# Elements of Scheduling

collected and edited by Jan Karel Lenstra and David B. Shmoys

This website presents the fragments of a book on machine scheduling. Work on the book started in 1977 but was never completed. The existing material is now made available for teaching purposes.

## PRELIMINARIES

## THE SINGLE MACHINE

## PARALLEL MACHINES

## MULTI-OPERATION MODELS

## MORE SCHEDULING

## BIBLIOGRAPHY

**SURVEY PAPERS**

Graham, Lawler, Lenstra, and Rinnooy Kan [1979]
Lawler, Lenstra, Rinnooy Kan, and Shmoys [1993]
Lenstra [1998]

---

© Centrum Wiskunde & Informatica — Jan Karel Lenstra

Back to top

# About

**Elements of Scheduling**

collected and edited by Jan Karel Lenstra and David B. Shmoys

This website presents the fragments of a book on machine scheduling. Work on the book started in 1977 but was never completed. The existing material is now made available for teaching purposes.

In the winter of 1976, Alexander Rinnooy Kan and Jan Karel Lenstra defended their PhD theses at the University of Amsterdam. Gene Lawler was on their committees. It was a natural idea to turn the theses into a textbook on scheduling. They set out to compile a survey with Ron Graham (1979), but progress on the book was hampered by the many research opportunities offered by the field. After David Shmoys joined the team in the mid 1980's, several chapters were drafted, and the survey was rewritten (1993). Gene passed away in 1994. Colleagues were asked to contribute chapters or to complete existing drafts. However, by the turn of the century the project was losing its momentum, and finite convergence to completion fell beyond our reach.

Over the years, several chapters have been used in the classroom. We continue to receive requests from colleagues who look for a text on the elements of scheduling at an advanced undergraduate or early graduate level. What exists is now available on this website. We have made a marginal effort in patching it up at some places but essentially put it up as it was written long ago. We did make an attempt to include most of the citations in the bibliography.

We owe many thanks and apologies to our colleagues who contributed chapters that waited years to be published. We hope that the material, in spite of all the gaps and omissions that the reader will encounter, will serve a useful purpose.

The help of Michael Guravage (Centrum Wiskunde & Informatica) and Shijiin Rajakrishnan (Cornell University) in formatting the chapters and realizing the website is gratefully acknowledged.

Below we review the status of each of the chapters and provide pointers to the pdf files.

## PRELIMINARIES

### 1. Determinictic machine scheduling problems
*Eugene L. Lawler, Jan Karel Lenstra, Alexander H.G. Rinnooy Kan, David B. Shmoys*
Chapter 1 defines the class of scheduling problems under consideration and introduces the three-field classification. Notes and references are up to date until about 1990.

### 2. Tools from algorithms and complexity theory
*David P. Williamson*
Chapter 2 reviews the tools and concepts from combinatorial optimization that are useful in designing scheduling algorithms: linear programming, network flows, dynamic programming, polynomial-time solvability, NP-hardness, and techniques for coping with NP-hardness. Notes and references are up to date until about 1997.

## THE SINGLE MACHINE

### 3. Minmax criteria
*Eugene L. Lawler, Jan Karel Lenstra, David B. Shmoys*
Chapter 3 considers two optimality criteria in a number of settings. For minimizing maximum cost, the focus is on polynomial-time optimization. For minimizing maximum lateness, the full spectrum of polynomial-time optimization, approximation and branch-and-bound is covered. Notes and references are up to date until 1992.

### 4. Weighted sum of completion times
*Eugene L. Lawler, Maurice Queyranne, Andreas S. Schulz, David B. Shmoys*
Chapter 4 centers around Smith's ratio rule. It follows two threads of work: the extension of the rule to other objective functions and various kinds of precedence constraints, and the design of approximation algorithms for NP-hard variants with general precedence constraints or release dates. Notes and references cover work up to 2006.

### 5. Weighted number of late jobs
*Eugene L. Lawler*
Chapter 5 deals with various problems with the objective of minimizing the unweighted or weighted number of late jobs, with dynamic programming as a leitmotiv. Notes and references are up to date until about 1990.

### 6. Total tardiness and beyond
A draft describing the branch-and-bound methods of the 1970's was made obsolete by later developments and never rewritten.

### 7. Nonmonotonic and multiple criteria
This is another blank spot.

## PARALLEL MACHINES

### 8. Minsum criteria
*Eugene L. Lawler*
Chapter 8 is a fragmented text, focusing on polynomial-time and pseudopolynomial-time optimization and providing pointers to results on NP-hardness, approximation and branch-and bound. Notes and references are up to date until about 1990.

### 9. Minmax criteria, no preemption
*David B. Shmoys, Jan Karel Lenstra*
Chapter 9 is about the design and performance analysis of approximation algorithms, including impossibility results and a brief excursion into probabilistic analysis. Notes and references are up to date until 1992.

### 10. Minmax criteria with preemption
*Eugene L. Lawler, Charles U. Martel*
Chapter 10 again focuses on polynomial-time optimization, using techniques ranging from the wrap-around rule to linear programming. Notes and references are up to date until 1990.

### 11. Precedence constraints
This is the most glaring omission.

## MULTI-OPERATION MODELS

### 12. Open shops
*Gerhard J. Woeginger*
Chapter 12 is based on a paper that was written for the 35th Symposium on Theoretical Aspects of Computer Science (2018). It deals with the minimization of makespan in nonpreemptive open shops. A pivotal role is played by a vector sum theorem of Steinitz (1913). Many open problems are listed. References are up to date until 2018.

### 13. Flow shops
*Jan Karel Lenstra, David B. Shmoys*
Chapter 13 discusses the design of approximation algorithms using Johnson's 2-machine flow shop algorithm and the vector sum theorem, and briefly reviews the empirical performance of heuristics. The literature on branch-and-bound and on flow shops with no wait in process or limited buffers is not covered. Notes and references cover work up to 1990.

### 14. Job shops
*Johann L. Hurink, Jan Karel Lenstra, David B. Shmoys*
Chapter 14 consists of three sections, presenting the disjunctive graph model, heuristics based on schedule construction and local search, and an approximation result based on the vector sum theorem. The many detailed complexity results for special cases and the advances in branch-and-bound are not covered. References are up to date until 2001.

## MORE SCHEDULING

### 15. Stochastic scheduling models
*Michael L. Pinedo*
Chapter 15 gives an overview of stochastic counterparts of the models considered in Chapters 3–14. Notes and references are missing.

16. **Scheduling in practice**
*Michael L. Pinedo*
Chapter 16 describes a diversity of practical applications of machine scheduling and discusses various aspects of scheduling systems. Figures, notes and references are missing.

## BIBLIOGRAPHY

The bibliography consists of the literature cited in the 1993 survey, supplemented with the references in Chapters 1–5, 8 10, 12 14. Citations are collected in bibliographic notes at the end of each chapter, with the exception of Chapters 12 and 14, where they occur in the main text.

The reader can download the surveys by Graham, Lawler, Lenstra, and Rinnooy Kan [1979] and Lawler, Lenstra, Rinnooy Kan, and Shmoys[1993], and a paper in memory of Gene Lawler by Lenstra [1998].

---

© Centrum Wiskunde & Informatica — Jan Karel Lenstra

Back to top

# Contents

# 1

# Deterministic Machine Scheduling Problems

Eugene L. Lawler
*University of California, Berkeley*

Jan Karel Lenstra
*Centrum Wiskunde & Informatica*

Alexander H.G. Rinnooy Kan
*University of Amsterdam*

David B. Shmoys
*Cornell University*

Scheduling theory is something of a jungle, encompassing a bewildering variety of problem types. In this book we shall be concerned with only a limited variety of the animals in this jungle, namely those which are *deterministic machine scheduling problems*. With the right techniques, some of these animals are easily domesticated. But others are quite intractable and submit to the taming techniques of *combinatorial optimization* with great difficulty. Before learning how to train these animals, we must be able to identify them and describe them. That is the purpose of this chapter.

## 1.1.   Machines, jobs, and schedules

The number of pans to be heated in the preparation of a gourmet dinner exceeds the number of burners available.  More ships are in a harbor than there are quays for unloading them.  Tasks assigned to a multiprogrammed computer system compete for processing on a timesharing basis. Planes requesting permission to use an airport runway are assigned places in a queue. Jobs in a job shop contend for processing by various machines.

Each of these situations suggests a problem in which limited resources must be allocated over time to a set of activities.  Adopting the terminology of the job shop, the resources can be described as *machines* and the activities as *jobs*.  Pans, ships, tasks and planes can be thought of as jobs, and burners, quays, central processing units and runways as machines.  Generally speaking, such problems involving machines and jobs are in the domain of *machine scheduling theory*.

Thus every instance of a scheduling problem involves a set of $m$ machines $M_1, M_2, ..., M_m$, comprising the *machine environment*, and a set of $n$ jobs $J_1, J_2, ..., J_n$, each requiring processing by one or more of the machines. As we shall explain more precisely later on, jobs have fixed *processing requirements*, which determine how long they must be processed. A processing requirement does not change in time and is unaffected by when the processing of a job is performed or by when the processing of other jobs occurs.

Also, as we shall describe, there may be constraints, in the form of *release dates* and *deadlines*, on the time period in which a given job is available for processing. There may be *precedence constraints* on the order in which jobs are processed. And one may allow *preemption*, or interruptions in the processing of jobs. But whatever the nature of these *job characteristics*, it is always understood that at any given point in time *no machine can process more than one job* and *no job can be processed by more than one machine*.

In this book we shall deal with only *deterministic* scheduling problems, in which all data are known precisely.  There will be no uncertainty about the processing requirements of the jobs or about the other job characteristics, which are all known in advance of scheduling.  Otherwise, we would be dealing with *stochastic* scheduling problems, which would involve us in *probability theory* and, most particularly, with *queueing theory* rather than *combinatorial optimization*.  In Chapter 15, we provide a small taste of stochastic scheduling.

The output of a scheduling procedure is a *schedule*, which specifies exactly what job, if any, each machine works on at each point in time.  A schedule can be represented by a *Gantt chart*, a device employed by managers and industrial engineers since the First World War.  In the chart shown in Figure 1.1, the horizontal axis indicates time and each band is identified with a machine.  The intervals in which a machine is assigned to no job are shaded; such periods are known as *idle time*.

Some schedules are better than others, and we seek to find a best schedule with respect to a specified *optimality criterion*.  In any schedule there is a well defined *completion time* $C_j$ for each job $J_j$, the time at which it is last processed.  A value $F(C_1, C_2, ..., C_n)$ is assigned to each possible schedule, and our goal is to find a schedule for which this value is minimized. We shall only be concerned with optimality criteria for which the function $F$ is *monotonic*. That is:

$$\text{if } C_j \leq C'_j (j = 1, ..., n), \text{ then } F(C_1, ..., C_n) \leq F(C'_1, ..., C'_n). \tag{1.1}$$

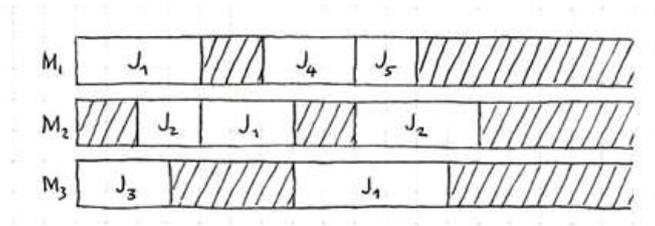

**Figure 1.1.** Gantt chart.

Now that we know something about what all the machine scheduling problems we shall deal with have in common, it is time to proceed with a systematic description and classification of problem types. As we shall elaborate in the remainder of this chapter, our description is based on three components: machine environment, job characteristics, and optimality criterion. Unless stated otherwise, all numerical data used in specifying a problem instance will be assumed to be integral.

## 1.2. Single machines and parallel machines

Every machine scheduling problem has a specified environment of $m$ machines $M_i$ ($i = 1, ..., m$) and a specified set of $n$ jobs $J_j$ ($j = 1, ..., n$), requiring processing by the machines. For each $J_j$ there is a specified *processing requirement* $p_j (j = 1, ..., n)$. (Later, in multi-operation models, we shall have a processing requirement for each of the separate operations comprising a job.) Recall that the processing requirement is independent of when the job is processed in a schedule. In particular, the processing requirement does not depend on the identity of the jobs that precede it and follow it. This means that we do not allow for the possibility of *sequence dependent setup and change-over times*. To do so would take us into the realm of the traveling salesman problem, which we now should like to avoid; see Exercise 1.1.

On occasion, we shall consider problems in which the processing requirements can take on only a restricted set of values. The simplest such case is that in which each $p_j = 1$. Clearly, imposing such a constraint can only make a problem easier to solve. Hence a *unit-time* scheduling problem is a *specialization* of the problem with *general* processing requirements, in that any solution prodedure for the latter can be used to solve the former. Ordinarily, we shall make the following assumption:

Processing requirements are arbitrary nonnegative integral values.    (1.2)

Condition (1.2) is an assumption that we shall make *by default* about any given problem, unless we explicitly state otherwise. Other default assumptions about job characteristics are:

All jobs are available for processing at time 0, but not before.    (1.3)

(All jobs have *release dates* equal to 0.)

It is feasible to process any job at any time after time 0.    (1.4)

(There are *no deadlines*.)

No constraints are imposed on the order in which jobs may be processed.    (1.5)

(Jobs are *independent*; there are *no precedence constraints*.)

Once a machine begins processing a job, it must process it to completion.    (1.6)

(Scheduling is *nonpreemptive*.)

The simplest machine environment is simply that of a *single machine* ($m = 1$), which is the subject of Chapters 3–7 of this book. Here the processing requirement $p_j$ simply indicates the amount of time for which the single machine must process job $J_j$.

Because of assumption (1.6), there is a well-defined sequence in which the jobs are processed in any given schedule, with each job in the sequence being completed before the next job is started. There are clearly an infinite number of schedules for the same sequence. But for *any given sequence* of the jobs, there is a *unique best schedule* in which the completion time of each and every job is as early as it is in any other schedule for the same sequence. We call this a *left-justified* schedule. For example, if the jobs are processed in the sequence $J_1, J_2, ..., J_n$, it is obtained by processing $J_1$ in the time interval $[0, p_1]$, $J_2$ in $[p_1, p_1 + p_2]$, ..., and $J_n$ in $[p_1 + ... + p_{n-1}, p_1 + ... + p_n]$. It follows from the monotonicity condition (1.1) that if any schedule for this sequence is optimal, this one is. Thus the single-machine scheduling problem reduces to a *sequencing* problem − that of determining which one of $n!$ possible sequences yields a schedule that is best with respect to the given optimality criterion.

The single-machine environment is a special case of *identical parallel machines*. In the case of this machine environment, a job may be performed on any one of the $m \geq 1$ available machines and each machine $M_i$ requires the same time $p_j$ to process job $J_j$. A computing center with $m$ identical computers provides an example of such a machine environment.

More generally, the $m$ parallel machines may have different *speeds* $s_i \geq 0$ ($i = 1, ..., m$), with $M_i$ requiring $p_j/s_i$ time units to perform $J_j$. These are known as *uniform parallel machines*. A computer center with $m$ computers, fully compatible and identical in all essential characteristics except for speed, provides an example of such a machine environment. Note that identical parallel machines are a special case of uniform parallel machines, in that it is understood that $s_i = 1$ for each $M_i$.

A still more general case of parallel machines is that of *unrelated parallel machines*, in which there are $mn$ specified parameters $s_{ij} \geq 0$ ($i = 1, ..., m$, $j = 1, ..., n$), with $M_i$ requiring $p_j/s_{ij}$ time units to perform $J_j$. A computer center with $m$ different computers, some good for certain tasks and less good for others, provides an example of such a machine environment. Identical, uniform and unrelated parallel machines are the subject of Chapters 8–11 of this book.

Note that the unit-time restriction does indeed provide meaningful specializations of identical and uniform parallel machine problems. However, in the case of unrelated parallel machines, the specialization is bogus: every unrelated parallel machine problem with arbitrary processing requirements $p_j$ and speeds $s_{ij}$ is equivalent to a unit-time problem with $p'_j = 1$ and $s'_{ij} = s_{ij}/p_j$. Unrelated parallel machine problems are in general much more difficult than identical and uniform machine problems and demand very different solution techniques. They also have a close relation with the multi-operation models we shall describe in Chapters 12–14, and some discussion of unrelated parallel machines is deferred to Chapters 8–11.

Again, because of condition (1.6), each schedule for a set of parallel machines provides a well-defined sequence in which the jobs are processed by any given machine. By carrying out the same kind of analysis we did for single machines, we see that parallel machine scheduling also reduces to a sequencing problem of sorts: we need consider only left-justified schedules, and there are as many of those as there are ways to assign the $n$ jobs to the $m$ machines and to sequence the jobs assigned to each of the machines. As we see in Exercise 1.2, there are exactly $(n+m-1)!/(m-1)!$ possibilities in the case of $m$ distinct machines.

**Exercises**

1.1. The legendary traveling salesman has to leave his home city and visit each of $n-1$ other cities exactly once, returning home at the end of his tour. He seeks to find a tour of minimum total length, given that the distance from city $j$ to city $k$ is a known positive value $c_{jk}$. Formulate the problem as a single-machine sequencing problem with $n+1$ jobs, where the time $p_{jk}$ required to perform a given job $J_k$ depends upon the identity of the job $J_j$ that precedes it. The objective should be to find a sequence of jobs that minimizes the total length of the schedule.

1.2. Show that there are $(n+m-1)!/(m-1)!$ ways to assign $n$ jobs to $m$ distinct machines and to sequence the jobs for each machine. (*Hint*: How many ways are there to permute a set of $n+m-1$ objects, $n$ of which are distinct and $m-1$ of which are identical?) How many possibilities are there if there are $m_i$ identical machines of each of $k$ different types (where $\sum_{i=1}^{k} m_i = m$)?

## 1.3. Release dates, deadlines, and precedence constraints

We have spoken of unit-time problems (all $p_j = 1$) as *specializations* that are obtained by departing from our default assumption (1.2). Now let us consider some *generalizations* that are obtained by departing from assumptions (1.3) and (1.4).

In (1.3) we assumed by default that all jobs are available for processing at time 0. Rather than restricting ourselves to such a *static* model, we might like to examine a *dynamic* model in which jobs are released for processing at various points in time. Each job $J_j$ can have a *release date* $r_j \geq 0$ specified for it; in order for a schedule to be feasible, it must satisfy the constraint that $S_j \geq r_j$, where $S_j$ is the *starting time* of $J_j$, the time at which it is first processed. We shall assume that all release dates are given to us at the time we are to prepare a schedule. In practice, of course, this may not be realistic since release dates may not be known with much certainty. Keeping this in mind, we shall occasionally investigate to what extent a solution procedure actually requires full information about release dates and processing requirements of jobs that are to become available in the future. A scheduling procedure that at any time $t$ does not require information about the $J_j$ with $r_j > t$ is said to be *on line* or *real time*.

Just as the starting time $S_j$ of $J_j$ may be constrained by a release date $r_j$, its completion time $C_j$ may be constrained by a *deadline* $\bar{d}_j$ with the requirement that $C_j \leq \bar{d}_j$, contrary to the default assumption (1.4). As we shall see, the presence of release dates and deadlines can only make problems more difficult to solve. However, any solution procedure that is effective for a scheduling problem with release dates and deadlines can be applied to the same problem when they are not present. So we are dealing with a strict generalization.

The observations we have made about the reduction of single and parallel machine scheduling problems remain valid under the imposition of release dates and deadlines. For any sequence there is a unique left-justified schedule, in which each job completion time is as early as in any other schedule consistent with the sequence. Thus, if $J_1, ..., J_n$ are to be processed by a certain machine in that order, each job should begin processing as early as possible, subject to the completion of its predecessor: $J_1$ should be processed in the time interval $[S_1, C_1]$, where $S_1 = r_1$ and $C_1 = S_1 + p_1$, and $J_j$ should be processed in the interval $[S_j, C_j]$, where $S_j = \max\{r_j, C_{j-1}\}$ and $C_j = S_j + p_j$, for $j = 2, ..., n$. The difficulty is that a sequence might be infeasible, because one or more of the completion times in this schedule may violate a deadline.

Additional feasibility constraints may arise if the jobs are not *independent* but related by *precedence constraints*, contrary to (1.5). For example, a sizeable project like building a house may be broken down into a number of jobs, such as laying the foundation, erecting the walls, building the roof, which must be done in a certain order. If $J_j$ must precede $J_k$, then we write $J_j -> J_k$. This is interpreted as meaning that $J_j$ must be completed before $J_k$ is started: $C_j \leq S_k$.

If precedence constraints exist, it is convenient to represent them by means of a *precedence digraph* $G$, with vertices $1, ..., n$ corresponding to jobs. The existence of an arc $(j, k)$ implies that $J_j \rightarrow J_k$. The number of arcs directed out of (into) a given vertex is said to be the *outdegree* (*indegree*) of the vertex. In order for the precedence constraints to be consistent, $G$ must be *acyclic*, i.e., contain no directed cycles.

A digraph can be represented by its *adjacency matrix* $A = (a_{jk})$, where $a_{jk} = 1$ if $(j, k)$ is an arc and $a_{jk} = 0$ otherwise. As is well known, a digraph is acyclic if and only if it is possible to number its vertices in such a way that each arc in the digraph extends from a lower numbered vertex to a higher numbered one. It follows that, for an appropriate numbering of the vertices, the adjacency matrix of a precedence digraph is *upper triangular*, with all 1's above the main diagonal.

A digraph is said to be *transitive* if the existence of arcs $(j, k)$ and $(k, l)$ implies the existence of an arc $(j, l)$. The *transitive closure* of a digraph $G$ is the digraph $G'$ obtained from $G$ by adding all arcs missing from $G$ that are implied by transitivity. It is easy to see that the transitive closure of any digraph is unique. On the other hand, $G'$ is said to be a *transitive reduction* if $G'$ can be obtained from $G$ by successively removing arcs that are implied by transitivity until no more such arcs remain. In general, a digraph may have several transitive reductions, resulting from various choices in the successive arcs that are removed. For acyclic digraphs this is not the case; as we will see in Exercise 1.3, the transitive reduction of an acyclic digraph is unique.

There are various restricted classes of precedence constraints that we shall want to consider (cf. Figure 1.2). We say that precedence constraints are in the form of *chains* if they can be represented by a (transitively reduced) digraph in which each vertex has indegree at most 1 and outdegree at most 1. Chain-type precedence constraints are a special case of *intree* constraints and of *outtree* constraints, which can be represented by (transitively reduced) digraphs in which each vertex has outdegree at most 1 and indegree at most 1, respectively.

Suppose a widget is to be assembled from its component parts. One job is to bolt parts $A$ and $B$ together to form subassembly $C$, another is to fit parts $D$, $E$ and $F$ together to form subassembly $G$. After this is done, there is the job of putting subassemblies $C$ and $G$ together to form a larger subassembly $H$, and so on. It is readily seen that the jobs in this assembly process are related by intree precedence constraints. Similarly, the jobs involved in disassembly are related by outtree constraints.

More generally, we shall simply say that precedence constraints are *tree* constraints if they are either intree or outtree constraints. That is, the class of tree digraphs is the union of the classes of intree and outtree digraphs, whereas the class of chain digraphs is their intersection. Later on, in Chapter 5, we shall define a still more general class of precedence constraints, called *series-parallel*.

Note that our observations about the reduction of single-machine scheduling problems to sequencing problems remain essentially valid, even if release dates, deadlines and precedence constraints are all imposed. It is just that certain sequences may be infeasible because they violate precedence constraints. And some of the sequences that are consistent with the precedence constraints may be infeasible because deadlines are violated, even when each successive job in the sequence is scheduled to start as early as possible. See Exercise 1.4.

**Exercises**

1.3. (a) Give an example of a digraph that has more than one transitive reduction. (Three vertices suffice.)

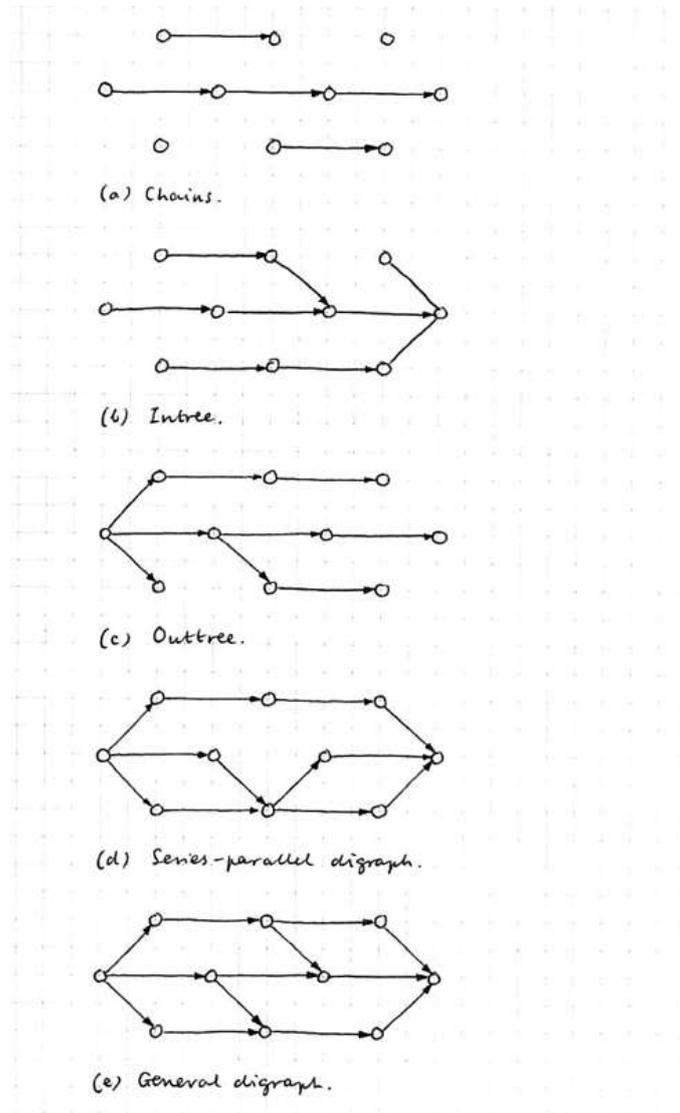

**Figure 1.2.**    Several types of precedence constraints

(b) Show that the transitive reduction of an acyclic digraph is unique.

1.4. We have seen in Exercise 1.2 that we may limit our search for the solution to a parallel machine scheduling problem to the $(n+m-1)!/(m-1)!$ possible ways in which it is possible to assign jobs to machines and to sequence the jobs assigned to each machine. Some of these possibilities may be infeasible because they violate precedence constraints.

(a) Show that precedence constraints are not necessarily satisfied if they are independently satisfied by each of the $m$ sequences of jobs that are to be performed by the machines.

(b) Show that, if a given set of $m$ sequences of jobs on machines satisfies the precedence constraints, left-justification yields a unique best schedule, in which each job completion time is as early as it is in any other schedule consistent with the sequences.

## 1.4. Preemption

Up to this point we have considered only *nonpreemptive scheduling*, in accord with default condition (1.6). Sometimes it is realistic to permit *preemption* or *job splitting*. If preemption is allowed, we will take that to imply that the processing of a job may be interrupted arbitrarily often and resumed at a later time on the same machine, or in the case of parallel machines at any time on a different machine. We assume there is no cost or loss of efficiency associated with preemption. All that matters is that the total processing requirement of a job be satisfied. Thus in the most general case of unrelated parallel machines, if $x_{ij}$ is the total amount of time that $M_i$ processes $J_j$, then it is necessary that

$$\sum_{i=1}^{m} s_{ij} x_{ij} = p_j \text{ for } j = 1, ..., n.$$

An obvious example of preemption is timesharing by jobs in a multiprogrammed computer system. In this case, the overhead involved in job swapping is nonneglible. But it is relatively small compared with the total amount of actual processing done, and for the purpose of scheduling it is not unreasonable to ignore this overhead.

Every feasible nonpreemptive schedule is, of course, also a feasible schedule when preemption is permitted. And there are scheduling problems for which there is no advantage to preemption, in the sense that there always exists a nonpreemptive schedule as good as any schedule involving preemption. For example, there is no advantage to preemption in a single-machine problem in which all release dates are 0; see Exercise 1.5. On the other hand, a problem may be such that no feasible nonpreemptive schedule exists, even though there are many feasible preemptive ones; see Exercise 1.6.

It follows that preemptive scheduling problems are not simply sequencing problems: optimization may require a search over a much larger number of possible schedules. There is no general rule for predicting whether the nonpreemptive version of a problem is either easier or harder than the preemptive version of the same problem. And certainly one cannot make a general statement that a nonpreemptive optimization procedure can be applied to solve a preemptive problem, or vice versa. Thus permitting preemption neither generalizes nor specializes the default assumption (1.6).

In general, it is a difficult problem to find, from among all optimal preemptive schedules, a schedule with the smallest possible number of preemptions. However, we shall be interested in keeping the number of preemptions down to a small number, in spite of our assumption that they have no cost. From a practical point of view this makes evident sense.

In the case that preemption is permitted, we will not consider the specialization to unit processing requirements. The reason for this is simply that preemptive unit-time models tend to be rather artificial and have not spawned any results of independent interest.

**Exercises**

1.5. Prove that there is no advantage to preemption in any single-machine problem in which all release dates are 0. Specifically, show that any feasible preemptive schedule with completion times $C_j$ can be modified to obtain a feasible nonpreemptive schedule with completion times $C'_j$ with $C'_j \leq C_j$ ($j = 1, ..., n$).

1.6. Give an example of a single-machine problem with release dates and deadlines in which all feasible schedules are preemptive.

## 1.5.   Optimality criteria

As we have noted, each schedule is assigned a value $F(C_1, ..., C_n)$, depending only on the job completion times, and we seek to find a feasible schedule for which this value is minimized. It should also be noted that we may restrict our attention to left-justified schedules: because of the monotonicity assumption (1.1), there will always be such a schedule among the optimal ones.

We shall deal with functions $F$ that result from cost functions $f_j$ assigned to the individual jobs $J_j$. We then distinguish between *minmax* problems, in which $F(C_1, ..., C_n)$ is given by

$$f_{\max} = \max_{j=1,...,n} f_j(C_j),$$

and *minsum* problems, in which $F(C_1, ..., C_n)$ is given by

$$\sum f_j = \sum_{j=1}^{n} f_j(C_j).$$

Both types of problems will be studied for general cost functions $f_j$, as well as for a number of special choices (cf. Figure 1.3).

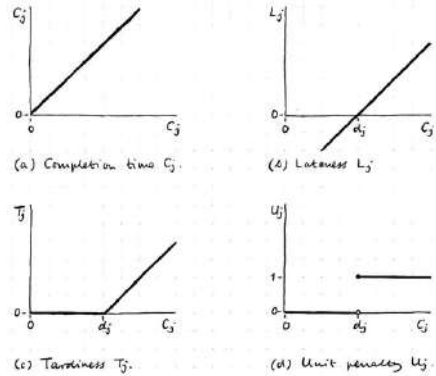

**Figure 1.3.** Four special choices of the cost function $f_j$.

The simplest choice for $f_j$ is $f_j(C_j) = C_j (j = 1, ..., n)$. The minmax problem then amounts to the minimization of *maximum completion time* $C_{\max} = \max_{j=1,...,n} C_j$. $C_{\max}$ is sometimes referred to as *makespan* or *schedule length*, since it represents the time required to process all jobs. This is a very natural and, not surprisingly, by far the most frequently encountered criterion.

The minsum problem in this case is to minimize *total completion time* $\sum C_j$, another natural choice. It is a special case of *total weighted completion time* $\sum w_j C_j$, where the *weight* $w_j$ of job $J_j$ reflects its relative importance. If $w_j = 1/n$ for all $j$, minimizing total weighted completion time reduces to minimizing *mean completion time*, which is equivalent to simply minimizing $\sum C_j$.

Now suppose each job $J_j$ is assigned a *due date* $d_j$, which serves as a yardstick by which the job completion cost is measured. A due date $d_j$ is quite distinct from a deadline $\bar{d}j$. If $C_j > \bar{d}_j$, the schedule is infeasible. If $C_j > d_j$, the schedule is not necessarily infeasible; it may just cost more. We say that $J_j$ is *early* if $C_j < d_j$ and *late* if $C_j > d_j$. A common cost function induced by due dates is *lateness*,

$$L_j = C_j - d_j.$$

For this choice, the minmax problem involves the minimization of *maximum lateness* $L_{\max} = \max_{j=1,...,n} L_j$, another popular criterion. Note that minimizing $C_{\max}$ is equivalent to minimizing $L_{\max}$ in the special case that all due dates are 0. So $L_{\max}$ is a proper generalization of $C_{\max}$.

We could generalize the $L_{\max}$ criterion by placing weights on the jobs. However, the minimization of maximum weighted lateness seems neither particularly attractive nor useful, so we shall ignore that possibility. Moreover, so far as minimizing total weighted lateness is concerned, observe that $\sum w_j L_j = \sum w_j C_j - \sum w_j d_j$. Since $\sum w_j d_j$ is a schedule-independent constant, $\sum w_j L_j$ and $\sum w_j C_j$ are equivalent, as are $\sum L_j$ and $\sum C_j$. It follows that we have no need to consider $\sum w_j L_j$ and $\sum L_j$ as separate criteria.

If $J_j$ is early, its lateness $L_j$ becomes negative. Sometimes a more realistic cost function to consider is *tardiness*,

$$T_j = \max\{0, C_j - d_j\}.$$

$T_j$ becomes nonzero only if $J_j$ is actually late. As we will see in Exercise 1.8, schedules that minimize $L_{\max}$ also minimize $T_{\max}$. However, the converse is not true: minimizing $T_{\max}$ does not necessarily minimize $L_{\max}$. We shall not be concerned with $T_{\max}$ as a criterion.

Minimization of *total tardiness* $\sum T_j$ and *total weighted tardiness* $\sum w_j T_j$ do provide interesting and challenging problems. Note that these criteria reduce to $\sum C_j$ and $\sum w_j C_j$ respectively if $d_j = 0$ for all $j$.

Finally, we may be interested in a different sort of measure of success in meeting due dates, in which we simply count the number of jobs that are late. In this case we assign a unit penalty for each late job:

$$U_j = \begin{cases} 0 & \text{if } C_j \le d_j, \\ 1 & \text{if } C_j > d_j. \end{cases}$$

We shall consider both the minimization of the *number of late jobs* $\sum U_j$ and of the *weighted number of late jobs* $\sum w_j U_j$.

### Exercises

1.7. Suppose that, in scheduling a single machine, we are interested in minimizing the total machine idle time prior to the completion of the last job. The minimization of total idle time is clearly equivalent to the minimization of $C_{\max}$. For which parallel machine environments is this observation also true?

1.8. (a) Prove that a schedule that minimizes $L_{\max}$ also minimizes $T_{\max}$.
(b) Construct a left-justified single-machine schedule showing that the converse is not true.

1.9. Prove that a schedule that minimizes $L_{\max}$ also minimizes $U_{\max}$.

1.10. Yet another cost function is *earliness* $E_j = d_j - C_j$. Do $E_{\max}$ and $\sum E_j$ satisfy the monotonicity assumption (1.1)?

1.11. Consider the cost function depicted in Figure 1.4 and show that it can be expressed as an appropriately weighted sum of completion time and tardiness.

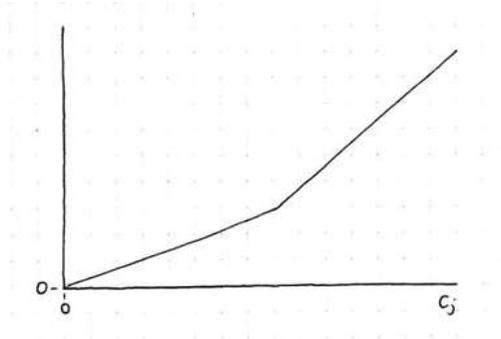

**Figure 1.4.** Cost function for Exercise 1.11.

## 1.6. Multi-operation models

Let us now consider *multi-operation* models, the subject of Chapters 12–14 of this book. In these models, each job $J_j$ consists of $\mu(j)$ *operations* $O_{1j}, O_{2j}, ..., O_{\mu(j)j}$, with the requirement that each operation $O_{ij}$ must be performed on a specified machine $M_{\iota(i,j)}$ $(1 \leq \iota(i,j) \leq m)$ for an amount of time equal to its processing requirement $p_{ij}$. No two of the operations of $J_j$ can be performed at the same time, and each operation must be completed before $J_j$ is considered to be finished. The completion time of a job is thus the maximum of the completion times of its operations.

In the case of the *open shop* model, each job $J_j$ has exactly $m$ operations, and each $O_{ij}$ must be performed on $M_i$. That is, $\mu(j) = m$ and $\iota(i,j) = i$ $(i = 1, ..., m, j = 1, ..., n)$. There is no restriction on the order in which the operations of a given job may be performed. As an example, consider a large automotive repair shop that is divided into several smaller shops: an engine shop, a radiator shop, a body shop, etc. A car is to have its engine tuned, its radiator repaired and its fender straightened. It is unimportant in which order these operations are performed, but no two of them can be carried out at the same time.

A *flow shop* is like an open shop except that the operations of each job $J_j$ must be carried out in the same fixed sequence: first $O_{1j}$, then $O_{2j}, ...,$ and finally $O_{mj}$. The completion time of $J_j$ is the completion time of $O_{mj}$. A small print shop, with a single typesetting machine, a single printing press and a single binding machine is an example of a flow shop.

The *job shop* is a generalization of the flow shop model, in the sense that the parameters $\mu(j)$ and $\iota(i,j)$ are unrestricted. Each job requires the services of some or all of the machines in its own fixed sequence, with repetitions allowed. For example, one job might require operations on $M_1$, $M_3$ in that order, and another might require operations on $M_3$, $M_2$, $M_1$, $M_2$, $M_3$ in that order. The completion time of $J_j$ is the completion time of $O_{\mu(j)j}$, its last operation. Machine shops provide the classical example of a job shop. One job might require the use of the lathe, then the drill press,

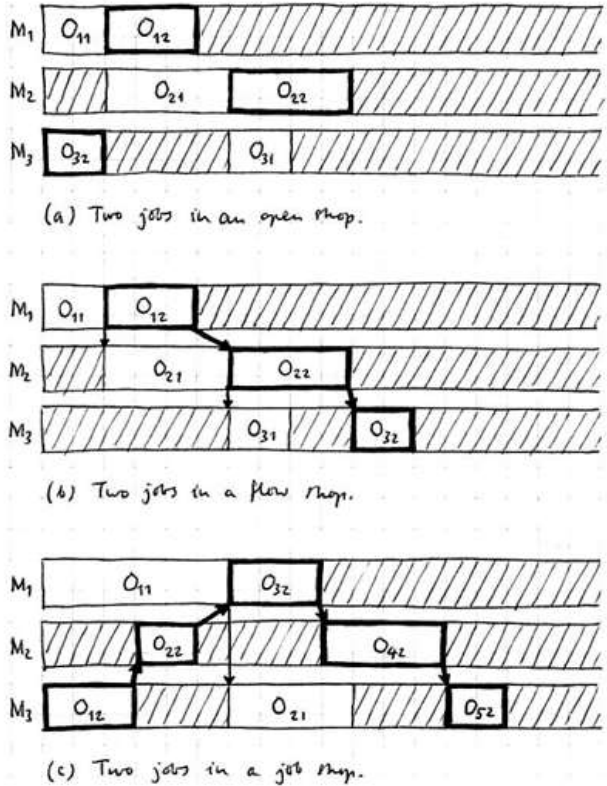

**Figure 1.5.**   Shop schedules.

then the grinder, but another job might require the same machines in a different order.

These three types of multi-operation models are illustrated in Figure 1.5. Their treatment in Chapters 12–14 will be restricted to models with general processing times and without release dates, deadlines and precedence constraints. Moreover, we shall only be concerned with the minimization of $C_{\max}$ (with a single excursion to $L_{\max}$ ). A few interesting results for problems outside this class are included as exercises.

It is easy to think of a common generalization of the parallel-machine model and the multi-operation model. Rather than assume the existence of a single machine that can perform a given operation, we could assume that a given operation can be performed on any one of a set of identical, uniform or unrelated parallel machines. However, we shall not pursue such a generalization.

## 1.7. Problem classification

We are now ready to explain our system of description and classification. Each problem type we shall consider will be described as $\alpha|\beta|\gamma$, where $\alpha$ denotes the machine environment, $\beta$ indicates job characteristics, and $\gamma$ specifies the optimality criterion.

First the $\alpha$-field. We shall use the following mnemonics to indicate machine environments:

| | |
|---|---|
| $P$ | identical parallel machines; |
| $Q$ | uniform parallel machines; |
| $R$ | unrelated parallel machines; |
| $O$ | open shop; |
| $F$ | flow shop; |
| $J$ | job shop. |

Each of these letters is followed by a constant, by an $m$, or by no symbol at all. In the first case, the constant indicates the number of machines, which is thereby specified as part of the problem type. For example, $F2|\beta|\gamma$ denotes a problem involving a two-machine flow shop. The second case indicates that the number of machines is an unspecified constant. Thus, $Fm|\beta|\gamma$ denotes a flow shop problem with some fixed number of machines. If there is no symbol following the letter, the number of machines is a variable, the value of which is part of the problem instance. Hence, $F|\beta|\gamma$ is a flow shop problem in which the number of machines will be specified together with the other numerical data. (We will discuss this distinction further in Chapter 3.) And what about single-machine problems? These are simply denoted by the numeral 1, i.e., $1|\beta|\gamma$. (Note that $P1$, $Q1$, $R1$, $O1$, $F1$ and $J1$ are all single-machine environments.)

Now for the $\beta$-field. This field indicates deviations from the default assumptions (1.2-1.6). So if nothing at all appears in this field, each of these assumptions applies. Thus $1||\gamma$ indicates a single-machine problem with all release dates equal to 0, no deadlines, no precedence constraints, and no preemption permitted. We shall employ the following mnemonics to indicate deviations from the default assumptions:

| | |
|---|---|
| *pmtn* | preemption is permitted; |
| *chain* | chain-type precedence constraints; |
| *intree* | intree-type precedence constraints; |
| *outtree* | outtree-type precedence constraints; |
| *tree* | tree-type precedence constraints; |
| *sepa* | series-parallel precedence constraints; |
| *prec* | general precedence constraints; |
| $r_j$ | release dates; |
| $\bar{d}_j$ | deadlines; |
| $p_j = 1$ | unit processing requirements. |

For example, $1|pmtn, prec, r_j|\gamma$ indicates a preemptive single-machine problem with general precedence constraints and release dates (but without deadlines and with general processing requirements). We will occasionally use the $\beta$-field to indicate other job characteristics, which, in certain situations, appear to merit a separate investigation. Some examples are:

$$\mu(j) \leq 2 \quad \text{each job has one or two operations;}$$
$$d_j = d \quad \text{all due dates are equal;}$$
$$p_j \in \{1, 2\} \quad \text{the processing requirements are equal to 1 or 2.}$$

We have introduced ten criteria that are of interest:

$$f_{max} \quad \text{general minmax criterion;}$$
$$C_{max} \quad \text{maximum completion time;}$$
$$L_{max} \quad \text{maximum lateness;}$$
$$\sum f_j \quad \text{general minsum criterion;}$$
$$\sum C_j \quad \text{total completion time;}$$
$$\sum w_j C_j \quad \text{total weighted completion time;}$$
$$\sum T_j \quad \text{total tardiness;}$$
$$\sum w_j T_j \quad \text{total weighted tardiness;}$$
$$\sum U_j \quad \text{number of late jobs;}$$
$$\sum w_j U_j \quad \text{weighted number of late jobs.}$$

Our notation and classification is summarized on the fold-out at the end of this book.

We conclude this chapter by repeating that not all problems generated by our classification scheme are of equal interest and that some combinations of choices are excluded *a priori*. As has been indicated above, in the case of unrelated parallel machines and in the case that preemption is permitted, we will not consider the specialization to unit processing requirements, and other restrictions apply to the investigation of multi-operation models. We also exclude the combination of deadlines with $C_{max}$ or $L_{max}$: the resulting problems are in some sense equivalent to the $L_{max}$ problem without deadlines.

## Exercises

1.12. The first paragraph of Section 1.1 mentions five practical scheduling situations. Formulate each of these in terms of the problem classification. In each case, consider the relevance of the possible machine environments, job characteristics, and optimality criteria.

## Notes

1.1. *Machines, jobs, and schedules.* Conway, Maxwell, and Miller [1967] wrote the first book on scheduling theory. Their text is still remarkable for the way it combines deterministic scheduling with queueing and simulation. Other books on deterministic scheduling include the undergraduate texts by Baker [1974] and French [1982], the collection of advanced expository reviews edited by Coffman [1976], and the dissertations of Rinnooy Kan [1976] and Lenstra [1977]. The proceedings volume edited by Dempster, Lenstra, and Rinnooy Kan [1982] provides surveys of the broader area of deterministic and stochastic scheduling and emphasizes developments on the interface between scheduling and queueing theory.

The present book is to some extent an outgrowth of the comprehensive survey papers by Graham, Lawler, Lenstra, and Rinnooy Kan [1979], Lawler, Lenstra, and Rinnooy Kan [1982], and Lawler, Lenstra, Rinnooy Kan, and Shmoys [1993]. More tutorial surveys are given by Lawler and Lenstra [1982], who consider the influence of precedence constraints, and Lawler [1983], who concentrates on polynomial algorithms and open problems. We further mention the *NP*-completeness column on multiprocessor scheduling by Johnson [1983], the annotated bibliography of the scheduling literature covering the period 1981-1984 by Lenstra and Rinnooy Kan [1985], the discussions of new directions in scheduling by Lenstra and Rinnooy Kan [1984], Blazewicz [1987] and Blazewicz, Finke, Haupt, and Schmidt [1988], and the overviews of single-machine scheduling by Gupta and Kyparisis [1987] and of multiprocessor and flow shop scheduling by Kawaguchi and Kyan [1988]. Graves [1981] reviews the broader area of production scheduling, which includes machine scheduling as well as lot sizing, with particular attention for practical aspects.

Henry Laurence Gantt (1861-1919) was a mechanical and industrial engineer. A disciple of Frederick W. Taylor, 'the father of scientific management', he held slightly more enlightened views on the social impacts of his work. He developed his famous charts during the First World War at the Ordnance Bureau of the United States Army, in order to provide methods for a simple, fast and accurate comparison between a production plan and its realization. The origin of the Gantt chart is reviewed by Porter [1968]. Gantt discussed the underlying principles in his paper 'Efficiency and democracy' [Gantt, 1919A], which he presented in December 1918 at the Annual Meeting of The American Society of Mechanical Engineers; see also his monograph *Organizing for Work* (i.e., for production, not for profit) [Gantt, 1919B]. His life and work are described by Alford [1934] and Rathe [1961], and his charts by Clark [1922] in a book that has been translated into twelve languages. It is fair to say that the Gantt charts as we use them are a gross simplification of the originals, both in purpose and design.

1.2. *Single machines and parallel machines.* The traveling salesman problem is dealt with in more detail by Lawler, Lenstra, Rinnooy Kan, and Shmoys [1985].

Left-justified schedules are also referred to as 'semi-active' schedules in part of the scheduling literature.

1.3. *Release dates, deadlines, and precedence constraints.* The investigation of precedence constraints presupposes an elementary knowledge of the theory of directed graphs, for which we refer to standard books such as those by Wilson [1972] and Bondy and Murty [1976]. Transitive closures and transitive reductions are discussed by Aho, Hopcroft, and Ullman [1974].

1.5. *Optimality criteria.* Objective functions that are monotonic in the job comple-
tion times are sometimes said to be 'regular'.

1.7. *Problem classification.* The classification scheme for deterministic machine
scheduling problems was developed by Graham, Lawler, Lenstra, and Rinnooy Kan
[1979]. It is based on the classification scheme for scheduling and queueing prob-
lems introduced by Conway, Maxwell, and Miller [1967].

# Contents

**i**

# 2

# Tools from algorithms and complexity theory

David P. Williamson
*Cornell University*

Given the taxonomy of deterministic scheduling problems in Chapter 1, we could immediately begin the discussion of their solution, but this would be a bit like a naturalist heading out into the forest with just a field guide and no equipment. Instead, we spend a brief moment in this chapter familiarizing ourselves with many tools and concepts from the field of combinatorial optimization that will be useful in designing scheduling algorithms in subsequent chapters. The field of combinatorial optimization is now sufficiently broad that any given topic we cover has already had a book written about it. Thus we will not cover any topic in depth, but rather cover the basic material that will be needed for the rest of this book. We will give references to more comprehensive treatments in the notes at the end of the chapter.

In the first part of the chapter we consider some well-studied algorithmic techniques: namely, linear programming, network flow, and dynamic programming. These techniques are sufficiently general that they can be used to solve several scheduling problems, and can be used as subroutines within algorithms for other scheduling problems. Furthermore, the techniques are efficient in practice. In the second part of the chapter, we turn to the concept of the complexity of algorithms and problems. Just as the naturalist, possessing a taxonomy of plants, would like indications of whether a plant is poisonous or not, the theory of NP-completeness helps determine whether a given scheduling problem is likely to have an efficient algorithm to solve it. We end the chapter by discussing a few techniques that can be brought to bear on the more poisonous varieties of scheduling problems.

**1**

### 2.1. Algorithmic techniques

**Linear programming.** One of the most useful tools from combinatorial optimization is *linear programming*. In linear programming, we find a non-negative, rational vector $x$ that minimizes a given linear objective function in $x$ subject to linear constraints on $x$. More formally, given an $n$-vector $c \in \mathbb{Q}^n$, an $m$-vector $b \in \mathbb{Q}^m$, and an $m \times n$ matrix $A = (a_{ij}) \in \mathbb{Q}^{m \times n}$, an optimal solution to the linear programming problem

$$\text{Min} \quad \sum_{j=1}^{n} c_j x_j$$

subject to:

$$(P) \qquad \sum_{j=1}^{n} a_{ij} x_j \geq b_i \quad \text{for } i = 1, \ldots, m \qquad (2.1)$$

$$x_j \geq 0 \qquad \text{for } j = 1, \ldots, n \qquad (2.2)$$

is an $n$-vector $x$ that minimizes the linear *objective function* $\sum_{j=1}^{n} c_j x_j$ subject to the *constraints* (2.1) and (2.2). The vector $x$ is called the *variable*. Any $x$ which satisfies the constraints is said to be *feasible*, and if such an $x$ exists, the linear program is said to be *feasible*. If there does not exist any feasible $x$, the linear program is called *infeasible*. The term "linear program" is frequently abbreviated to *LP*. Sometimes LPs are expressed more compactly in matrix/vector notation as follows:

$$\text{Min } c^T x$$

subject to:

$$Ax \geq b$$

$$x \geq 0,$$

where $c^T$ denotes the transpose of $c$. There are very efficient, practical algorithms to solve linear programs; LPs with tens of thousands of variables and constraints are solved routinely.

One could imagine variations and extensions of the linear program above: for example, maximizing the objective function rather than minimizing it, having equations in addition to inequalities, and allowing variables $x_j$ to take on negative values. However, the linear program $(P)$ above is sufficiently general that it can capture all these variations, and so is said to be in *canonical form*. To see this, observe that maximizing $\sum_{j=1}^{n} c_j x_j$ is equivalent to minimizing $-\sum_{j=1}^{n} c_j x_j$, and that an equation $\sum_{j=1}^{n} a_{ij} x_j = b_i$ can be expressed as a pair of inequalities $\sum_{j=1}^{n} a_{ij} x_j \geq b_i$ and $-\sum_{j=1}^{n} a_{ij} x_j \geq -b_i$. Finally, a variable $x_j$ which is allowed to be negative can be expressed in terms of two non-negative variables $x_j^+$ and $x_j^-$ by substituting $x_j^+ - x_j^-$ for $x_j$ in the objective function and the constraints.

Another variation of linear programming, called *integer linear programming* or *integer programming*, allows constraints requiring a variable $x_j$ to be an integer. For instance, we can require that $x_j \in \mathbb{N}$, or that $x_j$ be in a bounded range of integers, such as $x_j \in \{0, 1\}$. Unlike linear programming, there is currently no efficient, practical algorithm to solve general integer programs; in fact, many quite small integer programs are very difficult to solve. In Section 2.4, we will see evidence that it is unlikely that such an algorithm can exist. Nevertheless, integer programming remains a useful tool because it is a compact way to model problems in combinatorial optimization, and because there are several important special cases that do have efficient algorithms.

To illustrate the usefulness of linear programming in solving scheduling problems, consider the problem $R|pmtn|C_{\max}$. We will show that the problem of deciding how to allocate portions of each job to each machine can be formulated as a linear program, although we defer until Chapter 10 the problem of constructing a feasible schedule from the allocations. Let the variable $C$ denote the makespan of the schedule, which we wish to minimize, and let the variable $x_{ij}$ denote the fraction of the $j$th job to be allocated to the $i$th machine. Thus for any job $j$, $\sum_{i=1}^{m} x_{ij} = 1$. The total amount of processing to be performed by machine $i$ is then $\sum_{j=1}^{n} p_{ij} x_{ij}$, so that $C \geq \sum_{j=1}^{n} p_{ij} x_{ij}$. Finally, no job can be processed for more than $C$ units of time, so that $C \geq \sum_{i=1}^{m} p_{ij} x_{ij}$. In Chapter 10, we will see that any feasible values of $x$ and $C$ can be converted into a schedule of makespan $C$. Thus we can formulate the problem as the following linear program,

$$\text{Min} \quad C$$

subject to:

$(R)$

$$\sum_{i=1}^{m} x_{ij} = 1 \qquad j = 1, \ldots, n$$

$$C - \sum_{j=1}^{n} p_{ij} x_{ij} \geq 0 \qquad i = 1, \ldots, m$$

$$C - \sum_{i=1}^{m} p_{ij} x_{ij} \geq 0 \qquad j = 1, \ldots, n$$

$$x_{ij} \geq 0 \qquad i = 1, \ldots, m; \quad j = 1, \ldots, n,$$

and the optimal solution to the linear program, $C$, is the minimum possible makespan.

Observe that if we add the constraints $x_{ij} \in \{0, 1\}$ for all $i, j$ to the LP $(R)$, then this integer program *models* the problem $R||C_{\max}$, in the sense that the optimal solution to the integer program has the same value as the optimal solution to the problem $R||C_{\max}$, and the solution $x_{ij}$ indicates which machines should process which jobs. However, adding these constraints makes the problem much harder to solve than the linear program.

Linear programming has a very interesting and useful concept of *duality*. To explain it, we begin with a small example. Consider the following linear program in

canonical form:

Min     $6x_1 + 4x_2 + 2x_3$

subject to:

$$4x_1 + 2x_2 + x_3 \geq 5$$
$$x_1 + x_2 \geq 3$$
$$x_2 + x_3 \geq 4$$
$$x_i \geq 0 \qquad \text{for } i = 1, 2, 3.$$

Observe that because all variables $x_j$ are non-negative, it must be the case that the objective function $6x_1 + 4x_2 + 2x_3 \geq 4x_1 + 2x_2 + x_3$. Furthermore, $4x_1 + 2x_2 + x_3 \geq 5$ by the first constraint. Thus we know that the value of the objective function of an optimal solution to this linear program (called the *optimal value* of the linear program) is at least 5. We can get an improved lower bound by considering combinations of the constraints. It is also the case that $6x_1 + 4x_2 + 2x_3 \geq (4x_1 + 2x_2 + x_3) + 2 \cdot (x_1 + x_2) \geq 5 + 2 \cdot 3 = 11$, which is the first constraint summed together with twice the second constraint. Even better, $6x_1 + 4x_2 + 2x_3 \geq (4x_1 + 2x_2 + x_3) + (x_1 + x_2) + (x_2 + x_3) \geq 5 + 3 + 4 = 12$, by summing all three constraints together. Thus the optimal value of the LP is at least 12.

In fact, we can set up a linear program to determine the best lower bound obtainable by various combinations of constraints. Suppose we take $y_1$ times the first constraint, $y_2$ times the second, and $y_3$ times the third, where the $y_i$ are non-negative. Then the lower bound achieved is $5y_1 + 3y_2 + 4y_3$. We need to ensure that

$$6x_1 + 4x_2 + 2x_3 \geq y_1(4x_1 + 2x_2 + x_3) + y_2(x_1 + x_2) + y_3(x_2 + x_3),$$

which we can do by ensuring that no more than 6 copies of $x_1$, 4 copies of $x_2$, and 2 copies of $x_3$ appear in the sum; that is, $4y_1 + y_2 \leq 6$, $2y_1 + y_2 + y_3 \leq 4$, and $y_1 + y_3 \leq 2$. We want to maximize the lower bound achieved subject to these constraints, which gives the linear program

Max     $5y_1 + 3y_2 + 4y_3$

subject to:

$$4y_1 + y_2 \leq 6$$
$$2y_1 + y_2 + y_3 \leq 4$$
$$y_1 + y_3 \leq 2$$
$$y_i \geq 0 \qquad i = 1, 2, 3$$

This maximization linear program is called the *dual* of the previous minimization linear program, which is referred to as the *primal*. It is not hard to see that any feasible solution to the dual gives an objective function value that is a lower bound on the optimal value of the primal.

We can create a dual for any linear program; the dual of the canonical form LP $(P)$ above is

$(D)$

$$\text{Max} \quad \sum_{i=1}^{m} b_i y_i$$

subject to:

$$\sum_{i=1}^{m} a_{ij} y_i \leq c_j \qquad \text{for } j = 1, \ldots, n$$

$$y_i \geq 0 \qquad \text{for } i = 1, \ldots, m.$$

As in our small example, we introduce a variable $y_i$ for each linear constraint in the primal, and try to maximize the lower bound achieved by summing $y_i$ times the $i$th constraint, subject to the constraint that the variable $x_j$ not appear more than $c_j$ times in the sum. In matrix/vector notation this is

$$\text{Max} \quad y^T b$$

subject to:

$$y^T A \leq c$$

$$y \geq 0.$$

We now formalize our argument above that the value of the dual of the canonical form LP is a lower bound on the value of the primal. This fact is called *weak duality*.

**Theorem 2.1** [ Weak duality ]. *If $x$ is a feasible solution to the LP $(P)$, and $y$ a feasible solution to the LP $(D)$, then $\sum_{j=1}^{n} c_j x_j \geq \sum_{i=1}^{m} b_i y_i$.*

*Proof.*

$$\sum_{j=1}^{n} c_j x_j \quad \geq \quad \sum_{j=1}^{n} \left( \sum_{i=1}^{m} a_{ij} y_i \right) x_j \tag{2.3}$$

$$= \quad \sum_{i=1}^{m} \left( \sum_{j=1}^{n} a_{ij} x_j \right) y_i$$

$$\geq \quad \sum_{i=1}^{m} b_i y_i, \tag{2.4}$$

where the first inequality follows by the feasibility of $y$, the next equality by an interchange of summations, and the last inequality by the feasibility of $x$. $\square$

A very surprising, interesting, and useful fact is that when both primal and dual LPs are feasible, their values are exactly the same! This is sometimes called *strong duality*.

**Theorem 2.2** [ Strong duality ]. *If the LPs $(P)$ and $(D)$ are feasible, then for any optimal solution $x^*$ to $(P)$ and any optimal solution $y^*$ to $(D)$, $\sum_{j=1}^{n} c_j x_j^* = \sum_{i=1}^{m} b_i y_i^*$.*

As an example of this, for the small, three-variable LP and its dual we saw earlier, the optimal value is 14, achieved by setting $x_1^* = 0$, $x_2^* = 3$, and $x_3^* = 1$ in the primal, and $y_1^* = 0$, $y_2^* = 2$, and $y_3^* = 2$ in the dual. A proof of Theorem 2.2 is beyond the scope of this chapter, but one can be found in the textbooks on linear programming referenced in the notes at the end of the chapter.

An easy but useful corollary of strong duality is a set of implications called the *complementary slackness conditions*. Let $\bar{x}$ and $\bar{y}$ be feasible solutions to $(P)$ and $(D)$, respectively. We say that $\bar{x}$ and $\bar{y}$ obey the complementary slackness conditions if $\sum_{i=1}^{m} a_{ij}\bar{y}_i = c_j$ for each $j$ such that $\bar{x}_j > 0$ and if $\sum_{j=1}^{n} a_{ij}\bar{x}_j = b_i$ for each $i$ such that $\bar{y}_i > 0$. In other words, whenever $\bar{x}_j > 0$ the dual constraint that corresponds to the variable $x_j$ is met with equality, and whenever $\bar{y}_i > 0$ the primal constraint that corresponds to the variable $y_i$ is met with equality.

**Corollary 2.3** [ Complementary slackness ]. *Let $\bar{x}$ and $\bar{y}$ be feasible solutions to the LPs $(P)$ and $(D)$, respectively. Then $\bar{x}$ and $\bar{y}$ obey the complementary slackness conditions if and only if they are optimal solutions to their respective LPs.*

*Proof.* If $\bar{x}$ and $\bar{y}$ are optimal solutions, then by strong duality the two inequalities (2.3) and (2.4) must hold with equality, which implies that the complementary slackness conditions are obeyed. Similarly, if the complementary slackness conditions are obeyed, then (2.3) and (2.4) must hold with equality, and it must be the case that $\sum_{j=1}^{n} c_j\bar{x}_j = \sum_{i=1}^{m} b_i\bar{y}_i$. By weak duality, $\sum_{j=1}^{n} c_j x_j \geq \sum_{i=1}^{m} b_i y_i$ for any feasible $x$ and $y$ so therefore $\bar{x}$ and $\bar{y}$ must be optimal. □

So far we have only discussed the case in which the LPs $(P)$ and $(D)$ are feasible, but of course it is possible that one or both of them are infeasible. The following theorem tells us that if the primal is infeasible and the dual is feasible, the dual must be *unbounded*: that is, given a feasible $y$ with objective function value $z$, then for any $z' > z$ there exists a feasible $y'$ of value $z'$. Similarly, if the dual is infeasible and the primal is feasible, then the primal is unbounded: given feasible $x$ with objective function value $z$, then for any $z' < z$ there exists a feasible $x'$ with value $z'$. If an LP is not unbounded, we say it is *bounded*.

**Theorem 2.4.** *For primal and dual LPs $(P)$ and $(D)$, one of the following four statements must hold: (i) both $(P)$ and $(D)$ are feasible; (ii) $(P)$ is infeasible and $(D)$ is unbounded; (iii) $(P)$ is unbounded and $(D)$ is infeasible; or (iv) both $(P)$ and $(D)$ are infeasible.*

Sometimes in the design of scheduling algorithms it is helpful to take advantage of the fact if an LP is feasible, there exist feasible solutions of a particular form, called *basic* solutions. Furthermore, if an optimal solution exists, then there exists an optimal solution that is basic. Most linear programming algorithms will return a basic optimal solution. Suppose for a moment that in the canonical primal LP, there are more variables than constraints, that is, $n \geq m$. A basic solution to the LP is obtained by setting $n - m$ of the variables $x_j$ to zero, treating the inequalities as equalities, and solving the resulting $m \times m$ linear system (assuming the system is consistent and the given $m$ columns are linearly independent). In fact, the oldest and most frequently used linear programming algorithm, called the *simplex method*, works by moving from basic solution to basic solution, at each step swapping a variable set to zero for a variable in the linear system in a particular manner until an optimal solution is reached. When there are more constraints than variables ($m \geq n$), a basic solution is obtained by selecting $n$ of the constraints, treating them as equalities, and solving the resulting $n \times n$ linear system (assuming the system is consistent and the $n$ constraints are linearly independent). The solution obtained might not be feasible (since we ignored some constraints), but if an optimal solution exists, there will exist one of this form.

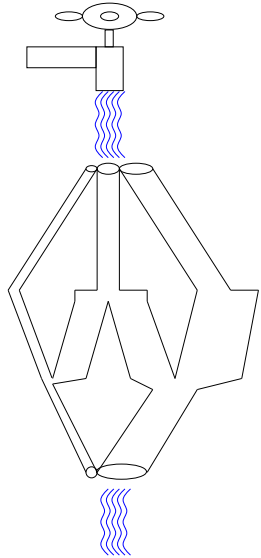

**Figure 2.1.** Example of a maximum flow problem.

**Network flow.** We now turn to another useful tool from combinatorial optimization, called *network flow*. An example of the most fundamental problem in this area is shown in Figure 2.1. We have a source of fluid and a destination for it joined by a network of pipes, each pipe with its own capacity. We would like to know at what rate we can send a flow of fluid from the source to the destination given the capacity of the pipes. The problem is usually abstracted as a directed graph $G = (V, E)$, with two distinguished nodes, a *source* node $s$ and a *sink* node $t$, such that no arc enters the source, and no arc leaves the sink. A *capacity* $u_{ij}$ is associated with each arc $(i, j)$ of the directed graph (see Figure 2.2). This *maximum flow problem* is used to model flow in pipes, traffic on streets, and the movement of goods via various modes of transportation, among other things.

It is not hard to see that the maximum flow problem can be modelled as a lin-ear program. Create a variable $x_{ij}$ for each arc $(i, j)$ to denote the flow on the arc. Then we wish to maximize the total flow out of the source, $\sum_{j \in V:(s,j) \in E} x_{sj}$, subject to two types of constraints. First, for any arc $(i, j)$ the flow on arc $(i, j)$ must not exceed its capacity; that is, $x_{ij} \leq u_{ij}$. Second, for any node $k \neq s, t$, the flow coming into node $k$ must be equal to the flow going out of node $k$; that is, $\sum_{i \in V:(i,k) \in E} x_{ik} - \sum_{j \in V:(k,j) \in E} x_{kj} = 0$. The first type of constraint is called a *capacity constraint*, and the second type is called a *flow conservation constraint*. The maximum flow problem can thus be modelled as the linear program

$$
\begin{aligned}
& \text{Max} && \sum_{j \in V:(s,j) \in E} x_{sj} \\
& \text{subject to:} \\
(MF) \quad && \sum_{i \in V:(i,k) \in E} x_{ik} - \sum_{j \in V:(k,j) \in E} x_{kj} = 0 && k \in V: k \neq s, t \\
& && 0 \leq x_{ij} \leq u_{ij} && (i, j) \in E,
\end{aligned}
$$

and solving the linear program gives a maximum flow.

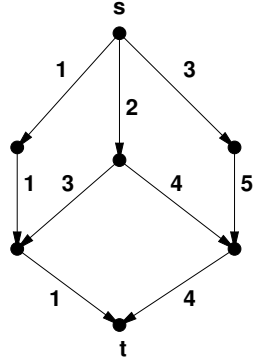

**Figure 2.2.** Abstraction of the maximum flow problem in Figure 2.1

Although the maximum flow problem and the other network flow problems considered here can all be modelled as linear programs, we consider them separately for two reasons: first, they form an extremely useful subclass of linear programs; and second, as we discuss at more length later on, there are special-purpose algorithms for network flow problems that are much more efficient than the general linear programming algorithms.

Unlike most linear programs, the maximum flow LP above has the property that if the capacities $u$ are integer, then the basic solutions $x$ are also integer. In particular, if an optimal solution exists, then there is an optimal solution such that the values of the $x_{ij}$ are integer. In other words, if the capacities $u$ are integer, then adding integrality constraints $x_{ij} \in \mathbb{Z}$ does not change the optimum value. This turns out to be true for all the network flow problems we discuss in this section, and, as we will see, this is a useful fact for designing scheduling algorithms.

Network flow problems also turn out to have interesting combinatorial dual problems, which are sometimes useful in their own right. For example, there is a natural combinatorial structure to the maximum flow problem that gives upper bounds on the amount of flow we can send from $s$ to $t$. Let $S$ be a set of vertices containing $s$ but not $t$. Let $u(S)$ denote the total capacity of all the arcs with their tails in $S$ and their heads in $\bar{S}$; that is, $u(S) = \sum_{i \in S, j \in \bar{S}} u_{ij}$. We call $S$ an *s-t cut* of the graph, and $u(S)$ the *value* of the *s-t* cut. It is not difficult to see that for any *s-t* cut $S$ the total amount of flow going from $s$ to $t$ cannot exceed $u(S)$. The strongest upper bound on the flow is then a *minimum s-t cut* – the *s-t* cut $S$ that minimizes $u(S)$.

Of course, the maximum flow problem also has a linear programming dual, which is as follows:

$$\text{Min} \quad \sum_{(i,j) \in E} u_{ij} z_{ij}$$

subject to:

$(MFD)$

$$
\begin{aligned}
z_{ij} + y_j - y_i \geq 0 & \qquad \text{for } (i,j) \in E; i \neq s,t; j \neq s,t \\
z_{sj} + y_j \geq 1 & \qquad \text{for } (s,j) \in E \\
z_{it} - y_i \geq 0 & \qquad \text{for } (i,t) \in E \\
z_{ij} \geq 0 & \qquad \text{for } (i,j) \in E.
\end{aligned}
$$

We will shortly prove that in fact this linear programming dual corresponds to the minimum *s-t* cut problem, so that the value of the linear programming dual and the combinatorial dual are precisely the same.

**Lemma 2.5.** *The value of the dual LP* $(MFD)$ *is exactly the value of a minimum s-t cut.*

By strong duality, this immediately implies the following celebrated *max flow/min cut theorem.*

**Theorem 2.6.** *The value of a maximum flow in a directed graph is the same as the value of its minimum s-t cut.*

This theorem can be extended to undirected graphs. The maximum flow in an undirected graph is the maximum flow in the directed graph obtained by replacing each undirected edge $(i, j)$ of capacity $u_{ij}$ with two directed arcs $(i, j)$ and $(j, i)$ of capacity $u_{ij}$. We remove arcs entering $s$ and leaving $t$. The capacity of an *s-t* cut $S$ in an undirected graph is simply the sum of the capacities of all edges with one endpoint in $S$ and the other not in $S$. Then the theorem holds as before.

We now prove Lemma 2.5.

*Proof of Lemma 2.5.* We prove equality by proving first that the value of LP is no greater than the value of a minimum *s-t* cut, and then the reverse. Given a minimum *s-t* cut $S$, we create a solution to the LP (*MFD*) of value $u(S)$ by setting $z_{ij} = 1$ if $i \in S, j \notin S$, and $z_{ij} = 0$ otherwise. We set $y_i = 1$ if $i \in S$ and $y_i = 0$ otherwise. It is easy to verify that $(y, z)$ is a feasible solution of value $u(S)$, so that the value of the LP must be no greater than that of a minimum *s-t* cut.

We now prove that there exists an *s-t* cut of value no greater than the value of an optimal solution $(y^*, z^*)$. For notational convenience, we add variables $y_s^* = 1$ and $y_t^* = 0$. We select a value $U$ uniformly at random from the interval $(0, 1]$, and use it to create an *s-t* cut $S$: if $y_i^* \geq U$, then $i \in S$, otherwise $i \notin S$. Obviously $s \in S$ and $t \notin S$. The probability that arc $(i, j)$ ends up in the cut is then $\max(0, \min(y_i^*, 1) - \max(y_j^*, 0))$, which is no greater than $z_{ij}^*$ by the feasibility of LP solution $y^*$. Thus the expected value of the *s-t* cut produced is at most $\sum_{(i,j) \in E} u_{ij} z_{ij}^*$. Therefore, there exists an *s-t* cut of value at most $\sum_{(i,j) \in E} u_{ij} z_{ij}^*$, and we are done. $\square$

We can use the maximum flow problem to determine whether or not a feasible solution exists for the problem $P|pmtn, r_j, \bar{d}_j|-$. In Chapter 10, we will see a rule by McNaughton that shows that a schedule of makespan $T$ can be constructed for the problem $P|pmtn|C_{\max}$ if and only if $T \geq \max\{\max_j p_j, \frac{1}{m}\sum_{j=1}^{n} p_j\}$, where $m$ is the number of machines. We now show how the maximum flow problem can be used to reduce the feasibility of $P|pmtn, r_j, \bar{d}_j|-$ to McNaughton's rule. First, sort the release dates $r_j$ and the deadlines $\bar{d}_j$ into increasing order, and obtain a list of times (without repetitions) $0 = t_0 < t_1 < t_2 < \cdots < t_q$, such that for every $r_j$ and $\bar{d}_j$ there exists some $k$ and $l$ such that $t_k = r_j$ and $t_l = \bar{d}_j$. Construct a network with a source node $s$, a sink node $t$, $n$ nodes $J_j$ (one for each job $j$), and $q$ nodes $T_k$ (one for each time interval $[t_{k-1}, t_k]$). For each job $j$, add an arc $(s, J_j)$ of capacity $p_j$ to the network, and for each node $T_k$, add an arc $(T_k, t)$ of capacity $m(t_k - t_{k-1})$. Finally, for each job $j$, let $k$ and $l$ be such that $t_k = r_j$ and $t_l = \bar{d}_j$, and for each $h$, $k + 1 \leq h \leq l$, add an arc $(J_j, T_h)$ of capacity $t_h - t_{h-1}$. We call this network $N$, and we can show the following theorem.

**Theorem 2.7.** *There is a feasible solution to an instance of $P|pmtn, r_j, \bar{d}_j|-$ if and only if the corresponding network $N$ has maximum flow value exactly $\sum_{j=1}^{n} p_j$.*

*Proof.* Given a schedule, we can construct a flow of the required value. On each arc $(s, J_j)$ put a flow of value of value $p_j$, while on each arc $(J_j, T_k)$ put a flow of value equal to the amount of time job $j$ was processed in time interval $[t_{k-1}, t_k]$. Clearly the capacity constraints for these arcs are obeyed, since the capacity of arc $(s, J_j)$ is $p_j$, and the capacity of arc $(J_j, T_k)$ is $t_k - t_{k-1}$. For each arc $(T_k, t)$ put a flow of value equal to the amount of processing performed by the $m$ machines in time interval $[t_{k-1}, t_k]$. This flow cannot have value more than $m(t_k - t_{k-1})$, and so the capacity constraint on arc $(T_k, t)$ is obeyed. The flow is of value $\sum_{j=1}^{n} p_j$, since that is the total amount of flow out of the source $s$. The flow conservation constraints are obeyed at each node: for each node $J_j$, $p_j$ units of flow enter from the source $s$, and $p_j$ units leave since $p_j$ units of time are scheduled for job $j$ overall. Similarly, for node $T_k$, the total flow entering and leaving $T_k$ is equal to the amount of processing performed during the corresponding time interval.

In a similar fashion, we can show that given a flow of value $\sum_{j=1}^{n} p_j$, we can construct a feasible schedule by using McNaughton's rule. Let $p_j^k$ be the amount of flow on arc $(J_j, T_k)$ (and 0 if the arc does not exist). Since the flow has value $\sum_{j=1}^{n} p_j$, there must be flow of value $p_j$ on each arc $(s, J_j)$, so that by flow conservation, $\sum_{k=1}^{q} p_j^k = p_j$ for all jobs $j$. By the capacity constraints on arcs $(J_j, T_k)$ we know that $p_j^k \leq t_k - t_{k-1}$ for each $j$ and $k$, and by capacity constraints on arcs $(T_k, t)$, we know that $\sum_{j=1}^{n} p_j^k \leq m(t_k - t_{k-1})$ for each $k$. Thus by McNaughton's rule, during the time interval $t_k - t_{k-1}$ we can construct a schedule in which $p_j^k$ units of job $j$ are feasibly scheduled, since $t_k - t_{k-1} \geq \max\{\max_j p_j^k, \frac{1}{m}\sum_{j=1}^{n} p_j^k\}$. Putting these schedules together gives an overall feasible schedule for the instance from time 0 to time $t_q$, since for each job $j$, $p_j$ units are scheduled, and by the construction of the network, they must be scheduled between time $r_j$ and $\bar{d}_j$. □

There are other network flow problems which are useful for solving scheduling problems. In Chapter 10 we will look at a variation on the maximum flow problem in which the capacities of the arcs leaving the source are a linearly increasing function of a parameter $\lambda$ (that is, $u_{sj} = a_{sj} + \lambda \cdot b_{sj}$, where $a_{sj}, b_{sj} \geq 0$), and the capacities of the arcs entering the sink are a linearly decreasing function of $\lambda$ (that is, $u_{it} = a_{it} - \lambda \cdot b_{it}$, where $a_{it}, b_{it} \geq 0$). In the *parametric maximum flow problem*, we would like to compute the value of the maximum flow for $k$ different non-negative values of $\lambda$. Of course, this can be done by computing a maximum flow $k$ times, but it turns out that there are more efficient algorithms. We discuss this further later in the chapter.

One of the most general network flow problems assigns costs to the arcs of the directed graph, so that sending a unit of flow through arc $(i, j)$ costs $c_{ij}$ units. The objective function is then to find the flow that minimizes the total cost, and so is called the *minimum-cost flow problem*. For each node $i$ in the directed graph there is a specified *supply* value $b_i$, which must be the difference between the flow leaving node $i$ and the flow entering $i$. Those nodes for which $b_i > 0$ are called *sources* and those for which $b_i < 0$ are called *sinks*. Note that in order for the problem to have a feasible solution, it must be the case that $\sum_{i \in V} b_i = 0$. Arcs $(i, j)$ can have lower bounds $l_{ij}$ in addition to capacities $u_{ij}$: that is, there must be at least $l_{ij}$ units of flow through arc $(i, j)$. The problem can then be formulated as the following linear program:

$$\text{Min} \quad \sum_{(i,j) \in E} c_{ij} x_{ij}$$

subject to:

$$\sum_{(k,j) \in E} x_{kj} - \sum_{(i,k) \in E} x_{ik} = b_k \qquad \forall k \in V$$

$$l_{ij} \leq x_{ij} \leq u_{ij} \qquad \forall (i,j) \in E.$$

As in the case of the maximum flow problem, whenever the supplies $b_i$, the capacities $u_{ij}$, and the lower bounds $l_{ij}$ are integer, the basic solutions $x$ of this linear program are also integer.

One particularly useful special case of the minimum-cost flow problem is when the network is a complete directed bipartite graph, with arcs from sources to sinks, each source with supply 1 and each sink with supply $-1$, and $l_{ij} = 0$ and $u_{ij} = \infty$ for each arc. The assumption that the supplies sum to zero implies that there must be the same number of sources and sinks. Since $b$, $l$, and $u$ are integer, a basic optimal solution $x$ must be integer, and thus for each source $i$, there is exactly one sink $j$ such that $x_{ij} = 1$ (for all other sinks $k \neq j$, $x_{ik} = 0$). That is, each source is assigned to exactly one sink, and hence this special case is called the *assignment problem*, and solutions to this problem are called *assignments*.

We can use the assignment problem to solve the scheduling problem $1|p_j = 1|\sum_j f_j$. For each of the $n$ jobs, we create a source and a sink node. The $j$th source represents the $j$th job, and the $k$th sink represents the $k$th position in the sequence of jobs. Since the job in the $k$th position will complete at time $k$, the cost of the $j$th job finishing in the $k$th position is $c_{jk} = f_j(k)$. Thus finding the minimum-cost assignment gives a sequence of the jobs that minimizes $\sum_{j=1}^{n} f_j(C_j)$.

Another variant of the minimum-cost flow problem is the *transportation problem*; although we will not prove it, this variant is not a special case, but is entirely equivalent to the minimum-cost flow problem. As in the assignment problem, the directed graph is a complete directed bipartite graph, with arcs from sources to sinks, and $l_{ij} = 0$ and $u_{ij} = \infty$ for each arc. In the transportation problem, however, the sources have arbitrary positive integer supplies, and the sinks have arbitrary negative integer supplies. A variation of this problem allows the amount of flow entering each sink $i$ to be at most $|b_i|$, rather than exactly $|b_i|$. In this case $\sum_{i \in V} b_i \leq 0$ in order for the problem to have a feasible solution.

We can use this variation of the transportation problem to extend our algorithm for the problem $1|p_j = 1|\sum_j f_j$ to an algorithm for $P|p_j = 1|\sum_j f_j$. As before, we create a source node of supply 1 for each of the $n$ jobs. If we have $m$ machines, then since the schedule will have no idle time, every job will complete by time $\lceil n/m \rceil$, and so we create a sink node for each of the $\lceil n/m \rceil$ possible times at which a job can complete. Since at most $m$ jobs can be scheduled to complete at a given time $k$, we set $b_k = -m$ for sink $k$, and enforce that at most $m$ units of flow can enter any sink. Since the cost of the $j$th job finishing at time $k$ is $f_j(k)$, we set $c_{jk} = f_j(k)$. A basic optimal solution $x$ to this transportation problem is integer, so for each $j$, $x_{jk} = 1$ for some sink $k$ (and 0 for all others). Thus we can construct a schedule of the same cost as the flow by scheduling job $j$ at time $k$ on some machine. Because $\sum_{j=1}^{n} x_{jk} \leq m$, the schedule uses at most $m$ machines at any given time $k$.

The area of network flows is very rich, and there are many other flow variants – such as generalized flow and multicommodity flow – whose details are not as pertinent to the topic of this book. The notes at the end of the chapter contain suggested further reading.

**Dynamic programming.** We now turn to *dynamic programming*, which is a technique for solving problems, rather than a class of problems, as in linear programming and network flows. In particular, it is a method for solving *multi-stage decision problems*, in which choices must be made in a sequence of stages. The underlying idea is that for such problems we can sometimes divide the decision in the $j$th stage into several subproblems. These subproblems should have the property that given the optimum solutions to the subproblems in the $(j-1)$st stage, we can easily compute the optimum solution to the subproblems in the $j$th stage. One of the subproblems of the $n$th stage is the overall problem we want to solve, and thus the optimum sequence of choices can be found.

Usually a dynamic programming algorithm for a problem is embodied in a multi-dimensional array $A(j,\cdots)$, with the additional coordinates indexing the subproblems for the $j$th stage. Boundary conditions specify the values of $A(0,\cdots)$, and given the values of $A(j-1,\cdots)$, the values of $A(j,\cdots)$ are easily computed. The desired value is one of the entries of $A(n,\cdots)$.

To illustrate this method, we will consider how it applies to the *knapsack problem*. In the knapsack problem, we are given $n$ items, each of which has a size $s_j \geq 0$ and an integer value $v_j \geq 0$. We are also given a knapsack of capacity $B$. We assume $s_j \leq B$ for all items $j$. The goal is to find the subset of items of the greatest total value that can be placed in the knapsack; that is, the sum of their sizes must not exceed the capacity of the knapsack. The problem can be viewed as a multi-stage decision problem, since we can consider the items in order, from 1 to $n$, and decide whether or not to include the $j$th item in the knapsack.

Let us see how we can apply the dynamic programming technique to solve this problem. The main difficulty is breaking the decision for the $j$th stage into useful subproblems. In this case we consider the subproblems $(j,s)$ of finding the most valuable set of items in $\{1,\ldots,j\}$ that use total size, or space, no more than $s$. We store the value of the most valuable such set in a two-dimensional array $A(j,s)$, where the first coordinate will range over the items 1 to $n$, and the second coordinate will range over the possible total sizes of the items in the knapsack (from 0 to $B$). Supposing that we can compute the values in the array $A(j,s)$, then the value of the optimum solution is the most valuable subset of all the items that fits in space at most $B$; that is, the optimum value is $A(n,B)$.

Now let us see how to set the boundary conditions and how to compute $A(j,\cdot)$ from $A(j-1,\cdot)$. Certainly $A(0,s) = 0$ for any $s$ between 0 and $B$; that is, by using no items, the most valuable set of items that uses space at most $s$ has value 0. Now suppose that we know the values of $A(j-1,s)$ for $0 \leq s \leq B$, and we want to compute $A(j,s)$. To achieve the most valuable subset of items from $\{1,\ldots,j\}$ that has size at most $s$, either the $j$th item is used or it is not used. If it is not used, the most valuable subset of items from $\{1,\ldots,j\}$ having space at most $s$ must be the same as the most valuable subset of items from $\{1,\ldots,j-1\}$ having space at most $s$; that is, $A(j,s) = A(j-1,s)$. If the $j$th item is used, then the value of the subset of items is $v_j$ plus the most valuable subset of items from $\{1,\ldots,j-1\}$ that occupies space at most $s - s_j$ if this amount of space is non-negative; that is, $A(j,s) = v_j + A(j-1,s-s_j)$ if $s - s_j \geq 0$. The optimum choice of whether to use item $j$ or not is whichever maximizes the overall value; that is, $A(j,s) = \min[A(j-1,s), v_j + A(j-1,s-s_j)]$

```
1        For s ← 0 to B
2            A(0,s) ← 0
3        For j ← 1 to n
4            For s ← 0 to B
5                if s_j ≤ s
6                    A(j,s) ← min(A(j−1,s), v_j + A(j−1, s−s_j))
7                else
8                    A(j,s) ← A(j−1,s).
```

**Figure 2.3.**  Dynamic programming algorithm for the knapsack problem.

(if $s − s_j ≥ 0$, otherwise $A(j,s) = A(j−1,s)$).

Given this discussion, the algorithm for computing the value of an optimal solution is straightforward, and we give it in Figure 2.3. If we would like to know the set of items that gives this value, we can compute it from the array $A(j,s)$ by backtracking through the decisions that were made. We leave this as an exercise for the reader.

We can use this algorithm to solve the scheduling problem $1|\bar{d}_j = d|\sum_j w_j U_j$, since this scheduling problem is just a knapsack problem in which the capacity of the knapsack is $B = d$, and each job corresponds to an item of size $s_j = p_j$ and value $v_j = w_j$. Minimizing the total weight of late jobs is equivalent to maximizing the total weight of jobs that complete before the deadline $d$, so that a maximum-weight set of jobs $S$ that complete by time $d$ corresponds to a maximum-weight set of items whose total size is no more than $d$. In Chapter 5 we will see that the dynamic programming algorithm for the knapsack problem can be extended to solve the more general problem $1||\sum_j w_j U_j$.

We can improve on the algorithm of Figure 2.3 by observing that for a given $j$, it is possible that $A(j,s)$ is the same for many consecutive values of $s$; that is, the most valuable subset of items of $\{1, \ldots, j\}$ that uses size at most $s$ is the same for sizes $s = s'$ up to $s''$. Then rather than storing an entry for every value of $s$, we can create a new array $A'$, where $A'(j)$ contains a list $(t_1, w_1), (t_2, w_2), \ldots, (t_k, w_k)$, with the understanding that $A(j,s) = w_i$ for $t_i ≤ s < t_{i+1}$ (where $t_{k+1} = B + 1$); in other words, for each pair $(t_i, w_i)$ there is a subset of items in $\{1, \ldots, j\}$ that has value $w_i$ and uses space at most $t_i$. Since it is the case that $t_1 < t_2 < \cdots < t_k$ and $w_1 < w_2 < \cdots < w_k$, the number of elements in the list is no more than $B + 1$ and no more than one plus the maximum possible value of the knapsack. Certainly the maximum possible value of the knapsack is no more than $V = \sum_{j=1}^n v_j$. So the length of the list is at most $\min(B, V) + 1$.

Given the list for $A'(j−1)$, it is easy to compute the list for $A'(j)$. For each pair $(t_i, w_i)$ in the list $A'(j−1)$, we create a new pair $(t_i + s_j, w_i + v_j)$ if $t_i + s_j ≤ B$, since if it is possible to have a knapsack of value $w_i$ using space at most $t_i$ using a subset of items from $\{1, \ldots, j−1\}$, one can have a knapsack of value $w_i + v_j$ using space at most $t_i + s_j$ using items from $\{1, \ldots, j\}$ by including item $j$ to the previous knapsack.

```
1        A'(0) ← {(0,0)}
2        for j ← 1 to n
3            for each pair (t, w) in A'(j−1)
4                if t + s_j ≤ B
5                    add new pair (t + s_j, w + v_j) to A'(j)
6            merge pairs from A'(j−1) into A'(j)
7            discard dominated pairs from A'(j)
```

**Figure 2.4.** Another dynamic programming algorithm for the knapsack problem.

We then merge the list of old pairs and new pairs to obtain a list $(t'_1, w'_1), \ldots, (t'_l, w'_l)$ with $t'_1 \leq t'_2 \leq \cdots t'_l$. We then check to see whether we can discard some pairs, since in the final list it must be the case that $t'_1 < t'_2 < \cdots < t'_l$ and $w'_1 < w'_2 < \cdots < w'_l$. Thus we discard any *dominated* pair $(t'_i, w'_i)$; that is, any pair $(t'_i, w'_i)$ such that there exists another pair $(t'_k, w'_k)$ of value at least as great that uses no more space ($w'_k \geq w'_i$ but $t'_k \leq t'_i$). Clearly the resulting list is correct: if $(t'_i, w'_i)$ is on the list, we can pack items from a subset of $\{1, \ldots, j\}$ of value $w'_i$ in space $t'_i$. If we can pack items from $\{1, \ldots, j\}$ of value $w$ in space $t$, then either the packing uses item $j$ or not; if not, the old list of $A'(j−1)$ implied that value $w$ could be packed in space $t$, and if so, value $w − v_j$ could be packed in space $t − s_j$ which was implied by the old list of $A'(j−1)$, and therefore one of the new pairs implies that value $w$ can be packed in space $t$. We summarize the new algorithm in Figure 2.4.

## 2.2. Analysis of algorithms

So far we have had little to say about the efficiency of the algorithmic techniques described above, other than to give assurances that the techniques are efficient in practice. In this section we would like to make those assurances somewhat more precise, to the extent possible.

The caveat "to the extent possible" is necessary. When we have given algorithms, they have been in English-like descriptions rather than specific computer programs, since the algorithm is independent of a particular implementation of it, just as the text of a play is independent of a particular staging. Yet of course the particular implementation (computer language, choice of data structures, etc.) will affect the efficiency of the algorithm. Even a particular implementation on a specific machine will be affected by the compiler used, and even a given compilation of a particular program targeted to a particular computer architecture will be affected by issues such as cache size, memory latency, disk speed, and so forth.

Thus although we could in principle state how many "instructions" – arithmetic operations, memory fetches and stores, comparisons, branches, etc. – need to be executed by the English-like statements of our algorithms on a particular input, this might not correspond precisely to the amount of time taken by a computer running some implementation of the algorithm on that input. Furthermore, the time taken by each type of instruction might differ. For these reasons, we discuss the efficiency of algorithms in terms of their *order of growth*, a somewhat cruder measure than precise instruction counts. The order of growth indicates how the running time of an algorithm varies as the size of its input varies (e.g. the number of jobs, the number of machines, the size of the jobs). The order of growth of an algorithm is expressed in terms of *big-oh notation*, which we define as follows.

**Definition 2.8** [ Big-oh notation ]. *Given two functions $f, g : \mathbb{N} \to \mathbb{N}$, we say that $f(n) = O(g(n))$ if there exist some positive constants $c, n_0$ such that $f(n) \leq c \cdot g(n)$ for all $n \geq n_0$.*

Thus if an algorithm for a scheduling problem with $n$ jobs and $m$ machines executes at most $f(n, m) = 8nm^2 + 5nm + 6n + 3m + 2$ instructions, we simplify this using big-oh notation by saying that it takes $O(nm^2)$ *time*, or that its *running time* is $O(nm^2)$. Sometimes we refer to the instruction counts as the *time complexity* of the algorithm.

When specifying the time complexity of an algorithm, we usually refer to its *worst-case* behavior. For instance, one might have an algorithm that executes $f(n)$ instructions, where

$$f(n) = \begin{cases} c_1 n & \text{if } n \text{ composite} \\ c_2 n^2 & \text{if } n \text{ prime.} \end{cases}$$

Then $f(n) = O(n^2)$, but it is not the case that $f(n) = O(n)$. Hence we say that the algorithm takes $O(n^2)$ time.

Let us now examine the time complexity of the algorithms in Figure 2.3 and 2.4. First, we consider the algorithm of Figure 2.3. Lines 1–2 initialize the array $A$; each time through the loop takes at most some constant $c_1$ number of instructions; hence these lines execute in $O(B)$ time. Lines 3–8 calculate the array. The instructions in lines 5–8 take at most some constant $c_2$ number of instructions, which are executed $nB$ times, so that lines 3–8 take $O(nB)$ time. The overall algorithm executes at most $c_1 B + c_2 nB$ instructions; thus the running time is $O(nB)$. Now consider the algorithm of Figure 2.4. Line 1 takes a constant $d_1$ number of instructions. Lines 2–7 calculate the array. Lines 4–5 take a constant $d_2$ number of instructions, and are executed $nL$ times, where $L$ is the maximum length of any list $A'(j)$. We argued earlier that $L \leq \min(B, V) + 1$, so that lines 2–5 take $O(n \min(B, V))$ time. It is possible to implement lines 6 and 7 in $d_3 L$ instructions, so that overall lines 2–7 take $O(n \min(B, V))$ time. Thus the overall running time of the algorithm in Figure 2.4 is $O(n \min(B, V))$ time.

Sometimes it is useful to give a lower bound on the order of growth of the running time of an algorithm. This can be done using *big-omega* notation.

**Definition 2.9** [ Big-omega notation ]. *Given two functions $f, g : \mathbb{N} \to \mathbb{N}$, we say that $f(n) = \Omega(g(n))$ if there exist some positive constants $c, n_0$ such that $f(n) \geq c \cdot g(n)$ for all $n \geq n_0$.*

The reader can verify that the knapsack algorithm in Figure 2.3 takes $\Omega(nB)$ time. Recall our hypothetical algorithm that takes $c_1 n$ instructions when $n$ is composite and $c_2 n^2$ when $n$ is prime. This algorithm takes $\Omega(n)$ time, but not $\Omega(n^2)$ time.

Sometimes it is useful to capture the fact that, as in the case of the knapsack algorithm of Figure 2.3, both $f(n) = O(g(n))$ and $f(n) = \Omega(g(n))$; we do this with *big-theta* notation.

**Definition 2.10** [ Big-theta notation ]. *Given two functions $f, g : \mathbb{N} \to \mathbb{N}$, we say that $f(n) = \Theta(g(n))$ if $f(n) = O(g(n))$ and $f(n) = \Omega(g(n))$.*

Since the knapsack algorithm in Figure 2.3 takes $O(nB)$ time and $\Omega(nB)$ time, it takes $\Theta(nB)$ time.

Big-oh notation is also used for the amount of memory (sometimes called *space*) that an algorithm uses. For instance, the space required by the knapsack algorithm given in Figure 2.3 is dominated by the space needed to store the array $A(j, s)$. Since there are $n(B + 1)$ items in this array, we say that the algorithm requires $O(nB)$ space. This amount is referred to as the *space complexity* of the algorithm.

**The complexity of the algorithmic techniques.** We now turn to a discussion of the time complexity of the algorithmic techniques discussed in the previous section. The most popular algorithm for solving linear programs is called the *simplex method*, and it has a number of variants. Recall that in our discussion of basic solutions to linear programs, we observed that the simplex method moves from basic solution to basic solution by swapping a variable set to 0 for a variable in the linear system. This step is called *pivoting*, and the means by which simplex decides which two variables to swap is called the *pivot rule*. Many textbook pivot rules can be shown to require $\Omega(2^n)$ pivots in the worst case, where $n$ is the maximum of the number of rows and columns, and thus the simplex method can take a large amount of time in the worst case. However, these examples are pathological; in practice, the simplex method is very fast, and problems with tens of thousands of variables and constraints are solved routinely.

*Interior-point methods* constitute another class of algorithms for solving linear programs. Interior-point algorithms have much better worst-case time complexity than the simplex method: the fastest algorithm currently known needs to solve $O(\sqrt{n}L)$ linear systems, each of which takes $O(n^3)$ time in the worst case, where $L$ is the "size" of the linear program (that is, the number of bits needed to encode in binary the matrix $A$, the right-hand side $b$, and the objective function $c$). More practical variants of interior-point methods have larger time complexity ($O(nL)$ iterations instead of $O(\sqrt{n}L)$), but in practice the number of iterations required for these variants seems to be essentially constant, and indeed these algorithms sometimes outperform the simplex method.

Many algorithms have been devised for solving the maximum flow problem. As of the writing of this chapter, the algorithm for the maximum flow problem with the lowest worst-case time complexity in most cases has a running time of $O(\min(n^{2/3}, m^{1/2})m \log(\frac{n^2}{m}) \log U)$, where $n$ is the number of vertices in the graph, $m$ is the number of edges, and $U$ is size of the largest capacity. However, a class of algorithms called *push-relabel* algorithms (also called *preflow-push* algorithms) have been shown to work very well in practice, and appear to be faster than algorithms whose theoretical worst-case complexity is lower. The lowest known worst-case complexity of a push-relabel algorithm is $O(nm \log(\frac{n^2}{m}))$, and hence is theoretically faster than the previously mentioned algorithm only when $U$ is very large. The notes at the end of this chapter contain pointers to descriptions of these algorithms and studies of them. Interestingly, it has been shown that the parametric maximum flow problem can be solved in the same worst-case running time as a push-relabel algorithm.

As is the case with maximum flow algorithms, there are many minimum-cost flow algorithms. One type that has had successful implementations uses the simplex method with a special pivot rule to solve the associated linear program; this type is called a *network simplex* algorithm. Another type that works as well or better in practice uses the push-relabel algorithm as a subroutine. Specialized algorithms have been developed for the assignment and transportation problems, and again, many different types of algorithms have been proposed. See the notes at the end of the chapter for references.

### 2.3.  Complexity theory and a notion of efficiency

Given the current set of algorithms for solving linear programming and network flow problems as described above, it is natural to ask whether better algorithms might exist. For example, is it inherent in the nature of the knapsack problem that all algorithms to solve it must take $\Omega(n \min(B, V))$ time? This asks whether the *complexity* of the knapsack problem is such that no faster running time is possible. The study of such questions is called *complexity theory*.

There is an immediate difficulty with posing such questions, since the time bound given will depend on our model of a computer. For instance, one can imagine a parallel computer executing the knapsack algorithm in Figure 2.3 with $O(n)$ processors. Such a computer could execute the loop in lines 1–2 in a constant number of instructions. Perhaps a parallel machine with a reasonable number of processors could solve the knapsack problem in $O(n)$ time.

There are two possible ways to make the question well-posed. One is to fix a model of a computer. We have so far implicitly considered a single processor machine in which each arithmetic operation, comparison, and memory access takes a single instruction, regardless of the size of the numbers involved. This model is known as the *random access machine*, or the *RAM* model of computation. We can consider the inherent time complexity of problems with respect to the RAM.

Another way to make the question well-posed is to broaden it to whether a given problem has an efficient algorithm or not, and define efficiency in such a way that it is invariant under any reasonable model of a computer. This line of thinking has led to an emphasis on the class of *polynomial-time* algorithms.

**Definition 2.11** [ Polynomial-time algorithm ]. *An algorithm for a problem is said to run in polynomial time, or said to be a polynomial-time algorithm, with respect* *to a particular model of computer (such as a RAM) if the number of instructions executed by the algorithm can be bounded by a polynomial in the size of the input.*

More formally, let $x$ denote an *instance* of a given problem; for example, an instance of the knapsack problem is the number $n$ of items, the numbers $s_j$ and $v_j$ giving the sizes and values of the items, and the number $B$ giving the size of the knapsack. To present the instance as an input to an algorithm $A$ for the problem, we must encode it in bits in some fashion; let $|x|$ be the number of bits in the encoding of $x$. Then $|x|$ is called the *size* of the instance or the *instance size*. Furthermore, we say that $A$ is a polynomial-time algorithm if there exists a polynomial $p(n)$ such that the running time of $A$ is $O(p(|x|))$.

We consider an algorithm to be efficient if it runs in polynomial time, and we consider a problem to be efficiently solvable if it has a polynomial-time algorithm. This is an imperfect measure: for instance, we have already noted that the simplex method has a worst-case running time that is exponential in the size of the instance, but is efficient in practice. However, the correspondence between theory and practice is good enough that the equivalence is useful.

One benefit of equating efficiency with polynomial time is that given reasonable models of a computer, efficiency is invariant: that is, an algorithm that runs in polynomial time on one will run in polynomial time on another. Of course, the two polynomials might be different for the two different models of computation. For instance, consider any parallel RAM with a number of processors at most a polynomial in the size of the instance. This parallel machine can achieve a speed-up factor of at most the number of processors it has; thus a polynomial-time algorithm on an ordinary RAM will run in polynomial-time on a parallel RAM and vice versa.

It will be useful in later sections to refer to the class of problems P. The class P contains all *decision problems* that have polynomial-time algorithms. A decision problem is one whose output is either "Yes" or "No". It is not difficult to think of decision problems related to optimization problems. For instance, consider a decision variant of the knapsack problem in which, in addition to inputs $B$, and $v_j$ and $s_j$ for every item $j$, there is also an input $C$, and the problem is to output "Yes" if the optimum solution to the knapsack instance has value at least $C$, and "No" otherwise. The instances of a decision problem can be divided into "Yes" instances and "No" instances; that is, instances in which the correct output for the instance is "Yes" (or "No").

Such a decision problem for the knapsack problem is clearly no harder than the optimization version, but also is no easier: if we had a polynomial-time algorithm for the decision problem, we would also be able to get a polynomial-time algorithm for the optimization problem. To see this, first observe that we can use an algorithm for the decision problem to determine the optimum value $Z$ by performing a bisection search over the range $[0, V]$ (recall that $V = \sum_{j=1}^{n} v_j$ so that $V$ is an upper bound on the value of the knapsack). Once we have determined the optimum value $Z$, we can loop through the items, using the algorithm for the decision problem to see if the value of the optimum solution remains $Z$ when the $j$th item is removed from the instance. If so, we omit item $j$ for the remaining iterations of the loop. The items remaining at the end of the loop must belong to an optimal knapsack, and thus we have determined an optimal knapsack. We use $O(n + \log V)$ calls to the decision algorithm, so if that algorithm runs in polynomial time, there is a polynomial-time algorithm for the optimization problem.

Let us now return to the question of the complexity of the knapsack problem: does it have a polynomial-time algorithm? At first glance, it would appear that the answer is yes, since we have an $O(nB)$ algorithm, where $n$ and $B$ are part of the input. However, this gets us into some of the subtlety of the definition. Usually the data put into a computer are encoded in binary, so that the size of a number $B$ is $\lceil \log_2 B \rceil$ bits. Thus the running time $O(nB)$ is exponential in the size of $B$, not polynomial. If we encode the numeric inputs to the knapsack problem in *unary* rather than binary (that is, we use $B$ bits to encode $B$), then the running time $O(nB)$ is polynomial in the instance size. Algorithms which have this property are called *pseudopolynomial*.

**Definition 2.12** [ Pseudopolynomial-time algorithm ]. *An algorithm for a problem is said to run in pseudopolynomial time, or said to be a pseudopolynomial-time algorithm, with respect to a particular model of computer (such as a RAM) if the number of instructions executed by the algorithm can be bounded by a polynomial in the size of the instance when the numeric data is encoded in unary.*

## 2.4. Complexity theory and a notion of hardness

So, does the knapsack problem have a polynomial-time algorithm? As of the writing of this chapter, the answer is unknown, but there are substantial reasons to think not. Similar evidence suggests that many scheduling problems of interest do not have polynomial-time algorithms. In this section, we present this evidence via the class of problems NP, and the theory of NP-completeness.

Roughly speaking, the class NP is the set of all decision problems such that for any "Yes" instance of the problem, there is a short, easily verifiable "proof" that the answer is "Yes". Additionally, for each "No" instance of the problem, no such "proof" is convincing. What kind of "short proof" do we have in mind? Take the example of the decision variant of the knapsack problem given above. For any "Yes" instance, in which there is a feasible subset of items of value at least $C$, a short proof of this fact is a list of the items in the subset. Given the knapsack instance and the list, an algorithm can quickly verify that the items in the list have total size at most $B$, and total value at least $C$. Note that for any "No" instance, then no possible list of items will be convincing.

We now attempt to formalize this rough idea as follows. A short proof is one whose encoding is bounded by some polynomial in the size of the instance. An easily verifiable proof is one that can be verified in time bounded by a polynomial in the size of the instance and the proof. This gives the following definition.

**Definition 2.13** [ NP ]. *A decision problem is said to be in the problem class* NP *if*

*there exists a verification algorithm $A(\cdot, \cdot)$ and two polynomials, $p_1$ and $p_2$, such that:*

1. *for every "Yes" instance $x$ of the problem, there exists a proof $y$ with $|y| \leq p_1(|x|)$ such that $A(x, y)$ outputs "Yes";*

2. *for every "No" instance $x$ of the problem, for all proofs $y$ with $|y| \leq p_1(|x|)$, $A(x, y)$ outputs "No";*

3. *the running time of $A(x, y)$ is $O(p_2(|x| + |y|))$.*

NP stands for *non-deterministic polynomial time*. Most of the scheduling problems in this book have decision variants that are in the class NP.

Observe that nothing precludes a decision problem in NP from having a polynomial-time algorithm. However, the central problem of complexity theory is whether *every* problem in NP has a polynomial-time algorithm. This is usually expressed as the question of whether the class P of decision problems with polynomial-time algorithms is the same as the class NP, or, more succinctly, as whether P = NP. Although this question has been a matter of intense research for almost thirty years, its answer is unknown as of the writing of this chapter.

To tackle this problem, in the early 1970s several researchers showed that there are problems in NP that are representative of the entire class, in the sense that if they have polynomial-time algorithms, then P = NP, and if they do not, then P $\neq$ NP. These are the NP-*complete* problems; we will be more precise about their definition in a moment. Since then, thousands of problems have been shown to be NP-complete, yet none of them is known to have a polynomial-time algorithm. Most complexity theorists take this as strongly suggestive evidence that P $\neq$ NP. However, complexity theory is not yet able to prove this inequality, and thus the strongest statement that can be made is that it seems likely that every NP-complete problem does not have a polynomial-time algorithm.

Let us now turn to the definition of NP-*completeness*. To do this, we will need the notion of a *polynomial-time reduction*.

**Definition 2.14** [ Polynomial-time reduction ]. *Given two decision problems $A$ and $B$, there is a polynomial-time reduction from $A$ to $B$ (or $A$ reduces to $B$ in polynomial time) if there is a polynomial-time algorithm that takes as input an instance of $A$ and produces as output an instance of $B$ and has the property that a "Yes" instance of $B$ is output if and only if a "Yes" instance of $A$ is input.*

We will use the symbol $\preceq$ to denote a polynomial-time reduction so that we write $A \preceq B$ if $A$ reduces to $B$ in polynomial time. Sometimes the symbol $\leq_m^P$ is used in the literature to denote a polynomial-time reduction. We can now give a formal definition of NP-completeness.

**Definition 2.15** [ NP-complete ]. *A problem $B$ is NP-complete if $B$ is in NP, and for every problem $A$ in NP, there is a polynomial-time reduction from $A$ to $B$.*

The following theorem is now easy to show.

**Theorem 2.16.** *Let B be an* NP-*complete problem. If B has a polynomial-time algorithm, then* P = NP.

*Proof.* Given any problem $A$ in NP, we can create a polynomial-time algorithm for it as follows: the algorithm takes an instance $x$ of $A$ as input, uses the polynomial-time reduction from $A$ to $B$ to transform it into an instance $y$ of $B$, then uses the polynomial-time algorithm for $B$ on that instance. The algorithm outputs "Yes" if and only if the algorithm for $B$ outputs "Yes" for $y$. First, observe that the algorithm outputs "Yes" on $x$ if and only if $x$ is a "Yes" instance of $A$, by the properties of the reduction. Second, we will show that $A$ runs in polynomial time. Suppose that the running time for the reduction algorithm is bounded by polynomial $p_1(|x|)$ and the running time of the algorithm for $B$ is bounded by polynomial $p_2(|y|)$. Certainly $|y| \leq p_1(|x|)$, since the running time for the reduction must include the time spent writing down the bits of $y$. Thus the overall running time is $p_1(|x|) + p_2(p_1(|x|))$, which is a polynomial in $|x|$. □

A useful property of NP-complete problems is that once we have an NP-complete problem $B$ it is often easy to prove that other problems are also NP-complete. As we will see, all we have to do is show that a problem $A$ is in NP, and that $B \preceq A$. This follows as an easy corollary of the transitivity of polynomial-time reductions.

**Theorem 2.17.** *Polynomial-time reductions are transitive: that is, if $A \preceq B$ and $B \preceq C$, then $A \preceq C$.*

**Corollary 2.18.** *If $A$ is in* NP*, $B$ is* NP-*complete, and $B \preceq A$, then $A$ is also* NP-*complete.*

*Proof.* All we need to show is that for each problem $C$ in NP, there is a polynomial-time reduction from $C$ to $A$. Because $B$ is NP-complete, we know that $C \preceq B$. By hypothesis, $B \preceq A$. By Theorem 2.17, $C \preceq A$. □

*Proof of Theorem 2.17.* Since $A \preceq B$, there is an algorithm such that for some polynomial $p_1$ the algorithm takes an instance $x$ of $A$ as input and produces an instance $y$ of $B$ in time no more than $p_1(|x|)$, and $y$ is a "Yes" instance of $B$ if and only if $x$ is a "Yes" instance of $A$. Furthermore, $|y| \leq p_1(|x|)$. Because $B \preceq C$, there exist an algorithm such that for some polynomial $p_2$ an instance $y$ of $B$ can be transformed into an instance $z$ of $C$ in time no more than $p_2(|y|)$, such that $z$ is a "Yes" instance of $C$ if and only if $y$ is a "Yes" instance of $B$. Additionally, $|z| \leq p_2(|y|)$. Thus there is an algorithm such that any instance $x$ of $A$ can be transformed into an instance $z$ of $C$ in time no more than $p_3(|x|)$ such that $x$ is a "Yes" instance of $A$ if and only if $z$ is a "Yes" instance of $C$, where $p_3(\cdot)$ is the polynomial $p_2(p_1(\cdot))$. Thus $A \preceq C$. □

Theorem 2.18 shows that given one NP-complete problem, it is possible to prove that other problems are NP-complete. How do we prove a problem is NP-complete in the first place? Ingenious proofs providing the first NP-complete problems were developed independently by Cook and Levin in the early 1970s, but it is beyond the scope of this chapter to provide any details of the proofs. Interested readers can consult the chapter notes for references. A paper of Karp then showed that many problems of interest are NP-complete, and since then thousands of problems have been shown to be NP-complete. We list a few of them below.

**Partition**
**Input:** Positive integers $a_1, \ldots, a_n$ such that $\sum_{i=1}^{n} a_i$ is even
**Question:** Does there exist a partition of $\{1, \ldots, n\}$ into sets $S$ and $T$ such that $\sum_{i \in S} a_i = \sum_{i \in T} a_i$?

**3-Partition**
**Input:** Positive integers $a_1, \ldots, a_{3n}, b$, such that $b/4 < a_i < b/2$ for all $i$, and $\sum_{i=1}^{3n} a_i = nb$.
**Question:** Does there exist a partition of $\{1, \ldots, 3n\}$ into $n$ sets $T_j$ such that $\sum_{i \in T_j} a_i = b$ for all $j = 1, \ldots, n$? (By the condition on the $a_i$, each $T_j$ must contain exactly 3 elements.)

**Clique**
**Input:** An undirected graph $G = (V, E)$, and a positive integer $K$
**Question:** Does there exist a set of vertices $S \subseteq V$ with $|S| \geq K$ such that for all $i, j \in S$ with $i \neq j$, then $(i, j) \in E$? (Such a set $S$ is called a *clique* or an $|S|$-clique.)

The decision version of the knapsack problem given at the beginning of the section is also NP-complete. However, we know that this problem has a pseudopolynomial-time algorithm. This brings up an interesting distinction among the NP-complete problems. Some NP-complete problems, such as the knapsack and partition problems, are NP-complete only when it is assumed that their numeric data is encoded in binary. As we've seen, the knapsack problem has a polynomial-time algorithm if the input is encoded in unary; so does the partition problem. Other problems, however, such as the 3-partition problem above, are NP-complete even when their numeric data is encoded in unary. We call such problems *strongly* NP-*complete*, or, sometimes, *unary* NP-*complete*. In contrast, problems such as the knapsack and partition problems are called *weakly* NP-*complete* or *binary* NP-*complete*.

Given the NP-completeness of the problems above, we give proofs showing the NP-completeness of three different scheduling problems. From here on we sometimes use the notation $\alpha|\beta|\gamma$ to refer to the decision version of the problem as well as the optimization version. The reader should be able to tell which version is intended from the context.

**Theorem 2.19.** *The problem $1|r_j|L_{\max}$ is NP-complete.*

*Proof.* The decision version of $1|r_j|L_{\max}$ has an input parameter $B$, and a "Yes" instance is one for which the maximum lateness of an optimal schedule is no more than $B$. Certainly this problem is in NP: a short proof for the problem is a list of the starting time of every job. Because each start time is at most $\max_j r_j + \sum_{j=1}^{n} p_j$, the starting time of each is a number whose size is at most a polynomial in the instance size. Since there are $n$ start times, the size of the proof can be bounded by a polynomial in the size of the instance. We can verify in polynomial time whether such a list is a valid schedule and whether $L_{max} \leq B$, and output "Yes" or "No" as appropriate.

To prove that the problem is NP-complete, we give a polynomial-time reduction to it from the partition problem. Given an instance of the partition problem $a_1, \ldots, a_n$, we construct a scheduling instance with $n+1$ jobs. Let $A = \sum_{i=1}^{n} a_i$. Job $j$, for $1 \leq j \leq n$, has processing time $p_j = a_j$, release date $r_j = 0$, and deadline $\bar{d}_j = A+1$. For job $n+1$, set $p_{n+1} = 1$, $r_{n+1} = A/2$, and $\bar{d}_{n+1} = A/2+1$. Finally, set $B = 0$. Certainly this reduction can be performed in polynomial time.

We now need to show that this is a "Yes" instance of $1|r_j|L_{max}$ if and only if $a_1, \ldots, a_n$ is a "Yes" instance of the partition problem. If $a_1, \ldots, a_n$ is a "Yes" instance of the partition problem, then there are two sets $S$ and $T$ that partition $\{1, \ldots, n\}$ such that $\sum_{i \in T} a_i = \sum_{i \in S} a_i = A/2$. Thus the jobs in $S$ can be scheduled from time 0 to time $A/2$, job $n+1$ from time $A/2$ to $A/2+1$, and the jobs in $T$ from time $A/2+1$ to $A+1$; in this case, each job will complete by its deadline. Hence the constructed instance of $1|r_j|L_{max}$ is a "Yes" instance.

Similarly, if the constructed instance of $1|r_j|L_{max}$ is a "Yes" instance, then job $n+1$ must be processed from time $A/2$ to $A/2+1$. Some set $S$ of jobs is processed from time 0 to $A/2$, and the remaining set $T$ is processed from time $A/2+1$ to $A+1$. Then $S$ and $T$ partition $\{1, \ldots, n\}$ and $\sum_{i \in S} a_i \leq A/2$ and $\sum_{i \in T} a_i \leq A/2$. Hence $\sum_{i \in S} a_i = \sum_{i \in T} a_i = A/2$, and $a_1, \ldots, a_n$ is a "Yes" instance of the partition problem. $\square$

**Theorem 2.20.** *The problem $P||C_{max}$ is* NP-*complete.*

*Proof.* The decision version of the problem $P||C_{max}$ has an input $B$, and the instance is a "Yes" instance if $C_{max} \leq B$ in an optimal schedule. It is easy to show that this problem is in NP: the short proof contains for each job the starting time of the job and machine on which the job executes. As argued above, this proof has size that can be bounded by a polynomial in the instance size. A polynomial-time algorithm can check whether this is a feasible schedule and whether $C_{max} \leq B$ and output "Yes" or "No" as appropriate.

To prove that the problem is NP-complete, we give a polynomial-time reduction to it from the 3-partition problem. Given an instance of the 3-partition problem, $a_1, \ldots, a_{3n}$, and $b$, construct a scheduling instance with $3n$ jobs and $n$ machines. Each job $j$ has processing time $p_j = a_j$, and $B$ is set to $b$. Certainly this reduction can be performed in polynomial time.

We now need to show that this a "Yes" instance of $P||C_{max}$ if and only if $a_1, \ldots, a_{3n}$, $b$ is a "Yes" instance of the 3-partition problem. If $a_1, \ldots, a_{3n}, b$ is a "Yes" instance of 3-partition problem, then there are $n$ sets $T_j$ that partition $\{1, \ldots, 3n\}$ such that $\sum_{i \in T_j} a_i = b$. For each set $T_j$, schedule the jobs $i \in T_j$ in any order on the $j$th machine from time 0 to time $\sum_{i \in T_j} a_i = b$. This implies that $C_{max} = b$, and thus the constructed instance of $P||C_{max}$ is a "Yes" instance.

Similarly, suppose the constructed instance of $P||C_{max}$ is a "Yes" instance. Let $T_j$ denote the set of indices of jobs processed on the $j$th machine. Since the scheduling instance is a "Yes" instance, it must be the case that $\sum_{i \in T_j} a_i \leq b$ for each $j$, $1 \leq j \leq n$. But since $\sum_{j=1}^{n} \sum_{i \in T_j} a_i = \sum_{i=1}^{3n} a_i = nb$, it must be the case that $\sum_{i \in T_j} a_i = b$ for each $j$, $1 \leq j \leq n$. Hence $a_1, \ldots, a_{3n}, b$ is a "Yes" instance of the 3-partition problem. $\square$

One consequence of the proof of Theorem 2.20 is that $P||C_{max}$ is strongly NP-complete. This follows since 3-partition is strongly NP-complete; that is, it is NP-complete even if the $a_i$ and $b$ are encoded in unary. We can use the reduction of the Theorem 2.20 to reduce instances of the 3-partition problem with data encoded in unary to instances of $P||C_{max}$ with data encoded in unary in polynomial time. Thus the problem $P||C_{max}$ with data encoded in unary is also NP-complete.

**Theorem 2.21.** *The problem $P|prec, p_j = 1|C_{max}$ is NP-complete.*

*Proof.* The decision version of $P|prec, p_j = 1|C_{max}$ has an input $B$, and an instance is a "Yes" instance if $C_{max} \leq B$ in an optimal schedule. It is easy to verify that this problem is in NP.

We now give a polynomial-time reduction from the clique problem to $P|prec, p_j = 1|C_{max}$. Given the input graph $G = (V, E)$ and the bound $K$ of the clique problem, create one job $J_v$ for each vertex $v \in V$ and one job $J_e$ for each edge $e \in E$. Set $J_v \to J_e$ if $v$ is one of the two endpoints of $e$. Set $N = |V|$, $M = |E|$, and $L = K(K-1)/2$, so that $L$ is the number of edges in a clique of size $K$. For the moment we leave the number of machines $m$ unspecified, and create extra "dummy" jobs: $m - K$ jobs $X_a$, $m - (L + N - K)$ jobs $Y_b$, and $m - (M - L)$ jobs $Z_c$. We set $X_a \to Y_b \to Z_c$ for every $a, b, c$. Set $m$ to guarantee that there will be at least one $X_a$ job, one $Y_b$ job, and one $Z_c$ job; that is, $m = \max\{K, L + N - K, M - L\} + 1$. Finally, set $B = 3$.

For this instance of $P|prec, p_j = 1|C_{max}$, first observe the following: in any schedule of length 3, it must be the case that the dummy jobs $X_a$ are scheduled at time 0, the jobs $Y_b$ at time 1, and the jobs $Z_c$ at time 2, leaving only $K$ machines free at time 0, $L + N - K$ free at time 1, and $M - L$ free at time 2.

We now show that this is a "Yes" instance of $P|prec, p_j = 1|C_{max}$ if there is a clique of size at least $K$ in $G$. If this is the case, then the non-dummy jobs can be completed by time 3 by scheduling $K$ of the jobs $J_v$ corresponding to $K$ vertices of the clique at time 0, scheduling $L$ of the jobs $J_e$ corresponding to the $L$ edges between the $K$ vertices of the clique at time 1, scheduling the remaining $N - K$ jobs corresponding to vertices at time 1, and scheduling the remaining $M - L$ jobs corresponding to edges at time 2.

Finally, this instance is a "No" instance of $P|prec, p_j = 1|C_{max}$ if there is no clique of size $K$ in $G$. In this case, then even if $K$ jobs $J_v$ are scheduled at time 0, at most $L - 1$ jobs $J_e$ can be scheduled at time 1. Even if the remaining $N - K$ jobs $J_v$ are scheduled at time 1, this means there are still $M - (L - 1)$ jobs $J_e$ left to schedule at time 2, but there is only enough room for $M - L$ jobs. Hence not all jobs can complete by time 3, and therefore the constructed instance of $P|prec, p_j = 1|C_{max}$ is a "No" instance. $\square$

We conclude this section by defining the term NP-*hard*, which can be applied to either optimization or decision problems. Roughly speaking, it means "as hard as the hardest problem in NP". To be more precise, we need to define an *oracle*. Given a decision or optimization problem *A*, we say that an algorithm has *A* as an oracle (or has *oracle access* to *A*) if we suppose that the algorithm can solve instance of *A* with a single instruction.

**Definition 2.22** [ NP-hard ]. *A problem A is* NP-*hard if there is a polynomial-time algorithm for an* NP-*complete problem B when the algorithm has oracle access to A.*

The term "NP-hard" is most frequently applied to optimization problems whose corresponding decision problems are NP-complete; it is easy to see that many such optimization problems are indeed NP-hard. It is also easy to see that if *A* is NP-hard and there is a polynomial-time algorithm for *A*, then P = NP.

## 2.5.   Coping with NP-completeness

We now turn from complexity theory back to algorithmic techniques. We have seen that the theoretical measure of whether an optimization problem is efficiently solvable is whether it has a polynomial-time algorithm, and have seen that many scheduling problems are NP-complete and thus unlikely to have such algorithms. Yet such problems continue to arise in practice and need solution, whether they have efficient algorithms or not.

What then can be done? In this section, we explore some of the possible answers to this question. One possibility is to give an efficient *heuristic* that does not always find an optimal solution to the problem, but does find one that is "good enough". An *approximation algorithm* is a type of heuristic that comes with a proven performance guarantee. Another possibility is to give an algorithm that finds the optimal solution, and although it is not guaranteed to run in polynomial time, it runs quickly enough for instances arising in practice. Some such algorithms use *branch-and-bound* techniques, which are discussed below.

One last possibility is to make probabilistic statements about the kinds of instances that arise in practice, and then design algorithms that with high probability run quickly and find a good or optimal solution. We will not discuss probabilistic techniques in this section, although this approach is considered in Chapter 9.

**Approximation algorithms.** Although there are many heuristics and metaheuristic techniques for scheduling problems (such as simulated annealing), in this chapter we will only consider *approximation algorithms*. An approximation algorithm for an optimization problem is one that runs in polynomial time, and is guaranteed to provide a solution whose value is close to the optimum value. If the algorithm produces a solution whose value is always within a factor of $\alpha$ of the value of an optimal solution, then the algorithm is called an $\alpha$-approximation algorithm. The value $\alpha$ is called the *performance guarantee* or the *approximation factor* of the algorithm. In this chapter, we will use the convention that $\alpha > 1$ for minimization problems, and $\alpha < 1$ for maximization problems. For example, a 2-approximation algorithm for a minimization problem produces solutions of value no more than twice the optimum value, while a $\frac{1}{2}$-approximation algorithm for a maximization problem produces a solution of value at least $\frac{1}{2}$ the optimum value. However, no standard convention exists for maximization problems, and in the literature it is common to see $1/\alpha$ used as the performance guarantee so that the $\frac{1}{2}$-approximation algorithm above would also be referred to as a 2-approximation algorithm.

Sometimes it is possible to give a family of approximation algorithms that can produce solutions whose value comes arbitrarily close to the optimum value. That is, for each $\varepsilon > 0$, we can give a $(1+\varepsilon)$-approximation algorithm $A_\varepsilon$ for a minimization problem or a $(1-\varepsilon)$-approximation algorithm for a maximization problem. The running time of the family of algorithms $\{A_\varepsilon\}$ may depend on $\varepsilon$ in some nasty way (e.g., it may be exponential in $1/\varepsilon$), but since $\varepsilon$ is a fixed constant for each algorithm $A_\varepsilon$ this does not matter. We call such a family a *polynomial-time approximation scheme*, sometimes abbreviated as *PTAS*.

**Definition 2.23** [ Polynomial-time approximation scheme ]. *A polynomial-time approximation scheme (PTAS) is a family of algorithms $\{A_\varepsilon\}$ for an optimization problem such that for each $\varepsilon > 0$, $A_\varepsilon$ is a $(1+\varepsilon)$-approximation algorithm (for a minimization problem) or a $(1-\varepsilon)$-approximation algorithm (for a maximization problem).*

If a family of algorithms is a polynomial-time approximation scheme, and it is also the case that the running time of the family is also polynomial in $1/\varepsilon$, we call the family a *fully polynomial-time approximation scheme*, sometimes abbreviated as *FPTAS* or *FPAS*.

**Definition 2.24** [ Fully polynomial-time approximation scheme ]. *If $\{A_\varepsilon\}$ is a polynomial-time approximation scheme such that the running time of the family of algorithms is polynomial in $1/\varepsilon$, then $\{A_\varepsilon\}$ is called a fully polynomial-time approximation scheme.*

To illustrate these ideas, we show how we can obtain a fully polynomial-time approximation scheme for the knapsack problem (and thus for the problem $1|\bar{d}_j = d|\sum w_j U_j$). In Figure 2.4 of Section 2.1, we gave a dynamic programming algorithm for the knapsack problem that requires $O(n \min(B,V))$ time, where $n$ is the number of items, $B$ is the size of the knapsack, and $V = \sum_{i=1}^{n} v_i$ is the maximum possible value of the knapsack. If we could transform any instance of the knapsack problem into one for which the optimum solution was not too different in value, and for which $V$ was polynomial in $n$ and $1/\varepsilon$, then in time polynomial in the instance size we could find the optimum of the transformed instance, and it would be a good approximation of the original optimum solution. This is precisely what we do in our fully polynomial-time approximation scheme for the knapsack problem. The algorithm shown in Figure 2.5 rounds the values $v_j$ down to the nearest multiple of $\varepsilon W/n$, where $W$ is the maximum value of any item; call these new values $v'_j$. Then $V' = \sum_{j=1}^{n} v'_j = \sum_{j=1}^{n} \lfloor \frac{v_j}{\varepsilon W/n} \rfloor \leq n^2/\varepsilon$, so that invoking the dynamic programming algorithm on the transformed instance takes $O(nV') = O(n^3/\varepsilon)$ time. We now only

```
1        W ← 0
2        For j ← 1 to n
3              If vⱼ > W
4                    W ← vⱼ
5        K ← εW/n
6        For j ← 1 to n
7              v'ⱼ ← ⌊vⱼ/K⌋
8        Run dynamic programming algorithm in Figure 2.4 on instance with
              sizes sⱼ and values v'ⱼ.
```

**Figure 2.5.**   Fully polynomial-time approximation scheme for the knapsack problem.

need to show that the value of the solution found on the transformed instance is close to the optimum value of the original instance.

**Theorem 2.25.** *The algorithm in Figure 2.5 produces a solution of value at least* $(1-\varepsilon)$ *times the optimum value.*

*Proof.*   Let $S$ be the set of items found by running the algorithm of Figure 2.4 with the modified values $v'_j = \lfloor v_j/K \rfloor$, where $K = \varepsilon W/n$. Let $O$ be an optimal set of items for the original instance, and let OPT denote its value. We know that $W \leq \text{OPT}$, since one possible knapsack is to take the most valuable item (recall that $s_j \leq B$ for all items $j$). We also know, by the definition of the $v'_j$, that

$$Kv'_j \leq v_j < K(v'_j + 1), \tag{2.5}$$

which implies that $Kv'_j > v_j - K$. Then

$$
\begin{aligned}
\sum_{j \in S} v_j &\geq K \sum_{j \in S} v'_j && \text{(by (2.5))} \\
&\geq K \sum_{j \in O} v'_j && \text{(since } S \text{ is an optimal solution for the values } v'_j) \\
&\geq \sum_{j \in O} v_j - |O|K && \text{(by (2.5))} \\
&\geq \sum_{j \in O} v_j - nK \\
&= OPT - \varepsilon W \\
&\geq (1-\varepsilon)OPT.
\end{aligned}
$$

□

Sometimes it is possible to show that a particular performance guarantee $\alpha$ cannot be achieved unless P = NP; that is, we can't even get an approximate solution whose value is within a factor $\alpha$ of the optimum in polynomial time unless we could solve the problem optimally in polynomial time. For instance, it is not difficult to show that under quite reasonable conditions, no optimization problem whose decision variant is strongly NP-complete can have an fully polynomial-time approximation scheme.

**Theorem 2.26.** *Given an optimization problem whose decision version is strongly* NP-*complete, let* $|x|_u$ *denote the length of the encoding of an instance* $x$ *with the numeric data encoded in unary. Let* OPT$(x)$ *denote the optimum value of the instance* $x$. *If* OPT$(x)$ *is an integer-valued function, and there exists a polynomial* $p$ *such that* OPT$(x) < p(|x|_u)$ *for all instances* $x$, *then there is no fully polynomial-time approximation scheme for the problem unless* P = NP.

*Proof.* Assume the problem is a minimization problem; a trivially modified proof will work for a maximization problem. The decision version of the problem asks whether or not OPT$(x) \leq B$ for an instance $(x, B)$. Suppose there is an fully polynomial-time approximation scheme for the problem. For a given instance $x$, set $\varepsilon = 1/p(|x|_u)$. Then the fully polynomial-time approximation scheme returns a solution of value between OPT$(x)$ and

$$(1+\varepsilon)\,\mathrm{OPT}(x) = \left(1 + \frac{1}{p(|x|_u)}\right)\mathrm{OPT}(x) < \mathrm{OPT}(x) + 1.$$

Since OPT$(x)$ is integer valued, the algorithm returns a solution of value $OPT(x)$. Thus we can correctly decide if OPT$(x) \leq B$. Furthermore, the algorithm takes time polynomial in $|x|$ and $\frac{1}{\varepsilon} = p(|x|_u)$, which is polynomial in $|x|_u$. However, the decision variant is NP-complete even if the instance is encoded in unary, so we have a polynomial-time algorithm to decide an NP-complete problem, implying P = NP. $\square$

As an example of the applications of this theorem, consider the problem $P||C_{\max}$. The makespan of an optimal schedule is at most $\sum_{j=1}^{n} p_j$. If the data for $P||C_{\max}$ is encoded in unary, then certainly the encoding will require at least $\sum_{j=1}^{n} p_j$ bits to represent all the processing times $p_j$. Thus OPT$(x) < |x|_u + 1$, and one condition of the theorem is met. The other condition of the theorem is met since the decision version of $P||C_{\max}$ is strongly NP-complete. Thus there is no FPTAS for $P||C_{\max}$ unless P = NP.

Sometimes a proof of NP-completeness will imply that no $\alpha$-approximation algorithm can exist for a given $\alpha$ unless P = NP. As an example, we can use the proof of Theorem 2.21 to show the following theorem.

**Theorem 2.27.** *There can be no* $\alpha$-*approximation algorithm for* $P|prec, p_j = 1|C_{\max}$ *with* $\alpha < 4/3$ *unless* P = NP.

*Proof.* In the proof of Theorem 2.21, we showed that a "Yes" instance of the clique problem could be mapped to an instance of $P|prec, p_j = 1|C_{\max}$ with makespan 3, and a "No" instance was mapped to an instance with makespan greater than 3. Since $p_j = 1$ for all jobs, an instance with makespan greater than 3 must have makespan at least 4. Now suppose we have an $\alpha$-approximation algorithm for $P|prec, p_j = 1|C_{\max}$ with $\alpha < 4/3$. Then a polynomial-time algorithm for the clique problem is as follows: given an instance of the clique problem, transform the instance into an instance of $P|prec, p_j = 1|C_{\max}$ as in Theorem 2.21 and run the $\alpha$-approximation algorithm. If the approximation algorithm gives a schedule of length 3, output "Yes", otherwise "No". The algorithm runs in polynomial time, and gives the correct answer since whenever it is given a "Yes" instance of the clique problem, the approximation algorithm creates a schedule of length less than $\frac{4}{3} \cdot 3 = 4$, which must be a schedule of length 3. Thus the algorithm correctly outputs "Yes". If the algorithm is given a "No" instance of the clique problem, the makespan of the schedule must be at least 4, and so the algorithm correctly returns "No". Thus we cannot have an $\alpha$-approximation algorithm for $P|prec, p_j = 1|C_{\max}$ for $\alpha < 4/3$ unless P = NP. $\square$

Similar theorems can be shown for other scheduling problems; see Chapters 12–14 for examples.

**Branch and bound.** Another approach to dealing with NP-complete problems is to guarantee optimality but not efficiency, rather than guaranteeing efficiency but not optimality. One technique of this type is called *branch-and-bound*. This technique is a combination of an algorithm which searches the space of all possible solutions to a problem (the "branch" part) and an algorithm that bounds the optimum value, to rule out parts of the search space (the "bound" part). More sophisticated variations of this technique have been developed (e.g. branch-and-cut, branch-and-price), but we will only discuss the most basic version.

We begin by showing how this technique can be applied to the knapsack problem. First, we formulate the problem as an integer program:

$$\text{Max} \quad \sum_{j=1}^{n} v_j x_j$$

subject to:

$$\sum_{j=1}^{n} s_j x_j \leq B$$
$$x_j \in \{0, 1\} \qquad\qquad 1 \leq j \leq n.$$

Replacing the constraints $x_j \in \{0, 1\}$ with constraints $0 \leq x_j \leq 1$ gives a linear programming relaxation. Solving this LP is our "bounding" algorithm, since the optimal value of this LP solution is an upper bound on the cost of an optimal solution. In fact, the optimal solution to this LP can be obtained easily: suppose the items are indexed such that $v_1/s_1 \geq v_2/s_2 \geq \cdots \geq v_n/s_n$, and $\sum_{j=1}^{k-1} s_j \leq B$, but $\sum_{j=1}^{k} s_j > B$. Then an optimal LP solution will set $x_1 = x_2 = \cdots = x_{k-1} = 1$, $x_{k+1} = x_{k+2} = \cdots = x_n = 0$, and $x_k = \frac{1}{s_k}\left(B - \sum_{j=1}^{k-1} s_j\right)$. If $x_k = 0$, then the solution is integral, and the solution is an optimal solution to the knapsack problem as well.

If $x_k \neq 0$, then we enter our "branching" algorithm. We create the two subproblems of our knapsack problem, one in which the $k$th item cannot be included in the knapsack (equivalent to setting $x_k = 0$) and the other in which the $k$th item must be included in the knapsack (equivalent to $x_k = 1$); this latter problem is equivalent to a knapsack problem with item $k$ omitted and a knapsack of capacity $B - s_k$. Observe that if we can find the solution to these two subproblems, then whichever solution has the maximum value is the optimal solution to the original knapsack problem, since the $k$th item must either be included or excluded. We say that the two subproblems *partition the solution space* of the original problem. Since both subproblems are knapsack problems, we can apply the same approach as above to find their solution; eventually either the LP relaxation returns an integer solution, or is infeasible, in which case we no longer need to consider the subproblem.

This process is usually described in terms of a tree. The tree's root corresponds to the original knapsack problem, the root's children to the subproblems of the original problem, and in general the children of a node correspond to the subproblems of that node.

This overall algorithm might be quite slow, but the following ideas can be used to speed it up. We can keep track of the value of the *incumbent* solution, the best integral solution found thus far. If we reach a subproblem such that the value of the LP relaxation for that subproblem is less than the value of the incumbent solution, then we do not need to find an integral solution to that subproblem; whatever it is, it will not be the solution to the original problem because it will have value less than that of the incumbent solution. So we can "cut off" this part of the solution space; we say that the node in the tree corresponding to this subproblem has been *fathomed*. Obviously the more nodes we can fathom, the less work we will have to do. Since we do not initially have any integral solution to the problem (and we might have to solve many subproblems before we find one), it is common to run a heuristic either before or immediately after solving initial LP relaxation of the original problem, so that we have some incumbent solution with which to cut off parts of the solution space. For example, we can run the polynomial-time approximation scheme for knapsack of the previous section to get a good initial incumbent solution.

The discussion above illustrates one way of applying branch-and-bound to a particular problem, the knapsack problem. The ideas there can be generalized in a number of different ways. For instance, we needn't use a linear programming relaxation of an integer programming formulation of the problem of interest; we can also use some other easily computed bound on the value of an optimal solution, as long as we know when the bound is equal to the value of an optimal solution. In the example above, we "branched" on a fractional variable $x$ from the linear programming solution, creating two subproblems in which $x$ was set to 0 and $x$ was set to 1. In general, any way of creating two or more subproblems that partition the solution space will do.

Other methods exist to speed up branch-and-bound. The running time of branch-and-bound depends on the amount of time it takes to compute a bound and the number of subproblems created. Getting good bounds and having a good incumbent solution cut down on the number of subproblems, but it is also helpful to speed up the bound algorithm. One way to do this is to use a bound which can be computed more quickly than an LP relaxation. In the case of the knapsack problem, we had an LP bound which could be computed by inspection, but this is not always the case.

One way to get a bound which can be computed more quickly is to further relax the linear programming relaxation. We now consider one technique for doing so, called *Lagrangean relaxation*.

**Lagrangean relaxation.** *Lagrangean relaxation* is a technique that can be used to speed up the "bound" computation in branch-and-bound. Usually it is applied to integer or linear programs in which there are some "nice" inequalities and some "nasty" inequalities, in the sense that if the "nasty" inequalities were not present we would be able to solve the integer or linear program more easily (perhaps using a combinatorial algorithm, such as a network flow algorithm). The basic idea of Lagrangean relaxation is that we drop the nasty constraints from the integer program or linear program, but add penalties for their violation to the objective function.

For example, suppose we wish to solve the following integer program

$$Z_A^* = \quad \text{Min} \quad \sum_{j=1}^{n} c_j x_j$$

subject to:

$$(A) \qquad \sum_{j=1}^{n} a_{ij} x_j \geq d_i \qquad\qquad i = 1, \ldots, p$$

$$\sum_{j=1}^{n} b_{ij} x_j \geq e_i \qquad\qquad i = 1, \ldots, q$$

$$x_j \in \{0, 1\} \qquad\qquad j = 1, \ldots, n,$$

where the integer program is easier to solve without the $q$ constraints $\sum_{j=1}^{n} b_{ij} x_j \geq e_i$. Then in Lagrangean relaxation, we introduce constants $\lambda_1, \ldots, \lambda_q \geq 0$ and create the following integer program

$$L^*(\lambda) = \quad \text{Min} \quad \sum_{j=1}^{n} c_j x_j + \sum_{i=1}^{q} \lambda_i \left( e_i - \sum_{j=1}^{n} b_{ij} x_j \right)$$

subject to:

$$\sum_{j=1}^{n} a_{ij} x_j \geq d_i \qquad\qquad i = 1, \ldots, p$$

$$x_j \in \{0, 1\} \qquad\qquad j = 1, \ldots, n.$$

Observe that we can rearrange terms in the objective function in the following way:

$$\begin{aligned}
L(x, \lambda) &= \sum_{j=1}^{n} c_j x_j + \sum_{i=1}^{q} \lambda_i \left( e_i - \sum_{j=1}^{n} b_{ij} x_j \right) \\
&= \sum_{j=1}^{n} \left( c_j - \sum_{i=1}^{q} \lambda_i b_{ij} \right) x_j + \sum_{i=1}^{q} \lambda_i e_i.
\end{aligned}$$

Thus if we define $\tilde{c}_j = c_j - \sum_{i=1}^{q} \lambda_i b_{ij}$, the Lagrangean relaxation is of the same form as the easily solvable integer program, except that it has a constant term in the objective function:

$$L^*(\lambda) = \quad \text{Min} \quad \sum_{j=1}^{n} \tilde{c}_j x_j + \sum_{i=1}^{q} \lambda_i e_i$$

subject to:

$$\sum_{j=1}^{n} a_{ij} x_j \geq d_i \qquad i = 1, \ldots, p$$

$$x_j \in \{0, 1\} \qquad j = 1, \ldots, n.$$

By hypothesis, we can solve this integer program easily. Furthermore, the value of its solution, $L^*(\lambda)$, is a lower bound on $(A)$, $Z_A^*$. To see this, note that for any feasible solution $x$ to $(A)$,

$$\sum_{i=1}^{q} \lambda_i \left( e_i - \sum_{j=1}^{n} b_{ij} x_j \right) \leq 0,$$

and so $L(x, \lambda) \leq \sum_{j=1}^{n} c_j x_j$. Since any $x$ feasible for $(A)$ is feasible for the Lagrangean relaxation and the Lagrangean relaxation is a minimization problem, $L^*(\lambda) \leq Z_A^*$.

The technique can be applied to linear programs as well as integer programs; the discussion above carries through if the constraints $x_j \in \{0, 1\}$ are replaced with $x_j \geq 0$. As a concrete example of how Lagrangean relaxation applies to linear programs, consider the problem $1 | prec | \sum_j w_j C_j$. For the problem without precedence constraints, there is a linear program

$$\text{Min} \quad \sum_{j=1}^{n} w_j C_j$$

subject to:

$$\sum_{j=1}^{n} a_{ij} C_j \geq d_i \qquad \forall i$$

$$C_j \geq 0 \qquad j = 1, \ldots, n,$$

which gives the optimal solution, where the variable $C_j$ gives the completion time of job $j$. There is a simple $O(n \log n)$ time algorithm to compute the solution (see Chapter 4 for details of the linear program and algorithm). We can get a better lower bound on $1 | prec | \sum_j w_j C_j$ by adding constraints $C_l - C_k \geq p_k$ whenever $J_k \rightarrow J_l$. However, adding these constraints to the linear program makes it much harder to solve. Applying Lagrangean relaxation gives the linear program

$$L^*(\lambda) = \quad \text{Min} \quad \sum_{j=1}^n w_j C_j + \sum_{k,l:J_k \to J_l} \lambda_{kl} (p_k - C_l + C_k)$$

subject to:

$$\sum_{j=1}^n a_{ij} C_j \geq d_i \qquad\qquad \forall i$$

$$C_j \geq 0 \qquad\qquad j = 1, \ldots, n.$$

Rearranging the objective function yields

$$\begin{aligned}
L^*(\lambda) &= \sum_{j=1}^n \left( w_j - \sum_{k:J_k \to J_j} \lambda_{kj} + \sum_{l:J_j \to J_l} \lambda_{jl} \right) C_j + \sum_{k,l:J_k \to J_l} \lambda_{kl} p_k \\
&= \sum_{j=1}^n \tilde{w}_j C_j + \sum_{k,l:J_k \to J_l} \lambda_{kl} p_k,
\end{aligned}$$

where $\tilde{w}_j = w_j - \sum_{k:J_k \to J_j} \lambda_{kj} + \sum_{l:J_j \to J_l} \lambda_{jl}$. We can then quickly obtain a lower bound on the cost of optimal solution for the instance of $1|prec|\sum_j w_j C_j$ by running the $O(n \log n)$ time algorithm on the $1||\sum_j w_j C_j$ instance with weights $\tilde{w}_j$, and adding $\sum_{k,l:J_k \to J_l} \lambda_{kl} p_k$ to the resulting value.

Of course, in this case and in general, we would like to compute the best bound possible; that is, we would like to compute the maximum of $L^*(\lambda)$ over all $\lambda \geq 0$. In some cases, the given problem has enough structure that we are able to find quickly the $\lambda$ that maximizes $L^*(\lambda)$. In general, however, we can use *subgradient optimization* to find a good value of $\lambda$. First, we claim that $L^*(\lambda)$ is a concave function of $\lambda$, so that any local maximum of $L^*(\lambda)$ is also a global maximum. We could thus use standard gradient ascent methods to find the global maximum, except that $L^*(\lambda)$ is not everywhere differentiable. Instead, we use a generalization of a gradient called a *subgradient*. A vector $\alpha \in \mathbb{R}^q$ is a subgradient of $L^*(\cdot)$ at $\lambda$ if

$$L^*(\lambda') \leq L^*(\lambda) + (\lambda' - \lambda)^T \alpha$$

for all $\lambda' \geq 0$. For a subgradient $\alpha$, the set $\{\lambda' : (\lambda' - \lambda)^T \alpha \geq 0\}$ contains any $\lambda'$ for which $L^*(\lambda') > L^*(\lambda)$. Thus the subgradient gives a direction of ascent. Fortunately, a subgradient is easy to find. Consider the LP $(A)$. Let $x^*$ be an optimal solution for $\lambda$ so that $L^*(\lambda) = L(x^*, \lambda)$. Then the vector $\alpha$ with $\alpha_i = e_i - \sum_{j=1}^n b_{ij} x_j^*$ is a subgradient. To see this, choose some $\lambda' \geq 0$ and let $x'$ be an optimal solution for $\lambda'$ so that $L^*(\lambda') = L(x', \lambda')$. Then it is the case that

$$
\begin{aligned}
L^*(\lambda') = L(x', \lambda') &\leq L(x^*, \lambda') \\
&= \sum_{j=1}^{n} c_j x_j^* + \sum_{i=1}^{q} \alpha_i \lambda_i' \\
&= L^*(x^*, \lambda) - \sum_{i=1}^{q} \alpha_i \lambda_i + \sum_{i=1}^{q} \alpha_i \lambda_i' \\
&= L^*(\lambda) + (\lambda' - \lambda)^T \alpha.
\end{aligned}
$$

Subgradient optimization generally works by producing a sequence of $\lambda^i \geq 0$ in the following way: we start from some $\lambda^0$, compute $L^*(\lambda^i)$, find a subgradient $\alpha^i$, and then set $\lambda^{i+1} = \lambda^i + \delta^i \alpha^i \geq 0$, where $\delta^i > 0$ is a scalar step length. Note that in this case,

$$
L^*(\lambda^{i+1}) \leq L^*(\lambda^i) + \delta^i \|\alpha^i\|^2,
$$

so potentially the value of $L^*(\lambda^{i+1})$ has increased from $L^*(\lambda^i)$. The following theorem states that this sequence $\lambda^i$ converges to the optimum under certain conditions.

**Theorem 2.28.** *If $\delta^i \to 0$ as $i \to \infty$ and $\sum_{i=1}^{\infty} \delta^i = \infty$, then*

$$
L^*(\lambda^i) \to \max_{\lambda \geq 0} L^*(\lambda).
$$

The notes at the end of the chapter give references about Lagrangean relaxation and subgradient optimization.

### Acknowledgments

I am very grateful to Eugene Lawler, whose existing outline and drafts of this chapter proved a solid foundation on which I was able to build in my own fashion. I hope he would be pleased by the results. I am grateful to the editors, Jan Karel Lenstra and David Shmoys, for allowing me to participate in this volume. In addition, David Shmoys gave me many helpful suggestions on how to structure (and restructure) the chapter. I am also grateful to Greg Sorkin for many useful comments.

### Notes

*Combinatorial optimization.* For overviews to the field of combinatorial optimization, consult the books of Bertsimas and Tsitsiklis [1997], Cook, Cunningham, Pulleyblank, and Schrijver [1998], Lawler [1976], Nemhauser and Wolsey [1988], and Papadimitriou and Steiglitz [1982].

*Linear programming.* There are many books on linear programming. A nice introduction for beginners is the book by Chvátal [1983]. Schrijver [1986] gives a technical, dense, and comprehensive overview of the area, although many of the recent developments in interior point methods for solving linear programs are not included.

Wright [1997] gives an accessible treatment of this topic.

*Network flows.* For further reading on network flows, we recommend the book of Ahuja, Magnanti, and Orlin [1993]. The books above on combinatorial optimization all include sections on network flows, as do many introductory texts on algorithms (see Cormen, Leiserson, and Rivest [1990], for example). The first proof of the maximum flow/minimum cut theorem was given by Ford and Fulkerson [1956]. The proof of Theorem 2.5 given here is due to Bertsimas, Teo, and Vohra [1995]. The algorithm for deciding the feasibility of $P|pmtn, r_j, \bar{d}_j|-$ given McNaughton's rule is due to Horn [1974].

*Dynamic programming.* The term "dynamic programming" was coined by Bellman [1957]. He considers problems in continuous optimization. Other examples of dynamic programming applied to discrete optimization problems can be found in standard textbooks in algorithms, such as Cormen, Leiserson, and Rivest [1990]. The knapsack algorithm given in Figure 2.3 is due to Bellman and Dreyfus [1962]. The algorithm in Figure 2.4 is due to Lawler [1979].

*Complexity theory.* For overviews of the field of complexity theory, consult the books of Papadimitriou [1994] and Sipser [1997].

*The complexity of the algorithmic techniques.* The first example showing that the simplex method requires an exponential number of pivots for a certain pivot rule is due to Klee and Minty [1972]; other researchers followed with examples for other pivot rules.

For further information about interior-point methods, consult the books of Wright [1997] and Ye [1997].

See Ahuja, Magnanti, and Orlin [1993] for discussions of various types of network flow algorithms and their running times. The maximum flow algorithm with running time $O(\min(n^{2/3}, m^{1/2}) m \log(\frac{n^2}{m}) \log U)$ is due to Goldberg and Rao [1998]. The more practical push-relabel algorithm was discovered by Goldberg and Tarjan [1988]. Implementations of various maximum flow algorithms are studied in the volume edited by Johnson and McGeoch [1993]. See also Cherkassky and Goldberg [1997] for an implementation of the push-relabel algorithm. The result showing that the parametric maximum flow problem can be solved in the same running time as the push-relabel algorithm is due to Gallo, Grigoriadis, and Tarjan [1989]. The successive approximation push-relabel algorithm for minimum-cost flows is described in Goldberg and Tarjan [1990] and evaluated in Goldberg [1997]. The implementation there is compared to network simplex codes of Grigoriadis [1986] and Kennington and Helgason [1980].

*Complexity theory and a notion of hardness.* A slightly dated, but still very worthwhile introduction to the theory of NP-completeness can be found in the book of Garey and Johnson [1979]. Perhaps the most useful feature of the book is an ap-pendix listing some 300 NP-complete problems. The notion of NP-completeness and the first NP-complete problems were given independently by Cook [1971] and Levin [1973]. However, Karp [1972] was the first to show that many problems from combinatorial optimization are NP-complete.

Lenstra, Rinnooy Kan, and Brucker [1977] gives the proof of the NP-completeness of $1|r_j|L_{max}$ in Theorem 2.19. The proof of the NP-completeness of $P||C_{max}$ in Theorem 2.20 is due to Garey and Johnson [1978]. The first proof of the NP-completeness of $P|prec, p_j = 1|C_{max}$ is due to Ullman [1975]. The proof we give in Theorem 2.21 is due to Lenstra and Rinnooy Kan [1978].

*Approximation algorithms.* Shmoys [1995] gives an excellent survey of this area. A collection of surveys edited by Hochbaum [1997] gives a more in-depth treatment of various areas in approximation algorithms.

The polynomial-time approximation scheme we give for the knapsack problem is due to Ibarra and Kim [1975].

*Branch-and-bound.* Surveys of Lagrangean relaxation are given by Geoffrion [1974] and Fisher [1981]. The example of applying Lagrangean relaxation to $1|prec|\sum_j w_j C_j$ is due to Van de Velde [1990]. Theorem 2.28 follows from work of Polyak [1967].

# Contents

# 3

# Minmax criteria

Eugene L. Lawler
*University of California, Berkeley*

Jan Karel Lenstra
*Centrum Wiskunde & Informatica*

David B. Shmoys
*Cornell University*

Pity poor Bob Cratchit, his in-basket piled high and the holidays upon him. The in-basket demands the completion of $n$ jobs, each with its own due date. Cratchit has no idea how long any of the jobs will take. But he knows full well that he must complete all the jobs on time, else his employment at Scrooge, Ltd. will be terminated. In what order should he do the jobs?

J. R. Jackson supplied the answer in 1955, too late for Cratchit, but early enough to qualify as one of the first optimization algorithms in scheduling theory. According to Jackson's Earliest Due Date (EDD) rule, if there is any way to complete all of the jobs on time, then it can be accomplished by performing the jobs in order of nondecreasing due dates. Furthermore, such an EDD order minimizes maximum lateness, whether or not it meets all of the due dates. Thus, Cratchit's $1||L_{\max}$ problem is solved by simply putting the jobs in EDD order.

The EDD rule, as given by Jackson, is extremely simple. But it has many implications that are quite nonobvious. With prior modification of due dates, the EDD rule can be applied to solve $1|prec|L_{\max}$. A preemptive extension of the rule solves $1|pmtn,r_j|L_{\max}$ and $1|pmtn,prec,r_j|L_{\max}$. Furthermore, a generalization of the EDD rule, the Least Cost Last rule, solves the general problems $1|prec|f_{\max}$ and $1|pmtn,prec,r_j|f_{\max}$ in $O(n^2)$ time.

Given these successes, one might suppose that other problems can be solved effi-ciently by some sort of elaboration of the EDD rule as well. Indeed, a complicated algorithm that involves an iterative application of the EDD rule solves the case of equal processing times, $1|r_j, p_j = p|L_{max}$, in polynomial time. This is probably as far as we can get, however: the unrestricted problem $1|r_j|L_{max}$ is strongly NP-hard. In later chapters, we shall have many occasions to reflect on the fact that preemptive scheduling problems are often easier than their nonpreemptive counterparts, and never known to be harder. In the case of single-machine minmax problems, nonuniform release dates are innocuous when preemption is permitted, but calamatous when it is not.

When faced with an NP-hard problem, there are various things we can do. First, we can try to find interesting and useful special cases that are solvable in polynomial time, such as the ones mentioned above. Second, we can devise approximation algorithms with performance guarantees, as we will do for the related 'head-body-tail problem'. Third, we can avail ourselves of enumerative methods.

### 3.1. Earliest Due Date and Least Cost Last rules

The *Earliest Due Date (EDD) rule* schedules the jobs without idle time in order of nondecreasing due dates; ties are broken arbitrarily. The resulting schedule rule is called an EDD schedule.

**Theorem 3.1** [ Jackson's EDD rule ]. *Any EDD schedule is optimal for the problem* $1||L_{max}$.

*Proof.* Let $\pi$ be the ordering of the jobs in an EDD schedule, i.e., $J_{\pi(j)}$ is the $j$th job scheduled, and let $\pi^*$ be an optimal ordering. Suppose that $\pi \neq \pi^*$. Then there exist jobs $J_j$ and $J_k$ such that $J_k$ immediately precedes $J_j$ in $\pi^*$, but $J_j$ precedes $J_k$ in $\pi$, with $d_j \leq d_k$. Interchanging the positions of $J_k$ and $J_j$ in $\pi^*$ does not increase $L_{max}$, and decreases the number of pairs of consecutive jobs in $\pi^*$ that are in the opposite order in $\pi$. It follows that a finite number of such interchanges transforms $\pi^*$ into $\pi$, so that $\pi$ is optimal. ∎

**Corollary 3.2.** *For any instance of* $1||L_{max}$, *all due dates can be met if and only if they are met in any EDD schedule.*

An EDD schedule can be found by sorting the due dates, which requires $O(\log n)$ time. Sometimes $O(n)$ time suffices, as in the special case of equal processing times (see Exercises 3.3 and 3.4).

It is perhaps surprising that the EDD rule also solves the problem $1|prec|L_{max}$, provided that the due dates are first modified to reflect the precedence constraints. Similar modifications schemes will be encountered in Chapter 11 as well, but they will not be as simple as this one.

If $d_j < d_k$ whenever $J_j \to J_k$, then any EDD schedule is consistent with the given precedence constraints and thereby optimal for $1|prec|L_{max}$. Observe that, if $J_j \to J_k$,

we may set

$$d_j := \min\{d_j, d_k - p_k\} \tag{3.1}$$

without increasing the value of $L_{\max}$ in any feasible schedule, since

$$L_k = C_k - d_k \geq C_j + p_k - d_k = C_j - (d_k - p_k).$$

Let $G = (\{1, \ldots, n\}, A)$ be an acyclic digraph that represents the precedence relation. Consider updating the due dates in the following systematic manner: in each iteration, select a vertex $k \in V$ of outdegree 0 (that is, a *sink*) and, for each arc $(j, k)$, update $d_j$ using (3.1); when all such arcs have been processed, delete them and vertex $k$ from $G$.

We claim that, after executing this algorithm, $d_j < d_k$ whenever $(j, k) \in A$. At some stage, vertex $k$ becomes a sink and is subsequently selected in a particular iteration. Immediately after that iteration, the update (3.1) implies that $d_j < d_k$; since $d_k$ remains unchanged from this point on and $d_j$ can only decrease, our claim is valid. Hence, by blindly applying the EDD rule to the modified due dates, we automatically obtain a feasible schedule, which must therefore be optimal.

The due date modification algorithm takes $O(n + |A|)$ time. Since the EDD rule takes $O(n \log n)$ time, we can solve $1 | prec | L_{\max}$ in $O(n \log n + |A|)$ time. Moreover, the problem can be solved without any knowledge of the processing times of the jobs (see Exercise 3.5).

There is an important symmetry between the problems $1 || L_{\max}$ and $1 | r_j | C_{\max}$. The former problem can be viewed as minimizing the amount by which the due dates must be uniformly increased so that there exists a schedule in which each job meets its (modified) due date. For the latter problem, we may also restrict attention to schedules in which the jobs are processed without idle time between them, since in each feasible schedule of length $C_{\max}$ the start of processing can be delayed until $C_{\max} - \sum_j p_j$. The problem can then be viewed as minimizing the amount by which the release dates must be uniformly decreased so that there exists a schedule of length $\sum_j p_j$ in which no job starts before its (modified) release date. These two problems are mirror images of each other; one problem can be transformed into the other by letting time run in reverse.

More precisely, an instance of $1 | r_j | C_{\max}$ can be transformed into an instance of $1 || L_{\max}$ by defining due dates $d_j = K - r_j$ for some integer $K$. (One may choose $K \geq \max_j r_j$ in order to obtain nonnegative due dates.) An optimal schedule for the former problem is found by reversing the job order in an optimal schedule for the latter. If we wish to solve $1 | prec, r_j | C_{\max}$, then we must also reverse the precedence constraints, i.e., make $J_j \rightarrow J_k$ in the $L_{\max}$ problem if and only if $J_k \rightarrow J_j$ in the $C_{\max}$ problem. It follows that the algorithm for $1 | prec | L_{\max}$ also solves $1 | prec, r_j | C_{\max}$, in $O(n \log n + |A|)$ time. In Section 3.5 we will make the symmetry between release dates and due dates more explicit by introducing the 'head-body-tail' formulation of $1 | r_j | L_{\max}$.

While $1 | prec | L_{\max}$ can be solved without knowledge of the processing times, this is definitely not true for the more general problem $1 | prec | f_{\max}$. Nevertheless, the latter problem can be solved in $O(n^2)$ time, by a generalization of the EDD rule that we shall call the *Least Cost Last rule*.

Let $N = \{1, 2, \ldots, n\}$ be the index set of all jobs, and let $L \subseteq N$ be the index set of jobs without successors. For any subset $S \subseteq N$, let $p(S) = \sum_{j \in S} p_j$, and let $f_{\max}^*(S)$ denote the cost of an optimal schedule for the subset of jobs indexed by $S$. We may assume that the machine completes processing by time $p(N)$. Since one of the jobs in $L$ must be scheduled last, we have

$$f_{\max}^*(N) \geq \min_{j \in L} f_j(p(N)).$$

Since omitting one of the jobs cannot increase the optimal cost, we also have

$$f_{\max}^*(N) \geq f_{\max}^*(N - \{j\}) \text{ for all } j \in N.$$

Now let $J_l$ with $l \in L$ be such that

$$f_l(p(N)) = \min_{j \in L} f_j(p(N)).$$

We have

$$f_{\max}^*(N) \geq \max\{f_l(p(N)), f_{\max}^*(N - \{l\})\}.$$

But the right-hand side of this inequality is precisely the cost of an optimal schedule subject to the condition that $J_l$ is processed last. It follows that there exists an optimal schedule in which $J_l$ is in the last position. Since $J_l$ is found in $O(n)$ time, repeated application of this Least Cost Last rule yields an optimal schedule in $O(n^2)$ time.

**Theorem 3.3.** *The Least Cost Last rule solves the problem* $1 \mid prec \mid f_{\max}$ *in* $O(n^2)$ *time.* □

Theorem 3.3 can also be proved by a straightforward interchange argument, but the method used here will be applied to the problem $1 \mid pmtn, prec, r_j \mid f_{\max}$ in Section 3.2.

**Exercises**

3.1. Define the *earliness* of $J_j$ as $E_j = d_j - C_j$. Use an interchange argument similar to the one in the proof of Theorem 3.1 to show that the maximum earliness $E_{\max}$ is minimized by scheduling the jobs $J_j$ in order of nonincreasing $p_j - d_j$.

3.2. Prove Corollary 3.2

3.3. Show that $1 \mid p_j \leq p \mid L_{\max}$ can be solved in $O(np)$ time by appropriate sorting techniques.

3.4. Show that $1 \mid p_j = p \mid L_{\max}$ can be solved in $O(n)$ time. (Hint: Find $d_{\min} = \min_j d_j$. For each interval $[d_{\min} + (k-1)p, d_{\min} + kp)$ $(k = 1, \ldots, n)$, count the number of jobs with due dates in the interval and record the largest due date in the interval. Use these values to compute the maximum lateness of an EDD schedule. Then find a schedule that meets this bound.)

3.5. Show that $1 \mid prec \mid L_{\max}$ can be solved without any knowledge of the $p_j$ values. Devise a two-job example to show that this is not the case for the maximum weighted lateness problem $1 \mid \mid wL_{\max}$, where each job $J_j$ has a weight $w_j$ and $wL_{\max} = \max_j w_j L_j$.

3.6. Prove Theorem 3.3 using an interchange argument.

3.7. Suppose that processing times are allowed to be negative. (This may seem less unnatural if one views $p_j$ as the amount of a resource that is either consumed, if $p_j > 0$, or produced, if $p_j < 0$.) Processing begins at time 0, and each completion time $C_j$ is the total processing time of the jobs scheduled up to (and including) $J_j$. Extend the Least Cost Last rule to solve this generalization of $1||f_{\max}$ in $O(n^2)$ time. (Hint: It is possible to partition an instance of the generalized problem into two instances of the ordinary kind.)

3.8. Prove that the precedence-constrained version of the problem posed in Exercise 3.7 is NP-hard.

## 3.2. Preemptive EDD and Least Cost Last rules

We wish to extend the EDD rule to deal with nonuniform release dates. In this section we will consider such an extension for the case in which preemption is permitted. A nonpreemptive extension of the EDD rule is introduced in the next section.

The *preemptive EDD rule* solves the problem $1|pmtn, r_j|L_{\max}$ by scheduling the jobs in time, with a decision point at each release date and at each job completion time. A job $J_j$ is said to be *available* at time $t$ if $r_j \leq t$ and it has not yet completed processing. At each decision point, from among all available jobs, choose to process a job with the earliest due date. If no jobs are available at a decision point, schedule idle time until the next release date.

Observe that the preemptive EDD rule creates preemptions only at release dates, but not at the first one. That is, if there are $k$ distinct release dates, the rule introduces at most $k-1$ preemptions. If all release dates are equal, then there are no preemptions and the rule generates an ordinary EDD schedule, as described in the previous section.

We shall prove that the preemptive EDD rule produces an optimal schedule. For $S \subseteq N$, let $r(S) = \min_{j \in S} r_j$, $p(S) = \sum_{j \in S} p_j$, and $d(S) = \max_{j \in S} d_j$.

**Lemma 3.4.** *For any instance of $1|pmtn, r_j|L_{\max}$ or $1|r_j|L_{\max}$, $L^*_{\max} \geq r(S) + p(S) - d(S)$ for each $S \subseteq N$.*

*Proof.* Consider an optimal schedule, and let $J_j$ be the last job in $S$ to finish. Since none of the jobs in $S$ can start processing earlier than $r(S)$, $C_j \geq r(S) + p(S)$. Furthermore, $d_j \leq d(S)$, and so $L^*_{\max} \geq L_j = C_j - d_j \geq r(S) + p(S) - d(S)$. □

**Theorem 3.5.** *The preemptive EDD rule solves the problem $1|pmtn, r_j|L_{\max}$, with $L^*_{\max} = \max_{S \subseteq N} r(S) + p(S) - d(S)$.*

*Proof.* We will show that the preemptive EDD rule produces a schedule that meets the lower bound of Lemma 3.4 for an appropriately chosen set $S$.

Consider the schedule produced by the preemptive EDD rule (cf. Figure 3.1). Let $J_c$ be a *critical* job, that is, $L_c = L_{\max}$. Let $t$ be the latest time such that each job $J_j$ processed in the interval $[t, C_c]$ has $r_j \geq t$, and let $S$ index the subset of jobs processed in this interval. The interval contains no idle time: if there were an idle period, then its end satisfies the criterion used to choose $t$ and is later than $t$; hence, $p(S) \geq C_c - t$. Further, $r_j \geq t$ for each $j \in S$ and, since the machine is never idle in $[t, C_c]$, some job starts at time $t$; hence, $r(S) = t$. Finally, we show that $d(S) = d_c$. Suppose that this is not true, and let $t'$ denote the latest time within $[t, C_c]$ at which some job $J_j$ with $d_j > d_c$ is processed. Hence, each $J_k$ processed in $[t', C_c]$ has $d_k \leq d_c < d_j$; since the preemptive EDD rule did not preempt $J_j$ to start $J_k$, we have $r_k \geq t'$. But this implies that $t'$ should have been selected instead of $t$, which is a contradiction. Therefore, $L_{\max} = L_c = C_c - d_c \leq r(S) + p(S) - d(S)$, which proves the theorem. □

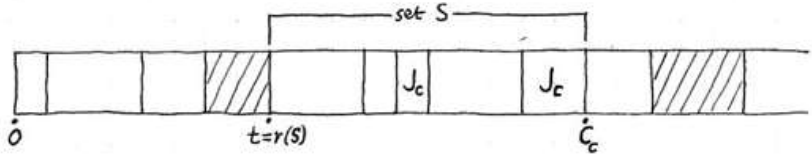

**Figure 3.1.**  Schedule obtained by the preemptive EDD rule.

The preemptive EDD rule is easily implemented to run in $O(n \log n)$ time with the use of two priority queues. Jobs are extracted from one queue in order of nondecreasing release dates. The second queue contains the jobs that are currently available. The available job with smallest due date is extracted from this queue, and reinserted upon preemption.

It is particularly important to note that the preemptive EDD rule is *on-line*, in the sense that it only considers available jobs for the next position in the schedule, without requiring any knowledge of their processing times and of the release dates of future jobs. This makes it an ideal rule for scheduling tasks with due dates on a real-time computer system.

After appropriate modification of both release dates and due dates, the preemptive EDD rule also solves the problem $1 \,|\, pmtn, prec, r_j \,|\, L_{\max}$. The condition that must be satisfied is that $r_j < r_k$ and $d_j < d_k$ whenever $J_j \rightarrow J_k$. In Section 3.1 we have already given a procedure for modifying the due dates. For the release dates, observe that, if $J_j \rightarrow J_k$, we may set

$$r_k := \max\{r_k, r_j + p_j\}$$

without changing the set of feasible schedules. We leave it to the reader to verify that $1 \,|\, pmtn, prec, r_j \,|\, L_{\max}$ can be solved in $O(n \log n + |A|)$ time, where $A$ is the arc set of the precedence digraph.

In some cases, the preemptive EDD rule produces a schedule without preemptions, which is therefore optimal for the nonpreemptive variant as well. This observa-tion applies to the cases of equal release dates ($1||L_{max}$), equal due dates ($1|r_j|L_{max}$), and unit processing times ($1|r_j, p_j = 1|L_{max}$); in the last case, our assumption that all release dates are integral is essential. The preemptive EDD rule solves a common generalization of the cases of equal release dates and equal due dates, which occurs when $d_j \le d_k$ whenever $r_j < r_k$; such release and due dates are said to be *similarly ordered* (see Exercise 3.9).

The preemptive EDD rule produces an optimal schedule with no *unforced idle time*; idle time is unforced if there is an available job. A significant consequence of this observation is the following.

**Theorem 3.6.** *For any instance of* $1|pmtn, prec, r_j|f_{max}$ *and* $1|pmtn, prec, r_j|\sum f_j$, *there exists an optimal schedule with no unforced idle time and with at most* $n - 1$ *preemptions.*

*Proof.* Given an optimal schedule in which each $J_j$ has a completion time $C_j^*$, create an instance of $1|pmtn, prec, r_j|L_{max}$ by setting $d_j = C_j^*$. These due dates satisfy the property that $d_j < d_k$ whenever $J_j \to J_k$. Modify the release dates so that they also conform to the precedence constraints, and apply the preemptive EDD rule. In the resulting schedule, each job $J_j$ completes no later than $C_j^*$. Hence, the new schedule is also optimal and the theorem is proved. $\square$

We shall now obtain a preemptive extension of the Least Cost Last rule that will enable us to solve $1|pmtn, r_j|f_{max}$ and even $1|pmtn, prec, r_j|f_{max}$ in $O(n^2)$ time.

Taking our cue from Theorem 3.6 which states that there is an optimal schedule with no unforced idle time, we first notice that we must work with a certain block structure. Consider running the following algorithm on an instance of $1|pmtn, r_j|f_{max}$: as long as jobs are available, choose one of them and schedule it to completion; otherwise, find the minimum release date of an unscheduled job, create an idle period until then, and continue from there. This algorithm partitions $N$ into *blocks*, where a block consists of a maximal subset of jobs processed continuously without idle time. Call this algorithm *Find Blocks*($N$). Recall that, for any subset $S \subseteq N$, $r(S) = \min_{j \in S} r_j$ and $p(S) = \sum_{j \in S} p_j$, and define $t(S) = r(S) + p(S)$. It follows that, if we find the blocks $B_1, \ldots, B_k$ in that order, then $B_i$ starts processing at time $r(B_i)$ and completes processing at time $t(B_i) < r(B_{i+1})$. Using this information, we obtain the following corollary of Theorem 3.6.

**Corollary 3.7.** *If* $B_1, \ldots, B_k$ *are the blocks of an instance of* $1|pmtn, r_j|f_{max}$, *then there exists an optimal schedule in which the jobs of* $B_i$ ($i = 1, \ldots, k$) *are scheduled from* $r(B_i)$ *to* $t(B_i)$ *without idle time.* $\square$

Corollary 3.7 implies that we can find an optimal schedule for each block, and then combine them to obtain an optimal schedule for the entire instance. Furthermore, the algorithm to solve $1|pmtn, r_j|f_{max}$ for a block can now follow a plan nearly identical to the one used for $1||f_{max}$. Let $f_{max}^*(B)$ denote the cost of an optimal schedule for the subset of jobs indexed by $B$. As above, we may assume that the machine completes processing block $B$ at time $t(B)$. Since some job in $B$ must finish last, we have

$$f_{\max}^*(B) \geq \min_{j \in B} f_j(t(B)).$$

Again, it is clear that

$$f_{\max}^*(B) \geq f_{\max}^*(B - \{j\}) \quad \text{for all } j \in B.$$

Now choose $J_l$ with $l \in B$ such that

$$f_l(t(B)) = \min_{j \in B} f_j(t(B)).$$

We have

$$f_{\max}^*(B) \geq \max\{f_l(t(B)), f_{\max}^*(B - \{l\})\}. \tag{3.2}$$

We will give a recursive algorithm that produces a schedule whose value $f_{\max}(B)$ meets this lower bound.

Algorithm  *Schedule Block*$(B)$
**if** $B = \emptyset$ **then** return the empty schedule;
$l := \operatorname{argmin} j \in B f_j(t(B)));$
**call** *Find Blocks*$(B - \{l\});$
schedule $J_l$ in the idle time generated between $r(B)$ and $t(B);$
**for** each block $B_i$ found **call** *Schedule Block*$(B_i).$

Note that, in the resulting schedule, job $J_l$ is processed only if no other job is available.

We shall prove by induction on the number of jobs in $|B|$ that *Schedule Block* constructs an optimal schedule. Clearly, this is true if $|B| = 0$. Otherwise, consider a *critical* job, i.e., one whose completion cost equals $f_{\max}(B)$. If $J_l$ is critical, then

$$f_{\max}(B) \leq f_l(t(B)),$$

since $J_l$ is certainly completed by $t(B)$. If one of the other jobs in $B$ is critical, then the correctness of the algorithm for smaller blocks, along with Corollary 3.7, implies that

$$f_{\max}(B) \leq f_{\max}^*(B - \{l\}).$$

It follows that the algorithm produces a schedule whose value matches the lower bound (3.2) on the optimal value. Hence, the schedule is optimal.

To analyze the running time of this algorithm, we must be a bit more careful about a few implementation details. Let $T(n)$ be the worst-case running time of *Schedule Block* on an input with $n$ jobs. We start by reindexing the jobs so that they are ordered by nondecreasing release dates. It then takes linear time to identify $J_l$, and also linear time to run *Find Blocks*, since the jobs are nicely ordered. In order to call *Schedule Block* recursively on each block found, we must make sure that the jobs in each block are numbered appropriately. However, we can separate the original sorted list of $n$ jobs into sorted lists for each block in linear time. Thus, if $n_1, \ldots, n_k$ are the sizes of the blocks found, then

$$T(n) \leq T(n_1) + \cdots + T(n_k) + O(n),$$

where $n_1 + \cdots + n_k = n - 1$, and $T(1) = O(1)$. This implies that $T(n) = O(n^2)$, as is easily verified by induction.

It is also important to analyze the number of preemptions generated by this algorithm. In the schedule constructed, $J_l$ can be preempted only at times $r(B')$ for the blocks $B'$ of $B - \{l\}$. This implies that the schedule contains at most $n - 1$ preemptions, and it is not hard to construct instances for which every optimal schedule has that many preemptions (see Exercise 3.10).

**Theorem 3.8.** *The problem* $1 \mid pmtn, r_j \mid f_{\max}$ *can be solved in* $O(n^2)$ *time in such a way that the optimal schedule has at most* $n - 1$ *preemptions.*

### Exercises

3.9. Show that the preemptive EDD rule solves $1 \mid r_j \mid L_{\max}$ in $O(n \log n)$ time in case the release and due dates are similarly ordered, i.e., $d_j \leq d_k$ whenever $r_j < r_k$.

3.10. Give a class of instances of $1 \mid pmtn, r_j \mid f_{\max}$ for which every optimal schedule has $n - 1$ preemptions.

3.11. Show how to extend Theorem 3.8 to apply to $1 \mid pmtn, prec, r_j \mid f_{\max}$. (Hint: After modifying the release dates in the usual way, the main difficulty is in identifying the correct job $J_l$, which is now restricted to be a job that has no successors within its block.)

3.12. Apply the algorithm of Theorem 3.8 to find an optimal schedule for the instance of $1 \mid pmtn, prec, r_j \mid f_{\max}$ given in Figure 3.2.

3.13. Show that the preemptive Least Cost Last rule solves $1 \mid prec, r_j, p_j = 1 \mid f_{\max}$ in $O(n^2)$ time.

3.14. Devise and prove a minmax theorem that generalizes Theorem 3.5 to $1 \mid pmtn, r_j \mid f_{\max}$.

### 3.3. A polynomial-time algorithm for jobs of equal length

We have seen that the preemptive EDD rule solves $1 \mid r_j \mid L_{\max}$ in $O(n \log n)$ time when all $r_j$ are equal, when all $d_j$ are equal, when all $p_j = 1$, and even when the $r_j$ and $d_j$ are similarly ordered. It is not easy to find additional special cases for which the preemptive EDD rule creates no preemptions.

We can modify the preemptive rule so that it is forced to produce a nonpreemptive schedule. With luck, some new special cases can be solved by the *nonpreemptive EDD rule*: schedule the jobs in time, with a decision point at the beginning of each block and at each job completion time. At each decision point, choose to process an available job with the earliest due date. If no jobs are available at a decision point, schedule idle time until the next release date.

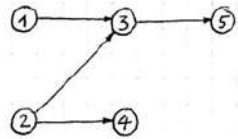

(a) Precedence constraints.

| j  | 1 | 2 | 3 | 4 | 5  |
|----|---|---|---|---|----|
| $r_j$ | 0 | 2 | 0 | 8 | 14 |
| $p_j$ | 4 | 2 | 4 | 2 | 4  |

(b) Release dates and processing times.

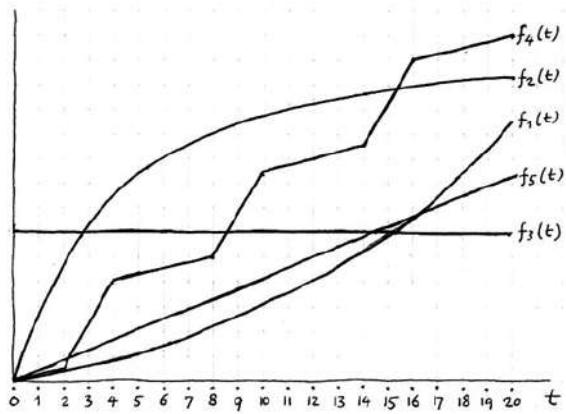

(c) Cost functions.

**Figure 3.2.**   Five-job instance of $1 \,|\, pmtn, prec, r_j \,|\, f_{\max}$ for Exercise 3.12

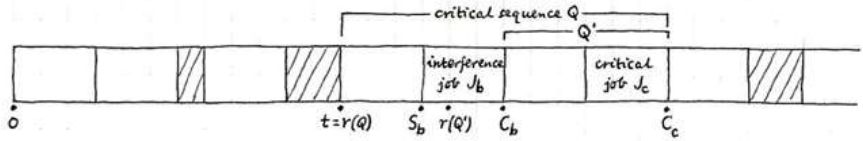

**Figure 3.3.** Schedule obtained by the nonpreemptive EDD rule.

Unfortunately, it is also not easy to find new cases for which the direct application of the nonpreemptive EDD rule can be guaranteed to yield an optimal schedule. However, the rule is invaluable as a component of more elaborate algorithms, for solving special cases, for obtaining near-optimal solutions, and for computing optimal schedules by enumeration. In this section, we shall develop an algorithm for the special case of equal processing times, $1|r_j, p_j = p|L_{max}$. This algorithm, unlike the simple EDD rule, is off-line, and unlike the special cases we have seen thus far, $1|r_j, p_j = p|L_{max}$ is not solvable by an on-line algorithm (see Exercise 3.15).

The reader should observe that the case in which all $p_j$ are equal to an arbitrary positive integer $p$ is very different from the case in which all $p_j = 1$. It is possible to rescale time so that $p = 1$. But then the release dates become rational numbers, contrary to the ground rule under which we asserted that the preemptive EDD rule solves $1|r_j, p_j = 1|L_{max}$.

There *are* instances of the general $1|r_j|L_{max}$ problem for which the solution delivered by the nonpreemptive EDD rule can be proved to be optimal. Recall that, for $S \subseteq N$, $r(S) = \min_{j \in S} r_j$, $p(S) = \sum_{j \in S} p_j$, and $d(S) = \max_{j \in S} d_j$. Consider a schedule delivered by the nonpreemptive EDD rule (cf. Figure 3.3). Let $J_c$ be a *critical* job, that is, $L_c = \max_j L_j$, and let $t$ be the earliest time such that the machine is not idle in the interval $[t, C_c]$. The sequence of jobs $Q$ processed in this interval is called a *critical sequence*. None of the jobs in $Q$ has a release date smaller than $t$; that is, $r(Q) = t$. Suppose that, in addition, none of the jobs in $Q$ has a due date larger than $d_c$; that is, $d(Q) = d_c$. In that case, the schedule must be optimal: its maximum lateness is equal to $r(Q) + p(Q) - d(Q)$, which, by Lemma 3.4, is a lower bound on the optimum.

This proof of optimality, however, is a happy turn of events, and many instances will not have the property that $d(Q) = d_c$. For those instances, there will be a job $J_b$ with $b \in Q$ and $d_b > d_c$. Any such job interferes with the optimality proof, and the one scheduled last in $Q$ is called an *interference* job. This notion will be useful in several aspects of algorithm design for $1|r_j|L_{max}$. We shall now apply it to derive a polynomial-time algorithm for $1|r_j, p_j = p|L_{max}$.

We will first consider the decision version of this problem and give a polynomial-time algorithm to decide whether, for a given integer $l$, there exists a schedule with value $L_{max} \leq l$. This is equivalent to the problem of deciding if every job can meet its due date, since a schedule satisfies $L_{max} \leq l$ if and only it satisfies $L_{max} \leq 0$ with respect to modified due dates $d_j + l$ $(j = 1, \ldots, n)$. Thus, we view the due dates as *deadlines* and call a schedule in which every job is on time a *feasible* schedule.

Suppose that the nonpreemptive EDD rule constructs an infeasible schedule. Find a critical job $J_c$ and its critical sequence $Q$. If there is no interference job in $Q$, then Lemma 3.4 implies that no feasible schedule exists. Suppose that there is an interference job $J_b$, and let $S_b$ be its starting time. Focus on that part of $Q$ that follows $J_b$; call this part $Q'$. The definition of an interference job implies that $d(Q') = d_c < d_b$. In spite of this, the nonpreemptive EDD rule selected $J_b$ for processing at time $S_b$. It follows that $r(Q') > S_b$. Consider any schedule in which a job starts in the interval $[S_b, r(Q'))$. Clearly, this cannot be a job in $Q'$. Since all processing times are equal, the jobs in $Q'$ are delayed at least as much as in the schedule produced by the nonpreemptive EDD rule. Hence, some job in $Q'$ must be late, and the schedule is infeasible. We conclude that no feasible schedule has any job starting in $[S_b, r(Q'))$, and therefore call that interval a *forbidden region*.

The main idea of the algorithm is to repeat this approach. We apply the nonpreemptive EDD rule, never starting jobs in any forbidden region that has been identified before. There are three possible outcomes: either a feasible schedule is found, or the instance is proved infeasible, or a new forbidden region is identified. In the first two cases, we are done; in the last case, we repeat. The running time analysis will rely on the fact that the number of forbidden regions is limited.

Before stating the algorithm in detail, we have to modify our definition of a critical sequence, since idle time that is caused by a forbidden region does not count. Given a critical job $J_c$, its critical sequence $Q$ consists of the jobs processed during the maximal interval $[t, C_c]$ such that all idle time within the interval occurs within forbidden regions. The decision algorithm is now as follows.

$\mathcal{F} := \emptyset$;                                     * initialize the set of forbidden regions *
**until** a feasible schedule is found                         * produce the next schedule *
$\quad N := \{1, \ldots, n\}$;                                     * $N$ is the index set of unscheduled jobs *
$\quad t := 0$;                                     * $t$ is the time at which the next job may start *
$\quad$ **while** $N \neq \emptyset$                                     * some job is not yet scheduled *
$\quad\quad$ **while** there exists $F \in \mathcal{F}$ such that $t \in F = [t_1, t_2)$
$\quad\quad\quad t := t_2$;                                     * advance $t$ beyond forbidden regions *
$\quad\quad A := \{j \in N | r_j \leq t\}$;                                     * $A$ is the set of available jobs at $t$ *
$\quad\quad$ select $j \in A$ for which $d_j$ is minimum;                                     * choose next job *
$\quad\quad S_j := t$;  $C_j := t + p$;                                     * schedule $J_j$ *
$\quad\quad N := N - \{j\}$;                                     * mark $J_j$ as scheduled *
$\quad\quad t := t + p$;                                     * advance $t$ to $C_j$ *
$\quad$ **if** some job is late
$\quad$ **then**                                     * prove infeasibility or find new forbidden region *
$\quad\quad$ find a critical job $J_c$;
$\quad\quad$ **if** there is no interference job  **then** output 'infeasible' and  **halt**;
$\quad\quad J_b := $ the interference job in the critical sequence ending with $J_c$;

$$Q' := \{j | S_b < S_j \le S_c\};$$
$$\mathcal{F} := \mathcal{F} \cup \{[S_b, r(Q'))\}.$$

The reader may wonder why this algorithm must stop. Observe that the upper endpoint of a forbidden region is always a release date. Hence, each job has a starting time of the form $r_j + sp$, where $s$ is an integer between 0 and $n-1$. Since the lower endpoints of the forbidden regions are starting times, there are $O(n^2)$ possible lower endpoints, and after that many iterations we cannot possibly generate a new forbidden region. The algorithm must terminate within $O(n^2)$ iterations.

In order to prove that the algorithm is correct, we need a lemma. Suppose that the jobs have identical release dates and identical processing times, and that we just wish to minimize schedule length; however, no job is allowed to start in any of a number of given forbidden regions. This problem can be solved in a simple way.

**Lemma 3.9** [ Simple algorithm ]. *Given a set $\mathcal{F}$ of forbidden regions, the problem $1|r_j = r, p_j = p|C_{\max}$ is solved by repeatedly choosing a job and starting it at the earliest possible time that is not contained in any interval in $\mathcal{F}$.*

*Proof.* Suppose that the lemma is incorrect. Choose a counterexample with the smallest number of jobs. Since the jobs are identical, we may assume that they are ordered $J_1, \ldots, J_n$ in both the schedule produced by the simple algorithm and the optimal schedule. Let $C_j$ and $C_j^*$, respectively, denote the completion time of $J_j$ in each of these schedules. By the choice of counterexample, $C_j \le C_j^*$ for $j = 1, \ldots, n-1$. In particular, since $C_{n-1} \le C_{n-1}^*$, any time at which the optimal schedule may start $J_n$ is also available for the simple algorithm. Hence, $C_n \le C_n^*$, which is a contradiction. $\square$

We claim that the decision algorithm has the following invariant property: *for any feasible schedule, no job starts at a time contained in a forbidden region $F \in \mathcal{F}$*. We shall first use this property to show that the algorithm is correct, and then establish its validity.

Assuming that the invariant property holds, it will be sufficient for us to show that no feasible schedule exists whenever the algorithm concludes that this is so. Suppose the algorithm halts without having produced a feasible schedule. In that case, the final schedule found has a critical sequence $Q$ with no interference job. By definition, the jobs in $Q$ are the only ones scheduled between $r(Q)$ and their maximum completion time, $\max_{j \in Q} C_j > d(Q)$. We shall invoke Lemma 3.9 to show that this set of jobs cannot be completed earlier than in the schedule found. Observe that the schedule of the critical sequence contains no idle time outside of the forbidden regions. Hence, this schedule could have been produced by the simple algorithm on a modified instance in which each release date is set equal to $r(Q)$. By Lemma 3.9, the maximum completion time of this schedule is optimal. However, it exceeds $d(Q)$. It follows that, in any schedule, the job in $Q$ that finishes last must be late.

We still must verify the invariant property. It certainly holds initially, when there are no forbidden regions, and we have already seen that it holds when the first forbidden region has been found. Assume that the invariant property holds for the first $k$ forbidden regions found. Suppose that, in iteration $k + 1$, an infeasible schedule is found, with a corresponding forbidden region $[S_b, r(Q'))$. Extract the schedule for the jobs in $Q' \cup \{b\}$. There is no idle time outside of a forbidden region in this schedule from $S_b$ until all jobs in $Q' \cup \{b\}$ are completed. Thus, if the release dates of these jobs were to be modified to $S_b$, then this schedule is a possible output of the simple algorithm. Its completion time exceeds $d(Q')$ and, by Lemma 3.9, no earlier maximum completion time is possible for this subinstance.

Suppose that there exists a feasible schedule with $S_j \in [S_b, r(Q'))$ for some job $J_j$. Of course, $j \notin Q'$. From this feasible schedule, extract the schedule for the jobs in $Q' \cup \{j\}$. By the invariant property, no job starts during any of the $k$ forbidden regions already identified. The schedule remains feasible if we change each release date to $S_b$. This makes $J_j$ identical to $J_b$. But it now follows that the completion time of the schedule must exceed $d(Q')$, which contradicts its feasibility. This completes the correctness proof of the algorithm.

As for the running time of the algorithm, we already know that there are $O(n^2)$ iterations. We still must figure out how to implement each iteration. The jobs are initially stored in a priority queue, ordered by their release dates. Whenever the available set $A$ is updated, we repeatedly test whether the minimum element in the queue is no more than the current time parameter $t$ and, if so, extract the minimum element from the queue. The available set is stored in another priority queue, ordered by due date. With these data structures, each iteration takes $O(n \log n)$ time. Hence, the decision algorithm runs in $O(n^3 \log n)$ time.

We now have solved the decision problem. In order to solve the optimization problem, we can use a naive bisection search to find $L_{\max}^*$ but there is a better way. Observe that any lateness is the difference between a starting time and a due date. Since there are only $O(n^2)$ possible starting times and $O(n)$ due dates, there are $O(n^3)$ possible values of $L_{\max}$. Even using brute force, we can calculate all of these, sort them, and then do a bisection search to find $L_{\max}^*$. As a result, $O(\log n)$ iterations suffice to find the optimal solution.

**Theorem 3.10.** *The problem $1|r_j, p_j = p|L_{\max}$ can be solved in $O(n^3 \log^2 n)$ time by an iterated version of the nonpreemptive EDD rule.*

We should make a final comment about implementing this algorithm. While we have stipulated that each forbidden region is derived from a critical job, the proof uses nothing more than that the job is late. Hence, whenever the algorithm schedules a job, it should check if that job is late. If it is, the current iteration ends, either by identifying an interference job relative to that job, or by concluding that the instance is infeasible.

The algorithm is easily extended to solve $1|prec, r_j, p_j = p|L_{\max}$ as well. As we have indicated in the previous sections, the release and due dates of a problem instance can be modified in such a way that $r_j < r_k$ and $d_j < d_k$ whenever $J_j \rightarrow J_k$, without changing the set of feasible schedules and their objective values. The nonpreemptive EDD rule, when run on the modified instance, will automatically respect the precedence constraints. Hence, any schedule generated by the algorithm

$n = 6$, $p = 3$

| $j$     | 1  | 2  | 3 | 4  | 5  | 6  |
|---------|----|----|---|----|----|----|
| $r_j$   | 0  | 1  | 2 | 6  | 13 | 15 |
| $\bar{d}_j$ | 12 | 22 | 5 | 10 | 20 | 19 |

**Figure 3.4.**  Instance of $1|r_j, p_j = p, \bar{d}_j|-$ for Exercise 3.16.

will satisfy the precedence constraints.

**Exercises**

3.15.  Show that there is no on-line algorithm to solve $1|r_j, p_j = p|L_{\max}$.

3.16.  Apply the algorithm of Theorem 3.10 to find a feasible schedule for the instance of $1|r_j, p_j = p, \bar{d}_j|-$ given in Figure 3.4.  What happens if $\bar{d}_6$ is changed to 18?

3.17.  Consider the special case of $1|r_j|L_{\max}$ in which $r_j + p_j \geq r_k$ for each pair $(J_j, J_k)$.

(a) Show that there exists an optimal schedule in which at least $n-1$ of the jobs are in EDD order.

(b) Describe how to find an optimal schedule in $O(n^2)$ time.

(c) Describe how to find an optimal schedule in $O(n \log n)$ time.

(d) Suppose that, instead of $r_j + p_j \geq r_k$, we have $d_j - p_j \leq d_k$ for each pair $(J_j, J_k)$. What can be said about this special case?

## 3.4.  NP-hardness results

We will now show that one cannot expect to obtain a polynomial-time algorithm for $1|r_j|L_{\max}$ in its full generality. More precisely, $1|r_j|L_{\max}$ is strongly NP-hard.

**Theorem 3.11.**  $1|r_j|L_{\max}$ *is NP-hard in the strong sense.*

*Proof.*  We shall show that the problem is NP-hard in the ordinary sense, by proving that the partition problem reduces to the decision version of $1|r_j|L_{\max}$. To establish strong NP-hardness, one can give an analogous reduction from 3-partition, but this is left to the reader (see Exercise 3.18).

An instance of partition consists of integers $a_1, \ldots, a_t, b$ with $\sum_{j=1}^{t} a_j = 2b$. The corresponding instance of $1|r_j|L_{\max}$ has $n = t + 1$ jobs, defined by

$$r_j = 0, \ \ p_j = a_j, \ \ d_j = 2b + 1, \ \ j = 1, \ldots, t;$$

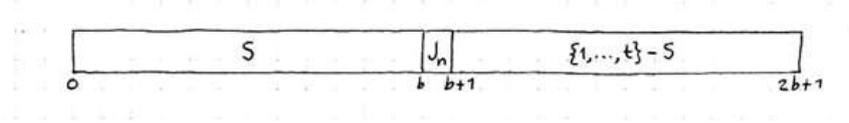

**Figure 3.5.** The reduction of partition to $1|r_j|L_{\max}$.

$$r_n = b, \quad p_n = 1, \quad d_n = b+1.$$

We claim that there exists a schedule of value $L_{\max} \leq 0$ if and only if the instance of partition is a yes-instance.

Suppose we have a yes-instance of partition; that is, there exists a subset $S \subset \{1,\ldots,t\}$ with $\sum_{j \in S} a_j = b$. We then schedule the jobs $J_j$ with $j \in S$ in the interval $[0,b]$, job $J_n$ in $[b, b+1]$, and the remaining jobs in $[b+1, 2b+1]$ (cf. Figure 3.5.) Note that no job starts before its release date, and that no job is late.

Conversely, consider a feasible schedule of value $L_{\max} \leq 0$. In such a schedule, $J_n$ must start at time $b$ and there cannot be any idle time. Hence, the jobs that precede $J_n$ have a total processing time of $b$. Since $p_j = a_j$ $(j = 1,\ldots,n)$, we conclude that we have a yes-instance of partition. □

**Exercises**

3.18. Prove that $1|r_j|L_{\max}$ is NP-hard in the strong sense.

3.19. Show that $1||max\{E_{\max}, L_{\max}\}$ is strongly NP-hard (cf. Exercise 3.1).

### 3.5. Approximation algorithms

Since $1|r_j|L_{\max}$ is NP-hard, it is natural to ask whether there are polynomial-time algorithms that find solutions guaranteed to be close to the optimum. Before answering this question, we must make sure that it is well posed. As discussed in Chapter 3, we are primarily concerned with the performance of an approximation algorithm in terms of its worst-case relative error. We would like to give an algorithm A such that, for example, $L_{\max}(A) \leq 2L_{\max}^*$ for every instance. This is a reasonable goal when $L_{\max}^*$ is positive. However, when $L_{\max}^* = 0$, we must find the optimum, and when $L_{\max}^* < 0$, we must find a solution that is better than optimal!

This difficulty is not avoided if we switch to the $T_{\max}$ objective. The maximum tardiness of a schedule is always nonnegative, but it can be zero. This has the interesting consequence that, for the problem $1|r_j|T_{\max}$, we cannot hope to find a polynomial-time algorithm with any finite performance bound.

**Theorem 3.12.** *If there exist a polynomial-time algorithm* A *and a constant $c > 1$ such that, for any instance of $1|r_j|T_{\max}$, $T_{\max}(A) < cT_{\max}^*$, then $P = NP$.* □

The proof follows directly from the reduction given in the proof of Theorem 3.11 (see Exercise 3.20).

For $L_{\max}$ problems, however, the whole notion of a $c$-approximation algorithm makes no sense. We can get around this difficulty by realizing that, unlike release dates and processing times, due dates have never been assumed to be nonnegative. By requiring that all due dates are *negative*, we ensure that each lateness is positive. And there do exist good approximation algorithms and, in fact, a polynomial approximation scheme for the problem $1\,|\,prec, r_j, d_j < 0\,|\,L_{\max}$.

A few words about the significance of this problem are in order. Any instance of $1\,|\,prec, r_j\,|\,L_{\max}$ can be transformed into an instance with negative due dates by subtracting a constant $K$ from each $d_j$, where $K > \max_j d_j$. Both instances are equivalent to the extent that a schedule is optimal for one if and only if it is optimal for the other. Still, a good approximation for the transformed instance may be a very poor approximation for the original, since the transformation adds $K$ to the maximum lateness of each schedule.

However, the problem $1\,|\,prec, r_j, d_j < 0\,|\,L_{\max}$ is not something we concocted for the sole purpose of being able to obtain good performance guarantees for approximation algorithms. The problem is equivalent to the *head-body-tail* problem, in which each job $J_j$ is to be processed on the machine during a period of length $p_j$, which must be preceded by a *release time* of length $r_j$ and followed by a *delivery time* of length $q_j$. A job $J_j$ can only start at time $S_j \geq r_j$; it then finishes at time $C_j = S_j + p_j$ and is delivered at time $L_j = C_j + q_j$. If $J_j \to J_k$, then it is required, as usual, that $C_j \leq S_k$. The objective is to minimize the maximum job delivery time. Obviously, the two problems correspond to each other via $d_j = -q_j$. A schedule is feasible for one problem if and only if it feasible for the other, and it has the same criterion values in both formulations.

We shall encounter the head-body-tail problem in Chapters 13 and 14, where it arises quite naturally as a relaxation of flow shop and job shop scheduling problems. In that context, it is convenient to view it as a three-machine problem, with each job $J_j$ having to be processed by $M_1, M_2, M_3$ in that order, for $r_j, p_j, q_j$ time units, respectively. Contrary to our usual assumptions, $M_1$ and $M_3$ are *non-bottleneck* machines, in that they can process any number of jobs simultaneously; $M_2$ can handle at most one job at a time and corresponds to the original single machine. Precedence constraints have to be respected on $M_2$ only. The objective is to minimize the maximum job completion time on $M_3$.

An appealing property of the head-body-tail problem is its symmetry. An instance defined by $n$ triples $(r_j, p_j, q_j)$ and a precedence relation $\to$ is equivalent to an *inverse* instance defined by $n$ triples $(q_j, p_j, r_j)$ and a precedence relation $\to'$, where $J_j \to' J_k$ if and only if $J_k \to J_j$. A feasible schedule for one problem instance is converted into a feasible schedule for its inverse by reversing the order in which the jobs are processed. The equivalence of $1\,|\,prec\,|\,L_{\max}$ and $1\,|\,prec, r_j\,|\,C_{\max}$ is a manifestation of this symmetry.

We shall use the language and notation of the head-body-tail problem in the rest of this chapter. Since $q_j = -d_j$ for each $J_j$, each variant of the EDD rule for $1|\,prec, r_j, d_j < 0|L_{\max}$ can be translated into its analogue for the head-body-tail problem by selecting the jobs in order of nonincreasing delivery times. Similarly, if $q(S) = \min_{j \in S} q_j$ for $S \subseteq N$, then we can restate Lemma 3.4 in the following way.

**Corollary 3.13.** *For any instance of* $1|r_j, d_j < 0|L_{\max}$, $L_{\max}^* \geq r(S) + p(S) + q(S)$ *for any* $S \subseteq N$. $\square$

### The nonpreemptive EDD rule (NEDD)

In Section 3.3 we have studied the structure of schedules produced by this rule. We now use that information to investigate its performance for instances with arbitrary processing requirements. We first derive two data-dependent bounds.

**Lemma 3.14.** *For any instance of* $1|r_j|L_{\max}$, *let* $J_c$ *be a critical job and let* $J_b$ *be an interference job in a critical sequence in a schedule produced by the nonpreemptive EDD rule; we then have*
*(i)* $L_{\max}(\text{NEDD}) - L_{\max}^* < q_c$;
*(ii)* $L_{\max}(\text{NEDD}) - L_{\max}^* < p_b$.

*Proof.* Consider a schedule produced by the nonpreemptive EDD rule. Let $Q$ be the critical sequence corresponding to $J_c$. By the definition of $Q$, the jobs in $Q$ start processing at $r(Q)$ and are processed without idle time for $p(Q)$ time units. Hence, $J_c$ completes processing at $r(Q) + p(Q)$, and

$$L_{\max}(\text{NEDD}) = r(Q) + p(Q) + q_c.$$

Corollary 3.13 implies

$$L_{\max}^* \geq r(Q) + p(Q) + q(Q) > r(Q) + p(Q).$$

By combining these inequalities, we obtain part (i) of the lemma.

Let $J_b$ be the interference job in $Q$, and let $Q'$ be the part of $Q$ that follows $J_b$. By the definition of $J_b$, $q_b < q_c = q(Q')$. Since $J_b$ was selected to start processing at time $S_b$, we have $S_b < r(Q')$, and

$$L_{\max}(\text{NEDD}) = S_b + p_b + p(Q') + q_c < r(Q') + p_b + p(Q') + q(Q').$$

Corollary 3.13 implies

$$L_{\max}^* \geq r(Q') + p(Q') + q(Q').$$

These inequalities together prove part (ii) of the lemma. $\square$

Since $L_{\max}^* \geq \max_j q_j$, part (i) of Lemma 3.14 implies that the nonpreemptive EDD rule is a 2-approximation algorithm for $1|r_j, d_j < 0|L_{\max}$. Alternatively, since $L_{\max}^* \geq p(N)$, part (ii) also implies this result.

In the precedence-constrained case, we first modify the data so that $r_j < r_k$ and $q_j > q_k$ whenever $J_j \to J_k$; each feasible schedule for the modified instance is feasible for the original one and has the same value. When the nonpreemptive EDD rule is applied to the modified instance without precedence constraints, the resulting schedule will still respect these constraints, and has a value strictly less than twice the optimum for the modified instance without precedence constraints. Clearly, this feasible solution for the original instance of $1 \mid prec, r_j, d_j < 0 \mid L_{\max}$ is also within a factor of 2 of the optimum for the more constrained problem.

**Theorem 3.15.** *For any instance of* $1 \mid prec, r_j, d_j < 0 \mid L_{\max}$, $L_{\max}(\text{NEDD})/L_{\max}^* < 2$. □

It is not hard to give a family of two-job examples that show that the bounds of Lemma 3.14 and Theorem 3.15 are tight (see Exercise 3.21).

**The iterated nonpreemptive EDD rule (INEDD)**

In Section 3.3, an iterated version of the nonpreemptive EDD rule did succeed where just the rule itself was not sufficient. We will adopt a similar strategy here.

When applied to an instance of $1 \mid prec, r_j, d_j < 0 \mid L_{\max}$, we view the nonpreemptive EDD rule as the procedure that first preprocesses the data to ensure that $r_j < r_k$ and $q_j > q_k$ whenever $J_j \to J_k$, and then builds a feasible schedule using the proper rule.

The iterated nonpreemptive EDD rule applies this nonpreemptive EDD rule to a series of at most $n$ instances, starting with the original instance. If the schedule obtained has a critical sequence with no interference job, then the algorithm terminates. Otherwise, there is a critical sequence $Q$, ending with the critical job $J_c$ and containing the interference job $J_b$. Since the small delivery time $q_b$ of $J_b$ interferes with the proof of optimality, we force $J_b$ to be processed after $J_c$ by increasing its release time $r_b$ to $r_c$ and reapplying the nonpreemptive EDD rule. The algorithm continues in this way, until either no interference job exists or $n$ schedules have been generated. From among the schedules generated, we select one with minimum $L_{\max}$ value. We claim that this is a 3/2-approximation algorithm.

Note that any schedule found is feasible for the original data. The original release dates are respected, since release dates are only increased throughout the algorithm. The precedence constraints are respected, since the data are modified in each iteration.

The following two lemmas deal with complementary cases.

**Lemma 3.16.** *For any instance of* $1 \mid prec, r_j, d_j < 0 \mid L_{\max}$, *if there does* not *exist an optimal schedule that is feasible with respect to the modified data used in the final iteration of the iterated nonpreemptive EDD rule, then* $L_{\max}(\text{INEDD})/L_{\max}^* < 3/2$.

*Proof.* Consider an optimal schedule $\sigma^*$ for the original instance. Focus on the last iteration in which $\sigma^*$ is still feasible with respect to the modified data. (These are the data *after* any preprocessing by the nonpreemptive EDD rule in that iteration.) Let $L_{\max}^*$ denote the optimum value for both the original instance and the modified instance considered in this iteration. In the schedule obtained in this iteration, let $J_c$ be a critical job, $Q$ its critical sequence, and $J_b$ the interference job, and let $L_{max}$ denote its value. We have

$$L_{max} = r(Q) + p(Q) + q_c \leq r_b + p_b + p(Q - \{b\}) + q_c.$$

As a result of this iteration, $r_b$ is increased to $r_c$, and $\sigma^*$ is no longer feasible. Thus, $J_b$ must precede $J_c$ in $\sigma^*$, so that

$$L_{max}^* \geq r_b + p_b + q_c.$$

It follows that

$$L_{max} - L_{max}^* \leq p(Q - \{b\}).$$

Lemma 3.14(ii) implies that

$$L_{max} - L_{max}^* < p_b.$$

Since $L_{max}^* \geq p(N)$, we have

$$\frac{L_{max}}{L_{max}^*} < 1 + \frac{\min\{p_b, p(Q - \{b\})\}}{L_{max}^*} \leq 1 + \frac{p(N)/2}{p}(N) = \frac{3}{2}. \quad \square$$

**Lemma 3.17.** *For any instance of* $1 \mid prec, r_j, d_j < 0 \mid L_{max}$, *if there* does *exist an optimal schedule that is feasible with respect to the modified data used in the final iteration of the iterated nonpreemptive EDD rule, then* $L_{max}(\text{INEDD})/L_{max}^* < 3/2$.

*Proof.* The algorithm terminates for one of two reasons: either there is no interference job, or $n$ schedules have been generated.

In the former case, the algorithm produces an optimal schedule for the final modified instance. This schedule must also be optimal for the original instance, since both instances have the same optimum value.

In the latter case, we will show that one of the schedules produced must meet the claimed bound. Observe that, in spite of the data modifications, the optimum value remains unchanged throughout the algorithm. Lemma 3.4 (ii) now implies that, in each iteration, $L_{max} - L_{max}^* < p_b$, where $J_b$ is the interference job. The proof would be complete if we could show that, in some iteration, $p_b \leq p(N)/2$, since $p(N) \leq L_{max}^*$. However, this is true for all but at most one job! Could the same job be the interference job for all $n$ iterations? No: in each iteration, another job (the critical one) is forced to precede it, and so after $n - 1$ iterations it would be the last job, which cannot be an interference job. $\square$

We combine Lemmas 3.16 and 3.17 to obtain the following theorem.

**Theorem 3.18.** *For any instance of* $1 \mid prec, r_j, d_j < 0 \mid L_{max}$, $L_{max}(\text{INEDD})/L_{max}^* < 3/2$.

It is not hard to show that this bound is tight (see Exercise 3.22).

**A polynomial approximation scheme**

The crucial fact in the proof of Theorem 3.18 is that there is at most one job with processing time greater than $p(N)/2$. Can we obtain better performance guarantees by using the more general observation that there are at most $k-1$ jobs with a processing time greater than $p(N)/k$? Yes, this is possible indeed, at least if we are willing to restrict attention to the problem without precedence constraints.

Let $k$ be a fixed positive integer. Call a job $J_l$ *long* if $p_l > p(N)/k$. Suppose that, somehow, we know the starting time $S_l^*$ of each long job $J_l$ in an optimal schedule $\sigma^*$, as well as the value $L_{max}^*$. We can use this information to produce a near-optimal schedule. Modify the release date and delivery time of each long job $J_l$ as follows:

$$r_l := S_l^*, \quad q_l := L_{max}^* - (S_l^* + p_l).$$

The modified instance has the same optimal value $L_{max}^*$ as the original instance, since $\sigma^*$ is still feasible and its value is unchanged. Also, for each long job $J_l$ we now have that $q_l \geq q_j$ for all of its successors $J_j$ in $\sigma^*$.

Now apply the nonpreemptive EDD rule to the modified instance to obtain a schedule $\sigma$. We claim that $\sigma$ has a value $L_{max} < (1 + 1/k)L_{max}^*$. Let $J_c$ be a critical job in $\sigma$, and let $Q$ be its critical sequence. If $Q$ does not have an interference job, then $\sigma$ is optimal. Otherwise, we will show that the interference job $J_b$ is not a long job; hence, $p_b \leq p(N)/k \leq L_{max}^*/k$, and Lemma 3.14(ii) implies the claimed performance bound. Suppose, for sake of contradiction, that $J_b$ is a long job. We have $r_b < r_c$ and $q_b < q_c$. The first inequality in combination with $S_b^* = r_b$ implies that $J_b$ precedes $J_c$ in $\sigma^*$. The second inequality in combination with the above observation regarding the successors of long jobs in $\sigma^*$ implies that $J_c$ precedes $J_b$ in $\sigma^*$. This is a contradiction.

Of course, we have no way of knowing when the long jobs start in an optimal schedule. We avoid this difficulty as follows. Suppose that, rather than the starting times of the long jobs in $\sigma^*$, we just know the positions at which they occur in $\sigma$. This knowledge is sufficient for us to reconstruct $\sigma$. All we need to do is to apply the nonpreemptive EDD rule to the $n$ jobs, subject to the condition that, when a long job occupies the next position in $\sigma$, then that job is scheduled next.

Once again, we have no way of knowing the positions of the long jobs in $\sigma$. However, there are at most $k-1$ long jobs, and hence there are $O(n^{k-1})$ ways to assign the long jobs to positions in a sequence of $n$ jobs. Our algorithm $A_k$ tests each of these possibilities and chooses the best schedule. The resulting schedule is guaranteed to be at least as good as $\sigma$. Since each application of the nonpreemptive EDD rule takes $O(n \log n)$ time, $A_k$ runs in $O(n^k \log n)$ time, which is polynomial for any positive constant $k$.

**Theorem 3.19.** *The family of algorithms $\{A_k\}$ is a polynomial approximation scheme for $1|r_j, d_j < 0|L_{max}$.* $\square$

This is the best result that we can reasonably expect to obtain for a strongly NP-hard problem, since the existence of a *fully* polynomial approximation scheme would imply that P=NP (cf. Chapter 2). It is not clear how the scheme should be extended to handle precedence constraints. A polynomial approximation scheme for $1|prec, r_j, d_j < 0|L_{\max}$ does exist, but is beyond the scope of this book.

**Exercises**

3.20. Prove Theorem 3.12. (Hint: How does algorithm A perform on the type of instance constructed in the proof of Theorem 3.11?)

3.21. Give a family of two-job instances showing that the performance bounds of the nonpreemptive EDD rule given in Lemma 3.14 and Theorem 3.15 are tight.

3.22. Prove that the performance bound of the iterated nonpreemptive EDD rule given in Theorem 3.18 is tight.

3.23. Give a tight performance analysis of the following algorithm for $1|r_j, d_j < 0|L_{\max}$: Run the nonpreemptive EDD rule on both the original instance and its inverse, and choose the better schedule.

3.24. Show that the following procedure is a 3/2-approximation algorithm for $1|r_j, d_j < 0$: Run the nonpreemptive EDD rule. If there is no interference job $J_b$, output this schedule. Otherwise, let $A = \{j : r_j \leq q_j, j \neq b\}$, $B = \{j : r_j > q_j, j \neq b\}$, and construct a second schedule by first scheduling the jobs indexed by $A$ in order of nondecreasing release dates, then job $J_b$, and finally the jobs indexed by $B$ in order of nonincreasing delivery times; output the better schedule.

## 3.6. Enumerative methods

Although $1|prec, r_j|L_{\max}$ is strongly NP-hard, computational experience has shown that it is not such a hard problem. There exist clever enumerative algorithms that perform remarkably well. In fact, this success has motivated the use of these algorithms for the computation of lower bounds for flow shop and job shop problems, which appear to pose greater computational challenges (see Chapters 13 and 14).

We will present two branch-and-bound algorithms to solve $1|r_j|L_{\max}$. Once again, it will be convenient to view the problem in its head-body-tail formulation.

**First branch-and-bound algorithm**

This is a relatively simple method. The *branching rule* generates all feasible schedules by making all possible choices for the first position in the schedule; for each of these choices it considers all possible choices for the second position, and so on. It is possible to exclude certain choices as being dominated by others. For example, it would be unfortunate to select as the next job one whose release date is so large that another job can be scheduled prior to it. More precisely, let $S$ denote the index set of jobs assigned to the first $l - 1$ positions, and let $t$ be the completion time of the last job in $S$; for the $l$th position, we need consider a job $J_k$ only if

$$r_k < \min_{j \notin S}\{\max\{t, r_j\} + p_j\}.$$

If this inequality does not hold, then the job minimizing the right-hand side could be made to precede $J_k$ without delaying the processing of $J_k$. Thus, we are sure that there exists an optimal schedule that satisfies this restriction.

We next consider the way in which a *lower bound* is computed for each node in the search tree. An attractive possibility is to schedule the remaining jobs from time $t$ onwards while allowing preemption. The preemptive EDD solves this problem in $O(n \log n)$ time (cf. Section 3.2).

Finally, we must specify a *search strategy*, which selects the next node of the tree for further exploration. A common rule is to select a node with *minimum lower bound* value. Whereas this rule helps to limit the number of nodes examined, the overhead in implementing the approach may overwhelm its advantages. A simple alternative is to do a *depth-first search* of the tree: the next node is a child of the current node (perhaps the one with minimum lower bound value); when all children of a node have been examined, the path towards the root is retraced until a node with an unexplored child is found.

A nice aspect of this approach is that it can be applied to solve other NP-hard $1|r_j|f_{\max}$ problems in an analogous way.

**Second branch-and-bound algorithm**

The second method makes a more extensive use of the mathematical structure that was developed in this chapter. The main idea is that each node in the search tree will correspond to a restricted instance of the problem, on which we will run the nonpreemptive EDD rule. The restrictions imposed on the instance are nothing more than precedence constraints between certain pairs of jobs. These constraints will be incorporated by modifying the release dates and the delivery times, as we have done throughout this chapter.

Consider a particular node in the tree and apply the nonpreemptive EDD rule to the corresponding instance. Suppose that the schedule obtained has a critical sequence with no interference job. This means that the schedule is optimal for the modified data or, in other words, optimal subject to the precedence constraints specified in that node.

On the other hand, suppose that there is an interference job $J_b$. Let $Q'$ be the index set of the jobs in the critical sequence that follow $J_b$, and let $L_{\max}$ be the value of the schedule. We know that

$$L_{\max} = S_b + p_b + p(Q') + q(Q') < r(Q') + p_b + p(Q') + q(Q').$$

Consider another schedule in which some job in $Q'$ precedes $J_b$ and another in $Q'$ follows it. The proof of Lemma 3.4 implies that the value of this schedule must be at least $r(Q') + p_b + p(Q') + q(Q')$, and so it is worse than the schedule just obtained. Hence, we may further restrict attention to those schedules in which $J_b$ either precedes all of the jobs in $Q'$ or follows all of them. This gives us our *branching rule*. Each node will have two children, each corresponding to one of these two additional constraints. Since we use the nonpreemptive EDD rule, we can enforce the first

constraint by setting

$$q_b := \max_{j \in Q'} q_j, \tag{3.3}$$

and the second one by

$$r_b := \max_{j \in Q'} r_j. \tag{3.4}$$

A simple way to compute a *lower bound* for a node is to take the maximum of $r(Q') + p(Q') + q(Q')$, $r(Q' \cup \{b\}) + p(Q' \cup \{b\}) + q(Q' \cup \{b\})$, and the lower bound of its parent. Note that, although $J_b$ and $Q'$ are determined by running the nonpreemptive EDD rule on the parent, the release dates and delivery times have been updated afterwards.

As in any branch-and-bound algorithm, a node is discarded if its lower bound is no smaller than the global *upper bound*, i.e., the value of the best schedule found thus far. An advantage of using the nonpreemptive EDD rule at each node is that each time we obtain a new feasible solution, which may improve the upper bound. For the same purpose, we also evaluate the schedule in which $J_b$ follows the jobs in $Q'$. The *search strategy* always selects a node with minimum lower bound.

The following trick can help to restrict the instance for a new node in the tree even further. Suppose there is a job $J_k$, $k \notin Q' \cup \{b\}$, for which $r(Q') + p_k + p(Q') + q(Q')$ exceeds the upper bound. By the same reasoning as above, we conclude that in any better schedule $J_k$ either precedes or follows all of the jobs in $Q'$. If also $r(Q') + p(Q') + p_k + q_k$ exceeds the upper bound, then $J_k$ must precede $Q'$, and we set $q_k := \max_{j \in Q'} q_j$. Similarly, if $r_k + p_k + p(Q') + q(Q')$ exceeds the upper bound, then $J_k$ follows $Q'$, and we set $r_k := \max_{j \in Q'} r_j$.

In comparing the two branch-and-bound methods, one may wonder how the two lower bound procedures relate. In this respect, Theorem 3.5 implies that the preemptive bound dominates the simple bound of the second algorithm. Why does the second algorithm neglect to run a superior bounding procedure? This is merely a question of balancing the sorts of work done by an enumerative method. It takes more time to obtain a better bound, and it is not clear *a priori* whether the improved lower bound justifies the additional work. There are procedures that in some cases even improve on the preemptive bound, but from an empirical point of view it appears to be preferable to use the simpler lower bound in case of the second branching strategy. Indeed, computational experiments suggest that this algorithm is among the current champions for solving $1|r_j|L_{\max}$.

**Exercises**

3.25. Prove that (3.3) and (3.4) enforce the desired precedence constraints not only in each of the two child nodes generated, but also in all of their descendants.

3.26. Construct an instance of $1|prec,r_j|L_{\max}$ which satisfies the property that $r_j < r_k$ and $d_j < d_k$ whenever $J_j \to J_k$, while its $L_{\max}^*$ value would decrease if the precedence constraints would be ignored.

3.27. How can the two branch-and-bound algorithms be adapted to solve $1|prec,r_j|L_{\max}$?

**Notes**

3.1. *Earliest Due Date and Least Cost Last rules.* These rules are due to Jackson [1955] and Lawler [1973].

Exercises 3.7 and 3.8 are from Monma [1980]. Hochbaum and Shamir [1989] gave an $O(n \log^2 n)$ algorithm for the maximum weighted tardiness problem, $1||wT_{\max}$. Fields and Frederickson [1990] gave an $O(n \log n + |A|)$ algorithm for $1|prec|wT_{\max}$, where $A$ is the arc set of the precedence digraph.

3.2. *Preemptive EDD and Least Cost Last rules.* Horn [1974] observed that $1|pmtn, r_j|L_{\max}$ and $1|r_j, p_j = 1|L_{\max}$ are solved by the preemptive EDD rule. Frederickson [1983] gave an $O(n)$ algorithm for $1|r_j, p_j = 1, \bar{d}_j|-$. Theorem 3.5 is due to Carlier [1982]; Nowicki and Zdrzalka [1986] observed that its proof is somewhat more elusive than originally believed.

The procedure for modifying release and due dates so that the several variants of the EDD rule automatically satisfy given precedence constraints was described by Lageweg, Lenstra, and Rinnooy Kan [1976]. Monma [1982] gave a linear-time algorithm for $1|prec, p_j = 1|L_{\max}$.

The generalization of the Least Cost Last rule for solving $1|pmtn, r_j|f_{\max}$ is due to Baker, Lawler, Lenstra, and Rinnooy Kan [1983]. Exercises 3.10 - 3.12 are also from their paper.

3.3. *A polynomial-time algorithm for jobs of equal length.* The algorithm for equal-length jobs is due to Simons [1978]. An alternative algorithm was proposed by Carlier [1979]. Garey, Johnson, Simons, and Tarjan [1981] gave an improved implementation of the decision algorithm, which runs in $O(n \log n)$ time.

3.4. *NP-hardness results.* Theorem 3.11 is due to Lenstra, Rinnooy Kan, and Brucker [1977].

3.5. *Approximation algorithms.* Schrage [1971] proposed the nonpreemptive EDD rule as a heuristic for $1|r_j|L_{\max}$, with the addition that ties on due date should be broken by selecting a job with maximum processing time. For the head-body-tail formulation, Kise, Ibaraki, and Mine [1979] showed that every left-justified schedule is shorter than three times the optimum. They considered six approximation algorithms, including the nonpreemptive EDD rule, and proved that all of them have a performance ratio of 2. Potts [1980B] proposed the iterated nonpreemptive EDD rule, which was the first method to achieve a better performance bound. Hall and Shmoys [1992] showed that the procedure which applies the iterated nonpreemptive EDD rule to an instance and its inverse and selects the better schedule is a 4/3-approximation algorithm. They observed that all approximation algorithms proposed thus far could easily be extended to handle precedence constraints. The polynomial approximation scheme presented is due to Lawler [-]. Hall and Shmoys [1992] gave a more efficient scheme: more precisely, they developed a family of algorithms $\{A_k'\}$ that guarantees $L_{\max}(A_k')/L_{\max}^* \leq 1 + 1/k$, where $A_k'$ runs in $O(n \log n + nk^{16k^2+8k})$ time. In a later paper, Hall and Shmoys [1990] extended their scheme to the precedence-constrained problem.

Exercises 3.23 and 3.24 are from Kise, Ibaraki, and Mine [1979] and Nowicki and Smutnicki [1994], respectively.

3.6. *Enumerative methods.* The first branch-and-bound algorithm is due to Baker and Su [1974], and the second one to Carlier [1982]. Exercise 3.25 is from Verkooijen [1991]. Carlier's method and the more recent branch-and-bound algorithm of Larson, Dessouky, and Devor [1985] are able to solve problem instances with up to 10,000 jobs, often without branching.

Earlier branch-and-bound algorithms were given by Dessouky and Margenthaler [1972] and by Bratley, Florian, and Robillard [1973]. McMahon and Florian [1975] proposed a lower bound that is not dominated by the preemptive bound. Lageweg, Lenstra, and Rinnooy Kan [1976] extended the algorithms of Baker and Su and of McMahon and Florian to the precedence-constrained problem, and demonstrated that, if for a given problem instance the range of the release times is smaller than the range of the delivery times, then it is computationally advantageous to apply the McMahon-Florian algorithm to the inverse instance. Exercise 3.26 is from their paper. Nowicki & Smutnicki [1987] discussed the relations between various lower bounds. Zdrzalka and Grabowski [1989] considered extensions of these enumerative methods to $1 \mid prec, r_j \mid f_{\max}$.

Dominance results among the schedules may be used in the obvious way to speed up enumerative procedures. Erschler, Fontan, Merce, and Roubellat [1982, 1983] considered the decision problem $1 \mid r_j, \bar{d}_j \mid -$, and introduced dominance based on the $[r_j, \bar{d}_j]$ intervals.

# Contents

# 4

# Weighted sum of completion times

Eugene L. Lawler
*University of California, Berkeley*

Maurice Queyranne
*University of British Columbia*

Andreas S. Schulz
*Technical University of Munich*

David B. Shmoys
*Cornell University*

Minimization of the mean completion time has always been an intuitively appealing objective. Although the origins of the *Shortest Processing Time (SPT) rule* are unknown, we do know that W. E. Smith, in one of the first publications in scheduling theory, showed that the *ratio rule*, a generalization of SPT, solves $1||\sum w_j C_j$. We also know that Smith pointed out in 1956 that the ratio rule begs for abstraction in the form of a "preference order."

During the 1960's and 70's, Smith's results on $1||\sum w_j C_j$ were extended to apply to precedence constraints of various kinds, first to parallel chains, then to rooted trees, and then to series-parallel networks, with rooted trees as a special case. During the same period, it was found that the preference order concept applies to a variety of sequencing problems, including the least cost fault detection problem, the two-machine permutation flow shop problem, and problems with maximum cumulative cost and total weighted discounted cost as their objectives. Moreover, it was observed

**1**

that the same O($n \log n$) algorithm solves any of these problems with series-parallel precedence constraints, provided the preference order has the property of applying to "sequences" and not simply to "jobs." This led to an elegant theory for dealing with precedence constraints in sequencing problems with a variety of scheduling objectives, of which total weighted completion time is but a prototypical example.

Later, the influence of polyhedral theory on combinatorial optimization made its impact on our understanding of a number of variations of the problem $1 | \; | \sum w_j C_j$. This understanding led not only to new perspectives on these older results, but also to the development of an extensive literature of approximation algorithms for these problems. In this chapter, we will highlight both of these threads of understanding.

## 4.1.  Smith's ratio rule

By applying Smith's ratio rule, the problem $1 | \; | \sum w_j C_j$ can be solved with nothing more than a simple O($n \log n$) sort of the jobs by their ratios $w_j / p_j$ (throughout this chapter, we assume all job weights are nonnegative and all processing times positive, i.e., $w_j \geq 0$, and $p_j > 0$, $j = 1, \ldots, n$).

**Theorem 4.1.** *A sequence is optimal for $1 | \; | \sum w_j C_j$ if and only if it places the jobs in order of nonincreasing ratios $w_j / p_j$.*

*Proof.* We first prove that having nonincreasing ratios is a necessary condition for a sequence to be optimal. Let $\pi$ be a sequence in which the jobs are not in ratio order. Then, in $\pi$, there is a job $i$ that immediately precedes a job $j$ and yet $w_i / p_i < w_j / p_j$. If job $j$ completes at time $C_j$, then job $i$ completes at time $C_j - p_j$. If we interchange these two jobs, this affects only *their* completion times, not those of other jobs. The result is a strict decrease in total cost, by

$$[w_i(C_j - p_j) + w_j C_j] - [w_j(C_j - p_i) + w_i C_j] = w_j p_i - w_i p_j$$
$$= p_i p_j \left( \frac{w_j}{p_j} - \frac{w_i}{p_i} \right) > 0,$$

from which it follows that $\pi$ is not optimal.

Conversely, we now prove that having nonincreasing ratios is a sufficient condition for a sequence to be optimal. Let $\pi$ be a sequence in which the jobs are in ratio order and let $\pi^*$ be an optimal sequence. If $\pi \neq \pi^*$, then in $\pi^*$ there is a job $i$ immediately preceding a job $j$, where $j$ precedes $i$ in $\pi$. But then $w_j / p_j \geq w_i / p_i$ (because in $\pi$ jobs are in order of nonincreasing ratios) and $w_i / p_i \geq w_j / p_j$ (by the first part of this proof because $\pi^*$ is optimal) and, therefore, $w_i / p_i = w_j / p_j$. Interchanging the jobs in $\pi^*$ creates a new sequence of equal cost. A finite number of such interchanges converts $\pi^*$ to $\pi$, demonstrating that $\pi$ is optimal. □

The ratio rule immediately specializes to the celebrated *Shortest Processing Time* or *SPT* rule.

**Corollary 4.2.** *A sequence is optimal for $1||\sum C_j$ if and only if it places the jobs in order of nondecreasing processing times $p_j$.*

The SPT rule is often applied to more complicated problems than $1||\sum C_j$, sometimes without much theoretical support for its performance. Although no additional proof of the SPT rule is needed, variations of the following proof turn out to be quite useful for parallel machine problems, as we shall see in Chapter 8. Suppose the jobs are executed in the order $1, 2, \ldots, n$. Then we have

$$
\begin{aligned}
C_1 &= p_1, \\
C_2 &= p_1 + p_2, \\
C_3 &= p_1 + p_2 + p_3, \\
&\cdots \\
C_n &= p_1 + p_2 + p_3 + \cdots + p_n,
\end{aligned}
$$

giving us

$$\sum_{j=1}^{n} C_j = n p_1 + (n-1) p_2 + (n-2) p_3 + \cdots + 2 p_{n-1} + p_n. \tag{4.1}$$

This means that the problem of minimizing $\sum_{j=1}^{n} C_j$ is equivalent to the problem of assigning the coefficients $1, 2, \ldots, n$ to the processing times $p_j$ in such a way that the weighted sum (4.1) is minimized. This is accomplished by assigning the coefficient 1 to the largest $p_j$, the coefficient 2 to the next-largest $p_j$, etc., and the coefficient $n$ to the smallest $p_j$, as can be verified by an interchange argument similar to that used in the proof of Theorem 4.1.

Smith's ratio rule produces an optimal schedule for the *nonpreemptive* problem $1||\sum w_j C_j$. The reader may wonder if the total cost could be further reduced by allowing preemption, or inserting idle time before all jobs are complete. The answer is "No," even for the most favorable model of preemption, whereby an interrupted job may be resumed at any date without any cost or time penalty. In fact, the same negative answer, "There is no advantage to preemption," applies to a broad class of single-machine scheduling problems. If we let $C = (C_1, C_2, \ldots, C_n)$ denote the completion time vector of a schedule, we say that an objective function $f(C)$ is *monotone* if $C \le C'$ (i.e., $C_j \le C'_j$ for each $j = 1, \ldots, n$) implies $f(C) \le f(C')$. We leave the proof of the following theorem as an exercise, since it can be proved easily with the machinery of the previous chapter.

**Theorem 4.3.** *There is no advantage to preemption or idle time for the single-machine problem $1| prec, \overline{d}_j, pmtn| f(C)$ whenever $f$ is monotone.*

Of course, this theorem can be applied to the problem $1| pmtn |\sum w_j C_j$, since we assume throughout this book that each $w_j \ge 0$, $j = 1, \ldots, n$. Hence, Theorem 4.3 implies that Smith's ratio rule is also optimal when preemption and/or idle time are allowed.

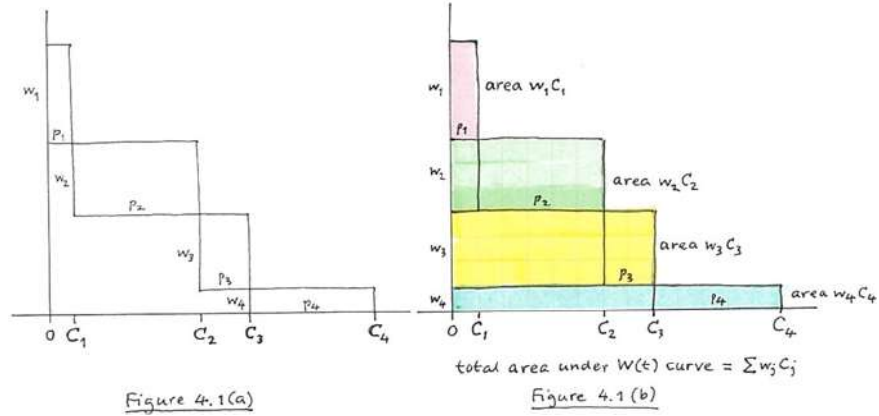

Many results in single-machine scheduling, starting with Smith's ratio rule, have a simple and intuitive geometric justification using *two-dimensional Gantt charts*. As in ordinary Gantt charts, the horizontal axis in a 2D Gantt chart represents time. The vertical axis represents total weight: at date $t \geq 0$, we plot the total weight $W(t)$ of all the jobs that have not yet been completed by date $t$. In a nonpreemptive schedule, if job $j$ is in process at date $t$, we may also plot $W(t) - w_j$, so the two horizontal lines $W(t)$ and $W(t) - w_j$ delimit a rectangle of length $p_j$ and height $w_j$, somewhat analogue to the rectangles that represent jobs in ordinary Gantt charts; see Figure 4.1(a).

For any nonpreemptive schedule, the area under the curve $W(t)$ is equal to $\sum_{j=1}^{n} w_j C_j$. This can be seen by identifying each term $w_j C_j$ with the area of the horizontal slabs in Figure 4.1(b). From this, it is clear that there is no advantage to inserting idle time.

We can draw a descending diagonal in each rectangle representing a job. The *slope* of job $j$ is the ratio $\rho(j) = w_j / p_j$; it is just the negative of the slope of its diagonal. Smith's ratio rule states that the area under the $W(t)$ curve is minimized when the jobs are sequenced with largest slopes first, that is, when the piecewise linear continuous curve $\overline{W}(t)$ defined by the slopes of the rectangles, is made convex. The adjacent pairwise argument used in the proof of Theorem 4.1 is visualized in Figure 4.2.

One interpretation of the $\sum w_j C_j$ objective is to consider $w_j$ to be the holding cost (euros per time unit) of a resource needed for the processing of each job $j$. We assume that we have an initial inventory of the resource that is exactly equal to the amount needed to process all the given jobs; so $W(0) = \sum_{j=1}^{n} w_j$ is the initial total holding cost. It is convenient to measure the resource inventory level at any time $t$ in monetary units (euros), and to identify it with the corresponding total holding cost $W(t)$. If all units of the resource used by job $j$ are consumed instantly at the completion of the job, then the inventory level $W(t)$ follows the piecewise constant curve plotted in the 2D Gantt chart.

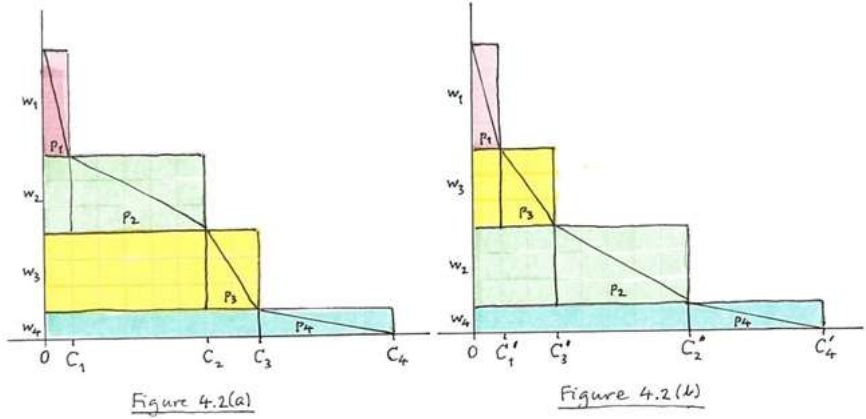

Figure 4.2(a)                                Figure 4.2(b)

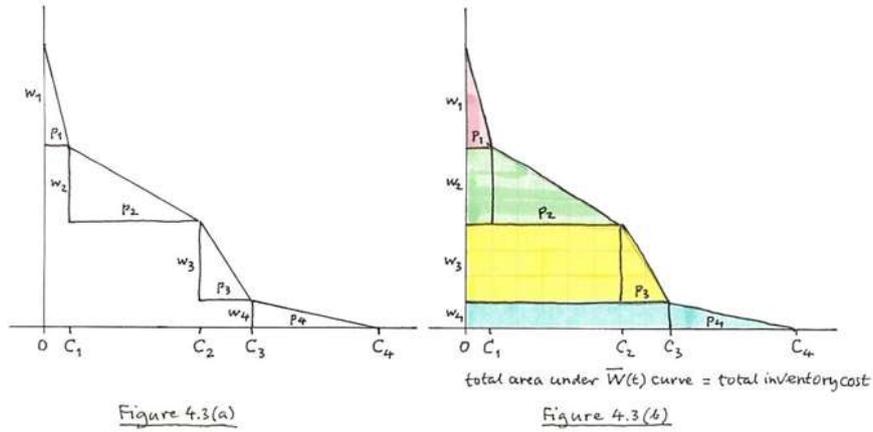

Figure 4.3(a)

total area under $\overline{W}(t)$ curve = total inventory cost

Figure 4.3(b)

In many situations, however, the resource is consumed at a constant rate $\rho(j) = w_j/p_j$ units per time unit during the processing of job $j$. In that case, the inventory level follows the curve $\overline{W}(t)$, as in Figure 4.3(a). The total inventory cost, $\int_0^{+\infty} \overline{W}(t)\,dt$ is the sum of the areas of the resulting horizontal trapezoids. Each trapezoid has the same area as a rectangle with length $(C_j - p_j + C_j)/2 = C_j - p_j/2$ and height $w_j$; see Figure 4.3(b). Alternatively, one can instead define the *mean busy time* $M_j$ of each job $j$ as the midpoint of its nonpreemptive processing: $[(C_j - p_j) + C_j]/2 = C_j - p_j/2$. Thus, we are also interested in finding schedules to minimize the total weighted mean busy time, $\sum w_j M_j$. In the nonpreemptive setting, since the difference between this new objective and $\sum w_j C_j$ is equal to $\sum w_j p_j/2$ (a constant) for any feasible schedule, the two objective functions lead to equivalent optimization problems. Throughout this chapter, we shall see that this alternative perspective significantly improves our understanding.

In the presence of release dates, idle time may be necessary, and we will see in Section 4.10 that there *is* an advantage to preemption, for both the $\sum w_j C_j$ and $\sum w_j M_j$ objectives. It will also be shown there that the preemptive weighted mean busy time problem $1 \mid r_j, pmtn \mid \sum w_j M_j$ is solvable to optimality by a very simple algorithm, whereas the preemptive weighted completion time problem $1 \mid r_j, pmtn \mid \sum w_j C_j$ is NP-hard.

Finally, whereas the nonpreemptive *optimization* problems with $\sum w_j C_j$ and $\sum w_j M_j$ objectives are equivalent (because of the constant difference $\frac{1}{2} \sum w_j p_j$), the corresponding *approximation* problems are *not* equivalent. In fact, an $\alpha$-approximation algorithm for the nonpreemptive $\sum w_j M_j$ objective is also an $\alpha$-approximation algorithm for the nonpreemptive $\sum w_j C_j$ objective, but the converse is not necessarily true.

**Exercises**

4.1. Prove Theorem 4.3.

4.2. Consider $1 \mid \overline{d}_j \mid \sum C_j$. Assume that there exists a sequence in which all jobs meet their deadlines. Show that the following algorithm produces an optimal feasible sequence:

> From among all jobs $j$ that are eligible to be sequenced last, i.e., are such that $\overline{d}_j \geq p_1 + \cdots + p_n$, put the job last which has the longest processing time. Repeat this procedure with the remaining $n - 1$ jobs.

4.3. Use 2D Gantt charts to show that an instance of $1 \mid prec \mid \sum w_j C_j$ with processing times $p_j$, weights $w_j$, and precedence constraints $\rightarrow$ is equivalent to an instance with the same set of jobs, $p'_j = w_j$, $w'_j = p_j$ for each job $j$, and $j \rightarrow' k$ if and only if $k \rightarrow j$. In particular, a sequence is optimal for the original instance if and only if the reverse sequence is optimal for the new instance, and both have the same objective function value.

## 4.2. Preference orders on jobs

A very general formulation of optimal sequencing problems is as follows: Given a real-valued function $f$ that assigns a cost $f(\pi)$ to each permutation $\pi$ of a set $N$ of $n$ jobs, find a permutation $\pi^*$ of $N$ such that

$$f(\pi^*) = \min\{f(\pi) : \pi \text{ is a permutation of } N\}.$$

If we know nothing at all about the structure of the function $f$, then we have no alternative but to evaluate $f$ for each of the $n!$ permutations of $N$. This occurs if $f$ is given to us by a "black box" subroutine, to which we can submit a permutation $\pi$, and from which we only receive the value $f(\pi)$ in return. However, we usually know quite a bit about the structure of the function $f$. We use knowledge of $f$ to solve problems like $1||L_{\max}$ and $1||\sum w_j C_j$ by Jackson's EDD rule and Smith's ratio rule, respectively. Perhaps other problems lend themselves to solution by similar rules. But, if so, what do we mean by "similar?" What is it we need to know about $f$ in order to infer such rules?

Throughout this section and Section 4.3, we will use slightly different notation and will represent a permutation or sequence as the concatenation of disjoint subsequences, e.g., a permutation $\pi$ of $n$ jobs may be represented as $\pi = (u, s, t, v)$, where each of the $n$ jobs appears in exactly one of the subsequences $u, s, t, v$. A single job corresponds to a sequence of length one. In the case of $1||L_{\max}$, we observed that $d_i \leq d_j$ implies that $f(u, i, j, v) \leq f(u, j, i, v)$ [Jackson's rule]. In the proof of Theorem 4.1 we showed that $w_i/p_i \geq w_j/p_j$ implies $f(u, i, j, v) \leq f(u, j, i, v)$ [Smith's rule].

**Definition 4.4.** *A transitive and complete relation $\leq_f$ on $N$ is said to be a* preference order on jobs*, relative to objective function $f$, if it satisfies the* adjacent pairwise interchange property on jobs*, i.e.,*

$$i \leq_f j \text{ implies that } f(u, i, j, v) \leq f(u, j, i, v), \tag{4.2}$$

*for all jobs $i, j$ and all subsequences $u, v$.*

Recall that a relation $\leq_f$ is *transitive* if, for each triple $i, j, k \in N$, $i \leq_f j$ and $j \leq_f k$ imply that $i \leq_f k$. A relation $\leq_f$ is *complete* if, for each pair $i, j \in N$, either $i \leq_f j$ or $j \leq_f i$. A relation that is both transitive and complete is sometimes called a *quasi total order*. Such a relation induces a linear ordering of equivalence classes, where $i$ and $j$ are in the same equivalence class if and only if both $i \leq_f j$ and $j \leq_f i$, in which case we may choose to write $i \equiv_f j$. If $i \leq_f j$, but it is not the case that $j \leq_f i$, we may write $i <_f j$.

**Theorem 4.5.** *Given a preference order $\leq_f$ on $N$, an optimal sequence can be found by sorting jobs according to $\leq_f$, with $O(n \log n)$ comparisons of jobs with respect to $\leq_f$.*

*Proof.* Let $\pi$ be any sequence consistent with the preference order, and let $\pi^*$ be an optimal permutation. If $\pi^*$ differs from $\pi$, then $\pi^*$ is of the form $(u, j, i, v)$, for some pair $i, j$ of jobs, where $i$ precedes $j$ in $\pi$ and hence $i \leq_f j$. From (4.2) it follows that $f(u, i, j, v) \leq f(\pi^*)$, and hence $(u, i, j, v)$ is also optimal. A finite number of such interchanges transforms $\pi^*$ into $\pi$ and shows that $\pi$ is optimal. $\square$

As already observed, Smith's ratio rule for $1||\sum w_j C_j$ and Jackson's EDD rule for $1||L_{\max}$ give rise to special cases of preference orders. Below we consider some other examples of sequencing problems for which there are preference orders on jobs.

**Total Weighted Discounted Completion Time.**

Sometimes jobs must be scheduled over a time period so long that inflation and interest charges must be taken into account. Suppose that we will be paid $w_j$ dollars upon the completion of job $j$. The present value of one dollar at time $t$ in the future is $\exp(-rt)$, where $r > 0$ is a fixed discount rate. Hence the present value of completing job $j$ at time $t$ is $w_j \exp(-rt)$. It follows that to maximize the present value of the payments we will receive we should minimize $\sum_j f_j(C_j)$ with respect to cost functions of the form

$$f_j(t) = -w_j \exp(-rt).$$

We assert (Exercise 4.6) that an optimal sequence is obtained by sequencing jobs in nondecreasing order of the ratios $w_j / [1 - \exp(rp_j)]$.

**Least Expected Cost Fault Detection.**

A system consisting of $n$ components is to be inspected by testing the components one at a time until either one fails (the system is found to be defective) or until all the components pass their tests (the system passes inspection). The cost of testing component $j$ is $c_j$ and the probability that it will pass its test is $q_j$. Tests are assumed to be statistically independent. Hence if the components are tested in the order $1, 2, \ldots, n$, the probability that it will be necessary to test component $j$ is

$$Q_j = q_1 \cdot q_2 \cdot \ldots \cdot q_{j-1},$$

where by convention $Q_1 = 1$. The expected cost of testing is then $\sum_j c_j Q_j$. We assert (Exercise 4.7) that it is optimal to test the components in nondecreasing order of the ratios $c_j / (1 - q_j)$.

**Weighted Monotone Cost Density.**

Suppose for each job $j$ we have

$$f_j(t) = w_j \int_{t-p_j}^{t} g(u)\, du,$$

where $g$ is a nondecreasing "cost density" function, and we want to minimize $\sum_j f_j(C_j)$. We assert (Exercise 4.9) that an optimal permutation is obtained by placing the jobs in order of nonincreasing $w_j$.

**Exercises**

4.4. Suppose someone gives you a function $f(\pi)$ in the form of a "black box" subroutine. But she also assures you that $f(\pi)$ is actually the weighted sum of completion times. You have total ignorance of the values of the parameters (i.e., the $p_j$'s and $w_j$'s) of the $n$ jobs in your problem instance. But you can still find an optimal sequence with $\mathrm{O}(n \log n)$ calls on the subroutine. How?

4.5. If $\leq_f$ is a preference order on jobs, does it follow that $i \leq_f j$ implies $f(u, i, v, j, w) \leq f(u, j, v, i, w)$, for all $i, j, u, v, w$? Prove or disprove.

4.6. Prove the validity of the ratio rule asserted for the total weighted discounted completion time problem.

4.7. Prove the validity of the ratio rule asserted for the least expected cost fault detection problem.

4.8. Any instance of the least expected cost fault detection problem can be transformed into an equivalent instance of the total weighted discounted completion time problem as follows. Let $r = 1$. For component $j$ with parameters $c_j$ and $q_j$, create a job $j$ with parameters $p_j = -\ln q_j$ and $w_j = -c_j/q_j$. Provide a similar transformation in the reverse direction, i.e., from the discounted completion time problem to the fault detection problem, showing that the two problems are equivalent.

4.9. In the case of a weighted monotone cost density function, prove that an optimal sequence is obtained by placing jobs in nonincreasing order of $w_j$.

4.10. Show that total weighted completion time is a special case of weighted monotone cost density.

4.11. Show that total weighted *discounted* completion time is a special case of weighted monotone cost density.

## 4.3. Preference orders on sequences & series-parallel precedence constraints

Let $\Pi$ be the set of all *feasible* permutations of a set $N = \{1, \ldots, n\}$ of $n$ jobs, and $f : \Pi \to \mathbb{R}$ be a cost function. The *constrained optimal sequencing problem* is to find a permutation $\pi^* \in \Pi$ such that $f(\pi^*) = \min\{f(\pi) : \pi \in \Pi\}$.

If the structure of the function $f$ is unknown, then there is no alternative but to evaluate the cost of each feasible permutation in $\Pi$. And although the number of feasible permutations may be much smaller than $n!$, this number may still be hopelessly large. In practice, we are likely to know quite a bit about both the function $f$ and the set $\Pi$. However, in order to successfully apply preference orders to constrained problems, our preference orders must satisfy stronger properties than before.

Let $\mathcal{N}$ denote the set of all *subpermutations* of the jobs $N$, i.e., the set of all sequences that can be formed from subsets of $N$. We shall call elements of $\mathcal{N}$ *sequences*, or *compound jobs*.

**Definition 4.6.** *A transitive and complete relation $\leq_f$ on $\mathcal{N}$ is said to be a* preference order on sequences*, relative to objective function $f$, if it satisfies the following* adjacent pairwise interchange property on sequences*:*

$s \leq_f t \quad$ *implies* $\quad f(u, s, t, v) \leq f(u, t, s, v)$ *for all disjoint sequences $u, s, t$ and $v$.*

Thus, by assumption, $\leq_f$ is a complete preorder or, in simpler terms, a total order with possible ties between sequences. We denote by $<_f$ the corresponding strict preorder, that is, $s <_f t$ when $s \leq_f t$ and $t \not\leq_f s$.

Some sequencing problems, like $1| \ |L_{\max}$, admit a preference order on jobs, but not on sequences. (See Exercise 4.13.) However, all of the other problems cited in Section 4.2 do admit preference orders on sequences, with the exception of problems with weighted monotone cost density functions. (See Exercise 4.14.) In particular, in the case of $1| \ |\sum w_j C_j$, the appropriate extension is $s \leq_f t$ if and only if $w(s)/p(s) \geq w(t)/s(t)$, where we extend our usual notation (slightly) to let $p(s)$ and $w(s)$ denote, respectively, the sum of the processing times and weights of the jobs in a sequence $s$. (See Exercise 4.12.)

In this section, we are primarily interested in the case when the set $\Pi$ of feasible permutations is specified by *precedence constraints*. We write $i \to j$ to denote the precedence constraint that job $i$ must appear before job $j$ in any feasible permutation $\pi$. One strategy for dealing with precedence constraints is to ignore them and simply sort the jobs by preference order. If we are lucky and the resulting sequence turns out to be feasible, then we are done. This follows from the fact that, by making all jobs independent, we have solved a relaxation of the original problem. If an optimal schedule for this relaxation happens to be feasible with respect to the precedence constraints, then it must also be optimal with respect to these constraints.

If the permutation $\pi$ obtained by sorting jobs by preference order is not feasible, it is due to the *collision* of one or more pairs of jobs $i$, $j$ where $i \to j$ and $j \leq_f i$. If it happens that $j$ and $i$ are consecutive in $\pi$ and $i \leq_f j$, the collision can be resolved immediately by interchanging $j$ and $i$ in $\pi$. Even if this is not the case, it may be possible to do something about the collision.

**Lemma 4.7.** *Let $i$ and $j$ be a pair of jobs such that $i \to j$ and $j \leq_f i$. Suppose that, for each $k \in N$ distinct from $i$ and $j$, either (i) $k \to i$; or (ii) $j \to k$; or (iii) $k$ is unrelated by the precedence constraints to $i$ and also unrelated to $j$. Then there exists an optimal feasible permutation in which $i$ immediately precedes $j$.*

*Proof.* Let $\pi^* = (t, i, u, j, v)$ be an optimal feasible permutation. If $u$ is empty, then we are done. Hence, assume that $u$ is not empty. Since each job $k$ in $u$ is unrelated to both $i$ and $j$, it follows that the precedence constraints are not violated by interchanging $i$ and $u$, or by interchanging $u$ and $j$. It must be the case that either $u \leq_f i$ or $j \leq_f u$, else we would have $i <_f u <_f j$, contradicting the hypothesis that $j \leq_f i$. Hence at least one of the two interchanges results in an optimal feasible permutation in which $i$ immediately precedes $j$. $\square$

When the hypotheses of the lemma are found to apply to a colliding pair of jobs $i$ and $j$, the pair can be replaced by the sequence $(i, j)$, with $(i, j)$ inheriting all of the precedence constraints of $i$ and $j$, e.g., if $j \to k$ then $(i, j) \to k$. The sequence $(i, j)$ can then be treated as a single job, or compound job, for the purpose of reapplying the lemma.

Precedence constraints that consist of *parallel chains* have the nice property that application of Lemma 4.7 is guaranteed to yield a set of sequences that is free of collisions. To see this, one need only note that if any pair of jobs is in collision, there is a colliding pair of jobs $i$, $j$ that are adjacent in a chain. But then $i$ and $j$ satisfy the hypotheses of the lemma and can be replaced by a single sequence. With no more than $n-1$ repeated applications of the lemma, a collision-free family of sets of sequences can be obtained, in the sense of the following definition. For this definition, let an ordering $(s_1, \ldots, s_\ell)$ of sequences be *consistent with* $\leq_f$ if $s_u \leq_f s_v$ for all $1 \leq u < v \leq \ell$; and *feasible with respect to the precedence constraints* if for any precedence relation $i \to j$, we have that $i \in s_u$, and $j \in s_v$, for some $u \leq v$.

**Definition 4.8.** *Let $X$ be a finite index set. A family $\mathcal{S} = \{S(x) : x \in X\}$ of sets of sequences is said to be* collision-free*, with respect to given precedence constraints and preference order $\leq_f$ on sequences, if*

1. *each job in $N$ is contained in exactly one of the sequences in $\cup_{x \in X} S(x)$; and*

2. *there exists an optimal feasible permutation of the jobs in $N$ in which the jobs in each sequence $s \in \cup_{x \in X} S(x)$ appear consecutively; and*

3. *any ordering of the sequences in each set $S(x)$ which is consistent with $\leq_f$ is also feasible with respect to the precedence constraints.*

In effect, Lemma 4.7 enables us to transform a problem instance with parallel-chains precedence constraints into an instance, consisting of sequences (or compound jobs) $s \in N'$, for which the family $\mathcal{S} = \{ N' \}$ is collision-free. Hence the resulting instance can be dealt with as if it was unconstrained and an optimal feasible permutation is obtained by simply sorting all these sequences in preference order $\leq_f$.

**Theorem 4.9.** *Let the precedence constraints form parallel chains. Given a preference order $\leq_f$ on $\mathcal{N}$, an optimal feasible permutation of $N$ can be found with $O(n \log n)$ comparisons of sequences with respect to $\leq_f$.*

*Series-parallel* partial orders are defined recursively as follows:

> any partial order $(N, \to)$, where $N$ is a singleton, is series-parallel.

Let $(N_1, \to)$ and $(N_2, \to)$ be disjoint partial orders (i.e., $N_1 \cap N_2 = \emptyset$) that are series-parallel. A partial order $(N_1 \cup N_2, \to)$ is also series-parallel, when relations between jobs in $N_1$ and jobs in $N_2$ are determined by either

> *series composition*, in which each job $i$ in $N_1$ precedes each job $j$ in $N_2$, i.e.,
> $i \to j$ for all $i \in N_1$ and $j \in N_2$,

or

> *parallel composition*, in which the jobs in $N_1$ and $N_2$ are unrelated, i.e.,
> $i \not\to j$ and $j \not\to i$ for all $i \in N_1$ and $j \in N_2$.

(Relations between pairs of jobs, both of which are in $N_1$ or in $N_2$, are unaffected.)

The structure of series-parallel precedence constraints is represented by a *composition* (or *decomposition*) tree in which each leaf of the tree is identified with a job and each internal node of the tree corresponds to a series or a parallel composition operation, and is accordingly labeled either "S" or "P." The left and right children of an S-node are respectively identified with the subsets $N_1$, $N_2$ of the series composition operation. The same is true of the children of a P-node, except that the left-right ordering of its children is immaterial.

Some examples of series-parallel precedence constraints and their composition trees are indicated in Figure 4.4. Note that parallel chains, in-trees, out-trees, and forests of in-trees and out-trees, are all special cases of series-parallel constraints. The smallest non-series-parallel partial order is the "Z" digraph shown in Figure 4.5. In fact, a partial order fails to be series-parallel if and only if it contains four elements in a "Z" relation.

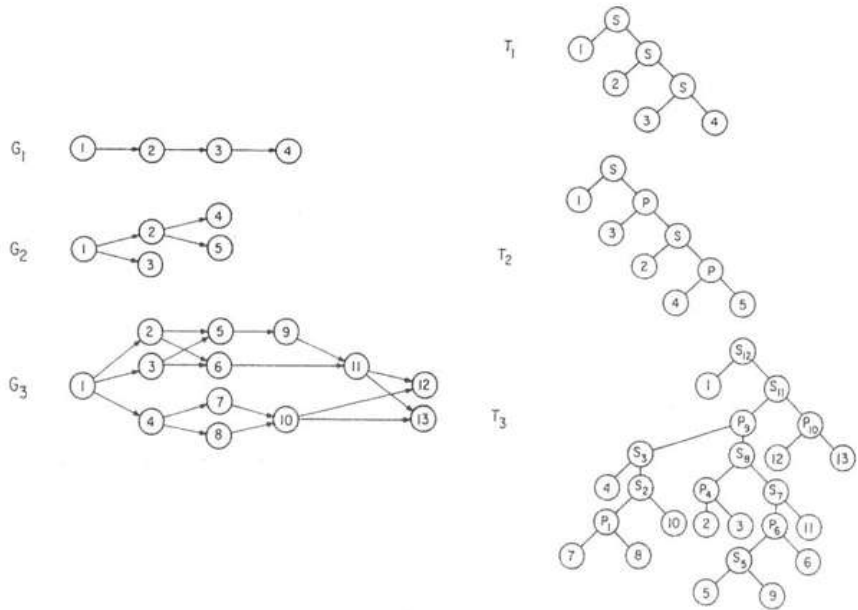

Figure 4.4:  Three series-parallel digraphs and their composition trees.

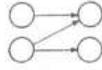

*Figure 4.5. The "Z" digraph.*

There are efficient algorithms for testing if precedence constraints are series-parallel. In particular, a digraph $G$ with $m$ arcs and $n$ nodes can be tested in $O(m+n)$ time to determine whether the partial order it induces is series-parallel. If the partial order is series-parallel, the algorithm "parses" it and returns a decomposition tree. If the partial order is not series-parallel, the algorithm returns a "Z," proving that it is not.

Given a decomposition tree for series-parallel precedence constraints and a preference order $\leq_f$ on sequences, a plausible strategy for finding an optimal feasible permutation is to work from the leaves of the tree toward the root, dealing with the subproblem at an internal node only after the subproblems at its children have been dealt with. At each node $x$ of the tree we propose to obtain a collision-free family $\mathcal{S}$ of sets of sequences, in which a set $S(x) \in \mathcal{S}$ contains sequences which contain every job corresponding to node $x$, and no other job. The algorithm will maintain the following two invariants after each node has been dealt with:

(1) the current family $\mathcal{S}$ of sets of sequences is collison-free; and

(2) the permutation obtained by sorting all sequences in $\mathcal{S}$ in preference order $\leq_f$ is optimal for the relaxed problem restricted only by the precedence constraints defined by the nodes dealt with so far.

Thus, at the root node $x_0$ the set $\mathcal{S}$ will consist of a single set $S(x_0)$, and an optimal feasible permutation will be obtained by simply sorting all these sequences in preference order $\leq_f$.

We initialize $\mathcal{S} = \{S(j) : j \in N\}$ where each $S(j)$ consists of the single sequence $j$. Thus all leaves have been dealt with, and invariants (1) and, by Theorem 4.5, (2) are verified. At a $P$-node $x$ with children $y$ and $z$, all that is necessary is to replace in $\mathcal{S}$ the sets $S(y)$ and $S(z)$ with the set $S(x) = S(y) \cup S(z)$, since none of the sequences in $S(x)$ collide. If $\mathcal{S}$ is collision-free before this replacement, then it also is collision-free afterwards. Moreover, invariant (2) continues to hold because we have not added any new precedence constraint. At an $S$-node, things are a bit more complicated. However, collisions can be resolved by repeated application of the following generalization of Lemma 4.7.

**Lemma 4.10.** *Let $\mathcal{S}$ be a collision-free family of sets of sequences. Let $S(a)$ and $S(b)$ be two sets in $\mathcal{S}$ such that, for every $s \in S(a)$ and $t \in S(b)$ there exist jobs $i \in s$ and $j \in t$ such that $i \to j$. Let $\alpha$ be a $\leq_f$-maximal sequence in $S(a)$ and $\beta$ be a $\leq_f$-* *minimal sequence in $S(b)$, such that $\beta \leq_f \alpha$. Assume that, for each $k \in N$ which is not in any sequence in $S(a) \cup S(b)$, either*

(i) *$k \rightarrow i$ for some job $i$ in every sequence in $S(a)$; or*

(ii) *$j \rightarrow k$ for some job $j$ in every sequence in $S(b)$; or*

(iii) *$k$ is unrelated by the precedence constraints to every job in every sequence in $S(a) \cup S(b)$.*

*Then there exists an optimal feasible permutation in which every sequence in $\mathcal{S}$ is consecutive and sequence $\alpha$ immediately precedes sequence $\beta$.*

*Proof.* Since $\mathcal{S}$ is collision-free, let $\pi^* = (t, \alpha, u, \beta, v)$ be an optimal feasible permutation in which each sequence appearing in $\mathcal{S}$ is consecutive, and for which the number $|u|$ of jobs between $\alpha$ and $\beta$ is as small as possible. If $|u| = 0$ we are done; hence assume that $|u| \geq 1$. If any sequence in $S(a)$ appears in $u$, let $s$ be the first such sequence and let $u = (u', s, u'')$ with no sequence in $u'$ contained in $S(a)$. If $u'$ is empty, then by the $\leq_f$-maximality of $\alpha$ in $S(a)$ we may interchange $s$ and $\alpha$, obtaining an optimal permutation with fewer than $|u|$ jobs between $\alpha$ and $\beta$, a contradiction to $u$ being of minimum size. If $u'$ is nonempty, $u'$ precedes $s \in S(a)$ in the feasible permutation $\pi^*$, and no sequence in $u'$ can be contained in $S(b)$ either. Therefore, by condition (iii) of the lemma, we can feasibly interchange $\alpha$ with $u'$, and $u'$ with $s$. The minimality of $|u|$ and the optimality of $\pi^*$ imply that $\alpha <_f u'$ and $u' \leq_f s$; hence $\alpha <_f s$, a contradiction to the $\leq_f$-maximality of $\alpha$ in $S(a)$. Therefore, no sequence in $S(a)$ can appear in $u$. A symmetric argument implies that no sequence in $S(b)$ can appear in $u$. But now we may feasibly interchange $\alpha$ with $u$, and $u$ with $\beta$, so we must have $\alpha <_f u <_f \beta$, a contradiction with the assumption $\beta \leq_f \alpha$.  $\square$

Let $x$ be an $S$-node and $y$ and $z$ be its left and right children. Let $\ell$ be a $\leq_f$-maximal sequence in $S(y)$ and $r$ be a $\leq_f$-minimal sequence in $S(z)$. If $\ell <_f r$ then we can simply merge $S(y)$ and $S(z)$ into a single set $S(x) = S(y) \cup S(z)$ and the resulting family is collision-free. Moreover, invariant (2) continues to hold. Otherwise, $r \leq_f \ell$ and we can apply Lemma 4.10: there exists an optimal feasible permutation in which every sequence in $\mathcal{S}$ is consecutive and sequence $\ell$ immediately precedes sequence $r$. Thus we concatenate the two sequences $\ell$ and $r$ into a single sequence $\lambda = (\ell, r)$. If $s <_f \lambda$ for all $s \in S(y) \setminus \{\ell\}$ and $\lambda <_f s$ for all $s \in S(z) \setminus \{r\}$, then we can replace $S(y)$ and $S(z)$ with $S(x) = (S(y) \setminus \{\ell\}) \cup (S(z) \setminus \{r\}) \cup \{\lambda\}$, and the resulting family is again collision-free. Otherwise, we form a collision-free family $\mathcal{S}'$ by replacing $S(y)$ and $S(z)$ with the three sets $S'(y) = S(y) \setminus \{\ell\}$, $\{\lambda\}$, and $S'(z) = S(z) \setminus \{r\}$. However, we aim to replace in turn these three sets with a single set $S(x)$ that comprises all of the corresponding jobs: for the resulting family to be collision-free, $S(x)$ must satisfy Property 3 of Definition 4.8.

If $\lambda \leq_f s$ for some $s \in S'(y)$ then we may again apply Lemma 4.10 with $S'(y)$ as $S(a)$ and $\{\lambda\}$ as $S(b)$: we conclude that there exists an optimal feasible permutation in which every sequence in $\mathcal{S}'$ is consecutive and a $\leq_f$-maximal sequence $\alpha$ in $S'(y)$ immediately precedes sequence $\lambda$. Thus replacing $\lambda$ with $(\alpha, \lambda)$ and $S'(y)$ with $S'(y) \setminus \{\alpha\}$ maintains the invariant that the resulting family is collision-free. Similarly, if $s \leq_f \lambda$ for some $s \in S'(z)$ then we may remove from $S'(z)$ a $\leq_f$-minimal sequence $\beta$ and merge it with $\lambda$ (so $(\lambda, \beta)$ now replaces $\lambda$). We may repeat these two operations until we obtain a collision-free family $\mathcal{S}'$ such that all sequences $s \in S'(y)$ satisfy $s <_f \lambda$ and all sequences $t \in S'(z)$ satisfy $\lambda <_f t$. We may now replace these three sets by their union $S(x) = S'(y) \cup \{\lambda\} \cup S'(z)$ and obtain the desired collision-free family. At this point, we have dealt with the $S$-node $x$ and may proceed to another tree node.

To summarize the preceding discussion, we now give a pseudo-code for the recursive computation of a set of sequences $S(x)$ at a node $x$ of a series-parallel decomposition tree, so as to maintain a collision-free family of sets of sequences. We suppose that the sets of sequences at each node are recorded in a priority queue supporting the operations of `findmin`, `findmax`, `deletemin`, `deletemax`, and `merge`. The `findmin` operation returns a sequence which is $\leq_f$-minimal in the set, while `deletemin` returns such a sequence and also deletes it from the set; similarly for `findmax` and `deletemax`. When $L$ is empty the `findmax` and `deletemax` operations return a dummy job such that $\max L <_f j$, for all jobs $j$; similarly, when $R$ is empty, $\min R >_f j$, for all jobs $j$. Finally, `merge` forms the union of two sets of sequences.

$S(x)$ :

**Case** ($x$ is a leaf):      **return** $S := \{j\}$, where $j$ is the job at $x$;
**Case** ($x$ is a P node):    **return** $S := \mathtt{merge}(S(\mathrm{left}(x)), S(\mathrm{right}(x)))$;
**Case** ($x$ is an S node):
    $L := S(\mathrm{left}(x));$     $R := S(\mathrm{right}(x));$
    **if**  $\mathtt{findmax}(L) <_f \mathtt{findmin}(R)$ **then return** $S := \mathtt{merge}(L, R)$;
    **else**
        $s := (\mathtt{deletemax}(L), \mathtt{deletemin}(R));$        * concatenation *
        **while** $(\mathtt{findmax}(L) \geq_f s)$ **or** $(\mathtt{findmin}(R) \leq_f s)$
            **if**  $\mathtt{findmax}(L) \geq_f s$ **then** $s := (\mathtt{deletemax}(L), s);$
                        **else** $s := (s, \mathtt{deletemin}(R));$
        **endwhile**
    $S := \mathtt{merge}(L, R);$
    **return** $S := \mathtt{merge}(S, \{s\})$;
    **endif**.

Each of the priority queue operations can be implemented to run in $O(\log n)$ time, and each operation is performed no more than $O(n)$ times. The final sort of the strings obtained at the root of the tree requires no more than $O(n \log n)$ comparisons, and invariant (2) implies that this is an optimal sequence for the entire instance. Hence we have the following result.

**Theorem 4.11.** *Let series-parallel precedence constraints be specified by a decomposition tree. Given a preference order $\leq_f$ on sequences, an optimal feasible permutation of $N$ can be found with $O(n \log n)$ comparisons of sequences with respect to $\leq_f$, and at most $O(n \log n)$ time for other operations.*

In particular, a variety of single-machine scheduling with precedence constraints can be solved in $O(n \log n)$ time; see the exercises below for examples.

### Exercises

4.12. Prove that the preference order on sequences defined for the $\sum w_j C_j$ criterion is correct.

4.13. Show that $1 | |L_{\max}$ does not admit a preference order on sequences.

4.14. Show that there is no preference order on sequences when $f$ is determined by weighted monotone cost density functions.

4.15. Find a preference order on sequences for each of the following problems:

(a) Total weighted discounted completion time problem.

(b) Least cost fault detection problem.

4.16. Provide pseudocode for obtaining a collision-free set in the case of parallel-chains precedence constraints. You should be able to achieve $O(n)$ running time.

### 4.4.  NP-hardness of further constrained min-sum problems

Unfortunately, it is relatively easy to move from the world of polynomial-time solvable problems to the world of NP-hard ones. In this section, we will present three NP-hardness results: $1 | r_j | \sum C_j$, $1 | r_j, pmtn | \sum w_j C_j$, and $1 | prec | \sum w_j C_j$.

**Theorem 4.12.** *The problem $1 | r_j | \sum C_j$ is NP-hard in the strong sense.*

*Proof.*  We will show that the 3-PARTITION problem (see Chapter 2) reduces to the decision version of $1 | r_j | \sum C_j$. Consider an instance of 3-PARTITION, consisting of positive integers $a_1, \ldots, a_{3t}, b$, with $b/4 < a_j < b/2$ for all $j$ and $\sum_j a_j = tb$. Recall that this is a yes-instance if and only if the index set $T = \{1, \ldots, 3t\}$ can be partitioned into $t$ mutually disjoint 3-element subsets $S_1, \ldots, S_t$ with $\sum_{j \in S_i} a_j = b$ for $i = 1, \ldots, t$. We will define an instance of $1 | r_j | \sum C_j$ and an integer $Z$ such that there exists a schedule of value $\sum C_j \leq Z$ if and only if the instance of 3-PARTITION is a yes-instance.

The scheduling instance has three types of jobs. First, for each $j \in T$, there is a job $J_j$ with release date 0 and processing time $a_j$. Second, for each $i \in \{1, \ldots, t-1\}$, there are $v$ jobs $K_k^{(i)}$ with release date $ib$ and processing time 0 ($k = 1, \ldots, v$). Finally, there are $w$ jobs $L_\ell$ with release date $tb$ and processing time 1 ($\ell = 1, \ldots, w$). The values of $Z$, $v$ and $w$ will be defined later.

Suppose we have a yes-instance of 3-PARTITION, and consider the following schedule (see Figure 4.6). The three jobs $J_j$ with $j \in S_i$ are processed in the in-terval $[(i-1)b, ib]$, for $i = 1, \ldots, t$; the sum of their completion times is bounded from above by

$$Z_J = \sum_{i=1}^{t} 3ib = \frac{3}{2}t(t+1)b.$$

The jobs $K_k^{(i)}$ are scheduled to start at their release dates; their total completion time is equal to

$$Z_K = \sum_{i=1}^{t-1} vib = \frac{1}{2}vt(t-1)b.$$

The jobs $L_\ell$ are processed consecutively in the interval $[tb, tb+w]$; their total completion time is given by

$$Z_L = \sum_{\ell=1}^{w}(tb + \ell) = wtb + \frac{1}{2}w(w+1).$$

We now define $v = Z_J$, $w = Z_J + Z_K$, and $Z = Z_J + Z_K + Z_L$. The schedule corresponding to the yes-instance of 3-PARTITION has a value $\sum C_j \leq Z$.

Conversely, consider a schedule satisfying $\sum C_j \leq Z$. We claim that, in any such schedule, all $J_j$ are completed by time $tb$. If this is not the case, then at least one $J_j$ as well as all $L_\ell$ finish after $tb$, so that

$$\sum C_j > \sum_{\ell=1}^{w+1}(tb + \ell) = Z_L + tb + w + 1 > Z.$$

It may be assumed that, for any i, all $K_k^{(i)}$ $(k = 1, \ldots, v)$ are processed consecutively; this follows from a straighforward interchange argument. We now also claim that, if $\sum C_j \leq Z$, then all $K_k^{(i)}$ start at their release dates. Otherwise, at least $v$ of these jobs are delayed by one time unit, so that

$$\sum C_j > Z_K + v + Z_L = Z.$$

We conclude that, in any schedule of value $\sum C_j \leq Z$, all jobs $J_j$ are processed in the interval $[0, tb]$, interrupted by zero-time jobs at each point in time $ib$ $(i = 1, \ldots, t-1)$. Hence, in each interval $[(i-1)b, ib]$, three jobs $J_j$ are processed for a total duration of $b$ time units. This implies that we have a yes-instance of 3-PARTITION.

Note that the number of jobs is proportional to $b$, so that the correctness of the reduction essentially depends upon the *strong* NP-completeness of 3-PARTITION. Finally, let us mention that one can modify the reduction such that all jobs have a positive length (Exercise 4.17). □

We did mention earlier that, in the presence of release dates, there can be advantage to preemption. In fact, for $1|r_j|\sum C_j$ there is; in addition, $1|r_j, pmtn|\sum C_j$ can be solved efficiently by the *Shortest Remaining Processing Time* (*SRPT*) rule; see

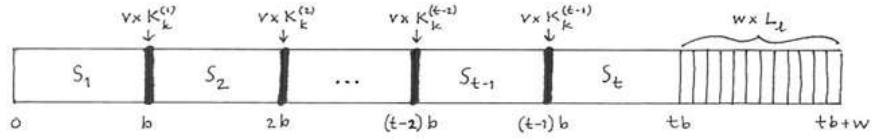

Figure 4.6

Theorem 4.25 below. However, the preemptive problem is NP-hard if jobs can have different weights.

**Theorem 4.13.** *The problem $1|r_j, pmtn| \sum w_j C_j$ is NP-hard in the strong sense.*

*Proof.* We will show that this problem is NP-hard in the ordinary sense, by a reduction from the PARTITION problem. It is not hard to extend this to a reduction from the 3-PARTITION problem, which implies NP-hardness in the strong sense; see Exercise 4.18.

An instance of PARTITION consists of positive integers $a_1, \ldots, a_t, b$ with $\sum_j a_j = 2b$. It is a yes-instance if and only if the index set $T = \{1, \ldots, t\}$ includes a subset $S$ with $\sum_{j \in S} a_j = b$.

Given an instance of PARTITION, we define $t$ jobs $j$ ($j = 1, \ldots, t$), with $r_j = 0$ and $p_j = w_j = a_j$. For these jobs, any nonpreemptive schedule without idle time is optimal: all $r_j = 0$, so that there is no advantage to preemption, and all $w_j/p_j = 1$, so that the ratio rule may choose any job order. The value of such a schedule is given by

$$Z_J = \sum_{1 \le j \le k \le t} a_j a_k.$$

(This can easily seen by considering 2-dimensional Gantt charts; see Section 4.5 below.) We define one more job, $K$, with release date $b$, processing time 1, and weight 2.

Suppose we have a yes-instance of PARTITION, and consider the following schedule. All jobs $j$ with $j \in S$ are processed in the interval $[0, b]$, job $K$ is scheduled in $[b, b+1]$, and the remaining jobs are processed in $[b+1, 2b+1]$ (see Figure 4.7(a)). Since these latter jobs are all delayed by one time unit, their contribution to $\sum w_j C_j$ increases by the sum of their weights, which is equal to the sum of their processing times. The value of this schedule is therefore equal to

$$Z = (Z_J + b) + 2(b+1) = Z_J + 3b + 2.$$

Now consider any feasible schedule, and suppose that job $K$ finishes at time $b+1+c$, for some $c \ge 0$. It may be assumed that there is no idle time in the interval $[0, 2b+1]$ and that job $K$ starts at time $b+c$. For the jobs that finish after job $K$, let $d$ be the total amount of processing done on them prior to $K$, for some $d \ge 0$ (see Figure 4.7(b)). Again, the contribution of these jobs to $\sum w_j C_j$ increases by the sum of their weights,

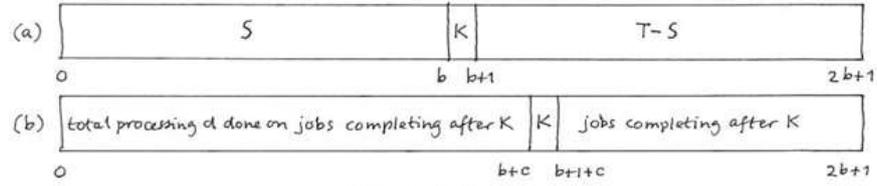

Figure 4.7

which is now equal to $d + b - c$. Hence, the value of this schedule is given by

$$(Z_J + d + b - c) + 2(b + c + 1) = Z_J + 3b + c + d + 2 = Z + c + d.$$

It follows that a schedule has value $\sum w_j C_j \leq Z$ if and only if $c = d = 0$; that is, job $K$ is processed in $[b, b+1]$ and each other job is entirely processed either before $K$ or after $K$. Such a schedule exists if and only if we have a yes-instance of PARTITION. □

Note that, alternatively, we could have given a very high weight to job $K$, thereby immediately fixing its starting time to $b$. The above technique, however, can be extended to yield similar and simple reductions to a number of related problems, including $1|\bar{d}_j|\sum w_j C_j$, $1||\sum w_j T_j$, and $P||\sum w_j C_j$.

We show the strong NP-hardness of the problem $1|prec|\sum w_j C_j$ by a two-step reduction from the LINEAR ARRANGEMENT problem, which is defined as follows: assign the vertices of an undirected graph $G = (V, E)$ to integer points on the real line so that the sum of edge lengths is minimized. More formally, given $G = (V, E)$ and a positive integer $Z$, is there a one-to-one mapping $f$ of $V$ to $\{1, \ldots, n\}$ such that $\sum_{\{u,v\} \in E} |f(u) - f(v)| \leq Z$? This problem is NP-complete, and we first reduce it to a version of $1|prec|\sum w_j C_j$ where jobs can have zero processing times and negative weights.

**Lemma 4.14.** *The optimal linear arrangement problem is polynomially reducible to the precedence-constrained single-machine scheduling problem with nonnegative processing times and arbitrary weights.*

*Proof.* Given an instance $G = (V, E)$ and $Z$ of the LINEAR ARRANGEMENT problem, we introduce a job for each vertex and one for each edge. Let $d_v$ be the degree of vertex $v$ in $G$. For each vertex $v \in V$, the corresponding job $v$ has weight $w_v = -d_v$ and processing time $p_v = 1$. For each edge $\{u, v\} \in E$, the corresponding job $\{u, v\}$ has weight $w_{\{u,v\}} = 2$ and processing time $p_{\{u,v\}} = 0$. Moreover, each edge job $\{u, v\}$ has exactly two predecessors, namely $u$ and $v$.

Suppose that we have a mapping $f$ of $V$ to $\{1, \ldots, n\}$ such that $\sum_{\{u,v\} \in E} |f(u) - f(v)| \leq Z$. Schedule the vertex jobs in the same order; i.e., $v$ is scheduled in position $f(v)$ among all vertex jobs. An edge job $\{u, v\}$ is scheduled immediately after its predecessor that completes later. The total weighted completion time of the resulting schedule is given by

$$\sum_{v \in V} w_v C_v + \sum_{\{u,v\} \in E} w_{\{u,v\}} C_{\{u,v\}}$$
$$= -\sum_{v \in V} d_v f(v) + \sum_{\{u,v\} \in E} 2 \max\{f(u), f(v)\}$$
$$= \sum_{\{u,v\} \in E} \left( 2 \max\{f(u), f(v)\} - f(u) - f(v) \right)$$
$$= \sum_{\{u,v\} \in E} |f(u) - f(v)| \ .$$

It is therefore at most $Z$. On the other hand, if there is a schedule of total weighted completion time at most $Z$, we may assume, without loss of generality, that each edge job $\{u,v\}$ is processed as soon as both jobs $u$ and $v$ are completed. Then, the same calculation as before implies that the order of the vertex jobs defines a solution of the given instance of the LINEAR ARRANGEMENT problem with value at most $Z$. □

Given an instance of $1 \mid prec \mid \sum w_j C_j$, adding a constant to the weight of every job not only changes the value of each schedule, but also can change the relative order of schedules with respect to their objective function values. However, if the constant is only added to the weights of jobs with processing time 1, and all other jobs have zero processing time, then the relative order of schedules is maintained. Let $d_{\max}$ be the maximal degree of a vertex in $G$. We can then add $d_{\max}$ to the weight of each vertex job in the proof of Lemma 4.14 to see that the scheduling problem remains NP-hard for instances with nonnegative weights.

One can also easily modify the reduction so that all jobs have positive processing times (Exercise 4.19). We have completed the proof of the following theorem.

**Theorem 4.15.** *The problem $1 \mid prec \mid \sum w_j C_j$ is NP-hard in the strong sense.*

**Exercises**

4.17. Modify the reduction in the proof of Theorem 4.12 so that all jobs in the resulting instance of $1 \mid r_j \mid \sum C_j$ have strictly positive processing times.

4.18. Give a reduction from the 3-PARTITION problem to show that the problem $1 \mid r_j, pmtn \mid \sum w_j C_j$ is indeed NP-hard in the *strong* sense.

4.19. Modify the reduction in the proof of Theorem 4.15 so that all jobs in the resulting instance of $1 \mid prec \mid \sum w_j C_j$ have *unit* processing time.

4.20. Modify the reduction in the proof of Theorem 4.15 so that all jobs in the resulting instance of $1 \mid prec \mid \sum C_j$ have unit weight and processing times 0 or 1.

### 4.5.  The ratio rule via linear programming

Linear programming methods have played a significant role in the development of combinatorial optimization; scheduling is no exception to this rule. We next present another proof of correctness for Smith's ratio rule using the tools of linear programming.

Let us draw a 2-dimensional Gantt chart where the "resource consumed" during the execution of every job $j$ is in fact the amount of processing, or *work* done on the job. The vertical axis may thus be interpreted as the *remaining work*; see Figure 4.8 for an illustration. Note that all jobs now have the same slope $\rho(j) = 1$. Recall that the mean busy time of a job $j$ in a nonpreemptive schedule is $M_j = C_j - p_j/2$.

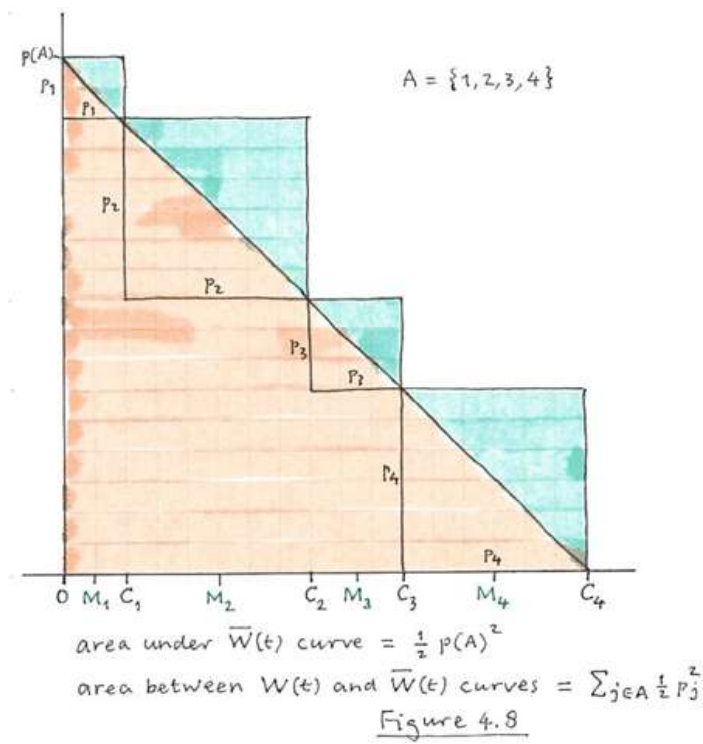

area under $\overline{W}(t)$ curve $= \frac{1}{2} p(A)^2$

area between $W(t)$ and $\overline{W}(t)$ curves $= \sum_{j \in A} \frac{1}{2} p_j^2$

Figure 4.8

Consider any subset $A \subseteq N$, where $N = \{1, \ldots, n\}$ is the set of all jobs to be processed. From the 2D Gantt chart, or from Smith's ratio rule, it follows that a schedule minimizes $\sum_{j \in A} p_j M_j$ if and only if all jobs in $A$ start at time zero and no idle time is incurred until the processing of all jobs in $A$ is complete. Jobs not in $A$ have a zero weight in the objective $\sum_{j \in A} p_j M_j$, and can be processed at any time after all jobs in $A$ are complete. The resulting minimum objective value is the area of the triangle below job set $A$, that is, $\frac{1}{2}\,p(A)^2 = \frac{1}{2}\left(\sum_{j\in A} p_j\right)^2$. This shows that for any subset $A \subseteq N$, the following so-called *parallel inequality*

$$\sum_{j\in A} p_j M_j \geq \frac{1}{2}\,p(A)^2 \tag{4.3}$$

holds for any feasible schedule in which no job can start before time 0 and the machine can process at most one job at a time. These inequalities are valid for *any* scheduling problem, whenever $N$ is a set of jobs or operations to be processed on a machine with unit speed and unit capacity.

The parallel inequalities may be written, using completion times instead of mean busy times, in the equivalent form

$$\sum_{j\in A} p_j C_j \geq \frac{1}{2}\,p(A)^2 + \frac{1}{2}\sum_{j\in A} p_j^2 \tag{4.4}$$

and may be interpreted as enforcing the requirement that no set of jobs can be completed too early on a machine with limited processing capacity.

The parallel inequalities (4.3) or (4.4), and variations or strengthenings thereof to take into account additional constraints or characteristics (such as precedence constraints, release dates, different processing speeds on different machines, etc.) play an important role in defining relaxations and approximation algorithms for more complicated scheduling problems.

In this light, we start by considering the problem that we know how to solve: $1|\,|\sum w_j M_j$. Suppose, without loss of generality, that we have reindexed the jobs so that $\rho(1) \geq \rho(2) \geq \cdots \geq \rho(n)$. Consider the linear program

$$\min\left\{\sum_{j\in N} w_j M_j : \sum_{j\in A} p_j M_j \geq \frac{1}{2}\,p(A)^2 \text{ for all } A \subseteq N\right\}. \tag{4.5}$$

We can show that the ratio rule is optimal by proving that if we schedule the jobs in the order $1, 2, \ldots, n$, this feasible solution is an optimal solution to this linear program. Consequently, this schedule must also be optimal for $1|\,|\sum w_j M_j$ (and also $1|\,|\sum w_j C_j$).

Let $\overline{M}_j$, $j = 1, \ldots, n$, denote the mean busy times of the jobs scheduled in the order $1, 2, \ldots, n$. We use linear programming duality to prove the optimality of this schedule. That is, we exhibit a feasible dual solution that satisfies the complementary slackness conditions with $\overline{M}_j$, $j = 1, \ldots, n$. The dual linear program has nonnegative variables $y_A$, for each $A \subseteq N$, and has a constraint for each $j = 1, \ldots, n$, $\sum_{A:j\in A} p_j y_A = w_j$, or equivalently, that $\sum_{A:j\in A} y_A = \rho(j)$.

The complementary slackness conditions for this pair of linear programs amount to $y_A > 0$ only if the corresponding parallel inequality is satisfied with equality. If we keep this in mind, then we can deduce a feasible solution for the dual as follows: the parallel inequalities corresponding to the sets $A_j = \{1, \ldots, j\}$ hold with equal-ity. Since $n$ is only in $A_n$, we set $y_{A_n} = \rho(n)$; working backwards, we see then that $y_{A_{n-1}} = \rho(n-1) - \rho(n)$, and in general, $y_{A_j} = \rho(j) - \rho(j+1)$, $j = 1, \dots, n-1$. The sorting of the jobs ensures that these values are all nonnegative. Moreover, $y_A = 0$ for all other $A \subseteq N$. Hence, we have feasible primal and dual solutions satisfying the complementary slackness conditions, and so they are optimal for their respective linear programs.

This proof of the correctness of the ratio rule actually says something much stronger, and truly remarkable. What it says, at its core, is that the parallel inequalities completely describe the feasible set of mean busy time vectors; more precisely, what we have shown is, in essence, that the feasible region of the linear program in $M_j$ variables is exactly the convex hull of vectors that correspond to mean busy times of feasible schedules. Of course, the same holds true for completion time vectors.

**Exercises**

4.21. If one would want to solve the linear program (4.5) by standard linear programming methods, one would have to deal with the exponential number of constraints. Because of the equivalence of optimization and separation (see Chapter 2), it suffices to solve the separation problem associated with the parallel inequalities $\sum_{j \in A} p_j M_j \geq \frac{1}{2} p(A)^2$, $A \subseteq N$. Given a vector $M^* \in \mathbb{Q}^N$, the separation problem is to decide whether $M^*$ satisfies all parallel inequalities and, if not, to produce a parallel inequality that is violated by $M^*$.

    (a) Defining the violation $v(A) = \frac{1}{2} p(A)^2 - \sum_{j \in A} p_j M_j^*$, $A \subseteq N$, compute $v(A \cup \{k\}) - v(A)$ for $k \notin A$, and $v(A) - v(A \setminus \{j\})$ for $j \in A$.

    (b) Assume, without loss of generality, that $M_1^* \leq M_2^* \leq \cdots \leq M_n^*$. Using (a), show that a parallel inequality most violated by $M^*$, if any, can be found among one of the consecutive sets $\{1\}, \{1,2\}, \dots, \{1,2,\dots,n\}$.

It follows that (4.5) can be solved in polynomial time. This should be hardly surprising, since it is solved by Smith's ratio rule. The real interest of the separation algorithm is in solving linear programming relaxations of more complicated scheduling problems, as will be seen later in this and other chapters.

4.22. Show that the mean busy times $M_j$, $j = 1, \dots, n$, of any feasible schedule for $1|r_j|\sum w_j C_j$ satisfy the inequalities $\sum_{j \in A} p_j M_j \geq (\min_{j \in A} r_j + \frac{1}{2} p(A)) p(A)$.

4.23. Solve the separation problem associated with the inequalities in Exercise 4.22.

## 4.6. Approximation algorithms for $1|prec|\sum w_j C_j$

In light of the NP-hardness of $1|prec|\sum w_j C_j$, it is natural to consider approximation algorithms. In Section 4.5, we used a linear program to give an alternative proof of optimality for Smith's ratio rule. Looking at this proof from a different angle, we see that sequencing jobs in nondecreasing order of their completion times in the solution to the linear program results in an optimal schedule for $1||\sum w_j C_j$. We now use the same idea to create a schedule whose cost is within a factor 2 of that of an optimum for the strongly NP-hard problem $1|prec|\sum w_j C_j$.

While the constraints of the linear program for $1||\sum w_j C_j$ consisted of the parallel inequalities (4.4) only, we add the following inequalities to make sure that the resulting schedule is consistent with the precedence constraints:

$$C_j - C_i \geq p_j \text{ for all pairs } i \rightarrow j. \qquad (4.6)$$

Let $C_j$, $j = 1, \ldots, n$, be a solution of the linear program to minimize $\sum_{j \in N} w_j C_j$ subject to the parallel inequalities (4.4) for all $A \subseteq N$ and the precedence constraints (4.6). Without loss of generality, we can reindex the jobs so that $C_1 \leq C_2 \leq \cdots \leq C_n$. Because of (4.6), $i < j$ whenever $i \rightarrow j$. In contrast to the case $1||\sum w_j C_j$, the values $C_j$ do in general not correspond to job completion times in an actual schedule, even when they correspond to a basic feasible solution; see Exercise 4.25. However, we can easily construct a feasible schedule by sequencing the jobs in the order $1, 2, \ldots, n$. The completion times $\overline{C}_j$ of this schedule are $\overline{C}_1 = p_1, \overline{C}_2 = p_1 + p_2, \ldots, \overline{C}_n = p_1 + p_2 + \cdots + p_n$. Note that this schedule does not violate any precedence constraints. We will now show that the total weighted completion time of this schedule is at most twice that of an optimal schedule.

Consider an arbitrary, but fixed job $k$. Recall that $(C_1, C_2, \ldots, C_n)$ satisfies the parallel inequalities (4.4); in particular, for $A = \{1, 2, \ldots, k\}$,

$$\sum_{j=1}^{k} p_j C_j \geq \frac{1}{2} \left( \sum_{j=1}^{k} p_j \right)^2,$$

where we have even dropped a nonnegative term from the right-hand side of inequality (4.4). Because of the ordering of jobs, the left-hand side of this inequality is bounded from above by $C_k \sum_{j=1}^{k} p_j$. We therefore obtain

$$\overline{C}_k = \sum_{j=1}^{k} p_j \leq 2 C_k .$$

Thus, $\sum_{j=1}^{n} w_j \overline{C}_j \leq 2 \sum_{j=1}^{n} w_j C_j$; given *any* feasible solution of the linear program to minimize $\sum_{j \in N} w_j C_j$ subject to all parallel inequalities and precedence constraints, we can construct a schedule with total weighted completion time at most twice its value. If we start with an optimal solution $C_j$, $j = 1, 2, \ldots, n$, of the linear program, then $\sum_{j \in N} w_j C_j$ is a lower bound on the cost of an optimal schedule; we have proved the following theorem.

**Theorem 4.16.** *Scheduling jobs in nondecreasing order of completion times in an optimal solution to the linear program* $\min\{\sum_{j=1}^{n} w_j C_j : C \text{ satisfies (4.4) and (4.6)}\}$ *is a 2-approximation algorithm for the problem* $1|prec|\sum w_j C_j$.

Actually, we still have to argue that the linear program in $C_j$ variables can be solved in polynomial time, because there are exponentially many parallel inequali-ties. While it follows from Exercise 4.21 that this is indeed the case, we here take a different route. Suppose there is another linear programming relaxation that only has polynomially many variables and constraints, so it can obviously be solved in polynomial time. Furthermore, suppose that every solution of the new linear program can be mapped to a feasible solution of the original linear program with no change of its objective function value. Then we could compute an optimal solution to the new linear program, map it to a solution of the old linear program, and apply the previous algorithm. We now describe such a linear program.

One way to derive another linear programming relaxation of the scheduling problem $1|prec|\sum w_jC_j$ is to formulate it as an integer program, and then relax the integrality constraints. Solving $1|prec|\sum w_jC_j$ is equivalent to determining, for each pair $i$ and $j$ of jobs, whether $i$ precedes $j$ in the solution, or not. For any pair $i \neq j$ of jobs, we introduce the variable $\delta_{ij} \in \{0,1\}$: $\delta_{ij} = 1$ indicates that job $i$ precedes job $j$, and $\delta_{ij} = 0$ indicates otherwise. Therefore,

$$\delta_{ij} + \delta_{ji} = 1 \text{ for all pairs } i,j \in N, i \neq j. \tag{4.7}$$

We can obviously represent every feasible schedule by a $0/1$-vector $\delta = (\delta_{ij})_{i \neq j}$ of this type; moreover, the resulting $0/1$-vectors satisfy further inequalities. The precedence constraints imply that

$$\delta_{ij} = 1 \text{ for all } i \rightarrow j. \tag{4.8}$$

In addition, when job $j$ precedes job $k$, and $k$ precedes $i$, then $j$ also precedes job $i$. This translates into the following inequalities:

$$\delta_{jk} + \delta_{ki} - \delta_{ji} \leq 1 \text{ for all triples } i,j,k \in N, i \neq j \neq k \neq i. \tag{4.9}$$

These inequalities are called *transitivity constraints*. On the other hand, every $0/1$-vector $\delta$ that satisfies the transitivity constraints (4.9), the precedence constraints (4.8), and (4.7) represents a feasible job sequence. Given such $\delta$, the completion time $C_j$ of job $j$ in the corresponding schedule is

$$C_j = \sum_{i \neq j} p_i \delta_{ij} + p_j. \tag{4.10}$$

So $1|prec|\sum w_jC_j$ is equivalent to minimizing $\sum_{j=1}^{n} \sum_{i \neq j} w_j p_i \delta_{ij} + \sum_{j=1}^{n} w_j p_j$ subject to the constraints $(4.7)-(4.9)$, and $\delta_{ij} \in \{0,1\}$ for all $i \neq j$. If we replace $\delta_{ij} \in \{0,1\}$ by

$$\delta_{ij} \geq 0 \text{ for all } i \neq j, \tag{4.11}$$

we obtain a linear program of polynomial size whose optimal value is a lower bound on the value of an optimal schedule. It remains to show that, if $\delta$ is a solution to this linear program, then the vector defined by (4.10), for $j = 1, 2, \ldots, n$, is a solution to the linear program in $C_j$ variables.

Let us consider the precedence constraints (4.6) first. Assume $i \to j$; then $\delta_{ij} = 1$, and (4.10) gives

$$C_j = \sum_{k \neq i,j} p_k \delta_{kj} + p_i + p_j \ .$$

Because of (4.7), $\delta_{ji} = 0$, and thus (4.10) yields

$$C_i = \sum_{k \neq i,j} p_k \delta_{ki} + p_i \ .$$

Let $k$ be a job with $k \neq i,j$. Since $\delta_{ji} = 0$, the transitivity constraint (4.9) for this triple implies that $\delta_{kj} = 1 - \delta_{jk} \geq \delta_{ki}$. Hence, $C_j \geq C_i + p_j$.

We now turn our attention to the parallel inequalities (4.4). Fix $A \subseteq N$. Then,

$$
\begin{aligned}
\sum_{j \in A} p_j C_j &= \sum_{j \in A} p_j \Big( \sum_{k \neq j} p_k \delta_{kj} + p_j \Big) = \sum_{\substack{j \in A, k \in N \\ j \neq k}} p_j p_k \delta_{kj} + \sum_{j \in A} p_j^2 \\
&\geq \sum_{\substack{j,k \in A \\ j \neq k}} p_j p_k \delta_{kj} + \sum_{j \in A} p_j^2 = \sum_{\substack{j,k \in A \\ j < k}} p_j p_k (\delta_{jk} + \delta_{kj}) + \sum_{j \in A} p_j^2 \\
&= \frac{1}{2} \, p(A)^2 + \frac{1}{2} \sum_{j \in A} p_j^2 \ .
\end{aligned}
$$

The last equality follows from (4.7).

**Corollary 4.17.** *Let $\delta$ be an optimal solution to the linear program $\min \{ \sum_{k \neq j} w_k p_j \delta_{jk} : \delta \geq 0 \ \text{satisfies} \ (4.7)-(4.9) \}$. Define $C_j$ according to (4.10), for $j = 1, 2, \ldots, n$, and sequence the jobs in nondecreasing order of $C_j$ values. Then, the total weighted completion time of the resulting schedule is at most twice that of an optimal schedule.*

The analysis of this 2-approximation algorithm is tight; i.e., there exist instances for which the resulting schedule has cost close to twice that of an optimal schedule. There are also instances for which the value of an optimal schedule is essentially twice that of the linear program; see Exercise 4.27.

The technique of scheduling jobs in order of their "completion times" in a linear programming relaxation of the problem is of use in more complicated problems than $1 \,|\, prec \,|\, \sum w_j C_j$, including parallel machine and open shop problems.

Interestingly, no approximation algorithm with a performance guarantee strictly less than 2 is known for the problem $1 \,|\, prec \,|\, \sum w_j C_j$. However, there is a generic way of designing a 2-approximation algorithm that does not require solving a linear program, which we describe next. For that, we need the concept of Sidney decomposition.

### Exercises

4.24. For an instance of $1 \,|\, | \sum w_j C_j$, consider the linear program $\min \{ \sum_{i,j:i \neq j} w_j p_i \delta_{ij} : \delta_{ij} + \delta_{ji} = 1, \delta_{ij} \geq 0 \ \text{for all} \ i \neq j \}$. Show that the optimal value of this linear program plus $\sum_j w_j p_j$ is equal to the value of an optimal schedule. How can one use this fact to give another proof of the optimality of Smith's ratio rule?

4.25. Consider the following instance of $1|prec|\sum w_j C_j$: there are three jobs of unit length each, job 1 precedes jobs 2 and 3, and $w_1 = 0$, $w_2 = 1$, and $w_3 = 1$. Show that $C_1 = 4/3$, $C_2 = 7/3$, and $C_3 = 7/3$ is an optimal basic feasible solution of the linear program $\min\{\sum_j w_j C_j : C \text{ satisfies (4.4) and (4.6)}\}$.

4.26. Show that the analysis that led to Theorem 4.16 is tight:

(a) Describe a family of instances for which the ratio of the objective function value returned by the algorithm to that of an optimal schedule converges to 2 with increasing values of $n$.

(b) Give a family of instances for which the ratio of the cost of an optimal schedule to that of an optimal solution to the linear program in completion time variables converges to 2 with increasing values of $n$.

4.27. Show that the analysis that led to Corollary 4.17 is tight.

## 4.7. Sidney decompositions

Recall that any optimal sequence for $1||\sum w_j C_j$ orders jobs according to nonincreasing ratios $w_j/p_j$. In the presence of precedence constraints, we already saw that this ordering can lead to collisions that cannot always be resolved. The Sidney decomposition offers one way to extend Smith's ratio rule to $1|prec|\sum w_j C_j$ by considering ratios of subsets. A subset $I \subseteq N$ that contains all its predecessors under the precedence constraints is said to be an *initial set* of $N$. That is, $I$ is an initial set if $j \in I$ and $i \to j$ imply $i \in I$. Note that a subset $I \subseteq N$ is an initial set of $N$ if and only if there is a feasible sequence that schedules all jobs in $I$ before all remaining jobs. A *ratio-maximal initial set* $I$ is an initial set of $N$ with maximum ratio $\rho(I) = w(I)/p(I)$. Here, as before, $w(I) = \sum_{j\in I} w_j$ and $p(I) = \sum_{j\in I} p_j$.

**Theorem 4.18.** *Let $I^*$ be a ratio-maximal initial set of $N$. There exists an optimal sequence of $N$ that schedules the jobs in $I^*$ before all remaining jobs.*

Before we prove this theorem, we discuss its implications and related aspects. First, note that an initial set is a closure in the directed graph defined by reversing all precedence constraints. A ratio-maximal initial set can therefore be computed in polynomial time; see Chapter 2. Second, we can apply Theorem 4.18 to the set $N \setminus I^*$ of remaining jobs. If we repeat this, then we eventually obtain a decomposition $(I_1, I_2, \ldots, I_k)$ of $N$ such that $I_i \cap I_j = \emptyset$, for all $1 \le i < j \le k$, $I_1 \cup I_2 \cup \ldots \cup I_k = N$, and $I_j$ is a ratio-maximal initial set of $N \setminus (I_1 \cup \ldots \cup I_{j-1})$. A decomposition of this kind is known as a *Sidney decomposition*. Note that each set $I_j$, for $j = 1, 2, \ldots, k$, is ratio-maximal for itself; i.e., $\rho(I) \le \rho(I_j)$ for all initial sets $I \subseteq I_j$. A sequence is *consistent with* a given Sidney decomposition if it first schedules the jobs in $I_1$, then the jobs in $I_2$, and so forth, until it eventually schedules all jobs in $I_k$. We can now reformulate Theorem 4.18 so that it encompasses Smith's ratio rule (Theorem 4.1).

**Corollary 4.19.** *A sequence $\pi = (\pi_1, \pi_2, \ldots, \pi_k)$ is optimal for $1 \mid prec \mid \sum w_j C_j$ if $\pi$ is consistent with a Sidney decomposition $(I_1, I_2, \ldots, I_k)$ of $N$, and each subsequence $\pi_j$ is optimal for $I_j$, $j = 1, \ldots, k$.*

Corollary 4.19 implies that it suffices to consider approximation algorithms that are consistent with a Sidney decomposition. In particular, if we had a 2-approximation algorithm for instances whose ground sets are ratio-maximal, then we could apply this algorithm to all sets $I_j$ in a Sidney decomposition, concatenate the resulting sequences in order, and thus obtain a 2-approximation algorithm for the entire instance. It turns out that we do not have to work hard to find such a 2-approximation algorithm.

An important structural property of instances with a ratio-maximal ground set $N$ is that sequencing the jobs in *any* feasible order constitutes a 2-approximation algorithm. Assume that $N$ is a ratio-maximal initial set of itself. That is, $\rho(I) \le \rho(N)$ for all initial sets $I \subseteq N$. Let us interpret this situation with the help of 2D Gantt charts. Consider an arbitrary feasible schedule, such as the one in Figure 4.9, and the line segment connecting the point $(0, w(N))$ on the vertical axis with the lower right corner of any rectangle representing a job $j$. The absolute value of the slope of this line segment is equal to the ratio $\rho(I)$ of the initial set $I$ of jobs defined by $I = \{k \in N : C_k \le C_j\}$. For any job $j$, the lower endpoint of this line segment is on or above the diagonal defined by the two points $(0, w(N))$ and $(p(N), 0)$, because $N$ is ratio-maximal and the negative value of the slope of this diagonal is $\rho(N)$. Recall that the area under the curve $W(t)$ is equal to $\sum_{j=1}^{n} w_j C_j$. So we have just argued that the total weighted completion time of any feasible schedule, especially that of an optimal one, is at least the area under the diagonal, which is $w(N)p(N)/2$. On the other hand, the cost of any schedule is at most $w(N)p(N)$. The proof of the following theorem is complete.

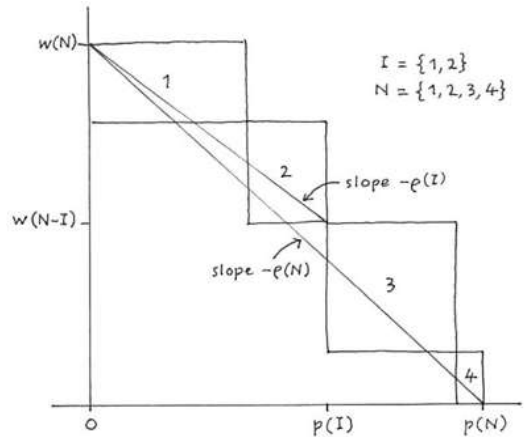

Figure 4.9

**Theorem 4.20.** *The total weighted completion time of any feasible sequence for*
$1 \mid prec \mid \sum w_j C_j$ *that is consistent with a Sidney decomposition is at most twice that of*
*an optimal schedule.*

So even if the only ratio-maximal initial set is $N$ itself, i.e., we cannot divide the
original problem into smaller pieces, we are at least assured that no feasible sequence
is too far from optimal.

It is time to prove Theorem 4.18. For notational convenience, we use the equiv-
alent integer programming formulation of $1 \mid prec \mid \sum w_j C_j$ to do so. We have to prove
that, if $I^*$ is a ratio-maximal initial set of $N$, then there exists an optimal solution
$\delta^*$ to the integer program defined by the constraints (4.7) – (4.9) and the objective
$\sum_{i \neq j} w_j p_i \delta_{ij}$ such that $\delta^*_{ij} = 1$ for all $i \in I^*, j \in N \setminus I^*$.

*Proof.* [Theorem 4.18] Let $\delta$ be an optimal solution of the integer program. Suppose
$\delta_{ij} < 1$ for some $i \in I^*, j \in N \setminus I^*$. For each $k \in I^*$, define $I_k = \{j \in N \setminus I^* : \delta_{jk} > 0\}$. If a job $i \in N \setminus I^*$ is a predecessor of $j \in I_k$, i.e., $i \to j$, then the transitivity
constraint (4.9) applied to the triple $i, j, k$, and $\delta_{ji} = 1 - \delta_{ij} = 0$ imply that $\delta_{ik} = 1 - \delta_{ki} \geq \delta_{jk} > 0$. Hence, $i \in I_k$, and $I_k$ is an initial set of $N \setminus I^*$. Similarly, for each
$k \in N \setminus I^*$, $F_k = \{i \in I^* : \delta_{ki} > 0\}$ is a final set of $I^*$. (That is, $I^* \setminus F_k$ is an initial set
of $I^*$.)

Let $\varepsilon = \min\{\delta_{ij} : i \in N \setminus I^*, j \in I^*, \delta_{ij} > 0\}$, and consider the vector $\delta^*$ defined as:

$$\delta^*_{ij} = \begin{cases} \delta_{ij} + \varepsilon & \text{if } i \in I^*, j \in N \setminus I^*, \text{ and } \delta_{ij} < 1; \\ \delta_{ij} - \varepsilon & \text{if } i \in N \setminus I^*, j \in I^*, \text{ and } \delta_{ij} > 0; \qquad \text{for } i, j \in N, i \neq j. \\ \delta_{ij} & \text{otherwise}; \end{cases}$$

Clearly, $\delta^*_{ij} + \delta^*_{ji} = 1$ for all $i \neq j$, and $\delta^*_{ij} = 1$ for all $i \to j$. Consider now a triple
$i, j, k \in N, i \neq j \neq k \neq i$. If $\{i, j, k\} \in I^*$ or $\{i, j, k\} \in N \setminus I^*$, then the associated tran-
sitivity constraint is satisfied because $\delta^*$ and $\delta$ coincide in the relevant components.
Otherwise, it is convenient to rewrite the transitivity constraint $\delta_{jk} + \delta_{ki} - \delta_{ji} \leq 1$ as
$\delta_{jk} + \delta_{ki} + \delta_{ij} \leq 2$. We may assume, because of symmetry, that $i \in I^*$ and $j \in N \setminus I^*$.
Suppose that $\delta^*_{jk} + \delta^*_{ki} + \delta^*_{ij} > 2$. There are two cases: either $k \in I^*$ or $k \in N \setminus I^*$. In the
first case, $\delta^*_{ki} = \delta_{ki}$, so we must have $\delta^*_{ij} = \delta_{ij} + \varepsilon$ and $\delta^*_{jk} > 0$. But then $\delta_{jk} = \delta^*_{jk} + \varepsilon$,
and, therefore, $\delta^*_{jk} + \delta^*_{ki} + \delta^*_{ij} = \delta_{jk} + \delta_{ki} + \delta_{ij}$, a contradiction. The second case is
handled analogously. Thus, $\delta^*$ also satisfies the transitivity constraints.

The difference in the objective function values of $\delta$ and $\delta^*$ can be calculated as
follows:

$$\begin{aligned} \sum_{\substack{i,j \in N \\ i \neq j}} p_i w_j \delta_{ij} - \sum_{\substack{i,j \in N \\ i \neq j}} p_i w_j \delta^*_{ij} &= \varepsilon \sum_{k \in N \setminus I^*} p_k w(F_k) - \varepsilon \sum_{k \in I^*} p_k w(I_k) \\ &= \varepsilon \sum_{k \in N \setminus I^*} p_k p(F_k) \rho(F_k) - \varepsilon \sum_{k \in I^*} p_k p(I_k) \rho(I_k) \ . \end{aligned}$$

Because $I^*$ is ratio-maximal, $\rho(I_k) \leq \rho(I^*) \leq \rho(F_k)$, for all $k$; see Exercise 4.28. Hence, the difference in objective function values can be bounded from below by

$$\varepsilon \rho(I^*) \Big( \sum_{k \in N \setminus I^*} p_k p(F_k) - \sum_{k' \in I^*} p_{k'} p(I_{k'}) \Big) \ .$$

Because $k' \in F_k$ if and only if $k \in I_{k'}$, this expression evaluates to zero. Therefore, the objective function value of $\delta^*$ is not worse than that of $\delta$. So $\delta^*$ is an optimal solution as well, and $\delta^*_{ij} = 1$ for all $i \in I^*$ and $j \in N \setminus I^*$. $\square$

Of course, in the proof $\varepsilon = 1$, and the reader might have wondered why we avoided using the integrality of $\delta$ and $\delta^*$. The reason is that the result holds true for the linear programming relaxation in $\delta$ variables as well. That is, if $I^*$ is a ratio-maximal initial set, then there exists an optimal solution $\delta^*$ to the linear program to minimize $\sum_{j \in N} w_j C_j$ subject to (4.7) – (4.11) such that $\delta^*_{ij} = 1$ whenever $i \in I^*$ and $j \in N \setminus I^*$. We can use the same proof; while $\varepsilon$ need not be equal to 1 anymore, the variable that determined the value of $\varepsilon$ will be reduced to 0 in $\delta^*$. If $\delta^*$ does not have the desired property yet, we can repeat the same procedure with $\delta = \delta^*$ until it does.

**Corollary 4.21.** *Let $(I_1, I_2, \ldots, I_k)$ be a Sidney decomposition of $1 \mid prec \mid \sum w_j C_j$. There exists an optimal solution $\delta^*$ of the linear program to minimize $\sum_{j \in N} w_j C_j$ subject to (4.7) – (4.11) such that $\delta^*_{ij} = 1$ for all $i \in I_h$, $j \in I_\ell$ whenever $1 \leq h < \ell \leq k$.*

### Exercises

4.28. Let $S$ be a ratio-maximal initial set of $N$, $I$ an initial set of $N \setminus S$, and $F$ a final set of $S$. Show that $\rho(I) \leq \rho(S) \leq \rho(F)$.

4.29. Let $I$ and $J$ be two ratio-maximal initial sets of $N$. Are $I \cap J$ and $I \cup J$ ratio-maximal initial sets as well? Prove or disprove.

4.30. Show that there exists a unique Sidney decomposition $(I_1, I_2, \ldots, I_k)$ such that $\rho(I_1) > \rho(I_2) > \cdots > \rho(I_k)$. Prove also that for any other Sidney decomposition $(J_1, J_2, \ldots, J_\ell)$ of the same instance and any index $i \in \{1, \ldots, \ell\}$, $J_i \subseteq I_j$ for some $j \in \{1, \ldots, k\}$.

4.31. Let $(N, \rightarrow)$ be a series-parallel partial order that is the parallel composition of $(N_1, \rightarrow)$ and $(N_2, \rightarrow)$. Show that there exists a ratio-maximal initial set $I$ of $N$ that is completely contained in either $N_1$ or $N_2$.

### 4.8. An integrality theorem for series-parallel precedence constraints

With Corollary 4.21 in place, the following result comes essentially for free.

**Theorem 4.22.** *When the precedence constraints are series-parallel, then the linear program to minimize $\sum_{j \in N} w_j C_j$ subject to (4.7) – (4.11) has an integer optimal solution.*

*Proof.* The proof is by induction on the number of jobs. The result is obviously true if there are just one or two jobs. Suppose it is true for all instances with series-parallel precedence constraints on $n$ jobs, and consider the case with $|N| = n+1$. The set $N$ is either a series or a parallel composition of two proper subsets $N_1$ and $N_2$. In the first case, all variables $\delta_{ij}$ with $i \in N_1$ and $j \in N_2$ have to be equal to 1, because of (4.8). Hence, the linear program for $N_1 \cup N_2$ decomposes into two separate linear programs, one for $N_1$ and one for $N_2$, each of which has an integer optimal solution, by induction. In the second case, $N$ is the parallel composition of $N_1$ and $N_2$. Let $I^*$ be an inclusion-minimal ratio-maximal initial set of $N$; it follows from Exercise 4.31 that $I^*$ is entirely contained in $N_1$ or $N_2$. By Corollary 4.21, there exists an optimal solution $\delta^*$ such that $\delta^*_{ij} = 1$ for $i \in I^*$ and $j \in N \setminus I^*$. The linear program for $N$ therefore decomposes again into two separate, smaller linear programs for $I^*$ and $N \setminus I^*$, respectively, which have integer optimal solutions. $\square$

Theorem 4.22 implies that the optimal value of the linear program is identical to the value of the optimal schedule. It does not necessarily imply that an optimal solution to the linear program is integer (and, therefore, a schedule) because there can be noninteger optimal basic feasible solutions, although this can be overcome by data perturbation (see Exercise 4.32 (a) and (b)). There is a related, seemingly weaker linear program in $\delta$ variables that has the same properties as the one that we have presented so far. It has the additional property that *all* its basic feasible solutions are integer if the precedence constraints are series-parallel. To prove this result, it will be convenient to work with the following characteristic of series-parallel precedence constraints.

**Lemma 4.23.** *If the precedence constraints are series-parallel, then there exists a total ordering $\pi$ of all jobs, which is consistent with the precedence constraints, such that, for all triples $i, j, k \in N$ with $i \to j$ and $k$ unrelated to $i$ and $j$, either $k$ precedes $i$ in $\pi$ or $j$ precedes $k$.*

*Proof.* Consider the decomposition tree of the precedence constraints, and the total ordering of jobs obtained by parsing the leafs from left to right. Note that this ordering is consistent with the precedence constraints; if $i \to j$, then the leaf corresponding to $i$ is to the left of that of $j$. Now consider a job $k$ that is unrelated to $i$ and $j$. The tree has an internal node representing a parallel composition where $i$ and $j$ are part of one of the two subtrees, while $k$ is in the other subtree. Hence, $k$'s leaf is to the left of the leafs of both $i$ and $j$, or to their right. $\square$

We use the total ordering $\pi$ from Lemma 4.23 to eliminate half of the $\delta$ variables from the previous linear program. Because of (4.7), i.e., $\delta_{ij} + \delta_{ji} = 1$, we know the value of $\delta_{ij}$ when we know that of $\delta_{ji}$, and vice versa. We only keep the variables $\delta_{ij}$ for which $i$ precedes $j$ in $\pi$. We also eliminate all variables whose values are fixed by the precedence constraints; i.e., we do not keep $\delta_{ij}$ with $i \to j$. Apart from the nonnegativity constraints, we are left with the transitivity constraints, some of which we shall drop as well. More precisely, we only keep the transitivity constraints

$$\delta_{jk} + \delta_{ki} - \delta_{ji} \leq 1$$

for triples $i, j, k \in N$, $i \neq j \neq k \neq i$, for which $i \to j$, and $k$ is unrelated to both $i$ and $j$. Depending on whether $j$ precedes $k$ or $k$ precedes $i$ in $\pi$, we derive one of the following two inequalities:

$$\delta_{jk} \leq \delta_{ik} \text{ or } \delta_{ki} \leq \delta_{kj} \ .$$

For convenience, let us assume that we (re)index the jobs so that they appear in the order $1, 2, \ldots, n$ in $\pi$. The resulting linear program,

$$\min \quad \sum_{\substack{i<j \\ i \neq j}} w_j p_i \delta_{ij} \tag{4.12a}$$

$$\text{s.t.} \quad \delta_{jk} \leq \delta_{ik} \qquad \text{for } i \to j, j < k, \tag{4.12b}$$

$$\delta_{ki} \leq \delta_{kj} \qquad \text{for } i \to j, k < i, \tag{4.12c}$$

$$\delta_{ij} \geq 0 \qquad \text{for } i < j, \tag{4.12d}$$

is a min-weight closure problem (see Chapter 2; in particular, its constraint matrix is totally unimodular, and thus all basic feasible solutions are integral.

Interestingly, the linear program (4.12) has the same optimal value as the previous one, although it has fewer constraints (Exercise 4.32 (c)). Because one can find an optimal sequence when one can determine the optimal value (Exercise 4.33), this yields another way of efficiently computing an optimal sequence for series-parallel precedence constraints. As the derivation of the linear program (4.12) only hinges on Lemma 4.23, we have actually proved that $1|prec|\sum w_j C_j$ can be solved in polynomial time for a much larger class of precedence constraints. The class of partial orders possessing a *nonseparating linear extension*, i.e., a total order $\pi$ with the property described in Lemma 4.23 coincides with that of *two-dimensional* partial orders. The *dimension* of a partial order $\to$ is the minimum number of linear orders whose intersection is $\to$. There are efficient algorithms for recognizing two-dimensional partial orders. If the partial order is two-dimensional, they return a nonseparating linear extension. We can conclude this section with the following theorem.

**Theorem 4.24.** *If the precedence constraints are two-dimensional, an optimal sequence for $1|prec|\sum w_j C_j$ can be found in polynomial time.*

### Exercises

4.32. Consider the linear program to minimize $\sum_{j \in N} w_j C_j$ subject to (4.7) – (4.11).

(a) Show that it can have nonintegral optimal basic feasible solutions, even if there are no precedence constraints.

(b) Let $0 < \varepsilon < 1/(2 \sum_{j \in N} p_j)$, and assume that the precedence constraints are series-parallel. Prove that the instance with redefined weights $\widetilde{w}_j = w_j + \varepsilon 2^j$ has a unique Sidney decomposition. Use this fact in the proof of Theorem 4.22 to show that the linear program has a unique optimal solution, which is a schedule.

(c) Show that any optimal solution to this linear program is also an optimal solution to the linear program (4.12).

**4.33.** Suppose someone gives you a "black box" subroutine that, given an instance of $1|prec|\sum w_jC_j$, returns its optimal value. Show that you can find an optimal sequence with a polynomial number of calls on the subroutine.

**4.34.** Show that a partial order is two-dimensional if and only if it has a nonseparating linear extension.

**4.35.** Present a two-dimensional partial order that is not series-parallel.

## 4.9. Approximation algorithms for $1|r_j|\sum C_j$

Once we add release dates to the problem of minimizing the average completion time, we again reach the realm of *NP*-complete problems. Since we are unlikely to have a polynomial-time algorithm to solve this problem, we turn our attention to approximation algorithms. Our approach will again be based on solving a relaxation to the problem; in this case, we relax the condition that the schedule must be nonpreemptive. We then show how to convert the optimal preemptive schedule to a nonpreemptive one, without increasing the objective function value of the schedule too much.

The first component of this approach is that the preemptive variant can be solved efficiently. The natural rule to consider is the *Shortest Remaining Processing Time* (*SRPT*) rule: at each moment in time, one should always be processing a job that could be completed earliest. This rule only preempts jobs when a new job is released. A straightforward interchange argument can be used to prove the following theorem; this will be left as an exercise.

**Theorem 4.25.** *Any SRPT schedule is optimal for the problem $1|r_j,pmtn|\sum C_j$.*

An important property of the SRPT rule is that it is an online algorithm, in the sense that the algorithm need only know of the existence of a job and its parameters at the moment in time that it is released.

A seemingly naive approach to converting a preemptive schedule into a nonpreemptive one is to schedule the jobs in the order in which they completed in the preemptive one. Yet, recall that a similar approach has worked well for $1|prec|\sum w_jC_j$. Let $C_j$, $j = 1\ldots,n$, denote the completion times of jobs in a preemptive schedule. We can assume, without loss of generality, that we have reindexed the jobs so that $C_1 \leq C_2 \leq \cdots \leq C_n$. We construct the minimal nonpreemptive schedule in which the jobs are processed in this order; job 1 is processed from $r_1$ to $r_1 + p_1$, and each job $j + 1$ is either scheduled to start at the time $\overline{C}_j$ that $j$ completes, or at its release date, $r_{j+1}$, whichever is later.

We can analyze the completion time $\overline{C}_j$ of job $j$ in the following way. Trace backwards from $\overline{C}_j$ in this schedule to find the latest idle time prior to the completion of job $j$; we see that

$$\overline{C}_j = r_k + \sum_{\ell=k}^{j} p_\ell,$$

where job $k$ is the first job processed after this idle time. Since the completion of job $k$ also precedes the completion of job $j$ in the preemptive schedule, we have that $r_k \leq C_k \leq C_j$. Furthermore, all of the jobs $\ell$, $\ell = k, \ldots, j$, have been completely processed in the preemptive schedule by $C_j$, and hence $\sum_{\ell=k}^{j} p_\ell \leq C_j$. Therefore $\overline{C}_j \leq 2C_j$.

Thus, we see that $\sum_{j=1}^{n} w_j \overline{C}_j \leq 2 \sum_{j=1}^{n} w_j C_j$; given *any* preemptive schedule, we can find a nonpreemptive schedule with total weighted completion time no more than twice that of the preemptive one. Of course, if we start with the optimal preemptive schedule, then we have found a nonpreemptive schedule with objective function value at most twice that of the preemptive optimum. The preeemptive optimum is always at most the nonpreemptive optimal value, and so we have just proved that the schedule found has total weighted completion time at most twice the optimal value.

**Theorem 4.26.** *Sequencing jobs in order of nondecreasing completion times in the SRPT schedule is a 2-approximation algorithm for the problem* $1|r_j|\sum C_j$.

In fact, we can strengthen this result to derive an algorithm that is an online 2-approximation algorithm. The idea behind this is quite simple. We maintain a queue of available jobs (which handles the jobs in a "first-in first-out" manner, and hence fixes the order in which they are processed in the schedule). Instead of making a job available at its release time, we will use the online SRPT rule to create a *shadow* schedule. This schedule is not used for the actual processing of the jobs, but only as a kind of side computation. Nonetheless, this side computation can be done in an online manner.

We make a job $j$ available only at date $C_j^{\text{SRPT}}$, the time that it is *completed* in the SRPT schedule. Clearly, this queueing mechanism ensures that the jobs are processed in the order of their completion in the preemptive schedule, but it most likely introduces additional idle time into the schedule. However, the analysis does not change much. Now, the latest idle time prior to $\overline{C}_j$ ends with the point in time that some job $k$ is made available (instead of being released). Hence,

$$\overline{C}_j = C_k^{\text{SRPT}} + \sum_{\ell=k}^{j} p_\ell \leq 2C_j^{\text{SRPT}},$$

and we still have that $\sum w_j \overline{C}_j \leq 2 \sum w_j C_j^{\text{SRPT}}$; we have proved the following theorem.

**Theorem 4.27.** *Scheduling jobs in the order of nondecreasing SRPT completion times with delayed starts is an online 2-approximation algorithm for the problem* $1|r_j|\sum C_j$.

Surprisingly, this result is best possible, in the sense that, for any $\alpha < 2$, there does not exist a deterministic online $\alpha$-approximation algorithm for $1|r_j|\sum C_j$ (Exercise 4.37). However, as we shall see next, if the algorithm is allowed to "toss coins," one can prove a better result, in expectation. For each input, the objective function value of the schedule found by such a randomized algorithm is a random variable; it is natural to analyze the performance of such a randomized algorithm by considering its expected value, and to show, for each possible input, that this expectation is no more than a factor of $\rho$ times the optimum value. We call an algorithm with this performance guarantee a *randomized $\rho$-approximation algorithm*.

We will now give an online randomized $e/(e-1)$-approximation algorithm called the *randomized $\alpha$-point algorithm*. The algorithm is based on a more general procedure to convert a preemptive schedule $\sigma$ into a nonpreemptive one. The algorithm first picks an $\alpha \in (0,1]$ according to some probability density function $f$. We define the *$\alpha$-point $C_j(\alpha)$* of job $j$ to be the first moment in time where a total of $\alpha p_j$ units of processing of job $j$ have been completed in $\sigma$. We now modify the online algorithm discussed above, so that each job is made available, and placed into the queue, only at its $\alpha$-point. This means that the algorithm now schedules the jobs in the order that the $\alpha$-points occur in $\sigma$, with the additional condition that $j$ is not allowed to be started prior to $C_j(\alpha)$. Subject to these two constraints, the schedule $\bar{\sigma}$ produced is minimal in the sense that each jobs starts as early as possible. Let $\overline{C}_j$ denote the completion time of job $j$ in $\bar{\sigma}$; that is, if $\pi$ is the order in which the jobs are processed, then $\pi(1)$ completes at time $\overline{C}_{\pi(1)} = C_{\pi(1)}(\alpha) + p_{\pi(1)}$, and in general, each job $\pi(j)$ starts at the maximum of its $\alpha$-point in $\sigma$ and $\overline{C}_{\pi(j-1)}$, and completes $p_{\pi(j)}$ time units later, $j = 2,\ldots,n$. We will analyze the expected objective function value $\sum_j \overline{C}_j$ as a function of the choice of $\alpha$.

We next introduce notation that will be convenient for the analysis of $\bar{\sigma}$. We will give, for each $j = 1,\ldots,n$, an upper bound on $\overline{C}_j$ that depends on our choice of $\alpha$. Let us focus on the completion time of some job $j$. Let $\beta_k(\alpha)$ be the fraction of job $k$ that has been completed in the preemptive schedule $\sigma$ by the $\alpha$-point of job $j$, that is, by $C_j(\alpha)$. For simplicity, let $\beta_k = \beta_k(1)$. Similarly, let $R(\alpha)$ denote the total idle time in the preemptive schedule $\sigma$ prior to the $\alpha$-point of $j$. Since the machine, at any point in time, is either processing some job, or idle, we get the following lemma.

**Lemma 4.28.** *In the preemptive schedule $\sigma$, for each $\alpha \in (0,1]$, the $\alpha$-point of job $j$ is equal to $C_j(\alpha) = R(\alpha) + \sum_{k=1}^{n} \beta_k(\alpha) p_k$.*

On the other hand, we show next that we can also express the completion time of $j$ in $\bar{\sigma}$ in terms of similar components.

**Lemma 4.29.** *In the nonpreemptive schedule $\bar{\sigma}$, job $j$ completes by time*

$$\overline{C}_j \leq C_j(\alpha) + \sum_{k:\beta_k(\alpha) \geq \alpha} (1 + \alpha - \beta_k(\alpha)) p_k.$$

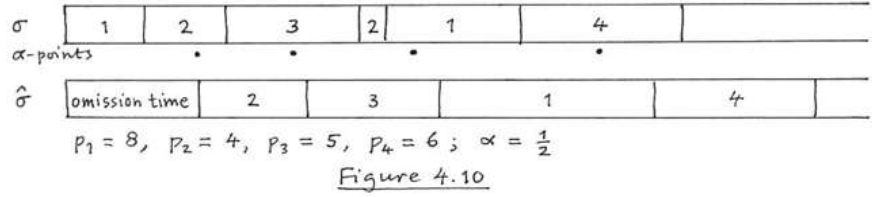

$p_1 = 8$, $p_2 = 4$, $p_3 = 5$, $p_4 = 6$ ; $\alpha = \frac{1}{2}$

Figure 4.10

*Proof.* We shall analyze a worse nonpreemptive schedule $\hat{\sigma}$, in the sense that for each $k = 1, \ldots, n$, we have that the completion time $\hat{C}_k \geq \overline{C}_k$.

Imagine starting out with the optimal schedule $\sigma$ for the preemptive relaxation, and write the Gantt chart for this schedule on a strip of ticker tape. We will modify this schedule by a number of "cut and paste" operations to convert it into a nonpreemptive schedule $\hat{\sigma}$.

Consider the schedule $\sigma$ "over time," starting at time 0 and advancing in time, until an $\alpha$-point of some job $k$ is reached. Cut the schedule at this point, and paste into this schedule a block of time of length $p_k$ that will be used for the nonpreemptive processing of job $k$. For the intervals of time that are spent processing job $k$ earlier in this schedule, replace the processing of job $k$ with new idle time, which we will call *omission time*. For each interval of time that is spent processing $k$ after this nonpreemptive block, cut that time interval out of the schedule, and then rejoin the schedule without any new idle time introduced at this point. Continue processing the schedule in this way until the $\alpha$-point of each job has been reached. Call this schedule $\hat{\sigma}$. See Figure 4.10 for an illustration.

We begin by making a number of observations. The effect of inserting the nonpreemptive block of length $p_k$ shifts the entire subsequent schedule exactly $p_k$ time units later. However, the deletion of the processing of job $k$ beyond this new block of length $p_k$ shifts earlier some portion of the schedule beyond this block. A total of length $(1 - \alpha)p_k$ is deleted, and so the end of the schedule is shifted $(1 - \alpha)p_k$ time units earlier by these deletions. However, the net effect of the changes prompted by reaching the $\alpha$-point of job $k$ does not shift any portion of the schedule earlier. This allows us to conclude, for example, that the schedule is feasible, since each job $k$ starts no earlier than $r_k$ before the modifications, and hence must start no earlier than $r_k$ after them. Furthermore, we also see that as we modify the schedule, when we find an $\alpha$-point of job $k$ in the schedule, it must currently correspond to a time that is at least $C_k(\alpha)$. Thus, we are scheduling the nonpreemptive block for each job $k$ to start no earlier than $C_k(\alpha)$. Finally, the jobs are processed in this schedule in the same order as their $\alpha$-points in $\sigma$. We have constructed a schedule $\hat{\sigma}$ in which the jobs are processed in the same order as in $\sigma$, and no job $k$ starts before $C_k(\alpha)$. Since $\overline{\sigma}$ is the minimal nonpreemptive schedule that is consistent with these constraints, we see that $\overline{C}_k \leq \hat{C}_k$, $k = 1, \ldots, n$.

We now analyze the schedule $\hat{\sigma}$ up to the completion time of job $j$. Each time interval on the ticker tape after these operations corresponds to one of the following:

- true idle time (i.e., in the preemptive schedule $\sigma$, not omission time): these add up exactly to $R(\alpha)$ since we have not altered true idle time.

- processing time: the jobs $k$ that are processed in $\hat{\sigma}$ by time $\hat{C}_j$ are exactly those that reach their $\alpha$-point in $\sigma$ no later than $C_j(\alpha)$; in terms of the notation that we have set up, this is equivalent to writing that $\beta_k(\alpha) \geq \alpha$. Hence this total is equal to $\sum_{k:\beta_k(\alpha)\geq\alpha} p_k$.

- omission time for each job $k$ that has reached its $\alpha$-point in $\sigma$ by time $C_j(\alpha)$ (including $j$ itself): for each job $k$, we replaced the first $\alpha p_k$ times unit of its processing by omission time (here all of this time occurs prior to the processing of $j$); hence this totals $\sum_{k:\beta_k(\alpha)\geq\alpha} \alpha p_k$.

- omission time for each job $k$ that reaches its $\alpha$-point in $\sigma$ after time $C_j(\alpha)$: in this case, considering just the part of schedule $\hat{\sigma}$ up to $\hat{C}_j$, only the fraction $\beta_k(\alpha)$ of the processing of $k$ that precedes $C_j(\alpha)$ is replaced by omission time; hence this contributes a total of $\sum_{k:\beta_k(\alpha)<\alpha} \beta_k(\alpha) p_k$.

This yields

$$\hat{C}_j = R(\alpha) + \sum_{k:\beta_k(\alpha)<\alpha} \beta_k(\alpha) p_k + (1+\alpha) \sum_{k:\beta_k(\alpha)\geq\alpha} p_k.$$

Applying Lemma 4.28, we see that this can be written as

$$\hat{C}_j = C_j(\alpha) + \sum_{k:\beta_k(\alpha)\geq\alpha} (1+\alpha-\beta_k(\alpha)) p_k.$$

Since $\overline{C}_j \leq \hat{C}_j$, for any given choice of $\alpha$,

$$\overline{C}_j \leq C_j(\alpha) + \sum_{k:\beta_k(\alpha)\geq\alpha} (1+\alpha-\beta_k(\alpha)) p_k.$$

This completes the proof of the lemma. $\square$

This also has the following immediate consequence.

**Corollary 4.30.** *In the nonpreemptive schedule $\hat{\sigma}$, job $j$ completes by time*

$$\overline{C}_j \leq C_j + \sum_{k:\beta_k\geq\alpha} (1+\alpha-\beta_k) p_k.$$

*Here, $C_j$ is the completion time of job $j$ in the preemptive schedule $\sigma$.*

*Proof.* From Lemma 4.29,

$$
\begin{aligned}
\overline{C}_j &\leq C_j(\alpha) + \sum_{k:\beta_k(\alpha)\geq\alpha} (1+\alpha-\beta_k(\alpha)) p_k \\
&= C_j(\alpha) + \sum_{k:\beta_k(\alpha)\geq\alpha} (\beta_k-\beta_k(\alpha)) p_k + \sum_{k:\beta_k(\alpha)\geq\alpha} (1+\alpha-\beta_k) p_k.
\end{aligned}
$$

The second term in this last expression corresponds to work done on jobs $k$ between $C_j(\alpha)$ and $C_j$, and hence the sum of the first two terms is at most $C_j$. $\square$

It is now just a straightforward calculation to compute an upper bound on the expected completion time $E[\overline{C}_j]$ of job $j$ in $\bar{\sigma}$: for any probability density function $f(x)$, we can compute the expectation of the upper bound just derived.

**Theorem 4.31.** *If $\alpha \in (0, 1]$ is selected according to the probability density function $f(x) = e^x/(e-1)$, then for the nonpreemptive schedule $\bar{\sigma}$, $E[\overline{C}_j] \leq \frac{e}{e-1} C_j$.*

*Proof.* We can bound $E[\overline{C}_j]$ by integrating the bound in Corollary 4.30. Note that $\alpha$ appears in two ways in the bound of this corollary: in the $(1 + \alpha - \beta_k)p_k$ term, and in the fact that we need that $\alpha \leq \beta_k$ for any job $k$ that contributes to the bound. Alternatively, for each job $k$, we can focus on those $\alpha$ such that $\alpha \leq \beta_k$. Hence, we have that

$$E[\overline{C}_j] \leq C_j + \sum_{k=1}^{n} p_k \int_0^{\beta_k} f(\alpha)(1 + \alpha - \beta_k) d\alpha.$$

Straightforward calculus (if one recalls that integrating $xe^x$ yields $e^x(x-1)$) shows that

$$\int_0^{\beta} f(x)(1 + x - \beta) dx = \frac{\beta}{e-1}.$$

Hence, since $\sum_{k=1}^{n} p_k \beta_k \leq C_j$ by Lemma 4.28,

$$E[\overline{C}_j] \leq C_j + \frac{1}{e-1} \sum_{k=1}^{n} p_k \beta_k \leq \frac{e}{e-1} C_j.$$

$\square$

The expectation of the total completion time is, by linearity of expectation, within a factor $e/(e-1)$ of the total completion time of the preemptive schedule $\sigma$. If we apply this randomized conversion to the optimal solution $\sigma^*$ for $1|r_j, pmtn|\sum C_j$ found by the SRPT rule, we see that the expected total completion time is at most $e/(e-1)$ times the preemptive optimum. The preemptive optimum is, of course, a lower bound on the nonpreemptive optimum and hence we have obtained the following theorem.

**Theorem 4.32.** *The randomized $\alpha$-point algorithm which draws $\alpha$ according to the density function in Theorem 4.31 provides an online randomized $e/(e-1)$-approximation algorithm for the problem $1|r_j|\sum C_j$.*

There are a few immediate implications of this result. Note that we have constructed a nonpreemptive schedule of total completion time within a factor of $e/(e-1) \approx 1.58$ of the preemptive optimum. This places a limit on the power of preemption: by allowing preemption, we can improve the objective function by at most a factor of 1.58. As this argument does not depend on our ability to solve the preemptive problem in polynomial time, the limit on the power of preemption is the same for the total weighted completion time objective.

Many people are troubled at first by the fact that our focus on approximation algorithms is in order to be guaranteed that the result is good, and yet now we only know that it is expected to be good. While that is true, it is important to realize that this theorem says that *for every input* we are assured that the expected result will be good, and the source of randomness is merely the coin tosses used by the algorithm.

However, one can convert this randomized algorithm into a deterministic algorithm (provided we are considering off-line algorithms, where we can see the entire input in advance). We now just compute the minimal schedule in which the jobs are ordered consistently with their α-points. (That is, $\overline{C}_1 = r_1 + p_1$ and $\overline{C}_j = \max\{\overline{C}_{j-1}, r_j\} + p_j$.) Since we have shown that by choosing α in this random manner, the α-point algorithm delivers a schedule with expected objective function value within a factor of 1.58 of the optimum, then there must exist some α with the property that if that α is used, then that schedule has objective function value within a factor of 1.58 of the optimum. So, if we could compute an α for which this α-point algorithm would deliver the best solution, then the schedule found must be within a factor of 1.58 of the optimum.

If we consider the schedule constructed for various values of α, the schedule only changes when the order in which the jobs reach their α-points in the SRPT schedule $\sigma^*$ changes. This can only happen when there exists some job that is preempted when exactly an α fraction of its processing has been completed. Recall that in $\sigma^*$ a job is only preempted if another job arrives whose processing time is shorter than the remaining one of the current job. In particular, there are at most $n-1$ preemptions. Hence there are at most $n$ potentially critical values of α, and we can compute these critical values directly from $\sigma^*$. If we run the α-point algorithm for all of these values of α and take the best schedule, then we have deterministically computed a schedule within a factor of 1.58 of optimal.

**Theorem 4.33.** *The best α-point algorithm is a (deterministic)* $\frac{e}{e-1}$*-approximation algorithm for the problem* $1|r_j|\sum C_j$.

The only reason that the same approach does not lead to an approximation algorithm of the same performance guarantee for $1|r_j|\sum w_j C_j$ is that its preemptive relaxation $1|r_j, pmtn|\sum w_j C_j$ is NP-hard (see Theorem 4.13). We will therefore discuss a similar conversion technique that starts from another, efficiently computable preemptive schedule in the next section.

**Exercises**

4.36. Prove Theorem 4.25.

4.37. Show that no deterministic approximation algorithm for $1|r_j|\sum C_j$ that works online, can have a performance guarantee smaller than 2.

4.38. Use Yao's minimax principle to show that no randomized approximation algorithm for $1|r_j|\sum C_j$ that works online, can have an expected performance guarantee smaller than $e/(e-1)$.

### 4.10. Approximation algorithms for $1|r_j|\sum w_j C_j$

For nonpreemptive schedules, the mean busy time of each job $j$ is a mere translation of the completion time by $p_j/2$. But it is also interesting to consider mean busy times for preemptive schedules. Let $I_j(t) = 1$ if job $j$ is in process at time $t$, and 0 otherwise. The function $I_j$ is called the *indicator function* of job $j$ in the given schedule. Because the schedule is feasible, all $p_j$ units of work for each job $j$ are processed, that is,

$$\int_0^{+\infty} I_j(t)\, dt = p_j.$$

Consider again the 2-dimensional Gantt chart, and the interpretation that $\rho(j)$ specifies the holding cost of job $j$ per unit of unprocessed work. Consider the curve defined by the function $\overline{W}(t)$, introduced in Section 4.1, which is piecewise linear; each piece corresponding to the (partial) processing of job $j$ has slope $-\rho(j) = -w_j/p_j$. As we have already seen, the area under this curve corresponds to the total cost of holding the inventory over the planning horizon. We can decompose this area by breaking it into horizontal trapezoidal pieces, each of which corresponds to a unique job $j$ being (partially) processed for an interval $[s, s']$; see Figure 4.11. The area in this trapezoid is

$$\rho(j)(s'-s)(s'+s)/2 = \int_s^{s'} \rho(j)\, t\, dt$$

and so the total area in all of the trapezoids corresponding to $j$ is $\int_0^{+\infty} \rho(j)\, t\, I_j(t)\, dt$. This is the total cost of holding job $j$ over the planning horizon; this is equivalent to $w_j$ times $M_j$, the *mean busy time* of job $j$, where

$$M_j = \frac{1}{p_j} \int_0^{+\infty} I_j(t)\, t\, dt\,. \tag{4.13}$$

Note that this is the natural extension of the definition of $M_j$ in the nonpreemptive case. Thus the mean busy time $M_j$ is the average time at which the machine is processing job $j$. Even in this preemptive setting, if we let $S_j$ denote the start time of job $j$ in a given schedule, we have

$$S_j + \frac{1}{2}\, p_j \le M_j \le C_j - \frac{1}{2}\, p_j\,, \tag{4.14}$$

where each of these inequalities holds with equality if and only if job $j$ is processed without preemption.

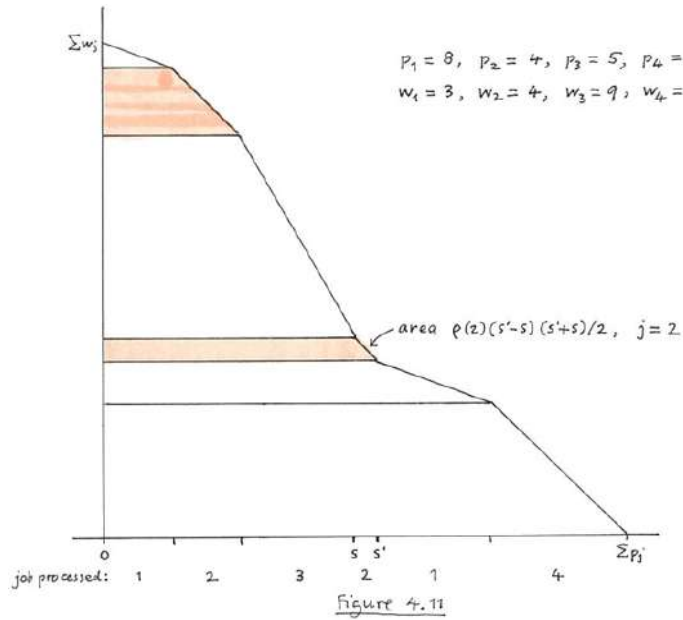

$p_1 = 8, \ p_2 = 4, \ p_3 = 5, \ p_4 = 6$
$w_1 = 3, \ w_2 = 4, \ w_3 = 9, \ w_4 = 6$

area $\rho(z)(s'-s)(s'+s)/2, \ j=2$

job processed:    1      2      3    2    1      4

Figure 4.11

The interpretation of $\sum_j w_j M_j$ as the area under the curve $\overline{W}(t)$ leads to an easy extension of the ratio rule to solve the problem $1|r_j, pmtn|\sum w_j M_j$. Intuitively, this area is made small by having the curve descend to 0 as sharply and as early as possible. In other words, consider the *preemptive ratio rule*: at each moment in time, among all available jobs choose a job $j$ for which the ratio $w_j/p_j$ is maximum. Thus, we need preempt a job $j$ only when a job $k$ with a bigger ratio is released.

We can prove the optimality of the preemptive ratio rule by an interchange argument. Suppose that the schedule $\sigma$ produced by the algorithm is not optimal, and consider an optimal schedule $\sigma^*$ whose earliest difference from $\sigma$ is as late as possible. Suppose that the two schedules agree up to time $t$, at which point $\sigma^*$ processes job $j$, whereas $\sigma$ processes job $k$. Of course, this means that $\sigma^*$ has not completed work on job $k$ at time $t$, and $k$ must still be processed later, next starting at some time $t'$. Consider these two fragments of processing jobs, job $j$ at $t$ and job $k$ at $t'$ in $\sigma^*$, and suppose that the jobs are processed in time periods $[t, t+\delta]$ and $[t', t'+\delta]$, respectively, for some $\delta > 0$. Both jobs have been released by $t$, and so we could swap the processing of $\delta$ time units of the two jobs. By the preemptive ratio rule, $\rho(k) \geq \rho(j)$. However, since the objective function value is the area under the $\overline{W}(t)$ curve, we see that unless $\rho(j) = \rho(k)$, this interchange will improve the objective function value, which contradicts the optimality of $\sigma^*$. But if $\rho(j) = \rho(k)$ then this interchange produces another optimal schedule, but one for which the ratio rule agrees even longer than for $\sigma^*$. This contradiction yields the following theorem.

**Theorem 4.34.** *A schedule is optimal for $1|r_j, pmtn| \sum w_j M_j$ if and only if it is a preemptive ratio rule schedule.*

In the previous section, we saw that the fact that $1|r_j, pmtn| \sum C_j$ could be solved in polynomial time led to a good approximation algorithm for the nonpreemptive variant. Next, we will see that Theorem 4.34 provides a similar engine for the weighted extension. In fact, it might seem natural to take the solution provided by this algorithm, and then process the jobs nonpreemptively in the order in which their mean busy times occur. Unfortunately, this algorithm does not perform well (see Exercise 4.40). Instead, we will use the second technique of the previous section: we will consider a randomized approach to converting the preemptive ratio rule schedule to a nonpreemptive one.

In fact, we will consider a somewhat more flexible algorithm, which we call the *randomized $\alpha_j$-point algorithm*:

(1) compute an optimal schedule for $1|r_j, pmtn| \sum w_j M_j$;
(2) for $j = 1, \ldots, n$, pick $\alpha_j$ independently and uniformly at
random from the interval $(0, 1]$;
(3) for each job $j$, compute its $\alpha_j$-point $C_j(\alpha_j)$
in the optimal preemptive schedule;
(4) schedule the jobs nonpreemptively in nondecreasing order of their
$\alpha_j$-points, always scheduling the next job as early as possible
(subject to this ordering constraint).

As we did in the previous section, our analysis of the algorithm is more general than we need for deriving an approximation algorithm; we will show that given any preemptive schedule in which job $j$ has mean busy time $M_j$, the expected value of $\sum w_j C_j$ for the schedule produced by this algorithm is at most $2 \sum_j w_j(p_j/2 + M_j)$.

We start with an easy observation, that follows directly from the definition of the mean busy time: choosing $\alpha_j$ uniformly in $(0, 1]$ is just a rescaled version of choosing a time uniformly within $(0, p_j]$.

**Lemma 4.35.** *If $\alpha_j$ is selected uniformly at random in $(0, 1]$, then $E[C_j(\alpha_j)] = M_j$.*

Let $\overline{C}_j$ denote the completion time of job $j$ in the schedule $\overline{\sigma}$ produced by the algorithm. Note that $\overline{C}_j$ is a random variable, since the ordering depends on the random choices $\alpha_j$, $j = 1, \ldots, n$. Nonetheless, we can write that $\overline{C}_j = P_j + R_j$, where the random variables $P_j$ and $R_j$ are, respectively, the total processing and total idle times prior to the completion of job $j$ in the schedule.

**Lemma 4.36.** *The randomized $\alpha_j$-point algorithm produces a schedule that, with probability 1, has $R_j \leq C_j(\alpha_j)$, $j = 1, \ldots, n$.*

*Proof.* We will show that for *any* realization of the values $\alpha_j$, $j = 1, \ldots, n$, the machine is never idle within the time interval $(C_j(\alpha_j), \overline{C}_j]$; clearly, this implies the lemma. Any job $k$ that is processed prior to job $j$ can have its release date no later than $C_j(\alpha_j)$ (since it has reached its $\alpha_k$-point by then). Suppose that there was idle time in the interval $(C_j(\alpha_j), \overline{C}_j]$; this must be due to the fact that the next job in the ordering is not yet released, but this is impossible, since all jobs processed before $j$ have been released by this point. $\square$

Next we analyze, for each job $j$, the total time $P_j$ the machine is busy prior to the completion of $j$.

**Lemma 4.37.** *The randomized $\alpha_j$-point algorithm produces a schedule for which $E[P_j] \leq p_j + M_j$.*

*Proof.* Focus on a job $j$; each job is processed prior to its own completion, and so we see that

$$E[P_j] = p_j + \sum_{k \neq j} Pr[k \text{ precedes } j \text{ in the schedule } \bar{\sigma}] p_k.$$

In order to analyze the summand, we first fix a value $\alpha \in (0, 1]$, and condition on the event that $\alpha_j = \alpha$. This fixes the $\alpha_j$-point of job $j$ at some time $t$. Once we have done this, it is easy to compute the conditional probability that job $k$ precedes job $j$; $k$ precedes $j$ if and only if $\alpha_k$ is chosen such that $C_k(\alpha_k)$ is at most $t$. If we let $\beta_k$ denote the fraction of job $k$ that is processed in the preemptive schedule prior to $t$, then the probability that $C_k(\alpha_k)$ is at most $t$ is exactly $\beta_k$. Note that this uses the fact that each $\alpha_k$ is chosen independently of $\alpha_j$. Thus, $\sum_{k \neq j} Pr[k \text{ precedes } j \text{ in } \bar{\sigma} \,|\, \alpha_j = \alpha] = \sum_{k \neq j} \beta_k p_k$. But this is exactly the total processing done prior to $t$ (in the preemptive schedule) on jobs other than $j$, and hence is at most $t = C_j(\alpha)$. Thus, $E[P_j \,|\, \alpha_j = \alpha] \leq p_j + C_j(\alpha)$.

Taking the expectation over all such choices $\alpha$ and applying Lemma 4.35, we see that $E[P_j] \leq p_j + M_j$. $\square$

It is now a simple matter of putting the pieces together to obtain the following theorem.

**Theorem 4.38.** *The randomized $\alpha_j$-point algorithm is a randomized 2-approximation algorithm for $1|r_j|\sum w_j C_j$.*

*Proof.* Let $C_j^*$, $j = 1, \ldots, n$, denote the completion times in an optimal schedule for $1|r_j|\sum w_j C_j$. This implies that the mean busy times for this schedule are $C_j^* - p_j/2$, and hence the preemptive ratio rule produces a schedule whose mean busy times $M_j$, $j = 1, \ldots, n$, satisfy the inequality

$$\sum_{j=1}^{n} w_j M_j \leq \sum_{j=1}^{n} w_j (C_j^* - p_j/2).$$

However, the randomized $\alpha_j$-point algorithm produces a schedule with completion times $\overline{C}_j$, $j = 1, \ldots, n$, such that, by Lemmata 4.36 and 4.37,

$$E\left[\sum_{j=1}^{n} w_j \overline{C}_j\right] = \sum_{j=1}^{n} w_j E[\overline{C}_j] = \sum_{j=1}^{n} w_j E[P_j + R_j] \leq \sum_{j=1}^{n} w_j(p_j + 2M_j) \leq 2\sum_{j=1}^{n} w_j C_j^*,$$

which concludes the proof of the theorem. □

Note that the randomized $\alpha_j$-point algorithm works online. One can also derandomize it to obtain a deterministic 2-approximation algorithm; however, for this we need to see the entire instance in advance, and so it does not work online. We conclude this chapter by describing a (deterministic) online algorithm that produces a solution that is within a factor 2 of the offline optimum.

The *delayed ratio rule* works as follows: At time $t$, let $j$ be an available job with highest $w_j/p_j$ ratio. If $t \geq p_j$, then start job $j$. Otherwise, do nothing until either time $p_j$ or a job with better ratio is released, whichever comes first.

Let $\sigma_0$ be the schedule returned by the delayed ratio rule. As any other feasible nonpreemptive schedule for $1|r_j|\sum w_j C_j$, $\sigma_0$ can be decomposed into blocks of consecutive jobs and idle time between the blocks. For the analysis, it will suffice to look at an arbitrary block. Let $I$ be the instance defined by the jobs in one such block, and let $\sigma$ be the restriction of $\sigma_0$ to $I$. We define another instance $I'$ from $I$ by changing the release dates of all jobs $j$ in $I$ to $r_j' = \min\{S_j, 2r_j\}$, where $S_j$ is the starting time of job $j$ in the schedule $\sigma$. Moreover, we add a new job $k$ to $I'$ with $r_k = 0$ and $p_k = t_I$, where $t_I = \min\{S_j : j \in I\}$. In order to define $w_k$, let us consider the situation in $\sigma_0$ right before time $t_I$, when the machine was idle (unless $t_I = 0$, in which case we set $w_k = 0$). This idle time can be caused by either the fact that no job is available or the job $h$ with the currently highest ratio $w_h/p_h$ is delayed because of $p_h \geq t_I$. If $h$ belongs to $I$, we define $w_k = w_h p_k/p_h$; otherwise $w_k = 0$. In either case, we have

$$\sum_{j \in I} w_j p_j \geq w_k p_k. \tag{4.15}$$

Finally, let $\sigma'$ be defined from $\sigma$ by assigning job $k$ to the time period $[0, t_I)$.

**Lemma 4.39.** *The schedule $\sigma'$ is an optimal preemptive schedule for the instance $I'$ and the weighted mean busy time objective.*

*Proof.* According to Theorem 4.34, we only need to show that $\sigma'$ processes at any point $t$ in time a job with the highest ratio among all the jobs that are not yet completed. We distinguish two cases.

In case $t < t_I$, $\sigma'$ is currently processing job $k$. Suppose there is another job $j \in I'$ available at that time; i.e., $r_j \leq r_j' \leq t$. Job $j$ is not started in $\sigma_0$ before time $t_I$, so $w_h/p_h \geq w_j/p_j$. If $h \in I'$, then $w_k/p_k = w_h/p_h$, and $k$ is a job of highest ratio available at time $t$. If $h \notin I'$, the algorithm did not wait until time $p_h$ before it started processing other jobs. By the definition of the delayed ratio rule, this implies $w_j/p_j > w_h/p_h$ for all jobs $j \in I$, a contradiction.

In case $t \geq t_I$, suppose that job $i$ is processed at time $t$ and job $j$ is available. Hence, $r'_j \leq t \leq S_i + p_i \leq 2S_i$. The last inequality follows from the fact that the delayed ratio rule only starts a job after its processing time has passed. Since $j$ does not start at its release time $r'_j$, we have $r'_j = 2r_j$. Therefore, $r_j = r'_j/2 \leq S_i$, and $j$ was available when job $i$ was started. Hence, $w_i/p_i \geq w_j/p_j$. $\square$

Let $\sigma^*$ be an optimal preemptive schedule for the instance $I$ and the weighted completion time objective. We denote the mean busy times and completion times of a job $j$ in a schedule $\pi$ by $M_j^\pi$ and $C_j^\pi$, respectively. The following equations are easily verified:

$$\sum_{j \in I} w_j C_j^\sigma = \sum_{j \in I} w_j M_j^\sigma + \frac{1}{2} \sum_{j \in I} w_j p_j, \tag{4.16}$$

$$\sum_{j \in I} w_j M_j^\sigma = \sum_{j \in I'} w_j M_j^{\sigma'} - \frac{1}{2} w_k p_k, \tag{4.17}$$

$$\sum_{j \in I} w_j M_j^{\sigma^*} \leq \sum_{j \in I} w_j C_j^{\sigma^*} - \frac{1}{2} \sum_{j \in I} w_j p_j. \tag{4.18}$$

From the preemptive schedule $\sigma^*$ we define a "pseudo-schedule" by letting the machine process all jobs at half speed. That is, if a fraction of job $j$ was previously scheduled in the interval $[a_j, b_j)$, it is now scheduled in $[2a_j, 2b_j)$. Consequently, the mean busy time of each job increases by a factor of 2. We also process job $k$ at half speed from time 0 to $2t_I$. This pseudo-schedule satisfies the release dates of the instance $I'$, and we use it to derive a feasible preemptive schedule $\hat\sigma$ for $I'$ with no further increase in any mean busy time. We can now put the pieces together:

$$\begin{aligned}
\sum_{j \in I} w_j C_j^\sigma &\overset{(4.16)}{=} \sum_{j \in I} w_j M_j^\sigma + \frac{1}{2} \sum_{j \in I} w_j p_j \\
&\overset{(4.17)}{=} \sum_{j \in I'} w_j M_j^{\sigma'} - \frac{1}{2} w_k p_k + \frac{1}{2} \sum_{j \in I} w_j p_j \\
&\overset{\text{Lemma 4.39}}{\leq} \sum_{j \in I'} w_j M_j^{\hat\sigma} - \frac{1}{2} w_k p_k + \frac{1}{2} \sum_{j \in I} w_j p_j \\
&\leq 2 \sum_{j \in I'} w_j M_j^{\sigma^*} + \frac{1}{2} \sum_{j \in I'} w_j p_j \\
&\overset{(4.18)}{\leq} 2 \sum_{j \in I} w_j C_j^{\sigma^*} - \frac{1}{2} \sum_{j \in I} w_j p_j + \frac{1}{2} w_k p_k \\
&\overset{(4.15)}{\leq} 2 \sum_{j \in I} w_j C_j^{\sigma^*}.
\end{aligned}$$

**Theorem 4.40.** *The delayed ratio rule is a deterministic online 2-approximation algorithm for $1|r_j|\sum w_j C_j$.*

**Exercises**

4.39. Define the mean busy time $M_S$ of a subset $S \subseteq N$ of jobs as the average point in time that the machine is busy scheduling a job from $S$. Show that $\sum_{j \in S} p_j M_S = \sum_{j \in S} p_j M_j$.

4.40. Give an example for $1|r_j|\sum w_j C_j$ that shows that scheduling the jobs nonpreemptively in order of nondecreasing mean busy times in the optimal preemptive ratio schedule can lead to a schedule that is arbitrarily worse than the optimal one.

4.41. Prove that the preemptive ratio rule is a 2-approximation algorithm for $1|r_j, pmtn|\sum w_j C_j$. Show that this analysis is tight.

4.42. Give an example that shows that the randomized $\alpha_j$-point algorithm can return an exponential number of different schedules.

4.43. Use the method of conditional probabilities to derandomize the randomized $\alpha_j$-point algorithm.

**Notes**

4.1. *Smith's ratio rule.* The ratio rule is, of course, due to W. E. Smith (1956). The observation

> given two nondecreasing sequences of numbers,
>
> $$a_1 \le a_2 \le \cdots \le a_n \text{ and } b_1 \le b_2 \le \cdots \le b_n,$$
>
> the permutation $\sigma = (n, n-1, \ldots, 2, 1)$ minimizes $\sum_{j=1}^{n} a_j b_{\sigma(j)}$ among all permutations of $(1, 2, \ldots, n)$ used in the alternate justification of the SPT rule, appears in Hardy, Littlewood and Pólya (1934). Two-dimensional Gantt charts go at least as far back as Eastman, Even and Isaacs (1964). Goemans and Williamson (2000) showed that many results in single-machine scheduling have a simple and intuitive geometric justification using 2D Gantt charts. Mean busy times were introduced by Goemans, Queyranne, Schulz, Skutella and Wang (2002) in the context of linear programming relaxations and approximation algorithms for $1|r_j|\sum w_j C_j$; see Sections 4.5 and 4.10 for further details. Exercise 4.3 is due to Goemans and Williamson (2000). The equivalence in this exercise was shown by Chudak and Hochbaum (1999) and, for the case all $p_j = 1$, von Arnim, Faigle and Schrader [1990].

4.2. *Preference orders on jobs.* When Smith (1956) defined a preference order, he did not require the relation to be transitive. This has the consequence that there might not exist a total order consistent with the relation. However, for relations for which there does exist a consistent total order, Smith's definition and the one given in this section are equivalent. Jackson (1955) showed that the earliest due date rule is optimal for $1||L_{\max}$. The total weighted discounted completion time criterion was introduced by Rothkopf (1966). The least cost fault detection problem was studied by Boothroyd (1960), Mitten (1960]), Mankekar and Mitten (1965), and Garey (1973), among others. Monotone cost density functions were introduced by Lawler and Sivazlian (1978). The cumulative cost problem was formulated by Addel-Wahab and Kameda (1978) and generalized by Monma (1980).

4.3. *Preference orders on sequences & series-parallel precedence constraints.* Monma and Sidney (1979) introduced the notion of preference relations on sequences. Parallel-chains precedence constraints were treated by Conway, Maxwell and Miller (1967). Horn (1972), Adolphson and Hu (1973), and Sidney (1975) gave algorithms for tree-like precedence constraints. Lawler (1978) presented the O($n \log n$) algorithm for series-parallel precedence constraints. Goemans and Williamson (2000) used 2D Gantt charts and linear programming duality to give an alternative proof of correctness of Lawler's algorithm for $1|prec|\sum w_j C_j$. The equivalent characterization of series-parallel partial orders in terms of a forbidden substructure is due to Duffin (1965). Valdes, Tarjan and Lawler (1982) gave a linear-time algorithm that recognizes series-parallel partial orders and creates a decomposition tree.

4.4. *NP-hardness of further constrained min-sum problems.* The strong NP-hardness of $1|r_j|\sum C_j$ was established by Lenstra, Rinnooy Kan and Brucker (1977). The proofs presented here for this result and the NP-hardness of $1|r_j, pmtn|\sum w_j C_j$ are due to Lenstra. The latter result was originally obtained by Labetoulle, Lawler, Lenstra and Rinnooy Kan (1984). Lawler (1978) showed that $1|prec|\sum w_j C_j$ is NP-hard, even if either all $w_j = 1$ or all $p_j = 1$. The proof of Theorem 4.15 follows that of Lenstra and Rinnooy Kan (1978).

4.5. *The ratio rule via linear programming.* The parallel inequalities were conceived by Wolsey (1985) and Queyranne (1993), who also showed that these inequalities completely describe the convex hull of feasible completion time vectors for $1|\,|\sum w_j C_j$. The relationship between Smith's ratio rule and optimizing a linear function over the polyhedron defined by the parallel inequalities is laid out further in Queyranne (1993) and Queyranne and Schulz (1994). The separation algorithm for the parallel inequalities discussed in Exercise 4.21 appeared in Queyranne (1993). The "shifted" parallel inequalities in Exercises 4.22 and 4.23 were introduced by Queyranne (1988).

4.6. *Approximation algorithms for $1|prec|\sum w_j C_j$.* Hall, Schulz, Shmoys and Wein (1997) devised linear programming based approximation algorithms to yield the first constant performance guarantees for a variety of minimum-sum single-machine scheduling problems, including Theorem 4.16, and gave the tight examples asked for in Exercise 4.26. Potts (1980C) suggested the integer programming formulation in $\delta$-variables, and Wolsey (1989) showed that any feasible solution to its linear programming relaxation satisfies the parallel inequalities and precedence constraints, which implies Corollary 4.17. A family of instances provided by Chekuri and Motwani (1999) demonstrates that the integrality gap of this linear programming relaxation is essentially 2 (Exercise 4.27). The first part of Exercise 4.24 is due to Peters (1988) and Nemhauser and Savelsbergh (1992). Potts (1980C) showed that the linear programming relaxation of his integer programming formulation with the transitivity constraints dropped can be used to obtain an alternative derivation of Smith's ratio rule, which corresponds to the second part of this exercise. 2-approximation algorithms for $1|prec|\sum w_j C_j$ that do not rely on solving a linear program were proposed by Chekuri and Motwani (1999), Chudak and Hochbaum (1999), and Margot, Queyranne and Wang (2003).

4.7. *Sidney decompositions.* The concept of Sidney decomposition is, of course, due to J. B. Sidney (1975). The proof of Theorem 4.18 presented herein is adapted from Correa and Schulz (2005). Chekuri and Motwani (1999) and Margot, Queyranne and Wang (2003) observed that each sequence that is consistent with a Sidney decomposition is within a factor of 2 from optimum. The 2D Gantt chart illustration of this fact is due to Goemans and Williamson (2000). Correa and Schulz (2005) noted that all known 2-approximation algorithms are of this type. Exercise 4.30 is based on an observation by Margot, Queyranne and Wang (2003).

4.8. *An integrality theorem for series-parallel precedence constraints.* This section is largely based on Correa and Schulz (2005). Chudak and Hochbaum (1999) suggested to drop all transitivity constraints associated with triples of unrelated jobs from Potts' linear programming relaxation. They observed that the resulting linear program is half-integral and can be solved by a min-cut computation, and used this insight to develop a combinatorial 2-approximation algorithm for the single-machine problem with general precedence constraints. Two-dimensional partial orders were introduced by Dushnik and Miller (1941), who also proved that a partial order is two-dimensional if and only if it has a nonseparating linear extension. The first polynomial-time recognition algorithm for 2D orders was presented by Pnueli, Lempel and Even (1971). Exercise 4.32, which is needed to complete the proof of Theorem 4.24, is based on a conjecture by Correa and Schulz (2005) that was proved by Ambühl and Mastrolilli (2006).

4.9. *Approximation algorithms for* $1|r_j|\sum C_j$. The optimality of the SRPT rule for $1|r_j, pmtn|\sum C_j$ was shown by Baker (1974). Phillips, Stein and Wein (1998) introduced the idea of scheduling jobs in order of their preemptive completion times into the design of approximation algorithms. Theorem 4.26 belongs to them. The lower bound of 2 on the performance guarantee of any deterministic online algorithm is due to Hoogeveen and Vestjens (1996). Chekuri, Motwani, Natarajan and Stein (2001) presented the randomized $\alpha$-point algorithm and proved Theorem 4.32. Goemans et al. (2002) showed that the sample space of this algorithm is O($n$), which makes the straightforward derandomization possible. The construction of a matching lower bound on the performance guarantee of any randomized online algorithm can be found in Vestjens (1997).

4.10. *Approximation algorithms for* $1|r_j|\sum w_j C_j$. The randomized $\alpha_j$-point algorithm was suggested by Goemans et al. (2002). They also showed that drawing the $\alpha_j$'s according to a truncated exponential distribution results in an expected performance guarantee of 1.69. Working with the same $\alpha$ for all jobs complicates the analysis, and the best density function known leads to an expected performance guarantee of 1.75. Theorem 4.40 is due to Anderson and Potts (2002). The proof presented here follows an interpretation by Sitters of a simplified proof suggested by Queyranne. Exercises 4.39, 4.42 and 4.43 are from Goemans et al. (2002). The tight example sought in Exercise 4.41 was presented by Schulz and Skutella (2002). Therein, they also proved the upper bound, which is due to Goemans, Wein and Williamson (2000).

# Contents

# 5

# Weighted number of late jobs

Eugene L. Lawler

*University of California, Berkeley*

Once it happened that two of the authors of this book found themselves circling an airport in a dense fog. When the fog broke, they could see many other aircraft. In fact, there were so many planes in the air it seemed unlikely that all of them could land before running out of fuel. While waiting to reach earth again, the authors amused themselves by applying their knowledge of scheduling theory to the situation. There were $n$ planes to be scheduled for landing on a single runway, where plane $J_j$ carried $w_j$ passengers, required $p_j$ time units to land, and would run out of fuel by time $d_j$. The problem of minimizing the number of crashed planes, $1||\sum U_j$, could be solved in polynomial time. However, the problem of minimizing the number of passenger fatalities, $1||\sum w_j U_j$, was NP-hard. Because time was limited, it seemed plausible that the flight controllers might choose to solve the easier problem. And since the authors were traveling in a very small plane, this observation was reassuring. It provided additional motivation to write this book.

In this chapter we first show that the NP-hard problem confronting the flight controllers, $1||\sum w_j U_j$, can be can be solved in $O(nW)$ time, where $W = \sum w_j$, with a minor modification of the knapsack algorithm described in Chapter 2. We then show that a much streamlined version of the algorithm, due to Moore and Hodgson, solves the problem $1||\sum w_j U_j$ in $O(n \log n)$ time, under the condition that the processing times and job weights are oppositely ordered. We also show that, although the problem $1|r_j|\sum U_j$ is strongly NP-hard, it can be solved in $O(n \log n)$ time, provided that the release dates and due dates are similarly ordered. Finally, we show that a considerable elaboration of the knapsack algorithm solves the general preemptive problem $1|pmtn, r_j|\sum w_j U_j$ in $O(nk^2W^2)$ time, where $k$ is the number of distinct release dates.

All of the computational results presented in this chapter are for independent jobs, because NP-hardness results concerning precedence constraints are very limiting. In particular, we show that even the problem $1|chains, p_j = 1|\sum U_j$ is strongly NP-hard.

**1**

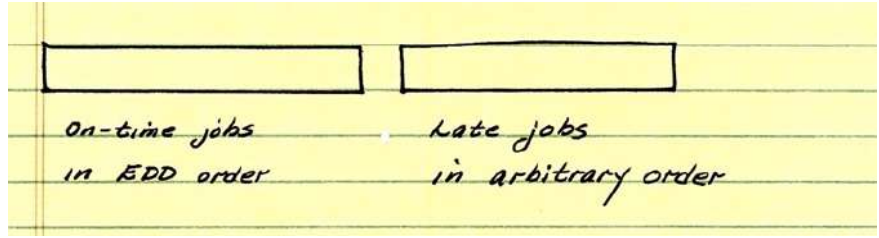

**Figure 5.1.** Form of an optimal schedcule

## 5.1. Some Preliminaries

We begin with some simple but fundamental observations that hold for $1||\sum w_j U_j$, $1|pmtn, r_j|\sum w_j U_j$, and for all special cases of these problems. First, we note the following: because of the character of the $\sum w_j U_j$ objective, if a given job $j$ is scheduled to be late, it might as well be arbitrarily late. This means that, in the absence of precedence constraints, there exists an optimal schedule in which all jobs that are on time precede all jobs that are late. Second, we observe that the jobs that are on time can be assumed to be scheduled by the following rule: always schedule the job with the earliest due date. In the presence of release dates, this is an *EDD rule* which may produce a schedule with preemptions. It follows that there must exist an optimal schedule like that shown in Figure 5.1.

From the previous observations it follows that the problems $1||\sum w_j U_j$ and $1|pmtn, r_j|\sum w_j U_j$ reduce to the problem of finding a maximum-weight feasible set of jobs, where by a *feasible* set we mean a set of jobs that are all completed on time when scheduled by the (extended) EDD rule, that is, this rule produces a feasible schedule.

Throughout this chapter we shall normally assume that jobs are numbered in EDD order, i.e.,

$$d_1 \leq d_2 \leq ... \leq d_n.$$

We shall also employ the following notation. As in Chapter 3, if $S \subseteq N = \{1, ..., n\}$, let $p(S) = \sum_{j \in S} p_j$, $r(S) = \min\{r_j | j \in S\}$, and similarly define $w(S) = \sum_{j \in S} w_j$ ; finally, let $C(S)$ denote the completion time of the last job in an (extended) EDD schedule for the jobs indexed by $S$.

### Exercises

5.1. Show that the following procedure computes a maximum-cardinality feasible index set $S$ for the problem $1|r_j, p_j = 1|\sum U_j$ : schedule the jobs in time, starting with $S$ empty. At each time $t$, from among all the unprocessed jobs $J_j$ such that $r_j \leq t < d_j$, if any, add to $S$ a job $J_j$ with the earliest possible due date. Show that

this algorithm can be implemented to run in $O(n \log n)$ time.

5.2. The problem $1|r_j, p_j = 1|\sum w_j U_j$ can be solved by applying the greedy matroid algorithm, as follows. Index the jobs in nonincreasing order of job weights. Then process the jobs in order, solving the recurrence relations

$$S^{(0)} = \emptyset,$$

$$S^{(j)} = \begin{cases} S^{(j-1)} \cup \{j\}, & \text{if } S^{(j-1)} \cup \{j\} \text{ is feasible,} \\ S^{(j-1)}, & \text{otherwise.} \end{cases}$$

The set $S^{(n)}$ then indexes a maximum-weight feasible set of jobs. Show that this algorithm can be implemented to run in $O(n^2)$ time.

5.3. Show that it is possible to solve the problem $1|p_j = 1|\sum U_j$ in $O(n)$ time. (Hint: Start with an $O(n)$ bucket sort of the due dates, by noting that any due date greater than $n$ can be reduced to $n$.)

## 5.2. Dynamic Programming Solution of $1||\sum w_j U_j$

The problem $1||\sum w_j U_j$ is NP-hard because the special case $1|d_j = d|\sum w_j U_j$ is equivalent to the NP-hard knapsack problem. But in Chapter 2 we showed that the knapsack problem can be solved by a dynamic programming computation within the pseudopolynomial time bound of $O(nW)$. We shall now show that a minor modification of the dynamic programming computation solves $1||\sum w_j U_j$ within the same time bound.

The problem of finding a maximum-weight feasible set for the $1||\sum w_j U_j$ problem can be formulated as an integer linear programming problem:

$$
\begin{aligned}
\text{maximize } &\sum w_j x_j \\
\text{subject to } &p_1 x_1 &&\leq d_1 \\
&p_1 x_1 + p_2 x_2 &&\leq d_2 \\
&p_1 x_1 + p_2 x_2 + p_3 x_3 &&\leq d_3 \\
&\quad\vdots &&\quad\vdots \\
&p_1 x_1 + p_2 x_2 + p_3 x_3 + ... + p_n x_n &&\leq d_n
\end{aligned}
$$

and

$$x_j \in \{0, 1\}, \text{ for } j = 1, 2, ..., n.$$

Note that when all the $d_j$'s are equal, the last inequality implies all the others and the problem reduces to the ordinary knapsack problem.

The triangular form of the inequality constraints enables us to employ a dynamic programming procedure essentially the same as that described for the knapsack problem in Chapter 2. Let $P^{(j)}(w)$ denote the minimum total processing time for any feasible subset of jobs $J_1,...,J_j$ that has total weight exactly $w$. Then we have

$$P^{(0)}(w) = \begin{cases} 0, & \text{if } w = 0, \\ +\infty, & \text{otherwise.} \end{cases}$$

$$P^{(j)}(w) = \begin{cases} \min\{P^{(j-1)}(w), P^{(j-1)}(w-w_j) + p_j\}, & \text{if } P^{(j-1)}(w-w_j) + p_j \leq d_j, \\ P^{(j-1)}(w), & \text{otherwise.} \end{cases}$$

$$(5.1)$$

As before, there are $O(W)$ equations to solve at each of $n$ iterations, $j = 1, 2, ..., n$. Each equation requires a constant number of arithmetic operations. It follows that the values $P^{(n)}(w)$ can be computed in $O(nW)$ time; the maximum weight of a feasible set is the largest value of $w$ such that $P^{(n)}(w)$ is finite.

As we described in Chapter 2, the dynamic programming recurrence equations can be solved by computing lists of dominant pairs $(w, P^{(j)}(w))$. At each iteration a list of candidate pairs is formed from the list existing at the end of the previous iteration. This list is merged with the existing list, with dominated pairs discarded in the course of the merge. In the case at hand, a pair $(w + w_j, P^{(j-1)}(w) + p_j)$ is discarded as infeasible at iteration $j$ whenever $P^{(j-1)}(w) + p_j > d_j$, whereas in the the knapsack problem a pair is discarded whenever $P^{(j-1)}(w) + p_j > d$, where $d$ is the knapsack capacity.

Recall that in Chapter 2 we also described how to compute upper bounds on the value of the solutions that can be obtained from pairs $(w, P)$ and how to use these bounds to eliminate pairs $(w, P)$ whose bounds do not exceed the value of a known feasible solution. We can compute comparable bounds for the problem $1||\sum w_j U_j$ by solving the linear programming relaxation of its integer programming formulation. This relaxation turns out to be an interesting scheduling problem in its own right: Let $V_j$ denote the amount of processing of job $j$ that is done after its due date $d_j$. Then the linear programming relaxation is equivalent to the problem of minimizing weighted *late work*, i.e., $1|pmtn|\sum w_j V_j$. In Exercise 5.4 we ask the reader to devise an efficient algorithm for solving this problem.

**Exercises**

5.4. Devise an $O(n \log n)$ algorithm for solving the weighted late work scheduling problem $1||\sum w_j V_j$. Recall that it is easy to solve the linear-programming relaxation of the knapsack problem, by including items in the knapsack in order of nonincreasing ratio $w_j/p_j$, until the knapsack is completely filled, using a fraction of the last item if necessary. Devise a similar algorithm for solving the linear programming relaxation of $1||\sum w_j U_j$. (Hint: Construct an optimal schedule by starting at the latest due date and working backward in time.)

### 5.3. A Fully Polynomial Approximation Scheme

In Chapter 2, we showed that it is possible to convert a pseudopolynomial-time algorithm for the knapsack problem into a fully polynomial approximation scheme. The $O(nW)$ algorithm for $1||\sum w_j U_j$ given in Section 5.2 is, of course, a pseudopolynomial-time algorithm. We shall now show how to construct a fully polynomial approximation scheme for the $1||\sum w_j U_j$ problem.

In the previous section, we have focused on the equivalent problem of finding a maximum-weight feasible set of jobs. These problems are not equivalent from the point of view of approximation: a $\rho$-approximation algorithm for finding a maximum-weight feasible set need not be a $\rho$-approximation algorithm for $1||\sum w_j U_j$. For example, the latter must produce an optimal schedule for instances in which all jobs can be scheduled on time, whereas this is certainly not the case for the former. In contrast to the analogous situation for $1|r_j|L_{\max}$, it is possible to obtain optimal solutions in this particular case, by using the EDD rule.

Let $W^*$ denote the minimum weight of a set of late jobs; hence $W - W^*$ is the maximum weight of a feasible set. The dynamic programming algorithm for $1||\sum w_j U_j$ given in Section 5.2 generates $O(W - W^*)$ dominant pairs $(w, P)$ in each of $n$ iterations, yielding a time bound of $O(n(W - W^*))$. For our purposes, we will need a slightly different algorithm, which has an $O(nW^*)$ running time. This can be obtained by a similar dynamic programming approach that maintains pairs $(w, P)$, where $w$ is the weight of a set of late jobs with processing time at most $P$, and the notion of dominance is reversed (see Exercise 5.5).

As in the case of the knapsack problem, the principal technique used is to round the data to have fewer significant digits, in order to make the dynamic programming algorithm take less time. This rounding and rescaling technique will be augmented by one further idea. We will first need to obtain a schedule of total weight that is within a factor of $n$ of the optimum. This will enable us to find an appropriate factor $\delta$, by which to rescale the data. Suppose that we find a schedule that minimizes the maximum weight of a late job. Since this is the special case of $1||f_{\max}$ with $f_j = w_j U_j$, $j = 1, \ldots, n$, Theorem 3.? implies that it can be solved using the least-cost-last rule in $O(n^2)$ time; let $w$ denote the maximum-weight late job in this schedule. Clearly, $w \leq W^*$, and the schedule just obtained has late jobs of total weight at most $nw$.

We now set

$$\delta = \frac{\varepsilon w}{n},$$

and compute the rounded weights

$$\bar{w}_j = \lfloor \frac{w}{\delta} \rfloor.$$

The claimed dynamic programming algorithm can be applied to the rounded instance in $O(nW^*/\delta)$ time.

Let the minimum-weight set of late jobs for the original instance be indexed by $S^*$, and let the set of late jobs in the schedule returned by the algorithm on the rounded data be indexed by $\bar{S}$. Note that $\delta \bar{w}_j \leq w_j \leq \delta(\bar{w}_j + 1)$, and so

$$\sum_{j \in \bar{S}} w_j \leq \delta |\bar{S}| + \sum_{j \in \bar{S}} \delta \bar{w}_j.$$

Using the optimality of $\bar{S}$ with respect to $\bar{w}_j$, $j = 1, ..., n$, and the inequality $|\bar{S}| \leq n$, we can further bound the right-hand side by

$$\leq \delta n + \sum_{j \in S^*} \delta \bar{w}_j \leq \varepsilon w + \sum_{j \in S^*} w_j \leq (1 + \varepsilon) \bar{W}^*.$$

Since $W^*/\delta \leq n^2/\varepsilon$, the algorithm runs in $O(n^3/\varepsilon)$ time. Thus we have shown the following theorem.

**Theorem 5.1.** *There exists a fully polynomial approximation scheme to solve the problem* $1||\sum w_j U_j$.

### Exercises

5.5. Give a dynamic programming algorithm for $1||\sum w_j U_j$ that runs in $O(nW^*)$ time.

5.6. Give a fully polynomial approximation scheme to find a maximum-weight feasible set of jobs.

### 5.4. The Moore-Hodgson Algorithm

A simple and elegant algorithm of Moore and Hodgson solves the unweighted problem $1||\sum U_j$ in $O(n \log n)$ time. More generally, it solves the problem $1||\sum w_j U_j$ under the condition that processing times and job weights are .ul oppositely ordered. By this we mean that $p_i < p_j$ implies $w_i \geq w_j$, for all $i$ and $j$. Note that opposite ordering necessarily holds if either all $p_j = 1$ or all $w_j = 1$. Hence the special case of opposite ordering of processing times and job weights is a common generalization of the problems $1|p_j = 1|\sum w_j U_j$ and $1||\sum U_j$.

The Moore-Hodgson algorithm is as follows: Starting with $S^{(0)} = \emptyset$, process the jobs in EDD order, constructing feasible sets $S^{(j)}$, $j = 1, 2, ..., n$, by the recurrence relations

$$S^{(j)} = \begin{cases} S^{(j-1)} \cup \{j\}, & \text{if } p(S^{(j-1)} \cup \{j\}) \leq d_j, \\ S^{(j-1)} \cup \{j\} - \{l\}, & \text{otherwise,} \end{cases} \tag{5.2}$$

where

$$l = \mathrm{argmax}\{p_i | i \in S^{(j-1)} \cup \{j\}\}$$
$$= \mathrm{argmin}\{w_i | i \in S^{(j-1)} \cup \{j\}\}.$$

The set $S^{(n)}$ is then a maximum-weight feasible set.

TThe Moore-Hodgson algorithm is efficiently implemented with a priority queue $S$ supporting the operations of *insert* and *deletemax*, as indicated below:

**Algorithm 1:** Moore-Hodgson algorithm

$S := \emptyset$;
$p(S) := 0$;
**for** $j = 1, 2, \ldots, n$ **do**
    $\quad insert(S, j)$;
    $\quad p(S) := p(S) + p_j$;
    $\quad$ **if** $p(S) > d_j$ **then**
        $\quad\quad l := deletemax(S)$;
        $\quad\quad p(S) := p(S) - p_l$;
    $\quad$ **end**
**end**

Each of the $n$ insertions and each of the at most $n$ deletions requires $O(\log n)$ time. Hence the overall running time required to generate a maximum-weight feasible set $S^{(n)}$ is bounded by $O(n \log n)$.

Although the Moore-Hodgson algorithm seems quite unrelated to the dynamic programming algorithm of Section 5.2, the two are closely related. We shall demonstrate this by deriving the Moore-Hodgson algorithm from the list-making version of the dynamic programming algorithm.

Recall that the input to iteration $j$ of the dynamic programming algorithm is the list of dominant pairs generated at iteration $j - 1$ :

$$(w(0), P^{(j-1)}(w(0))), (w(1), P^{(j-1)}(w(1))), \ldots, (w(k), P^{(j-1)}(w(k))),$$

where

$$0 = w(0) < w(1) < \ldots < w(k),$$
$$0 = P^{(j-1)}(w(0)) < P^{(j-1)}(w(1)) < \ldots < P^{(j-1)}(w(k)) \leq d_{j-1}.$$

Each pair $(w(i), P^{(j-1)}(w(i)))$ in the list of dominant pairs is realized by a feasible set $S^{(j-1)}(w(i)) \subseteq \{1, 2, ..., j-1\}$, with

$$w(S^{(j-1)}(w(i))) = w(i)$$
$$p(S^{(j-1)}(w(i))) = P^{(j-1)}(w(i)).$$

Ordinarily, we could not expect these feasible sets to be related in any special way. But when the processing times and job weights are oppositely ordered, a happy thing occurs: the feasible sets form a tower. Specifically, each set $S^{(j-1)}(w(i))$ contains exactly one job $J_{j(i)}$ that is not contained in the set $S^{(j-1)}(w(i-1))$ which is immediately below it in the tower. That is,

$$
\begin{aligned}
S^{(j-1)}(w(0)) = \emptyset &\subseteq S^{(j-1)}(w(1)) = \{j(1)\} \\
&\subseteq S^{(j-1)}(w(2)) = \{j(1), j(2)\} \subseteq ... \\
&\subseteq S^{(j-1)}(w(k)) = \{j(1), j(2), ..., j(k)\}.
\end{aligned}
\tag{5.3}
$$

from which it follows that

$$P^{(j-1)}(w(i)) - P^{(j-1)}(w(i-1)) = p_{j(i)}, \text{ and } w(i) - w(i-1) = w_{j(i)}, i = 1, 2, ..., k.$$

Furthermore,

$$p_{j(1)} \leq p_{j(2)} \leq ... \leq p_{j(k)} \tag{5.4}$$

and

$$w_{j(1)} \geq w_{j(2)} \geq ... \geq w_{j(k)}. \tag{5.5}$$

We shall refer to relations (5.3)-(5.5) as the *tower-of-sets* property of feasible sets.

At this point we may gain some insight from an example. Consider the following problem data:

| $j$   | 1 | 2 | 3 | 4 | 5 |
|-------|---|---|---|---|---|
| $p_j$ | 2 | 1 | 5 | 4 | 3 |
| $w_j$ | 4 | 5 | 1 | 2 | 3 |

Assume that all due dates are very large, so that they are of no consequence. At the end of iteration 4 we have the following list of dominant pairs,

$$(0, 0), (5, 1), (9, 3), (11, 7), (12, 12),$$

which are plotted as dots in Figure 5.2. With reference to the Gantt charts shown in Figure **??**, we see that these pairs are realized by a tower of feasible sets:

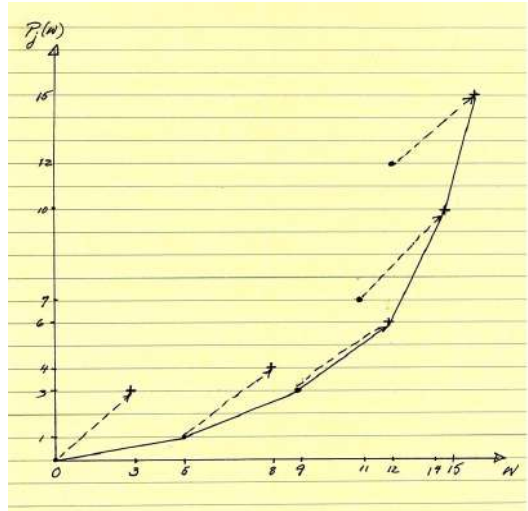

**Figure 5.2.** Graph of dominant pairs

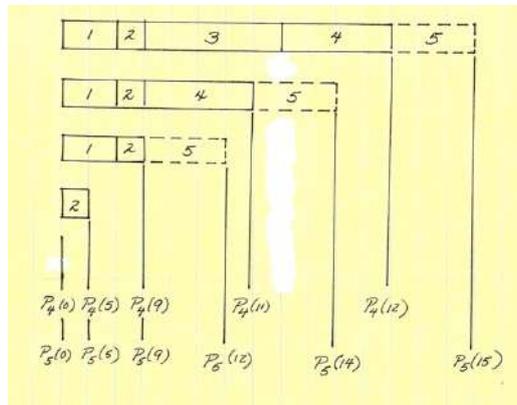

**Figure 5.3.** Gantt charts.

$$S^{(4)}(0) = \emptyset \subseteq S^{(4)}(5) = \{2\}$$
$$\subseteq S^{(4)}(9) = \{2, 1\}$$
$$\subseteq S^{(4)}(11) = \{2, 1, 4\}$$
$$\subseteq S^{(4)}(12) = \{2, 1, 4, 3\}$$

From the existing list of dominant pairs we form a list of candidates,

$$(3, 3), (8, 4), (12, 6), (14, 10), (15, 15),$$

which are plotted as +'s in Figure 5.2. When the two lists are merged, and dominated pairs discarded, the surviving pairs are

$$(0, 0), (5, 1), (9, 3), (12, 6), (14, 10), (15, 15).$$

Again with reference to Figure 5.3, we find that these pairs are realized by a tower of feasible sets:

$$S^{(5)}(0) = S^{(4)}(0) = \emptyset \subseteq S^{(5)}(5) = S^{(4)}(5) = \{2\}$$
$$\subseteq S^{(5)}(9) = S^{(4)}(9) = \{2, 1\}$$
$$\subseteq S^{(5)}(12) = S^{(4)}(9) \cup \{5\} = \{2, 1, 5\}$$
$$\subseteq S^{(5)}(14) = S^{(4)}(11) \cup \{5\} = \{2, 1, 5, 4\}$$
$$\subseteq S^{(5)}(15) = S^{(4)}(12) \cup \{5\} = \{2, 1, 5, 4, 3\}.$$

We shall prove by induction that the tower-of-sets property holds at each iteration. For the base case of the induction, note that the property holds at the end of iteration 0 when the only dominant pair, (0,0), is realized by the empty set. Assume, by inductive hypothesis, that the property holds at the end of iteration $j - 1$. At iteration $j$ the candidate pairs that are formed are

$$(w(0) + w_j, P^{(j-1)}(w(0)) + p_j), (w(1) + w_j, P^{(j-1)}(w(1)) + p_j), ..., (w(k) + w_j, P^{(j-1)}(w(k)) + p_j).$$

Let $h$ be the largest index $i$, $1 \leq i \leq k$, if any, such that $p_{j(i)} \leq p_j$ and $w_{j(i)} \geq w_j$. (The case in which there is no such $h$ is left as an exercise.) Recall that the existing pairs

$$(w(1), P^{(j-1)}(w(1))), ..., (w(h), P^{(j-1)}(w(h)))$$

can be rewritten as

$$(w(0) + w_{j(1)}, P^{(j-1)}(w(0)) + p_{j(1)}), ..., (w(h-1) + w_{j(h)}, P^{(j-1)}(w(h-1)) + p_{j(h)}).$$

Opposite ordering of processing times and job weights implies that these pairs respectively dominate the candidate pairs

$$(w(0) + w_j, P^{(j-1)}(w(0)) + p_j), ..., (w(h-1) + w_j, P^{(j-1)}(w(h-1)) + p_j).$$

Furthermore, by rewriting the other existing pairs as

$$(w(h) + w_{j(h+1)}, P^{(j-1)}(w(h)) + p_{j(h+1)}), ..., (w(k-1) + w_{j(k)}, P^{(j-1)}(w(k-1)) + p_{j(k)})$$

we see that they are, respectively, dominated by the candidate pairs

$$(w(h) + w_j, P^{(j-1)}(w(h)) + p_j), ..., (w(k-1) + w_j, P^{(j-1)}(w(k-1)) + p_j),$$

Hence the merged list at the end of iteration $j$ contains the pairs

$$(w(0), P^{(j-1)}(w(0))), ..., (w(h), P^{(j-1)}(w(h))),$$

followed by the pairs

$$(w(h) + w_j, P^{(j-1)}(w(h)) + p_j), ..., (w(k) + w_j, P^{(j-1)}(w(k)) + p_j).$$

These pairs are realized by a new tower of feasible sets

$$\begin{aligned}
S^{(j)}(w(0)) = S^{(j-1)}(w(0)) = \emptyset &\subseteq S^{(j)}(w(1)) = S^{(j-1)}(w(1)) = \{j(1)\} \\
&\subseteq S^{(j)}(w(2)) = S^{(j-1)}(w(2)) = \{j(1), j(2)\} \\
&\subseteq S^{(j)}(w(h)) = S^{(j-1)}(w(h)) = \{j(1), ..., j(h)\} \\
&\subseteq S^{(j)}(w(h) + w_j) = S^{(j-1)}(w(h)) \cup \{j\} = \{j(1), ..., j(h), j\} \\
&\subseteq S^{(j)}(w(k-1) + w_j) = S^{(j-1)}(w(k-1)) \cup \{j\} \\
&\qquad = \{j(1), ..., j(h), j, j(h+1), ..., j(k-1)\} \\
&\subseteq S^{(j)}((w(k) + w_j) = S^{(j-1)}(w(k)) \cup \{j\} \\
&\qquad = \{j(1), ..., j(h), j, j(h+1), ..., j(k-1), j(k)\}.
\end{aligned}$$

Of course, this is the case only if

$$P^{(j-1)}(w(k)) + p_j = p(S^{(j-1)} \cup \{j\}) \le d_j.$$

If not, the pair $(w(k) + w_j, P^{(j-1)}(w(k)) + p_j)$ is discarded and the last set in the new tower is either

$$S^{(j-1)}(w(k-1)) \cup \{j\}, \text{if } h < k,$$

or

$$S^{(j-1)}(w(k)), \text{if } h = k.$$

In either case, it follows that the relation (5.3) holds at the end of iteration $j$. Since properties (5.4) and (5.5) follow directly from the definition of $h$, we see that at the end of iteration $j$, the tower-of-sets property holds.

It is now evident that list-making and list-merging is totally unnecessary. If we know the largest set in the tower at any given iteration, then we know everything about the tower. Furthermore, it is now a small step to verify that the Moore-Hodgson recurrence relations are designed to compute the largest set $S^{(j)}$ at iteration $j$ from the largest set $S^{(j-1)}$ at iteration $j-1$. Thus the Moore-Hodgson algorithm is, in effect, a streamlined version of the dynamic programming computation for the problem $1||\sum w_j U_j$.

As a final note, observe that the algorithm does not require that either processing times or job weights be integers. Furthermore, the maximum-weight feasible set that is computed is invariant under changes in job weights, provided the relative ordering of the weights is unchanged (see Exercises 5.10 and 5.11).

**Exercises**

5.7. Describe how to determine in $O(n \log n)$ time whether or not processing times and job weights are oppositely ordered.

5.8. Prove that an instance of the knapsack problem with oppositely ordered $w_j$'s and $p_j$'s can be solved by filling the knapsack with items in nonincreasing order of the ratios $w_j/p_j$, until no further item can be added. (No item is fractionalized.)

5.9. Complete the proof of the Moore-Hodgson algorithm by supplying the argument for the case in which there is no index $h$, as defined in the inductive proof.

5.10. The following greedy procedure is an alternative to the Moore-Hodgson algorithm (but requires $O(n^2)$ running time; cf. Exercise 5.2). Index the jobs in nonincreasing order of job weights and in nondecreasing order of processing times. Then, starting with $S^{(0)} = \emptyset$, process the jobs in this order, solving the recurrence relations

$$S^{(j)} = \begin{cases} S^{(j-1)} \cup \{j\}, & \text{if } S^{(j-1)} \cup \{j\} \text{ is feasible,} \\ S^{(j-1)}, & \text{otherwise.} \end{cases}$$

Prove that the set $S^{(n)}$ computed by this procedure is the same as the set $S^{(n)}$ computed by the Moore-Hodgson algorithm.

5.11. Prove that the algorithm in the previous exercise computes a feasible set $S^{(n)}$ satisfying the following (very strong) optimality property: The weight of the $i$th weightiest job in $S^{(n)}$ is at least as great as the weight of the $i$th weightiest job in any other feasible set.

5.12. (a) Suppose that a certain feasible subset $R$ of jobs is required to be on time. Describe how to modify the Moore-Hodgson algorithm to accommodate this constraint. (b) Suppose that, in addition to a due date $d_j$, each job $j$ has a (hard) deadline $\bar{d}_j \geq d_j$. Suppose that our objective is to minimize $\sum U_j$, subject to satisfaction of all the deadlines. Show that the problem posed in part (a) is a special case of this problem. (Note: The problem $1|\bar{d}_j|\sum U_j$ has been shown to be NP-hard.)

5.13. Describe how to adapt the Moore-Hodgson algorithm to solve the problem $1|pmtn, r_j|\sum w_j U_j$, under the condition that processing times and job weights are oppositely ordered and, in addition, release dates and due dates are oppositely ordered.

5.14. Opposite ordering of release dates and due dates is a special case of nesting of release date-due date intervals. (Intervals $[r_j, d_j]$, $j = 1, 2, ..., n$, are said to be *nested* if for all pairs $j, k$, either $[r_j, d_j]$ and $[r_k, d_k]$, are disjoint (except possibly at an end point) or one interval is contained in the other. Generalize the algorithm derived for Exercise 5.13 to solve $1|pmtn, r_j|\sum w_j U_j$ in the special case that processing times and job weights are oppositely ordered and release date-due date intervals are nested. (Hint: Form a rooted tree (or a forest of trees) in which the nodes are identified with intervals and the children of a node are identified with the maximal intervals contained within it. Work from the leaves of the tree upward, computing a maximum weight feasible set at each node. You will probably need a data structure supporting the operations of *insert*, *deletemax*, and *merge*, each of which can be implemented to run in $O(\log n)$ time. The algorithm should run in $O(n \log n)$ time.)

## 5.5. Similarly Ordered Release Dates and Due Dates

Release dates make things much more difficult. The problem $1|r_j|\sum U_j$ is strongly NP-hard, as can be shown by straightforward transformation from 3-PARTITION (see Exercise 5.19). It follows that there is no polynomial-time algorithm for $1|r_j|\sum U_j$ and no pseudopolynomial-time algorithm for $1|r_j|\sum w_j U_j$ (unless $P = NP$). Nevertheless, it is possible to generalize the dynamic programming algorithm of Section 5.2 to solve $1|r_j|\sum w_j U_j$ in the case that release dates and due dates are .ul similarly ordered. By this we mean that $d_i < d_j$ implies $r_i \leq r_j$, for each pair of jobs $J_i$ and $J_j$. Accordingly, in this section we shall hereafter assume that the jobs have been given an EDD numbering such that

$$r_1 \leq r_2 \leq ... \leq r_n, \text{and}$$
$$d_1 \leq d_2 \leq ... \leq d_n.$$

Under this condition, there is no advantage to preemption and the problems $1|r_j|\sum w_j U_j$ and $1|pmtn, r_j|\sum w_j U_j$ are equivalent. Furthermore, the extended EDD rule produces schedules with no preemptions.

Let $C^{(j)}(w)$ denote the earliest completion time of a feasible subset of jobs $J_1, J_2, ..., J_j$

having total weight exactly $w$. Generalizing the recurrences (5.1), we have

$$C^{(0)}(w) = \begin{cases} 0, & \text{if } w = 0, \\ +\infty, & \text{otherwise,} \end{cases}$$

$$C^{(j)}(w) = \begin{cases} \min\{C^{(j-1)}(w), \max\{r_j, C^{(j-1)}(w - w_j) + p_j\}\}, & \text{if } \max\{r_j, C^{(j-1)}(w - w_j)\} + p_j \le d_j, \\ \text{otherwise.} \end{cases}$$

$$\text{(5.6)}$$

As in the case of (5.1), there are $O(W)$ equations to solve at each of $n$ iterations, and each equation requires a constant number of arithmetic operations. Hence the values $C^{(n)}(w)$ can be computed in $O(nW)$ time. The maximum weight of a feasible set is given by the largest value of $w$ such that $C^{(n)}(w)$ is finite.

It happens that when job weights are equal, a variation of the Moore-Hodgson algorithm computes a maximum-cardinality feasible set in $O(n \log n)$ time. Let $q_j \le p_j$ be an *effective* processing time that is imputed to $J_j$, as we shall describe later, and define

$$q(S) = \sum_{j \in S} q_j,$$

for any set $S \subseteq \{1, ..., n\}$. Starting with $S^{(0)} = \emptyset$, we process the jobs in EDD order, constructing feasible sets $S^{(j)}$, $j = 1, 2, ..., n$, by the recurrence relations

$$S^{(j)} = \begin{cases} S^{(j-1)} \cup \{j\}, & \text{if } q(S^{(j-1)} \cup \{j\}) \le d_j - r_j, \\ S^{(j-1)} \cup \{j\} - \{l\} & \text{otherwise.} \end{cases} \quad \text{(5.7)}$$

where

$$l = \operatorname{argmax}\{q_i | i \in S^{(j-1)} \cup \{j\}\}.$$

The set $S^{(n)}$ is then a maximum-cardinality feasible set.

The inductive proof argument that we shall make to justify the above recurrence relations for $1|r_j|\sum U_j$, under the condition of similarly ordered release dates and due dates, parallels that of the previous section. Before proceeding with this argument, let us try to gain some insight from a numerical example with the following problem data:

| $j$ | 1 | 2 | 3 | 4 | 5 |
|-----|---|---|---|---|---|
| $r_j$ | 1 | 2 | 3 | 5 | 6 |
| $p_j$ | 3 | 7 | 5 | 3 | 3 |

Assume that all due dates are very large and hence are of no consequence. The dominant pairs existing at the end of iteration 4 are of the form $(i, C^{(4)}(i))$, $i = 0, 1, ..., 4$. These dominant pairs, plotted as dots in Figure 5.4, are as follows:

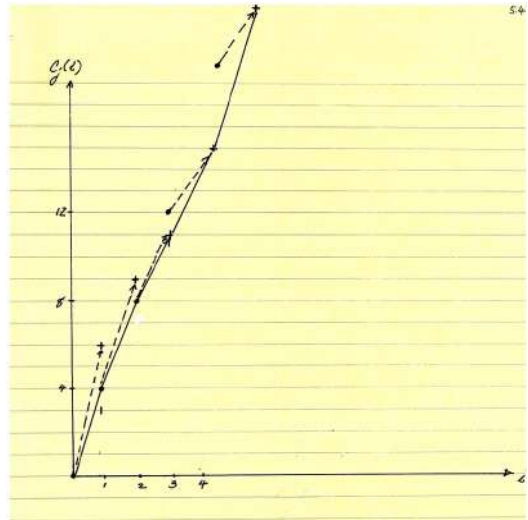

**Figure 5.4.**

$$(0,0), (1,4), (2,8), (3,12), (4,19).$$

With reference to Figure 5.5, we observe that these existing dominant pairs are realized by a tower of feasible sets:

$$S^{(4)}(0) = \emptyset \subseteq S^{(4)}(1) = \{1\}$$
$$\subseteq S^{(4)}(2) = \{1,4\}$$
$$\subseteq S^{(4)}(3) = \{1,4,3\}$$
$$\subseteq S^{(4)}(4) = \{1,4,3,2\}.$$

Candidate pairs $(i+1, \max\{r_5, C^{(4)}(i)\} + p_5)$, $i = 0,1,...,4$, are formed from the existing list. These are plotted as +'s in Figure 5.4, and are as follows:

$$(1,6), (2,9), (3,11), (4,15), (5,22).$$

The candidate pairs $(1,6)$ and $(2,9)$ are each dominated by the existing pair $(2,8)$. The existing pairs $(3,12)$ and $(4,19)$ are respectively dominated by the candidate pairs $(3,11)$ and $(4,15)$. Hence when the two lists of pairs are merged, the surviving

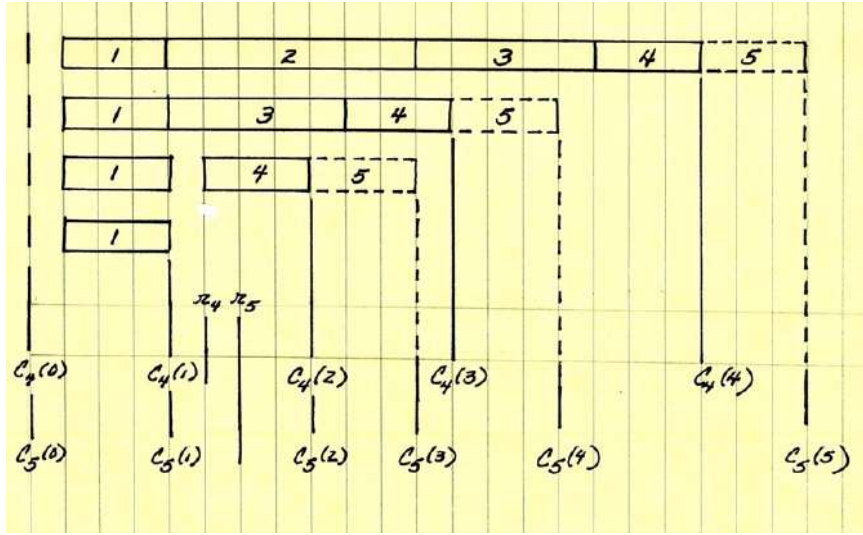

**Figure 5.5.**

dominant pairs are

$$(0,0),(1,4),(2,8),(3,11),(4,15),(5,22).$$

Again with reference to Figure 5.5, we see that these pairs are again realized by a tower of feasible sets:

$$S^{(5)}(0) = S^{(4)}(0) = \emptyset \subseteq S^{(5)}(1) = S^{(4)}(1) = \{1\}$$
$$\subseteq S^{(5)}(2) = S^{(4)}(2) = \{1,4\}$$
$$\subseteq S^{(5)}(3) = S^{(4)}(2) \cup \{5\} = \{1,4,5\}$$
$$\subseteq S^{(5)}(4) = S^{(4)}(3) \cup \{5\} = \{1,4,5,3\}$$
$$\subseteq S^{(5)}(5) = S^{(4)}(4) \cup \{5\} = \{1,4,5,3,2\}.$$

In general, the input to iteration $j$ is a list of dominant pairs,

$$(0,C^{(j-1)}(0)),(1,C^{(j-1)}(1)),...,(k,C^{(j-1)}(k)),$$

where each pair $(i,C^{(j-1)}(i))$ is realized by a feasible set $S^{(j-1)}(i)$, with

$$|S^{(j-1)}(i)| = i,$$
$$c(S^{(j-1)}(i)) = C^{(j-1)}(i).$$

(Recall that, by definition, $c(S)$ is the time of completion of the last job in an extended EDD schedule for $S$.) We assert that the feasible sets form a tower:

$$\begin{aligned}
S^{(j-1)}(0) = \emptyset \subseteq S^{(j-1)}(1) &= \{j(1)\} \\
&\subseteq S^{(j-1)}(2) = \{j(1), j(2)\} \\
&\quad ... \\
&\subseteq S^{(j-1)}(k) = \{j(1), j(2), ..., j(k)\}.
\end{aligned} \tag{5.8}$$

For each $j(i) \in S^{(j-1)}(k)$, $i = 1, 2, ..., k$, define

$$q_{j(i)} = \max\{r_j, C^{(j-1)}(i)\} - \max\{r_j, C^{(j-1)}(i-1)\}. \tag{5.9}$$

We assert that

$$q_{j(1)} \leq q_{j(2)} \leq ... \leq q_{j(k)}, \tag{5.10}$$

and refer to (5.8) and (5.10) as the *tower-of-sets* property for this problem.

It is useful to point out that (5.9) is equivalent to the following less compact way to determine the effective processing time:

$$q_{j(i)} = \begin{cases} 0, & \text{if } C^{(j-1)}(i) \leq r_j, \\ C^{(j-1)}(i) - C^{(j-1)}(i-1), & \text{if } r_j \leq C^{(j-1)}(i-1), \\ C^{(j-1)}(i) - r_j, & \text{if } C^{(j-1)}(i-1) < r_j < C^{(j-1)}(i). \end{cases}$$

The third condition holds for at most one of $q_{j(i)}$, $i = 1, ..., k$. It is not hard to verify that in each of the three cases, $q_{j(i)} \leq p_{j(i)}$: the additional processing of $J_{j(i)}$ can not delay the completion time of the extended EDD schedule by more than $p_{j(i)}$ time units beyond the maximum of the completion time for $J_{j(1)}, ..., J_{j(i-1)}$ and its release date.

We show next that the effective processing times play an important implicit role in determining the dominating pairs after iteration $j$ of the dynamic programming algorithm. To perform iteration $j$, we form a list of candidate pairs,

$$(i+1, \max\{r_j, C^{(j-1)}(i)\} + p_j), i = 0, 1, ..., k.$$

Let $h$ be the largest index $i$, $1 \leq i \leq k$, if any, such that $q_{j(i)} \leq p_j$. (As in Section 5.4, the case in which there is no such $h$ is left as an exercise.) For $i = 1, ..., h$,

$$p_j \geq q_{j(i)} = \max\{r_j, C^{(j-1)}(i)\} - \max\{r_j, C^{(j-1)}(i-1)\} \geq C^{(j-1)}(i) - \max\{r_j, C^{(j-1)}(i-1)\},$$

and so it follows that the existing pairs

$$(1, C^{(j-1)}(1)), ..., (h, C^{(j-1)}(h))$$

respectively dominate the candidate pairs

$$(1, \max\{r_j, C^{(j-1)}(0)\} + p_j), ..., (h, \max\{r_j, C^{(j-1)}(h-1)\} + p_j).$$

Furthermore, when $i > h$, we have that $q_{j(i)} > p_j \geq 0$, which implies that $q_{j(i)} = C^{(j-1)}(i) - \max\{r_j, C^{(j-1)}(i-1)\}$. As a result,

$$p_j < q_{j(i)} = C^{(j-1)}(i) - \max\{r_j, C^{(j-1)}(i-1)\},$$

and so the candidate pairs

$$(h+1, \max\{r_j, C^{(j-1)}(h)\} + p_j), ..., (k, \max\{r_j, C^{(j-1)}(k-1)\} + p_j).$$

respectively dominate the existing pairs

$$(h+1, C^{(j-1)}(h+1)), ..., (k, C^{(j-1)}(k)).$$

Hence the merged list at the end of iteration $j$ contains the pairs

$$(0, C^{(j-1)}(0)), ..., (h, C^{(j-1)}(h)),$$

followed by the pairs

$$(h+1, \max\{r_j, C^{(j-1)}(h)\} + p_j), ..., (k+1, \max\{r_j, C^{(j-1)}(k)\} + p_j).$$

These pairs are realized by a new tower of feasible sets

$$
\begin{aligned}
S^{(j)}(0) = S^{(j-1)}(0) = \emptyset &\subseteq S^{(j)}(1) = S^{(j-1)}(1) = \{j(1)\} \\
&\subseteq S^{(j)}(2) = S^{(j-1)}(2) = \{j(1), j(2)\} \subseteq ... \\
&\subseteq S^{(j)}(h) = S^{(j-1)}(h) = \{j(1), ..., j(h)\} \\
&\subseteq S^{(j)}(h+1) = S^{(j-1)}(h) \cup \{j\} = \{j(1), ..., j(h), j\} \subseteq ... \\
&\subseteq S^{(j)}(k) = S^{(j-1)}(k-1) \cup \{j\} = \{j(1), ..., j(h), j, j(h+1), ..., j(k-1)\} \\
&\subseteq S^{(j)}(k+1) = S^{(j)}(k) \cup \{j\} = \{j(1), ..., j(h), j, j(h+1), ..., j(k-1), j(k)\}.
\end{aligned}
$$

That is, this is the case if

$$\max\{r_j, C^{(j-1)}(k)\} + p_j \leq d_j.$$

If not, the pair $(k+1, \max\{r_j, C^{(j-1)}(k)\} + p_j)$ is discarded and the largest set in the new tower is either

$$S^{(j-1)}(k-1) \cup \{j\}, \text{if } h < k,$$

or

$$S^{(j-1)}(k), \text{if } h = k.$$

This shows that property (5.8) holds at the end of iteration $j$. Property (5.10) holds as well, as we indicate below.

It would be inefficient to compute the $q_{j(i)}$ values at each iteration directly from the definition (5.9). Instead, we shall derive a more efficient procedure. Let $j(1), j(2), ..., j(k)$ be the elements of $S^{(j-1)}(k)$ ordered to indicate the inclusions (5.8). Suppose that $h$ is the index computed above, so that the tower at the end of iteration $j$ is ordered

$$j(1), j(2), ..., j(h), j, j(h+1), ..., j(k);$$

of course, $j(k)$ might be deleted, but assume that it is not. (An identical argument proves the other case.) Let $j'(i)$, $i = 1, ..., k+1$, denote the $i$th element of this sequence.

At the start of iteration $j$, each job $J_k$ in this tower has effective processing time $q_k$. (For $J_j$, let $q_j = p_j$.) We have already seen that

$$C^{(j)}(i) = \begin{cases} C^{(j-1)}(i), & \text{if } i = 1, ..., h, \\ \max\{r_j, C^{(j-1)}(i-1)\} + p_j, & \text{if } i = h+1, ..., k+1. \end{cases}$$

This recurrence for $C^{(j)}(i)$ can be used to show that the effective processing times used in iteration $j$,

$$q_{j'(i)} = \max\{r_j, C^{(j)}(i)\} - \max\{r_j, C^{(j)}(i-1)\}, \tag{5.11}$$

by applying the recurrence separately in each of the three cases: (a) $i = 1, ..., h$; (b) $i = h+1$; and (c) $i = h+2, ..., k+1$. Therefore, to update these values for the next iteration, we need only update them to reflect the new release date $r_{j+1}$. If the release date is increased by one unit, the effect is to decrease by one the effective processing time of the first job in the tower that currently has a positive effective processing time. The same idea can be used to update the release date from $r_j$ to $r_{j+1}$ in (5.11), by repeating this $r_{j+1} - r_j$ times; this can clearly be made efficient by making the maximum allowed reduction to $q_{j'(i)}$ at once, for each $i = 1, ..., k$, in that order.

We can now show that (5.10) is maintained by the algorithm. In iteration $j$, $j$ is inserted into the tower between $j(h)$ and $j(h+1)$ according $q_j$'s place in the sorted order of effective processing times. Furthermore, after the update just described, the modified effective processing times are still sorted in nondecreasing order.

In order to implement the algorithm we propose to maintain two sets, $S$ and $S'$, where $S$ contains jobs whose effective processing times are strictly positive and $S'$ contains jobs with zero effective processing times. At the end of iteration $j$, the maximum feasible set in the tower is then $S^{(j)} = S \cup S'$. We propose to implement the set $S$ by a priority queue that supports the operations of *insert*, *deletemin*, and *deletemax*. For the set $S'$, we need only support the operation *insert*. At iteration $j$, the following subroutine is performed to modify the effective processing times.

```
modify( ):
r := r_j − r_{j−1} ;
while (r > 0 and S ≠ ∅) do
    i := deletemin(S) ;
    q(S) := q(S) − q_i ;
    if (q_i ≤ r) then
        insert(S', i) ;
        r := r − q_i ;
    else
        q_i := q_i − r ;
        insert(S, i) ;
        q(S) := q(S) + q_i ;
        r := 0 ;
    end
end
```

We can now implement the recurrence relations (5.7) as follows:

```
similar( ):
S := ∅ ;
S' := ∅ ;
q(S) := 0 ;
for j = 1 to n do
    call modify( );
    q_j := p_j ;
    insert(S, j) ;
    q(S) := q(S) + q_j ;
    if (q(S) > d_j − r_j) then
        l := deletemax(S) ;
        q(S) := q(S) − q_l ;
    end
end
return (S ∪ S') ;
```

Note that $O(n)$ of each of the operations *insert*, *deletemin*, and *deletemax* are performed in the course of the computation. Each of these operations takes $O(\log n)$ time. Hence the overall running time of the algorithm is bounded by $O(n \log n)$.

**Exercises**

5.15. Devise an example to show that the tower-of-sets property (5.8),(5.10), does not hold for $1|r_j|\sum w_j U_j$ with similarly ordered release dates and due dates and with oppositely ordered processing times and weights.

5.16. Show that the tower-of-sets property does hold for the problem $1|r_j, p_j = 1|\sum w_j U_j$ in the case of similarly ordered release dates and due dates. Devise an $O(n \log n)$ algorithm for solving this problem (cf. Exercise 5.2).

5.17. If $V_j$ is defined as in Section 5.4, show that the problem $1|pmtn, r_j|\sum w_j V_j$ can be solved in $O(n \log n)$ time, under the condition that $[r_j d_j]$ intervals are nested. Processing times and job weights may be arbitrary. (Hint: Modify the algorithm for Exercise 5.14, so as to remove jobs from an infeasible set $S^{(j)}$ in nonincreasing order of the ratios $w_i/p_i$, if necessary fractionalizing the last job removed and reinserting it, until $p(S^{(j)}) = d_j - r_j$.)

5.18. Devise an $O(nW^2)$ algorithm for $1|pmtn, r_j|\sum w_j U_j$ with nested release date-due date intervals (and arbitrary processing times and job weights).

5.19. Show that $1|r_j|\sum U_j$ is strongly *NP*-hard.

## 5.6. The General Problem $1|pmtn, r_j|\sum w_j U_j$

The problem $1|pmtn, r_j|\sum w_j U_j$ can be solved by dynamic programming in $O(nk^2 W^2)$ time, where $k$ is the number of distinct release dates. When release dates and due dates are similarly ordered, the dynamic programming recurrences specialize to (5.6), which can be solved in $O(nW)$ time. And when $k = 1$, the recurrences further specialize to (5.1).

Our dynamic programming recurrences iteratively compute two types of values, $C^{(j)}(r, w)$ and $P^{(j-1)}(r, r', w)$. In the following definitions, $j$ is an index, $1 \le j \le n$, $r$ and $r'$ are release dates, $r \le r'$, and $w$ is an integer, $0 \le w \le W$. Once again, we assume the jobs are indexed in EDD order.

We define

$$C^{(j)}(r, w) = \min\{c(S)\},$$

where the minimum is taken with respect to feasible sets $S$ such that

$$S \subseteq \{1, 2, ..., j\}, r(S) \ge r, \text{ and } w(S) \ge w.$$

If there is no such feasible set $S$, $C^{(j)}(r, w) = +\infty$. We also define $P^{(j-1)}(r, r', w)$ to be the minimum amount of processing *after time* $r_j$ in an extended EDD schedule, with respect to feasible sets $S$ such that

$$S \subseteq \{1, 2, ..., j-1\}, r(S) \ge r, c(S) \le r', \text{ and } w(S) \ge w.$$

If there is no such feasible set, $P^{(j-1)}(r, r', w) = +\infty$.

It follows from the above definitions that the maximum weight of a feasible set is given by the largest value of $w$ such that $C^{(n)}(r_{\min}, w)$ is finite, where $r_{\min} =$

$\min_j\{r_j\}$.

*Computation of* $C^{(j)}(r,w)$. We shall compute the values $C^{(j)}(r,w)$ in $n$ iterations, $j = 1, 2, ..., n$, starting with the initial conditions

$$C^{(0)}(r,w) = \begin{cases} r, & \text{if } w = 0, \\ +\infty, & \text{otherwise.} \end{cases}$$

Observe that $j$ cannot be contained in a feasible set $S$ with $r(S) > r_j$. Hence

$$C^{(j)}(r,w) = C^{(j-1)}(r,w) \text{ if } r > r_j,$$
$$\leq C^{(j-1)}(r,w), \text{ otherwise }.$$

It follows that at iteration $j$ we have only to compute the values of $C^{(j)}(r,w)$ for which $r \leq r_j$. So let $r \leq r_j$ and suppose $S \subseteq \{1, 2, ..., j\}$ realizes the value $C^{(j)}(r,w)$. We distinguish three cases, as follows.

*Case 1.* $j \notin S$. Then

$$C^{(j)}(r,w) = C^{(j-1)}(r,w).$$

*Case 2.* $j \in S$, and the processing of $J_j$ begins after all jobs in $S - \{j\}$ are completed in the extended EDD schedule. We then have

$$C^{(j)}(r,w) = \max\{r_j, c(S - \{j\})\} + p_j.$$

We may assume that $S - \{j\}$ is such that

$$c(S - \{j\}) = C^{(j-1)}(r, w - w_j).$$

(If not, replace $S - \{j\}$ by a feasible subset of $\{1, 2, ..., j - 1\}$ for which this is so.) Then

$$C^{(j)}(r,w) = \max\{r_j, C^{(j-1)}(r, w - w_j)\} + p_j.$$

*Case 3.* $j \in S$, and the processing of $J_j$ begins before all jobs in $S - \{j\}$ are completed in the extended EDD schedule of $S$. This means that there is idle time within the interval $[r_j, c(S - \{j\}))$. in the EDD schedule for $S - \{j\}$. Recall from Chapter 4 that a block indexes a maximal subset of jobs that are processed continuously without idle time. Let $S'$ be the .ul last block in the extended EDD schedule of $S - \{j\}$, i.e.,

$$r(S') = \max\{r(B)|B \text{ a block of the extended EDD schedule of } S - \{j\}\},$$

where $r(S') > r_j$. Let $r' = r(S')$ and $w' = w(S')$. It must be the case that

$$c(S') = C^{(j-1)}(r', w'),$$

else $S$ is not optimal. We may also assume that the total amount of processing done on jobs in $(S - \{j\}) - S'$ in the interval $[r_j, r']$ does not exceed $P^{(j-1)}(r, r', w - w_j - w')$. This means that the total amount time available for the processing of $J_j$ in the interval $[r_j, r']$ is

$$(r' - r_j) - P^{(j-1)}(r, r(S'), w - w_j - w(S')),$$

and the amount of processing done on $J_j$ after time $r'$ is

$$\max\{0, p_j - (r' - r_j) + P^{(j-1)}(r, r', w - w_j - w')\}.$$

Hence the completion time of the last job in $S$ is

$$C^{(j-1)}(r', w') + \max\{0, p_j - (r' - r_j) + P^{(j-1)}(r, r', w - w_j - w')\}. \tag{5.12}$$

Now observe that the expression (5.12) must be minimum with respect to sets $S'$, with $r(S') > r_j$, $w' = w(S') \le w - w_j$. In other words,

$$C^{(j)}(r, w) = \min_{r', w'}\{C^{(j-1)}(r', w') + max\{0, p_j - r' + r_j + P^{(j-1)}(r, r', w - w_j - w')\}\}$$

Putting Cases 1, 2, and 3 together, for $r \le r_j$ we have the recurrence relations

$$C^{(j)}(r, w) = \min \left\{ \begin{array}{l} C^{(j-1)}(r, w), \\ \max\{r_j, C^{(j-1)}(r, w - w_j)\} + p_j, \\ \min_{r', w'}\{C^{(j-1)}(r', w') + \max\{0, p_j - r' + r_j + P^{(j-1)}(r, r', w - w_j - w')\}\} \end{array} \right\} \tag{5.13}$$

where the inner minimization is taken over all distinct release dates $r' > r_j$ such that $r' \in \{r_1, r_2, ..., r_{j-1}\}$ and all integers $w'$, $0 < w' \le w - w_j$. It is important to note that (5.13) is valid only if the right hand side does not exceed $d_j$; if this is not so, set $C^{(j)}(r, w) = +\infty$.

*Computation of* $P^{(j-1)}(r, r', w)$. We shall now derive recurrence relations for computing $P^{(j-1)}(r, r', w)$, for all distinct release dates $r, r'$, with $r \le r'$.

We have as initial conditions

$$P^{(j-1)}(r, r', 0) = 0.$$

If $w > 0$, then $P^{(j-1)}(r, r', w)$ is realized by a nonempty set $S \subseteq \{1, 2, ..., j-1\}$. We distinguish two cases.

*Case 1.* $r(S) > r$. Then

$$P^{(j-1)}(r, r', w) \leq P^{(j-1)}(r^+, r', w),$$

where we define $r^+$ to be the smallest distinct due date larger than $r$.

*Case 2.* $r(S) = r$. Let $S' \subseteq S$ be the block of $S$ such that $r(S') = r$, and let w(S') = w'. We may assume that $c(S') = C^{(j-1)}(r, w')$). Hence the total amount of processing done on $S'$ in the interval $[r_j, r']$ is

$$\max\{0, C^{(j-1)}(r, w') - r_j\}.$$

Let $r''$ be the smallest release date greater than or equal to $C^{(j-1)}(r, w')$. It must be the case that the total amount of processing done on $S - S'$ in the interval $[r_j, r']$ is $P^{(j-1)}(r'', r', w - w')$. Hence the total amount of processing done on $S$ in the interval $[r_j, r']$ is

$$\max\{0, C^{(j-1)}(r, w') - r_j\} + P^{(j-1)}(r'', r', w - w'). \tag{5.14}$$

Now observe that the expression (5.14) must be minimum with respect to sets $S'$, with $r(S') = r$, $c(S') \leq r'' \leq r'$, and $w(S') \leq w$. That is,

$$P^{(j-1)}(r, r', w) = \min_{0 < w' \leq w} \{\max\{0, C^{(j-1)}(r, w') - r_j\} + P^{(j-1)}(r'', r', w - w')\},$$

where $r''$ is the smallest release date no less than $C^{(j-1)}(r, w')$.

Putting the above two cases together, we have

$$P^{(j-1)}(r, r', w) = \min \left\{ \begin{array}{l} P^{(j-1)}(r^+, r', w), \\ \min_{0 < w' \leq w}\{\max\{0, C^{(j-1)}(r, w') - r_j\} + P^{(j-1)}(r'', r', w - w')\} \end{array} \right\} \tag{5.15}$$

giving us the recurrences we need.

We shall now analyze the time and space complexity of the dynamic programming computation. At each of $n$ iterations, $j = 1, 2, ..., n$, there are $O(k^2 W)$ of the $P^{(j-1)}(r, r', w)$ values to compute, one for each combination of $r, r', w$. By (5.15), each $P^{(j-1)}(r, r', w)$ is found by minimization over $O(W)$ choices of $w' \leq w$. Hence the time required to compute the $P^{(j-1)}(r, r', w)$ values at each iteration is bounded by $O(k^2 W^2)$. There are $O(kW)$ of the $C^{(j)}(r, w)$ values to compute, one for each combination of $r$ and $w$. By (5.13), each $C^{(j)}(r, w)$ is found by minimization over $O(kW)$ choices of $r', w'$. Hence the time required to compute the $C^{(j)}(r, w)$ values at each iteration is bounded by $O(k^2 W^2)$. It follows that the overall time bound for these computations is $O(nk^2 W^2)$. Space requirements are clearly bounded by $O(k^2 W)$.

As we have observed, the maximum weight of a feasible subset can be obtained by finding the maximum value of $w$ such that $C^{(n)}(r_{\min}, w)$ is finite. (The $O(W)$ time required for this is dominated by the time required for other computations.) In practice, however, one wants to be able to construct a maximum-weight feasible set, not simply to find its weight. The most straightforward way to do this is to compute an incidence vector of the set realizing each $P^{(j-1)}(r, r', w)$ and $C^{(j)}(r, w)$ value. The computation of these incidence vectors can be carried out with an expenditure of $O(n^2 k^2 W)$ time, which is dominated by the $O(nk^2 W^2)$ time bound obtained above. However, because $O(k^2 W)$ $n$-vectors must be stored, this approach increases space requirements to $O(nk^2 W)$.

We claim that it is possible to use pointers to construct a maximum-weight feasible set, while maintaining the time and space bounds of $O(nk^2 W^2)$ and $O(k^2 W)$. We leave this as an exercise.

The EDD Rule creates preemptions only at release dates. Hence when the jobs in a maximum-weight feasible set are scheduled, at most $k-1$ preemptions are created, at most one at each distinct release date other than the first.

Observe that when release dates and due dates are similarly ordered, there are no release dates $r' > r_j$ over which the inner minimization in (5.14) can be, hence these recurrence relations simplify to

$$C^{(j)}(r, w) = \min \left\{ \begin{array}{l} C^{(j-1)}(r, w), \\ \max\{r_j, C^{(j-1)}(r, w - w_j)\} + p_j \end{array} \right\}.$$

Now let $C^{(j)}(w) = C^{(j)}(r_{\min}, w)$ and we have simply

$$C^{(j)}(w) = \min \left\{ \begin{array}{l} C^{(j-1)}(w), \\ \max\{r_j, C^{(j-1)}(w - w_j)\} + p_j \end{array} \right\},$$

and we have the recurrence equations (5.6).

When all release dates are equal,

$$\max\{r_j, C^{(j-1)}(w - w_j)\} = C^{(j-1)}(w - w_j),$$

and the recurrence further simplifies to

$$C^{(j)}(w) = \min\{C^{(j-1)}(w), C^{(j-1)}(w - w_j) + p_j\},$$

and we have the recurrence equations (5.1).

### Exercises

5.20. Show that it is possible to use pointers to construct a maximum-weight feasible set, while maintaining the time and space bounds of $O(nk^2 W^2)$ and $O(k^2 W)$.

### 5.7. Precedence Constraints

When the jobs are related by precedence constraints, severely restricted versions of the $1||\sum w_j U_j$ problem become *NP*-hard. We will show below that the case of unit weights and unit processing times, $1|prec, p_j = 1|\sum U_j$, is *NP*-hard. We will then refine this result and show that it still holds for the case of chain-type precedence constraints, where each job has at most one predecessor and at most one successor.

**Theorem 5.2.** $1|prec, p_j = 1|\sum U_j$ *is NP-hard in the strong sense.*

*Proof.* We will show that the clique problem reduces to the decision version of $1|prec, p_j = 1|\sum U_j$. An instance of the clique problem is given by a graph $G = (V, E)$ and an integer $k$. The question is if $G$ has a clique (i.e., a complete subgraph) on at least $k$ vertices.

Given an instance of clique, let $l = \binom{k}{2}$ denote the number of edges in a clique of size $k$. The corresponding instance of the scheduling problem will have $n = |V| + |E|$ unit-time jobs. We introduce a job $J_v$ for every vertex $v \in V$ and job $J_e$ for every edge $e \in E$, with a precedence constraint $J_v \rightarrow J_e$ whenever $v$ is an endpoint of $e$. Each "vertex job" $J_v$ has a due date $d_v = n$, and each "edge job" $J_e$ has a due date $d_e = k + l$. Note that no vertex job can be late in a schedule without idle time. We claim that there is a schedule with at most $|E| - l$ late jobs if and only if $G$ has a clique of size at least $k$.

Suppose that a clique on $k$ vertices exists. We first schedule the $k$ corresponding vertex jobs in the interval $[0, k]$. In view of the precedence constraints, we can then schedule the $l$ jobs corresponding to the clique edges in the interval $[k, k+l]$ ; they are on time. They are followed by the remaining vertex jobs in $[k+l, |V|+l]$ and, finally, the remaining edge jobs in $[|V|+l, n]$ ; these $|E| - l$ edge jobs are late. This schedule meets the claimed bound on the number of late jobs.

Conversely, suppose that there is a schedule in which at most $|E| - l$ jobs are late, or, equivalently, in which at least $l = \binom{k}{2}$ edge jobs are completed by time $k + l$. These jobs must be preceded by at least $k$ vertex jobs. It follows that there is a set of $k$ vertex jobs that releases $l$ edge jobs for processing or, in other words, that a clique of size $k$ exists. □

**Theorem 5.3.** $1|chain, p_j = 1|\sum U_j$ *is NP-hard in the strong sense.*

*Proof.* To simplify the exposition, we present the reduction in two stages. We first show that the exact 3-cover problem reduces to the decision version of $1|chain|\sum U_j$, with general processing times. We then show that this problem can be reduced to an equivalent problem with unit processing times.

An instance of exact 3-cover consists of a set $T = \{1, ..., 3t\}$ and a family $\mathcal{S} = \{S_1, ..., S_s\}$ of 3-element subsets of $T$. It is a yes-instance if $\mathcal{S}$ includes an exact cover, i.e., a subfamily of $t$ subsets whose union is $T$ (cf. Figure 5.6(a)).

The corresponding instance of $1|chain|\sum U_j$ will have a job $J_i$ for each subset $S_i$ and a job $J_{ij}$ for each occurrence of an element $j$ in a subset $S_i$. More precisely, for each $S_i = \{j, k, l\} \in \mathcal{S}$ $(i = 1, ..., s)$, we introduce four jobs, $J_i, J_{ij}, J_{ik}, J_{il}$, which are related by chain-like precedence constraints, $J_i \to J_{ij} \to J_{ik} \to J_{il}$. Each 'subset job' $J_i$ is of length 1, and each 'occurrence job' $J_{ij}$ is of length $js$. The sum of all of the processing times (and hence the length of the schedules we will be considering) is denoted by $d = s + \sum_{i=1}^{s} \sum_{j \in S_i} js$. All subset jobs $J_i$ have due date $d$ ; none of these can be late. Each occurrence job $J_{ij}$ is assigned a due date $d_{ij}$ defined by

$$d_{ij} = t + s + 2s + ... + js = t + j(j+1)s/2, j \in S_i, i = 1, ..., s.$$

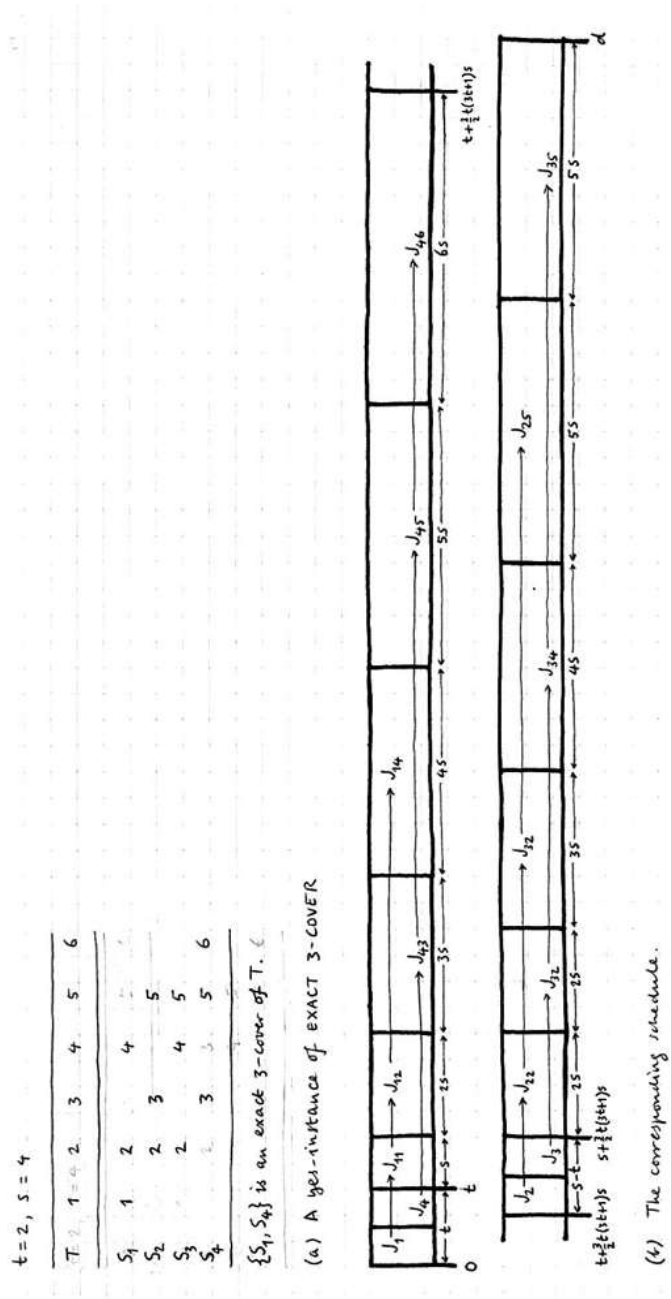

**Figure 5.6.** An exact 3-cover

We claim that there is a schedule with no more than $3(s-t)$ late jobs if and only if there is an exact 3-cover.

Suppose that $\mathcal{S}$ includes an exact 3-cover $\mathcal{S}'$. We then construct the following schedule (cf. Figure 5.6(b)). First, the $t$ jobs corresponding to the subsets $S_i \in \mathcal{S}'$ are scheduled in the interval $[0,t]$. Now note that each of these $t$ subset jobs heads a chain of three occurrence jobs such that the $3t$ elements to which these occurrences refer are all distinct. That is, for every element $j \in T$, there is exactly one occurrence job $J_{ij}$ for which the preceding subset job $J_i$ has been scheduled; this $J_{ij}$ is now scheduled in the interval $[d_{ij} - js, d_{ij}]$. In this way, $3t$ occurrence jobs occupy the interval $[t, t + 3t(3t+1)s/2]$ ; they are all on time. They are followed by the remaining $s-t$ subset jobs and, finally, the remaining $3(s-t)$ occurrence jobs; these latter jobs are late. This schedule meets the claimed bound on the number of late jobs.

Conversely, suppose that there exists a feasible schedule in which at most $3(s-t)$ jobs are late, or, equivalently, in which at least $3t$ occurrence jobs are on time. This implies that,

for every element $j \in T$, exactly one of its occurrence jobs $J_{ij}$ is on time.   (5.16)

The verification of this crucial implication is left to the reader (see Exercise 5.21). Assertion (5.16), in turn, implies that the amount of time available for processing the subset jobs that release these $3t$ occurrence jobs is bounded from above by

$$\max_{j \in T} d_{ij} - \sum_{j \in T} js = t + 3t(3t+1)s/2 - 3t(3t+1)s/2 = t.$$

Hence, $\mathcal{S}$ must include a subfamily of at most $t$ subsets whose union covers $T$. This subfamily constitutes an exact 3-cover.

This completes the first stage of the proof. It remains to be shown that the scheduling problem can be reduced to an equivalent problem with unit processing times. This is straightforward. We replace each occurrence job $J_{ij}$ by a chain of $js$ unit-time jobs, $J_{ij}^{(1)} \to ... \to J_{ij}^{(js-1)} \to J_{ij}^{(js)}$, with due dates $d_{ij}^{(1)} = ... = d_{ij}^{(js-1)} = d$, $d_{ij}^{(js)} = d_{ij}$ ($j \in S_i, i = 1, ..., s$). Given any feasible schedule in which the jobs of such a chain are not processed consecutively, we can obtain another schedule by moving the first $js-1$ jobs of the chain to the right, up to the $js$th one, thereby moving some other jobs to the left. The new schedule is still feasible, since no precedence constraints are violated, and it has no more late jobs, since the jobs that are moved to the right cannot be late. Hence, each chain $J_{ij}^{(1)} \to ... \to J_{ij}^{(js)}$ can be considered as a single job $J_{ij}$ with processing time $js$ and due date $d_{ij}$. As for the size of the final $1|chain, p_j = 1|\sum U_j$ instance, note that it has only $d < 9s^2t$ jobs; the entire transformation is polynomial because sufficiently small processing times have been chosen at the first stage. □

### Exercises

5.21. (a) Consider the $1||\sum U_j$ problem. Suppose that there are $t$ nonempty job sets, $FSJ_1,...,FSJ_t$, such that all jobs in $FSJ_j$ have a processing time $j$ and a due date $1 + 2 + ... + j (j = 1,...,t)$. Prove that in each optimal schedule exactly one job from $FSJ_j$ finishes at its due date, while the other jobs from $FSJ_j$ are late $(j = 1,...,t)$. (b) Use (a) to verify assertion (5.16) in the proof of Theorem 5.3.

### Notes

5.1. *Some preliminaries.* Each feasible set for an instance of the problem $1|r_j, p_j = 1|\sum w_j U_j$ is an independent set of a transversal matroid, hence the algorithm described in Exercise 5.2 is indeed an instance of the matroid greedy algorithm. The observation that $1|p_j = 1|\sum U_j$ can be solved in $O(n)$ time is due to Monma [1982].

5.2. *Dynamic programming solution of* $1||\sum w_j U_j$. Karp [1972] gave the reduction of the knapsack problem to $1|d_j = d|\sum w_j U_j$. The integer linear programming formulation of $1||\sum w_j U_j$ and the dynamic programming algorithm for solving this generalization of the knapsack problem are adapted from [Lawler & Moore, 1969]. Exercise 5.4 is due to Potts & Van Wassenhove [1988], who use linear relaxations as lower bounds within a branch and bound procedure for $1||\sum w_j U_j$.

5.3. *A fully polynomial approximation scheme.* Theorem 5.1 is due to Gens & Levner [1978]. By obtaining a preliminary upper bound on the optimum that is within a factor of 2, Gens & Levner [1981] improved the running time to $O(n^2 \log n + n^2 k)$. Exercise 5.6 is due to Sahni [1976]. Ibarra & Kim [1978] gave a polynomial approximation scheme to find a maximum feasible set for $1|d_j = d, tree|\sum w_j U_j$.

5.4. *The Moore-Hodgson algorithm.* Moore [1968] suggested a less elegant algorithm for the unweighted problem $1||\sum U_j$ but specifically credited the recurrences (5.2) to a suggestion by his colleague Hodgson, hence the appellation *Moore-Hodgson*. Maxwell [1970] provided an alternative derivation of the Moore-Hodgson algorithm based on ideas from linear and integer programming. The generalization of the algorithm to oppositely ordered processing times and job weights was noted by Lawler [1976A]. Sidney [1973] observed that the algorithm could be adapted to the case considered in Exercise 5.12. The NP-hardness of the problem $1|\bar{d}_j|\sum U_j$ was proved by Lawler [1982B]. The Moore-Hodgson algorithm and the algorithm of Lawler [1976A] were used to obtain lower bounds in the branch and bound procedure of Villareal & Bulfin [1983] to solve $1||\sum w_j U_j$ with arbitrary weights and processing times.

5.5. *Similarly ordered release dates and due dates.* The strong NP-hardness proof of $1|r_j|\sum U_j$ is due to Lenstra [–]. An $O(n^2)$ dynamic programming solution of $1|r_j|\sum U_j$, under the condition of similar ordering of release dates and due dates, is described by Kise, Ibaraki & Mine [1978]. The $O(n\log n)$ algorithm for this problem is due to Lawler [1982B]. The tower-of-sets property is known to hold for a number of special cases of $1|pmtn, r_j|\sum w_j U_j$, aside from those mentioned in the exercises, for example, as in the case that release dates and processing times are similarly ordered, and in opposite order to job weights, which is due to Lawler [–].

5.6. *The general problem $1|pmtn, r_j|\sum w_j U_j$.* The dynamic programming algorithm of this section is adapted from Lawler [1990], which improves on a less efficient algorithm given earlier by Lawler [1982B].

5.7. *Precedence constraints.* Theorem 6.2 is due to Garey and Johnson [1976], Theorem 6.3 to Lenstra and Rinnooy Kan [1980]. Ibaraki, Kise, and Mine [1976] proved that $1|chain, r_j, p_j = 1|\sum U_j$ is NP-hard.

# Contents

i

# 8

# Minsum criteria

Eugene L. Lawler

*University of California, Berkeley*

This is a fragmented chapter. Our focus is on the nonpreemptive and preemptive minimization of total completion time in polynomial time. We also present a pseudopolynomial-time algorithm for the minimization of total weighted completion time and related problems. We review NP-hardness results without providing proofs, and give pointers to the very few results on approximation and branch-and-bound.

## 8.1. Total completion time, nonpreemptive scheduling

In the absence of precedence constraints, we know that whatever jobs are performed on a given machine in an optimal schedule, they will be performed on that machine in SPT order. Recall formula (4.1) derived in Chapter 4:

$$\sum C_j = np_1 + (n-1)p_2 + \cdots + p_n.$$

We can solve the problem $P||\sum C_j$ by *coefficient matching*: There are $m$ 1's, $m$ 2's, ..., $m$ $(n-1)$'s, and $m$ $n$'s to match with $n$ $p_j$ values so as to obtain the smallest possible weighted sum. It is clear that an optimal solution to this matching problem is obtained by matching the $m$ 1's with the $m$ largest $p_j$'s, the $m$ 2's with the $m$ next-largest $p_j$'s, and so forth, ending up by matching one of the coefficients $\lceil n/m \rceil$ with the smallest $p_j$.

Thus an optimal schedule for $P||\sum C_j$ is as shown in Figure 8.1. One can also state an extended SPT rule for identical parallel machines, as follows: Schedule the jobs in time, with a decision point at the completion of a job. At each decision point choose to process the shortest job that has not yet been assigned to a machine.

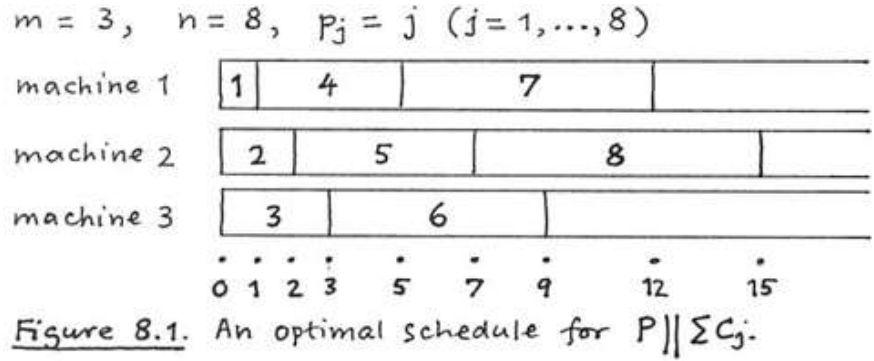

Figure 8.1. An optimal schedule for $P||\sum C_j$.

In the case of $Q||\sum C_j$, we notice that for the jobs assigned to machine $i$ the formula takes on the form:

$$\sum C_j = (n/s_i)p_1 + ((n-1)/s_i)p_2 + \ldots + (1/s_i)p_1.$$

Like $P||\sum C_j$, the problem $Q||\sum C_j$ can be solved by coefficient matching, except that now the coefficients are of the form $k/s_i$. The largest $p_j$ should be matched with the smallest of these coefficients ... and the smallest $p_j$ should be matched with the $n$th-largest of these coefficients.

Here is how to generate the coefficients $k/s_i$ in nondecreasing order of size: Place the fractions $1/s_i$, $i = 1, 2, \ldots, m$, into a priority queue $Q$ supporting the operations of **delete-min** and **insert**. Then

```
for h = 1, 2, ..., n do
    output k/s_i := delete-min(Q);
    insert (k + 1)/s_i into Q;
end
```

Clearly it is possible to match the coefficients and construct an optimal schedule in $O(n \log n)$ time.

The problem $R||\sum C_j$ can still be solved by coefficient matching. Note that, when job $j$ is scheduled in the $k$th last position on machine $i$, it contributes $kp_{ij}$ to the total completion time. We now describe schedules in terms of 0-1 variables $x_{(ik)j}$, where $x_{(ik)j} = 1$ if job $j$ is the $k$th last job processed on machine $i$, and $x_{(ik)j} = 0$, otherwise. The problem is then to minimize

$$\sum_{i,k} \sum_j k p_{ij} x_{(ik)j}$$

subject to

$$\sum_{i,k} x_{(ik)j} = 1, \quad \text{for } j = 1, \ldots, n;$$

$$\sum_j x_{(ik)j} \leq 1, \quad \text{for } i = 1, \ldots, m, \ k = 1, \ldots, n;$$

$$x_{(ik)j} \in \{0, 1\}, \quad \text{for } i = 1, ..., m, \ j, k = 1, ..., n.$$

The constraints ensure that each job is scheduled exactly once and that each position on each machine is occupied by at most one job. This is a weighted bipartite matching or assignment problem, so that we may replace the integrality constraints by nonnegativity constraints without altering the feasible set. It can be solved in $O(n^3)$ time.

**Exercises**
8.1. Let us say that two schedules for $P||\sum C_j$ are equivalent if they differ only by a renumbering of machines. Show that there exists at least

$$(m!)^{\lfloor (n/m) \rfloor - 1}$$

nonequivalent optimal schedules.
8.2. Show that the SPT rule solves $P||\sum w_j C_j$ for the special case in which processing times and weights are oppositely ordered.
8.3. Suppose that machine $i$ is not available until time $t_i \geq 0$, i.e., it must remain idle until then. Show how to incorporate this condition into the matching problem formulated for $R||\sum C_j$.
8.4. Show how to reduce the time required to set up the matching problem for $R||\sum C_j$ to $O(n^2 \log n)$ time. Hint: For each job $j$, one need only consider the smallest $n$ coefficients of the form $kp_{ij}$. For each job $j$ establish a priority queue $Q_j$ to generate the $n$ smallest coefficients of this form.

## 8.2. Total completion time, preemptive scheduling

We now turn to preemptive scheduling of parallel machines with respect to the $\sum C_j$ criterion. It turns out that there is no advantage to preemption for the problem $P|pmtn|\sum C_j$ (nor for $P|pmtn|\sum w_j C_j$ – see Section 8.3). That is, a schedule that is optimal for an instance of $P||\sum C_j$ is also optimal for a similar instance of $P|pmtn|\sum C_j$.

The problem $R|pmtn|\sum C_j$ is NP-hard in the strong sense. This is a surprising result, in view of the polynomial-time solvability of $R||\sum C_j$. There are very few problems for which allowing preemption makes the problem harder.

This leaves us with only the problem $Q|pmtn|\sum C_j$ to consider in this section. Knowing what we do about the $\sum C_j$ criterion, the SPT rule, and the solution of the problem $Q||\sum C_j$, what might we conjecture about the structure of an optimal schedule for $Q|pmtn|\sum C_j$? Our first, rather timid, guess might be that jobs will be completed in SPT order. This conjecture is indeed true, though its demonstration is nontrivial, as we shall see in proving Lemma 8.1. Our next guess might be to try the following greedy prescription:

*Preemptive SPT rule:* Taking the jobs in SPT order, preemptively schedule each successive job $j$ in the available time on the $m$ machines so as to minimize $C_j$.

This rule produces a schedule that looks like that shown in Figure 8.2. That is, job 1, the shortest job, is processed entirely on $M_1$, the fastest machine. Job 2, the second-shortest job, is processed on $M_2$, the second-fastest machine, until job 1 is completed. Then job 2 is processed on $M_1$ to completion. In general, job $j$ is scheduled for processing first on $M_m$, then on $M_{m-1}$,..., and finally on $M_1$. This makes $C_j$ as small as possible, given the time left available on the machines after jobs $1, 2, \ldots, j-1$ have been scheduled.

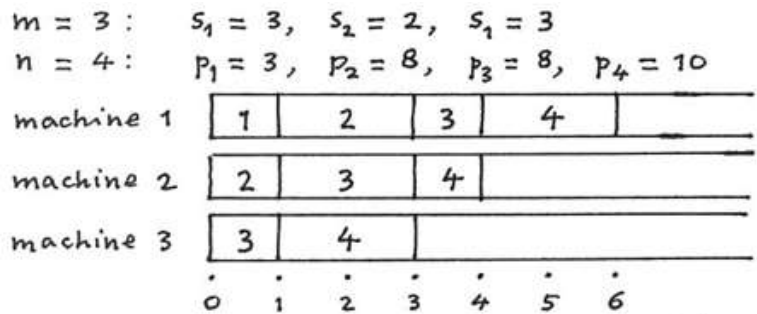

Figure 8.2.  Preemptive SPT schedule for $Q|pmtn|\sum C_j$.

The procedure we have described requires $O(n \log n)$ time for the initial sort of the jobs into SPT order and $O(mn)$ time to construct the schedule. The schedule has at most $(m-1)(n-m/2)$ preemptions (see Exercise 8.6), and no preemptions at all if the machines are identical. This latter observation proves our assertion that there is no advantage to preemption in the case of $P|pmtn|\sum C_j$ – provided the schedule constructed by the preemptive SPT rule is indeed optimal. We shall now prove the optimality of the schedule.

**Lemma 8.1.**  *For any instance of the $Q|pmtn|\sum C_j$ problem, there exists an optimal schedule in which the completion times of the jobs are in SPT order.*

*Proof.*  Let $S$ be an optimal schedule in which $C_i > C_j$ with $p_i < p_j$. Prior to time $C_j$ there are some periods during which job $i$ but not $j$ is processed, some periods in which job $j$ but not $i$ is processed, and some periods in which both jobs $i$ and $j$ are processed. Also, in the interval $[C_j, C_i]$ there are some periods in which job $i$ is processed. We propose to replace the processing of job $i$ in the interval $[C_j, C_i]$ with processing of job $j$, and to interchange some fraction $\lambda$, $0 < \lambda < 1$, of the processing of $i$ and $j$ prior to time $C_j$, thereby obtaining a feasible schedule $S'$ with completion times $C_i'$ and $C_j'$. Of each active period for job $i$ prior to $C_j$, a fraction $\lambda$ of that active period is given over to job $j$ in $S'$, and vice versa. Observe that $C_i' \leq C_j$ and $C_j' = C_i$.

We must now show that there is a value of $\lambda$ in the interval $(0, 1)$ such that the schedule $S'$ can be constructed as we have suggested. Let $t_i, t_j$ denote the total lengths of time, and let $\sigma_i, \sigma_j$ denote the effective rates at which jobs $i$ and $j$ are processed prior to time $C_j$ in schedule $S$. Let $t_i'$ be the total length of time and $\sigma_i'$ be the effective rate at which job $i$ is processed in the interval $[C_j, C_i]$. Then we have

$$\sigma_i t_i + \sigma_i' t_i' = p_i < p_j = \sigma_j t_j. \tag{8.1}$$

In order for $S'$ to be feasible, we must have

$$(1 - \lambda)\sigma_i t_i + \lambda \sigma_j t_j = p_i,$$

$$(1 - \lambda)\sigma_j t_j + \lambda \sigma_i t_i + \sigma_i' t_i' = p_j.$$

Solving for $\lambda$, we obtain

$$\lambda = \frac{\sigma_i' t_i'}{\sigma_j t_j - \sigma_i t_i}$$

It follows immediately from (8.1) that

$$\sigma_j t_j - \sigma_i t_i > \sigma_i' t_i' > 0,$$

and hence $0 < \lambda < 1$.

Note that $C_i' + C_j' \leq C_i + C_j$, and the completion times of all other jobs are the same in $S'$ as in $S$. It follows that a finite number of modifications of the type we have described will transform $S$ into an optimal schedule with completion times in SPT order. $\square$

**Theorem 8.2.** *The preemptive SPT rule computes an optimal schedule for* $Q|pmtn|\sum C_j$.

*Proof.* For convenience, assume that $n = m$, and that $s_1 \geq s_2 \geq \cdots \geq s_m$. (If $n < m$, discard the $m - n$ slowest machines, and if $n > m$, create $n - m$ zero-speed dummy machines.) Let $S$ be the schedule computed by the preemptive SPT rule, with $C_1 \leq C_2 \leq \cdots \leq C_n$. From the structure of the schedule $S$, it is apparent that

$$
\begin{aligned}
s_1 C_1 &= p_1, \\
s_2 C_1 + s_1(C_2 - C_1) &= p_2, \\
s_3 C_1 + s_2(C_2 - C_1) + s_1(C_3 - C_2) &= p_3,
\end{aligned}
$$

and so forth. Adding these equations yields

$$
\begin{aligned}
s_1 C_1 &= p_1, \\
s_2 C_1 + s_1 C_2 &= p_1 + p_2, \\
s_3 C_1 + s_2 C_2 + s_1 C_3 &= p_1 + p_2 + p_3,
\end{aligned}
$$

and so forth.

   Suppose $S^*$ is an optimal schedule. By Lemma 8.1 we may assume $C_1^* \leq C_2^* \leq \cdots \leq C_n^*$. For each $j$, $1 \leq j \leq n$, let us consider a lower bound on the value of $C_j^*$. The amount of processing done prior to $C_j^*$ is at least $p_1 + p_2 + \cdots + p_j$. The fastest way to perform this processing is to use machines $1, 2, \ldots, j$ before $C_1^*$, machines $1, 2, \ldots, j-1$ in the interval $[C_1^*, C_2^*]$, etc. This yields the system of inequalities of the form

$$
\begin{aligned}
&(s_1 + s_2 + \cdots + s_j)C_1^* + (s_1 + s_2 + \cdots + s_{j-1})(C_2^* - C_1^*) + \cdots + s_1(C_j^* - C_{j-1}^*) \\
&\geq p_1 + p_2 + \cdots + p_j.
\end{aligned}
$$

Adding these inequalities, we obtain,

$$
\begin{aligned}
s_1 C_1^* &\geq p_1, \\
s_2 C_1^* + s_1 C_2^* &\geq p_1 + p_2, \\
s_3 C^* 1 + s_2 C_2^* + s_1 C_3^* &\geq p_1 + p_2 + p_3,
\end{aligned}
$$

and so forth, which gives us

$$
\begin{aligned}
s_1 C_1^* &\geq s_1 C_1, \\
s_2 C_1^* + s_1 C_2^* &\geq s_2 C_1 + s_1 C_2, \\
s_3 C_1^* + s_2 C_2^* + s_1 C_3^* &\geq s_3 C_1 + s_2 C_2 + s_1 C_3,
\end{aligned}
$$

and so forth. Now let us multiply the $j$th inequality above by a positive value $\lambda_j$ and add the inequalities, thereby obtaining

$$
\begin{aligned}
&(\lambda_1 s_1 + \lambda_2 s_2 + \cdots + \lambda_n s_n)C_1^* + (\lambda_2 s_1 + \cdots + \lambda_n s_{n-1})C_2^* + \cdots + \lambda_n s_1 C_1^* \\
\geq\; &(\lambda_1 s_1 + \lambda_2 s_2 + \cdots + \lambda_n s_n)C_1 + (\lambda_2 s_1 + \cdots + \lambda_n s_{n-1})C_2 + \cdots + \lambda_n s_1 C_n.
\end{aligned}
$$

If the $\lambda_j$ values are chosen so that

$$
\begin{aligned}
\lambda_n s_1 &= 1, \\
\lambda_{n-1} s_1 + \lambda_n s_2 &= 1, \\
\lambda_{n-2} s_1 + \lambda_{n-1} s_2 + \lambda_n s_3 &= 1,
\end{aligned}
\tag{8.2}
$$

then we will have

$$
\sum C_j^* \geq \sum C_j,
$$

from which it follows that $S$, the schedule determined by the preemptive SPT rule, is optimal. Indeed, it follows from the fact that $s_1 \geq s_2 \geq \cdots \geq s_m$ that there is a solution to (8.2) in nonnegative $\lambda_j$'s. $\square$

### Exercises

8.5. Devise a simple example to show that an optimal schedule for $Q|pmtn|\sum C_j$ may require preemptions.

8.6. Show that the number of preemptions created by applying the preemptive SPT rule does not exceed $(m-1)(n-m/2)$. Suggestion: prove that the number of preemptions is equal to the number of active periods $-n$.

8.7*. Formulate a general class of problem instances for $Q|pmtn|\sum C_j$ to show that, for any $m$ and $n$, as many as $(m-1)(n-m/2)$ preemptions may be required for an optimal schedule. Hint: Find a class of problem instances with distinct machine speeds and job processing requirements for which the only schedule that is optimal is the one determined by the preemptive SPT rule.

8.8. Generalize Lemma 8.1 and its proof to apply to the $\sum w_j C_j$ criterion, in the special case that job processing times and weights are oppositely ordered.

8.9. Show that Lemma 8.1 and its proof remain valid for $Q|pmtn, \bar{d}_j|\sum C_j$, provided processing times and deadlines are similarly ordered.

8.10. Generalize Theorem 8.2 and its proof to apply to the $\sum w_j C_j$ criterion, in the special case that job processing times and weights are oppositely ordered.

8.11*. Obtain a closed form expression for $C_j$ when the preemptive SPT rule is applied to an instance of $Q|pmtn|\sum C_j$.

## 8.3. Total weighted completion time, complexity

A summary of the results of the previous two sections is as follows: $P||\sum C_j$ can be solved in $O(n \log n)$ time by a simple extension of the SPT rule, and since there is no advantage to preemption, the same rule solves $P|pmtn|\sum C_j$ as well. $Q||\sum C_j$ can be solved in $O(n \log n)$ time by matching the processing times $p_j$ in nonincreasing order with the coefficients $k/s_i$ in nondecreasing order. $R||\sum C_j$ can be solved in $O(n^3)$ time by a generalization of the coefficient-matching technique. $Q|pmtn|\sum C_j$ can be solved in $O(n \log n + mn)$ time by the preemptive SPT rule. $R|pmtn|\sum C_j$ is NP-hard.

A summary of the complexity status of generalizations of the problems studied in the previous sections is as follows:

- $\sum w_j C_j$ criterion – $P2||\sum w_j C_j$ is NP-hard. It can be shown that there is no advantage to preemption for $P|pmtn|\sum w_j C_j$. Hence, this NP-hardness result applies to $P2|pmtn|\sum w_j C_j$ as well. Dynamic programming provides a pseudopolynomial algorithm for $Qm||\sum w_j C_j$, as we show below, but this is of very limited practicality. Results on approximation and branch-and-bound for $P||\sum w_j C_j$ are scarce. (Eds: This chapter was written in 1990, but in the subsequent twenty years, the situation has changed significantly with respect to approximation algorithms.)

- Release dates and deadlines – In the absence of preemption, either release dates or deadlines induce NP-hardness; both $1|r_j|\sum C_j$ and $P2||C_{max}$ have already been shown to be NP-hard. Also, $P2|pmtn, r_j|\sum C_j$ has been shown to be NP-hard. The status of $P|pmtn, \bar{d}_j|\sum C_j$ is unknown. However, there are polynomial-time algorithms for solving $Q|pmtn, \bar{d}_j = \bar{d}|\sum C_j$.

- Precedence constraints – There is very little that can be done with precedence constraints in polynomial time, since even $P2|chains|\sum C_j$ is NP-hard. Also, $P|prec, p_j = 1|\sum C_j$ is NP-hard, but $P|outtree, p_j = 1|\sum C_j$ can be solved in polynomial time.

Let us now establish the existence of a pseudopolynomial algorithm for the problem $Qm||\sum w_j C_j$. The dynamic programming technique we shall describe also yields similar results for the problems $Rm||C_{max}$, $Rm||L_{max}$, $Rm||\sum w_j U_j$, and even $Rm|\bar{d}_j = \bar{d}|\sum w_j T_j$. What all these problems have in common is that there exists an optimal schedule in which the sequence of jobs performed by each of the $m$ machines is consistent with a permutation $\pi$ of the $n$ jobs that can be prescribed *a priori*. (For $\sum w_j C_j$, $\pi$ is a ratio order; for $L_{max}$ and $\sum w_j U_j$, $\pi$ is an EDD order; for $C_{max}$, $\pi$ is any order whatsoever. In the case of $\sum w_j U_j$, only the on-time jobs are in a sequence consistent with $\pi$.) Without loss of generality, let $\pi = (1, 2, ..., n)$. Define $F_j(t_1, ..., t_m)$ to be the minimum cost of a schedule for the jobs $1, 2, ..., j$, subject to the constraint that the last job on machine $i$ is completed at time $t_i$, for $i = 1, 2, ..., m$. Then, for $f_{max}$ criteria we have

$$F_j(t_1, ..., t_m) = \min_{i=1,...,m} \max\{f_j(t_i), F_{j-1}(t_1, ..., t_i - p_{ij}, ..., t_m)\},$$

and for $\sum f_j$ criteria,

$$F_j(t_1, ..., t_m) = \min_{i=1,...,m}\{f_j(t_i) + F_{j-1}(t_1, ..., t_i - p_{ij}, ..., t_m)\}.$$

In both cases we have initial conditions

$$F_0(t_1, ..., t_m) = \begin{cases} 0, & \text{if } t_i = 0, \text{ for } i = 1, 2, ..., m, \\ +\infty, & \text{otherwise.} \end{cases}$$

There are $O(nC^m)$ values of $F_j(t_1, ..., t_m)$ to compute, where $C$ is an upper bound on the completion time of any job in an optimal schedule. The computation of each value requires minimization over $m$ alternatives and hence $O(m)$ time. The cost of an optimal schedule is given by the minimum of the values $F_n(t_1, ..., t_m)$, where $t_i \leq C$. Hence, the cost of an optimal schedule can be computed in $O(mnC^m)$ time, which is pseudopolynomial for fixed $m$. For variations on this dynamic programming scheme, see the exercises below.

**Exercises**

8.7. Devise a simple example to show that an optimal schedule for $P2|pmtn, r_j|\sum C_j$ may require preemptions.

8.8. A time bound of $O(mnC^m)$ is indicated for the dynamic programming algorithm presented in this section, where $C$ is an upper bound on the completion time of any job in an optimal schedule. Show that $C \leq \sum p_j/m + p_{max} \leq \lceil(n/m)\rceil p_{max}$, and hence the time bound can be expressed as $O(n^2 p_{max} C^m)$.

8.9. Modify the dynamic programming procedure, as necessary, to solve $R||\sum w_j U_j$.

8.10. Show that in the case of uniform machines and the $C_{max}$, $L_{max}$, $\sum w_j C_j$ criteria, the dynamic programming running time bound of $O(mnC^m)$ can be reduced to $O(mnC^{m-1})$. Why can this not be done in the case of the $\sum w_j U_j$ criterion?

8.11. Consider the problems $Qm||\sum w_j C_j$, $Rm||C_{max}$, $Rm||L_{max}$, $Rm||\sum w_j U_j$, and $Rm|\bar{d}_j = \bar{d}|\sum w_j T_j$.

(a)    For which of these problems is it possible to adapt the dynamic programming algorithm to deal with nonuniform release dates? Explain.

(b)    Same question, with respect to deadlines.

(c)    Both release dates and deadlines?

8.12. Recall the recurrence relations for the dynamic programming solution of the problem $1||\sum f_j$:

$$F(S) = \min_{j \in S}\{f_j(p(S)) + F(S-j)\}. \tag{8.3}$$

As an alternative to the dynamic programming technique described in this section, let us consider two ways in which we might adapt the recurrence (8.3) to the solution of the problem $R||\sum f_j$.

(a)    Let $F_i(S)$ denote the minimum cost of a sequence for the subset $S$ on machine $i$. Use recurrences (8.3) to compute $F_i(S)$ for each $i = 1, 2, \ldots, m$, and for all $S \subseteq N$. Let $G_i(S)$ denote the minimum cost for a schedule of the subset $S$ on machines $1, 2, \ldots, i$, where

$$G_i(S) = \min_{S' \subseteq S}\{F_i(S') + G_{i-1}(S-S')\}.$$

Show that the time required to compute $G_m(N)$, the cost of an optimal schedule for all $n$ jobs on all $m$ machines, is $O(m3^n)$.

(b)    View the problem of finding an optimal schedule as that of constructing a single optimal sequence obtained by concatenating the m sequences for the individual machines. (A similar concatenation of sequences was contemplated for the solution of Exercise 1.2 in Chapter 1, in which the reader was asked to show that there are $(n+m-1)!/(m-1)!$ distinct schedules for $m$ parallel machines.) Define $F_i(S,t)$ to be the minimum cost of a sequence for the subset $S$ in which the last job in the sequence finishes at time $t$ on machine $i$. Then we have as our basic recurrence relations:

$$F_i(S,t) = \min_{j \in S}\{f_j(t) + F_i(S-j, t-p_{ij})\}.$$

Indicate how to obtain $F_i(S,0)$ from values for $F_i(S,t)$ and supply appropriate initial conditions. Show that the time required to compute the cost of an optimal schedule is $O(mnC2^n)$, where $C$ is an upper bound on the completion time of any job in an optimal schedule. Show that the time bound can be expressed as $O(n^2 p_{\max} 2^n)$.

(c)     For what value of $n$ can you reasonably expect $mnC2^n$ to be smaller than $m3^n$?

(d)     Describe the adaptations in the procedure necessary to solve $1|r_j,\bar{d}_j|\sum f_j$.

## 8.4.    Other minsum criteria, complexity

Having pretty well disposed of parallel machine problems with $\sum C_j$ and $\sum w_j C_j$ criteria, let us consider the complexity situation with respect to other minsum criteria. What we have is largely a litany of bad news:

- $\sum U_j$, $\sum w_j U_j$ – $P2||\sum U_j$ is NP-hard, as demonstrated by a simple transformation from SUBSET SUM. As we have seen in Section 8.3, dynamic programming yields a pseudopolynomial algorithm for $Rm||\sum w_j U_j$. $P|pmtn|\sum U_j$ turns out to be NP-hard in the ordinary sense; the proof is of interest because the problem is one for which preemption *is* advantageous, and the proof must take this into account. NP-hardness of the problem in the strong sense is an open question. It is possible to solve $Qm|pmtn|\sum w_j U_j$ in $O(n^3 m W^2)$ time, which is pseudopolynomial, but becomes polynomial, $O(n^3 m)$, when all $w_j = 1$.

- $\sum T_j$, $\sum w_j T_j$ – Since $1||\sum T_j$ and $1|pmtn|\sum T_j$ are NP-hard, all parallel machine problems with the $\sum T_j$ criterion are NP-hard as well. Furthermore, since $1||\sum w_j T_j$ and $1|pmtn|\sum w_j T_j$ are NP-hard in the strong sense, it is unreasonable even to hope for pseudopolynomial algorithms for parallel machine versions of these problems with fixed $m$. (Recall that in Section 8.3 the best we could offer was a pseudopolynomial algorithm for the special case of a common due date, $Rm|d_j = d|\sum w_j T_j$.)

- *Unit-time jobs* – Here is good news: The problem $Q|p_j = 1|\sum f_j$ can be solved in $O(n^3)$ time by a simple extension of the matching technique we applied to solve the problem $1|p_j = 1|\sum f_j$ in Section 2.1. Recall that we solved the single-machine unit-time problem by optimally matching the $n$ given jobs $j = 1, 2, \ldots, n$ with the $n$ time slots $t = 1, 2, \ldots, n$. This involved formulating and solving an assignment or matching problem for an $n \times n$ matrix with entries of the form $f_j(t)$. The only new wrinkle required to solve $Q|p_j = 1|\sum f_j$ in the same way is in the generation of the values of $t$ for the $n$ time slots to which the jobs matched.

  Because we assume the functions $f_j$ are monotone, we know there is an optimal schedule with no unforced machine idle time. This means that if $n_i$ unit-length are processed on machine $i$, those jobs will be completed at times $t = 1/s_i, 2/s_i, \ldots, n_i/s_i$. More generally, the completion times of the jobs in an optimal schedule can be assumed to be the $n$ smallest values of $t$ generated by the algorithm we described in Section 8.1:

**for** $h = 1, 2, \ldots, n$ **do**
  **output** $k/s_i :=$ **delete-min**$(Q)$;
  **insert** $(k + 1)/s_i$ into $Q$;
**end**

When this loop was used to generate coefficients for the problem $Q||\sum C_j$ in Section 8.1, $k$ denoted the $k$th-last position on machine $i$, whereas here $k$ denotes the $k$th-earliest time slot and $k/s_i$ is the completion time of the job processed in that slot.

For general $f_j$ functions, $O(n^2)$ time is required to construct the data for the matching problem and $O(n^3)$ time is required to solve it. However, for certain scheduling objectives, the matching problem has a trivial solution. For example, in the case of $Q|p_j = 1|\sum w_j C_j$, the problem is simply that of matching the largest weight $w_j$ with the smallest completion time, the second-largest $w_j$ with the second-smallest completion time, and so on. Thus the time required to solve $Q|p_j = 1|\sum w_j C_j$ is only $O(n \log n)$, the time required to sort the $w_j$ and to generate the values of $t$. For other examples, see the exercises.

### Exercises

8.13. Describe how to solve $Q|r_j, p_j = 1|\sum C_j$ in $O(n \log n)$ time.

8.14. Describe how to solve $Q|p_j = 1|\sum U_j$ in $O(n \log n)$ time.

8.15. Describe how to solve $Q|p_j = 1|f_{\max}$ in $O(n^2)$ time.

### Notes

8.1. *Total completion time, nonpreemptive scheduling.* The extension of the SPT rule to $P||\sum C_j$ is due to Conway, Maxwell & Miller [1967]. The solution of $Q||\sum C_j$ is due to Horowitz & Sahni [1976]. $R||\sum C_j$ was formulated as a matching problem by Horn [1973] and by Bruno, Coffman & Sethi [1974].

8.2. *Total completion time, preemptive scheduling.* Lemma 8.1 is due to Lawler [−]; the surveys by Graham et al. [1979] and Lawler et al. [1993] erroneously attribute the result to Lawler & Labetoulle [1978]. Theorem 8.2 was proved by Gonzalez [1977]. Sitters [2005] established NP-hardness of $R|pmtn|\sum C_j$ by a reduction from 3-dimensional matching.

8.3. *Total weighted completion time, complexity.* NP-hardness of $P2||\sum w_j C_j$ was proved by Bruno, Coffman & Sethi [1974]; Lenstra, Rinnooy Kan & Brucker [1977] gave a simpler proof. McNaughton [1959] showed that there is no advantage to preemption in $P|pmtn|\sum w_j C_j$.

For $P||\sum w_j C_j$, an obvious idea is to apply list scheduling with the jobs listed according to nondecreasing ratios $p_j/w_j$. Eastman, Even & Isaacs [1964] investigated the performance of this ratio rule and gave a lower bound on the optimum solution value. This lower bound has been the basis for the branch-and-bound algorithms of Elmaghraby & Park [1974], Barnes & Brennan [1977], and Sarin, Ahn & Bishop [1988]. Kawaguchi & Kyan [1986] refined the analysis of the ratio rule and gave a performance ratio of $(\sqrt{2}+1)/2$. Sahni [1976] gave algorithms $A_k$ with running time $O(n(n^2 k)^{(m-1)})$ and performance ratio $1 + 1/k$.

Gonzalez [1977] devised a complicated, but polynomial-time, algorithm for $Q|pmtn, \bar{d}_j = \bar{d}|\sum C_j$. As we note in Exercise 8.9, jobs are executed in SPT order for this problem. This makes it possible to give the problem a linear programming formulation, using ideas presented in Section 10.2. McCormick and Pinedo [1995] have investigated an LP formulation in which the objective is to minimize a linear combination of flow time and makespan, and gave an algorithm that minimizes flow time subject to a fixed makespan deadline. They showed how to generate the entire trade-off curve between total completion time and makespan. The schedules generated put the jobs with the shortest processing times on the fastest machines, except when it is necessary to fit a block of long jobs under the deadline.

NP-hardness of $P2|tree|\sum C_j$ and $P2|chain|\sum C_j$ was proved by Sethi [1977] and Du, Leung & Young [1991], respectively. The dynamic programming formulation for $Qm||\sum w_j C_j$ is due to Rothkopf [1966] and Lawler & Moore [1969].

**8.4.** *Other minsum criteria, complexity.* NP-hardness of $P|pmtn|\sum U_j$ was proved by Lawler [1983]. Also the pseudopolynomial-time algorithm for $Qm|pmtn|\sum w_j U_j$ is due to Lawler [1979A]. Lawler [−] and Dessouky et al. [1990] noticed that $Q|p_j = 1|\sum f_j$ and $Q|p_j = 1|f_{\max}$, and variations of these problems, can be formulated and solved as matching problems. Lawler [1976A] proposed an $O(n \log n)$ algorithm for the special case of $P|p_j = 1|\sum w_j U_j$.

# Contents

# 9

# Minmax criteria, no preemption

David B. Shmoys
*Cornell University*

Jan Karel Lenstra
*Centrum Wiskunde & Informatica*

Four authors wish to write a book, which will consist of fifteen chapters. Each chapter is to be written by a single author. The chapters differ in length, and the authors differ in expertise. The speed at which an author will be able to complete a chapter depends on his, or her, familiarity with its subject matter. The completion date of the manuscript is to be minimized. Who should do what?

Irrespective of the practical relevance of this model, it is one of the central problems in scheduling theory. In its general form, the authors are unrelated parallel machines, the chapters are jobs, and the problem is $R||C_{\max}$. The reader should recall the definitions of unrelated, uniform, and identical parallel machines (see Section 1.2). We have seen that, for these machine environments, there is a drastic difference in complexity between minimizing *maximum* and *total* completion time: $R||\sum C_j$ is solvable in polynomial time (see Section 8.1), but $P2||C_{\max}$ is already *NP*-hard (see Section 2.4). As long as the number of machines is a constant, dynamic programming can be applied to solve $Rm||C_{\max}$ in pseudopolynomial time (see Section 8.3); however, when $m$ is an input, even $P||C_{\max}$ is *NP*-hard in the strong sense (see Section 2.4).

This chapter, then, is about hard problems. The design of efficient approximation algorithms and of enumerative optimization methods is called for. We will emphasize the performance analysis of approximation algorithms. The empirical analysis of approximation and optimization procedures will not be discussed in detail. We have
chosen to summarize computational work of interest in the notes.

We start in Section 9.1 with some classical results for $P||C_{\max}$ due to R. L. Gra-
ham, whose work initiated the investigation of the worst-case performance of heuris-
tic methods for optimization problems. More recent results on performance guaran-
tees for identical, uniform, and unrelated machines are given in Sections 9.2, 9.3, and
9.4, respectively. Section 9.5 considers guarantees for unrelated machines which, in
all likelihood, cannot be achieved.

In Section 9.6, we switch to probabilistic analysis. We will see that algorithms
that can behave pretty badly in the worst case do quite well on average. From a prac-
tical point of view, the two approaches are complementary. A worst-case approach
provides a performance bound for every instance, but the bound may be pessimistic,
as it depends on anomalous cases that rarely occur. A probabilistic analysis has the
potential to provide a more accurate picture of the real world. However, the resulting
statements carry no guarantee for any individual instance, and the analysis is ulti-
mately no more realistic than the probability distribution over the class of instances
on which it is based.

## 9.1.   Identical machines: classical performance guarantees

In Graham's ground-breaking paper, the problem studied is $P||C_{\max}$, and the method
analyzed is the *list scheduling* (LS) rule: the jobs are listed in any fixed order, and
whenever a machine becomes idle, the next job from the list is assigned to start
processing on that machine. Surprisingly, this simple rule has a constant relative
performance bound.

**Theorem 9.1.** *For any instance of $P||C_{\max}$, $C_{\max}(LS)/C_{\max}^* \leq 2 - 1/m$.*

*Proof.*   Let $J_\ell$ be the last job to be completed in a list schedule (cf. Figure 9.1). Note
that no machine can be idle before the starting time $S_\ell$ of $J_\ell$. Intuitively, it is obvious
that a list schedule is no longer than twice an optimal one. Consider two parts of the
schedule, before and after $S_\ell$. The optimum is at least as long as the first part, since
no machine is idle between 0 and $S_\ell$, and the optimum is also at least as long as the
second part, since this part has the length of a single job.

More formally, the fact that no machine is idle prior to $S_\ell$ implies that $\sum_{j \neq \ell} p_j \geq mS_\ell$. Therefore,

$$C_{\max}(\text{LS}) = S_\ell + p_\ell \leq \frac{1}{m} \sum_{j \neq \ell} p_j + p_\ell = \frac{1}{m} \sum_j p_j + \frac{m-1}{m} p_\ell. \qquad (9.1)$$

By using the inequalities

$$C_{\max}^* \geq \frac{1}{m} \sum_j p_j, C_{\max}^* \geq p_\ell,$$

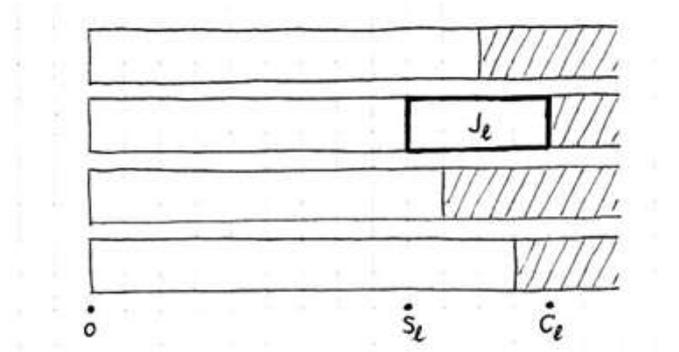

**Figure 9.1.** A list schedule

we obtain the claimed bound. □

The bound of Theorem 9.1 is tight for any value of $m$, as is shown by the following class of instances. Let $n = m(m-1) + 1$, $p_1 = ... = p_{n-1} = 1$, $p_n = m$, and consider the list $(J_1, ..., J_n)$. The list scheduling rule first assigns $m-1$ unit-length jobs to each of the $m$ machines and then is left scheduling $J_n$, so that $C_{\max}(\text{LS}) = 2m - 1$. The optimal schedule has length $C_{\max}^* = m$: it assigns $J_n$ to a machine by itself and balances the remaining $m(m-1)$ jobs among the $m-1$ other machines.

As the previous example suggests, list scheduling may perform poorly when the last job to finish is very long, and inequality (9.1) implies that this is the only way in which poor schedules are generated. A natural way to try to prevent this is to use a list in which the jobs are sorted in order of nonincreasing processing requirement. The next theorem shows that this *longest processing time* (LPT) rule performs significantly better than arbitrary list scheduling.

**Theorem 9.2.** *For any instance of $P||C_{\max}$, $C_{\max}(LPT)/C_{\max}^* \leq 4/3 - 1/3m$.*

*Proof.* Suppose that the theorem is false, and consider a counterexample with the minimum number of jobs. Let job $J_\ell$ be the job that completes last. If $\ell < n$, then the instance consisting of jobs $\{J_1, ..., J_\ell\}$ is a smaller counterexample, since the completion time of the LPT schedule is unchanged, whereas the optimal schedule length can only decrease by considering only a subset of the jobs. Hence, job $J_n$ is the last to finish. We will consider two cases separately: (*i*) $p_n \leq C_{\max}^*/3$; and (*ii*) $p_n > C_{\max}^*/3$. In case (*i*) it is easy to see that (9.1) immediately implies the theorem, since

$$C_{\max}(\text{LPT}) \leq \frac{1}{m} \sum_j p_j + \frac{m-1}{m} p_\ell \leq C_{\max}^* + \frac{m-1}{m} \frac{C_{\max}^*}{3} = \left(\frac{4}{3} - \frac{1}{3m}\right) C_{\max}^*.$$

We will show that if case (*ii*) holds, then the LPT rule delivers an optimal solution. Observe that in the optimal schedule it is impossible for three jobs to be assigned to one machine, since the total processing time of those jobs is at least $3p_n$, which exceeds $C^*_{\max}$. Hence $n \leq 2m$. Furthermore, we can assume that $n > m$, since otherwise, the LPT rule assigns each machine at most one job, which is clearly optimal.

Consider the following schedule $\pi$: we first show that it is optimal, and then show that LPT must output $\pi$.

For $\pi$, we first assign job $J_i$ to machine $M_i$, $i = 1, \ldots, m$. (Of course, we can assume without loss of generality that is also true for the LPT schedule.) Furthermore, we also assign the job $J_{m+i}$ to machine $M_{m+1-i}$, $j = 1, \ldots, n - m$. In other words, jobs $J_j$ and $J_{2m-j+1}$ are paired together on machine $M_j$, $j = 2m - n + 1, \ldots, m$.

Suppose, for a contradiction, that this schedule is not optimal; let $\pi'$ denote an optimal schedule. If $C^\pi_{\max} = p_j$ for some job $J_j$ (requiring of course that $n < 2m$), then we get a contradiction directly, since $p_j$ must be a lower bound on the makespan of any schedule. Hence $C^\pi_{\max} = p_j + p_{2m-j+1}$, for some $j = 1, \ldots, m$. In $\pi'$, for each $k = 1, \ldots, j - 1$, job $J_k$ cannot be paired on the same machine with some job $J_\ell$, where $\ell = 1, \ldots, 2m - j + 1$, since otherwise,

$$C^{\pi'}_{\max} \geq p_k + p_\ell \geq p_j + p_{2m-j+1} \geq C^\pi_{\max},$$

and hence $\pi'$ is not better than $\pi$. This means that in $\pi'$, the jobs $J_1, \ldots, J_{j-1}$ must all be assigned to distinct machines; index the machines so that $J_i$ is scheduled in $\pi'$ on $M_i$, $i = 1, \ldots, j - 1$. Furthermore, we know that jobs $J_j, \ldots, J_{2m-j+1}$ are not scheduled on machines $M_1, \ldots, M_{j-1}$. Hence, there are $(2m - j + 1) - (j - 1) = 2(m - j + 1)$ jobs scheduled on $m - j + 1$; since each machine gets at most 2 jobs, we can conclude that each of the machines $M_j, \ldots, M_m$ is assigned exactly 2 jobs in $\pi'$. Consider the machine with job $J_j$; this machine completes at time at least $p_j + p_{2m-j+1} = C^\pi_{\max}$, and hence $\pi'$ is not better than $\pi$. We have shown that $\pi$ is optimal.

Now we want prove that the LPT rule produces $\pi$. We have indexed the machines and jobs so that the LPT schedule coincides with $\pi$ for scheduling the first $m + 1$ jobs. (We will assume that when scheduling jobs $J_{m+1}, J_{m+2}, \ldots$, if two machines complete their jobs at the same time, we next assign a job to the higher indexed machine.) Consider the first time that LPT schedules in a way different from $\pi$, when scheduling some job $J_k$, where $k > m + 1$. In $\pi$, we assign job $J_k$ to machine $M_{2m-k+1}$. For LPT to select a different machine for $J_k$, it must be that some machine $M_j$, $j = 2m - k + 1, \ldots, m$ completes its second job no later than $M_{2m-k+1}$ completes its first (i.e., $J_{2m-k+1}$). Hence,

$$C_{2m-k+1} = p_{2m-k+1} \geq p_j + p_{2m-j+1} > 2C^*_{\max}/3.$$

But then, in $\pi$, job $J_k$ is assigned to the same machine as $J_{2m-k+1}$, and hence it completes in $\pi$ at time $p_{2m-k+1} + p_k$, which must be greater than $C^*_{\max}$, which contradicts the optimality of $\pi$. Hence, in case (*ii*), the schedule produced by the LPT rule is optimal. $\square$

Again, the bound of Theorem 9.2 is tight for any value of $m$; see Exercise 9.1.

Inequality (9.1) holds whenever there is no idle time prior to the start of the job that completes last. We will exploit this observation in the design of the following algorithm $A_k$ and its analysis. Let $k$ be a fixed positive integer. Partition the job set into two parts: the *long* jobs and the *short* jobs, where a job $J_s$ is said to be short if $p_s \leq (1/km) \sum_j p_j$. Note that there are less than $km$ long jobs, and hence less than $m^{km}$ possible schedules for the long jobs. Enumerate all of these schedules, and choose the shortest one. Extend this schedule by using list scheduling for the short jobs. As in the analysis of the LPT rule, we consider the last job $J_\ell$ to be completed, and distinguish between two cases. If $J_\ell$ is a short job, then inequality (9.1) implies that

$$C_{\max}(A_k) < C_{\max}^* + p_\ell \leq C_{\max}^* + \frac{1}{km} \sum_j p_j \leq (1 + \frac{1}{k}) C_{\max}^*.$$

If $J_\ell$ is a long job, then the schedule obtained is optimal, since $C_{\max}(A_k)$ is equal to the length of the optimal schedule for just the long jobs, which is clearly no more than $C_{\max}^*$. Algorithm $A_k$ can be implemented to run in $O((m^{km} + n) \log m)$ time, which is polynomial for any fixed $k$ and $m$. We have proved the following theorem.

**Theorem 9.3.** *The family of algorithms $\{A_k\}$ forms a polynomial approximation scheme for $Pm||C_{\max}$.*

This result has been improved into various directions. There is even a *fully* polynomial approximation scheme for $Pm||C_{\max}$. It is based on the same ideas as the more general scheme for $Rm||C_{\max}$, which is to be described in Section 9.4; see Theorem 9.16. Further, in Section 9.2 we will give a polynomial approximation scheme for $P||C_{\max}$, where $m$ is an input; see Corollary 9.7.

For the special case that $m = 2$, there is an algorithm with the same performance guarantee as the LPT rule (see Exercise 9.3), but with much better empirical behavior. The main idea of this *differencing* ($D$) method is that two jobs are assigned at a time, one to each machine. The decision as to which job is assigned to which machine is made by scheduling a smaller instance, where both jobs are replaced by a single artificial job with processing time equal to the difference of their processing times. More precisely, if there is only one job, assign it to $M_1$; otherwise, if $J_j$ and $J_k$ are the longest and second longest jobs, respectively, replace these by a new job $J_\ell$ with $p_\ell = p_j - p_k$ and call this procedure recursively; to convert the resulting schedule to one for the original instance, if $J_\ell$ is assigned to $M_i$, then instead assign $J_j$ to $M_i$ and $J_k$ to the other machine. We will return to the differencing method in Section 9.6.

**Exercises**

9.1. Prove that the bound given in Theorem 9.2 is tight for each $m \geq 1$.

9.2. Use the following approach to obtain a polynomial approximation scheme similar to the scheme $\{A_k\}$ of Theorem 9.3: schedule the $\ell$ longest jobs optimally and complete the schedule using list scheduling.

9.3. Prove that for any instance of $P2||C_{\max}$, $C_{\max}(D)/C_{\max}^* \leq 7/6$, and show that this bound is tight.

## 9.2. Identical machines: using bisection search for guarantees

The algorithms discussed in the previous section are only a first step towards understanding how to obtain good solutions for $P||C_{\max}$ efficiently. In this section, it will be useful to consider the *decision version* of $P||C_{\max}$: given a deadline $d$, does there exist a schedule in which all jobs are completed by time $d$? An equivalent way to pose this question is the following: given $n$ items of specified sizes and $m$ bins of capacity $d$ each, is it possible to pack the items in the bins so that no bin contains items of total size more than $d$?

The problem of finding the minimum number of bins in which a given set of items can fit is called the *bin-packing problem*. For this problem, there is an approximation algorithm, called *first fit decreasing* ( FFD ), which resembles the LPT rule: number the bins, list the items in order of nonincreasing size, and iteratively place the next item from the list into the lowest-index bin in which it will fit. It can be shown that, if $b^*$ is the minimum number of bins, then FFD uses at most $(11/9)b^* + 4$ bins.

The FFD procedure can also be used within an approximation algorithm for $P||C_{\max}$, which applies bisection search over the values of $d$. Recalling our assumption that all data are integral, we specify the *multifit* ( MF ) algorithm as follows:

> $S :=$ a list schedule ;
> $UB := C_{\max}(\text{LS})$;
> $LB := \max\{\max_j p_j, \frac{1}{m}\sum_j p_j\}$;
> **while** $LB \neq UB$ **do**
>> $d := \lfloor \frac{LB+UB}{2} \rfloor$;
>> run the $FFD$ algorithm to pack the jobs in bins of size $d$ (*);
>> **if** *more than m bins are used by the packing (\*\*)* **then**
>>> $LB := d+1$;
>>
>> **else**
>>> $UB := d;;$
>>> $S :=$ new schedule;
>>
>> **end**
>
> **end**

Multifit has a better performance guarantee than is known for any variant of list scheduling.

**Theorem 9.4.** *When multifit is run for $\ell$ iterations of bisection search, then for any instance of $P||C_{\max}$, $C_{\max}(MF)/C_{\max}^* \leq 13/11 + 2^{-\ell}$.*

This bound is tight. A surprising aspect of the proof of the theorem is that it does not make use of the performance of FFD as a bin-packing algorithm; the two guarantees seem to be unrelated.

In the multifit algorithm, the statements indicated by (*) and (**) can be viewed as approximately answering the decision version of $P||C_{\max}$. Another way to obtain such an approximation is by means of a $\rho$-*relaxed decision procedure*. This is an algorithm with the following properties: for each instance and any deadline $d$, (i) it either outputs 'no' or produces a schedule with $C_{\max} \leq \rho d$, and (ii) if the output is 'no', then there is no schedule with $C_{\max} \leq d$. Let $B_\rho$ denote the bisection search procedure in which statements (*) and (**) are replaced by:

run a $\rho$-relaxed decision procedure to schedule the jobs within the deadline $d$; (†)
**if** the output is 'no' (††)

We have now obtained a general framework for constructing approximation algorithms for $P||C_{\max}$ and, as we shall see, for other problems as well. The following lemma gives the main reason for its importance.

**Lemma 9.5.** *For any instance of $P||C_{\max}$, $C_{\max}(B_\rho)/C_{\max}^* \leq \rho$; furthermore, if the $\rho$-relaxed decision procedure runs in polynomial time, then $B_\rho$ runs in polynomial time.*

*Proof.* To prove the performance bound, we show that the algorithm maintains two invariants: at each iteration, (i) the schedule $S$ has length $C_{\max} \leq \rho UB$, and (ii) $C_{\max}^* \geq LB$. This suffices, since $UB = LB$ at termination, so that the final schedule has length $C_{\max} \leq \rho UB = \rho LB \leq \rho C_{\max}^*$.

Initially, both (i) and (ii) hold. Suppose that in a particular iteration $UB$ is updated to $d$. In this case, the decision procedure has produced a schedule with $C_{\max} \leq \rho d$, so that (i) still holds. Now suppose that $LB$ is updated to $d+1$. In this case, the decision procedure has output 'no' and so no feasible schedule completes by $d$; hence, $C_{\max}^* \geq d+1$ and (ii) still holds.

Furthermore, since the difference $UB - LB$ after $\ell$ iterations is an integer, bounded from above by $2^{-\ell} C_{\max}^*$, the algorithm must terminate after a polynomial number of iterations. Therefore, if the $\rho$-relaxed decision procedure runs in polynomial time, the entire bisection search is completed in polynomial time. □

As a consequence of this lemma, we need only design polynomial-time $\rho$-relaxed decision procedures in order to obtain efficient approximation algorithms. We will use this idea to obtain a polynomial approximation scheme for $P||C_{\max}$ (where $m$ is an input). By Theorem 2.26, this is the best we can hope for, since no fully polynomial approximation scheme exists unless $P = NP$.

For any fixed positive integer $k$, we will show how to construct a $(1 + 1/k)$-relaxed decision procedure. Consider an instance of $P||C_{\max}$, along with a deadline $d \geq$ $\max_j p_j$. Once again, we partition the job set into *short* jobs and *long* jobs: in this case, $J_s$ is called short if $p_s \le d/k$.

For the time being, suppose that we are fortunate and that there are no short jobs. In the polynomial approximation scheme for a fixed number of machines, we solved such an instance to optimality by complete enumeration. In order to reduce the running time to be polynomial in $m$, we will produce only a near-optimal schedule for the long jobs. To this end, we will use a dynamic programming algorithm to solve a special case of the bin-packing problem. We assert that, when the bin-packing problem is restricted to the class of instances where there are at most $c_1$ distinct item sizes and at most $c_2$ items may be packed in one bin, then it can be solved in $O((c_2 n)^{c_1})$ time (see Exercise 9.4). If $c_1$ is a constant, this is a polynomial bound.

The decision procedure works as follows. Round down each processing time to the nearest multiple of $d/k^2$. Use the claimed algorithm for the special case of bin-packing to find the minimum number of machines that suffices to complete all of the rounded jobs by time $d$, assigning at most $k-1$ jobs to each machine. If the optimal packing uses more than $m$ machines, then output 'no'; otherwise, consider the packing as a schedule for the original instance.

We want to show that this procedure satisfies the two properties of a $(1 + 1/k)$-relaxed decision procedure. Suppose that it produces a schedule. For any job, the difference between the rounded and the original processing time is less than $d/k^2$. Since each machine processes fewer than $k$ jobs, the cumulative effect of taking the original processing times is less than $k(d/k^2) = d/k$. Hence, the schedule completes by time $(1 + 1/k)d$. On the other hand, suppose that the original instance has a schedule of length at most $d$. Clearly, fewer than $k$ jobs are assigned to each machine. Hence, there must be a bin-packing for the rounded instance that uses at most $m$ machines, each processing at most $k-1$ jobs. Consequently, if the procedure outputs 'no', then there is no schedule for the original instance that completes by time $d$.

The rounding ensures that there are at most $k^2 + 1$ distinct processing times. Since $k$ is fixed, the bin-packing algorithm runs in polynomial time, and it follows that our procedure is a polynomial-time $(1 + 1/k)$-relaxed decision procedure.

Unfortunately, our discussion assumed that there were no short jobs. If there are short jobs, then first check if $\sum_j p_j > md$. If so, output 'no'. Otherwise, temporarily delete the short jobs and try to find a schedule for the long jobs with the procedure given above. Note that, if the original instance has a schedule that completes by $d$, then we will obtain a schedule for this subset of jobs that completes by $(1 + 1/k)d$. Once again, extend this schedule using list scheduling.

If this procedure outputs 'no', then clearly no schedule of length at most $d$ exists. Now we only need to show that, if a schedule is produced, then all short jobs are completed by time $(1 + 1/k)d$. Suppose that a short job $J_s$ finishes after $(1 + 1/k)d$. Since $p_s \le d/k$, it must have started after $d$, and thus all machines are processing jobs until after time $d$. But this contradicts the fact that $\sum_j p_j \le md$.

Summarizing, we have constructed the following $(1 + 1/k)$-relaxed decision procedure $D_k$:

**if** $\sum_j p_j > md$ **then**
   | output 'no';
**else**
   | temporarily focus on the subset of jobs $J_L := \{J_j | p_j > d/k\}$ ;
   | round down the processing time of each $J_j \in J_L$ to the nearest multiple of
     $d/k^2$ ;
   | using the rounded data, find the optimal packing of $J_L$ into bins of capacity
     $d$ where each bin is assigned fewer than $k$ jobs;
   | **if** *the number of bins used is more than $m$* **then**
     | output 'no';
   | **else**
     | interpret the packing as a schedule for the original data;
     | extend the schedule to include each $J_j \notin J_L$ by using list scheduling.
   | **end**
**end**

We have already seen that, for any fixed value of $k$, the procedure can be implemented to run in polynomial time. We have proved the following result.

**Theorem 9.6.** *For any fixed integer $k > 0$, the algorithm $D_k$ is a polynomial-time $(1 + 1/k)$-relaxed decision procedure for $P||C_{\max}$.*

**Corollary 9.7.** *If the algorithm $B_\rho$ uses the $\rho$-relaxed decision procedure $D_{\lceil c1/(\rho-1)rc\rceil}$, then the family of algorithms $\{B_\rho\}$ forms a polynomial approximation scheme for $P||C_{\max}$.*

**Exercises**

9.4. Give an $O((c_2 n)^{c_1})$ algorithm to solve the bin-packing problem in the case that there are at most at $c_1$ distinct item sizes and at most $c_2$ items may be packed in one bin.

9.5. Prove that the following algorithm is a 5/4-relaxed decision procedure for $P||C_{\max}$. The algorithm consists of six stages. In stage 1, check if $\sum_j p_j > md$; if so, output 'no' and halt. For stages 2 through 5, all $J_j$ with $p_j < d/4$ are temporarily set aside and the instance consisting of the remaining jobs is considered. In stage 2, while there exists a job $J_j$ with $p_j \geq d/2$, find the longest job $J_k$ such that $p_j + p_k \leq d$ and schedule $J_j$ and $J_k$ by themselves on one machine. Stages 3, 4 and 5 are executed $m$ times; if none of these attempts succeeds in scheduling all $J_j$ with $p_j \geq d/4$ on $m$ machines, then output 'no' and halt. For $k = 1, ..., m$, the $k$ th attempt is as follows: in stage 3, find the $2k$ unscheduled jobs with the longest processing times and schedule them two per machine in an arbitrary way; in stage 4, while there exists a job $J_j$ with $p_j > 5d/12$, schedule it on a machine with the two longest unscheduled jobs with processing time at most $3d/8$; in stage 5, schedule the remaining jobs three per machine in an arbitrary way. If a schedule has been constructed for all $J_j$ with $p_j \geq d/4$ using no more than $m$ machines, then stage 6 extends it for the remaining jobs by list scheduling.

9.6. Give a 6/5-relaxed decision procedure for $P||C_{\max}$ that runs in $O(mn)$ time. Can you improve this to $O(n \log n)$?

9.7. (a) Consider the following variant of $P||C_{\max}$, which we denote by $P|mem|C_{\max}$. Each job $J_j$ has both a processing requirement $p_j$ and a memory requirement $\rho_j$. Machine $M_i$ has a memory of size $\sigma_i$ and can only process job $J_j$ if $\sigma_i \geq \rho_j$. The aim is to find a schedule that minimizes the maximum completion time subject to the memory constraints. Consider the *largest memory first* ( *LMF* ) rule: the jobs are listed in order of nondecreasing memory requirement, and when a machine becomes idle, it chooses the next job from the list for which its memory is large enough. Prove that, for any instance of $P|mem|C_{\max}$, $C_{\max}(\text{LMF})/C^*_{\max} \leq 2 - 1/m$.

(b) Suppose that there is a total of $\bar{\sigma}$ units of memory. The memory can be partitioned among the $m$ machines in any way, but it must be done prior to scheduling the jobs. The aim is to find the optimal partition of the memory so that the schedule length for the resulting $P|mem|C_{\max}$ instance is minimized. Show how to compute a partition of the memory for which $C_{\max}(\text{LMF})/C^*_{\max} \leq 2 - 1/m$, where $C^*_{\max}$ is the optimal schedule length with respect to the optimal memory partition. (*Hint*: Suppose that the machines and jobs are indexed in order of nondecreasing memory size and requirement, respectively, and let *LB* be a lower bound on $C^*_{\max}$. For each $i = 1, ..., m$, let $k(i)$ be the smallest integer $k$ such that $\sum_{j=1}^{k} p_j > (i-1)LB$. Then we need $\sigma_i \geq \rho_{k(i)}$ in order to achieve the lower bound. If there is sufficient memory to assign $\rho_{k(i)}$ units to each $M_i$, then the algorithm terminates; if there is not, this fact can be used to increase *LB* and a new iteration is begun.)

9.8. It is not hard to extend the performance guarantees for variants of list scheduling to $P|r_j|L_{\max}$.

(a) Prove that, if list scheduling is used with the jobs given in *EDD* order, then, for any instance of $P|r_j|L_{\max}$, $L_{\max}(EDD) - L^*_{\max} \leq (2 - 1/m) \max_j p_j$.

(b) Prove that, in the delivery time model (as in Chapter 3), for any instance of $P|r_j|L_{\max}$, $L_{\max}(\text{LS})/L^*_{\max} < 2$.

## 9.3. Uniform machines: performance guarantees

Unfortunately, machines are not created equal, and may work at different speeds. Order the machines so that the speeds are nonincreasing; that is, $s_1 \geq s_2 \geq ... \geq s_m$. If $p_j$ denotes the processing requirement of job $J_j$, then processing $J_j$ on $M_i$ takes $p_j/s_i$ time units.

In this more general model, the simplest algorithms do not work quite as well. For example, consider the list scheduling rule. We can analyze this procedure using the same techniques as in the case of identical machines. Let $J_\ell$ be the job that completes last in some list schedule, and let $t$ denote the time at which $J_\ell$ begins processing; no machine is idle prior to time $t$. Since $M_i$ has the capacity to process $ts_i$ units by time

$t$, it follows that

$$\sum_{j\neq l} p_j \geq t \sum_i s_i.$$

Since there must be sufficient capacity to process all of the jobs by $C_{\max}^*$,

$$C_{\max}^* \sum_i s_i \geq \sum_j p_j.$$

The processing of $J_\ell$ takes at least $p_\ell/s_1$ time units, and so $C_{\max}^* \geq p_\ell/s_1$. On the other hand, $J_\ell$ is certainly processed in the list schedule within $p_\ell/s_m$ time units. Combining these pieces, we see that

$$C_{\max}(\text{LS}) \leq t + p_\ell/s_m \leq \frac{\sum_{j\neq l} p_j}{\sum_i s_i} + \frac{p_\ell}{s_m} = \frac{\sum_j p_j}{\sum_i s_i} + p_\ell\left(\frac{1}{s_m} - \frac{1}{\sum_i s_i}\right)$$
$$\leq C_{\max}^* + C_{\max}^* s_1\left(\frac{1}{s_m} - \frac{1}{\sum_i s_i}\right).$$

Thus, we have shown the following theorem.

**Theorem 9.8.** *For any instance of $Q||C_{\max}$, $C_{\max}(LS)/C_{\max}^* \leq 1 + s_1/s_m - s_1/\sum_i s_i$.*

In fact, this bound is tight, and by an appropriate choice of machine speeds may exceed any constant (see Exercise 9.9).

If we were to follow the approach adopted for identical machines, we would next focus on ways to construct a special list for which the list scheduling rule can then be run. However, unlike $P||C_{\max}$, there need not be a list for which the list scheduling rule gives the optimal schedule for $Q||C_{\max}$. The reason for this is that list scheduling always delivers a schedule in which no machine is idle while there is still an unscheduled job. It is easy to construct instances for which every optimal schedule contains such *unforced idle time* (see Exercise 9.10).

The following simple variant of list scheduling may create unforced idle time: schedule the jobs in the order of the list, always assigning the next job to the machine on which it would complete earliest. While this strategy has a better performance guarantee, it still does not deliver solutions for $Q||C_{\max}$ that are within a constant factor of the optimum. However, the LPT variant of this rule, which we shall call LPT′, works quite well.

**Theorem 9.9.** *For any instance of $Q||C_{\max}$, $C_{\max}(LPT')/C_{\max}^* \leq 2 - 1/m$.*

*Proof.* Suppose that the theorem is false, and consider a counterexample in which the sum of the number of jobs and the number of machines is minimum. Let $m$ and $n$, respectively, denote the number of machines and the number of jobs in this counterexample. We shall assume that the jobs are indexed so that $p_1 \geq p_2 \geq ... \geq p_n$.

Consider the schedule generated by LPT′. Suppose that no job is scheduled on machine $M_i$. Since $J_n$ is not scheduled on $M_i$, $p_n/s_i \geq C_{\max}(\text{LPT}') > C_{\max}^*$. Therefore, $M_i$ must be completely idle in any optimal schedule as well. But then, if we delete $M_i$, we have not changed either the LPT$'$ schedule or the optimal schedule, and we obtain a smaller counterexample, which is a contradiction. Similarly, suppose that $J_\ell$ is the last job to finish, where $\ell < n$. In that case, we can obtain a smaller counterexample by deleting all jobs $J_j$, $j > \ell$, since that does not change the length of the LPT$'$ schedule and does not increase $C^*_{\max}$. Thus, we may assume that $J_n$ is the last job to finish.

By the way in which $J_n$ is assigned to a machine, we see that for each $M_i$,

$$C_{\max}(\text{LPT}')s_i \leq p_n + \sum_{J_j \text{ on } M_i, j \neq n} p_j.$$

Summing this inequality for all $i = 1, ..., m$, we get

$$C_{\max}(\text{LPT}')\sum_i s_i \leq mp_n + \sum_{j \neq n} p_j = (m-1)p_n + \sum_j p_j.$$

Since no machine is idle, $m \leq n$, and thus $mp_n \leq \sum_j p_j$. From these inequalities, we see that

$$C_{\max}(\text{LP}T') \leq (2 - \frac{1}{m})\frac{\sum_j p_j}{\sum_i s_i} \leq (2 - \frac{1}{m})C^*_{\max},$$

which shows that our instance is not a counterexample. $\square$

Unlike all of the analyses that we have seen thus far, the bound given in Theorem 9.9 is not tight. In fact, by using additional structural information, one can show that the performance ratio is no more than 19/12. This is also not known to be tight; for the worst example known, $C_{\max}(\text{LPT}')/C^*_{\max} = 1.52$.

From the previous section, recall the notion of a $\rho$-relaxed decision procedure and the bisection search procedure $B_\rho$ in which it was used. The concept of a $\rho$-relaxed decision procedure can be applied equally well to $Q||C_{\max}$. Furthermore, we can adapt the initial lower and upper bounds to obtain the following analogue of $B_\rho$.

It is easy to see that the analogue of Lemma 9.5 is also valid for $Q||C_{\max}$. We will give a simple 3/2-relaxed decision procedure for $Q||C_{\max}$, which therefore yields a 3/2-approximation algorithm for $Q||C_{\max}$.

Our 3/2-decision procedure is a recursive algorithm; that is, it calls itself as a subroutine, but for a smaller instance. Let the procedure be called $Recurse(J, m, d)$ where $J$ denotes the set of jobs, $m$ indicates that there are machines $M_1, M_2, ..., M_m$ (where $s_1 \geq s_2 \geq ... \geq s_m$) and $d$ is the deadline being considered. Recall that the output must either be a schedule that completes by $(3/2)d$ or 'no', where 'no' is output only if there is no schedule that completes within the deadline $d$.

The procedure first checks if the capacity of the machines is sufficient for the given jobs: if $\sum_{J_j \in J} p_j > d \sum_{i=1}^m s_i$, then output 'no' and halt. Otherwise, if $m = 1$, all jobs in $J$ are assigned to machine $M_1$, and the procedure ends. If $m > 1$, let $S$ be the set of all jobs $J_j \in J$ such that $p_j \leq ds_m/2$. If there are no jobs $J_\ell$ in $J - S$ with $p_\ell \leq ds_m$, call $Recurse(J - S, m - 1, d)$. Otherwise, among all such jobs in $J - S$,

$S :=$ the LPT' schedule;
$UB := C_{\max}(\text{LPT}')$ ;
$LB := \max\{p_1/s_1, (\sum_j p_j)/(\sum_i s_i), C_{\max}(\text{LPT}')/2\}$ ;
**while** $LB \neq UB$ **do**
> $d := \lfloor \frac{LB+UB}{2} \rfloor$ ;
> run a $\rho$-relaxed decision procedure to schedule the jobs within the deadline $d$ ;
> **if** *the output is 'no'* **then**
>> $LB := d+1$;
>
> **else**
>> $UB := d$; $S := $ new schedule;
>
> **end**

**end**

let $J_r$ denote one of these with maximum processing requirement. Assign $J_r$ to be scheduled on $M_m$ and call $Recurse(J - S - \{J_r\}, m-1, d)$. If the procedure has not output 'no' and halted, extend the schedule for $J - S$ on $M_1, ..., M_m$ to include $S$ by using list scheduling; that is, order the jobs of $S$ arbitrarily, and assign the next job to be scheduled on the machine that is currently finishing earliest.

**Theorem 9.10.** *Recurse*$(\{J_1, ..., J_n\}, m, d)$ *is a 3/2-relaxed decision procedure for* $Q||C_{\max}$.

*Proof.* We first show that if there is a schedule that completes within time $d$, then the procedure outputs a schedule. To prove this, we will prove the following important claim: if the original instance has a schedule that completes by time $d$, then so must the smaller instance for which the recursive call is made. Given this, it is clear that no recursive call will ever output 'no' for an instance with a feasible deadline, and so a schedule will be output. If $d$ is a feasible deadline to schedule $J$ on the fastest $m$ machines, then $J - S$ can also be scheduled on these $m$ machines by time $d$. But for this subinstance, each job $J_j$ has $p_j > ds_m/2$, and so in any schedule that completes by time $d$, at most one of these jobs is scheduled on $M_m$. If all jobs have processing requirement more than $ds_m$, then clearly, $M_m$ must be completely idle. Otherwise, an interchange argument shows that there always exists a feasible schedule in which $J_r$ is scheduled on $M_m$. In either case, we have shown that the subinstance for which the recursive call is made has a schedule that completes within time $d$.

For the second half of the proof, we will show by induction on $m$ that if a schedule is produced, then it completes all jobs before $(3/2)d$. It is a simple exercise to verify the claim for $m = 1$. Suppose that the algorithm outputs a schedule. Then the recursive call must also output a schedule, and by the inductive hypothesis, all jobs in $J - S$ are completed before $(3/2)d$. The job $J_r$, if it exists, clearly completes by time $d$. Since the machines have sufficient capacity to schedule all jobs within $d$ time units, in any partial schedule, the machine that is currently finishing earliest must complete its jobs before time $d$. Each job $J_j \in S$ has $p_j \leq ds_m/2$ and therefore takes at most $d/2$ time units on *any* machine. But then each job in the list is assigned to a machine on which it completes before $(3/2)d$. □

There is an analogue of the multifit (MF) procedure for $Q||C_{\max}$ as well. This algorithm does still better.

**Theorem 9.11.** *When multifit is run for only $\ell$ iterations of bisection search, then for any instance of $Q||C_{\max}$, $C_{\max}(MF)/C^*_{\max} \leq 1.382 + 2^{-\ell}$.*

Finally, the approach that employs a $(1 + 1/k)$-relaxed decision procedure to construct a polynomial approximation scheme for $P||C_{\max}$ can also be extended to the case of $Q||C_{\max}$. The polynomial approximation scheme for $Q||C_{\max}$ is substantially more complicated, and beyond the scope of this book.

**Exercises**

9.9. Give a family of instances of $Q||C_{\max}$ that shows there does not exist a constant $c$ such that $C_{\max}(\text{LS})/C^*_{\max} \leq c$.

9.10. Give an instance of $Q||C_{\max}$ for which all optimal solutions have unforced idle time.

9.11. Show that the following algorithm is a 2-relaxed decision procedure for $Q||C_{\max}$. For each machine, construct a list of jobs so that these lists form a partition of the set of jobs. Assign each job $J_j$ to the list of the slowest machine $M_i$ such that $p_j \leq s_i d$. When a machine $M_i$ becomes idle, schedule the next job in its list on $M_i$; if its list is empty, find the the next slower machine with a non-empty list, and schedule the next job from *that* list on $M_i$. (When a job is scheduled, delete it from its list.) If the schedule constructed has $C_{\max} > 2d$, instead output 'no'.

## 9.4.   Unrelated machines: performance guarantees

When the machines are unrelated, it is possible that $J_1$ takes much longer on $M_2$ than on $M_1$, whereas $J_2$ takes much longer on $M_1$ than on $M_2$. This generalization makes it substantially more difficult to obtain good approximate solutions. We shall see that this statement can be made more precise. Throughout the next two sections, the time that it takes to process $J_j$ on $M_i$ will be denoted by $p_{ij} = p_j/s_{ij}$, which we assume to be integral.

The worst-case performance of list scheduling or any other known simple scheduling rule is pretty dismal. For example, the *greedy* (G) algorithm naively assigns each job to the machine on which it takes the least time. Clearly, this schedule completes within $T = \sum_j \min_i p_{ij}$. In addition, one can view $T$ as the minimum total requirement of the jobs, and since the best one could hope for is to balance this load evenly over all machines, $C^*_{\max} \geq T/m$.

**Theorem 9.12.** *For any instance of $R||C_{\max}$, $C_{\max}(G)/C^*_{\max} \leq m$.*

This bound is tight (see Exercise 9.12).

A clever variant of list scheduling is significantly better. Maintain a separate list for each machine, and sort all $n$ jobs for $M_i$ 's list in order of nondecreasing *relative speed*, $p_{ij}/\min_h p_{hj}$, $j = 1,...,n$. Whenever a machine is idle, assign the next unscheduled job in its list, unless the smallest ratio of an unscheduled job is more than *sqrtm*; in this case, no further jobs are scheduled on that machine. This *relative speed* ( RS ) rule can be shown to guarantee the following performance.

**Theorem 9.13.** *For any instance of* $R||C_{\max}$, $C_{\max}(RS)/C_{\max}^* \leq 2.5\sqrt{m}+1+1/(2\sqrt{m})$.

This bound is tight up to a constant factor.

Linear programming can be used to construct a much more effective procedure. One natural way to formulate $R||C_{\max}$ as an integer linear programming problem is as follows:

$$\min C_{\max}$$

subject to

$$\sum_j p_{ij}x_{ij} \leq C_{\max}, \qquad \text{for } i = 1,...,m, \qquad (9.2)$$

$$\sum_i x_{ij} = 1, \qquad \text{for } j = 1,...,n, \qquad (9.3)$$

$$x_{ij} \in \{0,1\}, \qquad \text{for } i = 1,...,m, j = 1,...,n, \qquad (9.4)$$

where $x_{ij}$ indicates whether job $J_j$ is assigned to $M_i$. If each integer constraint (9.4) is relaxed to the linear constraint $x_{ij} \geq 0$, then we can solve the resulting linear program $LP_1$ to obtain a lower bound on $C_{\max}^*$.

In fact, $LP_1$ can also be used to obtain good integer solutions as well. The main idea is to obtain an optimal solution $x^*$ to $LP_1$; then, if $x_{ij}^* = 1$, assign $J_j$ to machine $M_i$, and deal with the unassigned jobs in some other manner. We may assume that $x^*$ is an extreme point of $LP_1$, and we will use this in a critical way.

Consider any extreme point $\hat{x}$ of the feasible region of the linear program $LP_1$. There are $mn + 1$ variables in $LP_1$, and so there must be $mn + 1$ linearly independent constraints of $LP_1$ for which $\hat{x}$ is the unique feasible solution that satisfies these constraints with equality. However, other than those of the form $x_{ij} \geq 0$, there are only $m+n$ constraints. Therefore, all but $m+n-1$ of the components of $\hat{x}$ must equal 0. In order to satisfy the constraints (9.3), at least one component $\hat{x}_{ij}$ must be positive for each $j = 1,...,n$. There are at most $m-1$ other positive components, and so for all but at most $m-1$ of the constraints (9.3), exactly one variable is positive, and hence equal to 1. Therefore, the approach suggested above will be able to immediately assign all but at most $m-1$ jobs.

If the number of machines is small, then the schedule can be completed by enumerating all of the possible extensions and choosing the best one. This algorithm based on linear programming (LP) can be analyzed in the following way. Consider the schedule that is based in part on the integer assignments indicated by the optimal solution to $LP_1$, but is completed by assigning the remaining jobs to the machine on which they run in a particular optimal schedule. Of course, the schedule found by heuristic is at least as good as this new one. The new schedule can be split into two partial schedules in the obvious way. The partial schedule given by the linear programming solution has $C_{\max}$ no more than $C_{\max}^*$, as does the partial schedule for the remaining jobs. Since $C_{\max}(\mathrm{LP})$ is no more than the sum of these two parts, $C_{\max}(\mathrm{LP}) \leq 2C_{\max}^*$.

Unfortunately, there can be $m^{m-1}$ ways to assign each of the $m - 1$ jobs to one of $m$ machines, and so there is no apparent way to find the optimal extension in polynomial time. However, this takes only constant time if $m$ is fixed, and so we get the following result.

**Theorem 9.14.** *The algorithm LP is a polynomial-time algorithm for $Rm||C_{\max}$, and for any instance, $C_{\max}(LP)/C_{\max}^* \leq 2$.*

Once again, there exist examples that prove that this analysis is tight. Later in this section, we will derive a fully polynomial approximation scheme for $Rm||C_{\max}$.

This linear programming approach can be extended to yield a polynomial-time algorithm when $m$ is an input. Once again, we will switch to the perspective of finding a $\rho$-relaxed decision procedure. In this case, the bisection search procedure must be modified to initialize $S$, $UB$ and $LB$ as follows:

$S :=$ the greedy schedule;
$UB := C_{\max}(G)$ ;
$LB := \sum_j \min_i p_{ij}$;

Since $UB$ and $LB$ can be within a factor of $m$ initially, we might need to perform an additional $\log_2 m$ iterations until the bisection search terminates, but an analogue of Lemma 9.5 remains true.

We will show that the linear programming approach leads to a polynomial-time 2-relaxed decision procedure, which then yields a polynomial-time 2-approximation algorithm. Suppose that we wish to test if $C_{\max}^* \leq d$. We can view this as testing the feasibility of the following system of constraints:

$$\sum_j p_{ij}x_{ij} \leq d, \text{ for } i = 1,...,m, \qquad (9.5)$$

$$\sum_i x_{ij} = 1, \text{ for } j = 1,...,n, \qquad (9.6)$$

$$x_{ij} \in \{0,1\}, \text{ for } i = 1,...,m, j = 1,...,n. \qquad (9.7)$$

We will, as before, consider the linear relaxation of this, but in order to obtain a tighter relaxation, we add the constraints

$$x_{ij} = 0, \text{ if } p_{ij} > d. \qquad (9.8)$$

Let $\mathrm{LP}_2$ denote the linear relaxation of the system of constraints (9.5)–(9.8). We will show how to round any extreme point of $\mathrm{LP}_2$ into an *integer* solution that cor-responds to a schedule with $C_{\max} < 2d$. This rounding procedure can be done in polynomial time, and so we get the following polynomial-time 2-relaxed decision procedure: test if $LP_2$ is feasible; if not, output 'no'; otherwise, find an extreme point of $LP_2$ and round it to obtain the desired integer solution.

Suppose that $LP_2$ is feasible and we want to round the extreme point $\hat{x}$. One way to model the structure of this solution is to form a bipartite graph: $G(\hat{x}) = (M, J, E)$, where $M = \{M_1, ..., M_m\}$ and $J = \{J_1, ..., J_n\}$ are the sets of machines and jobs, respectively, and $E = \{(M_i, J_j) | \hat{x}_{ij} > 0\}$. As in the proof of Theorem 9.14, by observing that $LP_2$ has $mn$ variables and only $m + n$ constraints not of the form $x_{ij} \geq 0$, we see that $G(\hat{x})$ has no more edges than nodes. We now show that each connected component of $G(\hat{x})$ has this property.

Let $C$ be a connected component of $G(\hat{x})$. If $M_C$ and $J_C$ are the sets of machine and job nodes contained in $C$, let $\hat{x}_C$ denote the restriction of $\hat{x}$ to those $\hat{x}_{ij}$ for which $i \in M_C$ and $j \in J_C$, and let $\hat{x}_{\bar{C}}$ denote the remaining components of $\hat{x}$. For simplicity of notation, reorder the components so that $\hat{x} = (\hat{x}_C, \hat{x}_{\bar{C}})$. The connected component $C$ induces a smaller scheduling problem which restricts attention to the subset of machines $M_C$ and the subset of jobs $J_C$. We can formulate an analogous linear program to $LP_2$ for this subproblem, and denote it by $LP_C$.

We first prove that $\hat{x}_C$ is an extreme point of $LP_C$. Suppose not; then there exist distinct $y_1$ and $y_2$ such that $\hat{x}_C = (y_1 + y_2)/2$, where each $y_i$ is a feasible solution of $LP_C$. But now, $\hat{x} = ((y_1, \hat{x}_{\bar{C}}) + (y_2, \hat{x}_{\bar{C}}))/2$ where each $(y_i, \hat{x}_{\bar{C}})$ is a feasible solution of $LP_2$, which contradicts the fact that $\hat{x}$ was chosen to be an extreme point of $LP_2$. Hence, $\hat{x}_C$ is an extreme point of $LP_C$, and it follows that $G(\hat{x}_C) = C$ has no more edges than nodes.

Since each component of $G(\hat{x}_C)$ has no more edges than nodes, and is, by definition, connected, each component must be either a tree or a tree plus one additional edge. We now use this fact to round the corresponding extreme point $\hat{x}$. As before, for each edge $(M_i, J_j)$ with $\hat{x}_{ij} = 1$, we assign job $J_j$ to machine $M_i$. These jobs correspond to job nodes of degree 1, so that by deleting all of these nodes we get a graph of the same type, $G'(\hat{x})$, with the additional property that each job node has degree at least 2.

We show that $G'(\hat{x})$ has a matching that covers all of the job nodes. For each component that is a tree, root the tree at any node, and match each job node with any one of its children. (Note that each job node must have at least one child and that, since each node has at most one parent, no machine is matched with more than one job.) For each component that contains a cycle, take alternate edges of the cycle in the matching. (Note that the cycle must be of even length.) If the edges of the cycle are deleted, we get a collection of trees which we think of as rooted at the node that had been contained in the cycle. For each job node that is not already matched, pair it with one of its children. This gives us the desired matching. If $(M_i, J_j)$ is in the matching, assign job $J_j$ to be processed on machine $M_i$.

It is straightforward to verify that the resulting schedule has $C_{\max} \leq 2d$. For each machine $M_i$, at most one job $J_j$ is assigned to it based on the matching in $G'(\hat{x})$. Since the corresponding $\hat{x}_{ij}$ must be greater than 0, $p_{ij} \leq d$. For all of the remaining jobs assigned to $M_i$, the corresponding component of $\hat{x}$ is one, and so by (9.5), the total processing time of these jobs is at most $d$. Therefore, each machine is assigned jobs with total processing time at most $2d$. On the other hand, any schedule with $C_{\max} \le d$ corresponds to an (integer) feasible solution to $LP_2$, and so if $LP_2$ is infeasible, we are justified in answering 'no'.

By using this relaxed decision procedure within the usual framework, we have obtained the following result for this approximation algorithm based on linear programming ( $LP'$ ).

**Theorem 9.15.** *The algorithm $LP'$ is a polynomial-time algorithm for $R||C_{\max}$ and for any instance, $C_{\max}(LP')/C_{\max}^* \le 2$.*

It is not hard to construct a family of instances that show that this analysis cannot be improved to yield a better worst-case bound (see Exercise 9.13).

It is significant to note that the structure of $G(\hat{x})$ that was used in rounding $\hat{x}$ can also be used in finding an extreme point of $LP_2$. From a practical point of view, this characterization leads to a particularly efficient implementation of the simplex method. Alternatively, the structure of this linear program can be used to derive a special-purpose combinatorial algorithm that runs in polynomial time.

Consider again the special case when the number of machines is a fixed integer $m$. We will show that in this case, there is a fully polynomial approximation scheme. Clearly, this implies that such a scheme exists for the cases of a fixed number of identical machines or uniform machines. The heart of the scheme is a pseudopolynomial algorithm to find an optimal solution to $Rm||C_{\max}$. It is not hard to see that dynamic programming can be used to derive an algorithm that runs in $O(nmT^{m-1})$ time, where $T = \sum_j \min_i p_{ij}$ (see Exercise 9.18).

Given this pseudopolynomial-time algorithm, it is rather straightforward to complete the fully polynomial approximation scheme. For each positive integer $k$, we will construct a $(1 + 1/k)$-approximation algorithm. Approximate each processing time $p_{ij}$ by $\lfloor p_{ij}/\delta \rfloor$, where $\delta = T/(kmn)$. Note that this rescales $T$ to at most $kmn$. Find the optimal solution to this rescaled and rounded problem in $O(nm(kmn)^{m-1})$ time, and consider this schedule for the original processing times. The schedule obtained is also optimal for the instance with processing times given by $\delta\lfloor p_{ij}/\delta \rfloor$, $i = 1,...,m, j = 1,...,n$, and its value of $C_{\max}$ with these processing times is clearly a lower bound on the desired optimum. Since each $C_j$ is the sum of at most $n$ individual processing times, the difference between the value of $C_{\max}$ for this schedule with the original processing times and the value of $C_{\max}$ for this schedule with the rounded processing times is at most

$$n\delta = nT/kmn = \frac{1}{k}\frac{T}{m} \le \frac{1}{k}C_{\max}^*.$$

Therefore, this algorithm, which we shall call $H_k$, has the worst-case guarantee $C_{\max}(H_k)/C_{\max}^* \le 1 + 1/k$, and we have obtained the following result.

**Theorem 9.16.** *For any fixed $m$, the family of algorithms $\{H_k\}$ forms a fully polynomial approximation scheme for $Rm||C_{\max}$.*

**Exercises**

**9.12.** Give a family of examples that shows that for any number of machines, the analysis given for the greedy algorithm is tight.

**9.13.** Give a family of examples that shows that the analysis of LP′ given in Theorem 9.15 is tight. In addition, show that this bound is tight, even if $m = 2$.

**9.14.** Suppose that when job $J_j$ is assigned to machine $M_i$ a cost of $c_{ij}$ is incurred. Let $c^*$ denote the minimum total cost of any schedule with maximum completion time $C_{max}^*$. Use the linear programming approach to give an algorithm that delivers a schedule with $C_{max} \leq 2C_{max}^*$ with total cost no more than $c^*$.

**9.15.** Suppose that when job $J_j$ is scheduled on machine $M_i$, it may take anywhere between $l_{ij}$ and $u_{ij}$ time units. If job $J_j$ is scheduled on machine $M_i$ to be processed in $u_{ij}$ time units, a cost of $c_{ij}$ is incurred; job $J_j$ may be processed faster on $M_i$ by incurring an additional cost of $s_{ij}$ per unit decrease. For example, the cost of scheduling $J_j$ on $M_i$ to take $l_{ij}$ units is $c_{ij} + s_{ij}(u_{ij} - l_{ij})$. The objective is to schedule all jobs with minimum total cost, subject to the constraint that all jobs are completed within a deadline $d$. Use the linear programming approach to devise an algorithm that delivers a schedule that costs no more than this optimum, and completes all jobs within $2d$. (*Hint*: When job $J_j$ is processed for any time in the range between $l_{ij}$ and $u_{ij}$, this can be viewed as processing some fraction of the job on machine $M_i$ at the speed needed to process the entire job in $l_{ij}$ time units, and the complementary fraction is processed at the speed at which it would be processed in $u_{ij}$ time units.)

**9.16.** Consider the variant of the decision version of $R||C_{max}$ where each machine $M_i$ has its own deadline $d_i$. Modify the approach used in algorithm LP′ to give an algorithm that either outputs 'no', or else outputs a schedule in which each machine $M_i$ completes its assigned jobs by $d_i + \max_j p_{ij}$, where 'no' is output only when no schedule completes all jobs within the given deadlines.

**9.17.** The linear programming approach can also be used to design a polynomial approximation scheme for $Rm||C_{max}$, where the space required does not depend exponentially on $m$. Recall that in designing a $(1 + 1/k)$-approximation algorithm it is sufficient to find a $(1 + 1/k)$-relaxed decision procedure. To use this approach, one can define a suitable notion of a *bad* assignment, where the job takes a constant fraction of the time prior to the deadline if assigned to that machine. As a result, there are only a constant number of ways to make bad assignments of the jobs. Combine complete enumeration of these possibilities with the algorithm of Exercise 9.16, to design a $(1 + 1/k)$-relaxed decision procedure.

**9.18.** Construct a dynamic programming algorithm to solve $Rm||C_{max}$ in $O(nmT^{m-1})$ time, where $T = \sum_j \min_i p_{ij}$. (*Hint*: Recall Section 8.3.)

## 9.5. Unrelated machines: impossibilities

Unlike $P||C_{max}$ and $Q||C_{max}$, no polynomial approximation scheme is known for $R||C_{max}$. It is highly unlikely that such a scheme exists, since this would imply that

$P = NP$. In order to prove this assertion, we investigate the computational complexity of the decision version of $R||C_{max}$ with small integral deadlines.

**Theorem 9.17.** *For $R||C_{max}$, the question of deciding if $C_{max}^* \leq 3$ is NP-complete.*

*Proof.* We prove this result by a reduction from the 3-dimensional matching problem. We are given an instance of this problem, consisting of a family of triples $\{T_1, T_2, ..., T_m\}$ over the ground set $A \cup B \cup C$, where $A$, $B$ and $C$ are disjoint sets such that $|A| = |B| = |C| = n$; each $T_i$ satisfies $|T_i \cap A| = |T_i \cap B| = |T_i \cap C| = 1$. We construct an instance of the scheduling problem with $m$ machines and $2n + m$ jobs. Machine $M_i$ corresponds to the triple $T_i$, for $i = 1, ..., m$. There are $3n$ 'element jobs' that correspond to the $3n$ elements of $A \cup B \cup C$ in the natural way. In addition, there are $m - n$ 'dummy jobs'. (If $m < n$, we construct some trivial 'no' instance of the scheduling problem.) Machine $M_i$ corresponding to $T_i = (a_j, b_k, c_\ell)$ can process each of the jobs corresponding to $a_j$, $b_k$ and $c_\ell$ in one time unit and each other job in three time units. Note that the dummy jobs require three time units on each machine.

It is quite simple to show that $C_{max}^* \leq 3$ if and only if there is a 3-dimensional matching. Suppose there is a matching. For each $T_i = (a_j, b_k, c_\ell)$ in the matching, schedule the element jobs corresponding to $a_j$, $b_k$ and $c_\ell$ on machine $M_i$. Schedule the dummy jobs on the $m - n$ machines corresponding to the triples that are not in the matching. This gives a schedule with $C_{max} = 3$. Conversely, suppose that there is such a schedule. Each of the dummy jobs requires three time units on any machine and is thus scheduled by itself on some machine. Consider the set of $n$ machines that are not processing dummy jobs. Since these are processing all of the $3n$ element jobs, each of these jobs is processed in one time unit. Each three jobs that are assigned to one machine must therefore correspond to elements that form the triple corresponding to that machine. Since each element job is scheduled exactly once, the $n$ triples corresponding to the machines that are not processing dummy jobs form a matching. □

As an immediate corollary of this theorem, we get the following result.

**Corollary 9.18.** *For every $\rho < 4/3$, there does not exist a polynomial-time $\rho$-approximation algorithm for $R||C_{max}$, unless $P = NP$.*

*Proof.* Suppose there were such an algorithm. We will show that it yields a polynomial-time algorithm for the 3-dimensional matching problem. Given an instance $I$ of the 3-dimensional matching problem, map it into an instance of $R||C_{max}$ using the reduction given above, and then apply the presumed approximation algorithm. We have just seen that $I$ is a 'yes' instance if and only if the instance of $R||C_{max}$ does have a schedule of length 3. Therefore, when $I$ is a 'yes' instance, the approximation algorithm must output a schedule of length less than $(4/3)C_{max}^*$. But this length must be an integer, and so it must be at most 3. When $I$ is a 'no' instance, the algorithm produces a schedule of length at least 4. Therefore, the algorithm outputs a schedule of length at most 3 if and only if $I$ is a 'yes' instance. □

The technique employed in Theorem 9.17 can be refined to yield a stronger result.

**Theorem 9.19.** *For $R||C_{\max}$, the question of deciding if $C_{\max}^* \leq 2$ at most 2 is NP-complete.*

*Proof.* We again start from the 3-dimensional matching problem. We call the triples that contain $a_j$ *triples of type $j$*. Let $t_j$ be the number of triples of type $j$, for $j = 1, ..., n$. As before, machine $M_i$ corresponds to the triple $T_i$, for $i = 1, ..., m$. There are now only $2n$ element jobs, corresponding to the $2n$ elements of $B \cup C$. We refine the construction of the dummy jobs: there are $t_j - 1$ dummy jobs of type $j$, for $j = 1, ..., n$. (Note that the total number of dummy jobs is $m - n$, as before.) Machine $M_i$ corresponding to a triple of type $j$, say, $T_i = (a_j, b_k, c_\ell)$, can process each of the element jobs corresponding to $b_k$ and $c_\ell$ in one time unit and each of the dummy jobs of type $j$ in two time units; all other jobs require three time units on machine $M_i$.

Suppose there is a matching. For each $T_i = (a_j, b_k, c_\ell)$ in the matching, schedule the element jobs corresponding to $b_k$ and $c_\ell$ on machine $M_i$. For each $j$, this leaves $t_j - 1$ idle machines corresponding to triples of type $j$ that are not in the matching; schedule the $t_j - 1$ dummy jobs of type $j$ on these machines. This completes a schedule with $C_{\max} = 2$. Conversely, suppose that there is such a schedule. Each dummy job of type $j$ is scheduled on a machine corresponding to a triple of type $j$. Therefore, there is exactly one machine corresponding to a triple of type $j$ that is not processing dummy jobs, for $j = 1, ..., n$. Each such machine is processing two element jobs in one time unit each. If the machine corresponds to a triple of type $j$ and its two unit-time jobs correspond to $b_k$ and $c_\ell$, then $(a_j, b_k, c_\ell)$ must be the triple corresponding to that machine. Since each element job is scheduled exactly once, the $n$ triples corresponding to the machines that are not processing dummy jobs form a matching. □

**Corollary 9.20.** *For every $\rho < 3/2$, there does not exist a polynomial-time $\rho$-approximation algorithm for $R||C_{\max}$, unless $P = NP$.*

### Exercises

9.19. Prove for any integers $p < q$ such that $2p \neq q$, $R||C_{\max}$ is *NP*-hard even in the case that all $p_{ij} \in \{p, q\}$.

9.20. In contrast to Exercise 9.19, give a polynomial-time algorithm to solve the special case of $R||C_{\max}$ when each $p_{ij} \in \{1, 2\}$.

### 9.6. Two identical machines: probabilistic analysis

The results presented in the previous sections provide ample illustration of the power of a worst-case approach to the analysis of heuristics. While necessarily pessimistic, the outcome of a worst-case analysis at least yields an ironclad performance guarantee that will always be valid. This safety belt often comes at the expense of realism, in that computational experiments might indicate that the worst-case behavior is rarely registered in practice. On the contrary, the average performance of an approximation algorithm is usually strikingly better than its worst-case behavior would suggest. In this section, we shall consider a mathematical framework in which these empirical results can be analyzed, and reconsider several of the approximation algorithms already discussed, but from this perspective. As a caveat, however, the reader should note that we might have little understanding of what an average instance really is in practice; the results presented in this section might be no more realistic than our assumptions about the instances considered.

We first must outline the precise mathematical definitions that will capture the notion of the average performance of a heuristic. Probability theory provides an appropriate setting for this approach. We shall assume that the reader is familiar with the basic terminology of this area. In the spirit of empirical computational work, a problem instance will be regarded as being generated by a *random mechanism*. For example, for the scheduling problem $P||C_{max}$, one would typically assume that the processing times $p_j$, $j = 1, ..., n$, are *random variables* whose joint distribution is given in advance. Given a particular *realization* of these random variables, the heuristic solution is computed; its value is obviously a random variable as well, whose distribution can be analyzed and whose *expected value* informs us about the average behavior of the heuristic in question, especially when compared to the *expected value of the optimal solution*.

The *probabilistic analysis* of algorithms, then, starts from a probability distribution over the class of all problem instances, and focuses on the random variables describing algorithmic behavior on a randomly generated instance. The analysis can be technically demanding; frequently, it is *asymptotic* in nature, in that precise statements are only possible if the problem size is allowed to go to infinity. Hence, it is appropriate at this point to introduce the three modes of *stochastic convergence* that typically arise in such a situation.

If $\mathbf{y}_1, \mathbf{y}_2, ...$ is a sequence of random variables, then *almost sure (a.s.) convergence* of the sequence to a constant $c$ means that

$$\Pr\{\lim_{n \to \infty} \mathbf{y}_n = c\} = 1.$$

Sometimes this is referred to as convergence *with probability 1*. It implies the weaker condition *convergence in probability*, which requires that, for every $\varepsilon > 0$,

$$\lim_{n \to \infty} \Pr\{|\mathbf{y}_n - c| > \varepsilon\} = 0. \tag{9.9}$$

Finally, *convergence in expectation* means that

$$\lim_{n \to \infty} |Ex[\mathbf{y}_n] - c| = 0,$$

and also implies (9.9).

We shall illustrate some of these notions on the problem $P2||C_{max}$. In this case, a problem instance corresponds to the processing times $p_1, p_2, ..., p_n$; let us assume these to be independent random variables that are *uniformly distributed* over the interval [0,1], which is a model favored by analysts and experimenters alike.

Let us first consider the performance of the list scheduling (LS) algorithm from a probabilistic perspective. From (9.1), we deduce that in the case of $P2||C_{max}$,

$$C_{max}(\text{LS}) \leq \sum_j p_j/2 + p_{max}/2.$$

Since the first term on the right-hand side is a lower bound on $C_{max}*$, we see that

$$\mathbf{C}_{max}(LS)/\mathbf{C}_{max}^* \leq 1 + \mathbf{p}_{max}/\sum_j \mathbf{p}_j. \tag{9.10}$$

Obviously, $\mathbf{p}_{max} \leq 1$. In addition, the *strong* law of large numbers says that

$$(\sum_j \mathbf{p}_j)/(n/2) \to 1 \text{(a.s.)}.$$

Hence, the ratio $\mathbf{C}_{max}(LS)/\mathbf{C}_{max}^*$ itself also converges to 1 with probability 1. Put differently, the relative error of this simple heuristic almost surely vanishes for large problem sizes, and its depressing worst-case error of 50 percent occurs rather infrequently.

As we have seen above, *almost sure convergence* is just one of several ways to capture the asymptotic behavior of a sequence of random variables. Thus, there are other ways to capture the asymptotic optimality of these simple heuristics. They do, however, require a different probabilistic starting point.

By way of example, consider the inequality

$$\Pr\{\frac{\sum_j \mathbf{p}_j}{n} - \frac{1}{2} \geq t\} \leq e^{-2nt^2},$$

which is a special case of Hoeffding's inequality. From (9.10), with $t = 1/4$, one can easily deduce that

$$\Pr\left\{\frac{\mathbf{C}_{max}(LS) - \mathbf{C}_{max}^*}{\mathbf{C}_{max}^*} \leq \frac{4(m-1)}{3n}\right\} \geq 1 - e^{-n/8},$$

which, of course, confirms error convergence to 0 in probability, but which also provides precise information on the rate at which the relative error of list scheduling converges to 0. We shall later return to the third mode of convergence, convergence in expectation.

The entire analysis above has its implications for the optimal solution value $\mathbf{C}_{max}^*$ itself. Indeed, a trivial byproduct is that

$$\mathbf{C}_{max}^*/(n/4) \to 1 \text{( a.s.)}.$$

This is a typical example of what is usually referred to as *probabilistic* value analysis; for large $n$, the optimal solution value can be guessed with increasing relative accuracy. The *structure* of the optimal solution itself does not lend itself to a similar probabilistic convergence analysis, and so it is much easier to predict the solution value that to say anything about the solution itself.

Now that asymptotic optimality in the relative sense turns out to be within easy reach, it is tempting to examine the asymptotic behavior of the absolute error for $\mathbf{C}_{\max}(A) - \mathbf{C}_{\max}^*$ for several algorithms $A$. In the case of list scheduling, it converges to a value strictly greater than 0. Can we do better? The insight gained from the worst-case analysis leads us to consider the LPT rule, with the hope that this might demonstrate superior probabilistic performance too. Let us investigate this possibility from a probabilistic perspective by focusing on the absolute difference $\mathbf{d}_j(LPT)$ between the total processing time assigned to $M_1$ and to $M_2$ after $j$ jobs have been allocated. Clearly, $\mathbf{d}_n(LPT)/2$ is an upper bound on the absolute error.

Let $\mathbf{p}^{(1)} \leq \mathbf{p}^{(2)} \leq \cdots \leq \mathbf{p}^{(n)}$ denote the sorted list of processing times. The structure of the LPT rule immediately implies that

$$\begin{aligned}
\mathbf{d}_n(LPT) &\leq \max\{\mathbf{d}_{n-1}(LPT) - \mathbf{p}^{(1)}, \mathbf{p}^{(1)}\} \\
&\leq \max\{\mathbf{d}_{n-2}(LPT) - (\mathbf{p}^{(1)} + \mathbf{p}^{(2)}), \mathbf{p}^{(2)} - \mathbf{p}^{(1)}, \mathbf{p}^{(1)}\}.
\end{aligned}$$

Continuing recursively, we see that

$$\mathbf{d}_n(LPT) \leq \max_{k=1,\ldots,n} \{\mathbf{p}^{(k)} - \sum_{j=1}^{k-1} \mathbf{p}^{(j)}\}.$$

It is trivial to see that, for any fixed $\varepsilon \in (0,1)$,

$$\max_{k=1,\ldots,\lfloor \varepsilon n \rfloor} \{\mathbf{p}^{(k)} - \sum_{j=1}^{k-1} \mathbf{p}^{(j)}\} \leq \mathbf{p}^{(\lfloor \varepsilon n \rfloor)} \tag{9.11}$$

and

$$\max_{k=\lfloor \varepsilon n \rfloor+1,\ldots,n} \{\mathbf{p}^{(k)} - \sum_{j=1}^{k-1} \mathbf{p}^{(j)}\} \leq \mathbf{p}^{(n)} - \sum_{j=1}^{\lfloor \varepsilon n \rfloor} \mathbf{p}^{(j)}. \tag{9.12}$$

For the uniform distribution, it is known that $\mathbf{p}^{(\lfloor \varepsilon n \rfloor)}/\varepsilon \to 1$ (a.s.). In addition, it is not difficult to show that $\sum_{j=1}^{(\lfloor \varepsilon n \rfloor)} \mathbf{p}^{(j)}/n$ almost surely converges to a positive constant (which depends on $\varepsilon$). Since $\mathbf{p}^{(n)}$ is obviously at most 1, this implies that the right-hand side of (9.12) tends to $-\infty$, and hence the maximum of the right-hand sides of (9.11) and (9.12) tends to $\varepsilon$. This establishes the following theorem.

**Theorem 9.21.** *For* $P2||C_{\max}$, $\mathbf{d}_n(LPT) \to 0$ *almost surely.*

**Corollary 9.22.** *For* $P2||C_{\max}$, $\mathbf{C}_{\max}(LPT) - \mathbf{C}_{\max}^* \to 0$ *almost surely.*

We can use the same bounding technique to bound $\mathbf{d}_n(LPT)$ in expectation. From probability theory, we know that

$$Ex[\mathbf{p}^{(\lfloor \varepsilon n \rfloor)}] = \frac{\lfloor \varepsilon n \rfloor}{n+1}.$$

and

$$Ex\left[\mathbf{p}^{(n)} - \sum_{j=1}^{\lfloor \varepsilon n \rfloor} \mathbf{p}^{(j)}\right] = \frac{n}{n+1} - \sum_{j=1}^{\lfloor \varepsilon n \rfloor} \frac{j}{n+1}. \qquad (9.13)$$

For every fixed $\varepsilon \in (0,1)$, the right-hand side of (9.13) converges to $-\infty$, and hence the absolute error of the LPT rule also converges to 0 in expectation.

**Theorem 9.23.** *For $P2||C_{\max}$, $Ex[C_{\max}(LPT) - C_{\max}^*] \to 0$.*

This result can be extended for much more general models, such as $Q||C_{\max}$ (see Exercise 9.21).

Can we be still more ambitious? The above analysis for $P2||C_{\max}$ reveals that $\mathbf{d}_n(LPT)$ converges to 0 (a.s.), but does not provide any information on the *rate* at which this occurs. It is possible to estimate this rate; one finds that $\mathbf{d}_n(LPT)/(\log\log n/n)$ is almost surely bounded by a constant. At the same time, the smallest possible difference $\mathbf{d}_n^*$ is known to converge much faster to 0; $\mathbf{d}_n^*/(n^2 2^{-n})$ is almost surely bounded by a constant, so that here convergence occurs at an exponential rate. From that perspective, the best possible heuristic would be one that achieves a similarly fast convergence. Does such a heuristic exist?

The answer to this question is unknown, but an impressive improvement on the rate of convergence of the LPT rule is obtained by a modified version of the differencing method. The differencing method was introduced in studying the probabilistic performance of algorithms for this problem, but there is no known theoretical analysis of this method. As is often the case in the probabilistic analysis of algorithms, it seems to be easier to modify this algorithm to obtain one that is similar, and yet more amenable to analysis. The typical problem is one of dependencies among the data used at various stages of the algorithm. Initially, we have assumed that the processing requirements of the jobs are independent random variables, and this assumption is extremely important to the analyses that we have given. However, after even one iteration of the differencing method, we have lost this essential property. The basic idea is to modify the algorithm so that iterations of the algorithm can be analyzed as essentially independent steps. The technical details of this result are far too complicated to present here; however, we shall give an outline of the modified algorithm and try motivate the modifications made.

The modified differencing method (MD) works in a series of phases, where in each phase most of the jobs are paired, and each pair is replaced by a new job with processing requirement equal to the difference of their processing requirements. Suppose that in the current phase, each $p_j \in (0, U]$; we can subdivide this interval into a number of equal-length subintervals. For each pair, both jobs will be selected from the same interval, and so the size of the new job created will be bounded by the length of the subinterval. The crucial modification is to introduce *randomization* into the algorithm itself. For each subinterval, the algorithm randomly pairs all jobs (assuming that there are an even number of jobs). As a result, the processing requirements of the jobs created in this way will be independent random variables, and will be distributed according to a triangular distribution. Of course, not all subintervals will have an even number of jobs, and something more complicated must be introduced to handle this situation. The algorithm terminates when there are only a specified constant number of jobs remaining, and this small instance can be solved to optimality.

The crux of the analysis is to calculate the way in which the length of the upper bound $U$ changes from iteration to iteration, as compared to the number of jobs that remain. The value of $U$ at any stage is an upper bound on the value of $d_n(MD)$; thus, if we can show that with high probability, the value of $U$ at the end is very small, then we obtain the corresponding bound for the solution given by algorithm. Roughly speaking, it can be shown that the number of jobs in the $(i+1)$th phase is at least $c^{-i}n$ and the upper bound $U$ is at most $c^{(2^i)}n^{-i}$, where $c$ is a constant greater than 2. When the number of remaining jobs reaches the specified constant, the bound on the length of the interval is $n^{-O(\log n)}$.

**Theorem 9.24.** *For $P2||C_{\max}$, $C_{\max}(MD) - C_{\max}^* \leq d_n(MD)/2 = n^{-O(\log n)}$.*

In the probabilistic analysis of algorithms, it is often the case that the modified algorithm is justified by showing that, in some stochastic sense, the performance of the original algorithm dominates the modified one. It would be nice if this were true in this case, but this remains an important open question.

We have dealt with $P2||C_{\max}$ at length, since this example exhibits many of the ingredients typically encountered in a probabilistic analysis:

- a *combinatorial problem* which is *NP*-hard and hence difficult to solve;
- a *probability distribution* over all problem instances to generate problem data as realizations of independent and identically distributed random variables;
- a *probabilistic value analysis* that yields an asymptotic characterization of the optimal solution value as a simple function of the problem data;
- a *probabilistic error analysis* of a fast heuristic to prove that its relative or absolute error tends to 0 with increasing problem size in some stochastic sense; and
- a *rate of convergence analysis* that yields some indication of how large the problem size must be in order to demonstrate asymptotic behavior in practice, and this, moreover, allows for further differentiation among the heuristics.

**Exercises**

9.21. Using the approach suggested in the text, show that the absolute error of the LPT rule applied to $Q||C_{\max}$ a.s. converges to 0. (*Hint*: Rather than looking at the largest difference between the amounts of processing assigned to the various machines, consider the difference between largest and average amount.)

**Notes**
In these notes, a dagger (†) will be used to indicate that a performance bound is tight or asymptotically tight.

9.1. *Identical machines: classical performance guarantees.* The seminal paper of Graham [1966] contains the performance analysis of the list scheduling rule. In a later paper, Graham [1969] also analyzed the LPT rule as well as the polynomial approximation schemes for $Pm||C_{\max}$ given in Theorem 9.3 and Exercise 9.2. Sahni [1976] gave the first fully polynomial approximation scheme for $Pm||C_{\max}$. Garey, Graham, and Johnson [1978] give a tutorial introduction into the early work on performance guarantees for scheduling parallel machines. Karmarkar and Karp [1982] invented the differencing method; its worst-case analysis is due to Fischetti and Martello [1987].

9.2. *Identical machines: using bisection search for guarantees.* The FFD algorithm and other approximation algorithms for bin packing are reviewed by Coffman, Garey, and Johnson [1984]. Coffman, Garey, and Johnson [1978] proposed the multifit algorithm and proved that $C_{\max}(\text{MF})/C_{\max}^* \leq 1.22 + 2^{-\ell}$. Friesen [1984] improved the constant to 1.2 and gave instances that achieve a ratio of 13/11. Yue [1990] claimed that 13/11 is also an upper bound, but it appears that a complete proof was obtained later by Cao [1995]. Friesen and Langston [1986] gave a refined version of multifit which runs in time $O(n \log n + \ell \cdot n \log m)$ (where the constant embedded within the 'big Oh' notation is big indeed) and has a tight performance bound of $72/61 + 2^{-\ell}$.

The framework of using a relaxed decision procedure as well as the first polynomial approximation scheme for $P||C_{\max}$ are due to Hochbaum and Shmoys [1987]. They also gave the 5/4- and 6/5-relaxed decision procedures of Exercises 9.5 and 9.6, and a 7/6-relaxed decision procedure that runs in $O(n(m^4 + \log n))$ time. Exercise 9.7 is based on the work of Kafura and Shen [1977, 1978].

Several bounds are available that take into account the processing times of the jobs. Note that the probabilistic result of Theorem 9.21 relies on such a (worst-case) bound for list scheduling. Achugbue and Chin [1981] prove two results relating the performance ratio of list scheduling to the value of $\pi = \max_j p_j / \min_j p_j$. If $\pi \leq 3$, then

$$C_{\max}(\text{LS})/C_{\max}^* \leq \begin{cases} 5/3 & \text{if } m = 3, 4, \\ 17/10 & \text{if } m = 5, \\ 2 - \frac{1}{3\lfloor m/3 \rfloor} & \text{if } m \geq 6, \end{cases} \qquad (\dagger)$$

and if $\pi \leq 2$,

$$C_{\max}(\text{LS})/C_{\max}^* \leq \begin{cases} 3/2 & \text{if } m = 2, 3, \\ 5/3 - \frac{1}{3\lfloor m/2 \rfloor} & \text{if } m \geq 4. \end{cases} \qquad (\dagger)$$

For the case of LPT, Ibarra & Kim [1977] prove that

$$C_{\max}(\text{LPT})/C_{\max}^* \leq 1 + \frac{2(m-1)}{n} \text{ for } n \geq 2(m-1)\pi.$$

Much less is known about the worst-case performance of approximation algorithms for other minmax criteria. For $P|r_j|L_{\max}$, Exercise 9.8(a) is due to Gusfield [1984]; the result of Exercise 9.8(b) was observed by Hall and Shmoys [1989], who also developed a polynomial approximation scheme for this problem. For the case of equal processing times, $P|r_j, p_j = p|L_{\max}$, Simons [1983] extended Theorem 4.10 to obtain a polynomial algorithm, and Simons and Warmuth [1989] reduced the running time to $O(mn^2)$ by generalizing the approach of Garey, Johnson, Simons, and Tarjan [1981].

As for enumerative optimization methods, Bratley, Florian, and Robillard [1975] proposed an algorithm for $P|r_j, \tilde{d}_j|C_{\max}$ and tested it on problems with up to 3 machines and 25 jobs. Carlier [1987] gave an algorithm for $P|r_j|L_{\max}$, which performs well on problems with up to 8 machines and 100 jobs. Dell'Amico and Martello [1995] developed an algorithm for $P||C_{\max}$ and reported good results for random problems with up to 15 machines and 10,000 jobs; relatively small problems ($n \leq 50$ for $m = 10$ or 15) appear to be the hardest ones.

9.3. *Uniform machines: performance guarantees.* Theorem 9.8 is due to Liu and Liu [1974A, 1974B, 1974C], as are the observations given as Exercises 9.9 and 9.10. Morrison [1988] showed that LPT is better that LS, in that

$$C_{\max}(\text{LPT})/C_{\max}^* \leq \max\{\max_i s_i/(2\min_i s_i), 2\}. \qquad (\dagger)$$

Cho and Sahni [1980] considered the variant of list scheduling, LS′, which puts the next job in the list on the machine on which it will finish earliest, and proved that

$$C_{\max}(\text{LS}')/C_{\max}^* \leq \begin{cases} (1+\sqrt{5})/2 & \text{for } m = 2, \\ (1+(\sqrt{2m-2})/2 & \text{for } m > 2. \end{cases}$$

The bound is tight for $m \leq 6$ but, in general, the worst known examples have a performance ratio of $\lfloor (\log_2(3m-1)+1)/2 \rfloor$. This approach followed the work of Gonzalez, Ibarra, and Sahni [1977], who presented the LPT′ algorithm and the result given in Theorem 9.9. The improved upper and lower bounds for LPT′ were obtained by Dobson [1984] and Friesen [1987].

Friesen and Langston [1983] extended the multifit approach to uniform processors and proved that its performance bound is in between 1.4 and 1.341. They also showed that the decision to order the bins by increasing size is the correct one, since for decreasing bin sizes there exist examples with performance ratio 3/2. Chen [1991] proved the performance bound of Theorem 9.11.

Extending the work of Sahni [1976], Horowitz and Sahni [1976] gave a family of algorithms $A_k$ with running time $O(n^{2m}k^{m-1})$ and performance bound $1 + 1/k$, which is a fully polynomial approximation scheme for any fixed value of $m$. The 3/2-approximation algorithm implied by Theorem 9.10 and the polynomial approximation scheme for $Q||C_{\max}$ mentioned at the end of the section are due to Hochbaum and Shmoys [1988]. Williamson (private communication) observed the algorithm given in Exercise 9.11 and its analysis.

9.4. *Unrelated machines: performance guarantees.* The first paper on this subject was by Ibarra and Kim [1977], who considered the greedy method of Theorem 9.12 and Exercise 9.12 and other $m$-approximation algorithms. Davis and Jaffe [1981] analyzed a number of algorithms and obtained the bound for the RS rule given in Theorem 9.13. The linear programming approach of Theorem 9.14 is due to Potts [1985A]; it was extended by Lenstra, Shmoys, and Tardos [1990] to the algorithm LP' of Theorem 9.15 and Exercise 9.13. Trick [1990] proposed the model described in Exercise 9.15; Tardos (private communication) observed the results given in Exercises 9.14 and 9.15. For $Rm||C_{\max}$, the fully polynomial approximation scheme of Theorem 9.16 is due to Horowitz and Sahni [1976], and the polynomial approximation scheme derived in Exercises 9.16 and 9.17 is due to Lenstra, Shmoys, and Tardos [1990].

Hariri and Potts [1991] performed computational tests with approximation algorithms for $R||C_{\max}$ on problems with up to 50 machines and 100 jobs. Among five constructive heuristics, an implementation of the Lenstra-Shmoys-Tardos algorithm produced the best solutions but required nontrivial running times; a simple iterative improvement scheme, applying job reassignments and interchanges, was still able to achieve substantial improvements, in almost negligible amounts of time. Van de Velde [1993] developed optimization and approximation algorithms for $R||C_{\max}$, based on surrogate duality relaxation of the constraints (9.2). His branch-and-bound method performs reasonably well. His 'duality-based heuristic search' algorithm outperforms the constructive heuristics but, again, the method should be supplemented with some form of iterative improvement.

9.5. *Unrelated machines: impossibilities.* The results in this section are due to Lenstra, Shmoys, and Tardos [1990].

9.6. *Two identical machines: probabilistic analysis.* A basic text in probability theory is Feller [1968, 1971]. For the probabilistic analysis of scheduling algorithms, we refer to the monograph by Coffman and Lueker [1991], the surveys by Coffman, Lueker, and Rinnooy Kan [1988] and by Rinnooy Kan and Stougie [1989], and the annotated bibliography by Karp, Lenstra, McDiarmid, and Rinnooy Kan [1985]. The inequality of Hoeffding [1963] was applied by Coffman et al. in their survey paper to bound the rate of convergence of the relative error of the LS algorithm. Frenk and Rinnooy Kan [1987] proved that the absolute error of the LPT rule converges to 0 almost surely and in expectation, even for $Q||C_{\max}$; see also Coffman, Flatto, and Lueker [1984] and Loulou [1984]. The rate of convergence of this error was investigated by Boxma [1984] and Frenk and Rinnooy Kan [1986].  Karmarkar and Karp [1982] proposed and analyzed the MD method.

# Contents

# 10

# Minmax criteria with preemption

Eugene L. Lawler
*University of California, Berkeley*

Charles U. Martel
*University of California, Davis*

Good news! This chapter has no NP-hardness results to limit us, and hence no complicated approximation algorithms to consider. There are only polynomial algorithms to appreciate. In contrast to the nonpreemptive case where even $P2||C_{max}$ is NP-hard, we will be able to solve preemptive settings even for unrelated machines with release times and due-dates.

We begin with a simple linear time algorithm for $P|pmtn|C_{max}$ and then present the more complex algorithm of Gonzalez and Sahni that solves $Q|pmtn|C_{max}$ in essentially $O(n)$ time. We exploit the insights gained from this algorithm to derive efficient algorithms for $Q|pmtn|L_{max}$, $Q|pmtn, r_j|C_{max}$, $Q|pmtn, r_j|L_{max}$ and $Q|pmtn|C_{max}$ when processors have memory capacities and jobs have memory requirements. We conclude by showing that the problems $R|pmtn|C_{max}$, $R|pmtn|L_{max}$, $R|pmtn, r_j|C_{max}$ and $R|pmtn, r_j|L_{max}$ can be solved by linear programming.

## 10.1. Minimizing Makespan

McNaughton's 1959 solution of the problem $P|pmtn|C_{max}$ is probably the simplest and earliest instance of an approach that has been successfully applied to other preemptive scheduling problems: First provide an obvious lower bound on the cost of an optimal solution and then construct a schedule that meets this bound.

| $j$   | 1 | 2 | 3 | 4 | 5 | 6 | 7 | 8 |
|-------|---|---|---|---|---|---|---|---|
| $p_j$ | 5 | 8 | 6 | 2 | 3 | 1 | 3 | 4 |

**Table 10.1    Sample Data for McNaughton's Algorithm**

**Figure 10.1.**    Schedule obtained by McNaughton's algorithm.

In the case of $P|pmtn|C_{\max}$, let $p_{\max}$ be the largest $p_j$ value. We see that $C_{\max}$ must be at least

$$\max\{p_{\max}, (\sum_{j=1}^{n} p_j)/m\}. \tag{10.1}$$

A schedule meeting this bound can be constructed in $O(n)$ time: Schedule the jobs one at a time, in arbitrary order, filling up the available time on each successive machine before proceeding to the next, splitting the processing of a job whenever the above time bound on a given machine is met.

For example, consider the problem data in table 10.1:

Suppose there are four machines. Then the time bound given by (10.1) is

$C_{\max} \geq \max\{8, 32/4\} = 8$.

A schedule meeting this time bound is shown in Figure 10.1.

The number of preemptions occuring in a schedule constructed by McNaughton's algorithm is at most m-1, and it is possible to construct a class of problem instances for which an optimal schedule has at least this many preemptions. It is not hard to see that the problem of minimizing the number of preemptions is NP-hard. (Observe that our numerical example can be solved without preemptions.)

### 10.1.1. Uniform Machines: The Gonzalez-Sahni Algorithm

In the case of $Q|pmtn|C_{\max}$, generalizing the bound (10.1) shows that the length of an optimal schedule must be at least

$$\max\{\max(1 \leq k \leq m-1), \ \{(\sum_{j=1}^{k} p_j)/\sum_{i=1}^{k} s_i\}, (\sum_{j=1}^{n} p_j)/\sum_{i=1}^{m} s_i\}. \tag{10.2}$$

where $p_1 \geq \ldots \geq p_n$, and $s_1 \geq \ldots \geq s_m$. An algorithm of Gonzalez and Sahni enables us to construct a schedule meeting this bound. Their algorithm requires only $O(n)$ time if the jobs are given in in order of nonincreasing $p_j$ and the machines in order of nonincreasing $s_i$; without this assumption, the running time is $O(n+m\log m)$. The algorithm yields an optimal schedule with at most $2(m-1)$ preemptions, which is a tight bound.

To explain the algorithm, we find it convenient to generalize the usual definition of uniform parallel machines so as to allow the speeds of the machines to be time varying. Let $s_i(t)$ denote the speed of machine $i$ at time $t$ and assume that $s_1(t) \geq \ldots \geq s_m(t)$, for all $t$. The functions $s_i(t)$ may be discontinuous, but are integrable. The *processing capacity* of machine $i$ in the time interval $[t, t']$ is then

$$S_i(t, t') = \int_{t}^{t'} s_i(u) du. \tag{10.3}$$

In order for a job to be completed, it is necessary that the sum of the processing capacities in the time intervals in which the job is processed should equal its processing requirement. For example, if job $j$ is processed on machine 1 in the interval $[t_1, t_1']$ and on machine 2 in the interval $[t_2, t_2']$, then this processing is sufficient to complete the job if $S_1(t_1, t_1') + S_2(t_2, t_2') = p_j$.

Let $S_i, i = 1, \ldots, m$, denote the processing capacity of machine $i$ in the interval $[0, T]$. By a generalization of (10.2), we see that for there to exist a feasible preemptive schedule in the interval [0,T] it is necessary that

$$\begin{aligned}
S_1 &\geq p_1 \\
S_1 + S_2 &\geq p_1 + p_2
\end{aligned}$$

$$. \qquad\qquad . \tag{10.4}$$

$$\begin{aligned}
S_1 + S_2 + \ldots + S_{m-1} &\geq p_1 + p_2 + \ldots + p_{m-1} \\
S_1 + S_2 + \ldots + S_m &\geq p_1 + p_2 + \ldots + p_n.
\end{aligned}$$

Let T be the smallest value for which the inequalities (10.4) are satisfied. We shall construct a feasible schedule in the interval $[0, T]$ by scheduling the jobs one at a time, in arbitrary order. For each successive job j, we first find the machine with largest index k such that its (remaining) processing capacity $S_k \geq p_j$ and then consider three cases:

**Case 1** $S_k = p_j$, where $k \leq m - 1$. In this case we schedule job $j$ to be processed by machine $k$ for the entire period $[0, T]$. We then eliminate machine $k$ and job $j$ from the problem, leaving a problem with $m - 1$ machines and $n - 1$ jobs for which the inequalities (10.4) are again satisfied.

**Case 2** $S_k > p_j > S_{k+1}$, where $k \leq m - 1$. We assert that there exists a time $t, 0 < t < T$, such that

$p_j = \int_0^t s_k(u)du + \int_t^T s_{k+1}(u)du$.

To convince ourselves of this fact, we need only plot the curves of the continuous functions

$f(t) = \int_0^t s_k(u)du$, $g(t) = \int_t^T s_{k+1}(u)du$, and $f(t) + g(t)$.

We propose to schedule job $j$ for processing on machine $k$ in the interval $[0, t]$ and on machine $k + 1$ in the interval $[t, T]$. We then create a *composite* machine from the remaining available time on machines $k$ and $k + 1$. The capacity of this composite machine in the interval $[0, T]$ is

$$S_k + S_{k+1} - p_j = \int_0^t s_{k+1}(u)du + \int_t^T s_k(u)du \tag{10.5}$$

We then replace machines $k$ and $k + 1$ with this new composite machine, leaving us with a problem with $m - 1$ machines and $n - 1$ jobs, for which the inequalities (10.4) are again satisfied.

**Case 3** $S_m \geq p_j$. In this case we schedule job $j$ on machine $m$, thereby reducing its capacity to $S_m - p_j$. This leaves a problem with $m$ machines and $n - 1$ jobs for which the inequalities (10.4) are again satisfied.

It is now a simple matter to prove, by induction on the number of jobs, that inequalities (10.4) imply the existence of a feasible schedule. In other words, satisfaction of inequalities (10.4) is both necessary and sufficient for feasibility. We note that in fact the above algorithm has the strong property that after scheduling any job $j$ *each* of the sums $S_1, S_1 + S_2, \ldots, S_1 + \ldots + S_m$ is as large as possible for any legal scheduling of job $j$ on the set of processors which exist when $j$ is scheduled.

To illustrate the algorithm, consider a problem with jobs of $p_j = 28, 26, 16, 12, 10$. The total processing time of all jobs is 92. Thus if we were to schedule these jobs on machines of speeds 10, 8, 4 and 1, the value of equation (10.2) is the maximum of $26/10, 54/18, 70/22, 92/23$ which is $92/23 = 4$. The scheduling of job one (using Case 2 of the above algorithm) is shown in Figure 10.2 and the final schedule is shown in Figure 10.3.

### 10.1.2. Implementation of the algorithm

The idea of machines with time-varying speeds was introduced only to make it easy to describe the processing capacities of composite machines. At the beginning of the computation, before any composite machines are formed, all machines have constant speeds. In order to compute the minimum value of $C_{\max}$ by (10.2), we need only to determine the $m$ largest $p_j$ values and to order them. If the $p_j$ values are not already sorted, we can find the $m$ largest in $O(n)$ time by applying a selection algorithm, and then sort the $m$ largest values in $O(m \log m)$ time. Likewise, if the machine speeds are not already sorted, $O(m \log m)$ time suffices to sort the $s_i$ values.

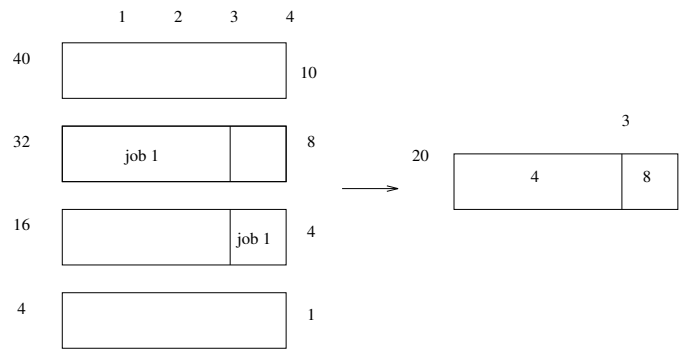

**Figure 10.2.** Scheduling job 1 using Gonzalex-Sahni algorithm.

**Figure 10.3.** Final schedule for $p_j = 28, 26, 16, 12, 10$.

We propose to maintain the (composite) machine capacities in a linked list $(S_1, ..., S_m)$. This means that each time a job is scheduled, only constant time is required to revise the list. Suppose, instead of scheduling the jobs in arbitrary order as described above, we schedule the jobs in nonincreasing order of processing time. Let $k(j)$ denote the largest value of $k$ such that $S_k \geq p_j$, with reference to the processing capacities existing at the time job $j$ is scheduled. Notice that, in all cases, $k(j) \geq k(j-1) - 1$. It follows that, by recording the value of $k(j-1)$ at the end of iteration $j-1$ for use in iteration j, only $O(n+m) = O(n)$ time is required to scan the list of machine capacities over the course of $n$ iterations. Moreover, after $m-1$ jobs have been scheduled, either only one composite machine remains, or Case 3 has occured. Once Case 3 occurs, the only machine capacities that must be checked are the last two capacities in the list.

It follows from the above analysis that $O(n + m \log m)$ time suffices for all operations required by the algorithm, except for the time required to actually schedule the jobs on the composite machines.

We propose to represent each composite machine by a doubly-linked list of triples $(i, [t_i, t_i'])$, where each triple gives the index of an elementary machine of speed $s_i$ and each time interval $[t_i, t_i']$ indicates a time interval of processing on that machine. By scanning the linked list for composite machine $k+1$ from the end and $k$ from the front in parallel (advancing by one interval on each machine until we find a pair of intervals which overlap), it is easy to determine the time t in Case 2. We can thus assure that we take $O(r+1)$ time to find the intervals used for job $j$ where $r$ is the number of elementary machine intervals which are completely used up by job $j$. It is not hard to verify that at most $O(n+m) = O(n)$ time is required for list scanning overall by the algorithm. Thus the total time for the algorithm is $O(n + m \log m)$. The number of preemptions introduced is discussed in the exercises.

The Gonzalez-Sahni algorithm is a an important building block for algorithms on uniform machines. In the next four sections we will use this algorithm as a subroutine to solve more complex uniform machine settings.

**Exercises**

10.1. Construct a class of $P|pmtn|C_{\max}$ problem instances for which any optimal schedule has at least $m-1$ preemptions.

10.2. Construct a class of $Q|pmtn|C_{\max}$ problem instances for which any optimal schedule has at least $2(m-1)$ preemptions.

10.3. Suppose each machine is available for processing in only certain specified time intervals. Describe how you would apply the Gonzalez-Sahni algorithm, subject to constraints of this type.

10.4. Show that, if jobs are scheduled in arbitary order, the Gonzalez-Sahni algorithm can be implemented to run in $O(m^2 + n)$ time.

10.5. Suppose that $k$ time units of idle time must elapse between the time when a job is preempted and the time when its processing is resumed on another machine.

(a) Prove that imposing this condition on identical machines increases $C_{\max}^*$ by at most $k-1$.

(b) Show that, for $k=1$, the $C_{\max}$ problem is solvable in polynomial time by a simple adaptation of McNaughton's rule. (For any fixed $k \geq 2$, the problem is NP-hard.)

10.6. Prove a bound on the on the maximum number of $(i,[t_i,t_i'])$ triples which will be used to represent the composite processors.

10.7. Use the result of problem 6 to show that the algorithm uses at most 2(m-1) preemptions.

## 10.2. Meeting Deadlines

The Gonzalez-Sahni algorithm gives us the insight necessary to develop an algorithm for solving the feasibility problem $Q|pmtn,\bar{d}_j|--$. We shall schedule the jobs in deadline order, so assume they are indexed with $\bar{d}_1 \leq \bar{d}_2 \leq \ldots \leq \bar{d}_n$. Let $S_i^{(j-1)}$ denote the capacity of (composite) machine $i$ in the time interval $[0,\bar{d}_{j-1}]$ after jobs $1,2,\ldots j-1$ have been scheduled. The capacities of the composite machines available for the processing of job $j$ will then be $S_i^{(j-1)} + s_i(\bar{d}_j - \bar{d}_{j-1})$, for $i=1,...,m$. The Gonzalez-Sahni (G-S) algorithm tells us that whatever the capacities of the composite machines, it is optimal to schedule job $j$ on the composite machines as determined by Cases 1-3 in the previous section (that is, this schedule maximizes each of the partial sums of the remaining machine capacities).

Concerning ourselves only with the composite machine capacities, and not with the actual scheduling of the jobs, the algorithm of Sahni and Cho is as follows:

Algorithm *Sahni-Cho*
**for** $i=1,...,m$;
   $S_i := 0$;
**for** $j=1,...,n$;
   **for** $i=1,...,m$;
      $S_i := S_i + s_i (\bar{d}_j - \bar{d}_{j-1})$;
Schedule job $j$ using G-S algorithm. Let $k$ be the largest index with $S_k \geq p_j$.
   Case $(S_k = p_j)$:          * schedule the job on composite machine $k$ from zero to $d_j$*
      **for** $i=k,..,m-1$
         $S_i := S_{i+1}$;
         $S_m := 0$;
   Case $(S_k > p_j > S_{k+1}$, and $k < m)$:  * schedule on composite machines $k, \quad k+1$ *
      $S_k := S_k + S_{k+1} - p_j$;
      **for** $i=k+1,...,m-1$
         $S_i := S_{i+1}$;
      $S_m := 0$;
   Case $(S_m \geq p_j)$                  * schedule the job on composite machine $m$ *

$S_m := S_m - p_j;$

The Sahni-Cho algorithm can be implemented to run in $O(mn + n\log n)$ time. The schedule it creates has at most $O(mn)$ preemptions, and there are settings which require $\Omega(mn)$ preemptions.

### 10.2.1.  Specialization to identical machines

Interestingly, the specialization of the Sahni-Cho algorithm to $P|pmtn, \overline{d}_j| - -$ is a good deal more efficient.

At each successive deadline $\overline{d}_j$, each (elementary) machine $i$ has been scheduled for the continuous processing of jobs in the interval $[0, a_i]$. We will now show that we can schedule each job using at most one preemption and maintaining the above property of continuous processing.

Suppose we maintain a nearly balanced tree of pairs $(i, a_i)$, sorted by $a_i$ value. Then, in $O(\log m)$ time, we will be able to locate the largest $a_i$ such that $\overline{d}_j - a_i \geq p_j$. We now schedule as follows.

If $\overline{d}_j - a_i = p_j$, we schedule job $j$ from $a_i$ to $\overline{d}_j$ on machine $i$ (as in case one of the G-S algorithm).

If this is the maximum $a_i$ value (thus the 'slowest' processor) we simply schedule job $j$ from $a_i$ to $a_i + p_j$ (this corresponds to case 3).

Otherwise we also find the smallest $a_r$ such that $\overline{d}_j - a_r < p_j$. Now schedule job $j$ on machine $r$ from $a_r$ to $\overline{d}_j$ and on machine $i$ for the remaining $p_j - (\overline{d}_j - a_r)$ time units starting at time $a_i$. This corresponds to case 2.

In each of the three cases above it is then easy to update $a_i$ (and $a_r$ in case 2) to its new value.

Finding $a_i$ and $a_r$ as well as updating their values can all be done in $O(\log m)$ time, and each job is then scheduled in O(1) time. Hence the overall running time of the algorithm is $O(n \log m)$. If time to sort the $\overline{d}_j$'s is included, the overall running time becomes $O(n \log n)$.

### 10.2.2.  Minimization of $L_{\max}$

We can apply Meggido's method to transform the Sahni-Cho feasibility algorithms to algorithms for solving $P|pmtn|L_{\max}$ and $Q|pmtn|L_{\max}$. What is needed is the smallest value of $\lambda$ such that the induced deadlines $\overline{d}_j = d_j + \lambda$ admit a feasible schedule. The Sahni-Cho feasibility algorithm has $n$ iterations. In the case of uniform machines, at each iteration a bisection search enables us to find the desired index $k$ with $O(\log m)$ comparisons of $p_j$ against the machine capacities, which are linear functions of $\lambda$. It follows that $L_{\max}$ can be minimized with $O(n \log m)$ calls on the feasibility algorithm, or $O(mn^2 \log m)$ time overall. In the case of identical machines, the feasibility algorithm also requires $O(\log m)$ comparisons at each iteration. The result of each comparison can be resolved by a call to the feasibility algorithm. Hence $L_{\max}$ can be minimized with $O(n \log m)$ calls on the feasibility algorithm, or $O(n^2 \log^2 m)$ time

overall.

Note: The problems $P|pmtn|L_{\max}$ and $Q|pmtn|L_{\max}$ are equivalent, by symmetry of release dates and due dates, to the problem $P|pmtn, r_j|C_{\max}$ and $Q|pmtn, r_j|C_{\max}$. Hence the algorithms described above solve these problems as well.

**Exercises**

10.8. Describe how to minimize *weighted* maximum lateness for identical and uniform machines. (Note: The relative order of the induced deadlines $\bar{d}_j = d_j + \lambda/w_j$ may change with $\lambda$.)

10.9. Give a family of $Q|pmtn, \bar{d}_j| --$ problems which require $\Omega(mn)$ preemptions.

## 10.3. The Staircase Algorithm for Release Dates

In this section we shall describe an algorithm for solving the problem $Q|pmtn, r_j|C_{\max}$. For reasons that will become apparent, we refer to this as the "staircase" algorithm. The details of the algorithm and its implementation are somewhat involved. However, the principal idea behind the algorithm is actually quite simple.

Assume the jobs are numbered in release date order, $r_1 \leq \ldots \leq r_n$. The algorithm creates a schedule by moving from one release date to the next, creating a subschedule for each successive interval $[r_k, r_{k+1}]$, $k = 1, 2, \ldots n-1$. At $r_k$, $k$ jobs have been released, and their remaining processing requirements in sorted order are $p_1^{(k)} \geq \ldots \geq p_k^{(k)} \geq 0$. (Included among these processing requirements is $p_k$, since job $k$ is released at $r_k$; no correspondence between the indexing of the the $p_j^{(k)}$'s and the release dates is intended.) The algorithm has the property that the remaining processing requirements are *evenly* distributed as possible. More specifically, there is no way that the jobs could have been processed before $r_k$ that would have yielded a smaller value for *any* of the partial sums

$p_1^{(k)}$,

$p_1^{(k)} + p_2^{(k)}$,

$\ldots$

$p_1^{(k)} + p_2^{(k)} + \ldots + p_k^{(k)}$.

At $r_k$, the algorithm utilizes the capacities of the $m$ machines in the interval $[r_k, r_{k+1}]$ so as to minimize each and every one of the new partial sums

$p_1^{(k+1)}$,

$p_1^{(k+1)} + p_2^{(k+1)}$,

$\ldots$

$p_1^{(k+1)} + p_2^{(k+1)} + \ldots p_k^{(k+1)}$.

where $p_1^{(k+1)} \geq \ldots \geq p_k^{(k+1)} \geq 0$. This means that when $r_n$ is reached, the remaining processing requirements $p_j^{(n)}$, $j = 1, \ldots, n$, are such that the length of the final interval $[r_n, C_{\max}]$ is minimized. (The length of this final interval is determined by the smallest value of $T$ such that inequalities (10.4) are satisfied, with respect to the

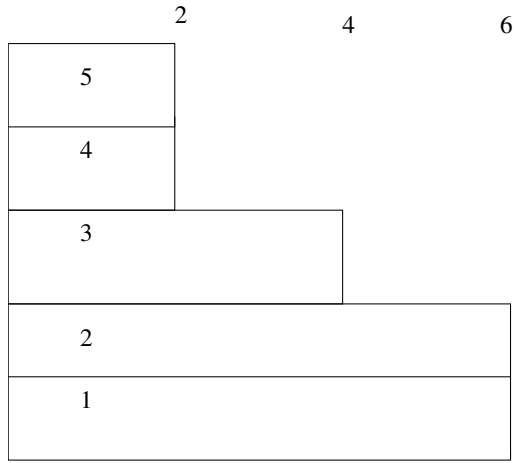

**Figure 10.4.**    staircase of remaining processing times

remaining processing requirements.)

The reader may wish to reflect on a certain duality between the staircase algorithm and the Sahni-Cho algorithm. As we have described, the staircase algorithm schedules several jobs at each iteration in such a way as to minimize each of the partial sums of the processing requirements remaining for the next iteration. By contrast, the Sahni-Cho algorithm schedules a single job at each iteration in such a way as to maximize each of the partial sums of the processor capacities remaining for the next iteration.

The staircase algorithm has the desirable property that it is *nearly on-line*. By this we mean that at each successive release date $r_k$ the algorithm does not require any knowledge of the jobs that are to be released at future times; all it requires is knowledge of the next release date $r_{k+1}$. (If it did not require knowledge of $r_{k+1}$, it would be truly on-line.) By contrast, the Sahni-Cho algorithm is off-line. When the Sahni-Cho algorithm is applied to an instance of the feasibility problem $Q|pmtn, r_j, \overline{d}_j = \overline{d}| - -$ (by applying it to a symmetrically equivalent instance of $Q|pmtn, \overline{d}_j| - -$), the algorithm iterates from the latest release date backward.

### 10.3.1.  Constructing the staircase

Let us consider how the algorithm determines the amount of processing to perform on each job in the interval $[r_k, r_{k+1}]$. For ease of notation, let us drop the superscripts and denote $p_j^{(k)}$ by $p_j$ and let $q_j$ denote the remaining processing time of the job associated with $p_j$ after processing in this interval. We also let $t = r_{k+1} - r_k$. For purposes of exposition, we assume for the time being that, if $m < k$, machines $m + 1, \ldots, k$, with $s_{m+1} = \ldots = s_k = 0$, are added to the model.

The $p_j$ can be viewed as defining a staircase pattern as shown in Figure 10.4. The $q_j$ will chosen in such a way that they form a similar pattern. Such a staircase can be characterized by grouping the remaining processing times into blocks of equal values. If there are $u$ different values remaining, we let $q(1)$ denote the maximum remaining value, $q(2)$ the second largest remaining value, ..., and $q(u)$ the smallest remaining value (with $q(u) \geq 0$). We then let $h(i)$ represent the *height* of the first $i$ blocks. Thus $h(1)$ is the number of jobs with $q(1)$ units of processing remaining, and in general $h(i)$ is the number of jobs with $q(i)$ or more units of processing remaining. In example 10.4 $q(1) = 6, q(2) = 4, q(3) = 2, h(1) = 2, h(2) = 3, h(3) = 5$. The staircase can be characterized by

$((h(1),q(1)),...,(h(u),q(u)),$
where $q_j = q(i)$ , for each job $j, h(i-1)+1 \leq j \leq h(i)$.
(Here $i = 1,...,u; h(0) = 0, h(u) = k$.) We always have:

$$q(i) > q(i+1), \quad i = 1,...,u-1. \tag{10.6}$$

The staircase is constructed in such a way that for each $i$, the capacities of machines $h(i-1)+1,...,h(i)$ are fully utilized to decrease $p_{h(i-1)+1},...,p_{h(i)}$ to $q(i)$. A second condition for feasibility is therefore that

$$\sum_{j=h(i-1)+1}^{l} q_j = (l - h(i-1))q(i) \geq \sum_{j=h(i-1)+1}^{l} p_j - t \sum_{j=h(i-1)+1}^{l} s_j \tag{10.7}$$

for $l = h(i-1)+1,...,h(i); i = 1,...,u$.
The corners of the staircase, except possibly the last one, correspond to strict equalities:

$$\sum_{j=h(i-1)+1}^{h(i)} q_j = (h(i) - h(i-1))q(i) = \sum_{j=h(i-1)+1}^{h(i)} p_j - t \sum_{j=h(i-1)+1}^{h(i)} s_j \tag{10.8}$$

for $i = 1,...,u-1$.
A third condition for feasibility is of course that

$$0 \leq q_j \leq p_j, j = 1,2,...,k. \tag{10.9}$$

We tentatively construct the first step of the staircase by setting
$h(1) := 1, q(1) := p_1 - ts_1$.
Generally, after constructing $i$ tentative steps, $(h(1),q(1)),...,(h(i),q(i))$, the tentative steps satisfy the feasibility conditions (10.6) to (10.9), except that possibly $q(i) < 0$, in violation of (10.9). We construct an $(i+1)st$ tentative step by setting
$h(i+1) := h(i) + 1, \qquad\qquad q(i+1) := p_{h(i+1)} - ts_{h(i+1)}$ .
If $q(i) > q(i+1)$ and $q(i) \geq 0$, we are finished with the construction of the new tentative step.
Suppose now that $q(i) < 0$, in violation of condition (10.9), or $q(i) \leq q(i+1)$, in violation of condition (10.6). In the first case, there is unused capacity on machines $h(i-1)+1, \ldots h(i)$; in both cases, some of the capacity of these machines must be used to process job h(i)+1, in order to satisfy conditions (10.6), (10.8) and (10.9). We therefore reconstruct the $i$-th step so as to include job $h(i)+1$ in that step, setting

$h(i) := h(i) + 1,$

and recalculating $q(i)$ according to (10.8),

$$q(i) := \left( \sum_{j=h(i-1)+1}^{h(i)} p_j - t \sum_{j=h(i-1)+1}^{h(i)} s_j \right) / (h(i) - h(i-1)) \tag{10.10}$$

Using the old $q(i), q(i+1)$ values, the formula for the total remaining processing of the jobs in the new combined step is:
$(h(i) - h(i-1) - 1)q(i) + p_{h(i)} - ts_{h(i)} = (h(i) - h(i-1) - 1)q(i) + q(i+1)$
Thus we can compute the new value of $q(i)$ as:

$$q(i) := [(h(i) - h(i-1) - 1)q(i) + q(i+1)]/(h(i) - h(i-1) \tag{10.11}$$

As a result, it may now be that $q(i-1) \leq q(i)$ ($q(i-1) < 0$ cannot occur). In this case, we reconstruct the $(i-1)$st step so as to include the i-th step: $h(i-1)$ is set to $h(i)$ and $q(i-1)$ is recalculated using an analagous approach to (10.11). We continue in this way until condition (10.6) is satisfied. The adjusted staircase includes one more job and may have fewer steps than before.

Pseudo-code for the staircase subalgorithm is as follows, where the interval length $t$ and processing times $p_1 \geq \ldots \geq p_k$ which remain at the start of interval $k$ are the inputs, and the resulting $q(i), h(i)$ values are the outputs:
staircase$(t, p_1, \ldots, p_k; q(1), \ldots, q(i); h(1), \ldots h(i))$:

```
h(0) := 0;
q(0) := ∞
i := 0;
for j = 1, .., k
    i := i + 1;
    h(i) := j;
    q(i) := pⱼ − tsₕ₍ᵢ₎;
    while (q(i − 1) ≤ q(i) or q(i − 1) < 0)
        q(i − 1) := [(h(i − 1) − h(i − 2))q(i − 1) + (h(i) − h(i − i))q(i)]/
        [h(i) − h(i − 2)];
        h(i − 1) := h(i);
        i := i − 1;
return q(1), . . . , q(i); h(1), . . . , h(i);
```

We have to verify that the resulting staircase $(h(1), q(1)), \ldots (q(k), h(k))$ and the corresponding remaining processing requirements $q(1), \ldots, q(u)$ indeed satisfy the the feasibility conditions (10.6–10.9). For (10.6) and (10.8) this is obvious. To see that (10.7) must be true, note that each $q(i)$ is initially defined by an equality constraint and can only increase thereafter. To verify condition (10.9), it suffices to show that $q(i) \leq p_{h(i)}$ (since if the new processing amount for jobs in the $ith$ step ($q(i)$) is equal to or less than the smallest intial processing amount ($p_{h(i)}$) then the condition holds for those jobs in the step which started with more processing). We initially set $q(i)$ to $p_{h(i)} - ts_{h(i)}$. When two steps are combined, the current value $h(i)$ becomes the height of the new step $i - 1$, and the new value of $q(i-1)$ is at most the old value of $q(i)$. Thus after combining the two steps, the new value $q(i-1)$ is still at most $p_{h(i-1)}$ which implies the desired result.

### 10.3.2. Complexity analysis

We now analyze the running time of the subalgorithm. The number of step constructions (first three lines of *for* loop) is exactly $k$. Each iteration of the *while* loop combines two steps, so this is done at most $k - 1$ times in total. Thus the entire *staircase* function runs in O($k$) time. This presupposes that the given $p_j$ values are ordered; but since the relative order of the remaining processing requirements does not change, we can maintain an ordered list of these values and insert the processing requirement $p_k$ of job $k$ that becomes available at $r_k$ in O($k$) time. Hence the subalgorithm determines the values $q(i)$ for each interval in O($k$) time. As has been indicated above, the Gonzalez-Sahni algorithm can be applied to construct an actual schedule for each interval in O($k$) time as well (since the machines and processing times are assumed to have been sorted). We thus have arrived at a nearly on-line algorithm that requires $O(n^2)$ time overall.

### 10.3.3. Correctness of the algorithm

We note first that not only does the relative order of the remaining processing requirements remain invariant, but also the following stronger property holds: as soon as two remaining processing requirements become equal, they remain equal. To see this, suppose that $p_j = p_{j+1}$ at time $r_k$, and they are in different steps, so, let $h(i) = j$. We set $q(i+1) = p_{j+1} - ts_{j+1}$. But $q(i) \leq p_j - ts_j \leq p_j - ts_{j+1} = q(i+1)$, and we have to reconstruct the $i$-th step so as to include job $j + 1$ as well.

This leads us to define the *rank* of an available job $j$ at time $r_k$ as the value $h(i)$ for which $h(i-1) + 1 \leq j \leq h(i)$ (where these are the final $h(i)$ values computed for the interval $[r_k, r_{k+1}]$). The rank of a job at time $r_n$ is defined analogously as its step height that would be found if the subalgorithm were to be applied in the interval $[r_n, C_{\max}^*]$. A job will be called *critical* if its rank is at most $m - 1$ and it hasn't been completed. Otherwise the job is *noncritical*. The rank of a job cannot decrease; in particular, once a job becomes noncritical, it never becomes critical again. It follows from (10.8) that in any interval the fastest $h(i)$ machines are exclusively processing the longest $h(i)$ critical jobs. A critical job is processed continuously from its release date until it either is completed or becomes noncritical.

These observations suggest the following correctness proof for the algorithm. First, suppose that the schedule ends at $C_{\max}^*$ with the simultaneous completion of $l$ critical jobs ($l < m$). At any time when $l'$ of these jobs are available, they are processed by the fastest $l'$ machines. Thus the maximum possible total processing has been devoted to these $l$ critical jobs prior to $r_n$, so, the schedule is clearly optimal.

Alternatively, suppose that the schedule ends with the simultaneous completion of $m$ noncritical jobs. If there is no idle time in the schedule it is clearly optimal. Otherwise, let $r_k$ be the last release date such that there is idle time in $[r_{k-1}, r_k]$ on some machine. All the jobs that are available but noncritical at time $r_{k-1}$ will thus be completed by time $r_k$. We conclude that the portion of the schedule for the remaining jobs has the following structure. Before $r_k$, the available critical jobs are processed by the fastest machines. Between $r_k$ and $C^*_{max}$, there is no idle time. It follows that the schedule is optimal for the jobs under consideration and thus that $C^*_{max}$ is the minimum time to complete all the jobs.

### 10.3.4.  A more efficient off-line implementation

Let us use the terminology of the prior section to describe a more efficient implementation of the staircase subalgorithm. We will reduce the running time by dealing more carefully with the noncritical jobs, circumventing the need to introduce machines of speed zero.

The noncritical jobs of lowest rank, i.e., jobs $h(i-1) + 1, \ldots h(i)$, where $h(i-1) + 1 \leq m \leq h(i)$, will be called *active*. In the interval $[r_k, r_{k+1}]$, their remaining processing requirements are reduced by machines $h(i-1) + 1, \ldots m$ to a common amount $q(i)$. The remaining processing requirements are not reduced at all, since they are assigned to dummy machines of speed zero.

As a first refinement, the subalgorithm does not have to deal with the active noncritical jobs individually, since their remaining processing requirements will remain equal throughout. They can easily be handled simultaneously by straightforward generalizations of (10.10) and (10.11). As a second refinement, the subalgorithm can be terminated as soon as either $h(i) = k$ or $h(i) \geq m$ and $q(i) > p_{h(i)+1}$.

Instead of maintaining an ordered list of all remaining processing requirements, we have only to do so for the largest $m - 1$ of them. We simply record the number of active noncritical jobs, their common remaining processing requirement, and the lowest index of any of them. Finally, we maintain a priority queue for the remaining requirements of the inactive noncritical jobs.

At each release date the processing requirement of the job(s) that becomes inactive is, depending on its size, inserted either in the ordered list in $O(m)$ time or in the priority queue in $O(\log n)$ time. The staircase computations for the longest $m - 1$ jobs and the active noncritical jobs require $O(m)$ time in each interval and $O(mn)$ time overall. The queue operations require $O(\log n)$ time when a job is inserted or deleted and $O(n \log n)$ time overall, since once an inactive job becomes active and is withdrawn from the queue, it remains active throughout. Hence successive applications of the modified subalgorithm determine the value $C^*_{max}$ in $O(n \log n + mn)$ time. As has been indicated above, the Sahni-Cho algorithm can be applied to construct an actual schedule in the interval $[r_1, C^*_{max}]$ in $O(n \log n + mn)$ time as well. We thus have arrived at an off-line algorithm that requires $O(n \log n + mn)$ time overall.

### Exercises

10.10. Devise an example to show that there can be no on-line algorithm for $Q|pmtn, r_j|C_{\max}$.

10.11. Suppose m uniform machines are capable of *processor sharing*. By this we mean that the processing capacity of a given subset of machines can be equally shared by a given subset of jobs (with processing occuring in infinitesimally small time slices, and preemption infinitely often). Show that under the assumption of processor sharing, an on-line algorithm does exist for $Q|pmtn, r_j|C_{\max}$. Describe how this algorithm is related to the staircase algorithm.

### 10.4. Network Flow Computations for Release Dates and Deadlines

Let us begin with the feasibility problem $P|pmtn, r_j, \bar{d}_j| - -$. Let $\{e_1, ..., e_{2n}\}, e_1 \leq ... \leq e_{2n}$, be the ordered collection of release dates $r_j$ and deadlines $\bar{d}_j$. If a release date and a deadline are equal, the smaller index is assigned to the release date. Let $E_k$ denote the time interval $[e_k, e_{k+1}]$, for $k = 1, 2, ..., 2n - 1$.

We shall construct a flow network with job nodes $j = 1, ..., n$, interval nodes $E_k$, $k = 1, ..., 2n - 1$, a source node $s$ and a sink node $t$. There is an arc $(j, k)$ of capacity $e_{k+1} - e_k$ from job node $j$ to each of the interval nodes $E_k$ such that $r_j \leq e_k$ and $\bar{d}_j \leq e_{k+1}$. In addition there is an arc $(s, j)$ of capacity $p_j$ from the source node to each job node $j$ and an arc $(k, t)$ of capacity $m(e_{k+1} - e_k)$ from each interval node to the sink node. We assert that there exists a feasible preemptive schedule for the given instance of the feasibility problem if and only if the flow network admits a flow which saturates all the arcs out of $s$ (and thus the total flow equals the sum of the $p_j$ values).

The reasoning is very simple. The flow value $f(j, k)$ for the arc $(j, k)$ represents the amount of processing of job $j$ that is done in interval $E_k$. The capacities $e_{k+1} - e_k$ assigned to the arcs $(j, k)$ assure that no job is scheduled for more than the length of $E_k$, and the capacity $m(e_{k+1} - e_k)$ of the arc $(k, t)$ assures that the total processing done by all jobs in interval $E_k$ can be completed. Thus, the amount of processing that is done on the various jobs in the interval $E_k$ satisfies the conditions (10.1).

If the network flow computation indicates that there exists a feasible schedule, McNaughton's algorithm can be applied to the arc flow values $f(j, k)$ to construct a feasible subschedule within a given time interval $E_k$ in $O(n)$ time. We thus can construct a feasible schedule in $O(p(n) + n^2)$ time, where $p(n)$ is the time required for the max flow computation.

Since any legal schedule can be converted to a feasible flow which saturates all arcs out of $s$, if the maximum flow is less than the sum of the $p_j$ values, we know that no legal schedule exists.

### 10.4.1.  $Q|pmtn, r_j, \bar{d}_j|-$

It requires a bit of cleverness to construct a flow network for the problem $Q|pmtn, r_j, \bar{d}_j|-$. In order for conditions (10.4) to be satisfied for the interval $E_k$, we must restrict the total processing done on any single job to $s_1(e_{k+1} - e_k)$, on any pair of jobs to $(s_1 + s_2)(e_{k+1} - e_k), \ldots$, on any $m-1$ jobs to $(s_1 + \ldots + s_{m-1})(e_{k+1} - e_k)$, and on all jobs to $(s_1 + \ldots + s_m)(e_{k+1} - e_k)$. What we shall do is modify the network of the prior subsection by replacing the node for $E_k$ by $m$ nodes $(i, E_k), i = 1, \ldots m$.

The capacity of each arc $(j, (i, E_k))$ will be $(s_i - s_{i+1})(e_{k+1} - e_k)$, and the capacity of each arc $((i, E_k), t)$ will be $i(s_i - s_{i+1})(e_{k+1} - e_k)$. (Here define $s_{m+1}$ to be zero.)

To validate the network construction we argue as follows. Consider the maximum amount of flow there can be from any set of $u$ job nodes to the $m$ nodes associated with interval $E_k$. We will now show that this total flow is at most $(s_1 + \ldots + s_u)(e_{k+1} - e_k)$, which is the maximum amount of processing we can do on $u$ jobs in this interval.

The maximum flow that can pass through the node $(i, E_k)$ is $\text{Min}\{u, i\}(e_{k+1} - e_k)$. Hence the total flow that can pass through all the nodes $(i, E_k)$ to $t$ is

$$
\begin{aligned}
&\phantom{+}\ (s_1 - s_2)(e_{k+1} - e_k) \\
&+\ 2(s_2 - s_3)(e_{k+1} - e_k) \\
&\phantom{+}\ \ldots \\
&+\ (u-1)(s_{u-1} - s_u)(e_{k+1} - e_k) \\
&+\ u(s_u - s_{u+1})(e_{k+1} - e_k) \\
&\phantom{+}\ \ldots \\
&+\ u(s_m)(e_{k+1} - e_k) \\
&=\ (s_1 + \ldots + s_u)(e_{k+1} - e_k),
\end{aligned}
$$

giving us the desired result.

Note that the network constructed for the $Q|pmtn, r_j, \bar{d}_j|-$ problem has $O(mn)$ nodes, hence the running time of the feasibility computation is $O(p(mn))$. However, the special structure of the flow network gives faster time bounds then for general networks of this size (this is partly explored in the exercises below).

**Exercises**

10.12. Show that the flow network for identical machines is actually a special case of the flow network for uniform machines. (That is, when $s_1 = \ldots = s_m$, and arcs of zero capacity are removed, the flow network for identical machines is obtained.)

10.13. Show that only $O(tn)$ nodes are needed for the uniform machine model when there are only t machine speeds.

10.14. Describe how to construct a network flow model for the case of 2 uniform machines, with only one node for each time interval $E_k$.

10.15. Suppose each job had a list of time intervals during which it could be processed. Describe how to modify the network flow formulation to deal with this setting.

### 10.5. Minimizing Maximum Lateness

We now apply the results of the previous section to solve the problems $P|pmtn,r_j|L_{\max}$ and $Q|pmtn,r_j|L_{\max}$. The network flow model enables us to test any given value of $L_{\max}$ for feasibility: for any given trial value $\lambda$, induce deadlines $\bar{d}_j = d_j + \lambda$ and perform a max-flow computation to test for the existence of a feasible schedule with respect to the induced deadlines. It follows that we can minimize $L_{\max}$ by finding the smallest value of $\lambda$ for which there is a feasible schedule.

Observe that the relative ordering of release dates and induced deadlines changes with $\lambda$, hence the topology of the flow network also changes with $\lambda$. There are at most $n^2$ *critical values* of $\lambda$ that are of particular concern, namely those values such that $d_j + \lambda = r_k$, for some $j$ and $k$. The node-arc structure of the flow network remains invariant for all values of $\lambda$ between two successive critical values; only the arc capacities change. Our first task is to compute the $n^2$ critical values of $\lambda$ and to carry out a bisection search over them, finding the largest *infeasible* critical value $\lambda_0$. This can be done in $O(n^2)$ time, plus the time required for $O(\log n)$ max-flow computations.

Having found $\lambda_0$, we are able to fix the topology of the flow network we shall be dealing with. The task that remains is finding the smallest increment $\delta$ such that the arc capacities induced for $\lambda_0 + \delta$ permit a flow value of P (the sum of the processing times). (In a degenerate case, it may be that $\lambda_0 + \delta$ equals the critical value of $\lambda$ next larger than $\lambda_0$, but this causes no problem.) We shall first consider how to minimize $\delta$ in the case of identical machines.

Let $E_k$, $k = 1, ..., 2n-1$, be the time intervals induced by $\lambda_0$. Each interval $E_k = [e_k, e_{k+1}]$ is of one of four types, depending upon whether $e_k$ and $e_{k+1}$ are release dates or induced deadlines; we shall refer to these four types as $[r,r]$, $[r,d]$, $[d,r]$, and $[d,d]$ intervals. When the trial value of $L_{\max}$ is increased from $\lambda_0$ to $\lambda_0 + \delta$, the length of an $[r,r]$ or $[d,d]$ interval remains unchanged, the length of an $[r,d]$ interval increases by $\delta$, and the length of a $[d,r]$ interval decreases by $\delta$. This means that when the arc capacities are expressed as a function of the parameter delta, we have the following results. Each arc $(j, E_k)$ has a capacity:

$$(e_{k+1} - e_k + \Delta) \tag{10.12}$$

and each arc $(E_k, t)$ has a capacity:

$$m(e_{k+1} - e_k + \Delta) \tag{10.13}$$

Where $\Delta = 0$ if $E_k$ is an $[r,r]$ or $[d,d]$ interval; $\Delta = \delta$, if $E_k$ is an $[r,d]$ interval; and $\Delta = -\delta$, if $E_k$ is a $[d,r]$ interval.

All arcs $(s, j)$ have capacity $p_j$.

Note that each arc $e_i$ in the flow network induced by $\lambda_0 + \delta$ has a capacity of the form $c_i + \mu_i \delta$ where $c_i$ is a constant and $\mu_i$ is a multiplier of value 0, 1, -1, or $m$.

### 10.5.1. First approach: apply Meggido's method

Because each arc capacity is a linear function of $\delta$, Meggido's method can be applied to find the minimum value of $\delta$ such that a flow of value $P$ (the sum of all $p_j$ values) can be achieved. A straightforward application of Meggido's Theorem (Chapter 2) yields a solution to $P|pmtn, r_j|L_{\max}$ in $O(p^2(n))$ time, where $p(n)$ is the running time of the max-flow algorithm chosen.

### 10.5.2. Second approach: iteration on trial values

The capacity of each $(s, t)$ cut in the flow network is also a linear function of $\delta$. Consider the capacity of a *minimum* cut. Each of the $2n - 1$ nodes $E_k$ is either on the source side or on the sink side of the cut. If $E_k$ is on the source side, arc $(E_k, t)$ contributes its capacity to the capacity of the cut as determined by (10.13). If $E_k$ is on the sink side, then at most $m$ arcs $(j, E_k)$ contribute their capacities to the capacity of the cut, where these capacities are determined by (10.12). (At most $m$ arcs $(j, E_k)$ can contribute their capacities to the capacity of a minimum cut, else the total capacity of these arcs would exceed that of the arc $(E_k, t)$.) Thus in the flow network with arc capacities induced by $\lambda_0$, each minimum cut $C'$ has a capacity as a function of $\delta$, which is $P' + \mu\delta$, where $P'$ is an integer constant less than $P$ and $\mu$ is an integer *multiplier* obtained by summing the $\mu_i$ values associated with the arcs in $C'$. Thus $\mu$ is no greater than $m(2n - 1)$. It follows that in order for the capacity of any given minimum cut $C'$ to be $P$, we must have $\delta = (P - P')/\mu$.

An iterative procedure for finding the optimal value of $\lambda$ is as follows. Find a minimum cut $C_0$, with capacity $P_0 < P$, in the flow network with arc capacities induced by $\lambda_0$. Set $\lambda_1 = \lambda_0 + (P - P_0)/\mu_0$. where $\mu_0$ is the multiplier for $C_0$. Find a minimum cut $C_1$, with capacity $P_1 > P_0$ in the network with arc capacities induced by $\lambda_1$. If $P_1 = P$, terminate. Otherwise iterate, setting $\lambda_i = \lambda_{i-1} + (P - P_{i-1})/\mu_{i-1}$, until a minimum cut $C_i$ is found with $P_i = P$.

Observe that in the network with arc capacities induced by $\lambda_i$, the capacities of all cuts with multipliers greater than or equal to $\mu_{i-1}$ are at least $P$. Hence at each iteration $i$, except possibly the last, $\mu_i > \mu_{i-1}$. Since there are at most $m(2n - 1)$ values for the multipliers, the procedure must terminate in $O(mn)$ iterations. We now have achieved a time bound of $O(mnp(n))$ for finding the optimal value of lambda.

### 10.5.3. Third approach: bisection search on $\delta$

The desired value of $\delta$ is $(P - P')/\mu$, for some cut $C'$ with capacity $P' + \mu\delta$. If the release times and due-dates are integers, $P'$ is a positive integer no greater than $P$ and $\mu$ is a positive integer no greater than $m(2n - 1)$. It follows that the optimum value of $\delta$ can be found by carrying out a bisection search over $O(mnP)$ ratios, which can be accomplished with $O(\log n + \log p_{\max})$ calls to the max-flow algorithm, yielding a time bound of $O(p(n)(\log n + \log p_{\max}))$.

### 10.5.4.  Uniform Machines

Now let us extend the above results to the case of uniform machines. Instead of the capacities (10.12), we have for each arc $(j,(i,E_k))$:

$(s_i - s_{i+1})(e_{k+1} - e_k)$, if $E_k$ is an $[r,r]$ or $[d,d]$ interval,

$(s_i - s_{i+1})(e_{k+1} - e_k + \delta)$, if $E_k$ is an $[r,d]$ interval,

$(s_i - s_{i+1})(e_{k+1} - e_k - \delta)$, if $E_k$ is a $[d,r]$ interval.

And instead of (10.13) we have for each arc $((i,E_k),t)$:

$i(s_i - s_{i+1})(e_{k+1} - e_k)$, if $E_k$ is an $[r,r]$ or $[d,d]$ interval,

$i(s_i - s_{i+1})(e_{k+1} - e_k + \delta)$, if $E_k$ is an $[r,d]$ interval,

$i(s_i - s_{1+1})(e_{k+1} - e_k - \delta)$, if $E_k$ is a $[d,r]$ interval.

Note that each arc $e_i$ in the flow network induced by $\lambda_0 + \delta$ still has a capacity of the form $c_i + \mu_i \delta$ where $c_i$ is a constant but now the multiplier $\mu_i$ is of the form $i(s_i - s_{i+1})$ where $i$ is an integer in the range $-m, \ldots, -1, 0, 1, \ldots, m$.

The capacity of each $(s,t)$ cut $C'$ is a linear function of the form $P' + \mu \delta$. For a min cut, $\mu$ is no greater than $(s_1 + s_2 + \ldots + s_m)n$ since we get at most this value by putting all of the $(i,E_k)$ nodes on the $s$ side of the cut.

A straightforward application of Meggido's method yields a running time of $O(p^2(mn))$. In order to apply the other two methods, we must assume that the machine speeds $s_i$ are integers. The iterative method then has a running time of $O((s_1 + \ldots + s_m)np(mn))$, and the bisection search method has a running time of $O(p(mn)(\log n + \log s_1 + \log p_{\max})$.

### Exercises

10.16.  Extend the ideas of this section to solve the *weighted* maximum lateness problems $P|pmtn,r_j|wL_{\max}$ and $Q|pmtn,r_j|wL_{\max}$. (Note: Each trial value of $\lambda$ in this setting induces deadlines $\overline{d}_j = d_j + \lambda/w_j$.)

10.17.  The bound of $m(2n-1)$ on $\mu$ for a min-cut in the indentical machine setting can be improved. Show that $\mu < mn$.

### 10.6.  Memory Constraints

We now consider a variation of $Q|pmtn|C_{\max}$ where we have the additional constraint that each processor $i$ has a *memory capacity* $c_i$, and each job $j$ has a *memory requirement* $q_j$. A job $j$ can be executed on machine $i$ if and only if

$$c_i \geq q_i \qquad (10.14)$$

This is a special case of $R|pmtn|C_{\max}$ which will be discussed in the next section.

Our approach will be to first describe a method for machines of identical speeds, then generalize this to a feasibility testing algorithm for uniform machines. Finally, we use search techniques to find the minimum possible completion time.

We assume that the $c_i$ values have been sorted so that $c_1 \geq \ldots \geq c_m$. We can then partition the jobs into sets $G_i$ where

$G_i = \{J_j \,|\, c_i \geq q_j > c_{i+1}\} \quad 1 \leq i \leq m-1$, and
$G_m = \{J_j \,|\, c_m \geq q_j\}$.

Thus $G_i$ is the set of all jobs which can be run on processor $i$ but not on processor $i+1$. We also define $F_i$ to be the set of all jobs which must be run on machines one to $i$:

$F_i = \cup_{j=1}^{i} G_j$.

We will also define $X_i$ to be the total processing of all jobs in $F_i$.

### 10.6.1.  Identical Machines

Kafura and Shen proved that when all machines have the same speed (thus speed one) the length of an optimal schedule is:

$$max\{max\{(1 \leq i \leq m) \ X_i/i\}, p_{\max}\} \tag{10.15}$$

The algorithm is quite simple: we form the sets $G_i$ and $F_i$ and use this to compute the value of $C_{\max}$ using (10.15). We then schedule the jobs in each set $G_i$ using McNaughton's rule.

We can form the $G_i$ sets on $O(n \log m)$ time and all other steps take $O(n)$ time, so the total time is $O(n \log m)$. Just as in McNaughton's algorithm there are at most $m-1$ preemptions.

### 10.6.2.  Uniform Machines

For this setting we know of no closed form expression for $C_{\max}$. Thus we start by considering the case where all jobs have a common deadline $D$ and try to find a schedule which completes all jobs by time $D$. At a high level our algorithm has the same structure as for identical machines: form the sets $G_i$ and then schedule the jobs in the sets $G_1, G_2, \ldots G_m$. The main difference is that each set is now scheduled using the Gonzalez-Sahni algorithm of Section 10.1.1. We will also have to be more careful about adding new processors.

Intuitively our goal is to schedule each set $G_i$ while leaving as much time on the fast processors as possible for future jobs. Fortunately that is exactly what the Gonzalez-Sahni algorithm does. Recall also that the Gonzalez-Sahni algorithm is set up to work on a composite processor system where each processor has a time varying speed.

Recall that our composite processors consist of *elementary processor intervals* which are just time intervals on our original processors. For example, if $D = 4$ composite processor one might have speed 5 in $[0, 2]$, speed 10 in $(2, 3]$ and speed 20 in $(3, 4]$. Note that the speeds increase as time increases. As we schedule jobs and add processors we will maintain the following properties: i) that within each composite processor speeds are nondecreasing as time increases; ii) no speed in processor $i$ is greater than the slowest speed in processor $i-1$. We can view this as lining up the composite processors from left to right starting with the last one on the left. As we go from left to right, the speeds of the elementary processor intervals are nondecreasing. We can view this as a sort of super composite processor of length $mD$ (with the early part possibly of speed zero). Let $R(T)$ denote the total processing capacity of this super processor using the last $T$ time units (e.g. $R(D)$ is the capacity of composite processor one, and $R(1.5\,D)$ is the total capacity of processor one and the fastest $D/2$ time units of processor two).

We now consider how to update a composite processor system. After scheduling a set of jobs $G_{i-1}$ we are left with a composite processor system $CP$ which has the remaining processing capacity of processors $1, 2, ..., i-1$. Before scheduling set $G_i$ we need to add processor $i$ to $CP$ to create a new composite processor system $CP'$. To add processor $i$, we splice it into CP as follows. First find the largest speed $s_r$ in CP such that $s_r < s_i$ (note $s_r$ could be zero). Let $k$ be the smallest index of a composite processor which has an elementary processor interval of speed $s_r$ and let $[t_1, t_2]$ be the last elementary processor interval of speed $s_r$ on composite processor $k$. We change composite processor $k$ to have speed $s_i$ in the interval $[0, t_2]$, and create a new composite processor $(k+1)$ which has speed $s_1$ in the interval $(t_2, D]$ and inherits the elementary processor intervals which were on machine $k$ in the interval $[0, t_2]$. The speeds of the other composite processors are unchanged, but those with index greater than $k$ all increase their index by 1. This maintains the property that speeds increase as we go later in time or to earlier indexed processors.

Let $CP_i$ be the composite processor system after our algorithm schedules the jobs in $G_i$. We define $R_i(T)$ as the total processing capacity using the fastest $T$ time units of processing in $CP_i$. The correctness of our feasibility algorithm follows from these facts:

i) Consider any alternate schedule $S$ for the jobs in $F_i$; let $R'(T)$ represent the total processing capacity of the $T$ fastest units of time which remains idle on processors $1, 2, ..., i$ in $S$. Then $R_i(T) \geq R'(T)$ for $0 \leq T \leq mD$.

ii) After splicing in processor $i+1$ as described above to get a new composite processor system, the $R$ function for this also dominates any alternate schedule's remaining processing capacity.

These facts are proved by showing that if these properties hold before a set of jobs is scheduled or a processor inserted, they must also hold afterwards.

**Theorem 10.1** [ Feasibility Algorithm ]. *The feasibility algorithm will schedule all jobs by time $D$ whenever such a schedule is possible.*

*Proof.* We will be able to schedule each set $G_i$ as long as $CP_i$ satisfies the inequalities of (10.4) for the Gonzalez-Sahni algorithm. If we ever reach a point where $G_i$ cannot be scheduled, the fact that each of the sums $S_1, S_1 + S_2, ..., S_1 + ... S_m$ is as large as possible (by facts i) and ii) above), shows that no schedule can complete all the jobs in $F_i$.

We now analyze the complexity of our algorithm. The sets $G_i$ can be constructed easily in $O(n \log m)$ time. To analyze the time for the Gonzalez-Sahni algorithm, note that each time we splice in a new processor we add at most two elementary processor intervals to our composite processor system. Since scheduling a job never creates additional elementary processor intervals, the total number of elementary processor intervals is at most $2m$. Each job can be scheduled using $O(\log m)$ time to find the correct composite processor(s) plus $O(l)$ time, where $l$ is the number of elementary processor intervals used up by this job. Thus the total time to schedule all jobs is $O(n \log m)$. The only remaining issue is adding a new processor. We can easily add a new processor in $O(m)$ time by simply scanning the doubly-linked list of elementary processor intervals which represents the appropriate composite processor. A more complex appraoch stores the speeds of the elementary processor intervals in a balanced binary search tree. This allows us to insert a new processor in $O(log m)$ time. Thus the overall running time is $O(n \log m)$.

The feasibility algorithm lets us test any common deadline $D$. To find $C_{max}$, the smallest feasible $D$, we can use Meggido's method using our feasibility routine as a subroutine. A simple application would result in a running time of $O(n^2 \log^2 m)$.

**Exercises**

10.18. Our description of the feasibility algorithm implicitly assumed that the memory sizes $c_i$ were all distinct. Describe how to modify the algorithm if there are ties among the $c_i$ values. In particular, describe how new processors are now inserted.

10.19. Describe the data structures needed to insert a new processor in $O(\log m)$ time. Also describe how to use and maintain these data structures when you schedule a new job.

10.20. Show that the feasibility algorithm introduces at most $3m - 3$ preemptions. Also show that this bound is tight by giving a family of problems which require $3m - 3$ preemptions to be scheduled.

## 10.7. Linear Programming Formulations for Unrelated Machines

Let $x_{ij}$ denote the total amount of time that machine $i$ processes job $j$. We shall now formulate some constraints that must be satisfied by any feasible schedule of length $C_{max}$.

minimize $C_{max}$
subject to

$$\sum_{j=1}^{n} x_{ij} \leq C_{max}, \quad i = 1, \ldots m, \tag{10.16}$$

$$\sum_{i=1}^{m} x_{ij} \leq C_{max}, \quad j = 1, \ldots n, \tag{10.17}$$

$$\sum_{i=1}^{m} (x_{ij}/p_{ij}) = 1, \quad j = 1, \ldots n, \tag{10.18}$$

and
$x_{ij} \geq 0$, for all $i = 1, \ldots m, j = 1, \ldots n$.

In this linear programming problem constraints (10.16) ensure that the total amount of processing done by any given machine does not exceed the time available, constraints (10.17) ensure that the total amount of processing done on any given job does not exceed the time available, and constraints (10.18) ensure that the processing done on any given job must be sufficient to complete it. Meeting these constraints is clearly necessary for any feasible schedule. But the converse is far from obvious. It is by no means clear that a feasible solution to this linear programming problem can always be transformed into a feasible schedule with the same value of $C_{\max}$.

To address this sufficiency question consider a matrix $X$ of $x_{ij}$ values which satisfy the LP constraints above. This matrix $X$ has exactly the same form as the constraints of an open shop problem: for each job $j$, the $x_{ij}$ values state the total amount of processing to be done for job $j$ on machine $i$. Thus we can achieve a legal schedule using the solution of the LP exactly when the preemptive open shop problem described by $X$ is feasible.

One classical theorem about open shop shows that any constraint matrix $X$ that satisfies (10.16) and (10.17) can be scheduled to complete all jobs by time $C_{\max}$. We will briefly describe that result below.

The problem $R|pmtn|L_{\max}$ can be solved by an elaboration of the above linear programming formulation. The problem $R|pmtn, r_j|L_{\max}$ requires calling on a linear programming computation as a subroutine, in essentially the same way as the network flow computation was called on in the previous section.

### 10.7.1. Minimizing makespan in preemptive open shops

The *open shop* scheduling environment will be the subject of Chapter 12. That chapter will be devoted to the non-preemptive setting, but here we will outline a simple result to find the minimum makespan preemptive schedule; that is, $O|ptmn|C_{\max}$, is solvable in polynomial time. The input to an open shop instance consists of the processing time $p_{ij}$ that job $j$ requires for its operation on machine $i$, for each $i = 1, \ldots, m, j = 1, \ldots, n$. At each point in time, each machine can process at most one operation, and each job can have at most one of its operations being processed by any machine. Let

$$C = \max\{\max_j \textstyle\sum_i p_{ij}, \max_i \textstyle\sum_j p_{ij}\}.$$

Call row $i$ *tight* if $\sum_j p_{ij} = C$, and *slack* otherwise. Similarly, column $j$ is *tight* if $\sum_i p_{ij} = C$, and is *slack* otherwise. We will give an algorithm that constructs a feasible schedule for which $C_{\max} = C$; hence, this schedule is optimal.

**Theorem 10.2.** *For any input $P = (p_{ij})$ to the open shop scheduling problem*

*O|ptmn|C*$_{\max}$*, there is an optimal solution of with makespan,*

$$C = \max\{\max_j \textstyle\sum_i p_{ij}, \max_i \textstyle\sum_j p_{ij}\},$$

*and this can be found in polynomial time*

Suppose that we can find a subset *S* of strictly positive elements of *P*, with exactly one element of *S* in each tight row and in each tight column, and at most one element of *S* in each slack row and in each slack column. We say that such a set *S* is a *decrementing set*, and use it to construct a *partial schedule* of length δ, for some δ > 0.

By construction, for this partial schedule, we can process all of the operations in *S* concurrently. We first require that δ ≤ $p_{ij}$ for each element $p_{ij} \in S$; this ensures that we have sufficient work remaining for each operation in *S* to process it throughout the full length δ of the partial schedule. However, there are rows and columns that do not have an element in *S* (and hence must be slack). For each slack row *i* (column *j*) of *P*, the total work remaining for that machine (job) is less than the maximum *C* computed; however, no work is being done on that machine (job) in this partial schedule. Hence, the larger we set δ for that partial schedule, the closer the slack total work remaining becomes to matching the remaining work in the tight rows and columns.

We want to ensure that the tight rows and columns remain tight. Hence, if row *i* has no element in *S*, we need that $C - \delta \geq \sum_j p_{ij}$ (or equivalently, that $\delta \leq C - \sum_j p_{ij}$); similarly, if column *j* has no element in *S*, we need $\delta \leq C - \sum_i p_{ij}$. Hence, we set δ to the minimum of these values (i.e., $p_{ij}$ for elements in *S*, and the corresponding differences for each row or column without an element in *S*).

We now have constructed a partial schedule of length δ, where the maximum of row and column sums for the remaining processing times decreases from *C* to *C* − δ, and yet there are either more elements of value 0, or else more tight rows or columns. We can iteratively build a schedule of length *C* for the input *P* by repeating this construction iteratively until the remaining processing times are all equal to 0. Since at least one row or column becomes tight or one element becomes 0, this must occur after *mn* + *m* + *n* iterations.

We must still show that such a decrementing set always exists. We have that

$$\sum_j p_{ij} \;=\; C, \quad \text{for each tight row } i;$$

$$\sum_i p_{ij} \;=\; C, \quad \text{for each tight column } j;$$

$$\sum_j p_{ij} \;\leq\; C, \quad \text{for each slack row } i;$$

$$\sum_i p_{ij} \;\leq\; C, \quad \text{for each slack column } j.$$

If we let $x_{ij} = p_{ij}/C$, then this is equivalent to:

$$\sum_j x_{ij} = 1, \quad \text{for each tight row } i;$$

$$\sum_i x_{ij} = 1, \quad \text{for each tight column } j;$$

$$\sum_j x_{ij} \leq 1, \quad \text{for each slack row } i;$$

$$\sum_i x_{ij} \leq 1, \quad \text{for each slack column } j.$$

Of course, we also know that $0 \leq x_{ij} \leq 1$, for each $i = 1, \ldots, m$, $j = 1, \ldots, n$. If we instead view $x$ as a variable, what we have argued is that $p/C$ is a feasible solution to this system of linear constraints. This system defines a bounded, non-empty polytope, and hence there must be an extreme point solution $x^*$. But this is nothing more than the assignment polytope, and we know that all of its extreme points are integer. If we let $S$ be the set of elements $p_{ij}$ corresponding to those $x_{ij}^* = 1$, we have obtained a decrementing set.

### Exercises

10.21.  Suppose "a little birdie" told you the optimal ordering of the completion times for an instance of the problem $R|pmtn|\Sigma\, C_j$. Formulate and solve as a linear programming problem.

### Notes

10.1. *Minimizing Makespan.* The algorithms of this section are adapted from Mc-Naughton [1959] and Gonzalez & Sahni [1978B].

Horvath, Lam & Sethi [1977] proved that the bound (10.2) can be met by a pre-emptive variant of the *LPT* rule, which, at each point in time, assigns the jobs with the largest remaining processing requirement to the fastest available machines. The algorithm runs in $O(mn^2)$ time and creates an optimal schedule with no more than $(m-1)n^2$ preemptions. In Chapter 15 we show that the same variant of the *LPT* rule solves a stochastic version of $Q|pmtn|C_{\max}$.

Exercise 10.3 concerning constraints on machine availability, is suggested by a problem formulation of Schmidt [1983]. Exercise 10.5 is from Rayward-Smith [1987B].

10.2. *Meeting Deadlines.* The algorithm of this section is adapted from [Sahni & Cho. 1980]. The Sahni-Cho algorithm is also used to solve $Q|pmtn|\sum w_j U_j$ [Lawler, 1979A]. See also Lawler & Martel [1989].

10.3. *The Staircase Algorithm for Release Dates.* Horn [1974] gives an $O(n^2)$ procedure for $P|pmtn|L_{\max}$ and $P|pmtn, r_j|C_{\max}$. The staircase algorithm for uniform machines was developed independently by Sahni & Cho [1979B] and Labetoulle, Lawler, Lenstra & Rinnooy Kan [1979]. The more efficient implementation for

$P|pmtn, r_j|C_{\max}$ is due to Sahni [ ]. Cf also Gonzalez & Johnson [1980].

**10.4.** *Network Flow Computations for Release Dates and Deadlines.* Horn [1974] showed that the feasibility problem $P|pmtn, r_j, \bar{d}_j|-$ can be formulated as a network flow problem. Bruno & Gonzalez [1976] noted that $Q2|pmtn, r_j, \bar{d}_j|-$ can be given a similar network flow formulation. Martel [1982] showed that $Q|pmtn, r_j, \bar{d}_j|-$ can be formulated as a special type of *polymatroidal* network flow problem, as described more generally by Lawler & Martel [1982]. Federgruen & Groenevelt [1986] showed that the problem can be reformulated as an ordinary network flow problem.

**10.5.** *Minimizing Maximum Lateness.* The algorithms of this section are adapted from Labetoulle, Lawler, Lenstra & Rinnooy Kan [1979]. Related ideas are discussed in Martel [1982] and Federgruen & Groenevelt [1986].

**10.6.** *Memory Constraints.* The algorithms of this section are adapted from Kafura & Shen [1977] and Martel [1985]. The second paper describes the feasibility algorithm and also shows that $C_{\max}$ can be found in $O(mn\log^2 m)$ time by looking more carefully at when the feasibility algorithm needed to be called using Megiddo's approach. Lai & Sahni [1984] showed how to minimize $L_{\max}$ for identical speed processors with memory constraints, and also considered settings with release times and due-dates [1983].

**10.7.** *Linear Programming Formulations for Unrelated Machines.* The linear programming formulations presented in this section are due to Lawler & Labetoulle [1978]. For fixed $m$, it seems possible that the linear program for $R|pmtn|C_{\max}$ can be solved in $O(n)$ time. Certainly this is true in the case of $m = 2$ [Gonzalez, Lawler & Sahni, 1990]. The algorithm for the preemptive scheduling of open shops is due to Gonzalez & Sahni [1976].

# Contents

# 12

# Open Shops

Gerhard J. Woeginger
*RWTH Aachen*

There are four blacksmiths working together. One of them has specialized in putting horseshoes on the left front leg of a horse, while the other three have specialized respectively in putting horseshoes on the left hind leg, the right front leg, and the right hind leg. If the work on one horseshoe takes five minutes, what is the minimum amount of time needed to put twenty-eight horseshoes on seven horses? (Note that a horse cannot stand on two legs.)

As each blacksmith has to handle 7 horseshoes, he needs at least 35 minutes of working time. The following picture with horses $A, B, C, D, E, F, G$ and blacksmiths $1, 2, 3, 4$ shows a schedule that meets this lower bound of 35 minutes. Note that each horse receives its four horseshoes during four different time slots (so that it never has to stand on two legs), and note that during each five minute time slot each blacksmith works for exactly five non-interrupted minutes on a single horse.

|  | minute 00–05 | minute 05–10 | minute 10–15 | minute 15–20 | minute 20–25 | minute 25-30 | minute 30-35 |
|---|---|---|---|---|---|---|---|
| Blacksmith 1 | A | B | C | D | E | F | G |
| Blacksmith 2 | B | C | D | G | F | E | A |
| Blacksmith 3 | C | D | G | E | A | B | F |
| Blacksmith 4 | D | A | F | B | C | G | E |

## 12.1. Problem statement and some definitions

An instance of the *open shop* scheduling problem consists of $m$ machines $M_1, \ldots, M_m$ and $n$ jobs $J_1, \ldots, J_n$. (Throughout, machines will be indexed by $i$ and jobs will be indexed by $j$.) Each job $J_j$ consists of $m$ independent operations $O_{i,j}$ with $i = 1, \ldots, m$. The operation $O_{i,j}$ of job $J_j$ has to be processed on machine $M_i$, which takes $p_{i,j}$ uninterrupted time units. For every job, the order in which its operations have to be processed is *not fixed* in advance but may be chosen arbitrarily by the scheduler; we stress that different jobs may receive different processing orders.

A *schedule* assigns every operation $O_{i,j}$ to a time interval of length $p_{i,j}$, so that no job is simultaneously processed on two different machines and so that no machine simultaneously processes two different jobs. The *makespan* $C_{\max}$ of a schedule is the largest job completion time. The optimal makespan is usually denoted by $C^*_{\max}$. This optimization problem is denoted by $O||C_{\max}$ (if the number $m$ of machines is given as part of the input) and by $Om||C_{\max}$ (if the number $m$ of machines is a fixed constant number).

In the blacksmiths and horseshoes puzzle, the four blacksmiths are four machines $M_1, M_2, M_3, M_4$. Each horse forms a job, and its four legs are the four operations of that job. All operations $O_{i,j}$ have length $p_{i,j} = 5$.

Here are some more notations. The length of the longest operation in an instance is denoted $o_{\max} = \max_{i,j} p_{i,j}$. The overall processing time of job $J_j$ will be denoted $p_j = \sum_{i=1}^{m} p_{i,j}$. The overall processing time assigned to machine $M_i$ is called the *load* $L_i = \sum_{j=1}^{n} p_{i,j}$ of the machine. The maximum job processing time is denoted $p_{\max} = \max_j p_j$ and the maximum machine load is denoted $L_{\max} = \max_i L_j$. As no job can be simultaneously processed on two different machines the makespan of any schedule satisfies $C_{\max} \geq p_{\max}$, and as no machine can simultaneously processes two different jobs the makespan satisfies $C_{\max} \geq L_{\max}$. This yields the following lower bound, which will be important throughout the chapter:

$$C^*_{\max} \geq \beta := \max\{L_{\max}, p_{\max}\} \tag{12.1}$$

We mention in passing that there are two other important shop scheduling problems that are closely related to the open shop problem: In a *flow shop*, every job must pass the machines in the same ordering $M_1, \ldots, M_m$. In a *job shop*, the ordering of the operations is fixed a priori for every job, and different jobs may have different orderings of operations. These variants will not be further discussed in the rest of this chapter.

## 12.2. Computational complexity: Fixed number of machines

Gonzalez & Sahni [1976] prove that the open shop on $m = 2$ machines allows a very simple polynomial time solution: There always exists a schedule whose makespan equals the lower bound $\beta$ in (12.1), and this schedule can be found in linear time.

**Theorem 12.1** (Gonzalez & Sahni, 1976)**.** *Problem $O2||C_{\max}$ is solvable in polynomial time.*

The algorithm in Theorem 12.1 is not hard to find (there are many possible approaches), and we leave it as a puzzle for the reader. A more general problem variant considers the completion time $C_1$ of the last operation on machine $M_1$ and the completion time $C_2$ of the last operation on $M_2$, and asks for a schedule that minimizes some objective function $f(C_1, C_2)$ of the two machine completion times. Based on extensive case distinctions, Shaklevich & Strusevich [1993] develop linear time algorithms for this variant, if the function $f(\cdot, \cdot)$ is non-decreasing in both arguments. Van den Akker, Hoogeveen & Woeginger [2003] provide a simpler proof of the same result. Sahni & Cho [1979A] prove strong NP-hardness of the no-wait problem. $O2|\text{no-wait}|C_{\max}$ in which the processing of the second operation of each job must start immediately after the completion of its first operation.

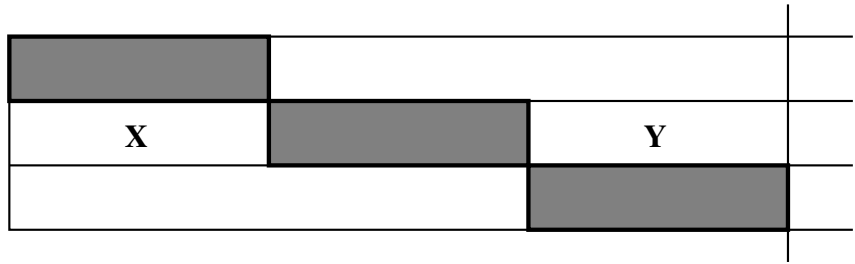

**Figure 12.1.** Illustration of the NP-hardness argument in Theorem 12.2.

Now let us turn to the cases of $Om||C_{\max}$ with $m \geq 3$ machines. As usual, the complexity jumps from easy to hard when we move from parameter 2 to parameter 3:

**Theorem 12.2** (Gonzalez & Sahni, 1976). *For every fixed $m \geq 3$, problem $Om||C_{\max}$ is NP-hard.*

*Proof.* We only show hardness for $m = 3$. The proof is a polynomial time reduction from the NP-hard PARTITION problem [Garey & Johnson, 1979]: *"Given $k$ positive integers $q_1, \ldots, q_k$ with $\sum_{i=1}^{k} q_i = 2Q$, does there exist an index set $I \subseteq \{1, \ldots, k\}$ with $\sum_{i \in I} q_i = Q$?"*

For $j = 1, \ldots, k$ we create a job $J_j$ with $p_{1,j} = p_{2,j} = p_{3,j} = q_j$. Furthermore, there is a job $J_{k+1}$ with $p_{1,k+1} = p_{2,k+1} = p_{3,k+1} = Q$. We claim that the PARTITION instance has answer YES, if and only if the constructed instance of $O3||C_{\max}$ has a schedule with makespan at most $3Q$. The (only if) part is straightforward. For the (if part), consider the three operations of job $J_{k+1}$ in a schedule with makespan $3Q$. By symmetry, we may assume that $J_{k+1}$ is first processed on machine $M_1$, then on $M_2$, and finally on $M_3$. Then the second operation generates two time intervals $X$ and $Y$ of length $Q$ on machine $M_2$; see Figure 12.1 for an illustration. The operations $O_{2,j}$ of the other jobs must be packed into intervals $X$ and $Y$, and thereby yield a solution for the PARTITION instance. □

As the PARTITION problem is NP-hard in the weak sense, the argument in Theorem 12.2 only yields NP-hardness in the weak sense for $Om||C_{\max}$. The precise com-plexity (strong NP-hardness versus pseudo-polynomial time solvability) of problem $Om||C_{\max}$ is unknown. This complexity question has been open since the 1970s, and it forms the biggest and most important gap in our understanding of open shop scheduling.

**Open problem 12.1.** *Prove that for every fixed number $m \geq 3$ of machines, problem $Om||C_{\max}$ is solvable in pseudo-polynomial time.*

### 12.3. Computational complexity: Arbitrary number of machines

Theorem 12.3 below formulates the main result on the complexity of problem $O||C_{\max}$, where the number of machines is specified as part of the input. Two fairly different proofs are known for this theorem. The first proof has been given by Lenstra [-] in the 1970s, and this proof might be one of the most-cited unpublished results in the scheduling literature. The second proof is implicit in the work of Williamson et al. [1997], who prove that $O||C_{\max}$ is NP-hard, even if all operations are of length $0, 1, 2$ and if the question is to decide whether there is a schedule with makespan 4.

**Theorem 12.3.** *Problem $O||C_{\max}$ is NP-hard in the strong sense.*

*Proof.* We present the original proof of Lenstra [-] and thereby save it for future generations. The proof is a polynomial time reduction from the 3-PARTITION problem [Garey & Johnson, 1979], which is NP-complete in the strong sense: *"Given an index set $I = \{1, \ldots, 3k\}$ and positive integers $q_1, \ldots, q_{3k}$ and $Q$ with $Q/4 < q_i < Q/2$ $(i \in I)$ and $\sum_{i \in I} q_i = kQ$, does there exist a partition $\{I_1, \ldots, I_k\}$ of $I$, so that $\sum_{i \in I_j} q_i = Q$ for $j = 1, \ldots, k$?"* (Note that each part $I_j$ must contain exactly three elements from $I$.)

Given an instance of 3-PARTITION, we define an instance of $O||C_{\max}$ with $2k-1$ machines $M_1, \ldots, M_{2k-1}$ and $6k-3$ jobs. For each $i \in I$, there is a job $J_i$ with processing time $q_i$ on $M_1$. For each $j = 1, \ldots, k-1$, there are three jobs:

- a red job $R_j$ with processing times 1 on $M_1$, $jQ + j - 1$ on $M_{2j}$, and $(k-j)Q + k - j + 1$ on $M_{2j+1}$;

- a green job $G_{2j}$ with processing time $(k-j)Q + k - j$ on $M_{2j}$;

- a green job $G_{2j+1}$ with processing time $jQ + j$ on $M_{2j+1}$.

Note that a job has operations of positive length on one or three machines; all undefined operations have zero length and can be ignored. We claim that the answer to the 3-PARTITION instance is YES, if and only if there exists a schedule of length at most $\beta = kQ + k - 1$; note that this value $\beta$ coincides with the bound stated in (12.1).

Suppose that we have a YES-instance of 3-PARTITION. We then construct a schedule of length $\beta$ by putting the jigsaw puzzle together as shown in Figure 12.2. For $j = 1, \ldots, k-1$:

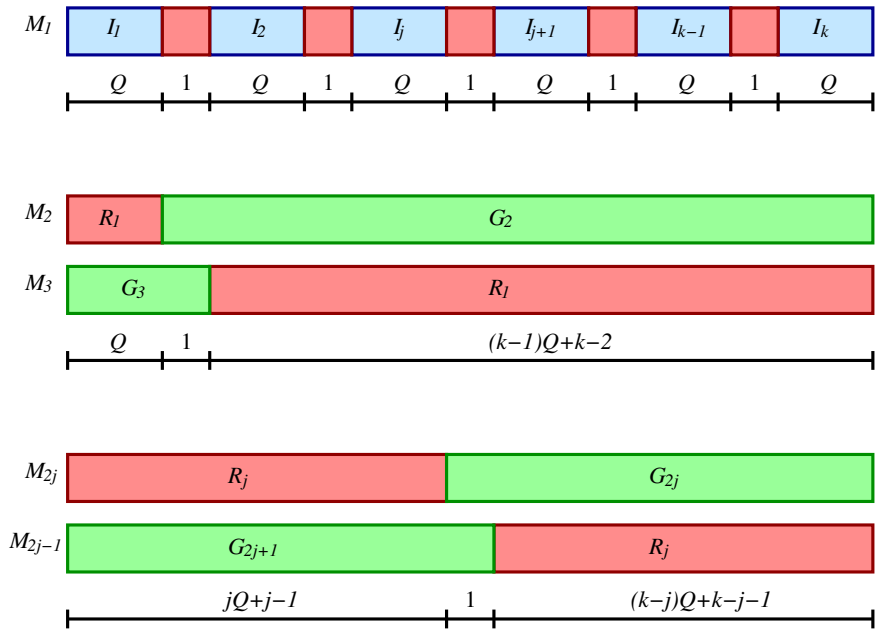

**Figure 12.2.** **An illustration for the schedule in the strong NP-hardness proof in Theorem 12.3.**

- job $R_j$ starts at 0 on $M_{2j}$, then visits $M_1$, and completes at $\beta$ on $M_{2j+1}$;

- job $G_{2j}$ completes at $\beta$ on $M_{2j}$;

- job $G_{2j+1}$ starts at 0 on $M_{2j+1}$.

The jobs $J_i$ ($i \in I$) are processed in the $k$ intervals of length $Q$ on $M_1$, according to the partition of $I$.

Now suppose that we have a schedule of length at most $\beta$. A simple calculation shows that the sum of all processing times is equal to $(2k-1)\beta$, which is the total machine capacity between 0 and time $\beta$. Hence, the schedule has length $\beta$ and no idle time.

On machine $M_2$, either job $R_1$ precedes job $G_2$ or $R_1$ follows $G_2$. If $R_1$ precedes $G_2$ on $M_2$, then $R_1$ must follow $G_3$ on $M_3$, and $R_1$ is processed on $M_1$ in the interval $(Q, Q+1)$. On the other hand, if $R_1$ follows $G_2$ on $M_2$, then $R_1$ must precede $G_3$ on $M_3$, and $R_1$ is processed on $M_1$ in the interval $(\beta - Q - 1, \beta - Q)$. It follows that in our schedule jobs $R_1$ and $R_{k-1}$ occupy the intervals $(Q, Q+1)$ and $(\beta - Q - 1, \beta - Q)$ on $M_1$.

Applying this argument to each pair $(R_j, R_{k-j})$ for $1 \le j < k/2$, and to $R_{k/2}$ in case $k$ is even, yields the conclusion that $M_1$ processes red jobs in the intervals $(j(Q+1) - 1, j(Q+1))$ for $j = 1, \dots, k-1$. The jobs $J_i$ ($i \in I$) must be packed into $k$ intervals of length $Q$ on $M_1$, and thereby yield a solution to the 3-PARTITION instance. $\square$

### 12.4.   A theorem on vector rearrangements

This section introduces an auxiliary problem and an auxiliary result that will be pivotal for the next section. Let $B \subset \mathbb{R}^d$ be the unit ball of a norm $\| \cdot \|$ on $\mathbb{R}^d$, that is, a $d$-dimensional closed convex body that is centrally symmetric about the origin. Suppose we are given $n$ vectors $v_1, \dots, v_n \in \mathbb{R}^d$ that satisfy

$$\sum_{i=1}^{n} v_i = 0 \qquad \text{and} \qquad \|v_i\| \le 1 \text{ for } 1 \le i \le n. \tag{12.2}$$

The goal is to find a permutation $v_{\pi(1)}, \dots, v_{\pi(n)}$ of these vectors, so that for every $k$ with $1 \le k \le n$ the norm $\|v_{\pi(1)} + v_{\pi(2)} + \dots + v_{\pi(k)}\|$ of the partial sum is small. Steinitz [1913] proved that the norms of these partial sums can be bounded by a constant that only depends on the unit ball $B$ (and Steinitz showed that the concrete constant $2d$ always works). The smallest such constant is called the *Steinitz constant* $c(B)$ of the norm.

**Theorem 12.4** (Grinberg & Sevastianov, 1980). *For any norm with unit ball $B \subset \mathbb{R}^d$, the Steinitz constant satisfies $c(B) \le d$.*

The proof of Theorem 12.4 by Grinberg & Sevastianov [1980] is an optimized and streamlined version of an earlier proof by Sevastianov [1978]. It is extremely elegant, and we will sketch it now. Hence, let us consider vectors $v_1, \ldots, v_n \in \mathbb{R}^d$ that satisfy (12.2). In a first step, we prove that there exists a system of subsets $V_d, V_{d+1}, \ldots, V_n$ of $\{v_1, \ldots, v_n\}$ that satisfies the following properties.

- $V_d \subseteq V_{d+1} \subseteq \cdots \subseteq V_n = \{v_1, \ldots, v_n\}$

- $|V_k| = k$ for $1 \leq k \leq n$

- There exist real numbers $\lambda_k(v) \in [0, 1]$ for $d \leq k \leq n$ and $v \in V_k$ with
  (A) $\sum_{v \in V_k} \lambda_k(v) = k - d$ for $d \leq k \leq n$, and
  (B) $\sum_{v \in V_k} \lambda_k(v) \cdot v = 0$ for $d \leq k \leq n$.

In other words, the coefficients $\lambda_k(v)$ constitute a linear dependency on $V_k$ where all coefficients add up to $k - d$. The subsets $V_k$ and the real numbers $\lambda_k(v)$ are constructed by a backward induction. For $k = n$, we have $V_n = \{v_1, \ldots, v_n\}$ and we define $\lambda_n(v) \equiv (n-d)/n$ for all $v$. These values satisfy condition (A) by definition, while condition (B) follows from (12.2).

Now assume that the sets $V_{k+1}, \ldots, V_n$ have already been defined together with the corresponding coefficients $\lambda_{k+1}(v), \ldots, \lambda_n(v)$. We consider the following system of linear constraints on $k + 1$ real variables $x(v)$ for $v \in V_{k+1}$.

$$\sum_{v \in V_{k+1}} x(v) = k - d \tag{12.3}$$

$$\sum_{v \in V_{k+1}} x(v) \cdot v = 0 \tag{12.4}$$

$$0 \leq x(v) \leq 1 \qquad \text{for } v \in V_{k+1} \tag{12.5}$$

Note that the system (12.3)–(12.5) is solvable, as can be seen for instance by setting

$$x(v) = \frac{k-d}{k+1-d} \lambda_{k+1}(v) \qquad \text{for } v \in V_{k+1}.$$

Hence the underlying $(k+1)$-dimensional polytope is non-empty. Any extreme point $x^*$ of this polytope must satisfy $k + 1$ of the linear constraints with equality. As constraint (12.3) yields one such equality and as constraint (12.4) yields $d$ such equalities (one per dimension), in an extreme point at least $k + 1 - (d + 1) = k - d$ of the inequalities in (12.5) must be tight. Because of (12.3), this implies that in an extreme point $x^*$ at least one of the variables $x^*(v)$ will be equal to zero. We construct the set $V_k$ by dropping the corresponding vector $v$ from $V_{k+1}$ and by setting $\lambda_k(v) = x^*(v)$. This completes the construction of the subset system.

In the second step, we transform the subset system into the desired permutation. The first $d$ vectors $v_{\pi(1)}, \ldots, v_{\pi(d)}$ are an arbitrary ordering of the vectors in set $V_d$. For $k \geq d + 1$, we choose vector $v_{\pi(k)}$ as the unique element of $V_k - V_{k-1}$. We claim that in the resulting permutation, the norm of every partial sum $\sum_{i=1}^{k} v_{\pi(i)}$ is at most $d$. Indeed, for $k \leq d$ the triangle inequality together with $\|v_i\| \leq 1$ implies $\|\sum_{i=1}^{k} v_{\pi(i)}\| \leq \sum_{i=1}^{k} \|v_{\pi(i)}\| \leq d$. For $d + 1 \leq k \leq n$, the claim follows from the following chain of equations and inequalities, which is based on properties (A) and (B) and on the triangle inequality.

$$
\begin{aligned}
\| \sum_{i=1}^{k} v_{\pi(i)} \| \;=\; & \| \sum_{v \in V_k} v \| \;=\; \| \sum_{v \in V_k} v - \sum_{v \in V_k} \lambda_k(v) \cdot v \| \\
\leq\; & \sum_{v \in V_k} (1 - \lambda_k(v)) \cdot \| v \| \;\leq\; \sum_{v \in V_k} (1 - \lambda_k(v)) \\
=\; & |V_k| - \sum_{v \in V_k} \lambda_k(v) \;=\; k - (k - d) \;=\; d
\end{aligned}
$$

This completes the proof of Theorem 12.4. Note that the constructed permutation does not depend on the underlying norm. Note furthermore that the entire construction can easily be implemented in polynomial time.

Banaszczyk [1987] slightly strengthened the bound in Theorem 12.4 on the Steinitz constants for norms in $d$-dimensional space to $c(B) \leq d - 1 + 1/d$. Bergström [1931] showed that the Steinitz constant of the Euclidean plane (2-dimensional space with Euclidean norm) equals $\sqrt{5}/2 \approx 1.118$. It is known (and easy to see) that the Steinitz constant of the $d$-dimensional Euclidean space is at least $\sqrt{d+3}/2$, and this might well be the correct value of the $d$-dimensional Euclidean Steinitz constant. However, at the current moment not even a sub-linear upper bound is known and the problem is wide open (even for $d = 3$).

### 12.5.  A tractable special case

Recall that $o_{\max}$ denotes the length of the longest operation, that $p_{\max}$ denotes the length of the longest job, and that $L_{\max}$ denotes the maximum machine load. Throughout this section we will assume that

$$
L_1 \;=\; L_2 \;=\; L_3 \;=\; \cdots \;=\; L_m \;=\; L_{\max} \qquad \text{and} \qquad o_{\max} = 1. \tag{12.6}
$$

The equality of all machine loads in (12.6) can be reached by adding dummy jobs, and $o_{\max} = 1$ can be reached by scaling. We will apply the machinery for vector rearrangements (as described in the preceding section) to the open shop scheduling problem $Om||C_{\max}$. We introduce a unit ball $B^*$ for a norm $\|\cdot\|_*$ in $(m-1)$-dimensional space, defined by

$$
B^* \;=\; \big\{ (x_1, \ldots, x_{m-1}) \in \mathbb{R}^{m-1} : |x_k| \leq 1 \text{ and } |x_k - x_\ell| \leq 1 \text{ for all } k \text{ and } \ell \big\}. \tag{12.7}
$$

Every job $J_j$ with processing times $p_{i,j}$ is translated into an $(m-1)$-dimensional vector

$$
v_j \;=\; (p_{1,j} - p_{m,j},\; p_{2,j} - p_{m,j},\; \ldots,\; p_{m-1,j} - p_{m,j}). \tag{12.8}
$$

Because of (12.6) we have $\sum_{j=1}^{n} v_j = 0$ and $\|v_j\|_* \leq 1$ for $1 \leq j \leq n$, so that the conditions (12.2) for the vector rearrangement Theorem 12.4 are satisfied. Consequently

**Figure 12.3.** The infeasible schedule that results from the vector rearrangement.

there exists a permutation $v_{\pi(1)}, \ldots, v_{\pi(n)}$ of these vectors, so that

$$\|v_{\pi(1)} + v_{\pi(2)} + \cdots + v_{\pi(k)}\|_* \leq m - 1 \qquad \text{for } k = 1, \ldots, n. \tag{12.9}$$

We construct an infeasible schedule that on each machine processes the jobs without idle time in the ordering $J_{\pi(1)}, \ldots, J_{\pi(n)}$; see Figure 12.3 for an illustration. This schedule is extremely infeasible, as it schedules all operations of every job into a short time interval; this is a consequence of (12.9) and the definition of norm $\|\cdot\|_*$. The positive effect of this type of infeasibility is that we have a good understanding of the global structure of this schedule. The completion time of operation $O_{i,j}$ in the infeasible schedule is denoted by $C_{i,j}$. Then for $k \geq 2$ we have

$$
\begin{aligned}
C_{1,\pi(k)} - C_{2,\pi(k-1)} &= \sum_{j=1}^{k} p_{1,\pi(j)} - \sum_{j=1}^{k-1} p_{2,\pi(j)} \\
&= \sum_{j=1}^{k-1} (p_{1,\pi(j)} - p_{2,\pi(j)}) + p_{1,\pi(k)} \leq (d-1) + 1 = d.
\end{aligned}
$$

In the inequality, we use (12.9) and $p_{1,\pi(k)} \leq o_{\max} = 1$ from (12.6). By applying analogous arguments, we derive

$$
\begin{aligned}
\Delta_1 &:= \max_{k \geq 2} C_{m,\pi(k)} - C_{1,\pi(k-1)} &\leq m \\
\Delta_2 &:= \max_{k \geq 2} C_{1,\pi(k)} - C_{2,\pi(k-1)} &\leq m \\
\Delta_3 &:= \max_{k \geq 2} C_{2,\pi(k)} - C_{3,\pi(k-1)} &\leq m \\
&\quad \cdots \quad \cdots \quad \cdots \\
\Delta_m &:= \max_{k \geq 2} C_{m-1,\pi(k)} - C_{m,\pi(k-1)} &\leq m
\end{aligned}
$$

This means that we can turn the infeasible schedule into a feasible schedule, by simply shifting all operations on every machine $M_i$ by $(i-1)m$ time units into the future. The makespan of the new schedule will be bounded by $L_{\max} + (m-1)o_{\max}$, which yields a reasonable approximation result. We will describe next how to get an even better result. We wrap the infeasible schedule around a cylinder with circumference $L_{max}$. Each of the individual machine schedules forms a ring around the cylinder that may be rotated. We freeze the ring for machine $M_1$, and we mark the starting time of job $J_{\pi(1)}$ as the zero-point.

We rotate the ring for machine $M_2$ by $\Delta_2$ time units and thereby shift the starting time of each operation by $\Delta_2$ into the future. By doing this, we resolve all collisions between operations on $M_1$ and operations on $M_2$: Every job has now disjoint processing intervals on $M_1$ and $M_2$. Then we rotate the ring for machine $M_2$ by $\varepsilon_2 \le o_{max}$ additional time units, so that one of the operations on $M_2$ is started at the marked zero-point. Next, we do a similar rotation of the ring for machine $M_3$ by $\Delta_2 + \Delta_3 + \varepsilon_2 + \varepsilon_3$ time units, so that all collisions between $M_2$ and $M_3$ are resolved and so that one of the operations on $M_3$ is started at the marked zero-point. And so on. The ring for machine $M_i$ is rotated by $\sum_{k=2}^{i} \Delta_k + \varepsilon_k$ time units, so that all collisions are resolved and so that some operation starts at the zero-point.

In the end, we cut the rings open at the marked zero-point and flatten them into a schedule for the considered open shop instance. If $L_{max} - \Delta_1$ is larger than the total length of all shifts, the resulting schedule will be feasible: Before the shifting, all operations of job $J_j$ were scheduled very close to each other in time. The first shift puts $O1,j$ and $O_{2,j}$ into disjoint processing intervals, and each of the later shifts puts another operation into a disjoint processing interval. As $L_{max}$ is sufficiently large, the later operations of job $J_j$ will not be shifted all the way around the cylinder and hence cannot cause collisions with the first operation $O1,j$ of that job. Since $\Delta_i \le m$ and since $\varepsilon_i \le o_{max} \le 1$, the total length of all shifts is at most $(m-1)(m+1)$, and this should be at most $L_{max} - \Delta_1 \ge L_{max} - m$.

**Theorem 12.5** (Fiala, 1983). *If an instance of $Om||C_{max}$ satisfies $L_{max} \ge (m^2 + m - 1)o_{max}$, then the optimal makespan is $L_{max}$. Furthermore, an optimal schedule can be computed in polynomial time.*

One consequence of Theorem 12.5 is that open shop problems in the real world are often easy to solve: If all jobs are drawn from the same distribution and if there is a sufficiently large number of jobs, then the condition $L_{max} \ge (m^2 + m - 1)o_{max}$ in Theorem 12.5 will hold true and the instances become easy to solve.

Belov & Stolin [1974] were the first to apply vector rearrangement methods in the area of scheduling (and they applied them to the flow shop problem). Fiala [1983] discovered the nice connection to the open shop problem; he actually proved a much stronger version of Theorem 12.5 where the factor $m^2 + m - 1$ is replaced by $8m' \log_2(m') + 5m'$ where $m'$ is the smallest power of 2 greater or equal to $m$. Bárány & Fiala [1982] improved Fiala's bound by a factor of 2, and Sevastianov [1992] improved it down to roughly $(16/3)m \log_2 m$. Sevastianov [1994] surveys and summarizes the history of vector rearrangement methods in the area of scheduling.

In the light of the above results, it is natural to ask for the smallest value $\eta(m)$, so that every instance of $Om||C_{max}$ with $L_{max} \ge \eta(m)o_{max}$ automatically satisfies $C^*_{max} = L_{max}$.

**Open problem 12.2.** *Derive good upper and lower bounds on $\eta(m)$ for $m \geq 3$.*

Sevastianov [1995] establishes the upper bound $\eta(m) \leq m^2 - 1 + 1/(m-1)$, which is useful for small values of $m$, and also establishes the lower bound $\eta(m) \geq 2m - 2$. Here is the bad instance for $m = 3$ machines that demonstrates $\eta(3) \geq 4$: There is one job with processing time 1 on each machine. Furthermore, for each machine $M_i$ ($i = 1, 2, 3$) there are three jobs with processing time $1 - \varepsilon$ on $M_i$ and processing time 0 on the other two machines. Then $L_{\max} = 4 - 3\varepsilon$ and $C^*_{\max} = 4 - \varepsilon$.

Sevastianov [1995] also shows that $Om||C_{\max}$ remains NP-hard, if it is restricted to instances with $L_{\max}/o_{\max} = \rho$ where $1 < \rho < 2m - 3$. It is not clear, what is going on for instances with $2m - 3 \leq \rho < \eta(m)$. Perhaps, the instances with $\rho < \eta(m)$ are all NP-hard; in that case $\eta(m)$ would be a threshold at which the complexity jumps from NP-hard to trivial.

**Open problem 12.3.** *Determine the computational complexity of the restriction of $Om||C_{\max}$ to instances with $L_{\max}/o_{\max} = 2m - 2$.*

## 12.6. Approximation: Arbitrary number of machines

Here is a simple greedy algorithm for $O||C_{\max}$: Start at time $t = 0$, and whenever some machine becomes idle and there is some job available that still needs processing on that machine then assign that job to that machine. Ties are broken arbitrarily. This greedy algorithm was formulated by Bárány & Fiala [1982] who attribute it to private communication with Anna Racsmány.

**Theorem 12.6** (Bárány & Fiala, 1982). *The greedy algorithm is a polynomial time approximation algorithm for $O||C_{\max}$ with worst case ratio at most 2.*

*Proof.* Consider a greedy schedule, and let $O_{i,j}$ be an operation that completes last. Then on machine $M_i$, the greedy schedule has busy time intervals and idle time intervals. The total length of the busy time intervals is $L_i \leq L_{\max}$. Whenever $M_i$ is idle, it is not processing job $J_j$ and the only possible reason for this is that job $J_j$ is being processed on one of the other machines. Therefore the total length of the idle time intervals is at most $p_j \leq p_{\max}$. This implies that the greedy makespan is at most $L_{\max} + p_{\max}$, which according to (12.1) is bounded by $2\beta \leq 2C^*_{\max}$. $\square$

The result in Theorem 12.6 can also be derived as a corollary to a more general result by Aksjonov [1988]. How good is the worst case analysis in this theorem? Consider the following instance with $m$ machines and $m^2 - m + 1$ jobs. For $i = 1, \ldots, m$ there are $m - 1$ dummy jobs that each need one unit of processing time on machine $M_i$ and zero processing time on all other machines. Furthermore, there is a job $J^+$ that needs one unit of processing time on every machine. There is a greedy schedule with makespan $2m - 1$ for this instance, in which from time 0 to time $m - 1$ all machines are busy with processing the dummy jobs, and from time $m - 1$ to time $2m - 1$ they process job $J^+$. As the optimal makespan is $C^*_{\max} = m$, the worst case ratio of the greedy algorithm is at least $2 - 1/m$; see Figure 12.4 for an illustration. Chen & Strusevich [1993] have settled the cases $m = 2$ and $m = 3$ of the following open problem by a tedious case analysis, and Chen & Yu [2001] have settled the case $m = 4$.

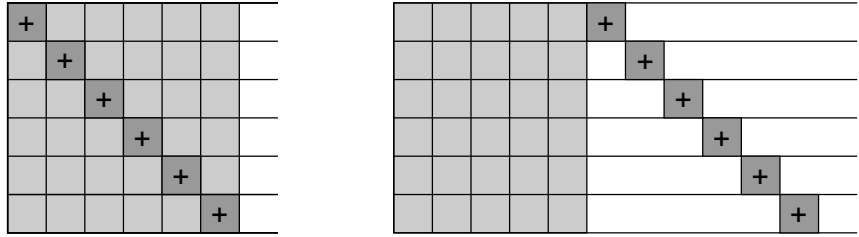

**Figure 12.4. A lower bound instance for the greedy algorithm on $m = 6$ machines. The dummy jobs are shown in light-gray, while the unit-time operations of job $J^+$ are shown in dark-gray and marked by +.**

**Open problem 12.4.** *Prove that the greedy algorithm for $Om||C_{\max}$ has worst case ratio at most $2 - 1/m$.*

A difficult open problem in this area asks whether there is a polynomial time approximation algorithm for $O||C_{\max}$ with worst case ratio strictly better than 2. One possible approach would work with the lower bound $\beta$ defined in (12.1). Sevastianov & Tchernykh [1998] have proved $C_{\max}^* \le 4\beta/3$ for problem $O3||C_{\max}$. Their proof is based on heavy case analysis and on case enumeration with the help of a computer program. As the computer program described by Sevastianov & Tchernykh [1998] takes more than 200 hours of computation time, this approach does not seem to be applicable to $m \ge 4$ machines.

**Open problem 12.5.** *Prove that any instance of $Om||C_{\max}$ satisfies $C_{\max}^* \le 3\beta/2$.*

Here is an instance that demonstrates that the factor $3/2$ in this open problem would in fact be best possible. The instance uses $m$ machines and $m + 1$ jobs. For $j = 1, \ldots, m$ the job $J_j$ consists of the operation $O_{jj}$ with processing time $p_{jj} = m - 1$ on machine $M_j$, and with processing times 0 on the other $m - 1$ machines. The final job $J_{m+1}$ has $m$ operations all with processing time 1. Then $\beta = m$ and $C_{\max}^* = \lceil m/2 \rceil + m - 1$. As $m$ becomes large, the ratio between $C_{\max}^*$ and $\beta$ tends to $3/2$.

Now let us turn to negative results on the worst case ratio of polynomial time approximation algorithms for $O||C_{\max}$. Williamson et al. [1997] prove that it is NP-hard to decide whether an $O||C_{\max}$ instance with integer processing times has optimal makespan at most 4. Since the optimal makespan of a NO-instance is at least 5, a polynomial time approximation algorithm with worst case ratio $5/4 - \varepsilon$ would allow us to distinguish the YES-instances from the NO-instances in polynomial time.

**Theorem 12.7** (Williamson et al., 1997). *Unless P=NP, problem $O||C_{\max}$ does not allow a polynomial time approximation algorithm with worst case ratio strictly better than* 5/4.

It might be possible to lift the hardness proof of Williamson et al. [1997] to get stronger inapproximability results.

**Open problem 12.6.** *Analyze the computational complexity of the $(a,b)$-versions of $O||C_{\max}$ instances with integer processing times: Decide whether the optimal makespan does satisfy $C_{\max}^* \leq a$ or whether it does satisfy $C_{\max}^* \geq b$.*

If this $(a,b)$-version turns out to be NP-hard for fixed integers $a$ and $b$, then $O||C_{\max}$ cannot have a polynomial time approximation algorithm with worst case ratio strictly better than $b/a$ unless $P = NP$. The result of Williamson et al. [1997] yields NP-hardness of the $(4,5)$-version, and they also show that the $(3,4)$-version is solvable in polynomial time. Hence the smallest interesting open cases would concern the $(5,7)$-version and the $(6,8)$-version.

## 12.7. Approximation: Fixed number of machines

For an arbitrary number of machines, polynomial time approximation algorithms cannot have worst case ratios very close to 1; see Theorem 12.7. For a fixed number of machines, the situation is much better and there is a polynomial time approximation scheme (PTAS).

**Theorem 12.8** (Sevastianov & Woeginger, 1998). *For every fixed $m \geq 3$, problem $Om||C_{\max}$ has a PTAS.*

We now show a proof of Theorem 12.8 for the special case $m = 3$. (The general case is based on exactly the same ideas, while some of the details become a bit messier.) So let us consider some instance of $O3||C_{\max}$, and let $\varepsilon$ with $0 < \varepsilon < 1$ be some small real number that indicates the desired precision of approximation. The running time of our algorithm will be polynomial in the instance size, but exponential in $1/\varepsilon$. The resulting makespan will come arbitrarily close to $C_{\max}^*$, if we let $\varepsilon$ tend to 0.

As often in approximation schemes for scheduling problems, the jobs are classified into *big* and into *small* jobs. We call a job *big*, if one of its operations has length $p_{i,j} \geq \varepsilon\beta$, where $\beta$ is the lower bound defined in (12.1). All other jobs are *small* jobs, and we want to assume for the moment that (***) all operations of all small jobs have length $p_{i,j} \leq \varepsilon^2\beta$; we will show later how to work around this assumption. Since the total processing time of all jobs is at most $3L_{\max} \leq 3\beta$ and as every big job has processing time at least $\varepsilon\beta$, there are at most $3/\varepsilon$ big jobs. The algorithm now works in two phases.

- In the first phase, we determine an optimal schedule for the big jobs. This can be done in $O(1)$ time, as the running time does only depend on $1/\varepsilon$ and does not depend on the instance size. In the resulting schedule, every machine processes at most $3/\varepsilon$ operations with at most $3/\varepsilon$ gaps of idle time between the operations; see Figure 12.5 for an illustration.

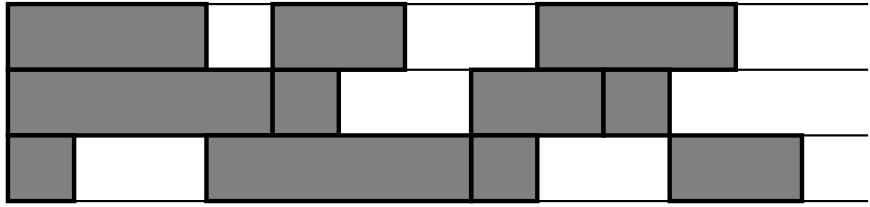

**Figure 12.5.**    An optimal schedule for the big jobs in the proof of Theorem 12.8.

- In the second phase, we pack the operations of the small jobs into the idle gaps. This is done greedily (as in Theorem 12.6). Start at time $t = 0$, and whenever some machine becomes idle at some time $t$, try to process an unprocessed small operation on that machine. There are two possible scenarios under which an unprocessed small operation $O_{i,j}$ cannot be started at time $t$: First another operation $O_{kj}$ of the same job might currently be processed on some other machine. Secondly, the remaining part of the current gap might be too small to accommodate $O_{i,j}$. If one of these scenarios occurs, we try to schedule some other small operation. If no operation can be scheduled, then the machine is left idle for the moment.

Now let us analyze the makespan $C_{\max}^A$ of the resulting approximating schedule. Let $O_{i,j}$ be an operation that completes last. In the first case assume that $O_{i,j}$ belongs to a big job. Then $C_{\max}^A$ coincides with the optimal makespan for the big jobs, and we actually have found an optimal schedule. In the second case assume that $O_{i,j}$ belongs to a small job. Then we consider the busy time intervals and the idle time intervals on machine $M_i$. The total length of all busy time intervals is the load $L_i \le \beta$. Whenever machine $M_i$ was idle, it could not process operation $O_{i,j}$. This means that either (i) job $J_j$ was being processed on one of the other machines, or that (ii) the remaining gap was too small to accommodate $O_{i,j}$. The total idle time of type (i) is bounded by the length of the small job $J_j$, which is at most $3\varepsilon^2\beta$. The total idle time of type (ii) is bounded by the number of gaps multiplied by the length of operation $O_{i,j}$, which is at most $(3/\varepsilon) \cdot (\varepsilon^2\beta) = 3\varepsilon\beta$. Altogether, this implies that the approximating makespan can be bounded as

$$C_{\max}^A = \text{Busy} + \text{Idle(i)} + \text{Idle(ii)} \le \beta + 3\varepsilon^2\beta + 3\varepsilon\beta \le (1 + 3\varepsilon^2 + 3\varepsilon)\,C_{\max}^*.(12.10)$$

As $\varepsilon$ tends to 0, the error factor $1 + 3\varepsilon^2 + 3\varepsilon$ tends to 1. This yields the desired PTAS modulo the assumption (***).

It remains to discuss what to do with assumption (***), which essentially postulates an empty no man's land between big operations (of length at least $\varepsilon\beta$) and small operations (of length at most $\varepsilon^2\beta$). In other words, under assumption (***) non-big jobs must not contain operations of intermediate length $\varepsilon^2\beta < p_{i,j} < \varepsilon\beta$. This assumption will be totally wrong for most instances, but we can come very close to it by playing around with the value of $\varepsilon$. This is done as follows.

For a real number $\delta$ with $0 < \delta < 1$, we say that a job is $\delta$-*big*, if one of its operations has length $p_{i,j} \geq \delta\beta$ and otherwise it is $\delta$-*small*. An operation $O_{i,j}$ is $\delta$-*nasty*, if it belongs to a $\delta$-small job and satisfies the inequality $\delta^2\beta < p_{i,j} < \delta\beta$. By $N(\delta)$ we denote the total length of all $\delta$-nasty operations. Now consider the real numbers $\delta_k = \varepsilon^{2^k}$ for $k \geq 0$. Then every operation $O_{i,j}$ is $\delta_k$-nasty for at most one choice of index $k$. We search for an index $k$ that satisfies the inequality

$$N(\delta_k) \leq \varepsilon\beta. \tag{12.11}$$

If some value $\delta_k$ violates (12.11), then the corresponding $\delta_k$-nasty operations consume at least $\varepsilon\beta$ of the total processing time of all operations (which is at most $3\beta$). Hence some $k \leq 3/\varepsilon$ will indeed satisfy (12.11). From now on we work with that particular index $k$ and with that particular value $\delta_k$.

The final approximation scheme works as follows. First we remove from the instance all the $\delta_k$-small jobs that contain some $\delta_k$-nasty operation. To the surviving jobs we apply the original approximation algorithm as described above with $\varepsilon := \delta_k$, and thereby find a schedule with makespan at most $(1 + 3\delta_k^2 + 3\delta_k)C_{\max}^*$ according to (12.10). In the end, we greedily add the previously removed jobs with $\delta_k$-nasty operations to this schedule. Since the overall processing time of all removed jobs is at most $3\varepsilon\beta$, this increases the makespan by at most $3\varepsilon\beta$. Since $\delta_k \leq \varepsilon$, this altogether yields a schedule of makespan at most $(1 + 3\varepsilon^2 + 6\varepsilon)C_{\max}^*$. This completes the proof of Theorem 12.8 for the special case $m = 3$.

An FPTAS (fully polynomial time approximation scheme) is a PTAS whose time complexity is also polynomially bounded in $1/\varepsilon$. The following open problem is closely linked to the existence of pseudo-polynomial time exact algorithms for $Om||C_{\max}$.

**Open problem 12.7.** *Prove that problem $Om||C_{\max}$ has an FPTAS for every fixed $m \geq 3$.*

## 12.8. Preemption and other optimality criteria

The term "open shop" is due to Gonzalez & Sahni [1976]. They give a polynomial-time algorithm for $O|pmtn|C_{\max}$, which finds a schedule of length $\beta$. Lawler & Labetoulle [1978] use the algorithm to construct optimal schedules for $R|pmtn|C_{\max}$ and $R|pmtn|L_{\max}$ from optimal solutions to linear programming formulations; see Section 10.7, where a sketch of the Gonzalez-Sahni algorithm is given. For $O|pmtn, r_j|L_{\max}$, Cho & Sahni [1981] observe that a trial value of $L_{\max}$ can be tested for feasibility by linear programming; bisection search is then appplied to minimize $L_{\max}$ in polynomial time.

Lawler et al. [1981, 1982] give a linear-time algorithm for $O2|pmtn|L_{\max}$, assuming that the due dates are preordered; they establish strong NP-hardness for $O2||L_{\max}$.

Liu & Bulfin [1985] provide an NP-hardness proof for $O3|pmtn|\sum C_j$; $O2|pmtn|\sum C_j$ remains open. For the nonpreemptive version $O2||\sum C_j$, Achugbue & Chin [1982A] prove strong NP-hardness and derive tight bounds on the length of arbitrary schedules and SPT schedules.

# Contents

**i**

# 13

# Flow Shops

Jan Karel Lenstra
*Centrum Wiskunde & Informatica*

David B. Shmoys
*Cornell University*

### 13.1. Johnson's algorithm for $F2||C_{max}$

In one of the first results in deterministic scheduling theory, Johnson gave an $O(n log n)$ algorithm to solve $F2||C_{max}$. The algorithm is surprisingly simple: first arrange the jobs with $p_{1j} \leq p_{2j}$, in order of nondecreasing $p_{1j}$ and then arrange the remaining jobs in order of nonincreasing $p_{2j}$.

The proof that this algorithm produces an optimal schedule is also straightforward. Notice that the algorithm produces a *permutation schedule*, where the order of the jobs on each machine is identical. An easy interchange argument shows that there exists an optimal schedule that is a permutation schedule. For a permutation schedule, it is easy to see that $C_{max}$ is determined by the processing time of some $\ell$ jobs on machine 1, followed by the processing time of $(n+1) - \ell$ jobs on machine 2. Note that this implies that if each $p_{ij}$ is decreased by the same value $p$, then for each permutation schedule, $C_{max}$ decreases by $(n+1)p$. Finally, observe that if $p_{1j} = 0$, then job $J_j$ is scheduled first in some optimal schedule, and similarly, if $p_{2j} = 0$, then $J_j$ is scheduled last is some optimal schedule. Putting these pieces together, we see that an optimal schedule can be constructed by repeatedly finding the minimum $p_{ij}$ value among the unscheduled jobs and subtracting this value from all processing times; if the minimum is zero, then the corresponding job is scheduled. This algorithm is clearly equivalent to the one given above.

### 13.2.  Approximation algorithms: simple results

Given that all but the most restrictive cases of the flow shop scheduling problem are *NP*-hard, it is natural to consider approximation algorithms for the problem. We will consider algorithms that produce permutation schedules, as well as algorithms that produce arbitrary schedules. In the former case, we must be careful to measure the performance relative to the optimal schedule rather than to the optimal permutation schedule.

In contrast to, for example, $O||C_{\max}$, simple algorithms for $F||C_{\max}$ do not give encouraging results. Consider the permutation schedule produced by a *list scheduling* (LS) algorithm; in other words, consider an arbitrary permutation schedule. In such a schedule, at least one operation is being performed at any time, so that $C_{\max}(\text{LS}) \leq \sum_{i=1}^{m} \sum_{j=1}^{n} p_{ij}$. On the other hand, the optimal schedule length $C_{\max}^*$ must be at least $\sum_{j=1}^{n} p_{ij}$, for any $M_i (i = 1, ..., m)$. There must be a machine for which this sum is at least $(\sum_{i=1}^{m} \sum_{j=1}^{n} p_{ij})/m$, since that is the average machine load. We have obtained the following result.

**Theorem 13.1.** *For any instance of $F||C_{\max}$, $C_{\max}(LS) \leq mC_{\max}^*$.*

The reader is invited to show that the analysis of algorithm LS is tight; see Exercise 13.1.

Johnson's algorithm for $F2||C_{\max}$ can be used to give simple algorithms with somewhat better performance. For example, suppose that the machines are grouped in $\lceil m/2 \rceil$ pairs: $(M_1, M_2)$, $(M_3, M_4)$, and so forth; if $m$ is odd, then $M_m$ is grouped by itself. We can find an optimal schedule for each pair; clearly, the length of each of these schedules is no more than $C_{\max}^*$. By concatenating the $\lceil m/2 \rceil$ two-machine schedules, we obtain an $m$-machine schedule of length at most $\lceil m/2 \rceil C_{\max}^*$. Note that this is not necessarily a permutation schedule.

A similar sort of *aggregation* algorithm (AG) produces a permutation schedule with the same performance guarantee. Aggregate $M_1, ..., M_{\lceil m/2 \rceil}$ into one virtual machine, and $M_{\lceil m/2 \rceil+1}, ..., M_m$ into another, with processing times $a_j = \sum_{i=1}^{\lceil m/2 \rceil} p_{ij}$ and $b_j = \sum_{i=\lceil m/2 \rceil+1}^{m} p_{ij}$. We can view this as an instance of $F2||C_{\max}$ and apply Johnson's algorithm to obtain a permutation $\pi$. This permutation can be viewed as a schedule for the original problem; how good is it?

One important fact for this analysis: the length of the permutation schedule $\pi$ for the original $m$-machine problem is no more than the length of $\pi$ for the aggregated two-machine problem. If we have an aggregated operation on the virtual machine 1 being processed for $a_j$ time units, then allocating this block of time across the first half of the machines, is sufficient to process each of the component operations of job $j$ in the corresponding interval.

We complete the analysis by showing that there is a schedule for the aggregated two-machine problem of length at most $\lceil m/2 \rceil C_{\max}^*$ (and hence the approximation for the $m$-machine problem is at most this much). Consider an optimal schedule for the $m$-machine problem. Group together $M_i$ and $M_{\lceil m/2 \rceil+i}$, for $i = 1, ..., \lfloor m/2 \rfloor$; if $m$ is odd, then $M_{\lceil m/2 \rceil}$ is grouped by itself. For each group, partition the schedule for its machines into time intervals in such a way that there is a one-to-one correspondence between the right endpoints of those intervals and the completion times of operations on machines in that group. If the intervals generated for all of the groups are sorted by their right endpoints and concatenated in that order, we obtain a preemptive schedule for the two-machine problem. Note that the sorting guarantees that the preemptive schedule is feasible for the two-machine problem; more precisely, no job is completed on the first virtual machine after it is started on the second. The length of this schedule is obviously no more than $\lceil m/2 \rceil C_{\max}^*$. But it is straightforward to argue that there is a nonpreemptive permutation schedule of the same length. Thus, we have proved the following theorem.

**Theorem 13.2.** *For any instance of $F||C_{\max}$, the aggregation algorithm AG delivers a permutation schedule with $C_{\max}(AG) \leq \lceil m/2 \rceil C_{\max}^*$.*

Again, the analysis of algorithm AG is tight; see Exercise 13.2. While Theorem 13.2 gives an upper bound on the quality of permutation schedules relative to the overall optimum, there are classes of instances for which the ratio between the lengths of the optimal permutation schedule and the overall optimal schedule is unbounded; see Exercise 13.3. In the next section, we will give a deeper result on the absolute difference between these two optima.

Many more approximation algorithms for flow shop scheduling have been proposed than the ones that we have analyzed in this section. In general, the objective has been to design methods that do well on a set of randomly generated instances, not to establish performance guarantees. In other words, the emphasis has been on empirical rather than worst-case analysis. In the rest of this section, we will mention some of these heuristics and summarize the experimental results that have been reported. We note that all of these methods produce permutation schedules.

We first discuss three algorithms that *construct* a schedule from scratch. The *slope index* rule orders the jobs $J_j$ in order of nonincreasing values $\sum_{i=1}^{m}[i-(m+1)/2]p_{ij}$. It is based on the same idea as Johnson's algorithm: a job whose operations tend to increase in length should be placed at the beginning of the schedule, and a job whose operations decrease in length should be at the end. The *generalized aggregation* rule uses Johnson's method in a more direct way: for $l = 1, ..., m-1$, apply the two-machine algorithm using processing times $\sum_{i=1}^{l} p_{ij}$ and $\sum_{i=m+1-l}^{m} p_{ij}$ ( $j = 1, ..., n$ ), and evaluate the resulting permutation as an $m$-machine schedule; choose the best of these $m-1$ schedules. Note that this algorithm is not a proper generalization of algorithm AG. The *longest insertion* rule is as follows: start with the empty partial schedule; at each step, select an unscheduled job with the maximum total processing requirement, and insert it in the current partial schedule in the position that minimizes the increase in schedule length.

There is ample empirical evidence to support the statement that the slope index rule is outperformed by the generalized aggregation rule, and that the longest insertion rule does still better. The latter algorithm is the current champion among constructive rules. These increases in solution quality come at the expense of increases in running time, as the reader can easily verify.

Another type of approximation algorithm applies *local search* to a given schedule. Here, one has to define a *neighborhood* for each schedule, i.e., a set of schedules that can be generated by a simple perturbation. Examples of perturbations are a transposition of two adjacent jobs, a transposition of two jobs that are not necessarily adjacent, and a shift of a job to a different position. In the basic variant of local search, the neighborhood of the initial schedule is searched for a better schedule. If such an improvement is found, the process restarts from there, and this continues until a schedule is found that is optimal in its neighborhood. A recent variant of local search is *simulated annealing*, which accepts deteriorations with a small and decreasing probability in an attempt to avoid bad local optima and to get settled in a global optimum. We will discuss this approach in more detail in Chapter 14.

Computational experiments suggest that the adjacent transposition neighborhood is too restrictive. Local search that starts with an arbitrary solution and applies adjacent transpositions tends to take more time and produce worse schedules than longest insertion. A longest insertion schedule can, however, often be improved by applying transpositions and shifts. Not surprisingly, simulated annealing with the shift neighborhood gives even better results but needs more time.

**Exercises**

13.1. Prove that the bound for list scheduling given in Theorem 13.1 is tight.

13.2. Prove that the bound for the aggregation algorithm given in Theorem 13.2 is tight.

13.3. Consider the following flow shop instance consisting of $2n$ machines and $n$ jobs: $p_{n-j+1,j} = p_{n+j,j} = 1, (j = 1, ..., n)$, and $p_{ij} = 0$ for all other $(i, j)$; here, 0 really denotes some arbitrarily small positive value.
(a) Prove that $C_{\max}^* = 2$.
(b) Prove that any permutation schedule has $C_{\max} \geq \sqrt{n}$. (*Hint:* A beautiful theorem of Erdös and Szekeres states that in any sequence of $n^2 + 1$ distinct integers, there is a decreasing subsequence of length $n + 1$ or an increasing subsequence of length $n + 1$. Use this theorem to identify a time-consuming subsequence of operations in any permutation schedule.)
(c) Use the insight gained in part (b) to construct a permutation schedule with $C_{\max} = 2\lceil \sqrt{n} \rceil$.

### 13.3. Approximation algorithms: geometric methods

The most significant results on approximation algorithms for flow shop scheduling are based on geometric methods. The main tool we will use is Theorem 12.4, the vector sum theorem, which was proved in the previous chapter.

The following result was already mentioned in Section 13.1. Let $p_{\max} = \max_{i,j} p_{ij}$.

**Theorem 13.3.** *For any instance of $F||C_{\max}$, a permutation schedule of length at most $C_{\max}^* + m(m-1)p_{\max}$ can be found in $O(m^2 n^2)$ time.*

*Proof.* Consider an arbitrary flow shop instance. For each $M_i$, let $\Pi_i = \sum_{j=1}^n p_{ij}$, and let $\Pi_{\max} = \max_i \Pi_i$. We will show that Theorem 12.4 provides a way to construct a permutation schedule of length at most $\Pi_{\max} + m(m-1)p_{\max}$. Since $\Pi_{\max} \le C_{\max}^*$, the theorem then follows.

Without loss of generality, we may assume that $\Pi_i = \Pi_{\max}$ for each $M_i$. If $\Pi_i < \Pi_{\max}$, we can iterate through the operations on $M_i$, increasing each $p_{ij}$ to at most $p_{\max}$, until the revised total reaches $\Pi_{\max}$.

Consider a permutation $\pi$ of $\{1, ..., n\}$, and suppose that each machine processes the jobs in the order $J_{\pi(1)}, ..., J_{\pi(n)}$. Let $I_{ij}$ denote the total idle time of $M_i$ up to the starting time of $O_{i\pi(j)}$. For a permutation schedule, it is easy to calculate all of the values $I_{ij}$. On $M_1$, we clearly have $I_{1j} = 0$ for all $j$. On $M_i$ $(i \ge 2)$, operation $O_{i\pi(j)}$ starts as soon as both $O_{i\pi(j-1)}$ and $O_{i-1,\pi(j)}$ are completed. In case the former operation finishes later, we have

$$I_{ij} = I_{i,j-1}.$$

In the latter case, $I_{ij} + \sum_{k=1}^{j-1} p_{i\pi(k)} = I_{i-1,j} + \sum_{k=1}^j p_{i-1,\pi(k)}$, or

$$I_{ij} = I_{i-1,j} + p_{i\pi(j)} + \sum_{k=1}^j (p_{i-1,\pi(k)} - p_{i\pi(k)}).$$

Suppose that $\pi$ is such that the following bound holds:

$$\sum_{k=1}^j (p_{i-1,\pi(k)} - p_{i\pi(k)}) \le (m-1)p_{\max}, j = 1, ..., n, i = 2, ..., m. \qquad (13.1)$$

This implies that the idle times are related by the recurrence

$$I_{ij} \le \max\{I_{i,j-1}, I_{i-1,j} + mp_{\max}\}.$$

With the initial conditions for $i = 1$, we conclude that $I_{mn} \le (m-1)mp_{\max}$. This yields the desired bound on the length of the permutation schedule corresponding to $\pi$:

$$C_{\max}(\pi) = \Pi_m + I_{mn} \le \Pi_{\max} + m(m-1)p_{\max}.$$

Fortunately, Theorem 12.4 is exactly what we need to compute a permutation that guarantees the inequality (13.1). Define the vectors

$$v_j = (p_{1j} - p_{2j}, p_{2j} - p_{3j}, ..., p_{m-1,j} - p_{mj}), j = 1, ..., n.$$

The assumption that $\Pi_i = \Pi_{\max}$ $(i = 1, ..., m)$ implies that $\sum_{j=1}^n v_j = 0$, so we can apply Theorem 12.2. Since $d = m - 1$ and $||v_j|| \le p_{\max}$, (13.1) immediately follows. This completes the proof. $\square$

It is natural to ask whether the bound given in Theorem 13.3 is tight. This is certainly not the case for $m = 2$, since there is a permutation schedule that is an overall optimal schedule. One might also wonder if $\Pi_{\max} + 2p_{\max}$ is a tight bound for $m = 2$; no, one can show that there always is a (permutation) schedule that achieves $\Pi_{\max} + p_{\max}$, and this is tight. For $m = 3$, there is a permutation schedule that is guaranteed to be no longer than $\Pi_{\max} + 3p_{\max}$, and this is tight; for $m = 4$, the best bound known is $\Pi_{\max} + 9p_{\max}$.

When the number of machines is fixed, Theorem 13.3 can also be used to obtain strong relative performance guarantees. Let $\varepsilon$ be an arbitrary positive constant. Call a job $J_j$ *big* if there is an operation $O_{ij}$ with $p_{ij} > \varepsilon\Pi_{\max}/m^2$, and *small* otherwise. Observe that there are less than $m^3/\varepsilon$ big jobs. (Otherwise, at least one machine would have to process at least $m^2/\varepsilon$ of the expensive operations, which would contradict the definition of $\Pi_{\max}$.) Hence, for fixed $m$ and $\varepsilon$, the number of big jobs is bounded by a constant.

Now, first use the algorithm of Theorem 13.3 to schedule all of the small jobs. The length of this partial schedule is at most

$$\Pi_{\max} + m(m-1)\varepsilon\Pi_{\max}/m^2 < (1+\varepsilon)\Pi_{\max} \le (1+\varepsilon)C_{\max}^*.$$

Next, consider all possible ways to schedule all of the big jobs at the conclusion of this partial schedule. Since there are only a constant number of big jobs and a constant number of machines, the number of such extensions is also constant (albeit possibly huge), and the best one can be selected in constant time. Since the completed schedule is at most $C_{\max}^*$ longer than the partial schedule, we obtain a schedule of length at most $(2+\varepsilon)C_{\max}^*$.

**Theorem 13.4.** *For any instance of $Fm||C_{\max}$ (i.e., with m fixed) and any $\varepsilon > 0$, a schedule of length at most $(2+\varepsilon)C_{\max}^*$ can be found in polynomial time.*

Note that the extension need not be a permutation schedule, and hence the completed schedule need not be one. However, if only permutation extensions are considered, then the same scheme delivers a solution of length at most $2 + \varepsilon$ times the length of the best permutation schedule. It remains an interesting open question to find algorithms with comparable performance whose running time is also polynomial in $m$.

### Exercises

13.4. Consider $F2||C_{\max}$. Give a polynomial-time algorithm for finding a permutation schedule of length at most $\Pi_{\max} + p_{\max}$. Show that this bound is tight.

### Notes

13.1. *The two-machine flow shop.* The two-machine flow shop algorithm is one of the most celebrated results in scheduling theory. It is due to Selmer M. Johnson [1954].

13.2. *Approximation algorithms: simple results.* Potts, Shmoys & Williamson [1991] exhibit a family of flow shop instances for which the best permutation schedule is longer than the true optimal schedule by a factor of more than $(1/2)\sqrt{m}$.

The analysis of algorithm LS is due to Gonzalez and Sahni [1978A]. They showed that the bound of Theorem 13.1 is tight even for LPT schedules, in which the jobs are ordered according to nonincreasing sums of processing times. They also gave the algorithm that consists of $\lfloor m/2 \rfloor$ applications of Johnson's algorithm. Algorithm AG was proposed and analyzed by Röck and Schmidt [1983].

The slope index rule is due to Palmer [1965], and the generalized aggregation rule to Campbell, Dudek, and Smith [1970]. Dannenbring [1977] compared eleven heuristics, including those of Palmer and Campbell et al. as well as two that apply adjacent transpositions. The longest insertion rule was proposed by Nawaz, Enscore, and Ham [1983] and tested by Turner and Booth [1987]. Osman and Potts [1989] report on experiments with traditional local search and simulated annealing.

We quote a few results that are not as simple as the title of this section suggests. In a series of papers, Nowicki & Smutnicki [1989, 1991, 1993] analyzed the performance of the algorithms mentioned in the previous paragraph. They are interested in the worst-case ratio of the length of the approximate schedule and the length of the optimal *permutation* schedule. The generalized aggregation rule has a ratio $\lceil m/2 \rceil$. For the slope index rule and for both adjacent transposition algorithms of Dannenbring, the ratio is about $m/\sqrt{2}$ (the precise value is a more complicated expression that depends on the parity of $m$). These bounds are tight. The worst-case ratio for the longest insertion rule is unknown, but instances exist that achieve a ratio of about $\sqrt{m}/2$.

Finally, Potts [1985B] investigated the performance of five approximation algorithms for $F2|r_j|C_{\max}$. The best one of these, called RJ', involves the repeated application of a dynamic variant of Johnson's algorithm to modified versions of the problem, and satisfies $C_{\max}(RJ')/C_{\max}^* \le 5/3$; this bound is tight.

13.3. *Approximation algorithms: geometric methods.* The connection between flow shop scheduling and the vector sum theorem, as stated in Theorem 13.3, was discovered independently by Belov & Stolin [1974], Sevast'yanov [1974], and Bárány [1981]. Theorem 13.4 is implicit in the work of Shmoys, Stein & Wein [1994].

# Contents

# 14

# Job Shops

Johann L. Hurink
*Twente University*

Jan Karel Lenstra
*Centrum Wiskunde & Informatica*

David B. Shmoys
*Cornell University*

## 14.1. The disjunctive graph model for $J||C_{\max}$

In the job shop scheduling problem, each job $J_j$ consists of a chain of operations (i.e., an ordered sequence of operations), where each operation is specified to be processed on a particular machine for a specified length of time (without interruption). Each machine can process at most one operation at a time. For $J||C_{\max}$, the objective is to minimize the makespan, that is, the time by which all jobs are completed.

The above description does not reveal much of the structure of this problem type. An illuminating problem representation is provided by the *disjunctive graph model* due to Roy & Sussmann [1964].

Given an instance of $J||C_{\max}$, the corresponding disjunctive graph is defined as follows. For every operation $O_{ij}$, there is a vertex, with a weight $p_{ij}$. For every two consecutive operations of the same job, there is a (directed) arc. For every two operations that require the same machine, there is an (undirected) edge. Thus, the arcs represent the job precedence constraints, and the edges represent the machine capacity constraints.

The basic scheduling decision is to impose an ordering on a pair of operations on the same machine. In the disjunctive graph, this corresponds to orienting the edge in question, in one way or the other. A schedule is obtained by orienting all of the

| $J_j$ | $m_j$ | $\mu_{1j}$ | $\mu_{2j}$ | $\mu_{3j}$ | $\mu_{4j}$ | $p_{1j}$ | $p_{2j}$ | $p_{3j}$ | $p_{4j}$ |
|---|---|---|---|---|---|---|---|---|---|
| $J_1$ | 3 | $M_1$ | $M_2$ | $M_3$ | — | 2 | 8 | 4 | — |
| $J_2$ | 4 | $M_2$ | $M_1$ | $M_3$ | $M_4$ | 7 | 3 | 6 | 3 |
| $J_3$ | 3 | $M_1$ | $M_2$ | $M_4$ | — | 5 | 9 | 1 | — |

(a)

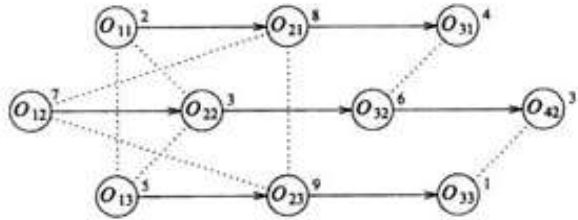

(b)

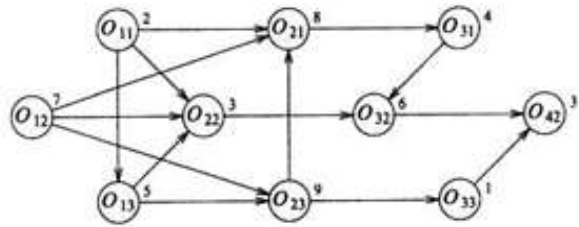

(c)

**Figure 14.1.  Job shop scheduling problem. (a) Instance. (b) Instance, represented as a disjunctive graph. (c) Feasible schedule, represented as an acyclic directed graph.**

edges.  The schedule is feasible if the resulting directed graph is acyclic, and its length is obviously equal to the weight of a maximum weight path in this graph.

The job shop scheduling problem has now been formulated as the problem of finding an orientation of the edges of a disjunctive graph that minimizes the maximum path weight. We refer to Figure 14.1 for an example.

## 14.2.  Approximation algorithms: computing good solutions

In this section we deal with the second class of approaches mentioned in Chapter 2 to cope with NP-complete problems: we present heuristic methods to solve the job shop problem.  These methods enter the scene if the computation time to get solu-tions is limited or the instances to be considered get larger. In both cases, branch-and-bound methods may fail. Depending on the available time or the size of the instances, simple constructive heuristics or more time consuming iterative methods may be used. The main focus of this section lies on the second class of heuristics. Starting from an initial solution achieved e.g., by a constructive heuristic, the iterative methods change the given solution slightly and repeat this process iteratively hoping that finally a good solution results. Based on the iterative process (searching) and the restriction to slight changes (local), these methods are called *local search* methods. The local search approaches presented here for the job shop problem work very well in practice, although they give no proven performance guarantee. However, getting approximation methods with a proven performance guarantee for the job shop problem is limited. Since it has been proven that determining whether or not a schedule of length $\leq 4$ exists is NP-complete, it follows that the problem of finding an approximation algorithm with a performance guarantee better than $\frac{5}{4}$ is NP-hard.

This section is organized as follows. In the following subsection we will give a brief description of constructive heuristics for the job shop problem. In addition to their value as stand alone methods, these methods can be used to calculate initial solutions for local search procedures, which form the contents of Subsection 14.2.2. Finally, in Subsection 14.2.3 some implementational issues are discussed.

### 14.2.1. Constructive methods

In this part we concentrate on constructive heuristics for the job shop problem. In addition to simple methods based on priority rules, we also discuss a more elaborate method called *shifting bottleneck* procedure.

As mentioned in Section 14.1, the main scheduling decision is to define processing orders on the machines leading to complete orientations. Given such a complete orientation, by longest path calculations we obtain a corresponding schedule in which an operation starts at the time where either its job or machine predecessor is finished or, if both do not exist, at time 0. Schedules that have this property (no operation can start earlier without changing at least one machine order) are called *semi-active* schedules. Two important subclasses of the semi-active schedules are the *active* schedules and the *non-delay* schedules. The first class contains only schedules (complete orientations) where it is not possible to start an operation earlier without delaying the start of another operation. The second class is a subset of the first and contains only schedules where a machine may be idle only if no jobs are available for processing. Whereas the set of active schedules is still a dominating set (i.e., it contains at least one optimal schedule), optimal schedules do not have to be a non-delay schedules (see Exercise 14.1).

Most of the priority driven heuristics for the job shop problem generate active or non-delay schedules. The basic principle of all these methods is the same and can also be found in the method to calculate upper bounds within branch-and-bound procedures: iteratively repeat the following steps

1. estimate the set $C$ of operations for which the job predecessors have already been scheduled,

2. select an operation in $C$ and schedule it next on its machine.

Thus, in each iteration one operation is added to the given partial schedule as the last operation on its machine.  In principle, Step 2 is always realized as follows:

2.1  determine a subset $\bar{C} \subset C$,

2.2  select an operation from $\bar{C}$ using a priority rule.

Depending on the realization of these two sub-steps, different constructive heuristics result. The priority rules in Step 2.2 may be simple rules like shortest processing time (SPT), longest processing time (LPT), longest remaining processing time (LRPT) etc., or may be more complex rules depending on a lower bound for the makespan assuming that the given operation is scheduled next.  The selection of the subset $\bar{C} \subset C$ in Sub-step 2.1 may be realized in three different ways:

- Select $\bar{C}$ as the subset of operations from $C$ that can start earliest (i.e., all operations in $\bar{C}$ can start at the same time if added to the current partial schedule and no other operation from $C$ can start earlier).
  In this case no idle time on machines occurs if operations are available for processing and, thus, the resulting schedule is a non-delay schedule.

- Let $u$ be an operation from $C$ that finishes earliest if added to the current partial schedule and let $i$ be the machine on which $u$ has to be scheduled.  Select $\bar{C}$ as the subset of operations from $C$ that have to be scheduled on machine $i$ and can start before $u$ would finish.
  In this case, no operation can be processed earlier without delaying another operation and, thus, the resulting schedule is an active schedule.

- Select the complete set $C$ (i.e., $\bar{C} = C$).
  In this case the resulting schedule will be a semi-active schedule.

From a theoretical point of view, the second choice is preferable since the set of non-delay schedules generated by the first choice in general does not have to contain an optimal schedule and since the additional solutions that can be generated by the third choice give no further improvements.  However, the first and third choice reflect the way planning in production plants is often carried out.

A more elaborate constructive heuristic for the job shop problem is the *shifting bottleneck heuristic* (SBH). In the remainder of this subsection we describe its basic version.

The basic idea of the SBH is to schedule the $m$ machines one by one.  In each stage the disjunctions belonging to a specific machine are fixed taking into account the selections already made for scheduled machines.  Furthermore, at the end of each stage, the partial schedule belonging to the fixed machines is re-optimized (see Figure 14.2 for the basic structure of SBH).

$M := \{M_1, \ldots, M_m\}$; $M_0 := \emptyset$;
**REPEAT**
      choose a machine $M_i$ from $M$;
      fix precedence relations for $M_i$;
      $M := M \setminus \{M_i\}$; $M_0 := M_0 \cup \{M_i\}$;
      re-optimize machines from $M_0$;
**UNTIL** $M = \emptyset$.

**Figure 14.2.**   Basic Structure of the Shifting Bottleneck Heuristic

The choice of the machine to be fixed next, the scheduling of this machine, and the re-optimization all rely on the same process: given a partial schedule via fixed disjunctions for a subset of machines, analyze the situation on a specific machine. The fixed disjunctions lead to earliest possible start times (release dates) and minimal delivery times (which can be transformed into due dates, see Chapter 3) for the operations to be scheduled on the specified machine. The resulting problem is a single-machine head-body-tail problem, for which an efficient branch-and-bound method exists to determine the minimal maximum lateness. The machine to be fixed next is chosen to be the machine with the largest maximum lateness of the head-body-tail problem (i.e., the bottleneck machine). On this machine, the disjunctions are fixed according to the optimal solution of the head-body-tail problem. For the re-optimization, the fixed machines (machines from $M_0$) are treated consecutively by first removing their fixed disjunctions and then rescheduling them according to the optimal solution of the head-body-tail problem that results after dropping the given schedule on the machine.

The SBH is a effective heuristic for the job shop problem originally developed by Adams, Balas and Zawack [1988]. Later on, several variants of the method and integrations of the method in other heuristics have been presented. The variations mainly differ in the way how the re-optimization is organized. Finally, it is worthwhile to mention that the basic structure of SBH also can be used to solve generalizations of the job shop problem.

### 14.2.2.   Local search

The methods presented in the previous subsection basically construct one feasible solution. However, as in the re-optimization step of SBH, it may be worthwhile to change this solution somehow and achieve other, hopefully better, feasible solutions. Local search methods form an important and widely used class of such approaches. For an overview of the literature on local search, we refer to Aarts and Lenstra [1997] and Osman and Laporte [1996].

In the following we first give a general overview of these methods and, afterwards, show how they can be adopted to the job shop problem.

choose an initial solution $s \in S$;
**REPEAT**
    choose a solution $s' \in N(s)$;
    $s := s'$;
**UNTIL** stopping criteria.

**Figure 14.3.**   Basic Structure of Local Search

**General overview.** Local search methods form a generic class of heuristics. Generally speaking, local search methods iteratively move through the set of feasible solution. Based on the current and maybe previously visited solutions, a new solution is chosen. The choice of the new solution is restricted to solutions that are somehow close to the current solution (in the 'neighborhood' of the current solution).

Looking at the general structure of local search, we may conclude that one of the basic ingredient of local search is the *neighborhood*. The neighborhood determines in an essential way how the search process will behave. To define a neighborhood on a set $S$ of feasible solutions, we have to specify a corresponding set of *neighbors* $N(s) \subset S$ for each solution $s \in S$. In principle, these sets $N(s)$ may be arbitrary subsets of $S$. However, following the idea of local search, these solutions should be somehow 'similar' to $s$. A systematic way of defining neighborhoods is to specify a set $OP$ of *operators* $op : S \to S$ which perturb solutions in some way. In this case, the set of neighbors of a solution $s$ is defined by $\mathcal{N}(s) = \{op(s) | op \in OP\}$.

Using the notion of neighborhoods, the basic structure of local search is given in Figure 14.3. This basic local search algorithm may be made concrete in various ways. Depending on the method of choosing solutions from the neighborhood of the current solution and the way in which the stopping criteria are defined, we get different local search methods.

A very intuitive and natural way to make concrete the choice of a solution from the neighborhood of the current solution and the stopping criteria in the algorithm of Figure 14.3 is to make in each step locally the best choice and to stop if this does not lead to an improvement. This method is called *iterative improvement* and has its roots in the year 1958. Its main characteristic is that it stops if a local optimum with respect to the given neighborhood has been reached. Generally, this local optima depends on the chosen initial solution and no information is available on how much the quality of this solution differs from that of the global optimum.

It took until the mid-1980s before alternative local search methods were developed that did not terminate at the first local optimum. The first method of this type was *simulated annealing*. This method has two stochastic elements. First, a solution is chosen from the set of neighbors of the current solution according to some given distribution (each neighbor usually has the same probability). Afterwards, depending on the difference between the objective values of the chosen and the current solution, it is decided whether we move to the chosen solution or stay with the current solution. If the chosen solution has a better objective value, we always move to this solution. Otherwise, we move to this solution with a probability that depends on the difference between the two objective values. More precisely, if $s_1$ denotes the current solution and $s_2$ the chosen solution, we move to $s_2$ with probability

$$p(s_1, s_2) = e^{\min\{c(s_1) - c(s_2), 0\}/c}, \tag{14.1}$$

where $c(s)$ denotes the objective value of a solution $s$. The parameter $c$ is a positive control parameter that decreases with increasing number of iterations and converges to 0. This has the effect that the probability of moving to a solution with larger objective value decreases in the course of time and that, finally, almost only improving solutions are accepted. Furthermore, the probability (14.1) has the property that large deteriorations of the objective function are accepted with lower probability than small deteriorations. Under certain conditions for the neighborhoods and the way in which the control parameter $c$ is decreased, it is possible to prove that simulated annealing is asymptotically an optimization algorithm. The proof is based on results on Markov chains.

Another local search method that allows nonimproving moves is the *tabu search* method. Its central idea is to deterministically guide the local search and to use information from the past to avoid cycling. The most straightforward realization of this idea is to always choose the best solution from the neighborhood (steepest descent or, if no descent is possible, mildest ascent), but to reduce the neighborhood by all solutions already visited. However, for practical reasons it is not possible to keep track of all solutions already visited and, thus, the process of reducing the set of neighbors is realized in a different way. Generally, the search for the next solution $s' \in N(s)$ is restricted to a subset $N_{red} \subset N(s)$, whereby either a set of operators from *OP* or solutions with certain properties are forbidden (tabu). These restrictions vary in time and are chosen so that it is not possible to revisit a certain solution within a given number of iterations. More precisely, depending on the last $k$ moves, a set of operators or a set of properties is defined as tabu. Since by this approach not only already visited solutions but also other solutions may be excluded from the search process, so-called *aspiration criteria* are used to overrule the tabu status of a solution. A simple aspiration criterion is that a solution has a better objective value than the best solution found thus far (such a solution has not been visited before and, thus, should not be tabu). However, more elaborate aspiration criteria can also be used.

Another type of solution approaches to solve combinatorial optimization problems is given by *genetic algorithms*. These methods are based on principles from population genetics and the theory of evolution. Roughly speaking, genetic algorithms start with some subset of feasible solutions (a population) and iteratively replace the current population by a next population. Thus, in principle, genetic algorithms are local search methods on subsets of solutions.

The process of building new populations is organized by two main operators. The first is called *mutation* and is applied with some given probability to each solution of the current population. A mutation changes the given solution slightly. Mostly, there are several alternatives in changing the solution and the choice for one of them is done randomly. Thus, a mutation is nothing but a random selection of one solution in the neighborhood of the given solution, whereby the neighborhood is defined by the possible changes mutations can perform. The second operator is called *recombination*. Recombination is used to produce a subset of new solutions (offsprings). To achieve this goal, pairs of solutions (parents) are selected from the current population and for each pair two new solutions (children) are constructed by 'shuffling' the given solutions. Again, there are several ways of shuffling solutions and the choice between these alternatives will be done randomly. For selecting pairs of solutions the quality of the solutions (objective values) plays an important role. Often, the parents are selected randomly using a distribution on the current population that depends on the quality of the solutions.

After applying mutations and recombinations a new population has to be determined. This new population will be a subset of the union of the population after mutation and the set of new solutions resulting from recombination. It is often chosen as the subset of the $k$ best solutions, where $k$ denotes the cardinality of the population.

**Application to the job shop problem.**  One main step for applying local search to a job shop problem is to decide on the underlying neighborhood. Most of the commonly used neighborhoods rely on the representation of solutions by complete orientations and change the orientation of one or more edges. In this context, two properties are helpful:

- reversing the orientation of an edge on a critical path of a feasible solution leads again to a feasible solution (see Exercise 14.2),

- only changes that effect the first or last operation of a block have the potential to lead to an improving solution.

The first property gives a sufficient condition to get feasible neighbors and the second indicates which solutions are non-promising with respect to the objective value. However, due to the nature of local search even a bad solution with respect to the objective value may be a good solution for the further search process. Furthermore, the given sufficient condition on feasibility may be much too restrictive for the search. Therefore, only a few of the following neighborhoods rely completely on these properties.

- the critical path interchange neighborhood $N_{inter}^{CP}$
  $N_{inter}^{CP}$ allows the interchange of two adjacent operations of a block on the critical path (i.e., the reversion of the corresponding edge) and, thus, is based on the first property.

- the end-of-block interchange neighborhood $N_{inter}^{end-block}$
  $N_{inter}^{end-block}$ is a sub-neighborhood of $N_{inter}^{CP}$, where only the interchange of the first two operations or the last two operations of a block are allowed. This reduction is based on the second property, since the interchange of two internal adjacent operations of a block cannot lead to an improved solution.

- the critical path permutation of three neighborhood $N_{3-perm}^{CP}$

  $N_{3-perm}^{CP}$ is an extension of $N_{inter}^{CP}$ and considers, in addition to the interchange of two adjacent operations of a block, the effect if after this interchange the machine predecessor or machine successor is interchanged with the two operations. More precisely, for four consecutive operations $p, v, w, s$ on machine with $v, w$ lying on a critical path, the following sequences are considered as neighbors if they lead to a feasible solution: $(p, w, v, s)$, $(w, p, v, s)$, $(w, v, p, s)$, $(p, w, s, v)$, and $(p, s, w, v)$.

- the end-of-block permutation of three neighborhood $N_{3-perm}^{end-block}$

  Again, $N_{3-perm}^{end-block}$ is the sub-neighborhood of $N_{3-perm}^{CP}$, where only neighboring operations $v, w$ are considered which are the first two operations or the last two operations of a block.

- the shift to the end of a block neighborhood $N_{shift}^{block}$

  $N_{shift}^{block}$ allows a shift of an operation of a block immediately in front of or after the other operations of its block, provided that the resulting solution is feasible. Otherwise, the operation is moved as far as possible to the beginning or end of the block resulting still in a feasible solution. Again, this neighborhood is based on the second property.

- the shift on the same machine neighborhood $N_{shift}^{machine}$

  $N_{shift}^{machine}$ executes shifts for two operations $v, w$ occurring on a critical path and scheduled on the same machine. If $v$ is scheduled before $w$ on the critical path, shifting $v$ directly after $w$ and shifting $w$ directly before $v$ are possible neighbors. However, if both the machine predecessor of $v$ and the machine successor of $w$ are on the critical path, the resulting neighbor cannot improve the current solution (see Exercise 14.3) and is not included in the neighborhood.

- the end-of-block three interchange neighborhood $N_{3-inter}^{end-block}$

  The basic idea of this neighborhood is the same as for $N_{inter}^{end-block}$. However, under certain conditions, not only the two adjacent operations $v, w$ are interchanged but also the job successor of the first operation ($v$) is interchanged with its machine successor and one of the job predecessors of the second operation ($w$) is interchanged with its machine predecessor.

- the shifting bottleneck neighborhood $N_{SBH}$

  A neighbored solution for $N_{SBH}$ is obtained by removing all orientations on a machine with at least one operation on a critical path and replacing them by any other orientation leading to a feasible solution.

- the several machine shifting bottleneck neighborhood $N_{SBH}^{several}$

  In contrast to $N_{SBH}$, not only the orientation on one machine but the orientations on $m - t$ machines are replaced for this neighborhood ($t$ is a small number depending on $m$).

Most of the presented neighborhoods change the processing order on only one machine. Only $N_{3-inter}^{end-block}$ and $N_{SBH}^{several}$ change orders on more than one machine.

For local search methods like iterative improvement, simulated annealing, and tabu search a choice for one of the above defined neighborhoods is sufficient to describe the search process. However, for genetic algorithms the situation is a bit more complex. Whereas the above neighborhoods may be used for mutations, also operators for the recombination have to be given. One possible realization of the recombination, which is somehow similar to the above defined neighborhoods is given as follows:

- recombination by interchanges $R_{inter}^{random}$
  Given two solutions $S_1$ and $S_2$ an offspring is constructed by repeating $\lceil nm/2 \rceil$ times: select randomly an edge $\{v, w\}$; if this edge is oriented in $S_1$ and $S_2$ in different ways, and if it belongs in $S_1$ to a critical path, interchange $v$ and $w$ in $S_1$. The resulting solution $S_1$ is the offspring.

Other realizations of the recombination rely on constructive heuristics for the job shop problem. The simplest one uses the priority driven heuristic presented in Subsection 14.2.1, which led to an active schedule:

- recombination by constructing active schedules $R_{heur}^{active}$
  Given two solutions $S_1$ and $S_2$ an offspring is constructed by a priority driven constructive heuristic using the following priority rule for the operations in $\bar{C}$ ($\bar{C}$ is determined by the second possibility given in Subsection 14.2.1): choose the operation which starts first in $S_1$ with probability $(1-\varepsilon)/2$, the operation which starts first in $S_2$ with probability $(1-\varepsilon)/2$, and randomly one of the other operations with probability $\varepsilon$ ($\varepsilon$ is a given small positive value).

All the presented neighborhoods and recombination methods may be realized using the mixed graph representation of the job shop problem. However, there also exists local search methods, especially genetic algorithms, which use different representations. One of these representations based on a constructive heuristic and perturbation of the data will be given in the notes.

Combining one of the sketched versions of local search with one of the above defined neighborhoods leads to a suitable heuristic for solving the job shop problem. However, to get efficient methods often some variations or adaptions of the methods are useful. In the following, some variations or adaptions of local search methods proposed for the job shop problem are given.

- Iterative improvement:

  This method stops in the first local optima and, therefore, evaluates in general only a few solutions. Thus, to make it somehow compatible to the other local search methods, a multi-start strategy (i.e., start the search for several different starting solutions) has to be applied.

- Simulated annealing:

Besides the 'pure' variant also backtracking strategies (after a certain number of iterations without improving the best found solution so far, move back to the best found solution and restart the search) or a combination with iterative improvement (do not evaluate a neighbor directly, but investigate its potential for the further search by applying (a few) iterations of iterative improvement to the neighbor before evaluating it) have been proposed.

- Tabu search:

  For tabu search several variations have been proposed in the literature. The first type of variation depends on the definition of the tabu status. In general a list (tabu list) keeps track of the last $k$ moves and the entries of the tabu list are used to define operators or solutions as tabu. On the one hand, the length of this list ($k$) may be fixed or may vary over time. On the other hand, the choice which operators/solutions are declared as tabu in dependence of the used operator may vary and also the strictness of the tabu status may be different (e.g. aspiration criteria).

  Another variation of tabu search results from the incorporation of backtracking strategies. Here one can vary the conditions for a backtracking move (e.g. maximum number of moves without improving the best found solution), the choice for the solution where the search is restarted (e.g. the best solution), and the restart conditions (e.g. use the next best neighbor or use the same solution but with a different length of the tabu list).

  Finally, within tabu search the search for the best non-tabu neighbor may be speeded up by replacing the exact evaluation of the objective value of neighbored solutions by quickly computed bounds on the change of the objective value.

- Guided local search:

  Guided local search uses elements from tabu search and back tracking and was first applied to the job shop problem by Balas and Vazacopoulos. In each step, a neighborhood tree of solutions is constructed, where the current solution serves as root. For each node of the tree, a set of descendants is created by generating neighbored solutions using some neighborhood structure. After creating the tree, one of its best solutions is chosen as the new current solution and, thus, serves as the root of the next tree.

  The main ingredient of guided local search is the construction of the neighborhood tree. Here the depth of the tree, the number of descendants per node, and the choice of the descendants play an important role. To bound the computation times, the depth and the number of descendants are kept small and the selection of descendants is based on lower bounds on the completion time of the neighboring solution and not on an exact evaluation of the makespan. Furthermore, within the generation process it is guaranteed that on a path to a node from the root, none of the neighborhood changes made is reversed.

- Genetic algorithms:

  Several proposed genetic algorithms for the job shop problem incorporate other local search methods. In the simplest variants, the solutions generated by mutations and recombinations are locally optimized by iterative improvement or improved by fast versions of simulated annealing.

  Other genetic algorithms used priority sequences for the machines as solution representation. For such a representation, a corresponding solution is achieved by applying some constructive heuristic based on priority rules. Somehow, these representations are similar to complete orientations. However, the main difference is that the set of sequences may not lead to a complete orientation since the resulting mixed graph may contain cycles. The advantage of the priority sequence based representation is that within the mutation and recombination process one does not have to deal with this feasibility question, since the application of the constructive heuristic acts as a repair mechanism. Some authors suggest that, after the application of the constructive heuristic, to replace the given priority sequence with the concrete sequences on the machines in the constructed schedule.

### 14.2.3. Implementation

Many constructive heuristics for the job-shop problem rely on priority-based dispatching rules. A survey of these methods is given by Haupt [1989]. However, the quality of the solutions achieved by these methods is not very good and these methods are mostly just used within other methods, e.g., to calculate initial solutions.

The more successful constructive method, the shifting bottleneck heuristic, was proposed by Adams et al. [1988]. In their original version, in each iteration they re-optimize all machines from $M_0$ three times in three cycles. In the first cycle the machines are rescheduled in the order they were inserted in $M_0$ and in last two cycles the machines are rescheduled in the order of their maximum lateness in the previous cycle. Note, that these rescheduling operations of the machines can be seen as one neighborhood step in $N_{SBH}$, where the current schedule on the machine is replaced by its best neighbor. Applegate and Cook [1991] present a variation of this method by not fixing the number of cycles for the re-optimization but repeating this step until for no machine an improvement is obtained.

For the job shop problem a large number of different local search methods has been proposed in the literature. It starts with more or less basic versions of the different search methods using a single neighborhood.

- Aarts et al. [1994] test iterative improvement with neighborhoods $N_{inter}^{CP}$ and $N_{3-inter}^{end-block}$.

- The first versions of simulated annealing are given by Matsuo et al. [1988] and van Laarhoven et al. [1992]. Matsuo et al. use neighborhood $N_{3-inter}^{end-block}$ but incorporate also some features of iterative improvement. Van Laarhoven et al. use neighborhoods $N_{inter}^{CP}$ and $N_{3-inter}^{end-block}$.

- Tabu search was first applied by Dell'Amico and Trubian [1993] and Taillard [1994]. Dell'Amico and Trubian use neighborhoods $N_{3-perm}^{end-block}$ and $N_{shift}^{block}$ and a tabu list of variable length, whereas Taillard uses neighborhood $N_{inter}^{CP}$ and tabu list of fixed length. Furthermore, Taillard does not estimate the makespan of all neighbored solutions exactly but uses lower bounds.

- Balas and Vazacopoulos [1998] proposed guided local search using the neighborhood $N_{shift}^{machine}$. They bound the size of the tree by limiting the number of descendants, by fixing part of the orientations, and by bounding the depth of the tree by a logarithmic function of the number of operations.

- The first genetic algorithms for the job shop problem were proposed by Davis [1985] and Falkenauer and Bouffouix [1991]. Both use priority sequences to represent solutions and a heuristic to produce a non-delay schedule from a given representations.

Besides these first, and often basic, local search approaches, various other, more elaborate, applications of local search to the job shop problem are given in the literature. Several of these approaches combine different local search methods or combine local search with the shifting bottleneck heuristic. For genetic algorithms besides the mentioned representation based on orientations or priority sequences, also other, less straightforward, representations have been proposed. E.g., Dorndorf and Pesch [1995] use a sequence of priority rules as representation. Using this sequence a schedule is built up by a simple priority driven heuristic, where in case of a conflict in the $i$-th iteration the priority rule on position $i$ of the sequence is used to solve the conflict. Ponnambalam et al. [2001] present a comparative evaluation of different representations for genetic algorithms for the job shop problem. They compare operation-based, job-based, priority list-based, and priority rule-based representations and come to the conclusion that priority list-based representations lead to the best quality of the solutions and that operation-based representation lead to the smallest computation times.

An overview on the application of local search to the job shop problem can be found in, e.g., Vaessens et al. [1996], Anderson et al. [1997], and Jain and Meeran [1999]. Besides a description of methods, these overviews contain also computational comparisons between approaches. A general conclusion from these comparisons is that hybrid approaches (combination of different methods) are mostly superior to pure approaches. In the remaining of this section we shortly describe the basic ideas behind the three most successful local search approaches for the job shop problem.

Balas and Vazacopoulos [1998] combine the shifting bottleneck heuristic with their guided local search approach. In the basic version they realize the re-optimization step of the shifting bottleneck heuristic by their guided local search approach using the current partial schedule as initial solution. The best schedule obtained in this processed is used to continue with the shifting bottleneck heuristic. Using the solution of this basic version as initial solutions two other variants are proposed. In the first, called *iterated guided local search*, iterations of re-optimization cycles are carried out until no improvement is found. Each cycle removes the orientation of one machine, applies guided local search for a limited number of trees, adds the removed machine again, and applies guided local search to the complete schedule for a limited number of trees. The second variant, called *reiterated guided local search*, uses the solution of the first variant as an initial solution and repeats a fixed number of cycles, where the orientation of $\sqrt{m}$ machines (randomly chosen) are removed, guided local search with a limited number of trees is applied to the resulting partial schedule, and the removed machines are added back by applying the basic version of the method.

Pezzella and Merelli [2000] also use elements from the shifting bottleneck heuristic and combine them with tabu search elements. The initial solution is achieved by a standard shifting bottleneck heuristic. Afterwards tabu search is applied using as neighborhood the neighborhood $N_{inter}^{CP}$ extended by some further neighbors achieved via shifts within blocks. Within the tabu search approach again an element of the shifting bottleneck heuristic is used: every time a new best solution has been found, a re-optimization as within the shifting bottleneck heuristic is applied to all machines involved in the critical path.

The third, and at the moment the best, approach is given by Nowicki and Smutnicki [1996]. Their approach is a tabu search approach using backtracking and the underlying neighborhood is $N_{inter}^{send-block}$. Each time a certain number of iterations yield no improvement to the best solution found, the method restarts from the best solution found with a different neighborhood move. Furthermore, if from one solution all possible restarts have been carried out, this best solution is no longer used for restarts, but the former best solution will act as base for the restarts.

**Exercises**

14.1.  Give an instance of the job shop problem, where no optimal schedule is a non-delay schedule.

14.2.  Show that reversing the orientation of an edge on a critical path of a feasible solution leads again to a feasible solution.

14.3.  For the neighborhood $N_{shift}^{machine}$ solutions resulting from shifts for operations $v, w$, where both the machine predecessor of $v$ and the machine successor of $w$ are on the critical path cannot improve the given solution.

### 14.3.  Approximation algorithms: geometric results

As was true for both open shops and flow shops, it is possible to derive a surprisingly strong result for the job shop problem using the vector sum theorem, which was proved in Chapter 12. The bound on the absolute error of the approximation algorithm will not be as good as it was for the flow shop problem, but it retains the crucial property: the absolute error is independent of the number of jobs, $n$, and is bounded by a function of the number of machines, $m$, the maximum number of operations of a job, $\mu_{max} = \max_j \mu(j)$, and the maximum processing time of an operation, $p_{max} = \max_{h,j} p_{hj}$. More precisely, we will prove the following result, due to Sevast'janov [1986].

**Theorem 14.1.** *For any instance of $J||C_{max}$, a schedule of length at most $C^*_{max} + O(m\mu^3_{max} p_{max})$ can be found in polynomial time.*

This theorem is significantly more difficult to prove than Theorem 13.3, the analogous result for flow shop scheduling. In fact, we will only prove that it holds under certain assumptions about the input. We assume that each job has the same number of operations and that each machine has the same load; that is, $\mu(j) = \mu$ for each $J_j$ and, if $\Pi_i = \sum_{t(h,j)=i} p_{hj}$ and $\Pi_{max} = \max_i \Pi_i$, then $\Pi_i = \Pi_{max}$ for each $M_i$. The algorithm will deliver a schedule of length $\Pi_{max} + O(m\mu^3 p_{max})$. Any instance can be modified to one satisfying these assumptions, without increasing the bound guaranteed; but this extension is left to the reader (see Exercise 14.4).

Since the proof is rather intricate, we will describe the main ideas before proceeding to the details. Focus on the set of the $h$th operations of jobs that must be performed on machine $M_i$, and let $\Pi_{hi}$ denote their total processing time. The key step in the algorithm will be to order the jobs in such a way that, for any $r$ consecutive jobs, the total processing time of their operations that belong to this set is approximately equal to $(r/n)\Pi_{hi}$. Note that a job need not have its $h$th operation on $M_i$, in which case it contributes 0 to the total. For simplicity of notation, assume that the jobs are indexed in this order.

We will use the ordering to partition the job set into $n/r$ sets of $r$ consecutive jobs each. The value of $r$ will be chosen so as to ensure that the proposed schedule is feasible on the one hand and sufficiently short on the other hand; for the time being, simply assume that $n/r$ is integral. Call the job set $\{J_{(g-1)r+1}, ..., J_{gr}\}$ *group $g$*, for $g = 1, ..., n/r$.

The schedule for $M_i$ will consist of a sequence of $n/r + \mu - 1$ *blocks* (cf. Figure 14.4). In the first block, we schedule the first operations of the jobs in group 1 that must go on $M_i$; in the second block, there are first operations from group 2 and second operations from group 1; and so on. A typical block $b$ contains first operations from group $b$, second operations from group $b - 1$, up to $\mu$th operations from group $b - \mu + 1$. The last group does not start its first operations until block $n/r$, and thus finishes only in block $n/r + \mu - 1$.

Note that, for any $b$ $(b = 1, ..., n/r + \mu - 1)$ and $j$ $(j = 1, ..., n)$, there is at most one machine that processes an operation of $J_j$ in its $b$th block. Now consider the total processing time in block $b$ on $M_i$, with $\mu \leq b \leq n/r$. The ordering ensures that the $h$th operations from group $b - h + 1$ that go on $M_i$ total roughly $(r/n)\Pi_{hi}$, for $h = 1, ..., \mu$. These sum, even more roughly, to $(r/n)\sum^\mu_{h=1} \Pi_{hi} = (r/n)\Pi_i = (r/n)\Pi_{max}$. However, this calculation does not apply to, for example, the second block, which contains only first and second operations, summing roughly to $(r/n)(\Pi_{1i} + \Pi_{2i})$. We correct this by adding idle time to certain blocks: for each block without $h$th operations, we introduce idle time of total length $(r/n)\Pi_{hi}$. For example, in the second block we introduce idle time totaling $(r/n)\sum_{h=3}^{\mu}\Pi_{hi}$. As a result, the length of each block on each machine is roughly equal to $(r/n)\Pi_{max}$.

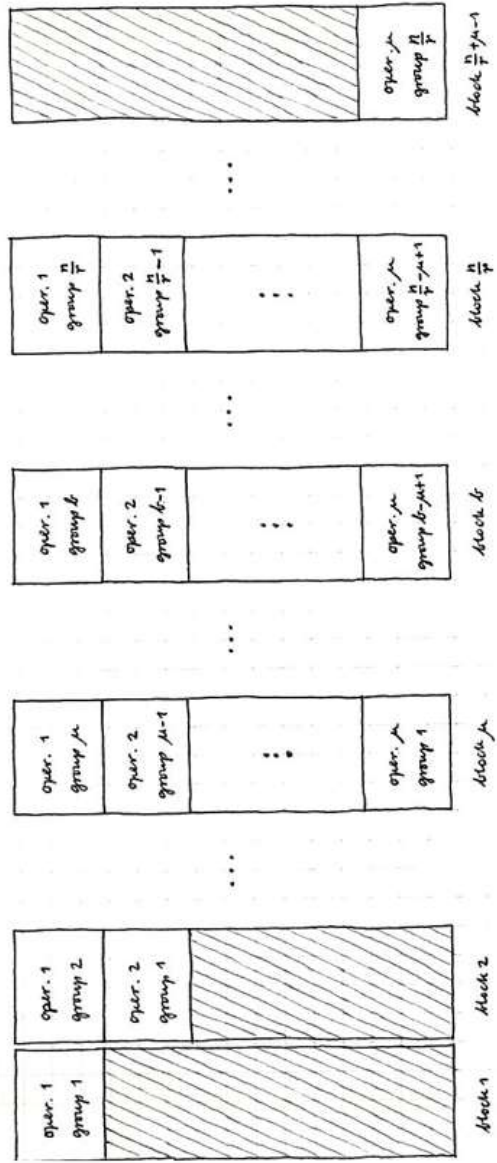

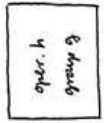

**Figure 14.4.**   Schedule for $M_i$

An intuitive interpretation of this schedule is that we try to group the jobs so that their operations can be synchronized and pipelined. The synchronization is achieved because each block takes roughly the same amount of time on each machine, and the pipelining is achieved because the starting times of the jobs are staggered by groups. What is the length of this schedule? Since each block is roughly of length $(r/n)\Pi_{max}$, and there are $n/r+\mu-1$ blocks, we get a schedule of length

$$\Pi_{max}+\frac{r}{n}(\mu-1)\Pi_{max}. \tag{14.2}$$

Unfortunately, there is a difference between blocks of roughly equal length, and blocks of equal length. The main problem with roughly equal length blocks is that it is possible that two consecutive operations of the same job are done simultaneously or even out of order. For example, block $b$ on $M_i$ may end somewhat later than block $b+1$ starts on $M_{i'}$. If $O_{hj}$ is scheduled late within block $b$ on $M_i$ and $O_{h+1,j}$ is scheduled early within block $b+1$ on $M_{i'}$, then $O_{h+1,j}$ may start before $O_{hj}$ is completed. This can be fixed by paying attention to the order in which the operations are performed within a block. We will subdivide each block into $r$ *phases*, and guarantee that all of the operations of the same job are scheduled within the same phase of subsequent blocks. As long as $r$ is large enough, two operations of a job will not overlap. In fact, we will choose $r$ to be roughly equal to $nm\mu^2 p_{max}/\Pi_{max}$. If this is substituted into expression (14.2) for the schedule length, we get the claimed result.

After this motivation, giving the proof is just a matter of writing out the precise equations and bounds that correspond to the intuition. We first make explicit the properties of the desired ordering of the jobs. The following notation for the processing times will be convenient:

$$p_{hj}^i=\begin{cases} p_{hj} & \text{if } \mathfrak{i}(h,j)=i,\\ 0 & \text{otherwise.}\end{cases}$$

**Lemma 14.2.** *For any instance of $J||C_{max}$, a permutation $\pi$ of $\{1,...,n\}$ such that*

$$\frac{k}{n}\Pi_{hi}-m\mu p_{max}\leq\sum_{j=1}^{k}p_{h\pi(j)}^i\leq\frac{k}{n}\Pi_{hi}+m\mu p_{max},h=1,...,\mu,i=1,...,m,k=1,...,n,$$

*can be found in polynomial time.*

*Proof.* This is a straightforward corollary of Theorem 12.4. For $j=1,...,n$, let $v_j$ be the $(m\mu)$-dimensional vector with components $p_{hj}^i-(\Pi_{hi}/n)$ ( $h=1,...,\mu$, $i=1,...,m$ ). By definition, $\sum_{j=1}^{n}p_{hj}^i=\Pi_{hi}$ for all $h$ and $i$, so that $\sum_{j=1}^{n}v_j=0$. Since $||v_j||\leq p_{max}$, Theorem 12.4 implies that a permutation $\pi$ can be found in polynomial

time such that

$$-m\mu p_{\max} \le \sum_{j=1}^{k} p_{h\pi(j)}^i - \frac{k}{n}\Pi_{hi} \le m\mu p_{\max}, h = 1, ..., \mu, i = 1, ..., m, k = 1, ..., n.$$

This is equivalent to the statement of the lemma. □

We will use this permutation $\pi$ to construct the schedule. For simplicity of notation, we reindex the jobs so that the identity permutation satisfies the property stated in the lemma. Focus on block $b$ on $M_i$. Recall that this block consists of the $h$th operations from group $b - h + 1$ that go on $M_i$, where $h$ ranges over all or part of the set $\{1, ..., \mu\}$ (cf. Figure 14.4). Recall also that group $b - h + 1$ consists of the jobs $J_{(b-h)r+1}, J_{(b-h)r+2}, ..., J_{(b-h+1)r}$. Hence, the operations processed in block $b$ on $M_i$ are contained in the following matrix:

$$\begin{bmatrix} O_{1,(b-1)r+1} & O_{1,(b-1)r+2} & \cdots & O_{1,br} \\ O_{2,(b-2)r+1} & O_{2,(b-2)r+2} & \cdots & O_{2,(b-1)r} \\ \vdots & \vdots & & \vdots \\ O_{\mu,(b-\mu)r+1} & O_{\mu,(b-\mu)r+2} & \cdots & O_{\mu,(b-\mu+1)r} \end{bmatrix} \quad (14.3)$$

To make things precise, we note that, for a given block index $b (1 \le b \le n/r + \mu - 1)$, the operation index $h$ ranges from $\max\{1, b - n/r + 1\}$ to $\min\{\mu, b\}$. An illegitimate value of $h$ implies a group index $g$ outside the range $\{1, ..., n/r\}$ and a job index outside the range $\{1, ..., n\}$.

Phase $s$ of block $b$ will contain the operations in column $s$ of matrix (14.3), for $s = 1, ..., r$. The schedule for this phase on $M_i$ is formed by considering the operations in column $s$ in order of increasing row index $h$. If such an operation $O_{hj}$ is not on $M_i$ or, in other words, if $p_{hj}^i = 0$, then continue to the next operation in the column. If an operation $O_{hj}$ is undefined or, in other words, if $h$ is outside its range, then let $M_i$ be idle for $\Pi_{hi}/n$ units of time. Otherwise, schedule $O_{hj}$ on $M_i$.

This completes the description of the schedule. We now have to show that it is a feasible schedule, and that its length is within the claimed bound. The following lemma is the key to both propositions. Let $z$ denote the total number of phases of the schedule, i.e., $z = r(n/r + \mu - 1) = n + r(\mu - 1)$, and let $C_{it}$ denote the time at which the first $t$ phases of the schedule are completed on $M_i$.

**Lemma 14.3.** *The schedule described above satisfies*

$$\frac{t}{n}\Pi_{\max} - m\mu^2 p_{\max} \le C_{it} \le \frac{t}{n}\Pi_{\max} + m\mu^2 p_{\max}, i = 1, ..., m, t = 1, ..., z.$$

*Proof.* Focus on a machine $M_i$ and a phase $t$. To obtain these bounds on $C_{it}$, we first fix $h$ $(1 \le h \le \mu)$ and compute the total time allocated to $M_i$ during the first $t$ phases, either to process $h$th operations or to be idle whenever $h$ is outside the range induced by the phase value. Let us extend the notation $p_{hj}^i$ to include these idle periods:

$$p_{hj}^i = \begin{cases} p_{hj}^i & \text{if } O_{hj} \text{ is defined,} \\ \Pi_{hi}/n & \text{otherwise.} \end{cases}$$

It is easily seen from the above matrix that the (possibly undefined) $h$ th operation that is slated for phase 1 is $O_{h,(1-h)r+1}$. Thus, the total time (both processing and idle) associated with $h$th operations on $M_i$ during the first $t$ phases is given by $\sum_{j=(1-h)r+1}^{(1-h)r+t} p_{hj}^i$. Suppose that $t'$ of these phases correspond to undefined operations, and that the remaining $t - t'$ correspond to legitimate operations (though not necessarily on $M_i$). We consider their contributions to this sum separately. Each of the former contributes an idle period of length $\Pi_{hi}/n$, which sum to $t'\Pi_{hi}/n$. The latter phases correspond to the first $t - t'$ jobs, and Lemma 14.2 implies that these sum to within $m\mu p_{\max}$ of $(t - t')\Pi_{hi}/n$. We see that

$$\frac{t}{n}\Pi_{hi} - m\mu p_{\max} \leq \sum_{j=(1-h)r+1}^{(1-h)r+t} p_{hj}^i \leq \frac{t}{n}\Pi_{hi} + m\mu p_{\max}.$$

Summing these inequalities over all possible $h$, we get

$$C_{it} = \sum_{h=1}^{\mu} \sum_{j=(1-h)r+1}^{(1-h)r+t} p_{hj}^i \geq \sum_{h=1}^{\mu} \left(\frac{t}{n}\Pi_{hi} - m\mu p_{\max}\right) = \frac{t}{n}\Pi_{\max} - m\mu^2 p_{\max}$$

and

$$C_{it} = \sum_{h=1}^{\mu} \sum_{j=(1-h)r+1}^{(1-h)r+t} p_{hj}^i \leq \sum_{h=1}^{\mu} \left(\frac{t}{n}\Pi_{hi} + m\mu p_{\max}\right) = \frac{t}{n}\Pi_{\max} + m\mu^2 p_{\max}.$$

$\square$

In order to ensure feasibility of the schedule, we choose $r$ so large that, irrespective of the machine, phase $br + s$ finishes no later than phase $(b+1)r + s$ starts. In other words, we want to guarantee that $C_{it} \leq C_{i't+r-1}$ for any pair $(M_i, M_{i'})$ and any $t$ ( $t = 1, ..., z - r$ ). By Lemma 14.3, it suffices to make sure that

$$\frac{t}{n}\Pi_{\max} + m\mu^2 p_{\max} \leq \frac{t+r-1}{n}\Pi_{\max} - m\mu^2 p_{\max}$$

or, equivalently,

$$2m\mu^2 p_{\max} \leq \frac{r-1}{n}\Pi_{\max}.$$

Therefore, if we set $r = lc2nm\mu^2 p_{\max}/\Pi_{\max}rc + 1$, then we get a feasible schedule.

As for the length of this schedule, we substitute this value of $r$ into the upper bound on $C_{iz}$ given by Lemma 14.3, recalling that $z = n + r(\mu - 1)$. We conclude that the schedule is no longer than

$$\frac{n + r(\mu - 1)}{n} \Pi_{\max} + m\mu^2 p_{\max}$$

$$= \Pi_{\max} + (lc\frac{2nm\mu^2 p_{\max}\Pi_{\max}}{r}c + 1)(\mu - 1)\frac{\Pi_{\max}n}{+} m\mu^2 p_{\max}$$

$$= \Pi_{\max} + O(m\mu^3 p_{\max}).$$

This completes the proof of Theorem 14.1.

**Exercises**

**14.4.** Show that Theorem 14.1 holds for any job shop instance. In particular, show how to pad the input so that each job has the same number of operations and each machine has the same load, without increasing the bound stated by the theorem. (*Hint*: To balance the load, consider increasing some of the $p_{hj}^i$ for which $\iota(h, j) \neq i$.) Does the proof really require that $n/r$ be integral?

**14.5.** Consider the special case of the job shop problem when both $m$ and $\mu_{\max}$ are fixed. Show that, for any $\varepsilon > 0$, there exists a $(2 + \varepsilon)$-approximation algorithm.

**14.6.** Consider the following generalization of the job shop problem: each job consists of a set of operations, whose order is constrained by a precedence relation, and yet no two of its operations may be processed simultaneously; the objective is to minimize the maximum completion time. (When the precedence relation of each job is a chain, we have the job shop problem.) Give a polynomial-time algorithm for this problem that delivers a solution of length $C_{\max}^* + O(m\mu_{\max}^3 p_{\max})$.

**14.7.** (a) Consider the special case $J|p_{ij} = 1|C_{\max}$. If one relaxes the capacity constraint of each machine, then there is a schedule of length $\mu_{\max}$, in which each $O_{ij}$ starts at time $i - 1$. Now, consider the schedules formed by delaying the start of each $J_j$ by some amount $t_j$, but then scheduling the operations without idle time; that is, $O_{ij}$ is started at time $t_j + i - 1$. Prove that, if each $t_j$ is chosen independently and uniformly at random in the range from 1 to $\Pi_{\max}$, then with high probability, no more than $4 \log(n\mu_{\max})$ operations are assigned to a machine at any time.

(b) Use this to give a randomized algorithm which, for every input, produces a schedule of length $O(\log(n\mu_{\max})C_{\max}^*)$ and is expected to run in polynomial time.

(c) By using Theorem 14.1, improve the bound in (b) to $O(\log(m\mu_{\max})C_{\max}^*)$.

# Contents

**i**

# 15

# Stochastic Scheduling Models

Michael L. Pinedo
*New York University*

## 15.1. Preliminaries

Production environments in real life are subject to many sources of uncertainty. These sources of uncertainty include machine breakdowns, unexpected releases of high priority jobs, i.e., jobs with large weights, and the unpredictability of processing times which are often not known in advance. Thus a good model of a scheduling problem would need to address these forms of uncertainty.

The goal of this chapter is not to give an exhaustive overview of the field of stochastic scheduling. It is rather meant to give an overview of stochastic counterparts of the specific deterministic models considered in the previous chapters and draw some comparisons between deterministic models and stochastic models. Because of space limitations, we present at times the known results not in their full generality, especially when an elaborate framework is needed to present such results.

The first section of this chapter goes over the notation, the classes of distributions, the various forms of stochastic dominance, and the different classes of scheduling policies.

In what follows, it is assumed that the *distributions* of the processing times, release dates and due dates are all known in advance, that is, at time zero. The actual *outcome* or *realization* of a random processing time only becomes known upon the completion of the processing; the realization of a release date or due date becomes known only at the point in time at which it actually occurs.

For this chapter we adopt the following notation. Random variables are capitalized, while the actual realized values are in lower case. Job $j$ has the following quantities of interest associated with it.

$X_{ij}$ = the random processing time of job $j$ on machine $i$; if job $j$ is only to be processed on one machine, or if it has the same processing times on each of the machines it may visit, the subscript $i$ is omitted.

$1/\lambda_{ij}$ = the mean or expected value of the random variable $X_{ij}$.

$R_j$ = the random release date of job $j$.

$D_j$ = the random due date of job $j$.

$w_j$ = the weight (or importance factor) of job $j$.

This notation is not completely analogous to the notation used for the earlier deterministic models. The reason $X_{ij}$ is used as the processing time in stochastic scheduling is due to the fact that $P$ usually refers to a probability. The weight $w_j$, similar to that in the deterministic models, is basically equivalent to the cost of keeping job $j$ in the system for one unit of time. In the queueing theory literature, which is closely related to stochastic scheduling, $c_j$ is often used for the weight or cost of job $j$. The $c_j$ and the $w_j$ are equivalent.

Distributions and density functions may take many forms. In what follows, for obvious reasons, only distributions of nonnegative random variables are considered.

A random variable from a continuous time distribution may assume any real nonnegative value within one or more intervals. A distribution function is typically denoted by $F(t)$ and its density function by $f(t)$, i.e.,

$$F(t) = P(X \le t) = \int_0^t f(t)\, dt,$$

where

$$f(t) = \frac{dF(t)}{dt}$$

provided the derivative exists. Furthermore,

$$\bar{F}(t) = 1 - F(t) = P(X \ge t).$$

An important example of a continuous time distribution is the *exponential* distribution. The density function of an exponentially distributed random variable $X$ is

$$f(t) = \lambda e^{-\lambda t},$$

and the corresponding distribution function is

$$F(t) = 1 - e^{-\lambda t},$$

which is equal to the probability that $X$ is smaller than $t$ (see *Figure* **??**). The mean or expected value of $X$ is

$$E(X) = \int_0^\infty t f(t)\, dt = \int_0^\infty t\, dF(t) = \frac{1}{\lambda}.$$

The parameter $\lambda$ is called the *rate* of the exponential distribution.

Another important distribution is the *deterministic* distribution. A deterministic random variable assumes a given value with probability one.

The *completion rate* $c(t)$ of a continuous time random variable $X$ with density function $f(t)$ and distribution function $F(t)$ is defined as follows:

$$c(t) = \frac{f(t)}{1 - F(t)}.$$

This completion rate is equivalent to the failure rate or hazard rate in reliability theory. For an exponentially distributed random variable $c(t) = \lambda$ for all $t$. That the completion rate is independent of $t$ is one of the reasons why the exponential distribution plays an important role in stochastic scheduling. This property is closely related to the *memoryless* property of the exponential distribution, which implies that the distribution of the *remaining* processing time of a job which already has been processed for an amount of time $t$, is still exponentially distributed with rate $\lambda$ and therefore identical to its processing time distribution at the very start of its processing.

Distributions can be classified based on their completion rate. An *Increasing Completion Rate (ICR)* distribution is defined as a distribution whose completion rate $c(t)$ is increasing in $t$, while a *Decreasing Completion Rate (DCR)* distribution is defined as a distribution whose completion rate is decreasing in $t$.

A subclass of the class of *ICR* distributions is the class of *Erlang(k, λ)* distributions. The *Erlang(k, λ)* distribution is defined as

$$F(t) = 1 - \sum_{r=0}^{k-1} \frac{(\lambda t)^r e^{-\lambda t}}{r!}.$$

The *Erlang(k, λ)* is a $k$-fold convolution of the exponential distribution and has a mean $k/\lambda$. Thus, if $k$ equals one, the distribution is an exponential with mean $1/\lambda$ and if $k$ and $\lambda$ both go to $\infty$, then the distribution becomes a constant, i.e., a deterministic distribution.

A subclass of the class of *DCR* distributions is the class of *mixtures of exponentials*. A random variable $X$ is distributed according to a mixture of exponentials if it is exponentially distributed with rate $\lambda_j$ with probability $p_j$, $j = 1, \ldots, n$, and

$$\sum_{j=1}^n p_j = 1.$$

The exponential as well as the deterministic distribution are special cases of *ICR*

distributions. The exponential distribution is *DCR* as well as *ICR*. The class of *DCR* distributions contains other special distributions. For example, let $X$ be distributed as follows: with probability $p$ it is exponentially distributed with rate $p$ and with probability $1 - p$ it is zero. Clearly, $E(X) = 1$. When $p$ is very close to zero this distribution will be referred to as an *Extreme Mixture of Exponentials (EME)*.

One way of measuring the variability of a distribution is through its coefficient of variation $C_v$. The coefficient of variation $C_v$ is defined as the variance of the distribution divided by the square of the mean, i.e.,

$$C_v = \frac{Var(X)}{(E(X))^2} = \frac{E(X^2) - (E(X))^2}{(E(X))^2}.$$

It can be verified easily that the $C_v$ of the deterministic distribution is zero and of the exponential distribution one. The $C_v$ of an extreme mixture of exponentials may be arbitrarily large (it goes to $\infty$ when $p$ goes to 0).

### 15.2. Stochastic Dominance and Classes of Policies

In stochastic scheduling random variables often have to be compared with one another. There are many ways in which comparisons between random variables can be made. Comparisons based on certain properties are typically referred to as *stochastic dominance*, i.e., a random variable *dominates* another with respect to some stochastic property.

**Definition 15.1.** *(i) The random variable $X_1$ larger in expectation than the random variable $X_2$ if $E(X_1) \geq E(X_2)$.*

*(ii) The random variable $X_1$ is said to be stochastically larger than the random variable $X_2$ if*

$$P(X_1 > t) \geq P(X_2 > t)$$

*or*

$$1 - F_1(t) \geq 1 - F_2(t)$$

*for all $t$. This ordering is usually referred to as stochastic ordering and is denoted by $X_1 \geq_{st} X_2$.*

*(iii) The random variable $X_1$ is almost surely larger than or equal to the random variable $X_2$ if $P(X_1 \geq X_2) = 1$. This ordering implies that the density functions $f_1$ and $f_2$ may overlap on at most one point. This ordering is denoted by $X_1 \geq_{a.s.} X_2$.*

Ordering in expectation is the crudest form of stochastic dominance. Stochastic ordering implies ordering in expectation since

$$E(X_1) = \int_0^\infty t f_1(t) dt = \int_0^\infty (1 - F_1(t)) dt = \int_0^\infty \bar{F}_1(t) dt.$$

The three forms of stochastic dominance described above all imply that the random variables being compared, in general, have different means. They lead to the following chain of implications.

$$\boxed{\text{almost surely larger} \implies \text{stochastically larger} \implies \text{larger in expectation}}$$

There are several other important forms of stochastic dominance that are based on the *variability* of the random variables assuming that the *means are equal*. In the subsequent definitions three such forms are presented. One of these is defined for density functions which are symmetric around the mean, i.e.,

$$f(E(X)+t) = f(E(X)-t)$$

for all $0 \leq t \leq E(X)$. Such a density function then has an upper bound of $2E(X)$.

**Definition 15.2.** *(i) The random variable $X_1$ is said to be larger than the random variable $X_2$ in the variance sense if the variance of $X_1$ is larger than the variance of $X_2$.*

*(ii) The random variable $X_1$ is said to be more variable than the random variable $X_2$ if*

$$\int_0^\infty h(t)dF_1(t) \geq \int_0^\infty h(t)dF_2(t)$$

*for all convex functions h. This ordering is denoted by $X_1 \geq_{cx} X_2$.*

*(iii) The random variable $X_1$ is said to be symmetrically more variable than the random variable $X_2$ if the density functions $f_1(t)$ and $f_2(t)$ are symmetric around the same mean $1/\lambda$ and $F_1(t) \geq F_2(t)$ for $0 \leq t \leq 1/\lambda$ and $F_1(t) \leq F_2(t)$ for $1/\lambda \leq t \leq 2/\lambda$.*

Again, the first form of stochastic dominance is somewhat crude. However, any two random variables with equal means can be compared with one another in this way.

From the fact that the functions $h(t) = t$ and $h(t) = -t$ are convex, it follows that if $X_1$ is "more variable" than $X_2$ then $E(X_1) \geq E(X_2)$ and $E(X_1) \leq E(X_2)$. So $E(X_1)$ has to be equal to $E(X_2)$. From the fact that $h(t) = t^2$ is convex it follows that $Var(X_1)$ is larger than $Var(X_2)$. Variability ordering is a partial ordering, i.e., not every pair of random variables with equal means can be ordered in this way. At times, variability ordering is also referred to as ordering *in the convex sense*.

It can be shown easily that symmetrically more variable implies more variable in the convex sense but not vice versa.

The forms of stochastic dominance described in *Definition 15.2* lead to the following chain of implications:

> symmetrically more variable $\implies$ more variable $\implies$ larger in variance

In stochastic scheduling, certain conventions have to be made which are not needed in deterministic scheduling. During the evolution of a stochastic process new information becomes available continuously. Job completions and occurrences of random release dates and due dates provide additional information that the decision-maker may wish to take into account while scheduling the remaining jobs. The amount of freedom the decision maker has in using this additional information is the basis for the various classes of decision making policies. In this section four classes of policies are defined.

The first class of policies is, in what follows, only used in scenarios where all the jobs are available for processing at time zero; the machine environments considered are the single machine, parallel machines and permutation flow shops.

**Definition 15.3.** *Under a nonpreemptive static list policy the decision maker orders the jobs at time zero according to a priority list. This priority list does not change during the evolution of the process and every time a machine is freed the next job on the list is selected for processing.*

Under this class of policies the decision maker puts at time zero the $n$ jobs in a list (permutation) and the list does not change during the evolution of the process. In the case of machines in parallel, every time a machine is freed, the job at the head of the list is selected as the next one for processing. In the case of a permutation flow shop the jobs are also put in a list in front of the first machine at time zero; every time the first machine is freed the next job on the list is scheduled for processing. This class of nonpreemptive static list policies is in what follows also referred to as the class of *permutation* policies. This class of policies is in a sense similar to the static priority rules often considered in deterministic models.

**Example 15.4.** *Consider a single machine and three jobs. All three jobs are available at time zero. All three jobs have the same processing time distributions, which is 2 with probability .5 and 8 with probability .5. The due date distributions are the same, too. The due date is 1 with probability .5 and 5 with probability .5. If a job is completed at the same time as its due date, it is considered to be on time. It would be of interest to know the expected number of jobs completed in time under a permutation policy.*

*Under a permutation policy the first job is completed in time with probability .25 (its processing time has to be 2 and its due date has to be 5); the second job is completed in time with probability .125 (the processing times of the first and second job have to be 2 and the due date of the second job has to be 5); the third job never will be completed in time. The expected number of on-time completions is therefore .375 and the expected number of tardy jobs is $3 - 0.375 = 2.625$.*

The second class of policies is a preemptive version of the first class and is in what follows only used in scenarios where jobs are released at *different* points in time.

**Definition 15.5.** *Under a preemptive static list policy the decision maker orders the jobs at time zero according to a priority list. This ordering includes jobs with nonzero release dates, i.e., jobs which are to be released later. This priority list does not change during the evolution of the process and at any point in time the job at the top of the list of available jobs is the one to be processed on the machine.*

Under this class of policies the following may occur. When there is a job release at some time point and the job released is higher on the static list than the job currently being processed, then the job being processed is preempted and the job released is put on the machine.

   Under the third and fourth class of policies, the decision-maker is allowed to make his decisions during the evolution of the process. That is, every time he makes a decision, he may take all information available at that time into account. The third class of policies does not allow preemptions.

**Definition 15.6.** *Under a nonpreemptive dynamic policy, every time a machine is freed, the decision maker is allowed to determine which job goes next. His decision at such a time point may depend on all the information available, e.g., the current time, the jobs waiting for processing, the jobs currently being processed on other machines and the amount of processing these jobs already have received on these machines. However, the decision maker is not allowed to preempt; once a job begins processing, it has to be completed without interruption.*

**Example 15.7.** *Consider the same problem as in* Example 15.4. *It is of interest to know the expected number of jobs completed in time under a nonpreemptive dynamic policy. Under a nonpreemptive dynamic policy the probability the first job is completed in time is again .25. With probability .5 the first job is completed at time 2. With probability .25 the due dates of both remaining jobs already occurred at time 1 and there will be no more on-time completions. With probability .75 at least one of the remaining two jobs has a due date at time 5. The probability that the second job put on the machine is completed in time is 3/16 (the probability that the first job has completion time 2 times the probability at least one of the two remaining jobs has due date 5 times the probability that the second job has processing time 2). Again, it is impossible to start the third job and complete it in time. The expected number of on-time completions is therefore .4375 and the expected number of tardy jobs is 2.5625.*

The last class of policies is a preemptive version of the third class.

**Definition 15.8.** *Under a preemptive dynamic policy, at any point in time, the decision maker is allowed to select the jobs to be processed on the machines. His decision at a time point may depend on all information available and may require preemptions.*

**Example 15.9.** *Consider again the problem of Example 15.4. It is of interest to know the expected number of jobs completed in time under a preeemptive dynamic policy. Under a preemptive dynamic policy, the probability that the first job is completed in time is again .25. This first job is either taken off the machine at time 1 (with probability .5) or at time 2 (with probability .5). The probability the second job put on the machine is completed in time is 3/8, since the second job enters the machine either at time 1 or at time 2 and the probability of being completed on time is 0.75 times the probability it has processing time 2, which equals 3/8 (regardless of when the first job was taken off the machine). However, unlike under the nonpreeemptive dynamic policy, the second job put on the machine is taken off with probability .5 at time 3 and with probability 0.5 at time 4. So now there is actually a chance that the third job that goes on the machine will be completed in time. The probability the third job is completed in time is 1/16 (the probability that the due date of the first job is 1 (=.5) times the probability that the due dates of both remaining jobs are 5 (=.25) times the probability that the processing time of the third job is 2 (=.5)). The total expected number of on-time completions is therefore $11/16 = 0.6875$ and the expected number of tardy jobs is 2.3125.*

It is clear that the optimal preemptive dynamic policy leads to the best possible value of the objective as in this class of policies the decision maker has the most information available and the largest amount of freedom. It is also clear that if all jobs are present at time zero and the environment is either a bank of machines in parallel or a permutation flow shop, then the optimal nonpreemptive dynamic policy is at least as good as the optimal nonpreemptive static list policy (see *Examples 15.4* and *15.7*).

There are several forms of minimization in stochastic scheduling. Whenever an objective function has to be minimized, it should be specified in what *sense* the objective is to be minimized. The crudest form of optimization is in the *expectation* sense, e.g., one wishes for example to minimize the *expected* makespan, that is $E(C_{max})$ and find a policy under which the expected makespan is smaller than the expected makespan under any other policy. A stronger form of optimization is optimization in the *stochastic* sense. If a schedule or policy minimizes $C_{max}$ stochastically, the makespan under the optimal schedule or policy is *stochastically* less than the makespan under any other schedule or policy. Stochastic minimization, of course, implies minimization in expectation. In the subsequent sections the objective is usually minimized in expectation. Frequently, however, the policies that minimize the objective in expectation minimize the objective stochastically as well.

### 15.3.  Single Machine Models

Stochastic models, especially with exponential processing times, may often contain more structure than their deterministic counterparts and lead to results which, at first sight, may seem surprising. Models that are NP-*hard* in a deterministic setting often allow a simple priority policy to be optimal in a stochastic setting.

In this section we first consider single machine models with arbitrary processing times in a nonpreemptive setting. Then we analyze models with exponentially distributed processing times.
For a number of stochastic problems, finding the optimal policy is equivalent to solving a deterministic scheduling problem. Usually, when such an equivalence relationship exists, the deterministic counterpart can be obtained by replacing all random variables with their means. The optimal schedule for the deterministic problem then minimizes the objective of the stochastic version in expectation.

One such case is when the objective in the deterministic counterpart is linear in $p_{(j)}$ and $w_{(j)}$, where $p_{(j)}$ and $w_{(j)}$ denote the processing time and weight of the job in the $j$-th position in the sequence.

This observation implies that it is easy to find the optimal permutation schedule for the stochastic counterpart of $1 \mid\mid \sum w_j C_j$, when the processing time of job $j$ is $X_j$, from an arbitrary distribution $F_j$, and the objective is $E(\sum w_j C_j)$. This problem leads to the stochastic version of the *WSPT* rule, which sequences the jobs in decreasing order of the ratio $w_j/E(X_j)$ or $\lambda_j w_j$. In what follows this rule is referred to either as the *Weighted Shortest Expected Processing Time first (WSEPT)* rule or as the "$\lambda w$" rule.

**Theorem 15.10.** *The WSEPT rule minimizes the expected sum of the weighted completion times in the class of nonpreemptive static list policies as well as in the class of nonpreemptive dynamic policies.*

*Proof.*  The proof for nonpreemptive static list policies is similar to the proof for the deterministic counterpart of this problem. The proof is based on an adjacent pairwise interchange argument identical to the one used for the deterministic counterpart of this problem. The only difference is that the $p_j$'s in that proof have to be replaced by the $E(X_j)$'s.

The proof for nonpreemptive dynamic policies needs an additional argument. It is easy to show that it is true for $n = 2$ (again an adjacent pairwise interchange argument). Now consider three jobs. It is clear that the last two jobs have to be sequenced according to the $\lambda w$ rule. These last two jobs will be sequenced in this order independent of what happens during the processing of the first job. There are then three sequences that may occur: each of the three jobs starting first and the remaining two jobs sequenced according to the $\lambda w$ rule. A simple interchange argument between the first job and the second shows that all three jobs have to sequenced according to the $\lambda w$ rule. It can be shown by induction that all $n$ jobs have to be sequenced according to the $\lambda w$ rule in the class of nonpreemptive dynamic policies: suppose it is true for $n-1$ jobs. If there are $n$ jobs it follows from the induction hypothesis that the last $n-1$ jobs have to be sequenced according to the $\lambda w$ rule. Suppose the first job is not the job with the highest $\lambda_j w_j$. Interchanging this job with the second job in the sequence, i.e., the job with the highest $\lambda_j w_j$, leads to an decrease in the expected value of the objective function. This completes the proof of the theorem. □

It can be shown that the nonpreemptive *WSEPT* rule is also optimal in the class of *preemptive* dynamic policies when all $n$ processing time distributions are *ICR*. This follows from the fact that any time when a preemption is contemplated, the $w_j/E(X_j)$ ratio of the job currently on the machine is actually higher than it was when first put on the machine (the expected remaining processing time of an *ICR* job decreases as processing goes on). If the ratio of the job was the highest among the remaining jobs when it was put on the machine, it remains the highest while it is being processed.

The same cannot be said about jobs with *DCR* distributions. The expected remaining processing time then *increases* while a job is being processed. So the weight divided by the expected remaining processing time of a job, while it is being processed, *decreases* with time. Preemptions may be thus advantageous with *DCR* processing times.

**Example 15.11.** *Consider n jobs with the processing time $X_j$ distributed as follows. The processing time $X_j$ is 0 with probability $p_j$ and it is distributed according to an exponential with rate $\lambda_j$ with probability $1-p_j$. Clearly, this distribution is DCR as it is a mixture of two exponentials with rates $\infty$ and $\lambda_j$. The objective to be minimized is the expected sum of the weighted completion times. The optimal preemptive dynamic policy is clear. All n jobs have to be tried out for a split second at time zero, in order to determine which jobs have zero processing times. If a job does not have zero processing time, it is taken immediately off the machine. The jobs with zero processing times are then all completed at time zero. After determining in this way which jobs have nonzero processing times, these remaining jobs are sequenced in decreasing order of $\lambda_j w_j$.*

The remaining part of this section focuses on due date related problems. Consider the stochastic counterpart of $1 \parallel L_{\max}$ with processing times having arbitrary distributions and deterministic due dates. The objective is to minimize the expected maximum lateness.

**Theorem 15.12.** *The EDD rule minimizes expected maximum lateness for arbitrarily distributed processing times and deterministic due dates in the class of nonpreemptive static list policies, the class of nonpreemptive dynamic policies and the class of preemptive dynamic policies.*

*Proof.* It is clear that the *EDD* rule minimizes the maximum lateness for any real-ization of processing times (after conditioning on the processing times, the problem is basically a deterministic problem and the results for the deterministic counterpart of this problem apply). If the *EDD* rule minimizes the maximum lateness for any realization of processing times then it minimizes the maximum lateness also in expectation (it actually minimizes the maximum lateness with probability 1). □

It can be shown that the *EDD* rule not only minimizes

$$E(L_{\max}) = E(\max(L_1, \ldots, L_n)),$$

but also $\max(E(L_1), \ldots, E(L_n))$. It is even possible to develop an algorithm for a stochastic counterpart of the more general $1 \mid prec \mid h_{\max}$ problem. In this problem the objective is to minimize the maximum of the $n$ expected costs incurred by the $n$ jobs, i.e., the objective is to minimize

$$\max \Big( E(h_1(C_1)), \ldots, E(h_n(C_n)) \Big),$$

where $h_j(C_j)$ is the cost incurred by job $j$ being completed at $C_j$. The cost function $h_j$ is nondecreasing in the completion time $C_j$. The algorithm is a modified version of the algorithm for the deterministic counterpart of this problem. The version here is also a *backward* procedure. Whenever one has to select a schedulable job for processing, it is clear that the distribution of its completion time is the convolution of the processing times of the jobs that have not yet been scheduled. Let $f_{J^c}$ denote the density function of the convolution of the processing times of the set of unscheduled jobs $J^c$. Job $j^*$ is then selected to be processed last among the set of jobs $J^c$ if

$$\int_0^\infty h_{j^*}(t) f_{J^c}(t) dt = \min_{j \in J^c} \int_0^\infty h_j(t) f_{J^c}(t) dt.$$

The *L.H.S.* denotes the expected value of the penalty for job $j^*$ if it is the last job to be scheduled among the jobs in $J^c$. This rule replaces one step in the algorithm for the deterministic counterpart of this problem. The proof of optimality is similar to the proof of optimality in the deterministic case. However, implementation of the algorithm is significantly more cumbersome as the evaluation of the integrals may not be easy.

We now discuss due date models with exponentially distributed processing times. Consider the stochastic version of $1 \mid d_j = d \mid \sum w_j U_j$ with job $j$ having an exponentially distributed processing time with rate $\lambda_j$ and a deterministic due date $d$. Recall that the deterministic counterpart is equivalent to the *knapsack* problem. The objective to be minimized is the expected weighted number of tardy jobs.

**Theorem 15.13.** *The WSEPT rule minimizes the expected weighted number of tardy*

*jobs in the classes of nonpreemptive static list policies, nonpreemptive dynamic policies and preemptive dynamic policies.*

*Proof.* First the optimality of the *WSEPT* rule in the class of nonpreemptive static list policies is shown. Assume the machine is free at some time $t$ and two jobs, with weights $w_1$ and $w_2$ and processing times $X_1$ and $X_2$, remain to be processed. Consider first the sequence $1, 2$. The probability that both jobs are late is equal to the probability that $X_1$ is larger than $d - t$, which is equal to $\exp(-\lambda(d - t))$. The penalty for being late is then equal to $w_1 + w_2$. The probability that only the second job is late corresponds to the event where the processing time of the first job is $x_1 < d - t$ and the sum of the processing times $x_1 + x_2 > d - t$. Evaluation of the probability of this event, through conditioning on $X_1$ (that is $X_1 = x$), yields

$$P(X_1 < d - t, X_1 + X_2 > d - t) = \int_0^{d-t} e^{-\lambda_2(d-t-x)} \lambda_1 e^{-\lambda_1 x} dx.$$

If $E(\sum wU(1, 2))$ denotes the expected value of the penalty due to jobs 1 and 2, with job 1 processed first, then

$$E\left(\sum wU(1, 2)\right) = (w_1 + w_2)e^{-\lambda_1(d-t)} + w_2 \int_0^{d-t} e^{-\lambda_2(d-t-x)} \lambda_1 e^{-\lambda_1 x} dx.$$

The value of the objective function under sequence $2, 1$ can be obtained by interchanging the subscripts in the expression above. Straightforward computation yields

$$E\left(\sum wU(1, 2)\right) - E\left(\sum wU(2, 1)\right) =$$

$$(\lambda_2 w_2 - \lambda_1 w_1)\frac{e^{-\lambda_1(d-t)} - e^{-\lambda_2(d-t)}}{\lambda_2 - \lambda_1}.$$

It immediately follows that the difference in the expected values is positive if and only if $\lambda_2 w_2 > \lambda_1 w_1$. Since this result holds for all values of $d$ and $t$, any permutation schedule that does not sequence the jobs in decreasing order of $\lambda_j w_j$ can be improved by swapping two adjacent jobs, where the first has a lower $\lambda w$ value than the second. This completes the proof of optimality for the class of nonpreemptive static list policies.

Induction can be used to show optimality in the class of nonpreemptive dynamic policies. It is immediate that this is true for 2 jobs (it follows from the same pairwise interchange argument for optimality in the class of nonpreemptive static list policies). Assume that it is true for $n - 1$ jobs. In the case of $n$ jobs this implies that the scheduler after the completion of the first job will, because of the induction hypothesis, revert to the *WSEPT* rule among the remaining $n - 1$ jobs. It remains to be shown that the scheduler has to select the job with the highest $\lambda_j w_j$ as the first one to be processed. Suppose the decision-maker selects a job which does not have the highest $\lambda_j w_j$. Then, the job with the highest value of $\lambda_j w_j$ is processed second. Changing the sequence of the first two jobs decreases the expected value of the objective function according to the pairwise interchange argument used for the nonpreemptive static list policies.

To show that *WSEPT* is optimal in the class of preemptive dynamic policies, suppose a preemption is contemplated at some point in time. The remaining processing time of the job then on the machine is exponentially distributed with the same rate as it had at the start of its processing (because of the memoryless property of the exponential). Since the decision to put this job on the machine did not depend on the value of $t$ at that moment or on the value of $d$, the same decision remains optimal at the moment a preemption is contemplated. A nonpreemptive policy is therefore optimal in the class of preemptive dynamic policies. □

This result is in marked contrast with the result for its deterministic counterpart, i.e., the knapsack problem, which is NP-*hard*.

The *WSEPT* rule does not necessarily yield an optimal schedule when processing time distributions are not all exponential.

*Theorem 15.13* can be generalized to consider breakdown and repair. Suppose the machine goes through "uptimes", when it is functioning and "downtimes" when it is being repaired. This breakdown and repair may form an arbitrary stochastic process. *Theorem 15.13* also holds under these more general conditions since no part of the proof depends on the remaining time till the due date.

*Theorem 15.13* can also be generalized to include different release dates with arbitrary distributions. Assume a finite number of releases after time 0, say $n^*$. It is clear from the results presented above that at the time of the last release the *WSEPT* policy is optimal. This may actually imply that the last release causes a preemption (if, at that point in time, the job released is the job with the highest $\lambda_j w_j$ ratio in the system). Consider now the time-epoch of the second last release. After this release a preemptive version of the *WSEPT* rule is optimal. To see this, disregard for a moment the very last release. All the jobs in the system at the time of the second to last release (*not* including the last release) have to be sequenced according to *WSEPT*; the last release may in a sense be considered a random "downtime". From the previous results it follows that all the jobs in the system at the time of the second last release should be scheduled according to preemptive *WSEPT*, independent of the time period during which the last release is processed. Proceeding inductively towards time zero it can be shown that a preemptive version of *WSEPT* is optimal with arbitrarily distributed releases in the classes of preemptive static list policies and preemptive dynamic policies.

The *WSEPT* rule also turns out to be optimal for other objectives as well. Consider the stochastic counterpart of $1 \mid d_j = d \mid \sum w_j T_j$ with job $j$ again exponentially distributed with rate $\lambda_j$. All $n$ jobs are released at time 0. The objective is to minimize the sum of the expected weighted tardinesses.

**Theorem 15.14.** *The WSEPT rule minimizes the expected sum of the weighted tardinesses in the classes of nonpreemptive static list policies, nonpreemptive dynamic policies and preemptive dynamic policies.*

*Proof.*   The objective $w_j T_j$ can be approximated by a sum of an infinite sequence of $w_j U_j$ unit penalty functions, i.e.,

$$w_j T_j = \sum_{l=0}^{\infty} w_j U_{jl}.$$

The first unit penalty $U_{j0}$ corresponds to a due date $d$, the second unit penalty $U_{j1}$ corresponds to a due date $d + \varepsilon$, the third corresponds to a due date $d + 2\varepsilon$ and so on (see *Figure ...*). From *Theorem 15.13* it follows that $\lambda w$ rule minimizes each one of these unit penalty functions. If the rule minimizes each one of these unit penalty functions, it also minimizes their sum.                                                    □

This theorem can be generalized along the lines of *Theorem 15.13* to include arbitrary breakdown and repair processes and arbitrary release processes, provided all jobs have due date $d$ (including those released after $d$).

Actually, a generalization in a slightly different direction is also possible. Consider the stochastic counterpart of the problem $1 \mid\mid \sum w_j h(C_j)$. In this model the jobs have no specific due dates, but are all subject to the *same* cost function $h$. The objective is to minimize $E(\sum w_j h(C_j))$. Clearly, $\sum w_j h(C_j)$ is a simple generalization of $\sum w_j T_j$ when all jobs have the same due date $d$. The function $h$ can again be approximated by a sum of an infinite sequence of unit penalties, the only difference being that the due dates of the unit penalties are not necessarily equidistant as in the proof of *Theorem 15.14.*

Consider now a stochastic counterpart of the problem $1 \mid\mid \sum w_j h_j(C_j)$, with each job having a *different* cost function. Again, all jobs are released at time 0. The objective is to minimize the total expected cost. The following ordering among cost functions is of interest: a cost function $h_j$ is said to be *steeper* than a cost function $h_k$ if

$$\frac{dh_j(t)}{dt} \geq \frac{dh_k(t)}{dt}$$

for every $t$, provided the derivatives exist. This ordering is denoted by $h_j \geq_s h_k$. If the functions are not differentiable for every $t$, the steepness ordering requires

$$h_j(t+\delta) - h_j(t) \geq h_k(t+\delta) - h_k(t),$$

for every $t$ and $\delta$. Note that a cost function being steeper than another does not necessarily imply that it is higher (see *Figure ??*).

**Theorem 15.15.** *If $\lambda_j w_j \geq \lambda_k w_k \Longleftrightarrow h_j \geq_s h_k$, then the WSEPT rule minimizes the total expected cost in the classes of nonpreemptive static list policies, nonpreemptive* *dynamic policies and preemptive dynamic policies.*

*Proof.* The proof follows from the fact that any increasing cost function can be approximated through the proper addition of a (possibly infinite) number of unit penalties at different due dates. If two cost functions, which may be at different levels, go up in the same way over an interval $[t_1, t_2]$, then a series of identical unit penalties go into effect within that interval for both jobs. It follows from *Theorem 15.13* that the jobs have to be sequenced in decreasing order of $\lambda w$ in order to minimize the total expected penalties due to these unit penalties. If one cost function is steeper than another in a particular interval, then the steeper cost function has one or more unit penalties going into effect within this interval, which the other cost function has not. To minimize the total expected cost due to these unit penalties, the jobs have to be sequenced again in decreasing order of $\lambda w$. □

The results in this section indicate that scheduling problems with exponentially distributed processing times allow for more elegant structural results than their deterministic counterparts. The deterministic counterparts of most of the models discussed in this section are NP-*hard*. It is intuitively acceptable that a deterministic problem may be NP-*hard* while its counterpart with exponentially distributed processing times allows for a very simple policy to be optimal. The reason is the following: all data being deterministic (that is, perfect data) makes it very hard for the scheduler to optimize. In order to take advantage of all the information available the scheduler has to spend an inordinately long time doing the optimization. On the other hand when the processing times are stochastic, the data are fuzzier. The scheduler, with less data at hand, will spend less time performing the optimization. The fuzzier the data, the more likely a simple priority rule minimizes the objective in expectation. Expectation is akin to optimizing for the average case.

## 15.4. Parallel Machine Models

This chapter deals with parallel machine models that are stochastic counterparts of the models discussed in *Chapter* ....**??** The body of knowledge in the stochastic case is considerably less extensive than in the deterministic case.

The results focus mainly on the expected makespan, the total expected completion time and the expected number of tardy jobs. In what follows the number of machines is usually limited to two. Some of the proofs can be extended to more than two machines, but such extensions usually require more elaborate notation. Since these extensions would not provide any additional insight, they are not presented here. The proofs for some of the structural properties of the stochastic models tend to be more involved than the proofs for the corresponding properties of their deterministic counterparts.

The first part of this section deals with nonpreemptive models; the results in this part are obtained through interchange techniques. The second part focuses on pre-emptive models; the results in this part are obtained through dynamic programming approaches. The third part deals with due date related models.

The first part of this section considers optimal policies in the class of nonpreemptive static list policies and in the class of nonpreemptive dynamic policies. Since preemptions are not allowed, the main technique for determining optimal policies is based on pairwise interchanges. The exponential distribution is considered in detail as its special properties makes the analysis relatively easy.

Consider *two* machines in parallel and $n$ jobs. The processing time of job $j$ is equal to the random variable $X_j$, that is exponentially distributed with rate $\lambda_j$. The objective is to minimize $E(C_{max})$. Note that this problem is a stochastic counterpart of $P2 \mid\mid C_{max}$, which is known to be NP-*hard*. However, in *Section 15.3* it already became clear that scheduling environments with exponentially distributed processing times often have structural properties which their deterministic counterparts do not have. It turns out that this is also the case with machines in parallel.

A nonpreemptive static list policy is followed. The jobs are put into a list and at time zero the two jobs at the top of the list begin processing on the two machines. When a machine becomes free the next job on the list is put on the machine. It is not specified in advance on which machine each job will be processed, nor is it known a priori which job will be the last one to be completed.

Let $Z_1$ denote the time when the second to last job is completed, i.e., the first time a machine becomes free with no jobs on the list to replace it. At this time the other machine is still processing its last job. Let $Z_2$ denote the time that the last job is completed on the other machine (i.e., $Z_2$ equals the makespan $C_{max}$). Let the difference $D$ be equal to $Z_2 - Z_1$. It is clear that the random variable $D$ depends on the schedule. It is easy to see that minimizing $E(D)$ is equivalent to minimizing $E(C_{max})$. This follows from

$$Z_1 + Z_2 = 2C_{max} - D = \sum_{j=1}^{n} X_j,$$

which is a constant independent of the schedule.

In what follows, a slightly more general two-machine problem is considered for reasons that will become clear later. It is assumed that one of the machines is not available at time zero and becomes available only after a random time $X_0$, distributed exponentially with rate $\lambda_0$. The random variable $X_0$ may be thought of as the processing time of an additional job which takes precedence and *must* go first. Let $D(X_0, X_1, X_2, \ldots, X_n)$ denote the random variable $D$, under the assumption that, at time zero, a job with remaining processing time $X_0$ is being processed on one machine and a job with processing time $X_1$ is being started on the other. When one of the two machines is freed a job with processing time $X_2$ is started, and so on (see *Figure* .....**??**). The next lemma, which we present without proof, examines the effect on $D$ of changing a schedule by swapping consecutive jobs 1 and 2.

**Lemma 15.16.** *For any $\lambda_0$ and for $\lambda_1 = \min(\lambda_1, \lambda_2, \ldots, \lambda_n)$*

$$E(D(X_0, X_1, X_2, \ldots, X_n)) \leq E(D(X_0, X_2, X_1, \ldots, X_n)).$$

This lemma constitutes a crucial element in the proof of the following theorem.

**Theorem 15.17.** *The LEPT rule minimizes the expected makespan in the class of nonpreemptive static list policies when there are two machines in parallel and exponentially distributed processing times.*

*Proof.* By contradiction. Suppose that a different rule is optimal. Suppose that according to this presumed optimal rule, the job with the longest expected processing time is not scheduled for processing either as the first or the second job. (Note that the first and second job are interchangeable as they both start at time zero.) Then an improvement can be obtained by performing a pairwise interchange between this longest job and the job immediately preceding this job in the schedule, as by *Lemma 15.16* this reduces the expected difference between the completion times of the last two jobs. Through a series of interchanges it can be shown that the longest job has to be one of the first two jobs in the schedule. In the same way it can be shown that the second longest job has to be among the first two jobs as well. The third longest job can be moved into the third position to improve the objective, and so on. With each interchange the expected difference, and thus the expected makespan, are reduced. □

The approach used in proving the theorem is basically an adjacent pairwise interchange argument. However, this pairwise interchange argument is not identical to the pairwise interchange arguments used in single machine scheduling. In pairwise interchange arguments applied to single machine problems, *no* restrictions were made on the relation between the two jobs to be interchanged and those that come after them. In *Lemma 15.16* jobs not involved in the interchange have to satisfy a special condition, viz., one of the two jobs being interchanged must have a larger expected processing time than all jobs following it. Requiring such a condition has certain implications. When no special conditions are required, an adjacent pairwise interchange argument actually yields two results: it shows that one schedule minimizes the objective while the *reverse* schedule maximizes that same objective. With a special condition like the one in *Lemma 15.16* the argument works only in one direction. It actually can be shown that the *SEPT* rule does *not* always maximize $E(D)$ among nonpreemptive static list policies.

The result presented in *Theorem 15.17* differs from the results obtained for its deterministic counterpart considerably. One difference is the following: minimizing makespan in a deterministic setting requires only an optimal *partition* of the $n$ jobs over the two machines. After the allocation has been determined, the set of jobs allocated to a specific machine may be sequenced in any order. With exponential processing times, a sequence is determined in which the jobs are to be *released* in order to minimize the expected makespan. No deviation is allowed from this sequence and it is not specified at time zero how the jobs will be partitioned between the machines. This depends on the evolution of the process.

In contrast with the results of *Section 15.3,* which do not appear to hold for distributions other than the exponential, the *LEPT* rule does minimize the expected makespan for other distributions as well.

Consider the case where the processing time of job $j$ is distributed according to a mixture of two exponentials, i.e., with probability $p_{1j}$ according to an exponential with rate $\lambda_1$ and with probability $p_{2j}$ $(= 1 - p_{1j})$ according to an exponential with rate $\lambda_2$. Assume $\lambda_1 < \lambda_2$. So

$$P(X_j > t) = p_{1j}e^{-\lambda_1 t} + p_{2j}e^{-\lambda_2 t}.$$

This distribution can be described as follows: when job $j$ is put on the machine a (biased) coin is tossed. Dependent upon the outcome of the toss the processing time of job $j$ is either exponential with rate $\lambda_1$ or exponential with rate $\lambda_2$. After the rate has been determined this way the distribution of the remaining processing time of job $j$ does not change while the job is being processed. So each processing time is distributed according to one of the two exponentials with rates $\lambda_1$ and $\lambda_2$.

The subsequent lemma again examines the effect on $D$ of an interchange between two consecutive jobs 1 and 2 on two machines in parallel. Assume again that $X_0$ denotes the processing time of a job 0 with an exponential distribution with rate $\lambda_0$. This rate $\lambda_0$ may be different from either $\lambda_1$ or $\lambda_2$.

**Lemma 15.18.** *For arbitrary $\lambda_0$, if $p_{11} \geq p_{12}$, i.e., $E(X_1) \geq E(X_2)$, then*

$$E(D(X_0, X_1, X_2, \ldots, X_n)) \leq E(D(X_0, X_2, X_1, \ldots, X_n)).$$

Note that there are no conditions on $\lambda_0$; the rate $\lambda_0$ may or may not be equal to either $\lambda_1$ or $\lambda_2$. Through this lemma the following theorem can be shown rather easily.

**Theorem 15.19.** *The LEPT rule minimizes the expected makespan in the class of nonpreemptive static list policies when there are two machines in parallel and when the processing times are distributed according to a mixture of two exponentials with rates $\lambda_1$ and $\lambda_2$.*

*Proof.* Any permutation schedule can be transformed into the *LEPT* schedule through a series of adjacent pairwise interchanges between a longer job and a shorter job immediately preceding it. With each interchange $E(D)$ decreases because of *Lemma 15.18.*                                                                                    □

Showing that *LEPT* minimizes the expected makespan can be done in this case without any conditions on the jobs that are not part of the interchange. This is in contrast with *Theorem 15.17,* where the jobs following the jobs in the interchange had to be smaller than the largest job involved in the pairwise interchange. The additional condition requiring the other expected processing times to be smaller than the expected processing time of the larger of the two jobs in the interchange, is *not* required in this case.

*Theorem 15.19* can be extended to include mixtures of three exponentials, with rates $\lambda_1$, $\lambda_2$ and $\infty$. The next example also shows that the *LEPT* rule does not necessarily minimize the expected makespan.

**Example 15.20.** *Let $p_{1j}$ denote the probability job $j$ is exponentially distributed with rate $\lambda_1$ and $p_{2j}$ the probability it is distributed with rate $\lambda_2$. Assume $\lambda_1 < \lambda_2$. The probability that the processing time of job $j$ is zero is $p_{0j} = 1 - p_{1j} - p_{2j}$. Through similar arguments as the ones used in* Lemma 15.18 *and* Theorem 15.19 *it can be shown that in order to minimize the expected makespan the jobs in the optimal nonpreemptive static list policy have to be ordered in decreasing order of $p_{1j}/p_{2j}$. The jobs with the zero processing times again do not play a role in the schedule. Clearly, the optimal sequence is not necessarily* LEPT.

The following example is a continuation of the previous example and an illustration of the *Largest Variance first (LV)* rule.

**Example 15.21.** *Consider the special case of the previous example with*

$$\frac{1}{\lambda_1} = 2$$

*and*

$$\frac{1}{\lambda_2} = 1.$$

*Let*

$$p_{0j} = a_j$$
$$p_{1j} = a_j$$
$$p_{2j} = 1 - 2a_j$$

*So*

$$E(X_j) = \frac{p_{1j}}{\lambda_1} + \frac{p_{2j}}{\lambda_2} = 1,$$

*for all $j$ and*

$$Var(X_j) = 1 + 4a_j.$$

*From the previous example it follows that sequencing the jobs in decreasing order of $p_{1j}/p_{2j}$ minimizes the expected makespan. This rule is equivalent to scheduling the jobs in decreasing order of $a_j/(1-2a_j)$. As $0 \le a_j \le 1/2$, scheduling the jobs in decreasing order of $a_j/(1-2a_j)$ is equivalent to scheduling in decreasing order of $a_j$, which in turn is equivalent to the* Largest Variance first *rule.*

The methodology used in proving that *LEPT* is optimal for the expected makespan on two machines does not easily extend to problems with more than two machines or problems with other processing time distributions. Consider the following generalization of this approach for *m* machines. Let $Z_1$ denote the time that the first machine becomes idle with no jobs waiting for processing, $Z_2$ the time the second machine becomes idle, and so on and let $Z_m$ denote the time the last machine becomes idle. Clearly $Z_m$ equals the makespan. Let

$$D_i = Z_{i+1} - Z_i \qquad i = 1, \dots, m-1.$$

From the fact that the sum of the processing times is

$$\sum_{j=1}^{n} X_j = \sum_{i=1}^{m} Z_i = mC_{\max} - D_1 - 2D_2 - \cdots - (m-1)D_{m-1},$$

independent of the schedule, it follows that minimizing the makespan is equivalent to minimizing

$$\sum_{i=1}^{m-1} iD_i.$$

A limited number of processing time distributions can be handled this way. For example, the settings of *Theorem 15.19* and *Examples 15.20* and *15.21* can be extended relatively easily through this approach. However, the scenario of *Theorem 15.17* cannot be extended that easily.

So far only the class of nonpreemptive static list policies has been considered in this section. It turns out, that most optimal policies in the class of nonpreemptive static list policies are also optimal in the classes of nonpreemptive dynamic policies and preemptive dynamic policies. The proof that a nonpreemptive static list policy is optimal in these other two classes of policies is based on induction arguments very similar to the ones described in the second and third parts of the proof of *Theorem 15.13*.

In what follows an entirely different approach is presented which first proves optimality in the class of preemptive dyanmic policies. As the optimal policy is a nonpreemptive static list policy, the policy is also optimal in the classes of nonpreemptive dynamic policies and nonpreemptive static list policies.

Only the expected makespan has been considered so far in this section. The total expected completion time $E(\sum C_j)$ in a nonpreemptive setting is a slightly more difficult objective to deal with than the expected makespan. Indeed, an approach similar to the one used to show that *LEPT* minimizes the expected makespan for exponential processing times, has not been found to show that *SEPT* minimizes the total expected completion time. However, if the processing times are distributed as in *Theorem 15.19*, it can be shown that *SEPT* minimizes the expected flow time and if the processing times are distributed as in *Example 15.21* it can be shown that *LV* minimizes the expected flow time.

Pairwise interchange arguments are basically geared to determine optimal policies in the class of nonpreemptive static list policies. After determining an optimal nonpreemptive static list policy it can often be argued that this policy is also optimal in the class of nonpreemptive dynamic policies and possibly in the class of preemptive dynamic policies.

In what follows an alternative proof for *Theorem 15.17* is presented. The approach is entirely different. A dynamic programming type proof is constructed within the class of *preemptive* dynamic policies. After obtaining the result that the nonpreemptive *LEPT* policy minimizes the expected makespan in the class of preemptive dynamic policies, it is concluded that it is also optimal in the class of nonpreemptive dynamic policies as well as in the class of nonpreemptive static list policies.

The approach can be used for proving that *LEPT* minimizes the expected makespan for *m* machines in parallel. It will be illustrated for 2 machines in parallel since the notation is much simpler.

Suppose $\lambda_1 \leq \lambda_2 \leq \cdots \leq \lambda_n$. Let $V(J)$ denote the expected value of the minimum remaining time needed (that is, under the optimal policy) to finish all jobs given that all the jobs in the set $J = j_1, \ldots, j_l$ already have been completed; if $J = \emptyset$, then $V(J)$ is simply denoted by $V$. Let $V^*(J)$ denote the same time quantity under the *LEPT* policy. Similarly, $V^*$ denotes the expected value of the remaining completion time under *LEPT* when no job has yet been completed.

**Theorem 15.22.** *The nonpreemptive LEPT policy minimizes the expected makespan in the class of preemptive dynamic policies.*

*Proof.* The proof is by induction on the number of jobs. Suppose that the result is true when there are less than *n* jobs. It has to be shown that it is also true when there are *n* jobs. That is, a policy which at time 0 (when there are *n* jobs waiting for processing) does not act according to *LEPT* but at the first job completion (when there are $n - 1$ jobs remaining to be processed) switches over to *LEPT* results in a larger expected makespan than when *LEPT* is adopted immediately from time zero on.

Conditioning on the first job completion yields

$$V = \min_{j,k} \left( \frac{1}{\lambda_j + \lambda_k} + \frac{\lambda_j}{\lambda_j + \lambda_k} V^*(\{j\}) + \frac{\lambda_k}{\lambda_j + \lambda_k} V^*(\{k\}) \right).$$

The expected time until the first job completion is the first term on the *R.H.S.*; the second (third) term is equal to the probability of job $j$ ($k$) being the first job to be completed, multiplied by the expected remaining time needed to complete the $n-1$ remaining jobs under *LEPT*. This last equation is equivalent to

$$0 = \min_{j,k} \Big( 1 + \lambda_j \Big( V^*(\{j\}) - V^* \Big) + \lambda_k \Big( V^*(\{k\}) - V^* \Big) + (\lambda_j + \lambda_k)\Big( V^* - V \Big) \Big).$$

Since $\lambda_1$ and $\lambda_2$ are the two smallest $\lambda_j$ values and supposedly $V^* \geq V$ the fourth term on the *R.H.S.* is minimized by $\{j,k\} = \{1,2\}$. Hence to show that *LEPT* is optimal it suffices to show that $\{j,k\} = \{1,2\}$ also minimizes the sum of the second and third term. In order to simplify the presentation let

$$A_j = \lambda_j \Big( V^*(\{j\}) - V^* \Big)$$

and

$$D_{jk} = A_j - A_k.$$

In order to show that

$$\lambda_j \Big( V^*(\{j\}) - V^* \Big) + \lambda_k \Big( V^*(\{k\}) - V^* \Big) = A_j + A_k$$

is minimized by $\{j,k\} = \{1,2\}$, it suffices to show that $\lambda_j < \lambda_k$ implies $A_j \leq A_k$ or, equivalently, $D_{jk} \leq 0$. To prove that $D_{jk} \leq 0$ is done in what follows by induction.

Throughout the remaining part of the proof $V^*$, $A_j$ and $D_{jk}$ are considered functions of the random variables $\lambda_1, \ldots, \lambda_n$. Define $A_j(J)$ and $D_{jk}(J)$, assuming jobs $j$ and $k$ are not in $J$, in the same way as $A_j$ and $D_{jk}$, e.g.,

$$A_j(J) = \lambda_j \Big( V^*(J \cup \{j\}) - V^*(J) \Big).$$

Before proceeding with the induction a number of identities have to be established. If $j$ and $k$ are the two smallest jobs not in the set $J$, then jobs $j$ and $k$ will, under *LEPT*, be processed first. Conditioning on the first job completion results in the identity

$$V^*(J) = \frac{1}{\lambda_j + \lambda_k} + \frac{\lambda_j}{\lambda_j + \lambda_k} V^*(J \cup \{j\}) + \frac{\lambda_k}{\lambda_j + \lambda_k} V^*(J \cup \{k\})$$

or

$$(\lambda_j + \lambda_k) V^*(J) = 1 + \lambda_j V^*(J \cup \{j\}) + \lambda_k V^*(J \cup \{k\}).$$

Similarly,

$$(\lambda_1 + \lambda_2 + \lambda_3)A_1 = \lambda_1(\lambda_1 + \lambda_2 + \lambda_3)V^*(\{1\}) - \lambda_1(\lambda_1 + \lambda_2 + \lambda_3)V^*$$
$$= \lambda_1\Big(1 + \lambda_1 V^*(\{1\}) + \lambda_2 V^*(\{1,2\}) + \lambda_3 V^*(\{1,3\})\Big)$$
$$- \lambda_1\Big(1 + \lambda_1 V^*(\{1\}) + \lambda_2 V^*(\{2\}) + \lambda_3 V^*\Big)$$
$$= \lambda_1\Big(\lambda_3 V^*(\{1,3\}) - \lambda_3 V^*(\{1\})\Big)$$
$$+ \lambda_2\Big(\lambda_1 V^*(\{1,2\}) - \lambda_1 V^*(\{2\})\Big) + \lambda_3 A_1$$

or

$$(\lambda_1 + \lambda_2)A_1 = \lambda_1 A_3(\{1\}) + \lambda_2 A_1(\{2\}).$$

The following identities can be established in the same way:

$$(\lambda_1 + \lambda_2)A_2 = \lambda_1 A_2(\{1\}) + \lambda_2 A_3(\{2\})$$

and

$$(\lambda_1 + \lambda_2)A_j = \lambda_1 A_j(\{1\}) + \lambda_2 A_j(\{2\}),$$

for $j = 3, \ldots, n$. Thus, with $D_{12} = A_1 - A_2$, it follows that

$$D_{12} = \frac{\lambda_1}{\lambda_1 + \lambda_2}D_{32}(\{1\}) + \frac{\lambda_2}{\lambda_1 + \lambda_2}D_{13}(\{2\}),$$

and

$$D_{2j} = \frac{\lambda_1}{\lambda_1 + \lambda_2}D_{2j}(\{1\}) + \frac{\lambda_2}{\lambda_1 + \lambda_2}D_{3j}(\{2\}),$$

for $j = 3, \ldots, n$.

Assume now as induction hypothesis that if $\lambda_j < \lambda_k$, and $\lambda_1 \leq \cdots \leq \lambda_n$, then

$$D_{jk} \leq 0$$

and

$$\frac{dD_{12}}{d\lambda_1} \geq 0.$$

In the remaining part of the proof, these two inequalities are shown by induction on $n$. When $n = 2$,

$$D_{jk} = \frac{\lambda_j - \lambda_k}{\lambda_j + \lambda_k}$$

and the two inequalities can be established easily.

Assume that the two inequalities of the induction hypothesis hold when there are less than $n$ jobs remaining to be processed. The induction hypothesis now implies that $D_{13}(\{2\})$ as well as $D_{23}(\{1\})$ are nonpositive when there are $n$ jobs remaining to be completed. It also provides

$$\frac{dD_{13}(\{2\})}{d\lambda_1} \geq 0.$$

This last inequality has the following implication: if $\lambda_1$ increases then $D_{13}(\{2\})$ increases. The moment $\lambda_1$ reaches the value of $\lambda_2$ jobs 1 and 2 become interchangeable. Therefore

$$D_{13}(\{2\}) \leq D_{23}(\{1\}) = -D_{32}(\{1\}) \leq 0.$$

From the fact that $\lambda_1 < \lambda_2$ it follows that $D_{12}$ is nonpositive. The induction hypothesis also implies that $D_{2j}(\{1\})$ and $D_{3j}(\{2\})$ are nonpositive, whereby $D_{2j}$ is nonpositive. This completes the induction argument for the first inequality of the induction hypothesis. The induction argument for the second inequality can be established by differentiating

$$\frac{\lambda_1}{\lambda_1 + \lambda_2} D_{32}(\{1\}) + \frac{\lambda_2}{\lambda_1 + \lambda_2} D_{13}(\{2\})$$

with respect to $\lambda_1$ and then using induction to show that every term is positive.  □

This proof shows that *LEPT* is optimal in the class of preemptive dynamic policies. As the optimal policy is a nonpreemptive static list policy it also has to be optimal in the class of nonpreemptive static list policies as well as in the class of nonpreemptive dynamic policies. In contrast with the first proof of the same result, this approach also works for an arbitrary number of machines in parallel. The notation, however, becomes significantly more involved.

The interchange approach described in the beginning of this section is not entirely useless. To show that the nonpreemptive *LEPT* policy is optimal when the processing time distributions are *ICR*, one has to adopt a pairwise interchange type argument. The reason is obvious. In a preemptive framework the remaining processing time of an *ICR* job that has received a certain amount of processing may become less (in expectation) than the expected processing time of a job that is waiting for processing. This then would lead to a preemption. The approach used in *Lemma 15.16* and *Theorem 15.17* can be applied easily to a number of different classes of distributions for which the approach used in *Theorem 15.22* does not appear to yield the optimal nonpreemptive schedule.

Consider minimizing the total expected completion time of the $n$ jobs. The processing time of job $j$ is exponentially distributed with rate $\lambda_j$. In *Chapter ....??* it was shown that the *SPT* rule is optimal for the deterministic counterpart of this problem. This gives an indication that in a stochastic setting the *Shortest Expected Processing Time first (SEPT)* rule may minimize the sum of the expected completion times under appropriate conditions. Consider again two machines in parallel with $n$ jobs. The processing time of job $j$ is exponentially distributed with rate $\lambda_j$. An approach similar to the one followed in *Theorem 15.22* for the makespan can be followed for the total expected completion time. The result then is that the nonpreemptive SEPT pol-icy minimizes the total expected completion time in the class of preemptive dynamic policies.

Actually, it turns out that a much more general result can be shown. Consider the more general setting where the $n$ processing times $X_1, \ldots, X_n$ come from arbitrary distributions $F_1, \ldots, F_n$ and $X_1 \leq_{st} X_2 \leq_{st} \cdots \leq_{st} X_n$.

**Theorem 15.23.** *The nonpreemptive* SEPT *policy minimizes the total expected completion time in expectation and even stochastically in the class of nonpreemptive dynamic policies.*

This result is more general than the corresponding result obtained for minimizing the expected makespan. Recall that *LEPT* does *not* minimize the expected makespan when the $X_1, \ldots, X_n$ are arbitrarily distributed and stochastically ordered.

In the remaining part of this section we consider the problem of two machines in parallel with *i.i.d.* job processing times distributed exponentially with mean 1, with precedence constraints in the form of an intree, and the expected makespan to be minimized in the class of preemptive dynamic policies (that is, a stochastic counterpart of $P2 \mid p_j = 1, intree \mid C_{\max}$). For the deterministic version of this problem the *Critical Path (CP)* rule (sometimes also referred to as the *Highest Level first (HL)* rule) is optimal. The *CP* rule is in the deterministic case optimal for an arbitrary number of machines in parallel, not just two.

For the stochastic version the following notation is needed. The root of the intree is level 0. A job is at level $k$ if there is a chain of $k-1$ jobs between it and the root of the intree. A precedence graph $G_1$ with $n$ jobs is said to be *flatter* than a precedence graph $G_2$ with $n$ jobs if the number of jobs at or below level $k$ in $G_1$ is larger than the number of jobs at or below level $k$ in graph $G_2$. This is denoted by $G_1 \prec_{fl} G_2$. In the following lemma two scenarios, both with two machines and $n$ jobs but with different intrees, are compared. Let $E(C_{\max}(i)(CP))$ denote the expected makespan under the *CP* rule when the precedence constraints graph takes the form of intree $G_i$, $i = 1, 2$.

Not that in the subsequent lemma and theorem, preemptions are allowed. However, it will become clear afterwards that for intree precedence constraints the *CP* rule does not require any preemptions. Also, recall that whenever a job is completed on one machine, the remaining processing time of the job being processed on the other machine is still exponentially distributed with mean one.

**Theorem 15.24.** *The nonpreemptive CP rule minimizes the expected makespan in the class of nonpreemptive dynamic policies and in the class of preemptive dynamic policies.*

As mentioned before, the results presented in *Theorems 15.17* and *15.22*, even though they were only proved for $m = 2$, hold for arbitrary $m$. The *CP* rule in *Theorem 15.24* is, however, not necessarily optimal for $m$ larger than two.

**Example 15.25.** *Consider three machines and 12 jobs. The jobs are all* i.i.d. *exponential with mean 1 and subject to the precedence constraints described in* Figure .....**??** *Scheduling according to the* CP *rule would put jobs 1, 2 and 3 at time zero on the three machines. However, straightforward algebra shows that starting with jobs 1, 2 and 4 results in a smaller expected makespan.*

In the deterministic setting discussed in *Chapter* ...**??** it was shown that the *CP* rule is optimal for $Pm \mid p_j = 1, intree \mid C_{\max}$ and $Pm \mid p_j = 1, outtree \mid C_{\max}$, for any $m$. One may expect the *CP* rule to be optimal when all processing times are exponential with mean 1 and precedence constraints take the form of an outtree. However, a counterexample can be found easily already in the case of 2 machines in parallel.

Consider again the problem of two machines in parallel with jobs having *i.i.d.* exponentially distributed processing times and subject to precedence constraints which take the form of an intree, but now with the expected flow time as the objective to be minimized. We present the following theorem without proof.

**Theorem 15.26.** *The nonpreemptive CP rule minimizes the total expected completion time in the class of nonpreemptive dynamic policies and in the class of preemptive dynamic policies.*

### 15.5. Stochastic Multi-Operation Models

Results for stochastic flow shop, open shop and job shop models are somewhat limited in comparison with the results for their deterministic counterparts.

For flow shops nonpreemptive static list policies, i.e., permutation schedules, are considered first. The optimal permutation schedules often remain optimal in the class of nonpreemptive dynamic policies as well as in the class of preemptive dynamic policies. For open shops and job shops, only the classes of nonpreemptive dynamic policies and preemptive dynamic policies are considered.

The results obtained for stochastic flow shops and job shops are somewhat similar to those obtained for deterministic flow shops and job shops. Stochastic open shops are, however, very different from their deterministic counterparts.

The first section discusses stochastic flow shops with unlimited intermediate storage and jobs not subject to blocking. The second section deals with stochastic flow shops with zero intermediate storage; the jobs are subject to blocking. The last section goes over stochastic open shops and stochastic job shops.

Consider two machines in series with unlimited storage between the machines and no blocking. There are $n$ jobs. The processing time of job $j$ on machine 1 is $X_{1j}$, exponentially distributed with rate $\lambda_j$. The processing time of job $j$ on machine 2 is $X_{2j}$, exponentially distributed with rate $\mu_j$. The objective is to find the nonpreemptive static list policy or permutation schedule that minimizes the expected makespan $E(C_{\max})$.

Note that this problem is a stochastic counterpart of the deterministic problem $F2 \mid\mid C_{\max}$. The deterministic two machine problem has a very simple solution, i.e., Johnson's rule. It turns out that the stochastic version with exponential processing times has a very elegant solution as well.

**Theorem 15.27.** *Sequencing the jobs in decreasing order of $\lambda_j - \mu_j$ minimizes the expected makespan in the class of nonpreemptive static list policies, the class of nonpreemptive dynamic policies and the class of preemptive dynamic policies.*

*Proof.* The proof of optimality in the class of nonpreemptive static list policies is in a sense similar to the proof of optimality in the deterministic case. It is by contradiction. Suppose another sequence is optimal. Under this sequence, there must be two adjacent jobs, say job $j$ followed by job $k$, such that $\lambda_j - \mu_j < \lambda_k - \mu_k$. It suffices to show that a pairwise interchange of these two jobs reduces the expected makespan. Assume job $l$ precedes job $j$ and let $C_{1l}$ ($C_{2l}$) denote the (random) completion time of job $l$ on machine 1 (2). Let $D_l = C_{2l} - C_{1l}$.

Perform an adjacent pairwise interchange on jobs $j$ and $k$. Let $C_{1k}$ and $C_{2k}$ denote the completion times of job $k$ on the two machines under the original, supposedly optimal, schedule and let $C'_{1j}$ and $C'_{2j}$ denote the completion times of job $j$ under the schedule obtained after the pairwise interchange. Let $m$ denote the job following job $k$. Clearly, the pairwise interchange does not affect the starting time of job $m$ on machine 1 as this starting time is equal to $C_{1k} = C'_{1j} = C_{1l} + X_{1j} + X_{1k}$. Consider the random variables

$$D_k = C_{2k} - C_{1k}$$

and

$$D'_j = C'_{2j} - C'_{1j}.$$

Clearly, $C_{1k} + D_k$ is the time at which machine 2 becomes available for job $m$ under the original schedule, while $C_{1k} + D'_j$ is the corresponding time after the pairwise interchange. First it is shown that the random variable $D'_j$ is stochastically smaller than the random variable $D_k$. If $D_l \geq X_{1j} + X_{1k}$, then clearly $D_k = D'_j$. The case $D_l \leq X_{1j} + X_{1k}$ is slightly more complicated. Now

$$P(D_k > t \mid D_l \leq X_{1j} + X_{1k}) = \frac{\mu_j}{\lambda_k + \mu_j} e^{-\mu_k t} + \frac{\lambda_k}{\lambda_k + \mu_j} \left( \frac{\mu_k}{\mu_k - \mu_j} e^{-\mu_j t} - \frac{\mu_j}{\mu_k - \mu_j} e^{-\mu_k t} \right).$$

This expression can be explained as follows. Since $D_l \leq X_{1j} + X_{1k}$, then, whenever job $j$ starts on machine 2, job $k$ is either being started or still being processed on machine 1. The first term on the *R.H.S.* corresponds to the event where job $j$'s processing time on machine 2 finishes before job $k$'s processing time on machine 1, which happens with probability $\mu_j/(\mu_j + \lambda_k)$. The second term corresponds to the event where job $j$ finishes on machine 2 after job $k$ finishes on machine 1; in this case the distribution of $D_k$ is a convolution of an exponential with rate $\mu_j$ and an exponential with rate $\mu_k$.

An expression for $P(D'_j > t \mid D_l \leq X_{1j} + X_{1k})$ can be obtained by interchanging the subscripts $j$ with the subscripts $k$. Now

$$P(D'_j > t \mid D_l \leq X_{1j} + X_{1k}) - P(D_k > t \mid D_l \leq X_{1j} + X_{1k}) =$$

$$\frac{\mu_j \mu_k}{(\lambda_j + \mu_k)(\lambda_k + \mu_j)} \frac{e^{-\mu_j t} - e^{-\mu_k t}}{\mu_k - \mu_j}(\lambda_j + \mu_k - \lambda_k - \mu_j) \leq 0.$$

So $D'_j$ is stochastically smaller than $D_k$. It can be shown easily, through a straightforward sample path analysis (i.e., fixing the processing times of job $m$ and of all the jobs following job $m$), that if the realization of $D'_j$ is smaller than the realization of $D_k$, then the actual makespan after the interchange is smaller than or equal to the actual makespan under the original sequence before the interchange. So, given that $D'_j$ is stochastically smaller than $D_k$, the expected makespan is reduced by the interchange. This completes the proof of optimality in the class of nonpreemptive static list (i.e., permutation) policies.

That the rule is also optimal in the class of nonpreemptive dynamic policies can be argued as follows. It is clear that the sequence on machine 2 does not matter. This is because the time machine 2 remains busy processing available jobs is simply the sum of their processing times and the order in which this happens does not affect the makespan. Consider the decisions which have to be made every time machine 1 is freed. The last decision to be made is at that point in time when there are only two jobs remaining to be processed on machine 1. From the pairwise interchange argument described above, it immediately follows that the job with the highest $\lambda_j - \mu_j$ value has to go first. Suppose that there are three jobs remaining to be processed on machine 1. From the previous argument it follows that the last two of these three have to be processed in decreasing order of $\lambda_j - \mu_j$. If the first one of the three is not the one with the highest $\lambda_j - \mu_j$ value, a pairwise interchange between the first and the second reduces the expected makespan. So the last three jobs have to be sequenced in decreasing order of $\lambda_j - \mu_j$. Continuing in this manner it is shown that sequencing the jobs in decreasing order of $\lambda_j - \mu_j$ is optimal in the class of nonpreemptive dynamic policies.

That the nonpreemptive rule is also optimal in the class of preemptive dynamic policies can be shown in the following manner. It is shown above that in the class of nonpreemptive dynamic policies the optimal rule is to order the jobs in decreasing order of $\lambda_j - \mu_j$. Suppose during the processing of a job on machine 1 a preemption is considered. The situation at this point in time is essentially no different from the situation at the point in time the job was started (because of the memoryless property of the exponential distribution). So, every time a preemption is contemplated, the optimal decision is to keep the current job on the machine. Thus the permutation policy is also optimal in the class of preemptive dynamic policies.    □

From the statement of the theorem, it appears that the number of optimal schedules in the exponential case is often smaller than the number of optimal schedules in the deterministic case. The following example makes this clear.

**Example 15.28.** *Consider n jobs with exponentially distributed processing times. One job has zero processing time on machine 1 and a processing time on machine 2 with a very large mean. Assume that this mean is larger than the sum of the expected processing times of the remaining $n-1$ jobs on machine 1. According to* Theorem 15.27 *these remaining $n-1$ jobs still have to be ordered in decreasing order of $\lambda_j - \mu_j$ for the sequence to minimize the expected makespan.*

*If all the processing times were deterministic with processing times equal to the means of the exponential processing times, it would not have mattered in what order the remaining $n-1$ jobs were sequenced.*

Although at first glance *Theorem 15.27* does not appear to be very similar to Johnson's result for its deterministic counterpart, the optimal schedule with exponential processing times is somewhat similar to the optimal schedule with deterministic processing times. If job $k$ follows job $j$ in the optimal sequence with exponential processing times, then

$$\lambda_j - \mu_j \geq \lambda_k - \mu_k$$

or

$$\lambda_j + \mu_k \geq \lambda_k + \mu_j$$

or

$$\frac{1}{\lambda_j + \mu_k} \leq \frac{1}{\lambda_k + \mu_j},$$

which, with exponential processing times, is equivalent to

$$E(\min(X_{1j}, X_{2k})) \leq E(\min(X_{1k}, X_{2j})).$$

This adjacency condition is quite similar to the condition for job $k$ to follow job $j$ in a deterministic setting, namely

$$\min(p_{1j}, p_{2k}) \leq \min(p_{1k}, p_{2j}).$$

There is another similarity between exponential and deterministic settings. Consider the case where the processing times of job $j$ on both machines are *i.i.d.* exponentially distributed with the same rate, $\lambda_j$, for each $j$. According to the theorem all sequences must have the same expected makespan. This result is similar to the deterministic proportionate flow shop, where all sequences also result into the same makespan.

We now focus on $m$ machine *permutation* flow shops. For these flow shops only the class of nonpreemptive static list policies is of interest, since the order of the jobs, once determined, is not allowed to change.

Consider an $m$ machine permutation flow shop where the processing times of job $j$ on the $m$ machines are *i.i.d.* according to distribution $F_j$ with mean $1/\lambda_j$. For such a flow shop it is easy to obtain a lower bound for $E(C_{\max})$.

**Lemma 15.29.** *Under any sequence*

$$E(C_{\max}) \geq \sum_{j=1}^{n} \frac{1}{\lambda_j} + (m-1) \max\left(\frac{1}{\lambda_1}, \ldots, \frac{1}{\lambda_n}\right)$$

*Proof.* The expected time it takes the job with the largest expected processing time to traverse the flow shop is at least $m \max(1/\lambda_1, \ldots, 1/\lambda_n)$. The time that this largest job starts on machine 1 is the sum of the processing times on the first machine of those jobs scheduled before the longest job. After the longest job completes its processing on the last machine, this machine remains busy for a time that is at least as large as the sum of the processing times on the last machine of all those jobs scheduled after the longest job. The lemma thus follows. □

One class of sequences plays an important role in stochastic permutation flow shops. A sequence $j_1, \ldots, j_n$ is called a *SEPT-LEPT* sequence, if there is a job $j_k$ in this sequence such that

$$\frac{1}{\lambda_{j_1}} \leq \frac{1}{\lambda_{j_2}} \leq \cdots \leq \frac{1}{\lambda_{j_k}}$$

and

$$\frac{1}{\lambda_{j_k}} \geq \frac{1}{\lambda_{j_{k+1}}} \geq \cdots \geq \frac{1}{\lambda_{j_n}}.$$

Both the *SEPT* and the *LEPT* sequence are examples of *SEPT-LEPT* sequences.

**Theorem 15.30.** *If $F_1 \leq_{a.s.} F_2 \leq_{a.s.} \cdots \leq_{a.s.} F_n$, then*
*(i) any SEPT-LEPT sequence minimizes the expected makespan in the class of non-preemptive static list policies and*

$$E(C_{\max}) = \sum_{j=1}^{n} \frac{1}{\lambda_j} + (m-1)\frac{1}{\lambda_n}.$$

*(ii) the SEPT sequence minimizes the expected flow time in the class of nonpreemptive static list policies and*

$$E(\sum_{j=1}^{n} C_j) = m \sum_{j=1}^{n} \frac{1}{\lambda_j} + \sum_{j=1}^{n-1} \frac{j}{\lambda_{n-j}}.$$

It is easy to find examples with $F_1 \leq_{a.s.} F_2 \leq_{a.s.} \cdots \leq_{a.s.} F_n$, where sequences which are not *SEPT-LEPT* are also optimal (it can be shown that when $F_1, F_2, \ldots, F_n$ are deterministic *any* sequence minimizes the makespan). However, in contrast with deterministic proportionate flow shops, when processing times are stochastic and $F_1 \leq_{a.s.} F_2 \leq_{a.s.} \cdots \leq_{a.s.} F_n$ not *all* sequences are always optimal.

**Example 15.31.** *Consider a flow shop with 2 machines and 3 jobs. Job 1 has a deterministic processing time of 11 time units. Job 2 has a deterministic processing time of 10 time units. The processing time of job 3 is zero with probability 0.5 and 10 with probability 0.5. It can be verified easily that* only *SEPT-LEPT sequences minimize the expected makespan. If the processing time of job 1 is changed from 11 to 20, then all sequences have the same expected makespan.*

Consider a two machine open shop where the processing time of job $j$ on machine 1 is the random variable $X_{1j}$, distributed according to $F_{1j}$, and on machine 2 the random variable $X_{2j}$, distributed according to $F_{2j}$. The objective is to minimize the expected makespan. As before, the exponential distribution is considered first. In this case, however, it is not known what the optimal policy is when $F_{1j}$ is exponential with rate $\lambda_j$ and $F_{2j}$ exponential with rate $\mu_j$. It appears that the optimal policy may not have a simple structure and may even depend on the values of the $\lambda$'s and $\mu$'s. The special case where $\lambda_j = \mu_j$ can be analyzed. In contrast with the results obtained for the stochastic flow shops the optimal policy now cannot be regarded as a permutation sequence, but rather as a policy which prescribes a given action dependent upon the state of the system.

**Theorem 15.32.** *The following policy minimizes the expected makespan in the class of preemptive dynamic policies as well as in the class of nonpreemptive dynamic policies: whenever a machine is freed, the scheduler selects from the jobs which have not yet undergone processing on either one of the two machines, the job with the largest expected processing time. If there are no such jobs remaining the decision-maker may take any job which only needs processing on the machine just freed. Preemptions never need to take place.*

It appears to be very hard to generalize this result to include a larger class of distributions.

**Example 15.33.** *Let the processing time of job $j$ on machine $i$, $i = 1, 2$, be a mixture of an exponential with rate $\lambda_j$ and zero with arbitrary mixing probabilities. The optimal policy is to process at time 0 all jobs for a very short period on both machines just to check whether their processing times on the two machines are zero or positive. After the nature of all the processing times have been determined, the problem is reduced to the scenario covered by* Theorem 15.32.

*Theorem 15.32* states that jobs which still have to undergo processing on both machines have priority over jobs which only have to be processed on one machine. In a sense, the policy described in *Theorem 15.32* is similar to the *Longest Alternate Processing Time first (LAPT)* rule for the deterministic $O2 \parallel C_{\max}$ problem.

From *Theorem 15.32* it follows that the problem is tractable also if the processing time of job $j$ on machine 1 as well as on machine 2 is exponentially distributed with rate 1. The policy that minimizes the expected makespan always gives priority to jobs that have not yet undergone processing on either machine. This particular rule does not require any preemptions. In the literature, this rule has been referred to in this scenario as the *Longest Expected Remaining Processing Time first (LERPT)* rule.

Actually, if in the two-machine case all processing times are exponential with mean 1 and if preemptions are allowed, then the sum of the expected completion times can also be analyzed. This model is an exponential counterpart of $O2 \mid p_{ij} = 1, pmtn \mid \sum C_j$. The total expected completion time clearly requires a different policy. One particular policy is appealing in the class of preemptive dynamic policies: consider the policy which prescribes the scheduler to process, whenever possible, on each one of the machines a job which already has been processed on the other machine. This policy may require the scheduler at times to interrupt the processing of a job and start with the processing of a job which just has completed its operation on the other machine. In what follows this policy is referred to as the *Shortest Expected Remaining Processing Time first (SERPT)* policy.

**Theorem 15.34.** *The preemptive SERPT policy minimizes the total expected completion time in a two machine open shop in the class of preemptive dynamic policies.*

*Proof.* Let $A_{ij}$, $i = 1, 2$, $j = 1, \ldots, n$, denote the time that $j$ jobs have completed their processing requirements on machine $i$. An idle period on machine 2 occurs if and only if

$$A_{1,n-1} \leq A_{2,n-1} \leq A_{1,n}$$

and an idle period on machine 1 occurs if and only if

$$A_{2,n-1} \leq A_{1,n-1} \leq A_{2,n}.$$

Let $j_1, j_2, \ldots, j_n$ denote the sequence in which the jobs leave the system, i.e., job $j_1$ is the first one to complete both operations, job $j_2$ the second, and so on. Under the *SERPT* policy

$$C_{j_k} = \max(A_{1,k}, A_{2,k}) = \max\left(\sum_{l=1}^{k} X_{1l}, \sum_{l=1}^{k} X_{2l}\right), \qquad k = 1, \ldots, n-1$$

This implies that the time epoch of the $k$ th job completion, $k = 1, \ldots, n-1$, is a random variable which is the maximum of two independent random variables, both with *Erlang(k)* distributions. The distribution of the last job completion, the makespan, is different. It is clear that under the preemptive *SERPT* policy the sum of the expected completion times of the first $n-1$ jobs that leave the system are minimized. It is not immediately obvious that *SERPT* minimizes the sum of all $n$ completion times. Let

$$B = \max(A_{1,n-1}, A_{2,n-1}).$$

The random variable $B$ is independent of the policy. At time $B$, each machine has at most one more job to complete. A distinction can now be made between two cases.

First, consider the case where, at $B$, a job remains to be completed on only one of the two machines. In this case, neither the probability of this event occurring nor the waiting cost incurred by the last job which leaves the system (at $\max(A_{1,n}, A_{2,n})$) depends on the policy. Since *SERPT* minimizes the expected sum of completion times of the first $n-1$ jobs to leave the system, it follows that *SERPT* minimizes the expected sum of the completion times of all $n$ jobs.

Second, consider the case where, at time $B$, a job remains to be processed on both machines. Either (i) there is one job left which needs processing on both machines or (ii) there are two jobs left, each needing processing on one machine (a different machine for each). Under (i) the expected sum of the completion times of the last two jobs to complete their processing is $E(B) + E(B+2)$, while under (ii) it is $E(B) + 1 + E(B) + 1$. In both subcases the expected sum of the completion times of the last two jobs is the same. As *SERPT* minimizes the expected sum of the completion times of the first $n-2$ jobs to leave the system, it follows that *SERPT* minimizes the expected sum of the completion times of all $n$ jobs. □

Unfortunately, no results have been reported in the literature with respect to stochastic open shops with more than 2 machines.

Consider now the two machine job shop with job $j$ having a processing time on machine 1 which is exponentially distributed with rate $\lambda_j$ and a processing time on machine 2 which is exponentially distributed with rate $\mu_j$. Some of the jobs have to be processed first on machine 1 and then on machine 2, while the remaining jobs have to be processed first on machine 2 and then on machine 1. Let $J_{1,2}$ denote the first set of jobs and let $J_{2,1}$ denote the second set of jobs. Minimizing the expected makespan turns out to be an easy extension of the two machine flow shop model with exponential processing times.

**Theorem 15.35.** *The following policy minimizes the expected makespan in the class of nonpreemptive dynamic policies as well as in the class of preemptive dynamic policies: when machine 1 is freed the decision-maker selects from $J_{1,2}$ the job with the highest $\lambda_j - \mu_j$ ; if all jobs from $J_{1,2}$ have received processing on machine 1 he may take any job from $J_{2,1}$. When machine 2 is freed the decision-maker selects from $J_{2,1}$ the job with the highest $\mu_j - \lambda_j$ ; if all jobs from $J_{2,1}$ have received processing on machine 2 he may take any job from $J_{1,2}$.*

The result described in *Theorem 15.35* is somewhat similar to the result obtained by Jackson for $J2 \parallel C_{\max}$. In deterministic scheduling the research on the more general $Jm \parallel C_{\max}$ has focused on heuristics and enumerative procedures. In stochastic scheduling less research has been done on job shops with more than two machines.

### 15.6.  Discussion

No framework or classification scheme has ever been introduced for stochastic scheduling problems. It is more difficult to develop such a scheme for stochastic scheduling problems than for deterministic scheduling problems. For example, it has to be specified which class of policies is considered, it has to be specified whether the processing times of the $n$ jobs are independent or correlated (e.g., equal to the same random variable), it may have to be specified that the processing times are of one type of distribution (e.g., exponential), while the due dates are of another (e.g., deterministic). For these reasons no framework has been introduced in this chapter either.

Table 15.1 outlines a number of scheduling problems of which stochastic versions are tractable. This list refers to most of the problems discussed in this chapter. In the distribution column the distribution of the processing times is specified. If the entry in this column specifies a form of stochastic dominance, then the $n$ processing times are arbitrarily distributed and ordered accordingly. The due dates in this table are considered fixed (deterministic).

Comparing Table 15.1 with the results described in earlier chapters of this book reveals that there are a number of stochastic scheduling problems that are tractable while their deterministic counterparts are NP-Hard. The four NP-Hard deterministic problems are:

$$
\begin{array}{ll}
\text{(i)} & 1 \mid r_j, pmtn \mid \sum w_j C_j, \\
\text{(ii)} & 1 \mid d_j = d \mid \sum w_j U_j, \\
\text{(iii)} & 1 \mid d_j = d \mid \sum w_j T_j, \\
\text{(iv)} & Pm \parallel C_{\max}.
\end{array}
$$

The first problem allows for a nice solution when the processing times are exponential and the release dates are arbitrarily distributed. The optimal policy is then the preemptive WSEPT rule. When the processing time distributions are anything but exponential it appears that the preemptive WSEPT rule is not necessarily optimal. The stochastic counterparts of the second and third problem also lead to the WSEPT rule when the processing time distributions are exponential and the jobs have a common due date which is arbitrarily distributed. Also here, if the processing times are anything but exponential the optimal rule is not necessarily WSEPT.

The stochastic counterparts of $Pm \parallel C_{\max}$ are slightly different. When the processing times are exponential the LEPT rule minimizes the expected makespan in all classes of policies. However, this holds for other distributions also. If the processing times are DCR (e.g., hyperexponentially distributed) and satisfy a fairly strong form of stochastic dominance, the LEPT rule is optimal as well. Note that if preemptions are allowed, and the processing times are DCR, the nonpreemptive LEPT rule remains optimal. Note also, that if the $n$ processing times have the same mean and are hyperexponentially distributed as in Example 15.21, then the LV rule minimizes the expected makespan.

Of course, there are also problems of which the deterministic version is easy and the version with exponential processing times is hard. Examples of such problems are:

(i)  $O2 \mid\mid C_{\max}$,
(ii)  $Pm \mid p_j = 1, tree \mid C_{\max}$.

For the $O2 \mid\mid C_{\max}$ problem the LAPT rule is optimal; when the processing times are exponential the problem appears to be very hard. For the deterministic problem $Pm \mid p_j = 1, tree \mid C_{\max}$ the CP rule is optimal. For the version of the same problem with all processing times i.i.d. exponential the optimal policy is not known and may depend on the form of the tree. One would expect that there are many scheduling problems of which the deterministic version with unit processing times is easy, and of which the stochastic version with all processing times i.i.d. exponential is hard.

**Table 15.1: Tractable Stochastic Scheduling Problems**

| DETERMINISTIC COUNTERPART | DISTRIBUTIONS | OPTIMAL POLICY |
|---|---|---|
| $1 \mid\mid \sum w_j C_j$ | arbitrary | WSEPT |
| $1 \mid r_j, pmtn \mid \sum w_j C_j$ | exponential | WSEPT (preemptive) |
| $Pm \mid\mid C_{\max}$ | exponential | LEPT |
| $Pm \mid pmtn \mid C_{\max}$ | exponential | LEPT |
| $Pm \mid\mid \sum C_j$ | $\geq_{st}$ | SEPT |
| $P2 \mid p_j = 1, intree \mid C_{\max}$ | exponential | CP |
| $P2 \mid p_j = 1, intree \mid \sum C_j$ | exponential | CP |
| $F2 \mid\mid C_{\max}$ | exponential | $(\lambda_j - \mu_j) \downarrow$ |
| $Fm \mid p_{ij} = p_j \mid C_{\max}$ | $\geq_{as}$ | SEPT-LEPT |
| $Fm \mid p_{ij} = p_j \mid \sum C_j$ | $\geq_{as}$ | SEPT |
| $O2 \mid p_{ij} = p_j \mid C_{\max}$ | exponential | Theorem 15.32 |
| $O2 \mid p_{ij} = 1, pmtn \mid \sum C_j$ | exponential | SERPT |
| $J2 \mid\mid C_{\max}$ | exponential | Theorem 15.35 |

# Contents

i

# 16

# Scheduling in Practice

Michael L. Pinedo
*New York University*

### 16.1. Introduction

In this chapter we focus on how the results presented in the earlier parts of this book are made useful in the real world. Before we go into the applications we do have to describe a number of the differences between the deterministic models considered in this book and the scheduling problems in the real world.

In practice, there are often not just $n$ jobs. At a certain point in time there may be indeed $n$ jobs to be scheduled, but additional jobs arrive at regular or random intervals.

There may be multiple objectives, that are subject to various weights. These weights often vary from day to day, making a parametric analysis necessary.

In the real world, it is often a *rescheduling* problem that has to be solved. That is, there is already a schedule, but there is a need to reschedule, because of a random perturbation (in the form of an arrival of a rush job, or the change in the priority of a job). This is one of the reasons why, in addition to the standard objectives described in this book, one more consideration is important, namely the "robustness" of the schedule. A schedule is called robust if the schedule, when subject to a random perturbation, does not need any major changes.

In the models considered in the earlier parts of this book, it is typically assumed that the machines are available continuously. However, in real life, machines are never available at all times. There may be many reasons why a machine may become unavailable. For example, the machine may be subject to (random) breakdowns or the machine may be subject to preventive maintenance. The availability of a machine may also depend on the shift schedules of the operators.

This chapter is organized as follows. The second section describes application of single machine models and the third section considers applications of parallel machine models. The fourth section focuses on multi-operation models. The fifth section gives an overview of the procedures that have proven to be popular in the scheduling engines and schedule generators developed and under development in industry.

## 16.2.  Application of Single Machine Models

When in a given environment one machine is the bottleneck, and it is so consistently, then a decomposition procedure may be appropriate. In such a decomposition procedure there is always a module that contains a procedure for solving a single machine scheduling problem. So the scheduling of the entire environment is done by tackling the bottleneck first. Since the bottleneck determines the overall throughput, it makes sense to schedule all other machines in such a way that the life of the bottleneck is made easy.

Such a situation occurs in many industrial settings.

**Example 16.1** (A Factory in the Packaging Industry). *Consider a factory that makes carton boxes for breakfast cereals. Each order is a request for a given number of cartons (e.g., 50,000) of a certain type. There is a committed shipping date at which time the order should be delivered. The raw materials to produce the carton are available at a certain date (the raw material includes the board, the ink needed in the printing process, the dies for the cutting, etc.). This data is typically provided by a Material Requirements Planning system. In this environment different jobs often may have different priorities, implying that each job has its own weight.*

*Suppose that in the facility there is a single printing machine for printing the paper board. This printing machine feeds into various cutters, that cut the boxes from the printed sheets. After the cutting process, the material moves to the gluers that produce the final box. Suppose that this single printing machine is the bottleneck. In order to do the scheduling of the entire factory, it makes sense to schedule the printing machine first and analyze it as a single machine model. There may be sequence dependent setup times on such a printing machine.*

*Before scheduling such a bottleneck, preliminary computations have to be done that yield "local" release dates and "local" due dates for the jobs at the bottleneck. The local release dates are estimated by retrieving from a Material Requirements Planning (MRP) system the date at which the job can start its route in the factory and estimating the time it will take the job to reach the bottleneck machine (in this example, where the bottleneck is at the first stage of the process, this transit time is typically negligible). The local due dates are estimated by taking into consideration the committed shipping dates and then estimating the amount of time it takes each job to traverse the facility after it has completed its processing on the bottleneck machine.*

   In the example above there may be sequence dependent setup times on the single machine that is being analyzed. These sequence dependent setup times may or may not be negligible. In the next example, however, sequence dependent setup times for sure play an important role.

**Example 16.2** (A Routing Problem). *Consider a warehouse with a single truck that has to bring merchandise to a number of clients. The delivery of merchandise to a client represents a job and the time that the truck spends at the client's site is the processing time. There may be release dates and due dates for the processing of that job, since the client may want the goods to be delivered within a given time period. However, the time it takes the truck to travel from one client to another represent a form of sequence dependent setup time that cannot be disregarded.*

   *There are a couple of possible objectives that could be minimized when scheduling (or, equivalently, routing) the truck. One objective is to minimize the total weighted tardiness, i.e., have the truck deliver the goods as much as possible within the intervals requested by the customers. Another objective is to minimize the total time it takes the truck to do all its deliveries. This implies that the the sum of the distances travelled (or, equivalently, the sum of the sequence dependent setups) has to be minimized. Such a problem is, of course, equivalent to the Travelling Salesman Problem.*

   From the examples above it appears that single machine models of two types in particular are of interest:

$$1 \mid r_j, setups \mid \sum w_j f_j(d_j, C_j),$$
$$1 \mid r_j, setups \mid \max(w_j f_j(d_j, C_j)).$$

There may be, in addition to a tardiness penalty, also an earliness penalty. A number of special cases of these problems have been considered in previous chapters, i.e., single machine problems with objectives $\sum w_j U_j$, $\sum w_j T_j$, and $L_{\max}$. These special cases are all unary NP-hard. The special case $\sum U_j$ is of some importance in practice as well, since one measure according to which plant managers often are judged is the percentage of on-time shipments (which is equivalent to $\sum U_j$).

   There are a number of ways in which these problems are dealt with in practice. First, one may establish a priority rule. The priority rules of interest are, of course, more complicated than the EDD or WSPT rules described in Chapter ...**??** However, they typically have somewhere an EDD or WSPT flavor in them.

   Second, one may attempt to apply local search procedures to these problems, i.e., either simulated annealing, tabu-search or genetic algorithms.

   Third, in applied problems one may at times apply rolling horizon procedures. Since jobs come in regularly over time, it makes sense to do some form of temporal decomposition.

   It is not often that one would apply branch and bound procedures to such problems in practice. The reason is the following: The inaccuracy in the data may often be in the order of 10 or 20%. It may not pay to try to do a perfect optimization and achieve a solution that is 1% better if the data are not sufficiently exact.  However, if one does apply branch and bound, then many of the dominance rules that have been established in the literature for these problems are useful.  For example, if $p_j < p_k$, $d_j < d_k$, and $w_j > w_k$, then it is known that, provided both jobs $j$ and $k$ are available, job $j$ must precede job $k$ (see Chapter 6).

More general single machine models have been considered in practice.  These more general models often include multiple objectives including earliness costs.

### 16.3.    Application of Parallel Machine Models

Parallel machine applications occur in practice when the bottleneck is a workcenter with a number of machines in parallel.  The remaining part of the factory is again disregarded in the same way as in the previous section. Such a setting is prevalent in many industries as well.

In industrial settings these parallel machines often have different speeds or are unrelated. Unrelated machines occur in practice often: certain jobs cannot be processed on just any one of the machines in parallel but rather only on machines that belong to a specific subset (making in essence the processing times on the other machines infinity).

The local release dates of all the jobs for the parallel machine workcenter have to be estimated and so have the due dates. The due date related objectives described in the previous section are applicable here as well. This implies that

$$Pm \mid r_j \mid \sum w_j f_j(d_j, C_j),$$
$$Qm \mid r_j \mid \sum w_j f_j(d_j, C_j),$$
$$Rm \mid r_j \mid \sum w_j f_j(d_j, C_j).$$

are problems of interest.

Consider the same factory as the one described in Example 16.1.  However, suppose now that the printing stage consists of $m$ identical machines in parallel and suppose that this stage is again the bottleneck.  It makes sense then to apply a decomposition procedure in which this stage is first considered as an isolated parallel machines model.

Consider the same warehouse as the one described in Example 16.2.  However, suppose that instead of a single truck, there is a fleet of $m$ trucks.  Again, the $n$ clients have to receive their goods, and each truck has to be assigned a number of clients. This implies that this problem is equivalent to a parallel machine scheduling problem.

The machines in a bank of parallel machines (or the trucks in a fleet of trucks) are very often not exactly identical. They may operate, for example, at different speeds. However, the processing time of job $j$ on machine $i$ may have a certain structure. For example, suppose job $j$ is associated with an order for a given quantity of items, say

$q_j$. The processing time of job $j$ on machine $i$ could then be of the form

$$p_{ij} = s_{ij} + q_j/v_{ij},$$

where $s_{ij}$ represents the setup time needed at machine $i$ for job $j$ (this setup time may depend on the machine as well as on the type of job and may or may not be sequence dependent). If the setup times on machine $i$ are sequence dependent, then the setup time has to be $s_{ikj}$, i.e., if job $k$ is followed by job $j$, then machine $i$ requires an amount of time $s_{ikj}$ for setup. The $v_{ij}$ is the speed at which machine $i$ can process job $j$. The speed $v_{ij}$ may be zero; if this is the case, then machine $i$ cannot process job $j$ (possibly for technological reasons).

**Example 16.3** (An Airport Terminal). *Consider an airport terminal with m gates. The gates are not all identical. Some are so close to others that they are not capable of serving wide-body planes. Some gates are so close to the terminal that planes have to be towed in, implying an additional setup time.*

*The arrival of a plane at the terminal is equivalent to the release date of a job. The deplaning of arriving passengers, the servicing of the plane and the boarding of the departing passengers constitute the job. This job has a processing time that is subject to a certain amount of randomness. The due date is the scheduled departure time of the plane and the completion time is the actual departure time of the job.*

*Jobs have to be assigned to machines in such a way that the total weighted tardiness is minimized. The weight $w_j$ of job $j$ depends on ......*

**Example 16.4** (A Hotel). *Consider a hotel with m rooms. There are n customers arriving over a given horizon and customer $j$ has an arrival date $r_j$ and a departure date $d_j$. The sojourn time of customer $j$ is $p_j = d_j - r_j$. If the hotel decides to give the customer a room for that period, then the hotel makes a profit $w_j$. The objective of the hotel is to maximize its profit. It is clear that the scheduling problem the hotel faces is equivalent to $Pm \mid r_j \mid \sum w_j U_j$.*

*This particular problem, where there is no freedom in the timing of the processing (i.e., $p_j = d_j - r_j$ ), is often referred to as a fixed interval scheduling problem, or simply an interval scheduling problem.*

Even in the case of a hotel it is typical that the "machines", i.e., the hotel rooms, are not all identical. There are singles, doubles and suites (with or without a view). So any given customer cannot just be assigned to any room; he may only be assigned to a room that belongs to a given subset. However, the "processing time" of the customer, i.e., his length of stay, does not depend on the room, i.e., in this case the machines do not have different speeds.

One objective that is also important in the parallel machine setting is the $C_{\max}$ objective. Minimizing the makespan in essence balances the workload over the ma-chines and balancing the workload enhances the capacity utilization. Workload balance is often an important objective in practice. (Of course, this objective did not play a role in the single machine models). As we have seen in Chapter 9 a very useful rule for assigning jobs to machine is based on the LPT heuristic. The next example shows an application of the LPT rule.

**Example 16.5** (A Manufacturing Facility for Printed Circuit Boards). *This example originated in a manufacturing facility of IBM in the U.S. The facility has a flexible flow line for the insertion of components onto printed circuit cards. A card is transported through the line on a "magazine", which contains 100 cards of the same type. From the scheduling point of view, such a magazine with a 100 cards is equivalent to one job. A job visits up to three banks of machines. The first bank consists of the so-called "DIP" inserters, the second bank of the so-called "SIP" inserters, and the third bank consists of a couple of robots. Each machine has a buffer that can hold one job, but additional storage is available, if necessary. Each job must be processed by at most one machine at each bank, but some jobs may skip some banks. Each job must visit the three banks in the same order. The assignment of the jobs to the specific machines at the various banks has to be done in advance, i.e., it is not possible to have the jobs arrive at a bank, join a single queue, and, when the job is at the head of the line, take the machine that becomes available first.*

*There are a number of objectives. First, one goal is to minimize the time to complete one day's production. This objective is basically equivalent to minimizing the makespan in a multi-processor flow shop. A second objective is to minimize the queueing in the buffers, since the buffers have a finite capacity.*

*The heuristic developed for this problem consists of three modules, that have to be executed one after another. Since one of the objectives is to minimize the queueing, it is advisable that at any given bank of machines the workload is balanced over the various machines. In order to balance the workload over the various machines at any given bank, the Longest Processing Time first (LPT) heuristic is used. Recall that the LPT heuristic basically establishes an assignment of jobs to machines. After a set of jobs has been assigned to a particular machine, the sequence in which the set of jobs are processed on the machine is immaterial, i.e., it does not affect the workload balance.*

The techniques applied to these parallel machine problems in practice are very similar to the techniques applied to the single machine scheduling problems. With machines in parallel there are often also multiple objectives. The multiple objectives include due date related objectives as well as workload balancing objectives.

## 16.4. Application of Multi-Operation Models

Multi-operation machine settings are prevalent in many industries. Actually, the examples in the single machine and the parallel machine settings were already extracted from setups that resemble multi-processor flow shops. Important applications of multi-operation models occur in the micro-electronics industries. The most important objectives in these industries are due date related, i.e., objectives typically more general than $L_{\max}$ (Uzsoy), $\sum w_j T_j$ and $\sum w_j U_j$.

**Example 16.6** (Manufacturing of Printed Circuit Boards). *Flow shops are prevalent in Printed Circuit Board (PCB) Manufacturing. The PCB manufacturing process consists of a number of steps that have to be applied sequentially to a panel that is typically made of epoxy. These steps may include:*

- (i) *drilling of holes,*
- (ii) *metallization*
  *(e.g., processing through a copper immersion bath),*
- (iii) *panel preparation*
  *(applying Dry Film Photo Resist (DFPR)),*
- (iv) *lamination and exposition to UV light,*
- (v) *developing to wash away the DFPR not exposed to UV light,*
- (vi) *electroplating of copper conductors,*
- (vii) *chemical stripping of DFPR that has been exposed to UV light,*
- (vii) *chemical etching of the copper layer,*
- (viii) *stripping of the tin,*
- (ix) *soldermask silkscreening,*
- (x) *solder dipping,*
- (xi) *punch pressing,*
- (xii) *inspection.*

*A job basically consists of an order for a batch of PCB's. The batch size could be anywhere from half a dozen to several thousands. The fact that job $j$ is basically associated with a given quantity of boards $q_j$ implies that the processing times of a job at the various stages are positively correlated. The processing time of job $j$ at machine $i$ could be of the form*

$$p_{ij} = s_{ij} + q_j/v_i,$$

*where $s_{ij}$ represents the setup time needed at machine $i$ for job $j$ (this setup time may depend on the machine as well as on the type of board); the $v_i$ is the speed of machine $i$ and gives an indication of how fast the machine can produce a board.*

*The main objective is typically to meet as many of the committed shipping dates as possible.*

The setting in PCB manufacturing is often a flow shop. However, if at any one of the stages there are two or more machines in parallel instead of a single machine, then the setting is referred to as a multi-processor flow shop (or flexible flow shop, or hybrid flow shop).

Multi-processor flow shops are common in many other industries as well.

**Example 16.7** (A Paper Mill). *Paper mills typically give rise to multi-processor flow shop models. A paper mill often has more than one paper machine (it may have 2, 3, or even more) which produce the rolls of papers. These rolls of papers may have to go to an adjacent converting facility that produces cut size paper (8.5 × 11). This converting facility has a number of cutters in parallel. The jobs that have to be processed have given committed shipping dates or due dates. The objective is to minimize some due date related penalty function.*

Actually, the problem in the real world is slightly more complicated because of the fact that the parallel machines at any one of the two stages are not exactly identical. This implies that certain jobs can only be processed on a subset of the machines at anyone of the stages.

The paper mill example is a typical example of a multi-processor flow shop without any recirculation. The next example is an example of a job shop with recirculation.

**Example 16.8** (Integrated Circuit Manufacturing). *Job shops are common in integrated circuit manufacturing (wafer fabs). The fabrication process in wafer fabs consists also of a number of steps.*

     *(i)   cleaning*
    *(ii)   oxidation, deposition, and metallization*
  *(iii)   lithography*
   *(iv)   etching*
    *(v)   ion implementation*
   *(vi)   photoresist stripping*
 *(vii)   inspection and measurement.*

*However, since wafers often have various layers of of circuits one on top of another, a wafer may have to go through these sequence of operations a number of times. One could refer to this environment as a flow shop with recirculation, or, in effect, a job shop.*

*The main objective here often depends on the type of wafer fab. If the facility is geared towards the mass production of DRAMs, then maximizing throughput, and therefore minimizing setup times, is important. If the facility is designed to produce more specialized products, then meeting customer due dates is more important.*

The next example is another example of a job shop with recirculation.

**Example 16.9** (Manufacturing of Nuclear Control Rods). *A manufacturing facility that produces nuclear control rods produces rods of a special alloy about 1.5 inch in diameter and 20 feet long for controlling nuclear reactions. The manufacturing procedure goes roughly as follows. Each incoming rod is extruded four times. The prototype rod is drawn through dies four times to slowly reduce its diameter. After being extruded four times the rod is cut into pieces of the appropriate length. Each time the diameter is reduced the process is said to have made a pass. Such a pass consists of pilgering, cleaning, and then tempering in an annealing surface to reduce the stresses induced by extrusion. The pilgering and the cleaning in the different passes is done at different workstations. However, there is only one annealing furnace. So at the end of each one of the first three passes a rod has to go through the same annealing furnace. The entire production process is depicted in Figure 2??. Because of the fact that that each job has to go three times through the same annealing furnace, this is an example of a job shop with recirculation. The main objective in this case is the minimization of the sum of the weighted tardinesses. So the problem can be modelled as $Jm \mid r_j \mid \sum w_j T_j$.*

*A number of different techniques have been developed for this problem, branch and bound techniques as well as shifting bottleneck techniques. When these techniques are used it is usually assumed that the weight of job $j$, $w_j$, does not change while the job traverses the shop.*

*However, another scheduling technique that is often used in practice in such a setting requires the use of a dispatching or priority rule at each workstation. Such rules may be used at the different stations independently from one another. There is actually an interesting phenomenon that often takes place in practice when priority rules are used. The weight or priority of a job, on its route through the shop, goes up. This implies that the weight of a job that is waiting at the furnace on its third pass is higher then the weight of a job that is waiting at the furnace on its first pass. The rationale for this higher priority is based on the fact that at the end of the third pass more value has been added to the job.*

## 16.5. General Principles of Scheduling System Design

Analyzing a scheduling problem and developing a procedure for dealing with the problem on a regular basis is, in the real world, only part of the story. The procedure has to be embedded in a system that enables the scheduler to actually apply it. The scheduling system has to be integrated into the information system of the organization, which can be a formidable task.

The information system within an organization or company typically is a very large Enterprise Resource Planning (ERP) system that serves as a backbone for the entire corporation. Every division of the organization has access to the system and many decision support systems are plugged into it, e.g., forecasting systems, inventory control systems, MRP systems, and planning and scheduling systems. The flow chart of such an ERP system is depicted in Figure 16...

A scheduling system consists of a number of different modules. Those of fundamental importance are:

  (i)    the schedule generation modules,
  (ii)   the user interface modules, and
  (iii)  the database, object base, and knowledge-base modules.

The interactions between these different modules are depicted in Figure 16.4.

## 16.6.  Schedule Generation Techniques in Scheduling Systems

A schedule generation module contains a suitable model with objective functions, constraints and rules, as well as heuristics and algorithms.

Current schedule generation techniques are an amalgamation of several approaches that have been converging in recent years. One approach, predominantly followed by industrial engineers and operations researchers, may be referred to as the *algorithmic* approach. Another, that is often followed by computer scientists and artificial intelligence experts, may be referred to as the *knowledge-based* approach. Recently, these two approaches have started to converge and the differences have become blurred. Some recent hybrid systems combine a knowledge base with fairly sophisticated heuristics. Certain segments of the procedure are designed according to the algorithmic approach, while other segments are designed according to the knowledge-based approach.

**Example 16.10** (Architecture of a Scheduling System in a Wafer Fab)**.** *A hybrid scheduling system has been designed for a particular semiconductor wafer fabrication unit. The system consists of two levels. The higher level operates according to a knowledge-based approach. The lower level is based on an algorithmic approach; it consists of a library of algorithms.*

*The higher level performs the first phase of the scheduling process. At this level, the current status of the environment is analyzed. This analysis takes into consideration due date tightness, bottlenecks, and so on. The rules embedded in this higher level determine the type of algorithm to be used at the lower level in each situation.*

The algorithmic approach usually requires a mathematical formulation of the problem that includes objectives and constraints.

A number of generic algorithmic procedures have proven to be very useful in practice. In this section we give a very short overview of the generic procedures that are popular.

Many industrial scheduling systems use priority or dispatching rules. The most popular priority rules are WSPT, EDD, CP, etc. These priority rules are based on a simple sort and operate therefore in $O(n \log n)$. They are therefore very useful for providing a first crack at the problem.

Many procedures in practice are based on some form of decomposition. Instances in practice are typically very large, involving hundreds of jobs and tens of machines. There are several forms of decomposition. One form of decomposition is based on machine decomposition. The machines are scheduled one at a time. Every time a machine is scheduled, the machine is scheduled throughout. One famous form of machine decomposition is often referred to as the shifting bottleneck procedure. Another form of decomposition is based on a partitioning of the time axis. These forms of decomposition are often called rolling horizon procedures.

Other techniques that have proven to be very popular in practice are the local search techniques, e.g., simulated annealing, tabu-search, and genetic algorithms. These techniques have proven to be very versatile, in particular for nonpreemptive scheduling problems, and easy to code.

There are cases of scheduling in practice that can be formulated very precisely and are not subject to uncertainty and randomness. Such cases often can be formulated as integer or disjunctive programs. Such formulations may lead then to enumeration techniques such as branch and bound. If the solution of such a problem is not time constrained, then a complete enumeration may be possible. If there are constraints with regard to the computation time, then beam search may be more appropriate.

Algorithmic schedule generation may consist of multiple phases (see Figure 4 (Young-Hoon Lee)). In the first phase, a certain amount of *preprocessing* is done, where the problem instance is analyzed and a number of statistics are compiled, e.g., the average processing time, the maximum processing time, the due date tightness. A second phase may consist of the actual algorithms and heuristics, whose structure may depend on the statistics compiled in the first phase. A third phase may contain a *postprocessor*. The solution that comes out of the second phase is fed into a procedure, such as simulated annealing or tabu-search, to see if improvements can be obtained. This type of schedule generation is usually coded in a procedural language such as Fortran, Pascal or C.

The knowledge-based approach is different from the algorithmic approach in various respects. This approach is often more concerned with underlying problem structures that cannot easily be described in an analytical format. In order to incorporate the scheduler's knowledge into the system, rules or objects are used. This approach is often used when it is only necessary to find a *feasible* solution given the many constraints or rules; however, as some schedules are ranked "more preferable" than others, heuristics may be used to obtain a "more preferred" schedule. Through a so-called *inference engine,* the approach attempts to find schedules that do not violate prescribed rules and satisfy stated preferences as much as possible. The inferencing techniques are usually so-called *forward chaining* and *backward chaining* algorithms. A forward chaining algorithm is knowledge driven. It first analyzes the data and the rules and, through inferencing techniques, attempts to construct a feasi-ble schedule.  A backward chaining algorithm is result oriented.  It starts out with a promising schedule and attempts to verify whether it is feasible.  Whenever a satisfactory solution does not appear to exist or when the scheduler judges that it is too difficult to find, the scheduler may reformulate the problem through a relaxation of the constraints.  The relaxation of constraints may either be done automatically (by the system itself) or by the user.  Because of this aspect, the approach has been at times also referred to as the *reformulative* approach.

The programming style used for the development of knowledge-based systems is different from the ones used for systems based on algorithmic approaches.  The programming style may depend on the form of the knowledge representation.  If the knowledge is represented in the form of *IF-THEN* rules, then the system can be coded using an expert system shell such as *OPS5*.  The expert system shell contains an inference engine that is capable of doing forward chaining or backward chaining of the rules in order to obtain a feasible solution. This approach may have difficulties with conflict resolution and uncertainty. If the knowledge is represented in the form of logic rules (see Example **??**), then an ideal programming language is Prolog. If the knowledge is represented in the form of frames, then a language with object oriented extensions is required, e.g., LISP or C++.  These languages emphasize user-defined objects that facilitate a modular programming style.

Algorithmic approaches as well as knowledge-based approaches have their advantages and disadvantages. An algorithmic approach has an edge if

(i)    the problem allows for a crisp and precise mathematical formulation,
(ii)   the number of jobs involved is large,
(iii)  the amount of randomness in the environment is minimal,
(iv)  some form of optimization has to be done frequently and in real time and
(v)   the general rules are consistently being followed without too many exceptions.

A disadvantage of the algorithmic approach is that if the scheduling environment changes, (for example, certain preferences on assignments of jobs to machines) the reprogramming effort may be substantial.

The knowledge-based approach may have an edge only if feasible schedules are required.  Some system developers believe that changes in the scheduling environment or rules can be more easily incorporated in a knowledge-based system than in a system based on the algorithmic approach. Others, however, believe that the effort required to modify any system is mainly a function of how well the code is organized and written; the effort required to modify does not depend that much on the approach used.

A major disadvantage of the knowledge-based approach is that obtaining a reasonable schedule may take substantially more computer time than an algorithmic approach. In practice certain scheduling systems have to operate in near-real time (it is very common that schedules must be generated within several minutes).

The amount of available computer time is an important issue with the selection of a schedule generation technique. The time allowed for schedule generation varies from application to application. Many applications require real time performance, where the schedule has to be generated in seconds or minutes on the computer at hand. This may be the case if rescheduling is required throughout the day, due to substantial schedule deviations. It would also be true if the scheduler runs iteratively, requiring human interaction between iterations (perhaps for adjustments of workcenter capacities). However, some applications do allow for overnight number crunching. For example, the scheduler at the end of the afternoon executes the program and wants an answer by the time he or she arrives at work the next day. A few applications require extensive number crunching. When, in the airline industry, quarterly flight schedules have to be determined, the investments at stake are such that a week of number crunching on a mainframe is fully justified.

As said before, the two approaches have been converging and most recent scheduling systems have elements of both. One language of choice is now C++ as it is an easy language for coding algorithmic procedures and it also has object-oriented extensions.

### 16.7.    User Interfaces in Scheduling Systems

User interface modules are important, especially with regard to the implementation process. Without an excellent user interface there is a good chance that, regardless of its capabilities, the system will be too unwieldy to use.

The user interfaces are very important parts of the system. These interfaces may determine whether the system is going to be used or not. Most user interfaces, whether the system is based on a workstation or PC, make extensive use of window mechanisms. The user often wants to see several different sets of information at the same time. This is the case not only for the static data that is stored in the database, but also for the dynamic data that depend on the schedule.

Some user interfaces allow for extensive user interaction. The scheduler can change the current situation or the current information. Other user interfaces do not allow the scheduler to change anything. For example, an interface that displays the values of all the relevant performance measures would not allow the scheduler to change any of the numbers. The scheduler may be able to change the schedule in another interface that then automatically changes the values of the performance measures, but the scheduler cannot change the performance measures directly.

User interfaces for database modules often take a fairly conventional form and may be determined by the particular database package used. These interfaces must allow for user interaction, as data such as due dates often have to be changed during a scheduling session.

The schedule generation module may provide the user with a number of computational procedures and algorithms. Such a library of procedures within the schedule generation module will require its own user interface, enabling the scheduler to select the appropriate algorithm or even design an entirely new procedure.

User interfaces that display schedule information take many different forms. In-terfaces for schedule *manipulation* determine the basic character of the system, as these are the ones used most extensively by the scheduler. The different forms of schedule manipulation interfaces depend on the level of detail as well as on the planning horizon being considered. Two such interfaces are described in detail, namely:

  (i)   the Gantt Chart interface,
  (ii)  the Dispatch List interface.

The first, and probably most popular, form of schedule manipulation interface is the Gantt chart (see Figure 5). The Gantt chart is the usual horizontal bar chart, with the x-axis representing the time and the y-axis, the various machines. A color and/or pattern code may be used to indicate a characteristic or an attribute of the corresponding job. For example, jobs that are completed after their due date under the current schedule may be colored red. The Gantt chart usually has a number of scroll capabilities that allow the user to go back and forth in time or focus on particular machines, and is usually mouse driven. If the user is not entirely satisfied with the generated schedule, he may wish to perform a number of manipulations on his own. With the mouse, the user can "click and drag" a job from one position to another. Providing the interface with a click and drag capability is not a trivial task for the following reason. After changing the position of a particular operation on a machine, other operations on that machine, which belong to other jobs, may have to be pushed either forward or backward in time to maintain feasibility. The fact that other operations have to be processed at different times may also have an effect on other machines. This is often referred to as *cascading* or *propagation* effects. After the scheduler repositions an operation of a job, the system may call a reoptimization procedure embedded in the scheduling routine to control the cascading effects in a proper manner.

**Example 16.11** (Cascading Effects and Reoptimization). *Consider a three machine flow shop with unlimited storage space between the successive machines and therefore no blocking. The objective is to minimize the total weighted tardiness. Consider a schedule with 4 jobs as depicted by the Gantt chart in Figure 6.a. If the user swaps jobs 2 and 3 on machine 1, while keeping the order on the two subsequent machines the same, the resulting schedule, because of cascading effects, takes the form depicted in Figure 6.b. If the system has reoptimization algorithms at its disposal, the user may decide to reoptimize the operations on machines 2 and 3, while keeping the sequence on machine 1 the way he constructed it. A reoptimization algorithm then may generate the schedule depicted in Figure 6.c. To obtain appropriate job sequences for machines 2 and 3, the reoptimization algorithm has to solve an instance of the two machine flow shop with the jobs subject to given release dates at the first machine.*

Gantt charts do have disadvantages, especially when there are many jobs and machines. It may be hard to recognize which bar or rectangle corresponds to which job. As space on the screen (or on the printout) is rather limited, it is hard to attach text to each bar. Gantt chart interfaces usually provide the capability to click the mouse on a given bar and open a window that displays detailed data regarding the corresponding job. Some Gantt charts also have a filter capability, where the user may specify the job(s) that should be exposed on the Gantt chart while disregarding all others. (The Gantt chart interface depicted in Figure 5 is from the LEKIN system developed at New Jersey Institute of Technology and New York University).

The second form of user interface displaying schedule information is the *dispatch-list* interface. (see Figure 7). Schedulers often want to see a list of the jobs to be processed on each machine in the order in which they are to be processed. With this type of display schedulers also want to have editing capabilities so they can change the sequence in which jobs are processed on a machine or move a job from one machine to another. This sort of interface does not have the disadvantage of the Gantt chart, since the jobs are listed with their job numbers and the scheduler knows exactly where each job is in a sequence. If the scheduler would like more attributes (e.g., processing time, due date, completion time under the current schedule, and so on) of the jobs to be listed, then more columns can be added next to the job number column, each one with a particular attribute. The disadvantage of the dispatch-list interface is that the scheduler does not have a good view of the schedule relative to time. The user may not see immediately which jobs are going to be late, which machine is idle most of the time, etc. (The dispatch-list interface in Figure 7 is also from the LEKIN system.)

Clearly, the different user interfaces for the display of schedule information have to be strongly linked with one another. When the scheduler makes changes in either the Gantt chart interface or in the dispatch-list interface, the dynamic data may change considerably due to the cascading effects or reoptimization routines. Changes made in one interface, of course, have to be shown immediately in the other interfaces as well.

User interfaces for the display of schedule information have to be linked with other interfaces also, e.g., database management interfaces and schedule generation interfaces. For example, the scheduler may modify an existing schedule in the Gantt chart interface by clicking and dragging; then he may want to freeze certain jobs in their respective positions. After doing this, he may want to reoptimize the remaining jobs, which are not frozen, using an algorithm in the schedule generation module. These algorithms are similar to the algorithms described in Part 2 for situations where machines are not available during given time periods (because of breakdowns or other reasons). The schedule manipulation interfaces have to be, therefore, strongly linked with the interfaces for algorithm selection.

Schedule manipulation interfaces may also have a separate window that displays the values of all relevant performance measures. If the user has made a change in the schedule the values before and after the change may be displayed. Typically, performance measures are displayed in plain text format. However, more sophisticated graphical displays may also be used.

Some schedule manipulation interfaces are sophisticated enough to allow the user to split a job into a number of smaller segments and schedule each of these separately. Splitting an operation is equivalent to (possibly multiple) preemptions. The more sophisticated schedule manipulation interfaces also allow different operations of the same job to overlap in time. In practice, this may occur in many settings. For example, a job may start at a downstream machine of a flow shop before it has completed its processing at an upstream machine. This occurs when a job is a large batch of identical items. Before the entire batch has been completed at an upstream machine, parts of the batch may already have been transported to the next machine and may have already started their processing there.

## 16.8. Database Issues in Scheduling Systems

The database modules play a crucial role in the functionality of the system. Significant effort is required to make a factory's database suitable for input to the scheduling system.

The database management subsystem may be either custom-made or a commercial system. A number of the commercial database systems available on the market have proven to be useful for scheduling systems. These are usually relational databases incorporating *Stuctured Query Language (SQL)*. Examples of such database management systems are Oracle, Sybase and Ingres.

Whether a database management subsystem is custom made or commercial, it needs a number of basic functions, which include multiple editing, sorting and searching routines. Before generating a schedule, the scheduler may want to see certain segments of the order masterfile and collect some statistics with regard to the orders and the related jobs. Actually, at times, he may not want to feed all jobs into the scheduling routines, but rather a subset.

Within the database a distinction can be made between *static* and *dynamic* data. Static data are all job and machine data that do *not* depend on the schedule. These include the job data that are specified in the customer's order form, such as the ordered product quantity (which is proportional to the processing times of all the operations associated with the job), the committed shipping date (the due date), the time at which all necessary material is available (the release date) and possibly some processing (precedence) constraints. The priorities (weights) of the jobs are also static data as they are not schedule dependent. Having different weights for different jobs is usually a necessity, but determining their values is not all that easy. In practice, it is seldom necessary to have more than three priority classes; the weights are then, for example, 1, 2 and 4. The three priority classes are sometimes described as "hot", "very hot" and "hottest" dependent upon the level of manager pushing the job. These weights actually have to be entered manually by the scheduler into the information system database. To determine the priority level, the person who enters the weight may use his own judgement, or may use a formula that takes certain data from the information system into account (for instance, total annual sales to the customer or some other measure of customer criticality). The weight of a job may also change from one day to another; a job that is not urgent today, may be urgent tomorrow. The scheduler may have to go into into the file and change the weight of the job before generating a new schedule. Static machine data include machine speeds, scheduled maintenance times, and so on. There may also be static data that are both job and machine dependent, e.g., the setup time between jobs $j$ and $k$ assuming the setup takes place on machine $i$.

The dynamic data consists of all the data that are schedule dependent: the starting and completion times of the jobs, the idle times of the machines, the times that a machine is undergoing setups, the sequences in which the jobs are processed on the machines, the number of jobs that are late, the tardinesses of the late jobs, and so on.

The calendar function is often also part of the database system. It contains information with regard to factory holidays, scheduled machine maintenance, number of shifts available, and so on. Calendar data are sometimes static, e.g., fixed holidays, and sometimes dynamic, e.g., preventive maintenance shutdowns.

Some of the more modern scheduling systems may rely on an object base in addition to (or instead of) a database. One of the main functions of the object base is to store the definitions of all object types, i.e., it functions as an object library and instantiates the objects when needed. In a conventional relational database, a data type can be defined as a schema of data. For example, a data type "job" can be defined as in Figure 9.a and an instance can be as in Figure 9.b. Object types and corresponding instances can be defined in the same way. For example, an object type "job" can be defined and corresponding job instances can be created. All the job instances have the same type of attributes.

There are two crucial relationships between object types, namely, the "is-a" relationship and the "has-a" relationship. An is-a relationship indicates a generalization and the two object types have similar characteristics. Often the two object types are referred to as a subtype and a supertype. For example, a "machine" object type is a special case of a "resource" object type and a "tool" object type is another special case of a resource object type. A "has-a" relationship is an aggregation relationship; one object type contains a number of other object types. A "workcenter" object is composed of several machine objects. A "plant" object comprises a number of workcenter objects. A "routing table" object consists of job objects as well as of machine objects.

Object types related by is-a or has-a relationships have similar characteristics with regard to their attributes. In other words, all the attributes of a supertype object are used by the corresponding subtypes. For example, a machine object has all the attributes of a resource object and it may also have some additional attributes. This is often referred to as inheritance. A hierarchical structure that comprises all object types can be constructed. Objects can be retrieved with commands that are similar to SQL commands in relational databases.

While virtually every scheduling system relies on a database or an object base, few scheduling systems have a module that serves specifically as a knowledge-base.

However, knowledge-bases may become more important in the future.

The design of a knowledge-base, in contrast with the design of a database or object base, has a significant impact on the overall architecture of the system, in particular on the schedule generator. The most important aspect of a knowledge-base is the knowledge *representation.* One form of knowledge representation is through *rules.* There are several formats for stating rules. A common format is through an *IF-THEN* statement. That is, *IF* a given condition holds, *THEN* a specific action has to be taken.

Another format for stating rules is through *predicate logic* that is based on propositional calculus. An appropriate programming language for dealing with rules in this format is Prolog.

A second form of knowledge representation is through so-called *frames* or *schemata.* A frame or schema provides a structured representation of an object or a class of objects. A frame is a collection of slots and values. Each slot may have a value class as well as a default value. Information can be shared among multiple frames by inheritance. Frames lower in the hierarchy are applicable to more specific operations and resources than the more generic frames at higher levels in the hierarchy.

Just like activities, constraints and rules can be represented by schemata as well.

## 16.9.  Scheduling Systems in Practice

The last three decades has seen the design and implementation of many scheduling systems. Some of these systems were application-specific and others were generic. Some were developed for research and experimentation and others were commercial.

A number of scheduling systems have been designed and developed in academia over the last three decades. Several universities developed research systems or educational systems that were often based on ideas and algorithms that were quite novel. An example of such a system is the Lekin system which can be downloaded free of charge from the web. Some of the academic systems have been handed over to industry and have led to the start-up of software companies.

The last two decades have witnessed the development of scores of commercial scheduling systems. There were a few major trends in the design and development of these commercial scheduling systems.

One trend started in the 1980s when a number of companies began to develop sequencing and scheduling software. Most of these companies tended to focus in the beginning only on sequencing and scheduling. They started out with the development of generic scheduling software that was designed to optimize flow lines or other types of machine environments. Some of these companies have grown significantly since their inception, e.g., ILOG, I2, Manugistics.

These companies, whenever they landed a contract, had to customize their software to the specific applications. Since they realized that customization of their software customization is a way of life, they usually tried to keep their schedule gen-erators as generic as possible. The optimization methodologies they adopted often included:

>   (i)    shifting bottleneck procedures,
>   (ii)   local search procedures,
>   (iii)  mathematical programming procedures.

These companies, which at the outset were focussing primarily on sequencing and scheduling, began to branch out in in the 1990s; they started to develop software for Supply Chain Management. This diversification became necessary since clients typically had a preference for dealing with vendors that could provide a suite of software modules capable of optimizing the entire supply chain; clients did not like to have to deal with different vendors and face all kinds of integration problems.

A second major trend in the development of sequencing and scheduling software had its source in a another corner of the software industry. This second trend started to take place in the beginning of the 1990s. Scheduling software started being developed by companies which at the outset specialized in ERP systems, e.g., SAP, Baan, J.D. Edwards, and PeopleSoft. These ERP systems basically are huge accounting systems that serve as a backbone for all the information requirements in a company. This backbone could then be used to feed information into all kinds of decision support systems, such as forecasting systems, inventory control, and sequencing and scheduling. The software vendors that specialized in ERP systems realized that it was necessary to branch out and develop decision support systems as well. A number of these companies either bought a scheduling software company (e.g., Baan bought Berclain), or started their in house scheduling software development (e.g., SAP), or established partnerships with scheduling software vendors.

Currently there are more than a hundred scheduling software vendors. Most of these are relatively small. The bigger players are I2, Cybertec, and Manugistics, all of them offering software for the entire supply chain. The main ERP vendors, e.g., SAP, Baan, PeopleSoft, J.D. Edwards, all offer sequencing and scheduling packages. Some of their scheduling modules had been developed internally, whereas other modules were developed through acquisitions of smaller software companies specializing in scheduling.

## 16.10. Discussion

From this chapter it should be clear that in practice it is extremely common that the procedures implemented are of a hybrid nature. Only the most application-specific system can rely on a very specific solution procedure. However, even those systems may need at times various solution techniques, because often instances of different sizes of the same problem have to be solved on different days requiring different approaches.

Some of the most popular techniques used in industrial systems are local search techniques. These techniques often have the important advantage that they are relatively easy to program and to implement. However, they have one disadvantage: usually they are not that easy to implement in settings where preemptions are allowed.

Industrial systems typically have one component that is very important in practice and that hardly plays any role in the theoretical models studied in the literature. That component is the calendar. The calendar is often a very large component in many scheduling systems. It plays a role in preventive maintenance scheduling, workforce (or shift) scheduling, and so on. The shift scheduling module may at times have to interact with the job (or machine) scheduling module.

There are dozens of software companies that have developed scheduling systems for the various industries. The costs of these systems range anywhere from $10,000 to $500,000. Often they require an amount of customization that costs even more.

One can make a distinction between generic scheduling systems and application-specific systems. The general architecture may be very different.

Lately academic researchers have been toying with the ideas of developing web enabled (Internet) scheduling systems. Designing scheduling systems have a number of important advantages. ...

# Elements of Scheduling